\documentclass[twoside,12pt,a4paper]{book}
\usepackage[headings]{fullpage}
\usepackage{verbatim}
\usepackage[utf8]{inputenc}
\usepackage[T1]{fontenc}
\usepackage{lmodern}
\usepackage[english]{babel} 
\usepackage{amssymb}
\usepackage{pifont} 
\usepackage{longtable} 
\usepackage{tabularx} 
\usepackage{graphicx} 
\usepackage{amsmath}
\usepackage{amsthm}
\usepackage{braket}

\usepackage{amssymb}
\usepackage{bbold}

\counterwithout{footnote}{chapter}
\numberwithin{equation}{section}

\usepackage[top=2.5cm, bottom=2.5cm, inner=3.cm, outer=2.cm]{geometry}
\usepackage{relsize}
\usepackage{xcolor}
\usepackage{soul}
\usepackage{endnotes}
\usepackage[mode=text,per-mode=symbol,exponent-product=\cdot ,input-uncertainty-signs=\pm, output-open-uncertainty=\pm]{siunitx}
\usepackage{csquotes}
\usepackage[colorlinks=true]{hyperref}
\definecolor{dark-gray}{gray}{0.13}
\hypersetup{urlcolor=blue,linkcolor=blue,citecolor=blue}
\usepackage{todonotes}
\usepackage{bookmark}
\usepackage{doi}
\usepackage{titlesec, blindtext}

\def\cJ{\mathcal J}
   
\usepackage[numbers, square, comma, sort&compress]{natbib}
\usepackage[nottoc]{tocbibind}
\usepackage{mathtools}

\usepackage{dsfont}
\usepackage{pst-all}
\usepackage{upgreek}
\usepackage{mathrsfs}
\usepackage{bigdelim}
\usepackage{multirow}
\usepackage{floatflt}
\usepackage{enumitem}
\usepackage{fancyhdr}
\usepackage{wrapfig}

\usepackage[Rejne]{fncychap}

\usepackage{graphicx}
\newlength{\neededheight}

\usepackage{IEEEtrantools}
\usepackage{mathrsfs}
\usepackage{slashed}

\usepackage{calc}
\usepackage{extarrows}

\usepackage{makeidx}
\makeindex

\usepackage[titletoc]{appendix}

\usepackage{afterpage}

\makeatletter
\newcommand{\DESCRIPTION@original@item}{}
\let\DESCRIPTION@original@item\item
\newcommand*{\DESCRIPTION@envir}{DESCRIPTION}
\newlength{\DESCRIPTION@totalleftmargin}
\newlength{\DESCRIPTION@linewidth}
\newcommand{\DESCRIPTION@makelabel}[1]{\llap{#1}}%
\newcommand{\DESCRIPTION@item}[1][]{%
  \setlength{\@totalleftmargin}%
       {\DESCRIPTION@totalleftmargin+\widthof{\textbf{#1 }}-\leftmargin}%
  \setlength{\linewidth}
       {\DESCRIPTION@linewidth-\widthof{\textbf{#1 }}+\leftmargin}%
  \par\parshape \@ne \@totalleftmargin \linewidth
  \DESCRIPTION@original@item[\textbf{#1}]%
}

\makeatother

\usepackage{cancel}

\usepackage{mathdots}
\usepackage{bigstrut}
\usepackage{dsfont}
\usepackage{ulem}
\usepackage{hhline}
\usepackage[percent]{overpic}

\usepackage{notoccite}

\usepackage{tikz} 
\usetikzlibrary{shapes,arrows,positioning,automata,backgrounds,calc,er,patterns}
\usepackage[compat=1.0.0]{tikz-feynman}

\usepackage{tikz-feynhand}

\newcommand{\beq}{\begin{equation}}
\newcommand{\eeq}{\end{equation}}
\newcommand{\bea}{\begin{eqnarray}}
\newcommand{\eea}{\end{eqnarray}}

\newcommand{\pslash}[1]{\rlap{/}\kern-0.8pt #1}
\newcommand{\Dslash}{\rlap{/}\kern-1.5pt D}

\def\today{\number\day\space\ifcase\month\or
January\or February\or March\or April\or May\or June\or
July\or August\or September\or October\or November\or December\fi
\space\number\year}
\newcount\mins \newcount\hours
\def\now{\hours=\time \mins=\time
	\divide\hours by60 \multiply\hours by60 \advance\mins by-\hours
	\divide\hours by60 
	\number\hours:\ifnum\mins<10 0\fi\number\mins~}

\def\stacksymbols #1#2#3#4{\def\theguybelow{#2}
    \def\verticalposition{\lower#3pt}
    \def\spacingwithinsymbol{\baselineskip0pt\lineskip#4pt}
    \mathrel{\mathpalette\intermediary#1}}
\def\intermediary #1#2{\verticalposition\vbox{\spacingwithinsymbol
    \everycr={}\tabskip0pt
    \halign{$\mathsurround0pt#1\hfil##\hfil$\crcr#2\crcr
             \theguybelow\crcr}}}

\makeatother

\renewcommand\L{\mathcal{L}}
\newcommand\D{\mathcal{D}}

\usepackage{color}
\usepackage{braket}

\definecolor{firebrick}{RGB}{178,34,34}
\definecolor{darkorange}{RGB}{255,140,0}
\definecolor{darkgreen}{RGB}{0,100,0}
\definecolor{seagreen}{RGB}{46,139,87}
\definecolor{lightseagreen}{RGB}{32,178,170}
\definecolor{forestgreen}{RGB}{34,139,34}
\definecolor{midnightblue}{RGB}{25,25,112}
\definecolor{navyblue}{RGB}{0,0,128}
\definecolor{cornflowerblue}{RGB}{100,149,237}
\definecolor{mediumblue}{RGB}{0,0,205}
\definecolor{lightgray}{RGB}{190,190,190}
\definecolor{slategray}{RGB}{112,138,144}

\newcommand{\raw}{\rightarrow}
\newcommand{\ee}{\mathrm{e}}
\newcommand{\I}{\mathrm{i}}
\newcommand{\bC}{\mathbb{C}}
\newcommand{\cP}{\mathcal{P}}
\newcommand{\cT}{\mathcal{T}}
\newcommand{\cO}{\mathcal{O}}
\newcommand{\cL}{\mathcal{L}}
\newcommand{\cM}{\mathcal{M}}

\newcommand{\cH}{\mathcal{H}}
\newcommand{\bR}{\mathbb{R}}
\newcommand{\bZ}{\mathbb{Z}}
\newcommand{\tr}{\mathrm{Tr}}
\newcommand{\nn}{\nonumber}

\newcommand{\kom}{\, ,\quad}
\newcommand*{\dif}{{\,\rm d}}
\newcommand*{\p}{\mathop{}\!\mathrm \partial}

\usepackage{mathrsfs} 

\definecolor{light-gray}{gray}{0.85}
\definecolor{tgray}{gray}{0.2}

\newtheoremstyle{AS}{\topsep}{\topsep}{\itshape\color{tgray}}{}{\bfseries}{.}{1pt}{}          
\theoremstyle{AS}

\usepackage[framemethod=TikZ]{mdframed}
\mdfsetup{skipabove=\topskip, innertopmargin=-5pt, skipbelow=\topskip}

\usepackage{lipsum} 

\newcounter{equ}[section]
\newenvironment{equ}[1][]{%
\stepcounter{equ}%
\ifstrempty{#1}%
{\mdfsetup{%
frametitle={%
\tikz[baseline=(current bounding box.east),outer sep=0pt]
\node[anchor=east,rectangle,fill=blue!20]
{\strut };}}
}%
{\mdfsetup{%
frametitle={%
\tikz[baseline=(current bounding box.east),outer sep=0pt]
\node[anchor=east,rectangle,fill=blue!20]
{\strut ~#1};}}%
}%
\mdfsetup{innertopmargin=5pt,linecolor=blue!20,%
middlelinewidth=2pt,topline=true,
frametitleaboveskip=\dimexpr-\ht\strutbox\relax,}
\begin{mdframed}[]\relax%
}{\end{mdframed}}

\newcounter{Boxequ}[section]
\newenvironment{Boxequ}[1][]{%
\mdfsetup{innertopmargin=10pt,linecolor=blue!20,%
middlelinewidth=2pt,topline=true,}
\begin{mdframed}[]\relax%
\vspace{-0.25cm}
}{\end{mdframed}}

\newcounter{summary}[section]

\definecolor{ocre}{RGB}{243,102,25}
\definecolor{mygray}{RGB}{243,243,244}
\definecolor{myred}{RGB}{245,73,73}

\newtheoremstyle{definition}
  {\topsep}
  {}
  {\normalfont}
  {}
  {\bfseries}
  {.}
  {.5em}
  {}
\theoremstyle{definition}
\newmdtheoremenv{definition}{Definition}[section]

\newmdtheoremenv[ 
skipabove=10pt,
roundcorner=5pt,
 middlelinewidth=2pt,
  backgroundcolor=blue!5,
  linecolor=blue!30,
  leftmargin=20pt,
  innerleftmargin=5pt,
  innerrightmargin=5pt,
  ]{satz}{Satz}[section]

  \newmdtheoremenv[
  skipabove=10pt,
roundcorner=5pt,
  middlelinewidth=2pt,
  backgroundcolor=blue!5,
  linecolor=blue!30,
  leftmargin=20pt,
  innerleftmargin=5pt,
  innerrightmargin=5pt,
  ]{DEF}[satz]{Definition}
  
    \newmdtheoremenv[
  skipabove=10pt,
roundcorner=5pt,
  middlelinewidth=2pt,
  backgroundcolor=blue!5,
  linecolor=blue!30,
  leftmargin=20pt,
  innerleftmargin=5pt,
  innerrightmargin=5pt,
  ]{claim}[satz]{Claim}
  
    \newmdtheoremenv[
    skipabove=10pt,
roundcorner=5pt,
  middlelinewidth=2pt,
  backgroundcolor=blue!5,
  linecolor=blue!30,
  leftmargin=20pt,
  innerleftmargin=5pt,
  innerrightmargin=5pt,
  ]{theo}[satz]{Theorem}
  
    \newmdtheoremenv[
    skipabove=10pt,
roundcorner=5pt,
 middlelinewidth=2pt,
  backgroundcolor=blue!5,
  linecolor=blue!30,
  leftmargin=20pt,
  innerleftmargin=5pt,
  innerrightmargin=5pt,
  ]{EX}[satz]{Exercise}
  
 \newmdtheoremenv[
 skipabove=10pt,
 roundcorner=5pt,
 middlelinewidth=2pt,
  backgroundcolor=blue!5,
  linecolor=blue!30,
  leftmargin=20pt,
  innerleftmargin=5pt,
  innerrightmargin=5pt,
  ]{lemma}[satz]{Lemma}
  
 \newmdtheoremenv[
 skipabove=10pt,
 roundcorner=5pt,
 middlelinewidth=2pt,
  backgroundcolor=blue!5,
  linecolor=blue!30,
  leftmargin=20pt,
  innerleftmargin=5pt,
  innerrightmargin=5pt,
  ]{koro}[satz]{Korollar}
  
   \newmdtheoremenv[
   skipabove=10pt,
 roundcorner=5pt,
 middlelinewidth=2pt,
  backgroundcolor=blue!5,
  linecolor=blue!30,
  leftmargin=20pt,
  innerleftmargin=5pt,
  innerrightmargin=5pt,
  ]{coro}[satz]{Corollary}
  
   \newmdtheoremenv[
   skipabove=10pt,
   roundcorner=5pt,
 middlelinewidth=2pt,
  backgroundcolor=blue!5,
  linecolor=blue!30,
  leftmargin=20pt,
  innerleftmargin=5pt,
  innerrightmargin=5pt,
  ]{propo}[satz]{Proposition}
  
\usepackage{amsfonts}
\usepackage{array}
\usepackage{tikz}
\usetikzlibrary{decorations.markings}

\usetikzlibrary{decorations.markings}
\usetikzlibrary{arrows}

\newcommand{\VEC}[1]{\vec{#1}} 

\newcommand*{\SUN}{\mathrm{SU}(N)}
\newcommand*{\SUTw}{\mathrm{SU}(2)}
\newcommand*{\SUTh}{\mathrm{SU}(3)}

\newcommand*{\UO}{\mathrm{U}(1)}

\newcommand{\cD}{\mathcal{D}}

\newcommand*{\UTh}{\mathrm{U}(3)}

\usepackage{setspace}

\let\emph\relax 
\DeclareTextFontCommand{\emph}{\em}

\renewcommand{\emph}[1]{\textit{#1}}
\usepackage{simpler-wick}

\newcommand\summaryname{\normalsize Abstract}
\newenvironment{Abstract}%
    {\normalsize\begin{center}%
    \bfseries{\summaryname} \end{center}}

\begin{document}
\setcounter{tocdepth}{4}

\pagenumbering{roman}

\thispagestyle{empty} 

\begin{flushright}
        \text{\small CERN-TH-2024-106}
\end{flushright}

\begin{center}

\vspace*{2cm}

{\LARGE \textbf{Cambridge Lectures on The Standard Model}}

\vspace*{2cm}

{\large {Fernando Quevedo$^{a,b,c}$ and Andreas Schachner$^{a,d,e}$}}

\vspace{0.5 cm}
{\textsl{$^{a}${\small DAMTP, CMS, Wilberforce Road, Cambridge, CB3 0WA, UK}}\\
\textsl{$^{b}${\small Department of Theoretical Physics, CERN, 1211 Meyrin, Switzerland}}\\
\textsl{$^{c}$ {\small New York University Abu Dhabi, PO Box 128199, Saadiyat Island, Abu Dhabi, UAE}}\\
\textsl{$^{d}${\small ASC for Theoretical Physics, LMU Munich, 80333 Munich, Germany}}\\
\textsl{$^{e}${\small Department of Physics, Cornell University, Ithaca, NY 14853, USA}}\\}

\vspace*{1.1cm}

\end{center}

\begin{Abstract}
\begin{changemargin}{0.75cm}{0.75cm}

These lecture notes cover the Standard Model (SM) course for \emph{Part III of the Cambridge Mathematical Tripos}, taught during the years 2020-2023. The course comprised 25 lectures and 4 example classes. Following a brief historical introduction, the SM is constructed from first principles. We begin by demonstrating that essentially only particles with spin/helicity $0, \frac{1}{2}, 1, \frac{3}{2}, 2$ can describe matter and interactions, using spacetime symmetries, soft theorems, gauge redundancies, Ward identities, and perturbative unitarity. The remaining freedom lies in the choice of the Yang-Mills gauge group and matter representations. Effective field theories (EFTs) are a central theme throughout the course, with the 4-Fermi interactions and chiral perturbation theory serving as key examples. Both gravity and the SM itself are treated as EFTs, specifically as the SMEFT (Standard Model Effective Field Theory). Key phenomenological aspects of the SM are covered, including the Higgs mechanism, Yukawa couplings, the CKM matrix, the GIM mechanism, neutrino oscillations, running couplings, and asymptotic freedom. The discussion of anomalies and their non-trivial cancellations in the SM is detailed. Simple examples of calculations, such as scattering amplitudes and decay rates, are provided. The course concludes with a brief overview of the limitations of the SM and an introduction to the leading proposals for physics beyond the Standard Model.

\end{changemargin}
\end{Abstract}

~\vfill

\noindent Version: \today ~(\now\!\!).

\newpage

{
\onehalfspacing
\tableofcontents
}

\pagenumbering{arabic} 


\chapter{\bf Introduction and History}
\label{chap:intro}

\vspace{0.5cm}
\begin{equ}[The Guideline]
{\it Our purpose in theoretical physics is not to describe the world as we find it, but to explain -- in terms of a few fundamental principles -- why the world is the way it is.}\\

\rightline{\it Steven Weinberg}
\end{equ}
\vspace{0.5cm}

The Standard Model is one of the greatest scientific achievements of all time.
It consistently describes all known fundamental particles and their interactions with the exception of gravity that is still properly described at low energies. In this sense we can now explain any fundamental physical phenomenon at the smallest distances that can be probed experimentally with spectacular success. We may therefore claim that, so far, the Standard Model is the fundamental theory of Nature.

The Standard Model is the most successful application of quantum field
theory when it comes to experimental verification. It is the final conclusion of many decades of intense research both on the theoretical and experimental side. Its structure was finally completed after the celebrated discovery of the Higgs particle in  2012. 
Over the  decades
since its ingredients were combined, thousands of measurements have been
made at energies $E\leq 1$ TeV, all consistent with the Standard Model.

The Standard Model describes the physics of the building blocks of all visible matter: spin $1/2$ quarks and leptons interacting via three fundamental forces, each
mediated by  spin 1 particles known as \emph{gauge bosons}.
Electrically charged particles feel the electromagnetic force by exchanging \emph{photons} as described by Quantum Electrodynamics (QED). The electromagnetic interactions are of long range due to the fact that photons are massless.
In contrast, the short range weak force
is responsible for certain radioactive decays such as the neutron $\beta$-decay and plays a crucial role in the thermonuclear interactions within stars.  The mediators of this interaction are the massive $W$ and $Z$ bosons. Their large mass is responsible for the  short range of the interaction.
The strong force binds
quarks into nucleons (protons and neutrons) and indirectly nucleons into nuclei; the carriers of the strong force
are appropriately called the \emph{gluons}. Leptons, such as electrons and neutrinos do not feel the strong force. Particles made out of quarks are called \emph{hadrons} which can be either \emph{baryons} (made up of 3 quarks) and $mesons$ (a quark-antiquark pair).

\

The aim of these lectures is threefold. First, to let the students appreciate all the twists and turns that drove scientists in the past century to discover and develop the Standard Model.
A historical perspective is important to appreciate the magnitude of the achievements, but also the surprises and human drama that came with the development of new ideas in particle physics.
Most importantly, it demonstrates that research is not a straight line of well developed arguments as usually presented in textbooks and lecture notes,
but rather a windy road with occasionally unforeseen twists and turns before a proper understanding emerges. 

The second goal is for students to internalise that, despite the Standard Model being only one in an infinite number of possible field theories, its structure is extraordinary rigid and compelling. Following general principles within our basic theories, \emph{Special Relativity}\footnote{We emphasise that it is Special Relativity and not General Relativity. The reason is that Special Relativity applies in general for all interactions whereas General Relativity describes only one of the interactions and we will see this may be obtained from Special Relativity and Quantum Mechanics applied to particles of helicity $\pm 2$. } and \emph{Quantum Mechanics}, we take a constructive approach arguing that it is essentially unavoidable that elementary particles are determined
by unitary representations of the Poincar\'e group limiting their nature to only a handful of possibilities, namely spins $0,1/2,1,3/2, 2$ out of an infinite number of possible spins.
Further, we will see why gauge invariance defining the three non-gravitational interactions is only a redundancy needed to properly describe the interactions.
Lastly, even though Field Theory is the basic formalism to describe interactions, the fundamental objects are actually the particles themselves, whereas fields are a necessary \emph{tool} to describe local interactions among particles. 

The third target of these lectures is for students to get familiar with the physical details of the Standard Model and be able to reproduce some of the key calculations and results that led to its successful completion. Throughout the lectures emphasis will be given to basic principles and potential loopholes that may be important to guide us towards the unknown physics beyond the Standard Model.
That is, we provide crucial methods to build new theories (or models) of nature -- a skill that is vital for any theoretical physicist.
We follow the guideline as in the quote of Steven Weinberg above with the key word being to {\it explain} rather than describe. We emphasise the explanatory power of the Standard Model towards all the experiments, but also towards some of the approximate or accidental symmetries such as isospin, flavour, baryon number, etc.
Just like the Standard Model provides a UV description of \emph{Effective Field Theories} (EFTs) such as the Fermi theory of weak interactions and the pion dynamics of Yukawa, it should itself only be regarded as an EFT once gravity or other UV physics is included.

\

These lecture notes are based on the course \emph{The Standard Model} in \emph{Part III of the Mathematical Tripos} of the University of Cambridge taught from 2020 to 2023. The subject is vast and some selection had to be made since we were limited to $24+1$ lectures. For complementary material we refer to the several excellent books on the subject. For a partial list see \cite{Weinberg:1995mt, Weinberg:1996kr, Burgess:2006hbd, Schwartz:2014sze, Peskin:1995ev, Burgess:2020tbq}. In particular, we follow the general structure and logic of the presentation of Weinberg's books \cite{Weinberg:1995mt, Weinberg:1996kr}.

Previous versions of this course over the past years by B. Allanach, C. Thomas, M. Wingate, and other colleagues before them provided an alternative, more phenomenological presentation of the subject.
They influenced the discussions of decay rates and cross sections in Appendix~\ref{app:drc}.
More recent also excellent lectures given by David Tong take yet another emphasis on this rich subject \cite{TongSM}.

The course only assumes a basic knowledge of group theory and a first exposure to quantum field theory although an effort is made to be as self-contained as possible. Subjects such as  path integrals, quantisation of non-abelian gauge theories and renormalisation group are discussed in coordination with a parallel course on Advanced Quantum Field Theory (AQFT), but they are briefly summarised for those students who were not taking AQFT.
The presentation aims at preparing students to think about the fundamental ideas underlying the Standard Model that could eventually be questioned once they start working on the subject to search for the physics that may lie beyond the Standard Model.

\

\newpage

\section{Brief History of the Standard Model}\label{sec:histSM} 

Before we begin with a careful treatment of the Standard Model itself, we present here a brief sketch of the historical developments that led to the Standard Model.
\

\noindent{
\small
\begin{tabular}{ccl}
\textbf{1600's} & & Classical {\bf Gravity}. First unification of interactions (Newton).\\
\textbf{1800's} & 1861 & Classical {\bf Electromagnetism}. Second unification  (Maxwell \cite{maxwell2010physical}).\\
 &1869& Periodic table (Mendeleyev). Discrete nature of matter not established.\\ 
\textbf{} & 1896& Radioactivity (Becquerel, P. \& M. Curie, Rutherford).  \\ 
 & &$\alpha$-, $\beta$- \& $\gamma$-decay as hint for instabilities in nature \&\\ 
  & & hint for new  {\bf Weak Interactions} \& {\bf Strong Interactions}.\\ 
   &1897& {\bf Electron} discovered (J. J. Thomson \cite{Thomson:1897cm}), first computation of $e/m_{e}$.\\ 
    && \emph{Beginning of particle physics!}\\ 
    \hline
\hline
\textbf{1900's} &1900-1930 & Quantum Mechanics developed \& established (e.g.  {\bf Photons} as particles).\\ 
 &1905& Special Relativity (Einstein \cite{Einstein:1905ve}, e.g.  $c=\text{const.}$ \& spacetime structure).\\ 
  && $\Rightarrow$ \emph{The two basic theories of nature.} \\ 
  \hline
\hline
\textbf{1910's} &1911& Rutherford \cite{Rutherford:1911zz} formulated a model for atoms ({\bf Proton} nucleus of $H$). \\ 
  & & First cloud chamber is constructed (Wilson). \\ 
  & 1912& Cosmic rays discovered (Hess \cite{Hess:1912srp,Hess:2018twh}).\\ 
& 1915& Einstein General Relativity \cite{Einstein:1911vc,Einstein:1914bt,Einstein:1915ca,Einstein:1915by,Einstein:1916vd}.\\ 
 & 1919& F. Aston \cite{aston1920xliv} postulates the ``whole numbers rule'' $\raw$ \textbf{proton}. \\ 
 \hline
\hline
 \textbf{1920's} && Bose \cite{Bose:1924mk} \& Fermi \cite{1999cond.mat.12229Z} statistics. \\ 
 & & Beginning of Quantum Field Theory: Dirac, Jordan, Heisenberg, ... \cite{weinberg1977search}.\\ 
  & &Dirac equation \cite{Dirac:1928hu,Dirac:1930ek} with solutions of charge $\pm 1$.\\ 
  \hline
\hline
 \textbf{1930's} &1930& Pauli predicts {\bf Neutrino} (energy \& momentum conservation in $\beta$-decay). \\ 
&1931& Dirac predicts {\bf Positron} $e^{+}$ as anti-particle of $e^{-}$ \cite{Dirac:1931kp}. \\ 
&1932& Anderson \cite{Anderson:1933mb} discovered positron $e^{+}$. \\ 
&& Chadwick \cite{Chadwick:1932ma} discovered {\bf Neutron}. \\ 
&& Heisenberg \cite{Heisenberg:1932dw} introduces isospin as symmetry between $n$ \& $p$. \\ 
&1934& Fermi theory of weak interactions \cite{fermi1934versuch}, e.g. $\beta$-decay: $n\raw p+e^{-}+\bar{\nu}$. \\ 
&1935& Yukawa theory of strong interactions \cite{Yukawa:1935xg}. \\ 
&& Scalar mediators {\bf Pions} $\pi$ predicted. \\ 
&& Short range potential $V\sim\ee^{-m_{\pi}r}/r$ with $m_{\pi}\sim 100$MeV. \\ 
&1936& Anderson, Neddermeyer \cite{Neddermeyer:1937md} discovered {\bf Muon} $\mu$ with $m_{\mu}\sim 100$MeV. \\ 
&& Condon et al. \cite{Cassen:1936dg} include isospin as d.o.f. in wave function. \\ 
&1939& Wigner's description of particles as representations of Poincar\'e group \cite{Wigner:1939cj}. \\
\hline
\hline
\end{tabular} 
}

\vspace*{-0.1cm}

\noindent{
\small
\begin{tabular}{ccl}
\hline
\hline
  \textbf{1940's}&1947 & Lamb shift \cite{Lamb:1947zz} (vacuum polarisation becomes relevant). \\ 
   && {\bf QED} (Schwinger \cite{Schwinger:1948iu,Schwinger:1948yj,Schwinger:1948yk,Schwinger:1949ra}, Feynman \cite{Feynman:1949zx,Feynman:1949hz,Feynman:1950ir}, \\
   &&\hphantom{{\bf QED} (} Tomonaga \cite{Tomonaga:1946zz,Koba:1947rzy}, Dyson \cite{Dyson:1949bp,Dyson:1949ha}). \\ 
&& Pions $\pi$ are discovered (charged 1947 \cite{Lattes:1947mw}, neutral 1950 \cite{Bjorklund:1950zz}). \\ 
\hline
\hline
  \textbf{1950's} && Particle accelerators and bubble chambers ($E\geq $MeV). \\ 
 && dozens of new particles discovered (mostly strongly interacting). \\ 
&& {\bf Hadrons}: Kaons, hyperons, ... \\ 
&& 2 classes: {\bf Mesons} (bosonic) \& {\bf Baryons} (fermionic). \\ 
&& Classification: charge (Q), baryon number (B), lepton number (L),...,\\ && strangeness (S) (Gell-Mann 1956 \cite{Gell-Mann:1956iqa}, ...) as new charge. \\ 
&1954& Yang-Mills (\& Shaw) theory generalising QED \cite{Yang:1954ek}.\\ 
&1956& Parity violation theoretically conjectured (Lee \& Yang \cite{Lee:1956qn}, Salam \cite{Salam:1957st}) \\ 
&& implies that parity is \emph{not} a fundamental symmetry of nature! \\ 
&& Discovery of {\bf (Anti-)Neutrino} (Cowan, Reines \cite{Cowan:1992xc}). \\ 
&1957& Wu discovered parity violation experimentally \cite{Wu:1957my}. \\ 
&& Neutrino oscillations proposed (Pontecorvo \cite{Pontecorvo:1957cp}). \\ 
&& V-A structure of weak interactions (Marshak \& Sudarshan \cite{Sudarshan:1958vf},\\
&&  \hphantom{V-A structure of weak interactions (} also Feynman, Gell-Mann).\\ 
\hline 
\hline
  \textbf{1960's} &1961& Eightfold Way (Gell-Mann \cite{GellMann:1961ky} \& Ne'eman \cite{Neeman:1961jhl}), see Fig.~\ref{fig:EFW}. \\ 
 &&Symmetry breaking (Nambu \cite{Nambu:1961tp,Nambu:1961fr},  Goldstone, Weinberg, Salam \cite{Goldstone:1961eq, Goldstone:1962es}).\\ 
&1962& Cabibbo mixing \cite{Cabibbo:1963yz}. \\ 
&& {\bf Muon Neutrino} $\nu_{\mu}$ discovered (Steinberger et al. \cite{Danby:1962nd}). \\ 
&& Preliminary Electroweak unification (Glashow \cite{Glashow:1961tr}, Salam-Ward \cite{Salam:1964ry}).\\ 
&1964& {\bf Quarks $u$, $d$, $s$}  proposed (Gell-Mann \cite{GellMann:1964nj} \& Zweig \cite{Zweig:1981pd,Zweig:1964jf}). \\ 
&& {\bf Higgs Mechanism} (Higgs \cite{Higgs:1964ia,Higgs:1964pj,Higgs:1966ev}, Brout, Englert \cite{Englert:1964et},\\
&&  \hphantom{{\bf Higgs Mechanism} (} Guralnik et al.\cite{Guralnik:1964eu}, Kibble \cite{Kibble:1967sv}).  \\ 
&& $\Omega^{-}$ discovery \cite{Barnes:1964pd}, CP violation \cite{Wolfenstein:1964ks,Wu:1964qx,Matumoto:1964uy}, Kaon decay \cite{Christenson:1964fg}.  \\ 
&& Colour introduced (Greenberg \cite{Greenberg:1964pe}, Han-Nambu \cite{Han:1965pf}). \\ 
&& {\bf Charm} quark predicted (Glashow, Bjorken \cite{Bjorken:1964gz}). \\ 
&1967& {\bf Electroweak Unification} (Weinberg \cite{Weinberg:1967tq}, Salam \cite{Salam:1968rm}).\\
&1968& Deep inelastic scattering (Friedman, Kendall, Taylor et al. \cite{Bloom:1969kc,Breidenbach:1969kd}). \\
&& Parton (quarks, gluons) composition of hadrons (Bjorken, Feynman). \\
&1968& Solar neutrino puzzle (Bahcall \& Davis \cite{Davis:1968cp}). \\
&1969& Anomalies (Bell, Jackiw \cite{Bell:1969ts}, Adler \cite{Adler:1969gk}). \\ 
\hline
\hline
\end{tabular} 
}

\noindent{
\small
\begin{tabular}{ccl}
\hline
\hline
  \textbf{1970's}& 1970& Glashow-Iliopoulos-Maiani (GIM) mechanism \cite{Glashow:1970gm}: existence of charm \\ 
 & &  quarks explains suppression of flavour-changing neutral currents.\\ 
&1971& Renormalisability of Weinberg-Salam model (t'Hooft \cite{tHooft:1971qjg}). \\ 
&1973& {\bf Asymptotic Freedom} (Gross-Wilczek \cite{Gross:1973ju}, Politzer \cite{Politzer:1973fx}). \\ 
&& {\bf QCD} (Fritzsch, Leutwyler, Gell-Mann \cite{Fritzsch:1973pi}).\\
&& Weak Neutral Currents measured at CERN \cite{Hasert:1973ff}. \\ 
&& Kobayashi-Maskawa mixing ($3$-families, CP-violation) \cite{Kobayashi:1973fv}. \\
&1974& $J/\Psi$ discovery (Richter et al. \cite{Augustin:1974xw,Aubert:1974js}) proves existence of {charm} quark. \\ 
&& Effective Field Theory (Wilson, Weinberg).\\ 
&1975&{\bf Tau Lepton} $\bf \tau$ (M. Perl et al. \cite{Perl:1975bf}). \\ 
&& Quark Jets (hadronisation) \& $2$-jet events ($e^{+}e^{-}\raw q\bar{q}\raw 2\,$jets). \\  
&1977& Upsilon discovered alongside the {\bf Bottom Quark} (Fermilab \cite{Herb:1977ek}). \\ 
&1979& {\bf Gluon} evidence  jets $e^{+}e^{-}\raw q\bar{q}q\raw 3\,$jets \cite{Barber:1979yr}. \\ 
\hline
\hline
  \textbf{1980's} &1983& Discovery of $\bf Z^{0}$, $\bf{W^{\pm}}$ (Rubbia et al. \cite{Arnison:1983rp} at CERN with $170$GeV collisions). \\ 
  \textbf{1990's} &1995& {\bf Top Quark} discovery (Fermilab \cite{Abe:1995hr,Abachi:1994td}). \\ 
 && LEP precision tests of SM ($<3$ light neutrinos). \\ 
  &1998& Neutrino Oscillations \cite{Fukuda:1998mi,Mikaelyan:1999pm}. \\ 
  \textbf{2000's} &2001& {\bf Tau Neutrino} $\nu_\tau$ discovery \cite{Kodama:2000mp}. \\ 
  \textbf{2010's} &2012& {\bf Higgs} discovery \cite{Chatrchyan:2012xdj,Aad:2012tfa}. \\ 
  &2016 & First detection of {\bf Gravitational Waves} by LIGO \cite{LIGOScientific:2016aoc}. \\ 
      \hline
\hline
\end{tabular} 
}

\begin{figure}[t!]
\centering
 \includegraphics[width=0.95\linewidth]{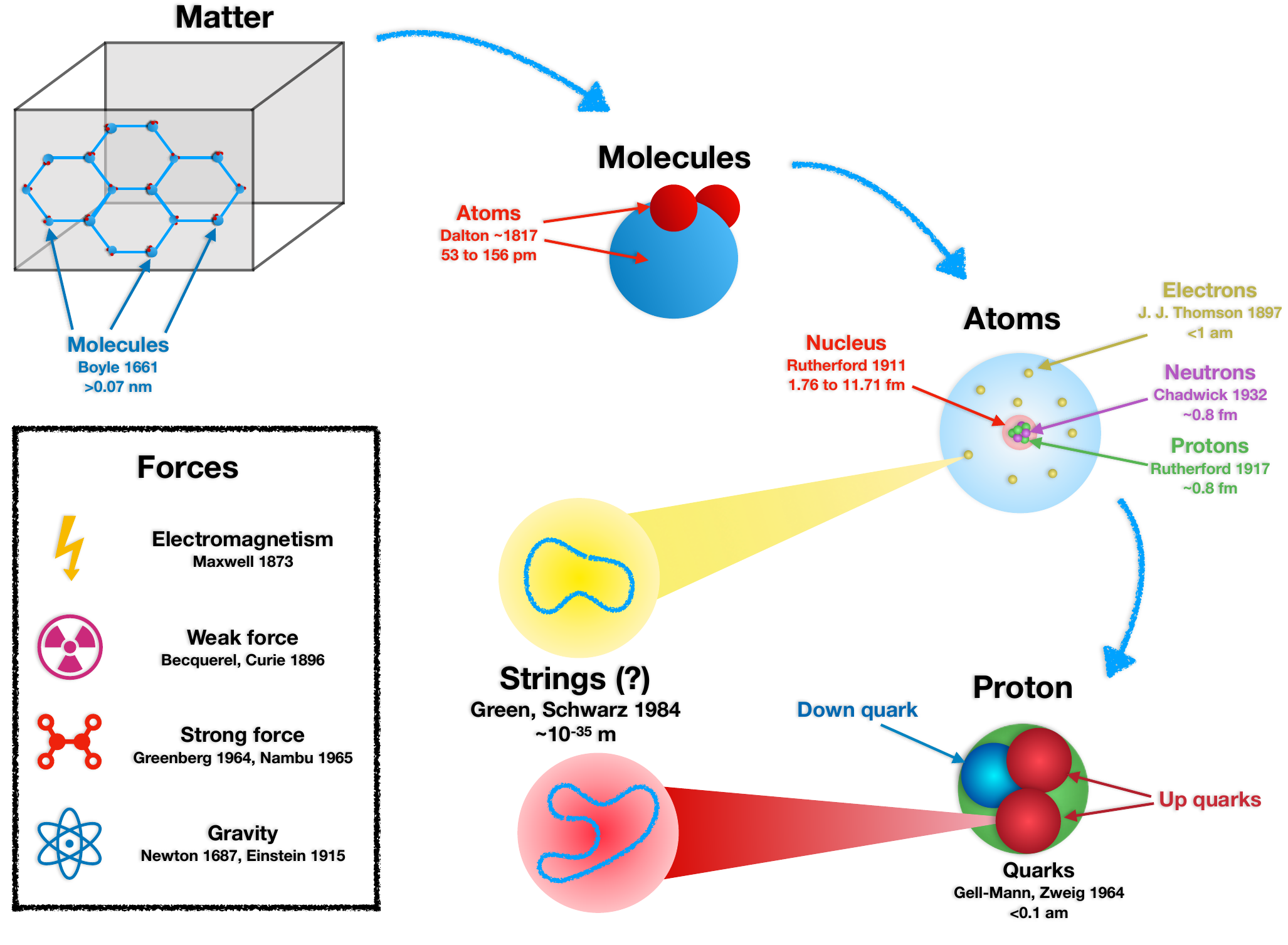}
\caption{An historical account of our understanding of matter and its constituents. The complementary forces are summarised in the box on the left. The dates either refer to the earliest major contribution towards a theoretical understanding or to the first experimental evidence.}\label{fig:HSM}
\end{figure}

\vspace*{0.5cm}

From the above,
we find that the history of the Standard Model involves many interesting scientific developments, great creativity, very hard work with ups and downs with incredible achievements. As usual, the human factor plays an important role as the following anecdotes establish:
\begin{itemize} 
\item How a combination of great theoretical ideas with ingenious and brave experimental initiatives managed to unlock the deepest mysteries of the elementary particles. 
\item In particular, how  the Cavendish laboratory played such a crucial role in the early part of the 20th century with some of the major discoveries that helped identify the structure of the nucleus and different elementary particles. It is interesting to notice that  the discoverer of the neutron (Chadwick) was a student of the discoverer of the proton (Rutherford) who in turn was a student of the discoverer of the electron (Thomson), completing the composition of all atoms. All of them have followed the steps of the previous Cambridge giants: Newton and  Maxwell who had performed the great unifications of the past, namely the gravitational forces on Earth and  space, and electricity, magnetism (and optics)\footnote{Without counting the equivalent unifications in Biology with Darwin and his Theory of Evolution and Crick and Watson's double helix structure of DNA. Great discoveries made also in Cambridge (including also the football rules!).}, which can be considered the start of the Standard Model. This can make us proud and humble to address these questions in this very same place where so many developments were made.
\item The fact that a few years after Thomson discovered the first elementary particle (the electron) his son managed to prove that the electron was also a wave  identifying the electron diffraction patterns.
\item The original reluctance to explain experimental results by introducing new particles (it delayed the identification of the neutron as an independent new particle instead of an electron-proton composite).
\item Pauli's bold proposal of the existence of a totally new class of particles, neutrinos, based on arguments of conservation laws.
\item Dirac's contrived prediction of the positron and anti-particles, while being again reluctant for some time to accept the positron as a new particle.
\item The several independent discoveries of the positron, but most failed to appreciate it or report on time. 
\item Yukawa's prediction of pions as mediators of the (strong) interactions among protons and neutrons to keep them together within the nucleus dominating the electromagnetic repulsion among protons. Contrary to the electromagnetic interactions which are long range, these strong interactions had to be only at the nuclear scale and Yukawa concluded that the mediators of the interactions were massive particles. Almost immediately  the muon was discovered at very similar mass as predicted by Yukawa creating confusion since the muons only interact by weak and electromagnetic interactions. The confusion finished when the pions were later discovered with a mass similar to the muons but with the properties Yukawa had predicted. It took many years for people to understand that Yukawa's theory was only an approximation of the fundamental strong interactions mediated by gluons. In current terminology Yukawa's theory is an Effective Field Theory (EFT) that is completed in the UV by QCD.
\item Fermi's theory to describe weak interactions such as $\beta$ decay in terms of four-particle interactions being very accurate at low energies but failing at larger energies. This is today also understood as an EFT that is completed in the UV by the exchange of the massive $Z$ and $W$ particles of the Standard Model. 
\item The role of a relatively simple issue in atomic physics such as the Lamb-shift leading to the full theoretical development of QED. 

\begin{figure}[t!]
\centering
 \includegraphics[scale=0.3]{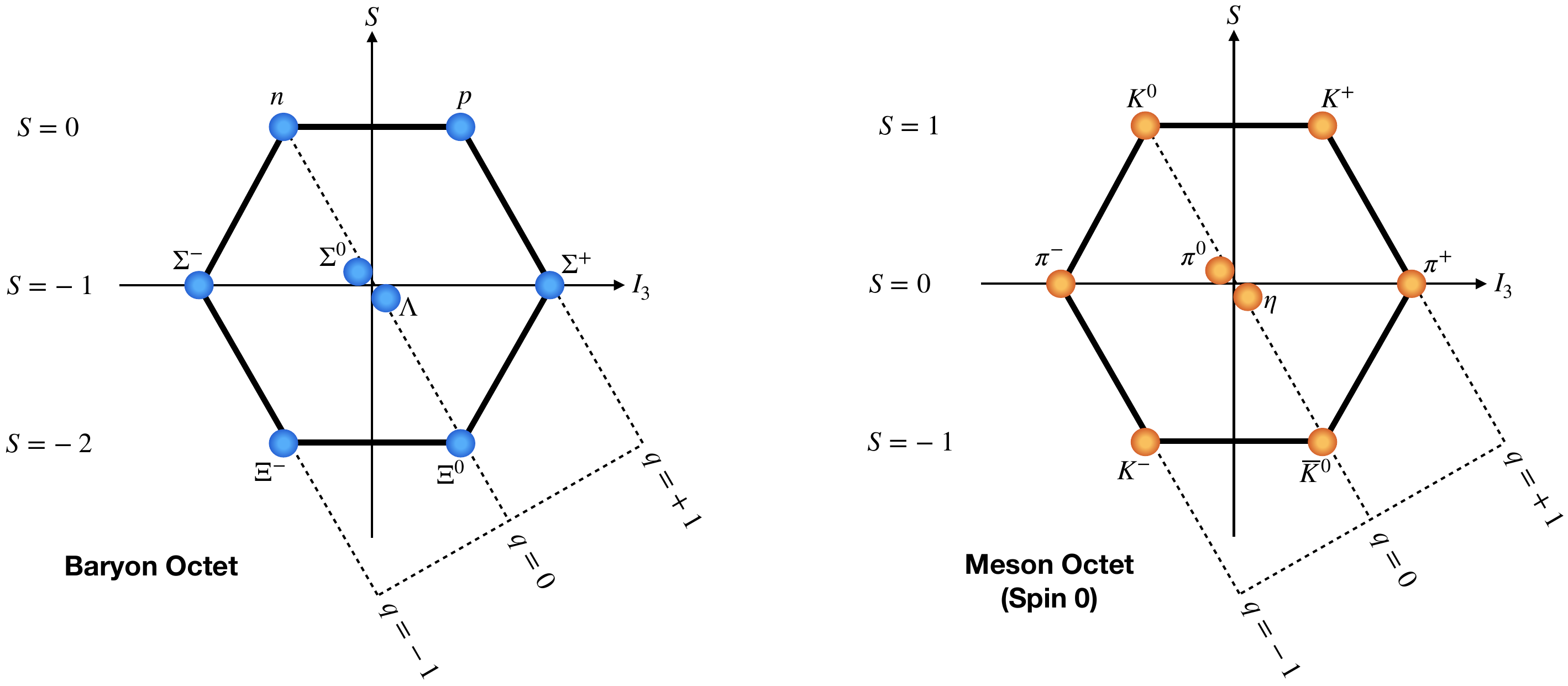}
  \includegraphics[scale=0.3]{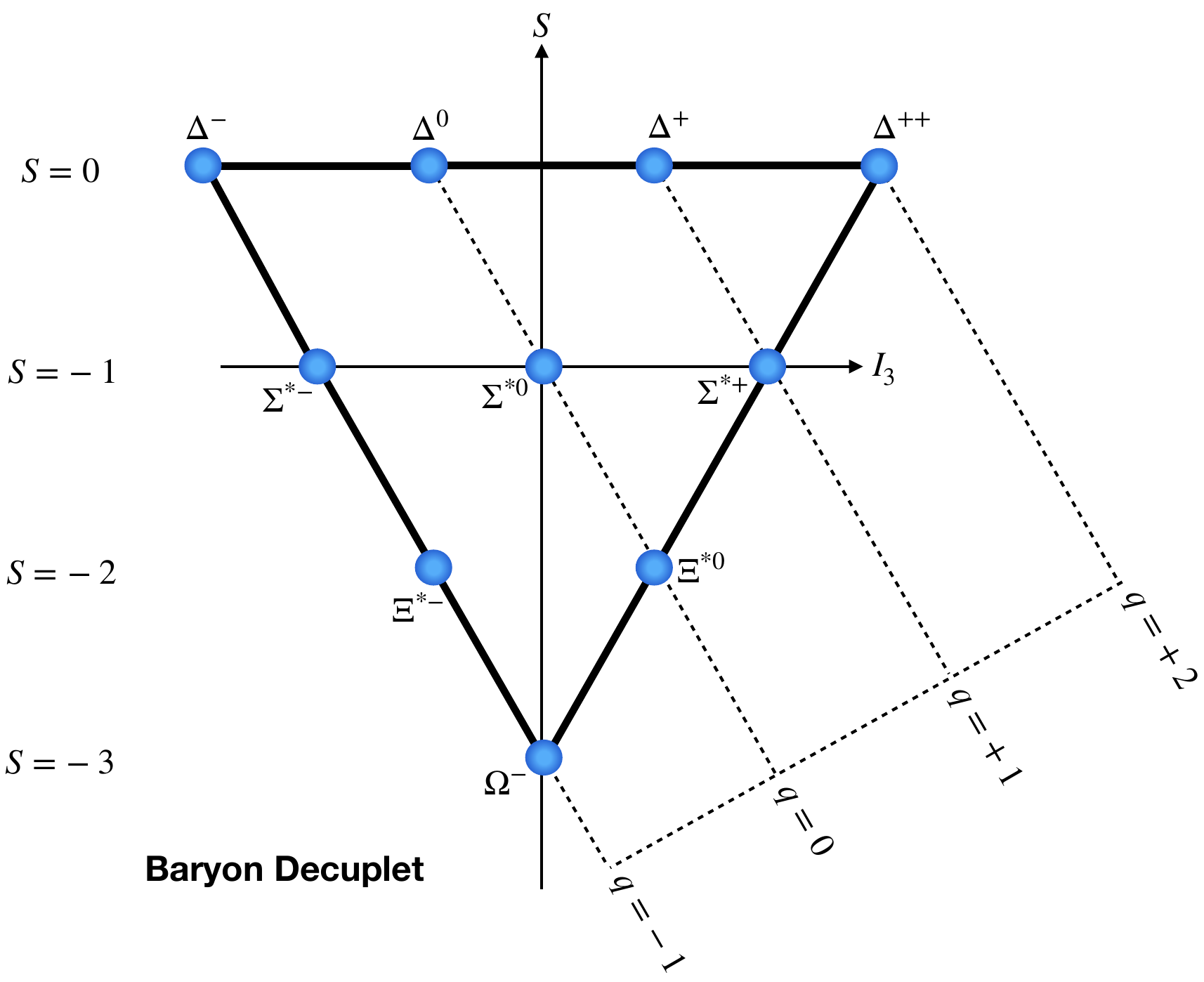}
\caption{\index{Eightfold way}The \emph{eightfold way}. The weight lattice of hadrons in their respective $\mathrm{SU}(3)_{f}$  {representations\protect\footnotemark} with the baryon octet (top left) corresponding to $\mathbf{8}$ of $\mathrm{SU}(3)_{f}$, pseudo-scalar meson octet (top right) again corresponding to $\mathbf{8}$ of $\mathrm{SU}(3)_{f}$ and the baryon decuplet (bottom) corresponding to $\mathbf{10}$ of $\mathrm{SU}(3)_{f}$. Here, $S$ denotes 'strangeness' and $I_{3}$ isospin. The prediction of the $\Omega^-$ particle and its subsequent discovery lead strong credibility to this approximate symmetry.}\label{fig:EFW}
\end{figure}
\footnotetext{Recall that for $\mathrm{SU}(3)$, $\mathbf{3}\otimes\bar{\mathbf{3}}=\mathbf{8}\oplus \mathbf{1}$ and $\mathbf{3}\otimes \mathbf{3}\otimes\mathbf{3}=\mathbf{10}\oplus\mathbf{8}\oplus\mathbf{8}\oplus\mathbf{1}$.}

\item The introduction of approximate symmetries such as strangeness and eightfold way (see Fig.~\ref{fig:EFW}) leading to the prediction of quarks.
\item The original reluctance to question the validity of parity violation and the great inspiration of Lee and Yang to prove it and Wu to confirm it experimentally in such a short time.
\item How physicists became involved in world's politics. For instance, the discoverer of the neutron, Chadwick, was a first world war prisoner for 4 years where he still managed to perform some experiments (similar to Karl Schwarzschild who found the black hole solution \cite{Schwarzschild:1916uq} of Einstein's equations while fighting in the war in 1915). Although the second world war somehow delayed fundamental scientific progress since some of the leading scientists were involved in projects such as the Manhattan project, after the war, scientists quickly returned to ask and answer fundamental questions and even  overcame political differences by having collaborations from both sides  during  the cold war. Although cases, like Pontecorvo, took sides and escaped from the West (after a mysterious disappearance) and others moved in the other direction. 
Like Harald Fritzsch, one of the pioneers of QCD, escaping in a folding boat from East Germany via Bulgaria to Turkey after organising a major protest.
\item The disappearance of one of the greatest minds of the 20th century (Majorana) that still remains a mystery.
\item Creating international institutions like CERN allowed scientists to show how a united effort among different countries can lead to  great achievements, much  earlier than politicians realised the same and proposed the European Union. CERN is still the best example of international scientific collaborations and defines the recently coined term of \emph{science diplomacy}.
\item Pauli's  dismissal of Salam's ideas on parity violation (that were later confirmed by Lee and Yang). 
\item Pauli's also criticism of Yang and Mills since, as he correctly pointed out, their theory predicted massless particles that should have been  observed and were not. It so happens that Pauli had obtained the same theory as Yang and Mills starting from a six-dimensional theory but decided not to publish his results because of the massless particles problem. He gave such a difficult time to Yang in a seminar at the Institute for Advanced Studies in Princeton that Yang decided to stop the seminar and sat down until Oppenheimer convinced him to continue \cite{O'Raifertaigh}.
\item How, in turn,  Salam discouraged his PhD student Ronald Shaw to publish the results of his thesis which were identical to those of Yang and Mills based on the same argument of Pauli. When years later people realised that Yang-Mills theories were the basis to describe the electroweak and strong interactions in the Standard Model (after the massless problem was solved by the Higgs mechanism and gluon confinement), Salam called the theory Yang-Mills-Shaw. But it was too late. 
\item How the V-A (vector minus axial vector) theory of Marshak and Sudarshan (followed by Gell-Mann and Feynman) played a key role in deciphering the weak interactions but originally contradicted four different experiments that ended up being wrong. 
\item How Weinberg (and the independent work by Salam) trying to understand strong interactions led him towards understanding the weak interactions and unifying them with the electromagnetic interactions in a 3-page paper \cite{Weinberg:1967tq} where he predicted neutral currents, the $W^\pm$ and $Z^0$ particles as well as the Standard Model Higgs. All these predictions were later confirmed experimentally. It is hard to find any written material in history carrying such an amount of information and successful predictions in such a  few number of words.
\item How the intuition of Bjorken and Feynman combined to extract the relevant information of the deep inelastic scattering experiments that determined the composite nature of protons and neutrons and finally are the best evidence for the existence of quarks.  
\item The fact that quarks for  several years were only considered mathematical objects (even by Gell-Mann himself) since they did not exist in liberty, but theorists and experimentalists were clever enough to find evidence for them and also for gluons despite the fact that they are confined within hadrons. 
\item The origin of colour (one of the most fundamental properties of the SM) just to address a problem with Pauli exclusion principle in the $\Omega^-$ (and other) particles. 
\item The importance of path integral techniques that allowed 't Hooft to prove the renormalisability of spontaneously broken gauge theories and how it suddenly gave credibility to the Weinberg-Salam model that had been totally ignored for 4 years.
\item The relevance of a simple minus sign that allowed Gross, Wilczek and Politzer to explain and understand the strong interactions (\emph{asymptotic freedom}). How several groups failed to identify it or recognise the importance and how the two groups finally agreed with this important sign.
\item The prediction of the charm quark based first on general symmetry arguments by Glashow and Marshak and later on a way to understand the suppression of  flavour changing neutral currents (GIM Mechanism of Sheldon Glashow, John Iliopoulos and Luciano Maiani)\index{GIM Mechanism} and subsequent discovery of charm in terms of the $J/\Psi$ particle. 
\item The fact that several groups missed the $J/\Psi$ discovery (one of the most surprising and exciting discoveries that helped to confirm the quark theory and particularly the existence of charm) but one of them (Lederman) was lucky enough to later find the upsilon particle leading to the unexpected discovery of the bottom quark. 
\item The different stages of trust in quantum field theory to describe elementary particles from the early attempts in the 1930s to excitement after QED in 1948 to almost rejection before asymptotic freedom and back to life again after that. 
\item How a very simple $\mathrm{SU}(3)$ extension  of the mixing ideas of Cabibbo for $\mathrm{SU}(2)$ by Kobayashi and Maskawa had the important information about CP violation in the Standard Model and how the whole scientific community was surprised and disappointed for Cabibbo to be left out of the Nobel prize.
\item How several groups simultaneously came up with the idea of the Higgs mechanism but failed to identify the importance for the weak interactions and did not even mention the Higgs particle (except only for Higgs but only apparently after the suggestion of the referee). 
\item The persistence for decades of one experimentalist (Ray Davis) and one theorist (John Bahcall) to insist that the solar neutrino problem (the fact that a much smaller number of neutrinos are detected than predicted) was a fundamental rather than astrophysical issue. This can be solved by noticing that the different types of neutrinos can oscillate converting from one type to another (as we will see in this course) and the ones that are produced in the Sun are changed to another kind of neutrino while leaving the core of the Sun explaining why less are detected when they arrive on Earth.
\item The impressive international efforts towards the search and final discovery of the top quark, the W,Z and Higgs particles, etc.
\item Before the Higgs discovery, how the CERN international efforts gave rise, essentially by accident,  to the world-wide-web (WWW) that made internet accessible to the general public, one of the most influential developments  of the past decades, confirming the importance to study basic science, even for its impact on spin-off applications.
\end{itemize}
These are just a few of the highlights for this  beautiful set of events that illustrate, as well as anything else, how science, as a human endeavour,  is made and how there is usually a large amount of confusion before things get properly  understood and then written in a logical way in textbooks and lecture notes like these. This history is also a lesson for current times in which there are many open questions and no clear guidance for the future rather than the knowledge acquired that led to the Standard Model and some open questions. Knowing how scientists have addressed and solved questions of the past is a good guidance for how to address new challenges. 

Even though this historical tour was brief and many of the concepts are unknown to the students, we would like to encourage the students to come back to some of these events after we have discussed the fundamental aspects of the Standard Model in a systematic rather than historical way to appreciate the greatness of the achievements and the combination of consistency, rigour, creativity, imagination and luck that is needed  to do fundamental research. As Weinberg emphasised, this is one of his four gold rules for scientists: \emph{you can get great satisfaction by recognising that your work in science is a part of history} \cite{weinberggold}.

\newpage

\section{Summary and Motivation}

Let us start with a first overall glimpse at the Standard Model just to introduce the concepts that will be developed during the subsequent lectures.

\subsection{A Brief Introduction to the Standard Model}

\begin{equ}[Definition:]
The Standard Model (SM) is a construction (or model or theory) that describes all the known elementary particles and their interactions in terms of relativistic quantum field theories.
\end{equ}

\textbf{Ingredients:}
\begin{enumerate}[label=(\arabic*)]
\item \emph{Spacetime}. The spacetime is $(3+1)$-dimensional \emph{Minkowski spacetime} with (global) symmetry group
\begin{equation}
G_{\text{gl}}=\mathbb{R}^{3,1}\rtimes \mathrm {O} (3,1)
\end{equation}
that is the semi-direct product of spacetime translations and the Lorentz group, corresponding to the \emph{Poincar{\'e} group}\index{Poincar{\'e} group} of special relativity.
\item \emph{Matter}. The particle content can be classified by the spin $s$ (or helicity), i.e., there are the \emph{Higgs} $H$ with $s=0$ as well as $3$ families of \emph{quarks} and \emph{leptons} with $s=1/2$, see also table~\ref{tab:SM}.
\item \emph{Interactions}. The interactions are given by $3$ gauge interactions with associated gauge bosons of spin $s=1$ and, in general, also gravitational interactions transmitted by a particle of spin $s=2$ known as the \emph{graviton}. The gauge forces are encoded in the gauge (or local) symmetry group
\begin{equation}\label{eq:GaugeGroupSM} 
G_{\text{SM}}=\underbrace{\underbrace{\mathrm{SU}(3)_{C}}_{\text{color}}}_{\text{strong}}\times \underbrace{\underbrace{\mathrm{SU}(2)_{L}}_{\text{left}}\times \underbrace{\mathrm{U}(1)_{Y}}_{\text{hypercharge}}}_{\text{electroweak}}
\end{equation}
where the subindex $C$ refers to \emph{colour} with $\mathrm{SU}(3)_C$ determining the strong interactions. The strong force binds
quarks into nucleons and nucleons into nuclei; the carrier of the strong force
is appropriately called the \emph{gluon}. The $L$ in $\mathrm{SU}(2)_L$ refers to \emph{left-handed} in the sense that weak interactions only act on left-handed particles. Finally $Y$ refers to \emph{hypercharge}. The gauge group in \eqref{eq:GaugeGroupSM} is broken by the Higgs boson through a non-zero vacuum expectation value, to a subgroup, namely
\begin{equation}
G_{\text{SM}}\xrightarrow{\,\;\text{SSB}\;\,} \underbrace{\mathrm{SU}(3)_{C}}_{\text{QCD}}\times \underbrace{\mathrm{U}(1)_{EM}}_{\text{QED}}\, .
\end{equation}
In the process referred to as \emph{spontaneous symmetry breaking} (SSB), the corresponding gauge bosons of the broken group ($W^{\pm}$ and $Z^{0}$) receive a mass, but there is also a remaining massless boson corresponding to the unbroken $\mathrm{U}(1)_{EM}$ with $EM$ standing for \emph{electromagnetic}.
This is the familiar photon of the electromagnetic interactions.
To reiterate, after the breaking of the symmetry, only the gluons and photons remain massless. Photons are free to move but gluons together with quarks are confined within the particles of strong interactions such as protons and neutrons.
The representations\footnote{We work in a particular representation where $\alpha\in \mathrm{U}(1)_{Y}$ acts on $\psi\in\bC$ in such a way that $\psi\raw\alpha^{6iy}\psi$, i.e., weak hypercharges appear in integer multiples of $1/6$. Keep in mind that different definitions are commonly used in the literature!} of the particles involved are summarised in table~\ref{tab:SM}. 

\item {\it Three families}. For the quarks and leptons, there are $3$ distinct families coming with the same copies of the representation:
\begin{equation*}
\text{Leptons:} \quad \begin{pmatrix}\nu_e \\ e \end{pmatrix},
\begin{pmatrix}\nu_\mu \\ \mu \end{pmatrix},
\begin{pmatrix}\nu_\tau \\ \tau \end{pmatrix} \qquad \qquad \text{Quarks:} \quad  \begin{pmatrix} u \\ d\end{pmatrix}, 
\begin{pmatrix} c \\ s\end{pmatrix}, 
\begin{pmatrix} t \\ b\end{pmatrix} \, .
\end{equation*}
Only the first family (with electron, its neutrino and up and down quarks) are enough to make the matter we know. The second (muon, its neutrino, charm and strange quarks) and third (tau-lepton, its neutrino, top and bottom quarks) are more massive and the corresponding particles are unstable having the particles of the first family as end results of their decay.

\end{enumerate}

\begin{table}
\centering
\begin{tabular}{|c|c|c|c|}
\hline 
Name & Label & {\small $\mathrm{SU}(3)_{C}$, $\mathrm{SU}(2)_{L}$, $\mathrm{U}(1)_{Y}$} &Spin/Helicity\\
\hline 
\hline
  & $Q^i_{L}=\bigl (
 u^i_{L},
 d^i_{L}
  \bigl )$  & $\bigl (\mathbf{3},\mathbf{2},+\frac{1}{6}\bigl )$ &$\frac{1}{2}$\\ [0.6ex]
  Quarks & $u^i_{R}$ &$\bigl (\bar{\mathbf{3}},\mathbf{1},\frac{2}{3}\bigl )$&$\frac{1}{2}$ \\ [0.6ex]
   & $d^i_{R}$  & $\bigl (\bar{\mathbf{3}},\mathbf{1},-\frac{1}{3}\bigl )$&$\frac{1}{2}$ \\ [1ex]
\hline 
 & $L^i_{L}=\bigl (
 \nu^i_{L},
 e^i_{L}
  \bigl )$ &   $\bigl (\mathbf{1},\mathbf{2},-\frac{1}{2}\bigl )$&$\frac{1}{2}$ \\  [0.6ex]
Leptons & $e^i_{R}$ & $(\mathbf{1},\mathbf{1},-1)$ &$\frac{1}{2}$\\  [0.6ex]
  & $\nu^{i*}_{R}$ & $(\mathbf{1},\mathbf{1},0)$ &$\frac{1}{2}$\\  [1ex]
\hline 
Higgs  & $H$& $\bigl (\mathbf{1},\mathbf{2},+\frac{1}{2}\bigl )$ &$0$\\  [0.6ex]
\hline 
Gluons &$g_{\alpha}$ & $\bigl (\mathbf{8},\mathbf{1},0\bigl )$ &$1$\\  [0.6ex]
$W$/$Z$-Bosons&$W^{\pm}, Z^{0}$ & $\bigl (\mathbf{1},\mathbf{3},0\bigl )$ &$1$\\  [0.4ex]
Photon&$\gamma$  & $\bigl (\mathbf{1},\mathbf{1},0\bigl )$ &$1$\\  [0.6ex]
\hline 
Graviton$^*$  & $h_{\mu\nu}$& $\bigl (\mathbf{1},\mathbf{1},0\bigl )$ &$2$\\  [0.5ex]
\hline
\end{tabular} 
\caption{Particle content of the Standard Model and the corresponding group representations.  The right-handed neutrino and the graviton are included here for completeness with the understanding that the couplings of the graviton to all other particles can be studied as long as the energies are small enough in terms of an effective QFT. }\label{tab:SM} 
\end{table}

\

It is remarkable that these simple ingredients are enough to account for the structure of the Universe as we know it including every single experience and measurement we make. There are some comments in place:
\begin{itemize}
\item \emph{Chirality}\index{Chirality}. Since the right-handed quarks $u_{R}$ and $d_{R}$ transform under the trivial representation of $\mathrm{SU}(2)_{L}$ in table~\ref{tab:SM}, that is to say they are $\mathrm{SU}(2)_{L}$-\emph{singlets}, they do not feel the $\mathrm{SU}(2)_{L}$-interactions. This is why we call the Standard Model a \emph{chiral} gauge theory.
Physically, this implies that left- and right-handed fermions feel certain gauge interactions differently, i.e., they couple non-democratically to the mediators (gauge fields) of a given force.
Mathematically speaking, left- and right-handed fermions transform in different $\mathrm{SU}(2)_{L}$-representations. Thus, the weak interaction is not parity invariant under exchange of left- and right-handed particles.
\item \emph{Charge quantisation}\index{Charge quantisation}. The electric charge is defined as
\begin{equation}
Q=T_{3}+Y
\end{equation}
where $T_{3}$ is the third generator of $\mathrm{SU}(2)_{L}$ which is the diagonal matrix with entries $(1/2,-1/2)$ and $Y$ the hypercharge under $\mathrm{U}(1)_{Y}$. For instance to compute the electric charge of the left-handed electron we observe that its hypercharge is $Y=-1/2$ and its value of $T_3$ is $-1/2$ giving $Q=-1$. For the right-handed electron the corresponding $T_3$ value is zero because it is a singlet and then charge and hypercharge are the same (here $+1$ for the positron). Computing the electric charges of quarks give multiples of $1/3$ instead of integers as we are familiar for electrons and protons.

\item \emph{Consistency conditions}. We observe that the assignment of these numbers such as hypercharge and the different representations of the Standard Model particles is not arbitrary. For instance it is easy to verify the following conditions for the hypercharges
\begin{equation}
\sum_{Left}\, Y-\sum_{Right}\, Y=\sum_{Left}\, Y^{3}-\sum_{Right}\, Y^3=0\, .
\end{equation}
Also the total number of particles transforming as a $\mathbf{3}$ of $\mathrm{SU}(3)_C$ equals the number of $\bar{\mathbf{3}}\, $ ($\#\mathbf{3}=\#\bar{\mathbf{3}}$) and the total number of $\mathrm{SU}(2)_L$ doublets  ($\#\mathbf{2}$) is even. Any modification of these numbers would render the theory mathematically inconsistent.  
This will be crucial in ensuring \emph{anomaly cancellation}\index{Anomaly cancellation} within the Standard Model. If the Standard Model was not chiral, these conditions would be trivially satisfied. It is the chiral structure of the Standard Model that makes it subject to potential inconsistencies and therefore makes it more interesting when they are satisfied. 
\item \emph{Coleman-Mandula theorem}\index{Coleman-Mandula theorem} \cite{Coleman:1967ad}. The total symmetry of the Standard Model is given by a direct product between a spacetime and an internal symmetry 
\begin{equation}
\text{spacetime}\otimes\text{internal gauge}=(\mathbb{R}^{3,1}\rtimes \mathrm {SO} (3,1))\otimes (\mathrm{SU}(3)_{C}\times \mathrm{SU}(2)_{L}\times \mathrm{U}(1)_{Y})\, .
\end{equation}
The Coleman-Mandula  theorem states that this structure is the most general\footnote{In fact, there is a loophole for this theorem,  the only possibility of extending the Poincar{\'e} group is to introduce \emph{supersymmetry}\index{Supersymmetry} which is ensured by the \emph{Haag-Lopuszanski-Sohnius theorem}\index{Haag-Lopuszanski-Sohnius theorem} \cite{Haag:1974qh}, see the corresponding \href{http://www.damtp.cam.ac.uk/user/fq201/susynotes.pdf}{Part III SUSY lecture notes} \cite{Quevedo:2010ui} for details.} for the full symmetry group, i.e., a direct product of the Poincar{\'e} group and an internal (gauge) group.
\item \emph{Gravity as an effective field theory}. We can treat gravity only as  what is called an effective QFT (EFT) by working with energies $E$ well below the \emph{Planck scale}\index{Planck scale}:
\begin{equation}
E\ll M_{P}=\sqrt{\dfrac{\hbar c}{G}}\sim 10^{18}\text{GeV}\, .
\end{equation}
At energies $E\sim M_{P}$, quantum effects of gravity become important and the EFT has to be replaced by a more fundamental theory that is ultra-violet complete. But for energies well  below $E\ll M_{P}$ working with quantum aspects of gravity as an EFT are predictable and reliable.
We will have to say more about the role of gravitational interactions within the Standard Model later in the course.
\item {\it Accidental symmetries}. There are  accidental symmetries known as \emph{Baryon number} $B$ and \emph{Lepton number} $L$. That is, the total number of baryons, such as the neutron and proton, and the total number of leptons such as the electron and neutrino are conserved in every interaction.
\item {\it Approximate symmetries}. The three families of quarks and leptons in which the members
 of each family behave the same as the other families except that the particles are heavier for each generation (e.g., the muon is like a heavier copy of the electron, the top quark of the up quark, etc.) implies that there are approximate symmetries known as \emph{flavour symmetries} such as $\mathrm{SU}(3)_{f}$ known as the \emph{eightfold way}\index{Eightfold way}, see Fig.~\ref{fig:EFW}. This flavour $\mathrm{SU}(3)_{f}$ should not  be confused with the colour $\mathrm{SU}(3)_C$ which is the symmetry describing the strong interactions. 
\item {\it Phases of the Standard Model}. The Standard Model is relatively simple, although not the simplest model we can imagine. Actually, it is rich enough to illustrate the  $3$ main phases of gauge theories: The {\it Coulomb phase}\index{Coulomb phase}  for $\mathrm{U}(1)_{EM}$ meaning that the corresponding gauge boson, the photon, moves freely; the {\it confining phase}\index{Confining phase} for $\mathrm{SU}(3)_{C}$ meaning that the interactions are so strong that the corresponding gauge bosons, the gluons, and the quarks are confined within hadrons; and the  {\it Higgs phase}\index{Higgs phase} for the weak interactions in  $\mathrm{SU}(2)_{L}\times \mathrm{U}(1)_{Y}$ meaning that the corresponding force is short range since the gauge bosons $W,Z$ are heavy after symmetry breaking. 
\end{itemize}

\subsection{Motivation for the Standard Model}

Why do we need to learn about the Standard Model?
\begin{itemize}
\item \emph{It is fundamental.} This is currently the most fundamental theory in science describing the nature and interactions of the building blocks of nature. 
\item \emph{Robustness}. It is based on self-consistent elegant and robust mathematical principles based on the two fundamental theories of physics, namely quantum mechanics and relativity, that can be used to explain the observable world from basic principles. Symmetries and their whole mathematical structure play a crucial role in the implementation of these basic principles.
\item \emph{It is true!} The SM is one of the greatest achievements in science history.
It is mathematically consistent and agrees with all experimental tests so far. In fact, many experiments have spectacularly confirmed predictions such as the existence of the $W^{\pm}$, $Z^{0}$ bosons, the top quarks, the Higgs particle, etc. On top of that, the observable physical quantities have been measured with unprecedented precision. For instance, the anomalous magnetic dipole moment of the electron is
\begin{equation}
a=\dfrac{g-2}{2}=(1159.65218091\pm0.00000026)\times 10^{-6}
\end{equation}
The agreement between theory and experiment is within one  part in a trillion, which is probably the best precision test of any scientific theory. Also, the fine structure constant (measured at small energies $E\ll 10^{2}$GeV)
\begin{equation}
\alpha^{-1}=\dfrac{\hbar c}{e^{2}}=137.035999084\,(21)
\end{equation}
has been tested to one part in a billion. These are only a few of the many experimental tests that the Standard Model has passed successfully over many years and the experimental results have been reproduced by independent experiments many times adding to the robustness of the theory.
\item \emph{It is the best test of validity of QFT.} Relativistic QFT is a very general framework, but with only a handful of general experimental predictions such as the existence of anti-particles, the CPT theorem, the spin-statistics connection and the running of couplings. The main reason that QFT is trusted is through its big success in describing the Standard Model. Only one in an infinite number of possible QFTs that happens to describe our world. 
\item \emph{Cosmology}. The Standard Model is the main theoretical framework to  successfully describe the early history of the universe known so far. The great success of the past decades to test cosmological theories with strong precision uses the properties of the Standard Model and provides a further experimental way to test the Standard Model. For instance, the excellent agreement between theory and observation on the abundance of the different elements coming from the big-bang for light elements and from the core of stars for the heaviest can all be traced to properties of the Standard Model. Furthermore,  the fact that no more than three light neutrinos are expected fits extraordinarily well with precision tests of the Standard Model as well as cosmological observations of the cosmic microwave background that put stringent bounds on what is known as \emph{dark radiation}\index{Dark radiation}, essentially ruling out further neutrino species.
\item \emph{It is incomplete!} We know that the Standard Model cannot be the final theory. For example, there is no description of some key observational facts such as the explanation of baryogenesis (why we are made up of baryons and not anti-baryons), the identity of dark matter for which there is overwhelming evidence, the explanation of the nature and origin of dark energy, responsible for the current acceleration of the universe and most importantly a fully quantum description of gravity. But whatever physics will replace it, the SM will remain as the valid description of the world at low energies and understanding the basic principles behind the Standard Model is a prerequisite to look for alternative theories to modify it and/or generalise it.
We summarise some of these directions in chapter~\ref{chap:probs}.
\end{itemize}

\newpage

\section{Outline for the lectures}


As is common in many textbooks, the structure of these lecture notes diverges from the historical development of the Standard Model summarised earlier. The primary reason for this is pedagogical clarity, which at times may obscure some of the unexpected discoveries and existential challenges the theory faced. Ultimately, the Standard Model's final form is the result of numerous breakthroughs and innovative ideas. While these lectures often present the definitive answers to foundational questions in particle physics upfront, it is important to remember that many physicists wrestled with these same issues for years -- this is the nature of research. Therefore, students should not be discouraged by the seemingly ad-hoc choices made in presenting some of the material.

\


In chapter~\ref{chap:stsym}, we revisit Wigner's classification of elementary one-particle states through unitary irreducible representations of the Poincar\'e group. We begin with an overview of the Poincar\'e algebra and its representations, giving special attention to spinor representations, where the concepts of left- and right-handedness -- crucial for chiral gauge theories -- are introduced. Finally, we explore in detail how both massless and massive elementary particles are described through Poincar\'e group representations.

After this classification of all ``relevant'' unitary representations of the Poincar{\'e} group that potentially play a role in nature in form of elementary one-particle states,
chapter~\ref{chap:fields} is concerned with the question: what do we do with them?
Traditionally, we pick one of two options by either building an \emph{on-shell} formalism for amplitudes (see \cite{Eden:1966dnq,Elvang:2013cua,Benincasa:2013faa,Cheung:2017pzi} for excellent reviews on this approach) or we introduce \emph{off-shell} objects that transform covariantly under Lorentz transformations.

In these lectures we pick the second option.
To describe interactions among many particle states,
we formulate axioms for a proper quantum theory in section~\ref{sec:particlestofields} such as unitarity and locality.
We define \emph{fields} as the superposition of one-particle states of fixed mass and spin/helicity.
Subsequently,
we can formulate an action principle for these fields which allows to systematically introduce interactions among particles in terms of operators in the Lagrangian, while producing the correct on-shell conditions for corresponding particles in the absence of interactions.
Next, we revisit standard arguments for organising physics by energy scales in section~\ref{sec:EFTs} which will be crucial in describing physical phenomena at low energies.
The organisation of interactions and their relevance at low energies are both heavily determined by the notion of symmetries.
We summarise the various types of such symmetries in section~\ref{sec:symmetriesQFT} most of which will play important roles in the Standard Model.

The advent of fields comes however also at a cost.
Among others, it introduces unphysical auxiliary degrees of freedom for particles with spin/helicity $\geq  1$.
This ultimately demands gauge redundancies as will be explained in chapter~\ref{chap:sym}.
The story is quite simple: to describe the interactions of many species of massless fields of helicity $1$,
the operators in the action have to follow certain rules dictated by an algebra underlying certain Lie groups.
This can in fact be derived from only requiring Lorentz invariance and unitarity of scattering amplitudes (in form of Ward identities) as detailed in App.~\ref{app:compton_sQED}.
These so-called \emph{gauge theories} play an outstanding role in the Standard Model since they describe the dynamics of the mediators of elementary forces like the photon in QED.
We collect their properties relevant for these lectures and also review soft theorems making statements about e.g. charge conservation in Sect.~\ref{sec:softtheorems}.

The next chapter is concerned with the breaking of symmetries.
We distinguish mainly two ways in which symmetries get broken, namely \emph{explicitly} or \emph{spontaneously}.
In the former case, a symmetry that exists under certain assumptions gets spoiled once other interactions are allowed.
In contrast, the spontaneous breaking of a symmetry just means that the ground state of a theory does not respect the full symmetry group, but merely some sub-group.
In chapter~\ref{chap:ssb}, we explain how this idea solves unitary problems in EFTs with massive spin-1 bosons.
In particular, the spontaneous breakdown of gauge symmetries will be key in understanding why the weak interactions are short ranged as well as how matter fields like the electron receive their masses in the Standard Model.

Yet another important aspect of symmetries is their manifestation in quantum theories.
The notion of anomalies introduced in section~\ref{sec:sym_gauge_anom} explains which classical symmetries can be promoted to quantum theories.
In the presence of non-vanishing anomalies, global symmetries are explicitly broken in the quantum theory, whereas for gauge (=local) symmetries the theory is rendered inconsistent.
The discussion of anomalies in the Standard Model will be deferred to Sect.~\ref{sec:anomalies_SM} once the full content of the Standard Model has been described.

A first step towards building up the Standard Model is taken in the subsequent chapter. There, we introduce the \emph{Electroweak Theory} as the unification of the weak interactions mediated by massive spin-1 bosons $W^{\pm},Z^{0}$ and the electromagnetic force due to the photon.
We will see the Higgs mechanism at play giving masses to $W^{\pm},Z^{0}$ as well as the matter fields.
Chapter~\ref{chap:ew} will also analyse in great detail how the Higgs particle cancels the dangerous contributions in the scattering of massive vector bosons, thereby preventing the otherwise expected loss of perturbative unitarity.
We explicitly write down the interactions of matter fields with the Higgs and vector bosons.
We show that in a basis of mass eigenstates some of the interactions are not flavour diagonal which leads us to the CKM mixing matrix.
A significant part of chapter~\ref{chap:ew} is devoted to neutrino physics which remains a very active field of research on the Standard Model.
For example, we illustrate how the decay of the $Z$-boson into lepton-antilepton pairs constrains the number of light species, i.e., neutrinos in Sect.~\ref{sec:fermioncouplingsEW}.
Lastly, we show how the electroweak interactions can be approximated by the $4$-Fermi theory at low enough energies.
As an application of these results, we compute the decay $\mu^{-}\rightarrow e^{-}+\nu_{\mu}+\bar{\nu}_{e}$ and compare our theoretical results with experimental measurements.

Chapter~\ref{chap:qcd} concerns the strong interactions which is theoretically described by \emph{Quantum Chromodynamics} (QCD).
At its heart, QCD is an $\mathrm{SU}(3)$ gauge theory providing additional charges, so-called \emph{colours}, for the quarks.
Ultimately, this was key in explaining the classification of hadrons through the eightfold way in Fig.~\ref{fig:EFW}.
The arguably most important property of QCD is however asymptotic freedom which ensures that the interactions becomes strong at low energies, while at the same time being well behaved in the high energy limit.
Ultimately, this result explains e.g. why we cannot see quarks and gluons in isolation -- the farther away a quark from a hadron, the stronger the interactions.
The last part of chapter~\ref{chap:qcd} discusses chiral Lagrangians obtained when treating the light quarks as effectively massless. In this way, the resulting approximate symmetry helps us classifying hadrons, while its breakdown quantifies the mass hierarchies observed in e.g. the baryon octet in Fig.~\ref{fig:EFW}.

In chapter~\ref{chap:SM},
we provide a short summary of parameters in the Standard Model with a particular focus on the $\Theta$-term and its relevance for quark masses.
As it turns out, this term combines in an interesting way the weak and strong interactions descriptions within the Standard Model.
Furthermore, we study potential anomalies of local and global symmetries in the Standard Model which also combine the strong and electroweak sectors of the model. We show explicitly that, despite being a chiral theory, quite remarkably the Standard Model is indeed free of gauge anomalies and therefore quantum mechanically consistent in a highly non-trivial way, while there are harmless anomalous global symmetries like lepton and baryon number.

The subsequent chapter addresses some of the open questions in the Standard Model and possible extensions.
First, we list the major open problems within the Standard Model and classify them according to their nature.
In section~\ref{sec:BSM}, we provide a short account of physics Beyond the Standard Model such as supersymmetry, grand unification or axions. We emphasise also bottom-up approaches. In this sense the power of effective field theories again play a major role in order to learn from a model independent way what may lie beyond the Standard Model.

The final chapter~\ref{chap:final_remarks} summarises the most important concepts derived over the course of these lectures and provides some concluding remarks.


\newpage


\def\beq{\begin{equation}}
\def\eeq{\end{equation}}

\newcommand{\vecb}{\left(\begin{array}{c}}
\newcommand{\vece}{\end{array}\right)}
\newcommand{\ccb}{\left(\begin{array}{cc}}
\newcommand{\cce}{\end{array}\right)}
\newcommand{\cccb}{\left(\begin{array}{ccc}}
\newcommand{\ccce}{\end{array}\right)}
\newcommand{\ccccb}{\left(\begin{array}{cccc}}
\newcommand{\cccce}{\end{array}\right)}
\newcommand{\cccccb}{\left(\begin{array}{ccccc}}
\newcommand{\ccccce}{\end{array}\right)}


\newcommand{\vph}{\varphi}
\newcommand{\vth}{\vartheta}
\newcommand{\ve}{\vec}
\newcommand{\pa}{\partial}
\newcommand{\kb}{k_{B}}
\newcommand{\hb}{\hbar}
\newcommand{\al}{\alpha}
\newcommand{\be}{\beta}
\newcommand{\ga}{\gamma}
\newcommand{\de}{\delta}
\newcommand{\ep}{\epsilon}
\newcommand{\vep}{\varepsilon}
\newcommand{\la}{\lambda}
\newcommand{\ka}{\kappa}
\newcommand{\om}{\omega}
\newcommand{\Ga}{\Gamma}
\newcommand{\De}{\Delta}
\newcommand{\Si}{\Sigma}
\newcommand{\La}{\Lambda}
\newcommand{\Om}{\Omega}
\newcommand{\Th}{\Theta}
\newcommand{\Psib}{\overline{\Psi}}
\newcommand{\mto}{\rightarrow}
\newcommand{\te}{\textrm}
\newcommand{\lap}{{\cal 4}} 
\newcommand{\inn}{ \ {\cal 3} \ } 
\newcommand{\eq}{  =  }
\newcommand{\co}{\ , \ \ \ \ \ \ }
\newcommand{\trf}{ \ \ \longmapsto \ \ }
\newcommand{\geg}{ \ \ \longrightarrow \ \ }
\newcommand{\thb}{\bar{\theta}}
\newcommand{\vac}{|\te{vac} \rangle}

\chapter{\bf Spacetime Symmetries}
\label{chap:stsym}

\vspace{0.5cm}
\begin{equ}[Importance]
{\it  If it were not for these symmetries, the work of science would have to be redone in every new laboratory and in every passing moment.}\\

\rightline{\it Steven Weinberg}
\end{equ}
\vspace{0.5cm}

In this chapter, we review basic techniques for constructing suitable representations of the Poincar\'e group -- the symmetry group of (Minkowski) spacetime\footnote{QFT in curved spacetime is a far more difficult endeavour, see e.g. Prof. Enrico Pajer's notes on field theory in cosmology (see this \href{https://www.damtp.cam.ac.uk/user/ep551/FTC.html}{link}) and books like \cite{parker2009quantum}.} -- which defines Special Relativity and is relevant for constructing the Standard Model.
We begin with a general discussion of the Poincar\'e algebra and its properties before introducing spinor representations of the Lorentz group $\mathrm{SO}(3,1)$. Subsequently, we will detail Wigner's classification of irreducible representations of the Poincar\'e group which define for us elementary particles.
We discuss in detail how these particles have to transform under discrete spacetime transformations which allows us to count all of the relevant polarisation states.
Among others, we will find that massless particles with helicity $\geq 0$ have always just two degrees of freedom which will severely constrain the field theories to be studied in the reminder of these lectures.

\vfill

\newpage

\section{Poincar\'e symmetry and spinors}
\label{sec:PoincareSymmetryAndSpinors}

For the vast majority of this lecture,
we will be interested in describing relativistic processes involving particles moving in $(3+1)$-dimensional Minkowski space $\bR^{3,1}$.
Any theory describing such phenomena must respect the symmetry inherited from the spacetime geometry -- heuristically, physical processes should not depend on the observer's initial frame.
This is the \emph{principle of relativity} stating that the laws of physics are the same in all viable frames of reference.
For particle physics in $\bR^{3,1}$, the spacetime symmetry group in question is the \emph{Poincar{\'e} group} $\cP(3,1)$.

The Poincar{\'e} group\index{Poincar{\'e} group $\cP(3,1)$} corresponds to the basic symmetries of special relativity, it acts on the Minkowski spacetime coordinates $x^{\mu}$ via
\begin{equation}x^{\mu} \ \ \mapsto \ \ x'^{\mu} \eq \underbrace{\La^{\mu}\,_{\nu}}_{\te{Lorentz}} x^{\nu} \ + \ \underbrace{a^{\mu}}_{\te{translation}}\kom \mu,\nu=0,1,2,3\, .
\end{equation}
It is sometimes convenient to use a shorthand notation $\lbrace\cdot |\cdot\rbrace$ for such a transformation\index{Poincar{\'e} transformations} where
\begin{equation}
 x'^{\mu} \eq\lbrace\Lambda|a\rbrace\, x^{\mu}\equiv \La^{\mu}\,_{\nu}\, x^{\nu}+a^{\mu}\, .
\end{equation}
Formally, the Poincar{\'e} group $\cP(3,1)$ or $\mathrm{ISO}(3,1)$ is the isometry group of Minkowski spacetime. It is a semidirect product of spacetime translations and the transformations corresponding to the Lorentz group\index{Lorentz group} of special relativity
\begin{equation}\label{eq:DefPoincareSymSp} 
\cP(3,1)=\bR^{3,1}\rtimes\mathrm{O}(3,1)\, ,
\end{equation}
which is just a fancy way of saying that $\cP$ leaves $\bR^{3,1}$ invariant and every Poincar{\'e} transformation can be decomposed into the product of the form
\begin{equation}
\lbrace\Lambda|a\rbrace=\lbrace\mathds{1}_{4\times 4}|a\rbrace\lbrace\Lambda|0\rbrace\, .
\end{equation}

The Lorentz transformations belong to the orthogonal $\mathrm{O}(3,1)$ group that leaves the metric tensor
\begin{equation}
\eta_{\mu \nu} = \te{diag}(1, \ -1 , \ -1 , \ -1)
\end{equation}
in the line element
\begin{equation}
\dif s^{2}=\eta_{\mu\nu}\dif x^{\mu}\dif x^{\nu}
\end{equation}
invariant, i.e.,
\begin{equation}\label{eq:InvMetricLorTraf} 
    \La^\mu\,_\rho \eta_{\mu\nu}\La^\nu\,_\sigma=\eta_{\rho\sigma} \qquad {\rm or} \qquad \La^{T} \, \eta \, \La \eq \eta\, .
\end{equation}
From this equation we can easily see that
\begin{equation} \det \La=\pm 1\end{equation}
and taking the $00$ component
\begin{equation}\left(\La^0\,_0\right)^2-\left(\La^1\,_0\right)^2-\left(\La^2\,_0\right)^2-\left(\La^3\,_0\right)^2=1\qquad \Longrightarrow \qquad |\La^0\,_0|\geq 1\, .
\end{equation}

Therefore the Lorentz group has 4 disconnected components according to the signs of $\det \La$ and $\La^0\,_0$.
We will mostly discuss those transformations $\La$ connected to the identity, i.e., the \emph{proper orthochronous group}\index{Lorentz group!proper}\index{Lorentz group!orthochronous} $\mathrm{SO}(3,1)^{\uparrow}$ for which $\det \Lambda=1$ (proper) and $\Lambda^0\,_0\geq 1$ (orthochronous).
All $\mathrm{O}(3,1)$ transformations can be obtained by combining the  $\mathrm{SO}(3,1)^{\uparrow}$ transformations with:
\begin{equation}
\left\{\mathds{1},\, \La_P,\, \La_T,\, \La_{PT}\right\}
\end{equation}
where 
\begin{itemize}
\item $\mathds{1}$ is the identity matrix,
\item $\La_P =\te{diag}(1,-1,-1,-1)$ is the \emph{parity}\index{Parity} transformation,
\item $\La_T=\te{diag}(-1,1,1,1)$ is \emph{time reversal}\index{Time reversal}, and
\item $\La_{PT}=\La_P\times \La_T$ is combined parity and time reversal.
\end{itemize} 
These four elements form a group known as \emph{Klein's four-group}\index{Klein's four-group}. From now on we will concentrate on those transformations connected to the identity and drop the arrow on $\mathrm{SO}(3,1)$ for simplifying the notation. For the same reason, we work from now on with the proper orthochronous Poincar{\'e} group and simply write $\cP(3,1)\equiv \cP^{\uparrow}_{+}=\bR^{3,1}\rtimes\mathrm{SO}(3,1)^{\uparrow}$.
 
 \subsection{The Poincar\'e Algebra}
 
 Let us consider infinitesimal Poincar{\'e} transformations\index{Poincar{\'e} transformations!Infinitesimal} $\lbrace\Lambda |a\rbrace\in\cP(3,1)$ for which
 \begin{equation} 
 \La^\mu\,_\nu=\delta^\mu\,_\nu+\omega^\mu\,_\nu\, ; \qquad a^\mu=\epsilon^\mu\qquad \omega^\mu\,_\nu,\, \epsilon^\mu\ll 1\, .
\end{equation}
Plugging this back in \eqref{eq:InvMetricLorTraf}, we find
\begin{align}
\La^\mu\,_\rho \eta_{\mu\nu}\La^\nu\,_\sigma= \eta_{\rho\sigma}+\omega_{\sigma\rho}+\omega_{\rho\sigma}+\cO(\omega^{2})\overset{!}{=}\eta_{\rho\sigma}\, .
\end{align}
To linear order, we deduce that
\begin{equation}
\omega_{\sigma\rho}=-\omega_{\rho\sigma}
\end{equation}
is an anti-symmetric tensor which has $6$ free parameters. Together with the $4$ translations $\epsilon^{\mu}$, a general Poincar\'e transformations must have $10$ parameters. This is the dimensionality of the Poincar\'e  group.
 
 In order to determine the algebra we can exponentiate the group elements. We will do this considering the action of operators acting on the Hilbert space $\cH$ relevant in quantum mechanics. A Poincar{\'e} transformation will be represented by a unitary operator $U(\La,a)$ acting on the Hilbert space vectors $|\Psi\rangle\in\cH$
\begin{equation}
|\Psi\rangle\rightarrow U(\La,a) |\Psi\rangle\qquad U=U^\dagger\, .
\end{equation}
Near the identity, we can expand to linear order
\begin{equation}
U(1+\omega,\epsilon)=\mathds{1}\, -\, \frac{\I}{2}\omega_{\mu\nu} M^{\mu\nu}+i\epsilon_\mu P^\mu 
\end{equation}
where  $M^{\mu\nu}=-M^{\nu\mu}$ and $P^\mu$ are the generators of the group. Since $U$ is unitary, both $M^{\mu\nu}$ and $P^\mu$ are Hermitian. As usual in group theory,
the above can be used to determine the algebra satisfied by the generators.
 
 First, since translations commute, their generators also commute with each other
\begin{equation} 
\left[P^\mu\, ,\, P^\nu \right] =0\, .
\end{equation}
Let us now consider the commutator $[P^\sigma, M^{\mu\nu}]$ by analysing how $P^\sigma$ transforms under Lorentz transformations.  For this we can consider the dual role of $P^\mu$. On the one hand, it is a vector that should transform (to leading order in $\omega_{\mu\nu}$) as
 \begin{align}
 P^\sigma&\rightarrow \La^{\sigma}\,_\rho \, P^\rho \nn\\
 &= \left(\delta^\sigma\,_\rho+\omega^\sigma\,_\rho\right) P^\rho\nn\\
 &= P^\sigma+\frac{1}{2}\left(\omega_{\alpha \rho}-\omega_{\rho\alpha}\right)\eta^{\sigma\alpha} P^\rho\nn\\
 &= P^\sigma+\frac{1}{2}\omega_{\alpha \rho}\left(\eta^{\sigma\alpha} P^\rho-\eta^{\sigma\rho} P^\alpha\right)
 \end{align} 
On the other hand, $P^\sigma$ as an operator transforms as
 \begin{align} 
 P^\sigma&\rightarrow U^\dagger P^\sigma U \nn\\
 &= \left(\mathds{1}+\frac{\I}{2}\omega_{\mu\nu}M^{\mu\nu}\right) P^\sigma \left(\mathds{1}-\frac{\I}{2}\omega_{\mu\nu}M^{\mu\nu}\right) \nn\\
 &= P^\sigma-\frac{\I}{2}\omega_{\mu\nu}\left(P^\sigma M^{\mu\nu}- M^{\mu\nu} P^\sigma\right)
\end{align}
Comparing both expressions we find the commutator $[P^\sigma, M^{\mu\nu}]$:
  \begin{equation}
  \bigl[P^\sigma\, , \, M^{\mu \nu} \bigr]  =  -\I \, \bigl(P^{\mu} \, \eta^{\nu \sigma} \ - \ P^{\nu} \, \eta^{\mu \sigma} \bigr)
  \end{equation}
 A similar argument can be used for the commutators of $M^{\mu \nu}$. Therefore the generators of the Poincar\'e group are $M^{\mu \nu}$ and $P^{\sigma}$ with algebra\index{Poincar\'e algebra}:
 \begin{equ}[Poincar\'e algebra]
 \vspace*{-0.6cm}
\begin{align}
\bigl[P^{\mu} \ , \ P^{\nu} \bigr] \ \ &= \ \ 0 \\
\bigl[M^{\mu \nu} \ , \ P^{\sigma} \bigr] \ \ &= \ \ \I \, \bigl(P^{\mu} \, \eta^{\nu \sigma} \ - \ P^{\nu} \, \eta^{\mu \sigma} \bigr)\\
\bigl[M^{\mu \nu} \ , \ M^{\rho \sigma} \bigr] \ \ &= \ \ \I\, \bigl(M^{\mu \sigma} \, \eta^{\nu \rho} \ + \ M^{\nu \rho} \, \eta^{\mu \sigma} \ - \ M^{\mu \rho} \, \eta^{\nu \sigma} \ - \ M^{\nu \sigma} \, \eta^{\mu \rho} \bigr)
\end{align}
\end{equ}
As an example, a 4-dimensional matrix representation for the $M^{\mu \nu}$ is
\begin{equation}
(M^{\rho \sigma})^{\mu}\,_{\nu} \eq  -\I\, \bigl(\eta^{\mu \sigma} \, \de^{\rho}\,_{\nu} \ - \ \eta^{\rho \mu} \, \de^{\sigma}\,_{\nu} \bigr) \ .
\end{equation}

Also, the definition of the operators
\begin{equation}\label{eq:LorGenFunctions} 
(M^{\rho\sigma})^{\mu}\,_{\nu}=\I\left (x^{\mu}\p_{\nu}-x_{\nu}\p^{\mu}\right )
\end{equation}
is a representation of the Lorentz generators acting on the space of functions.
Similarly, the operator
\begin{equation}
P^{\mu}=\I\p^{\mu}
\end{equation}
is the generator of translations in the representation defined by \eqref{eq:LorGenFunctions}. It is left as an exercise to verify that these operators satisfy the Poincar\'e algebra.

\subsection{Properties of the Poincar\'e group}
\label{sec:PropertiesOfLorentzGroup}

Let us summarise  the basic properties of the Poincar\'e group.

\subsubsection*{Conservation laws}

Recall that $P^{0}=H$ corresponds to the Hamiltonian and thus
\begin{equation}
[P^{0},P^{\mu}]=0
\end{equation}
implies \emph{conservation of energy and momentum}, whereas
\begin{equation}
[P^{0},M^{ij}]=0\quad i,j=1,2,3
\end{equation}
amounts to \emph{conservation of angular momentum}. There is no conservation law associated to the $M^{0i}$ generators since they do not commute with $P^0$.

\subsubsection*{Correspondence of $\mathrm{SO}(3,1)$ and $\mathrm{SU}(2) \otimes \mathrm{SU}(2) $}

There is a correspondence between the algebras of $\mathrm{SO}(3,1)$ and $\mathrm{SU}(2)\otimes \mathrm{SU}(2)$,\footnote{This is true only locally since $\mathrm{SO}(3,1)$ is not compact. This distinguishes global aspects of Lorentzian geometry from those of Riemannian manifolds. For instance, even if a spacetime is connected and geodesically complete, there exist points which cannot be connected by a geodesic \cite{Hawking:1973uf}.}
namely
\begin{equation}
 \mathrm{SO}(3,1) \  \leftrightarrow \ \mathrm{SU}(2) \otimes \mathrm{SU}(2) \, .
\end{equation}
This means that the representations of the $ \mathrm{SO}(3,1)$ algebra can be determined by those of $ \mathrm{SU}(2)\oplus  \mathrm{SU}(2)$.
This works as follows.
The generators $J_{i}$ of rotations and $K_{i}$ of Lorentz boosts can be defined as
\begin{equation}
J_{i} = \frac{1}{2} \; \ep_{ijk} \, M_{jk} \co K_{i} = M_{0i} \ ,
\end{equation}
therefore, using the Poincar\'e algebra we can easily derive the commutation relations
\begin{equation}
\bigl[J_{i},J_{j}\bigr] = \I\ep_{ijk}J_{k}\kom \bigl[J_{i},K_{j}\bigr] = \I\ep_{ijk}K_{k}\kom \bigl[K_{i},K_{j}\bigr] = -\I\ep_{ijk}J_{k}\, .
\end{equation}
In order to identify the different representations of the Lorentz group, it is instructive to consider the linear combinations
\begin{equation}
A_{i} =\frac{1}{2} \; \bigl(J_{i} \, + \, \I K_{i} \bigr) \co B_{i} = \frac{1}{2} \; \bigl(J_{i} \, - \, \I K_{i} \bigr)
\end{equation}
which are neither Hermitian nor anti-Hermitian if $J_i$ and $K_i$ are Hermitian. They satisfy $\mathrm{SU}(2)$ commutation relations,
\begin{equation}
\bigl[ A_{i} \ , \ A_{j} \bigr] =\I \ep_{ijk} \, A_{k} \co \bigl[B_{i} \ , \ B_{j} \bigr] = \I \ep_{ijk} \, B_{k} \co \bigl[A_{i} \ , \ B_{j} \bigr] = 0\, .
\end{equation}
These are two independent copies of the $\mathrm{SU}(2)$ algebra\index{$\mathrm{SU}(2)$}, but keeping in mind that the operators $A_i$ and $B_i$ are not Hermitian.
In contrast, the combination 
$\mathbf{J} = \mathbf{A} + \mathbf{B}$ is Hermitian and corresponds to the physical spin. 
We can then interpret $\mathbf{J} $ as the physical spin which itself  generates an $\mathrm{SU}(2)$ group. Recall that irreducible representations $R_{\Lambda_{j}}$ of $\mathrm{SU}(2)$ are labelled by 
 $j=0,\frac{1}{2},\ldots$ with $j$ determined from the eigenvalues of the \emph{quadratic Casimir Operator}\index{$\mathrm{SU}(2)$!Casimir operator} which is nothing but the ``total angular momentum'' $J^{2}=J_{1}^{2}+J_{2}^{2}+J_{3}^{2}$.
It commutes with the three generators $J_i$ and satisfies
\begin{equation}
J^{2}\ket{j}=j(j+1)\ket{j}\, \qquad j=0,\frac{1}{2}, 1, \cdots .
\end{equation}
Hence, we can use those eigenvalues to label the irreducible representations of $\mathrm{SU}(2)$. In the case of $\mathrm{SO}(3,1)$, we denote representations of $\mathrm{SU}(2)\otimes \mathrm{SU}(2)$ as
\begin{equation}
(A,B)\kom \qquad {\rm with} \, \, A,B=0,\dfrac{1}{2},1,\ldots
\end{equation}
Since $\mathbf{J} = \mathbf{A} + \mathbf{B}$,
we can see using the standard addition of angular momenta that the representation $(A,B)$ corresponds to spins $j=|A-B|\oplus (|A-B|+1)\cdots \oplus (A+B)$.

Under parity $P$ with $x^{0} \mapsto x^{0}$ and $\mathbf{x} \mapsto -\mathbf{x}$, we have
\begin{equation}
    J_{i} \ \ \mapsto \ \ J_{i} \co K_{i} \ \ \mapsto \ \ -K_{i} \ \ \ \Longrightarrow \ \ \  A_{i} \ \ \ \leftrightarrow \ \ \ B_{i} \ .
\end{equation}
Therefore, 
$A$ and $B$ are interchanged under parity transformation
\begin{equation}
	(A,B)\xleftrightarrow{P}(B,A)\, .
\end{equation}
We then call $A$ the \emph{left-handed} and $B$ the \emph{right-handed} component of $(A,B)$. 
Below, we will use this notion to label representations of SO(3,1) and to define their handedness (or chirality).

\subsubsection*{Universal cover of $\mathrm{SO}(3,1)$}

There is a homomorphism (not an isomorphism)\index{$\mathrm{SL}(2,\mathbb C)$}
\begin{equation}
\mathrm{SO}(3,1) \ \cong \ \mathrm{SL}(2,\mathbb C) \, ,
\end{equation}
where $ \mathrm{SL}(2,\mathbb C)$ is the group of $2\times 2$ complex matrices with unit determinant. To see this, take a 4 vector $X$ and a corresponding $(2 \times 2)$-matrix $\tilde{x}$,
\begin{equation}
X = x_{\mu} \, e^{\mu} = (x_{0} , x_{1} , x_{2} , x_{3}) \kom \tilde{x} = x_{\mu} \, \sigma^{\mu} = \ccb x_{0} + x_{3} &x_{1} - \I x_{2} \\ x_{1} + \I x_{2} &x_{0} - x_{3} \cce \ ,
\end{equation}
where $\sigma^{\mu}$ is the 4 vector of \emph{Pauli matrices}\index{Pauli matrices}
\begin{equation}\label{eq:PauliMatrices} 
\sigma^{\mu} \eq \left\{ \ccb 1 &0 \\ 0 &1 \cce \ , \ \ccb 0 &1 \\ 1 &0 \cce \ , \ \ccb 0 &-\I \\ \I &0 \cce \ , \ \ccb 1 &0 \\ 0 &-1 \cce \right\} \ .
\end{equation}
Transformations $X \mapsto \La X$ under $\mathrm{SO}(3,1)$ leave the modulus square
\begin{equation}
|X|^{2} = x_{0}^{2} \ - \ x_{1}^{2} \ - \ x_{2}^{2} \ - \ x_{3}^{2}
\end{equation}
invariant, whereas the action of $\mathrm{SL}(2,\mathbb C)$ mapping 
\begin{equation}
\tilde{x} \mapsto N\tilde{x} N^\dag
\end{equation}
with $N \in \mathrm{SL}(2,\mathbb C)$ preserves the determinant
\begin{equation}
\det \tilde{x} = x_{0}^{2} \ - \ x_{1}^{2} \ - \ x_{2}^{2} \ - \ x_{3}^{2} \ .
\end{equation}
This equals $|X|^2$. The map between $\mathrm{SL}(2,\mathbb C)$ and $\mathrm{SO}(3,1)$ is 2-1, since $N = \pm \mathds{1}_2$ both correspond to $\La = \mathds{1}_4$, but $\mathrm{SL}(2,\mathbb C)$ has the advantage of being simply connected, that means that  $\mathrm{SL}(2,\mathbb C)$\index{$\mathrm{SL}(2,\mathbb C)$!Universal cover}\index{$\mathrm{SL}(2,\mathbb C)$!Simply connectedness} is the \emph{Universal Covering Group} of the Lorentz group. This is important since it is the simply connected group manifold that is continuously connected to the identity operator.
Furthermore, since the map between $\mathrm{SL}(2,\mathbb C)$ and $\mathrm{SO}(3,1)$ is 2-1, a rotation by an angle $\theta$ in $\mathrm{SO}(3,1)$ is mapped to the matrix $\rm{diag}(e^{i\theta/2}, e^{-i\theta/2})$ in $\mathrm{SL}(2,\mathbb C)$. This in turn implies that it is only rotations by $\theta=4\pi$ (and not $2\pi$) that give the identity in $\mathrm{SL}(2,\mathbb C)$. This will turn out to be a crucial observation in order to describe particles of \emph{half-integer} spin as we will see later.

Let us briefly see why the manifold of $\mathrm{SL}(2,\mathbb C)$ is simply connected. By the polar decomposition of matrices, an $\mathrm{SL}(2,\mathbb C)$ matrix  $N$ can be written as $N=e^H U$ where $e^H$ is  a positive Hermitian matrix and $U$ is unitary. Since  $e^H$ has positive determinant then $\det N=1$ implies $\rm{Tr} \, H=0$ and $\det U=1$. Therefore $H$ and $U$ can be written as 
\begin{equation} H=\ccb a & b+\I c \\ b- \I c & -a \cce \, \qquad U=\ccb x+\I y & z+\I w \\ -z+\I w & x-\I y \cce
\end{equation}
with $a,b,c$ real parameters and $x,y,z,w$ constrained by $x^2+y^2+z^2+w^2=1$ therefore the manifold of $\mathrm{SL}(2,\mathbb C)$ is $\mathbb R^3\times \mathbb{S}^3$ which are simply connected, whereas the manifold for $\mathrm{SO}(3,1)$ is $\mathbb R^3\times \mathbb{S}^3/\mathbb{Z}_2$ which is doubly connected\index{$\mathrm{SL}(2,\mathbb C)$!Double cover}. Being the covering group, it is $\mathrm{SL}(2,\mathbb C)$ that is reached by exponentiating the algebra and therefore we are let to consider the representations of $\mathrm{SL}(2,\mathbb C)$ that we discuss next.

\section{Spinor representations of the Lorentz group}

Above, we already established that $\mathrm{SL}(2,\mathbb C)$ is the universal covering group of the Lorentz group $\mathrm{SO}(3,1)$. This should ring a bell: in quantum mechanics, we learned that $\mathrm{SU}(2)$ is the double cover of $\mathrm{SO}(3)$ and therefore it is representations of $\mathrm{SU}(2)$  that are the ones to be considered since  $\mathrm{SU}(2)$  is simply connected and $\mathrm{SO}(3)$ is not.
We observe a similar phenomenon for the Lorentz group in the sense that  representation theory of $\mathrm{SL}(2,\mathbb C)$ is the relevant one to study. Concentrating only on representations of  $\mathrm{SO}(3,1)$ we would miss the fundamental representations which are the spinor representations which we define next.

\subsection{Representations and invariant tensors of $\mathrm{SL}(2,\mathbb C)$}
\label{sec:RepresentationsAndInvariantTensorsOfSL2C}

To begin, we define the basic representations of $\mathrm{SL}(2,\mathbb C)$\index{$\mathrm{SL}(2,\mathbb C)$!Representations}.
Let $N\in \mathrm{SL}(2,\mathbb C)$, then we have:
\begin{itemize}
\item The fundamental representation $\psi_{\beta}$ transforming as
\begin{equation}\psi'_{\al} \eq N_{\al}\,^{\be} \, \psi_{\be} \co \al,\be = 1,2
\end{equation}
The elements of this representation $\psi_\al$ are called \emph{\bf left-handed Weyl spinors}\index{Weyl spinors}\index{Weyl spinors!Left-handed}.
\item The conjugate fundamental representation $\overline{\chi}_{\dot{\beta}}$ transforming as
\begin{equation}
\bar{\chi}'_{\dot{\al}} \eq N^{*}_{\dot{\al}}\,^{\dot{\be}} \, \bar{\chi}_{\dot{\be}} \co \dot{\al},\dot{\be} = 1,2
\end{equation}
Here $\bar{\chi}_{\dot{\be}}$ are called \emph{\bf right-handed Weyl spinors}\index{Weyl spinors!Right-handed}.
\item The contravariant representations $\psi^{\beta}$ and $\bar{\chi}^{\dot{\beta}}$
\begin{equation}
\psi'^{\al} \eq \psi^{\be} \, (N^{-1})_{\be}\,^{\al} \co \bar{\chi}'^{\dot{\al}} \eq \bar{\chi}^{\dot{\be}} \, (N^{*-1})_{\dot{\be}}\,^{\dot{\al}}
\end{equation}
as the dual representations of the two above.
\end{itemize}
The fundamental and conjugate representations are the basic representations of
$\mathrm{SL}(2, \mathbb C)$ and the Lorentz group, giving then the importance to spinors as the basic objects of special relativity, a fact that could be missed by not realising the connection of the Lorentz group and $\mathrm{SL}(2, \mathbb C)$.

We will show now that the contravariant representations are however not independent by explicitly showing how indices can be raised (or lowered) using specific tensors.
To see this, we consider three different ways to raise and lower indices.
\begin{itemize}
\item The metric tensor $\eta^{\mu \nu} = (\eta_{\mu \nu})^{-1}$ is invariant under $\mathrm{SO}(3,1)$ and is therefore used to lower and raise spacetime indices.
\item The analogue for $\mathrm{SL}(2,\mathbb C)$ is
\begin{equation}
\ep^{\al \be} \eq \ep^{\dot{\al} \dot{\be}} \eq \ccb 0 &1 \\ -1 &0 \cce \eq -\ep_{\al \be} \eq - \ep_{\dot{\al} \dot{\be}} \ ,
\end{equation}
since it is invariant under $\mathrm{SL}(2,\mathbb C)$ transformations
\begin{equation}\ep'^{\al \be} \eq \ep^{\rho \sigma} \, \left(N^{-1}\right)_{\rho}\,^{\al} \, \left(N^{-1}\right)_{\sigma}\,^{\be} \eq \ep^{\al \be} \cdot \det N^{-1} \eq \ep^{\al \be} \ .
\end{equation}
That is why $\ep^{\rho \sigma}$ is used to raise and lower indices
\begin{equation}\psi^{\al} \eq \ep^{\al \be} \, \psi_{\be} \co \bar{\chi}^{\dot{\al}} \eq \ep^{\dot{\al} \dot{\be}} \, \bar{\chi}_{\dot{\be}} \ ,
\end{equation}
so contravariant representations are not independent.
\item To handle mixed $\mathrm{SO}(3,1)$- and $\mathrm{SL}(2,\mathbb C)$-indices, recall that the transformed components $x_{\mu}$ should look the same, whether we transform the vector $X$ via $\mathrm{SO}(3,1)$ or the matrix $\tilde{x} = x_{\mu} \sigma^{\mu}$
\begin{equation}(x_{\mu} \, \sigma^{\mu})_{\al \dot{\al}} \ \ \mapsto \ \ N_{\al}\,^{\be} \, (x_{\nu} \, \sigma^{\nu})_{\be \dot{\ga}} \, N^{*}_{\dot{\al}}\,^{\dot{\ga}} \eq \left(\La^{-1}\right)_{\mu}\,^{\nu} \, x_{\nu} \, \sigma^{\mu} \ ,
\end{equation}
so the right transformation rule is
\begin{equation}(\sigma^{\mu})_{\al \dot{\al}} \eq N_{\al}\,^{\be} \, (\sigma^{\nu})_{\be \dot{\ga}} \, (\La)^{\mu}\,_{\nu} \, N^{*}_{\dot{\al}}\,^{\dot{\ga}} \ .
\end{equation}
This may be interpreted by saying that the Pauli matrices are invariant under a combined $\mathrm{SO}(3,1)$ transformation on its spacetime index with a  $\mathrm{SL}(2,\mathbb C)$ on its matrix indices. Similar relations hold for the quantity
\begin{equation}(\bar{\sigma}^{\mu})^{\dot{\al}\al}  =  \ep^{\al \be} \, \ep^{\dot{\al} \dot{\be}} \, (\sigma^{\mu})_{\be \dot{\be}} \eq (\mathds{1} , \ -\boldsymbol{\sigma}) \ .
\end{equation}
Note that this is the definition of $\bar\sigma$ and no other connection with $\sigma$ such as complex conjugation should be assumed despite the notation. Note in particular the chosen location of the dotted and undotted indices which differ between $\sigma^\mu$ and $\bar\sigma^\mu$. The order is conventional and keeps track of how the corresponding quantity transforms under  $\mathrm{SL}(2,\mathbb C)$. Both $\sigma$ and $\bar\sigma$ will play an important role next. In fact, we can already deduce that the \emph{Clifford algebra}\index{Clifford algebra}
\begin{equation}
\sigma^{\mu}\bar{\sigma}^{\nu}+\sigma^{\nu}\bar{\sigma}^{\mu}=2\eta^{\mu\nu}\mathds{1}_{2}
\end{equation}
appears naturally in our analysis which will give rise to \emph{Dirac spinors}\index{Dirac spinors} further below.
\end{itemize}

\subsection{Generators of $\mathrm{SL}(2,\mathbb C)$ and Weyl spinors}
\label{sec:GeneratorsOfSL2C}

Let us define tensors $\sigma^{\mu \nu}$, $\bar{\sigma}^{\mu \nu}$ as anti-symmetrised products of $\sigma$ matrices
\begin{align*}
(\sigma^{\mu \nu})_{\al}\,^{\be} \ \ &:= \ \ \frac{\I}{4} \; \bigl(\sigma^{\mu} \, \bar{\sigma}^{\nu} \ - \ \sigma^{\nu} \, \bar{\sigma}^{\mu} \bigr)_{\al}\,^{\be} \\
(\bar{\sigma}^{\mu \nu})^{\dot{\al}}\,_{\dot{\be}}  \ \ &:= \ \ \frac{\I}{4} \;  \bigl(\bar{\sigma}^{\mu} \, \sigma^{\nu} \ - \ \bar{\sigma}^{\nu} \, \sigma^{\mu}  \bigr)^{\dot{\al}}\,_{\dot{\be}}
\end{align*}
which satisfy the Lorentz algebra
\begin{equation}
\bigl[ \sigma^{\mu \nu} \ , \ \sigma^{\la \rho} \bigr] = \I \, \bigl( \eta^{\mu \rho} \, \sigma^{\nu \la} \ + \ \eta^{\nu \la} \, \sigma^{\mu \rho} \ - \ \eta^{\mu \la} \, \sigma^{\nu \rho} \ - \ \eta^{\nu \rho} \, \sigma^{\mu \la} \bigr)
\end{equation}
and similarly for $ \bar{\sigma}^{\mu \nu}$.
Then $\sigma^{\mu\nu}$ and $\bar{\sigma}^{\mu \nu}$ correspond to the generators of the Lorentz algebra in the spinor representation.

\noindent
Under a finite Lorentz transformation with parameters $\om_{\mu \nu}$, Weyl spinors transform as follows:
\begin{align*}
\psi_{\al} \ \ &\mapsto \ \ \exp\left(-\frac{\I}{2} \om_{\mu \nu} \sigma^{\mu \nu}\right)_{\al}\,^{\be} \, \psi_{\be} &\te{(left-handed)} \\
\bar{\chi}^{\dot{\al}} \ \ &\mapsto \ \ \exp\left(-\frac{\I}{2} \om_{\mu \nu} \bar{\sigma}^{\mu \nu}\right)^{\dot{\al}}\,_{\dot{\be}} \, \bar{\chi}^{\dot{\be}} &\te{(right-handed)}
\end{align*}
Now consider the spins with respect to the $\mathrm{SU}(2)$s spanned by the $A_{i}$ and $B_{i}$:
\begin{align*}
&J_{i} = \frac{1}{2} \sigma_{i} \co K_{i} = -\frac{\I}{2} \sigma_{i} \quad\Longrightarrow \quad\psi_{\al}: \;(A, \ B) = \left(\frac{1}{2} , \ 0\right)\kom \text{left-handed} \\
&J_{i} = \frac{1}{2}  \sigma_{i} \co K_{i} = +\frac{\I}{2}  \sigma_{i} \quad\Longrightarrow\quad\bar{\chi}^{\dot{\al}}: \;(A , \ B) = \left(0, \ \frac{1}{2}\right)\kom \text{right-handed} \, .
\end{align*}
Recall the the Pauli matrices correspond to generators of the rotation group in the $j=\frac{1}{2}$ representation since 
\begin{equation}
\sum_i\left(\frac{\sigma_i}{2}\right)^2=\frac{3}{4}=j(j+1) \qquad {\rm for } \qquad  j=\frac{1}{2} \, .
\end{equation}
The expressions above also justify the aforementioned term left- and right-handed components for $A$, $B$. The difference and independence between the left- and right-handed representations of  $\mathrm{SL}(2,\mathbb C)$ indicates that there is no reason to assume that parity is a fundamental symmetry and will be the reason behind the fact that the Standard Model is chiral in the sense that left- and right-handed representations are not the same. The believe that physicists before Yang and Lee had assuming parity should be an inherent symmetry of Nature is not justified and it is not surprising then that the laws of Nature are not invariant under parity as we will see later.
The concept of \emph{chirality}\index{Chirality} is ubiquitous not only in the Standard Model, but also more generally in various areas of modern  physics (and biology\footnote{DNA and aminoacids (and humans) are chiral and for some reason life on Earth is of one chirality. This has intrigued scientists for some time. Salam was known to have tried to find a connection between chirality in physics and biology, with no success.}). 

Some useful identities concerning the $\sigma^{\mu}$ and $\sigma^{\mu \nu}$ can be found in \cite{muller2010introduction}. For now, let us just mention the identities
\begin{align}
\sigma^{\mu \nu} = \frac{1}{2\I} \; \ep^{\mu \nu \rho \sigma} \, \sigma_{\rho \sigma} \kom \bar{\sigma}^{\mu \nu} =  -\frac{1}{2\I} \; \ep^{\mu \nu \rho \sigma} \, \bar{\sigma}_{\rho \sigma} \ ,
\end{align}
known as \emph{self duality}\index{Self duality} and \emph{anti-self duality} respectively. They are important because naively $\sigma^{\mu\nu}$ being antisymmetric seems to have
$\frac{4 \times 3}{2}$ components, but the self duality conditions reduces this by half. We then need the two sets of generators $\sigma^{\mu\nu}$ and $\bar{\sigma}^{\mu \nu}$ to complete the $6$ independent generators of $\mathrm{SL}(2,\mathbb C)$. A reference book illustrating many of the calculations for $2$-component spinors is \cite{muller2010introduction}.

\subsubsection*{Products of Weyl spinors}

We define the product of two Weyl spinors as
\begin{align}
\chi \psi = \chi^{\al} \, \psi_{\al}  = -\chi_{\al} \, \psi^{\al} \kom \bar{\chi} \bar{\psi} =  \bar{\chi}_{\dot{\al}} \, \bar{\psi}^{\dot{\al}} =  -\bar{\chi}^{\dot{\al}} \, \bar{\psi}_{\dot{\al}} \ ,
\end{align}
particularly,
\begin{equation}\psi \psi \eq \psi^{\al} \, \psi_{\al} \eq \ep^{\al \be} \, \psi_{\be} \, \psi_{\al} \eq \psi_{2} \, \psi_{1}\ - \ \psi_{1} \, \psi_{2} \ .
\end{equation}
Choose the $\psi_{\al}$ to be \emph{anticommuting Grassmann numbers\index{Grassmann numbers}}, $\psi_{1} \psi_{2} = - \psi_{2} \psi_{1}$, so $\psi \psi = 2\psi_{2} \psi_{1}$.

\noindent
From the definitions
\begin{equation}
\psi_{\al}^\dag  =  \bar{\psi}_{\dot{\al}} \kom \bar{\psi}^{\dot{\al}}  =  \psi^{*}_{\be} \, (\sigma^{0})^{\be \dot{\al}}
\end{equation}
it follows that
\begin{equation}
(\chi \psi)^\dag  =  \bar{\chi} \bar{\psi} \co (\psi \, \sigma^{\mu} \, \bar{\chi})^\dag  =  \chi \, \sigma^{\mu} \, \bar{\psi}
\end{equation}
which justifies the $\nearrow$ contraction of dotted indices in contrast to the $\searrow$ contraction of undotted ones.
 
\noindent
In general we can generate all higher dimensional representations of the Lorentz group by products of the fundamental representation $(\frac{1}{2}, \, 0)$ and its conjugate $(0, \, \frac{1}{2})$. The computation of tensor products
\begin{equation}
\left (\frac{r}{2}, \, \frac{s}{2}\right ) = \left (\frac{1}{2}, \, 0\right )^{\otimes r} \otimes \left (0, \, \frac{1}{2}\right )^{\otimes s}
\end{equation}
can be reduced to successive application of the elementary $\mathrm{SU}(2)$ rule (for $j\neq 0$)
\begin{equation}
\left (\frac{j}{2}\right ) \otimes \left (\frac{1}{2}\right ) = \left (\frac{j-1}{2}\right ) \oplus \left (\frac{j+1}{2}\right )\, .
\end{equation}

\noindent
Let us give two examples for tensoring Lorentz representations:
\begin{itemize}
\item $(\frac{1}{2}, \, 0)\otimes (0, \, \frac{1}{2})\ = \ (\frac{1}{2}, \, \frac{1}{2}) $

\noindent
Bispinors with different chiralities can be expanded in terms of the $\sigma^\mu_{\al \dot \al}$. Actually, the $\sigma$ matrices form a complete orthonormal set of $2 \times 2$ matrices with respect to the trace Tr$\{ \sigma^\mu \bar \sigma^\nu \} =2 \eta^{\mu \nu}$:
\begin{equation} \psi_\al \, \bar{\chi}_{\dot{\al}} =  \frac{1}{2} \; \left(\psi \, \sigma_\mu \, \bar{\chi}\right)\, \sigma^\mu_{\al\dot\al}  \end{equation}
Hence, two spinor degrees of freedom with opposite chirality give rise to a Lorentz vector $\psi  \sigma_\mu \bar{\chi}$.
\item $(\frac{1}{2}, \, 0)\otimes (\frac{1}{2}, \, 0)\ = \ (0,0) \oplus (1,0) $

\noindent
Alike bispinors require a different set of matrices to expand, $\ep_{\al \be}$ and $(\sigma^{\mu \nu})_{\al} \,^\ga \ep_{\ga \be} =: (\sigma^{\mu \nu} \ep^T)_{\al \be}$. The former represents the unique antisymmetric $2 \times 2$ matrix, the latter provides the symmetric ones. Note that the (anti-)self duality reduces the number of linearly independent $\sigma^{\mu \nu}$'s (over $\mathbb C$) from 6 to 3:
\begin{equation}\psi_\al \, \chi_\be  =  \frac{1}{2}\; \ep_{\al\be} \, \left(\psi\chi\right) \
+ \ \frac{1}{2} \; \left(\sigma^{\mu\nu} \, \ep^T\right)_{\al\be}\, \left(\psi \, \sigma_{\mu\nu} \, \chi\right) \end{equation}
The product of spinors with alike chiralities decomposes into two Lorentz irreducibles, a scalar $\psi\chi$ and a self-dual antisymmetric rank two tensor $\psi \, \sigma_{\mu\nu} \, \chi$. The counting of independent components of $\sigma^{\mu\nu}$ from its self-duality property precisely provides the right number of three components for the $(1,0)$ representation. Similarly, there is an anti-self dual tensor $\bar{\chi} \bar{\sigma}^{\mu \nu} \bar{\psi}$ in $(0,1)$.
\end{itemize}

\subsection{Dirac and Majorana spinors}
\label{sec:Dirac}


Here, we give the dictionary connecting the ideas of Weyl spinors with the more standard \emph{Dirac} theory in $D=4$ dimensions. 

\begin{equ}[Dirac spinor]

A \emph{Dirac spinor}\index{Dirac spinor} $\Psi_D$ is defined to be the direct sum of two Weyl spinors $\psi, \bar \chi$ of opposite chirality.
It therefore falls into a reducible representation of the Lorentz group,
\begin{equation}
\Psi_{D} =  \vecb \psi_{\al} \\ \bar{\chi}^{\dot{\al}} \vece = \left (0,\dfrac{1}{2}\right ) \oplus \left (\dfrac{1}{2},0\right )\ .
\end{equation}

\end{equ}

\noindent The Dirac analogue of the Weyl spinors' sigma matrices are the $4 \times 4$ gamma matrices $\ga^\mu$ subject to the \emph{Clifford algebra}\index{Clifford algebra}
 \begin{equation}
 \ga^{\mu} = \ccb 0 &\sigma^{\mu} \\ \bar{\sigma}^{\mu} &0 \cce \co \bigl\{ \ga^{\mu} \kom \ga^{\nu} \bigr\} = 2 \, \eta^{\mu \nu} \, \mathds{1} \ .
\end{equation}
Due to the reducibility, the generators of the Lorentz group take block diagonal form
\begin{equation}
\Si^{\mu \nu} = \frac{\I}{4} \, \ga^{\mu \nu} = \ccb \sigma^{\mu \nu} &0 \\ 0 &\bar{\sigma}^{\mu \nu} \cce
\end{equation}
and naturally obey the same algebra like the irreducible blocks $\sigma^{\mu \nu}$, $\bar \sigma^{\mu \nu}$
\vspace*{0.1cm}
\begin{equation} 
\bigl[ \Si^{\mu \nu} \ , \ \Si^{\la \rho} \bigr] = \I \, \left (\eta^{\mu \rho}\, \Si^{\nu \la}  +\eta^{\nu \la}\, \Si^{\mu \rho} -\eta^{\mu \la}\, \Si^{\nu \rho} -\eta^{\nu \rho}\, \Si^{\mu \la} \right )\, .
\end{equation}
\vspace*{0.1cm}
To disentangle the two inequivalent Weyl representations, one defines the chiral matrix $\ga^{5}$ as
\begin{equation}
\ga^{5} = \I\ga^{0} \, \ga^{1} \, \ga^{2} \, \ga^{3} = \ccb -\mathds{1} &0 \\ 0 &\mathds{1} \cce \ ,
\end{equation}
such that the $\psi (\chi)$ components of a Dirac spinors have eigenvalues (chirality) $-1 \, (+1)$ under $\ga^5$,
\begin{equation}
\ga^{5} \, \Psi_{D} = \ccb -\mathds{1} &0 \\ 0 &\mathds{1} \cce \, \vecb \psi_{\al} \\ \bar{\chi}^{\dot{\al}} \vece \eq \vecb -\psi_{\al} \\ \bar{\chi}^{\dot{\al}} \vece \ .
\end{equation}
Hence, one can define projection operators\index{Chirality!Projection operators} $P_{L}$, $P_{R}$,
\begin{equation}
P_{L}  =  \frac{1}{2} \; \bigl(\mathds{1} \ - \ \ga^{5} \bigr) \co P_{R}  =  \frac{1}{2} \; \bigl(\mathds{1} \ + \ \ga^{5} \bigr) \ ,
\end{equation}
eliminating one part of definite chirality, i.e.,
\begin{equation}
P_{L} \, \Psi_{D} \eq \vecb \psi_{\al} \\ 0 \vece \co P_{R} \, \Psi_{D} \eq \vecb 0 \\ \bar{\chi}^{\dot{\al}} \vece \ .
\end{equation}
The fact that Lorentz generators preserve chirality can also be seen from $\{ \ga^5, \ga^\mu \} = 0$ implying $[ \ga^5 , \Si^{\mu \nu} ] = 0$.

\noindent 
Finally, define the \emph{Dirac conjugate}\index{Dirac spinor!Dirac conjugate} $\Psib_{D}$ and \emph{charge conjugate}\index{Charge conjugation}\index{Charge conjugation!Charge conjugate spinor}\index{Dirac spinor!Charge conjugate} spinor $\Psi_{D}\,^{C}$ by
\begin{align}
\Psib_{D} =  (\chi^{\al} , \ \bar{\psi}_{\dot{\al}}) \eq \Psi_{D}^\dag \, \ga^{0} \kom \Psi_{D}\,^{C} = C \, \Psib_{D}^{T} \eq \vecb \chi_{\al} \\ \bar{\psi}^{\dot{\al}} \vece \ ,
\end{align}
where $C$ denotes the \emph{charge conjugation matrix}\index{Charge conjugation!Charge conjugation matrix}
\begin{equation}
C  =  \ccb \ep_{\al \be} &0 \\ 0 &\ep^{\dot{\al} \dot{\be}} \cce \ .
\end{equation}

There is a third basic type of spinors called
{\emph{Majorana spinors}}\index{Majorana spinors} $\Psi_{M}$ which have the property $\psi_{\al} = \chi_{\al}$,
\begin{equation}
\Psi_{M} \eq \vecb \psi_{\al} \\ \bar{\psi}^{\dot{\al}} \vece \eq \Psi_{M}\,^{C} \ ,
\end{equation}
which are neutral under charge conjugation.
A general Dirac spinor (and its charge conjugate) can be decomposed in terms of Majorana spinors as
\begin{equation}
\Psi_{D} \eq \Psi_{M1} \ + \ \I\Psi_{M2} \co \Psi_{D}\,^{C} \eq \Psi_{M1} \ - \ \I\Psi_{M2} \ .
\end{equation}
Note that there can be no spinors in 4 dimensions which are both Majorana and Weyl. This is a dimension dependent property. It can be shown that in dimensions $2\, {\rm mod} \, 8$ it is possible to have spinors which are both Majorana and Weyl,\footnote{This happens to be relevant in string theory for which the worldsheet dimension is $d=2$ and the target space dimension is $D=10$.} see for instance~App.~B.1 in \cite{Polchinski:1998rr}.

\section{Unitary Representations of the Poincar\'e group}
\label{sec:RepresentationsOfPoincareGroup}

We now will combine the $2$ fundamental theories of special relativity and quantum mechanics to find the unitary representations of the Poincar{\'e} group on quantum states. As usually unitarity\index{Unitarity} is required in order to have invariant observables (such as matrix elements).
Being non-compact, the Poincar\'e group does not have finite dimensional unitary representations.\footnote{Notice also that, when we labelled representations of the Lorentz group in terms of $(A,B)$ that are finite dimensional, since they correspond to the algebra of $\mathrm{SU}(2)\oplus \mathrm{SU}(2)$ the generators $A,B$ are not Hermitian so the corresponding representations are not unitary.}

\subsubsection*{Recap: the rotation group in Quantum Mechanics}

Before we consider the Poincar\'e group, let us reiterate some facts about unitary representations of the rotation group $\mathrm{SU}(2)$ in quantum mechanics.
Recall that the rotation group $\mathrm{SU}(2)$\index{$\mathrm{SU}(2)$} has generators $\{J_{i}: \ i=1,2,3 \}$ satisfying the algebra
\begin{equation}
    \bigl[J_{i} \ , \ J_{j} \bigr] \eq \I\ep_{ijk}\, J_{k} \ .
\end{equation}
Next, we define the Casimir operator
\begin{equation}
 J^{2} \eq \sum^{3}_{i = 1} J_{i}^{2}\, .
\end{equation}
In general, for a given group $G$, the Casimir operators are operators that commute with all the generators. They are important because Schur's Lemma guarantees that they are proportional to the identity within a given representation and therefore their eigenvalues can be used to label the representations. For compact Lie groups, the number of Casimir operators equals the rank of the group.\index{$\mathrm{SU}(2)$!Casimir operator}
In the present case of $\mathrm{SU}(2)$, the Casimir operator $J^{2}$ commutes indeed with all the $J_{i}$,
\begin{equation}
[J^{2},J_{i}]=0\quad \forall i=1,2,3\, ,
\end{equation}
and labels irreducible representations by eigenvalues $j(j+1)$ of $J^{2}$, that is,
\begin{equation}
J^{2}\ket{j;\lambda}=j(j+1)\ket{j;\lambda}\, .
\end{equation}
Within these representations, $\lambda$ parametrises the degeneracy of states in the same representation obtained by acting with the ladder operators on the highest weight state. Thus, we can make a choice and diagonalise the states with respect to $J_{3}$ with eigenvalues $\lambda=j_{3}$ so that
\begin{equation}
J_{3}\ket{j;j_{3}}=j_{3}\ket{j;j_{3}}\kom j_{3}= -j,-j+1,...,j-1,j\, .
\end{equation}
Hence, the corresponding states are labelled like $\ket{j;j_{3}}$.
These are in fact unitary representations and, since $\mathrm{SU}(2)$ is compact, they are finite-dimensional.
The latter will cease to be true for the Poincar\'e group.

\subsubsection*{The Poincar\'e group}

Now, let us consider the Poincar\'e group. The takeaway message from our recap about labelling irreducible representations of $\mathrm{SU}(2)$ is that we simply need to find the corresponding Casimir operators.\footnote{In general, Casimir operators form a basis of the center of the associated universal enveloping Lie algebra. For a semi-simple Lie group, the number of independent Casimir operators is given by the rank. But, since the Poincar{\'e} group is not semi-simple and, in particular, a semi-direct product of two groups, there is no direct theorem determining the number of independent Casimirs.}
In this case there are  two Casimir operators\index{Poincar{\'e} group!Casimir operators}. The first one corresponds to the square of the momenta: $C_{1}=P^{\mu}P_{\mu}$ which can be easily checked that it commutes with all the generators $P^\mu, M^{\mu\nu}$. The second one involves the \emph{Pauli-Ljubanski vector}\index{Pauli-Ljubanski vector} $W_{\mu}$,
\begin{equation}
W_{\mu} := \frac{1}{2} \; \ep_{\mu \nu \rho \sigma} \, P^{\nu} \, M^{\rho \sigma}
\end{equation}
where $\ep_{0123} = -\ep^{0123} = -1$. This operator satisfies the following commutation relations
\begin{align}
[W_{\mu},P_{\nu}]&=0\, ,\\[0.5em]
[W_{\mu},M_{\rho\sigma}]&=\I\left (\eta_{\mu\rho}W_{\sigma}-\eta_{\mu\sigma}W_{\rho}\right )\, ,\\[0.5em]
[W_{\mu},W_{\nu}]&=-\I\varepsilon_{\mu\nu\rho\sigma}W^{\rho}P^{\sigma}\, .
\end{align}
From these commutations relations we can check that a second Casimir corresponds to $C_{2}=W^{\mu}W_{\mu}$. Notice that at this level the Pauli-Ljubanski vector only provides a short
way to express the second Casimir. Even though $W_\mu$ has standard
commutation relations with the generators of the Poincar\'e group $
M_{\mu\nu}, P_\mu$ stating that it transforms as a vector under
Lorentz transformations and commutes with $P_\mu$ (invariant under
translations), the commutator $[W_\mu,W_\nu]\sim
\epsilon_{\mu\nu\rho\sigma}W^\rho P^\sigma$ implies that the $W_\mu$'s by
themselves are not generators of any algebra since the right hand side is quadratic and not linear in the corresponding operators. 

\noindent

Summarising, one can show that the Casimir operators\footnote{Notice that $C_{2}$ is a \emph{quartic} Casimir since it basically involves products of four generators.} for the Poincar{\'e} group are given by
\begin{equ}[Casimir operators of the Poincar\'e group]
\begin{equation}
C_{1}=P^{\mu}P_{\mu}\kom C_{2}=W^{\mu}W_{\mu}\, .
\end{equation}
\vspace*{-0.4cm}
\end{equ}
It is easy to verify that
\begin{equation}
[C_{1,2},P^{\mu}]=[C_{1,2},M^{\mu\nu}]=0\, .
\end{equation}
Poincar\'e multiplets\index{Poincar\'e multiplets} are therefore labelled $\ket{m,\omega;\lambda_{i}}$ so that
\begin{equation}
C_{1}\ket{m,\omega;\lambda_{i}}=m^{2}\ket{m,\omega;\lambda_{i}}\kom C_{2}\ket{m,\omega;\lambda_{i}}=f(m,\omega)\,\ket{m,\omega;\lambda_{i}}\, ,
\end{equation}
that is, $m^{2}$ are the eigenvalues of $C_{1}$ and $f(m,\omega)$ the ones of $C_{2}$. We have to work a little harder to determine the labels $\omega$ and the exact expression for the $f(m,\omega)$.

\noindent
As above, states within those irreducible representations carry extra labels $\lambda_{i}$ corresponding to all operators that can be diagonalised simultaneously (such as $J_{3}$ for $\mathrm{SU}(2)$). One of the $\lambda_{i}$ corresponds to the eigenvalue $p^{\mu}$ of the generator $P^{\mu}$ as a label. To find more labels, take the eigenvalue $p^{\mu}$ of $P^{\mu}$ as given and look for all elements of the Lorentz group that commute with $P^{\mu}$. This defines the \emph{Little} or \emph{Stability group}\index{Stability group}\index{Little group}\index{Poincar{\'e} group!Little group} which we denote as $L(p^{\mu})$.

Note that within a multiplet, at fixed momentum, the operator $P^\mu$ can be replaced by its eigenvalue $p^\mu$ and then the Pauli-Ljubanski vector can be seen as the combination of the generators of the Lorentz group that commutes with the momentum operator and its commutation relations determine the algebra of the Little group (now the right hand side of $[W_\mu,W_\nu]$ is a linear combination of the $W_\rho$'s since the $P^\mu=p^\mu$'s are just numbers within  the multiplet). 

\noindent
Our ultimate goal is to obtain unitary irreducible representations of the Poincar{\'e} group. This can be achieved using the arguments above that can be summarised as the following theorem 

\begin{theo}[see theorem 10.13 in \cite{Tung:1985na}]

Let ${p}^{\mu}$ be some fixed $4$-vector.
\begin{enumerate}
\item On the orbit\footnotemark\index{Orbit} $\cO({p}^{\mu})$, the independent components of $W^{\mu}$ form a Lie algebra $\cL(L({p}^{\mu}))$ of the Little group $L({p}^{\mu})$.
\item For every \emph{unitary} irreducible representation of $L({p}^{\mu})$, there exists an {\it induced} representation of the Poincar{\'e} group $\cP(3,1)$.
\item The unitary irreducible representations of $\cP(3,1)$ are characterised by eigenvalues of the Casimirs $P^{2}$ and $W^{2}$.
\end{enumerate}

\end{theo}
\footnotetext{{The orbit $\cO({p}^{\mu})$ consists of all $4$-vectors $q^{\mu}$ for which there exists $\Lambda\in \mathrm{SO}(3,1)$ so that $\Lambda^{\mu}\,_{\nu}q^{\nu}=p^{\mu}$.}}

We will now consider the different representations determined by fixing the momenta for different values and signs of $P^\mu P_\mu$.\index{Poincare{\'e} group!Particle representations}
\begin{itemize}
\item $P^\mu P_\mu=m^2>0$ ({\bf Massive particles}).

Valid choices of eigenvectors include $p^{\mu} = (m , 0  , \ 0  , \ 0)$ which have rotations as their little group since $p^{i}=0$, $i=1,2,3$, i.e., $L(p^{\mu})=\mathrm{SO}(3)$. Due to the completely antisymmetric tensor $\ep_{\mu \nu \rho \sigma}$ in the definition of $W_{\mu}$, it follows
\begin{equation}
W_{0} \eq 0 \co W_{i} \eq -m \, J_{i} \quad\Longrightarrow\quad C_{2}=m^{2}J^{2}\ .
\end{equation}
Thus, we have
\begin{equation}
C_{1}\ket{m,\omega;\lambda_{i}}=m^{2}\ket{m,\omega;\lambda_{i}}\kom C_{2}\ket{m,\omega;\lambda_{i}}=m^{2}j(j+1)\,\ket{m,\omega;\lambda_{i}}\, ,
\end{equation}
This identifies $\omega$ with $j$, while the remaining labels are specified as $\lambda_{i}\in\lbrace p^{\mu},j_{3}\rbrace$. Note that once the $p^\mu$ are fixed within a representation the components of the Pauli-Ljubanski vector do satisfy an algebra (since within one representation we can replace $P^\mu$ by $p^\mu$ and in this case the algebra is essentially the same as the rotation group since $W_i\propto J_i$). This algebra defines the Little group that has the well known finite dimensional representations. Hence, every particle with nonzero mass is an irreducible representation of the Poincar\'e  group with labels $\ket{m,j;p^{\mu},j_{3}}$. This {\it defines} a one-particle state and, in particular, an \emph{elementary particle} of mass $m$ and spin $j$. 

It is important to emphasise that the existence of these quantum states corresponding to elementary particles is a general consequence of the two basic theories, quantum mechanics and special relativity. We may then {\bf define elementary particles as unitary irreducible representations of the Poincar\'e group}. This is a remarkable result since it is a way to mathematically define the basic building blocks of nature.\footnote{Next time when someone asks you ``What are we made of?'', you may simply answer: ``We and everything else we know in nature are made of  unitary representations of the Poincar\'e group!''}

\item $P^\mu P_\mu=0$ ({\bf Massless particles}).

The simplest realisation is selecting the origin $p^\mu=(0,0,0,0)$ which is Lorentz invariant. Even though this seems like a trivial case, it  corresponds to a state  with no particles, the vacuum state $|0\rangle$. We will see the importance of this state later on. 

In order to have non-trivial representations corresponding to particle states, we can  take the momentum of the form $p^{\mu} = (E  , \ 0  , \ 0  , \ E)$ which implies
\begin{equation}
(W_{0}, \ W_{1} , \ W_{2}, \ W_{3}) \eq E \, \bigl( J_{3}, \ -J_{1} \, + \, K_{2}, \ -J_{2} \, - \, K_{1}, \ -J_{3} \bigr)
\end{equation}
\begin{equation}
 \bigl[W_{1} \ , \ W_{2} \bigr] = 0 \kom \bigl[W_{3} \ , \ W_{1} \bigr] = -\I E \, W_{2} \kom \bigl[W_{3} \ , \ W_{2} \bigr] = \I E \, W_{1} \ .
\end{equation}
These commutation relations are those for the Euclidean group in two dimensions (translations generated by $W_1,W_2$ and rotations generated by $W_3$, acting on an abstract two dimensional space) and again define the Little group for massless particles. This group, contrary to the massive case, has infinite dimensional unitary representations known as {\it continuous spin representations}\index{Poincar{\'e} group!Continuous spin representations}. 

A simple way to see this is to realise that $W_1$ and $W_2$ commute with each other and can be simultaneously diagonalised with eigenvalues $w_1, w_2$. If $w_1,w_2\neq 0$, then 
\begin{equation}
W^\mu W_\mu =-(w_1^2+w_2^2):=-\rho^2 \quad \Rightarrow \quad w_1=\rho \cos\theta\kom w_2=\rho \sin \theta\, .
\end{equation}
Hence, the representation can be labelled as $\ket{0,\rho; p^\mu,\theta}$. Therefore, the existence of these representations would imply particles with an extra continuous label on top of the momenta $p^\mu$. Since particles with these extra continuous labels have not been seen in nature, in order to proceed we concentrate only on the finite dimensional representations.\footnote{This is the argument given in \cite{Weinberg:1995v1}. Originally Wigner \cite{Wigner:1939cj} had argued that these states should be ignored since their existence  would require the need of infinite heat capacity.} 

This is not entirely satisfactory since, contrary to the massive case in which we extracted the most general implications of special relativity and quantum mechanics without any further assumptions, here we have to make an {\it ad-hoc} restriction to concentrate only on \textbf{finite dimensional} representations. This may be one of the points that may need further study.\footnote{There has been recent interest to extract physical information of hypothetical physical states belonging to the continuous spin representations (see for instance \cite{Schuster:2023xqa} and references therein). Furthermore, it has been argued that these states should not be present in perturbative string theory \cite{Font:2013hia}. Any information that can be extracted about these states in either direction may be relevant in the future extensions of the Standard Model.} 

Restricting to finite dimensional representations, $\mathrm{SO}(2)$ is the relevant subgroup of the Little group generated by $W_3$ as $w_{1}$, $w_{2}$ vanish. In that case, $W^{\mu} = \la P^{\mu}$ and states are labelled as $|0,0;p^{\mu},\la \rangle := |p^{\mu} , \la \rangle$, where $\la$ is called \emph{helicity}\index{Helicity} and corresponds to the component of angular momentum in the direction of motion of the particle.  Since we have seen that it is only rotations by $4\pi$ and not $2\pi$ that leave the physics invariant, we should expect
\begin{equation}
\ee^{2\pi \I \la} \, |p^{\mu} , \la \rangle \eq \pm |p^{\mu} , \la \rangle
\end{equation}
which requires $\la$ to be integer or half integer $\la = 0,\frac{1}{2},1,\ldots$.
Notice that contrary to the massive case in which the integer or half-integer nature of spin was dictated by 
group theory, i.e., the representations of $\mathrm{SU}(2)$, for the massless case we need to use a topological argument related to the simply connected nature of   $\mathrm{SL}(2,\mathbb C)$. We will see that essentially all the particles of the Standard Model will come from these massless representations of the Poincar\'e  group, e.g., $\la = 0$ (Higgs), $\la = \frac{1}{2}$ (quarks, leptons), $\la = 1$ ($\ga$, $W^{\pm}$, $Z^{0}$, $g$) and $\la = 2$ (graviton).
We will see that parity transforms states of helicity $\lambda$, $|p^{\mu} , \la \rangle$ to $|p^{\mu} , -\la \rangle$ and therefore, if parity is conserved, states such as the photon and graviton have two degrees of freedom
corresponding to $\lambda=\pm 1$, $\lambda=\pm 2$ respectively.

\item $P^\mu P_\mu=-m^2<0$ ({\bf tachyons})\index{Tachyons}.

A typical momentum can be $p^\mu=(0,m,0,0)$. This would correspond to a particle moving in a space-like trajectory (moving faster than light).
In particular, it would contradict causality. In some cases, these particles appear in physical theories when instead of expanding around a minimum of the energy we expand around a maximum and their presence would only indicate that we are expanding around the 'wrong' vacuum. Once a minimum is identified and the expansion is done around the minimum of the energy the particle would correspond to a normal massive particle as described above.
We encounter such a situation further below when discussing the electroweak phase transition in Chapter~\ref{chap:ew}.
\end{itemize}

All in all, we deduce that the states for massless and massive particles are finite dimensional representations of SO$(3)$ (massive) or SO$(2)$ (massless).\footnote{Similar observations can be made for one-particle states in higher-dimensional theories where representations of SO$(D)$ for some $D$ play a crucial role.}
It is important to emphasise that the existence of the aforementioned quantum states corresponding to elementary particles is a general consequence of the two basic theories, quantum mechanics and special relativity. We may then define elementary particles as unitary irreducible representations of the Poincaré group. This is a remarkable result since it is a way to mathematically define the basic building blocks of nature.
Below, we use these results to introduce \emph{quantum fields} in an attempt to build up an off-shell framework to describe particle interactions.
But before we get there, we have to briefly discuss the effects of discrete spacetime transformations which, as we will see throughout these lectures, will also have important consequences for the Standard Model.

\section{Discrete spacetime transformations}

So far we have only considered the representations of the proper orthochronous Lorentz group.
Let us now consider the disconnected components of the Lorentz group and consider the action of parity and time reversal that, as we saw before, together with the identity and their product define the Klein group.\footnote{As we will see in the next few chapters these transformations are not symmetries of the Standard Model. The combination $CPT,$ where $C$ is the charge conjugation operator, is however an exact symmetry.}
In the following, we denote operators representing a general Poincar{\'e} transformation $\lbrace\Lambda |a\rbrace\in \cP(3,1)$ as $U(\Lambda,a)$.

The transformation matrices for parity $P$ and time reversal $T$ can be written
\begin{equation}
\Lambda_{P}=\left (\begin{array}{cccc}
1 & 0 & 0 & 0 \\ 
0 & -1 & 0 & 0 \\ 
0 & 0 & -1 & 0 \\ 
0 & 0 & 0 & -1
\end{array} \right )\kom\Lambda_{T}=\left (\begin{array}{cccc}
-1 & 0 & 0 & 0 \\ 
0 & 1 & 0 & 0 \\ 
0 & 0 & 1 & 0 \\ 
0 & 0 & 0 & 1
\end{array} \right )\, .
\end{equation}
As operators acting on the Hilbert space of quantum states, we denote them as
\begin{equation}
\cP=U(\Lambda_{P},0)\kom\cT=U(\Lambda_{T},0)\, .
\end{equation}
For a general Poincar{\'e} transformation $\lbrace\Lambda |a\rbrace\in \cP(3,1)$,
we have\index{Parity!Operator}\index{Time reversal!Operator}
\begin{align}
\cP^{-1} U\cP&=U(\Lambda_{P}^{-1}\Lambda\Lambda_{P},\Lambda_{P}a)\kom \cT^{-1} U\cT=U(\Lambda_{T}^{-1}\Lambda\Lambda_{T},\Lambda_{T}a)\, .
\end{align}
Expanding as before around the identity using
\begin{align}
\La^\mu\,_\nu&=\delta^\mu\,_\nu+\omega^\mu\,_\nu\kom a^\mu=\epsilon^\mu\kom \omega^\mu\,_\nu\kom \epsilon^\mu\ll 1\nn\\
U(\Lambda,a)&=\mathds{1}-\dfrac{\I}{2}\omega_{\mu\nu}M^{\mu\nu}+\I\epsilon_{\mu}P^{\mu}\, ,
\end{align}
and recalling that $M^{\mu\nu}$ transforms as a tensor under Lorentz transformations, i.e.,
\begin{equation}
\dfrac{\I\omega_{\rho\sigma}}{2}\cP^{-1} M^{\rho\sigma}\cP=\dfrac{\I\omega_{\rho\sigma}}{2}\Lambda^{\rho}_{P}\,_{\mu}\Lambda^{\sigma}_{P}\,_{\nu}M^{\mu\nu}\, ,
\end{equation}
we compute
\begin{align}
&\cP^{-1} J_{i}\cP=J_{i}\kom\cP^{-1} K_{i}\cP=-K_{i}\, ,\\
&\cP^{-1} P_{i}\cP=-P_{i}\kom\cP^{-1} P^{0}\cP=P^{0}\, .
\end{align}

This is as expected since under parity we expect that the $0$-th component of the vector $P^\mu$ ($P^0=E$) is invariant whereas the spatial components change sign. Also the angular momentum $J_i$ should be invariant (being an axial vector as in classical mechanics ${\bf{J=r\times p}}$). It is straightforward to see that the parity operator is indeed unitary.
However, if we follow the same procedure to obtain the transformations under time reversal we encounter a problem. If the time reversal operator is also unitary it would map
$ \cT^{-1} P^{0}\cT=-P^{0}$. That means it would change positive energies to negative energies that seems unphysical.
To interpret this result, we recall the following theorem due to Wigner:

\begin{theo}[Wigner]\index{Wigner's theorem}\index{Unitary operators}\index{Anti-unitary operators}\index{Wigner's theorem}

Transformations on a Hilbert space preserving probabilities are either
\begin{itemize}
\item unitary and linear, i.e.,
\begin{equation}
\braket{U\Phi|{U\Psi}}=\braket{\Phi|\Psi}\kom U(\alpha\ket{\Phi}+\beta\ket{\Psi})=\alpha U\ket{\Phi}+\beta U\ket{\Psi}\, ,
\end{equation}
\item \textbf{or} anti-unitary and anti-linear, that is,
\begin{equation}
 \braket{U\Phi|{U\Psi}}=\braket{\Phi|\Psi}^{*}\kom U(\alpha\ket{\Phi}+\beta\ket{\Psi})=\alpha^{*} U\ket{\Phi}+\beta^{*} U\ket{\Psi}\, .
 \end{equation} 
\end{itemize}

\end{theo}

Now, in order to preserve positive energies, $\cT$ must be an anti-unitary and anti-linear operator,
\begin{equation}
\cT(\I\ket{\Phi})=-\I\cT\ket{\Phi}\, ,
\end{equation}
so that
\begin{align}
&\cT^{-1} J_{i}\cT=-J_{i}\kom\cT^{-1} K_{i}\cT=K_{i}\, ,\\
&\cT^{-1} P_{i}\cT=-P_{i}\kom\cT^{-1} P^{0}\cT=P^{0}\, .
\end{align}
Then, the time reversal operator $\cT$ maps positive energies to positive energies.

Let us come back to our one-particle states defined in section~\ref{sec:RepresentationsOfPoincareGroup} and understand their transformation behaviour under parity and time reversal. One can show that (for more details, see section 2.5 in \cite{Weinberg:1995v1}):
\begin{itemize}
\item For massive particles, we have the transformation properties
\begin{equation}
\ket{m,j;p^{\mu},j_{3}}\begin{cases}
\xrightarrow{\cP}&\eta_{P}\;\ket{m,j;p'^{\mu},j_{3}}\, ,\\[0.5em]
\xrightarrow{\cT}&\eta_{T}\;(-1)^{j-j_{3}}\;\ket{m,j;p'^{\mu},-j_{3}}
\end{cases}
\end{equation}
with $|\eta_{P}|=|\eta_{T}|=1$ and $p'^{\mu}$ is the result of the corresponding transformation acting on $p^\mu$ (under parity the spatial components change sign, under time reversal only the time component change sign, etc.).
\item For massless particles, one finds
\begin{equation}
\ket{p^{\mu},\lambda}\begin{cases}
\xrightarrow{\cP}&\eta_{P}^{\prime}\;\ee^{\mp\I\pi\lambda}\;\ket{p'^{\mu},-\lambda}\, ,\\[0.5em]
\xrightarrow{\cT}&\eta_{T}^{\prime}\;\ee^{\pm\I\pi\lambda}\;\ket{p'^{\mu},\lambda}\, .
\end{cases}
\end{equation}
Note that for $\lambda\neq 0$ there has to exist the opposite helicity states. In particular,
\begin{itemize}
\item the photon and the graviton have $\lambda=\pm 1$, $\lambda=\pm 2$ respectively which means that for each of them the $\pm \lambda$ represent two states of the same particle, since both gravitation and electromagnetism are invariant under parity (both graviton and photon are their own antiparticle).
\item if the neutrino were massless, then $\lambda=\pm 1/2$ may have a different interpretation, for instance $\lambda=+1/2$ could be identified with the neutrino and $\lambda=-1/2$ with the antineutrino, since the weak interactions are {\emph not} invariant under parity. Even though, as we will see, the neutrinos are expected to have a mass, it is still an open question if the neutrinos are or are not their own anti-particles.
\end{itemize}
\end{itemize}
We conclude this section with some comments that will be very relevant in the next chapters:
\begin{itemize}
\item Massive particles of spin $j=1$ have $2j+1=3$ polarisation states, namely $j_{3}=-1,0,1$. In contrast, massless particles of helicity $\lambda=1$ have only $2$ polarisation states with $\lambda=\pm 1$.
\item Massive particles of spin $j=2$ have $2j+1=5$ polarisation states, namely $j_{3}=-2,-1,0,1,2$. In contrast, massless particles of helicity $\lambda=2$ have \textbf{still} only $2$ polarisation states with $\lambda=\pm 2$.
\end{itemize}


\chapter{\bf From Particles to Fields}
\label{chap:fields}

\vspace{.5cm}
\begin{equ}[Particles vs Fields]
{\it  If it turned out that some physical system could not be described by a quantum field theory, it would be a sensation; if it turned out that the system did not obey the rules of quantum mechanics and relativity, it would be  a cataclysm.}\\

\vspace{-0.3cm}

\rightline{\it Steven Weinberg}
\end{equ}

\vspace{0.45cm}

In the previous chapter we stressed that only assuming two fundamental theories of nature, special relativity and quantum mechanics, the fundamental physical entities are the elementary particles labeled by the quantum number specified by the representations of the Poincar\'e group $|m,j;p^\mu, j_3\rangle$ and $|p^\mu,\lambda\rangle $ describing massive and massless particles.  

This chapter is devoted to the study of interactions among these particles. The general requirements of Poincar\'e invariance, locality and unitarity  will let us to introduce \emph{fields} as ``functions'' of spacetime which are operators made out of creation and annihilation operators that create and destroy the corresponding particles. We emphasise that fields are only a tool to describe interactions among the particles. Their introduction include more conditions than just the assumptions of special relativity and quantum mechanics. However, fields are the key objects to describe interactions among particles and their  use goes beyond the study of interactions among particles. They are key ingredients in any interacting theory which requires local interactions such as condensed matter systems. So their use is across different disciplines. They are often presented as the basic objects of high energy physics with the particles appearing as their excitations. The two descriptions are somehow manifestations of the wave-particle duality of quantum mechanics. 

Our emphasis on particles rather than fields as the fundamental objects resides on the fact that it may be possible that some of the ingredients assumed in the introduction of fields may be overcome in future descriptions of nature beyond the Standard Model. The spirit of this course is not only to introduce the basic tools to describe the Standard Model, but also to identify the key ingredients and assumptions that may help in shaping formulations {\it beyond} the Standard Model. However, for the rest of the course, we will use the powerful tool of field theory.

\section{Particle interactions and  fields}\label{sec:particlestofields} 

In the preceding section, we learned about the concept of one-particle states.
Here, our aim is to describe interactions among many particles.
Putting together Poincar\'e invariance with the extra assumptions of unitarity and locality will lead us to superpositions of the aforementioned one-particle states corresponding to \emph{fields}.
These objects allow us to develop a formalism known as \emph{Quantum Field Theory} (QFT) that is suitable to describe local interactions among particle states.
 
\subsection{Many particle states}

Let us begin by trying to understand how we can describe relativistic processes in a quantum mechanical theory.
Clearly,
$E=mc^{2}$ tells us that mass and energy are on equal footing which further implies that particles can be annihilated into energy.
To describe such processes, we aim at combining, as we said several times before, Lorentz invariance of special relativity with the notion of quantum mechanics and add further conditions such as locality to describe interactions. To this end, we initially need to introduce the space of multi-particle states.

The Hilbert space $\cH$ of all particle states can be decomposed as
\begin{equation}
\cH=\cH_{0}\oplus\cH_{1}\oplus\cH_{2}\oplus\ldots
\end{equation}
where
\begin{itemize}
\item $\cH_{0}$ encodes $0$-particle states, i.e., the vacuum $\ket{0}$
\item $\cH_{1}$ includes $1$-particle states, e.g.,those generated by the creation operator $a^{\dagger}(p,\lambda)$ from the vacuum state,
\begin{equation}
\ket{p^{\mu},\lambda}=a^{\dagger}(p,\lambda)\ket{0}\, .
\end{equation}
\item $\cH_{2}$ includes $2$-particle states, e.g., 
\begin{equation}
\ket{p_{1}^{\mu},\lambda_{1};p_{2}^{\mu},\lambda_{2}}=a^{\dagger}(p_{2},\lambda_{2})\ket{p_1^{\mu},\lambda_1}=\pm\ket{p_{2}^{\mu},\lambda_{2};p_{1}^{\mu},\lambda_{1}}\, .
\end{equation}
Here, the $+$ sign refers to integer spin/helicity states ({\bf bosons}), whereas the $-$ sign\footnote{Note that this is one of the most important $-$ signs in science since it is the origin of the Pauli exclusion principle that implies that nuclei, atoms, molecules and therefore all matter, have a non-trivial structure.} to half-integer spin/helicity states ({\bf fermions}).
\item $\ldots$
\end{itemize}
As usual the creation and annihilation operators satisfy
\begin{align}\label{eq:CommutationRelAnGenOps} 
[a(p,\lambda),a(p^{\prime},\lambda^{\prime})]_\pm=[a^{\dagger}(p,\lambda),a^{\dagger}(p^{\prime},\lambda^{\prime})]_\pm&=0\, ,\nn\\
 [a(p,\lambda),a^{\dagger}(p^{\prime},\lambda^{\prime})]_\pm&\propto \delta(p-p')\delta_{\lambda\lambda'} 
\end{align}
for bosons $[\cdot\, ,\cdot]_-=[\cdot\, ,\cdot]$ which are the standard commutators, whereas for fermions we change commutators for anti-commutators which may be written as $[\cdot\, ,\cdot]_+=\{\cdot\, ,\cdot\}$.
In this way, we construct the full Hilbert space of particle states that we would like to describe with a dedicated quantum theory.

\subsection{Interactions and Fields}

\begin{figure}[t!]
\centering
\vspace*{0.5cm}
\begin{tikzpicture}[scale=1.4]
\begin{feynhand}
\vertex (a0) at (0,0.75); 
\vertex (a1) at (-1.75,0) {$|\alpha^{\text{in}}\rangle$ at $t=-\infty$}; 
\vertex (a10) at (0,0) {$\mathbf{\vdots}$}; 
\vertex (a3) at (0,-0.75); 
\vertex [NWblob] (b) at (2,0) {};
\vertex (c0) at (4,0.75);
\vertex (c1) at (5.75,0.) {$\langle\beta^{\text{out}}|$ at $t=+\infty$}; 
\vertex (c10) at (4,0.) {$\vdots$}; 
\vertex (c22) at (4,-0.75); 
\propag [fer] (a0) to (b);
\propag [fer] (a3) to (b);
\propag [fer] (b) to (c0);
\propag [fer] (b) to (c22);
\end{feynhand}
\end{tikzpicture}
\vspace*{0.5cm}
\caption{\index{Scattering}A cartoon representation of a scattering process.}\label{fig:scattering_cartoon} 
\end{figure}
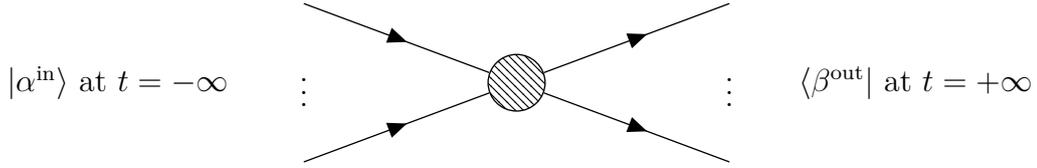

Interactions among many particle states are determined by computing the \emph{S-matrix}\index{S-matrix}. Typically, we can think of scattering processes as starting from an initial state $|\alpha^{\text{in}}\rangle$ at $t=-\infty$ and ending up with an out-state $\ket{\beta^{\text{out}}}$ at $t=+\infty$, see Fig.~\ref{fig:scattering_cartoon}. In between, particles interact in a complicated way by, e.g., colliding with each other or splitting up in a bunch of new particles. We define the S-matrix as
\begin{equation}
S_{\beta\alpha}=\braket{\beta^{\text{out}}|\alpha^{\text{in}}}=\delta_{\beta\alpha}+(2\pi)^4\delta(p_\beta-p_\alpha)\cM_{\beta\alpha}\, .
\end{equation}
The first term stands from the trivial event of no interactions at all. So the interesting physics is encoded in the \emph{Matrix Elements} $\cM_{\beta\alpha}$.

The standard questions we can ask for particle interactions are:
\begin{enumerate}
\item {\emph{\bf Decay Rates}.} The probability of decay of one particle to several particles. This is the simplest case in which the  $|\alpha^{\text{in}}\rangle$ state is one single particle. The $S$-matrix reduces to the probability of decay of the original particle to its daughter states, see Fig.~\ref{fig:decay_cartoon}. It is usually represented as $\Gamma (\alpha^{\text{in}}\rightarrow \beta^{\text{out}})$.  
The  decay rate per unit of phase space volume of the final states can be explicitly computed via
\begin{equation}
d\Gamma=\frac{d\Pi_{\text{LIPS}}}{2E_\alpha}\, |{\mathcal M}_{\beta\alpha} |^2
\end{equation} 
where $d\Pi_{\text{LIPS}}$ stands for Lorentz invariant phase space volume
\begin{equation}
d\Pi_{\text{LIPS}}\equiv (2\pi)^4 \delta^4(p^\mu_\alpha-\sum p)
\end{equation}
The important point for us is that $\Gamma$ is determined by $ |\cM_{\beta\alpha}|^2$ integrated and summed over all final momentum and spin states, see Appendix~\ref{app:drc} for details.

\begin{figure}[t!]
\centering
\vspace*{0.5cm}
\begin{tikzpicture}[scale=1.4]
\begin{feynhand}
\vertex (a1) at (-1.5,0) {$|\alpha^{\text{in}}\rangle$ at $t=-\infty$}; 
\vertex [NWblob] (b) at (2,0) {};
\vertex (c0) at (4,0.75);
\vertex (c1) at (5.75,0.) {$\langle\beta^{\text{out}}|$ at $t=+\infty$}; 
\vertex (c10) at (4,0.) {$\vdots$}; 
\vertex (c22) at (4,-0.75); 
\propag [fer] (a1) to (b);
\propag [fer] (b) to (c0);
\propag [fer] (b) to (c22);
\end{feynhand}
\end{tikzpicture}
\vspace*{0.5cm}
\caption{\index{Decay}A cartoon representation of a decay process.}\label{fig:decay_cartoon} 
\end{figure}
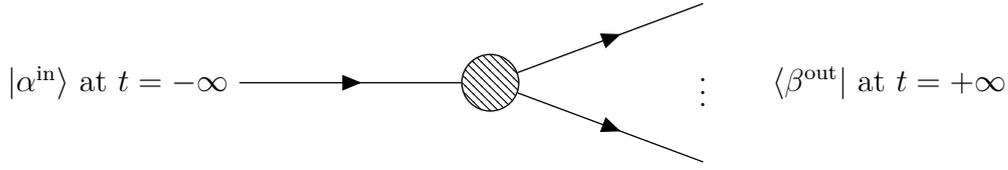

\item {\emph{\bf Cross Sections}.} As we know from Classical Mechanics, cross sections are the quantities that measure how often a scattering process between several particles happens. Cross sections are usually labelled as $\sigma$ and are also proportional to  $ |\cM_{\beta\alpha}|^2$, see Appendix~\ref{app:drc} for details.
\end{enumerate}
Therefore our goal is to find a way to determine $ |\cM_{\beta\alpha}|^2$ given a configuration of initial and final states interacting in a particular way.

In order to determine the matrix elements $ |\cM_{\beta\alpha}|^2$, in general, we require the following conditions on interactions:\index{Axioms of QFT}
\begin{enumerate}
\item \textbf{Poincar{\'e} invariance}\index{Poincar{\'e} invariance} of amplitudes. The S-matrix is invariant under Poincar{\'e} transformations.
\item \textbf{Unitarity}\index{Unitarity}: Probabilities add up to $1$ and are preserved under time evolution by $U=\ee^{-\I Ht}$. For the S-matrix, unitarity implies
\begin{equation}
\int\dif\beta\, S_{\beta\gamma}^{*}S_{\beta\alpha}=\int\dif\beta\, \braket{\gamma^{\text{in}}|\beta^{\text{out}}}\braket{\beta^{\text{out}}|\alpha^{\text{in}}}=\braket{\gamma^{\text{in}}|\alpha^{\text{in}}}=\delta(\alpha-\gamma)\, .
\end{equation}
or as an operator equation
\begin{equation}
S^\dagger S=1\, .
\end{equation}
\item \textbf{Stability}: This is the requirement that the energy should be bounded from below and there is a state of minimum energy, the vacuum $|0\rangle$.
\item \textbf{Locality}\index{Locality} (cluster decomposition)\index{Cluster decomposition}: 
Observables $\cO_1, \cO_2$ commute at space-like distances, i.e.,
\begin{equation}\label{eq:Locality} 
[\cO_1(x), \cO_2(y)]=0, \qquad (x-y)^2>0
\end{equation}
which is usually referred to as \emph{microcausality}\index{Microcausality}.

\end{enumerate}

As a consequence of the last requirement of locality,
we deduce that
the Hamiltonian $H$ is determined by a local function, the \emph{Hamiltonian density} $\cH(\mathbf{x},t)$, which is defined at each space point.
More precisely, we define the Hamiltonian as
\begin{equation}
H=\int\,\cH(\mathbf{x},t)\dif^{3}x
\end{equation}
corresponding to the sum of energies at each point in space. Similarly, the \emph{Lagrangian} $L$ is obtained from the \emph{Lagrangian density} $\cL(\mathbf{x},t)$ via
\begin{equation}
L=\int\, \cL(\mathbf{x},t)\dif^{3}x
\end{equation}
with associated \emph{action}
\begin{equation}
 S=\int\,\cL(\mathbf{x},t)\dif^{4}x\, .
\end{equation}
This locality requirement is crucial to determine interactions and is known as {\it cluster decomposition} which means that experiments performed at large enough distances do not affect each other. 

Now, we arrive at a conundrum: $\cH$ and $\cL$ are operators in position space, but particle states as derived above are defined in momentum space.
The way out is pretty obvious: we need to apply Fourier transformations to describe the corresponding states in terms of ``functions''\footnote{To be more precise, fields are not ordinary functions in the sense of conventional analysis. They are rather operator-valued objects or distributions, see e.g. \cite{Glimm:1987ylb}.} of spacetime coordinates -- objects that we call {\bf fields}. For any particle of given momentum and spin/helicity, we define a field $\Phi_{\alpha}$ as
\begin{equation}\label{eq:DefFieldInTermsOfAB} 
\Phi_{\alpha}(x^{\mu})=A_{\alpha}(x^{\mu})+B_{\alpha}^{*}(x^{\mu})
\end{equation}
in terms of
\begin{align}
A_{\alpha}(x^{\mu})&=\sum_\lambda\int\dif p\, u_{\alpha}(p,\lambda)\, a(p,\lambda)\,\ee^{\I p_{\mu}x^{\mu}}\, ,\\
B_{\alpha}^{*}(x^{\mu})&=\sum_\lambda\int\dif p\, v_{\alpha}(p,\lambda)\, b^{\dagger}(p,\lambda)\,\ee^{-\I p_{\mu}x^{\mu}}\, .
\end{align}
We use a multi-index notation where $\alpha = \mu_{1}\ldots \mu_{n}$ are spacetime indices of the corresponding representation under $\mathrm{SO}(3,1)$.
Here, the operators $a(p,\lambda)$ and $b^{\dagger}(p,\lambda)$ are raising and lowering operators as defined in the previous section with the commutation relations \eqref{eq:CommutationRelAnGenOps}.
The object $A$ is the field annihilating the corresponding particle, whereas $B$ is the field creating the anti-particle. 
Stated otherwise, fields are always of the form
\begin{equ}[Quantum Fields]
\begin{equation}\label{eq:FieldExpansionCreationAnnihi} 
\Phi_{\alpha}(x^{\mu})=\sum_\lambda\int\dif p\,\left ( u_{\alpha}(p,\lambda)\, a(p,\lambda)\, \ee^{\I p_{\mu}x^{\mu}}+ v_{\alpha}(p,\lambda)\,  b^{\dagger}(p,\lambda)\, \ee^{-\I p_{\mu}x^{\mu}} \right )\, .
\end{equation}
\end{equ}
For integer spin or helicity $n$, the object on the left transforms as some rank $n$ tensor under $\mathrm{SO}(1,3)^{\uparrow}$, that is, it transforms under the $(n/2,n/2)$ representation of the Lorentz algebra.
The integral and sum on the right is a superposition of all unitary irreducible representations of one-particle states as classified above via the little groups.
The parameter $\lambda$ labels spins for massive particles and helicity for massless particles, while the integration over momenta is performed using the invariant measure
\begin{equation}
\int\dif p=\int\dif^{4}p\, \delta(p^{2}-m^{2})=\int\dif^{3}p\dfrac{1}{2E_{p}(2\pi)^{3}} \kom E_p=p_0\kom E_p^2={\bf p}^2+m^2\, .
\end{equation}
The wave functions $u_{\alpha}(p,\lambda)$ and $v_{\alpha}(p,\lambda)$ in momentum space describe the dynamics of fields in spacetime given that they carry Lorentz indices $\alpha$. 
As we will in the next chapter, the functions $u_{\alpha}(p,\lambda)$ and $v_{\alpha}(p,\lambda)$ need to satisfy certain constraints in order to write down (off-shell) actions in terms of the fields.
The relation between the two sides in \eqref{eq:FieldExpansionCreationAnnihi} is determined by the \emph{coefficient functions} $\ee^{\I p_{\mu}x^{\mu}}\, u_{\alpha}(p,\lambda)$ and $\ee^{-\I p_{\mu}x^{\mu}}\, v_{\alpha}(p,\lambda)$ which carry both Lorentz indices $\alpha$ and $x$ as well as Poincar{\'e} representation labels ${p}$ and $\lambda$.

One import remark concern the form \eqref{eq:DefFieldInTermsOfAB} that we started with which ensures
the existence of anti-particles.
Let us explain this in more detail.
First, we note that writing the field $\Phi_{\alpha}(x^{\mu})$ in terms of $A_{\alpha}(x^{\mu})$ and $B_{\alpha}(x^{\mu})$ as in \eqref{eq:DefFieldInTermsOfAB} is essentially required by causality.
Above, we stated that all operators should commute at spacelike separations, cf.~Eq.~\eqref{eq:Locality}. That is, for a fixed time and two different locations, we demand that
\begin{equation}\label{eq:CausaField} 
\left[\Phi_{\alpha}(x^{i}, t),\Phi^*_{\alpha}(y^{i},t)\right]=0 \, .
\end{equation}
However, it is impossible for both $A_{\alpha}(x^{\mu})$ and $B_{\alpha}(x^{\mu})$ to satisfy this condition by themselves because
\begin{equation}
\left[A_{\alpha}(x^{i}, t),A^*_{\alpha}(y^{i},t)\right]\neq 0\, .
\end{equation}
These commutators can be explicitly derived from those for the creation and annihilation operators $a(p,\lambda)$ and $a^{\dagger}(p,\lambda)$ as defined in \eqref{eq:CommutationRelAnGenOps}.
Thus, both objects are needed to build fields $\Phi_{\alpha}(x^{\mu})$ satisfying \eqref{eq:CausaField}.
This is the requirement for the existence of \emph{anti-particles}.
If $a(p,\lambda)=b(p,\lambda)$, the particle is simply its own anti-particle.

\subsection{Field theories and their actions}

So far, we have seen that putting special relativity and quantum mechanics together lead us to classifying one-particle states in terms of their masses and spins.\footnote{Helicity is the appropriate term for massless particles. Nonetheless, one usually talks about \emph{spin} even in the case of massless states keeping in mind that the degrees of freedom are counted differently.}
Interactions lead us to the concept of locality and to use fields rather than particles to describe our theory. In finding all unitary irreducible representations of the Poincar{\'e} group, we defined states for fixed $p^{\mu}$ over which we have to integrate to get a suitable superposition of eigenstates, i.e., we found fields of the form
\begin{equation}
\Phi_{\alpha}(x^{\mu})=\sum_\lambda\int\dif p\,\left ( u_{\alpha}(p,\lambda)\, a(p,\lambda)\, \ee^{\I p_{\mu}x^{\mu}}+ v_{\alpha}(p,\lambda)\,  b^{\dagger}(p,\lambda)\, \ee^{-\I p_{\mu}x^{\mu}} \right )\, .
\end{equation}
The action $S$ becomes a function of these fields $\Phi_{\alpha}$ and their derivatives, i.e.,
\begin{equation}
S[\Phi_{\alpha},\p\Phi_{\alpha}]= \int\,\dif^{4}x\, \cL[\Phi_{\alpha},\p\Phi_{\alpha}]
\end{equation}
where $\cL[\Phi_{\alpha},\p\Phi_{\alpha}]$ is the Lagrange density or simply \emph{Lagrangian} of the theory.
Here, translation invariance forbids the explicit dependence of $\cL$ on the coordinates $x^{\mu}$, i.e., $\cL\neq \cL[\Phi_{\alpha},\p\Phi_{\alpha},x^{\mu}]$.
The Lagrangian is typically written as a sum of individual terms of the form
\begin{equation}\label{eq:LagGen} 
\cL[\Phi_{\alpha},\p\Phi_{\alpha}]=\sum_{i}\, c_{i}\, \cO_{i}(\Phi_{\alpha},\p\Phi_{\alpha})\, .
\end{equation}
Here, $c_{i}$ are some ``constant'' coefficients and $\cO_{i}$ are referred to as \emph{operators} since they are functions of the fields  $\Phi_{\alpha}$ which are themselves operators (as it can be seen from their dependence on the creation and annihilation operators). 
The equations of motion for $\Phi_{\alpha}$ are obtained as usual from the Euler-Lagrange equations, i.e.,
\begin{equation}\label{eq:EulerLagrangeEquation} 
\p_{\mu} \dfrac{\p\cL}{\p \p_{\mu} \Phi_{\alpha}}-\dfrac{\p\cL}{\p\Phi_{\alpha}} = 0\, .
\end{equation}
When the field configuration satisfies these classical equations, we say that the field is on its mass shell or {\it on-shell}. Otherwise we say it is {\it off-shell}. 
Quantisation of field theories proceeds most easily through the path integral approach where e.g. the partition function can be written as
\begin{equation}\label{eq:PathIntegral} 
\mathcal{Z} = \int\cD\Phi_{\alpha}\, \ee^{-S[\Phi_{\alpha},\p\Phi_{\alpha}]}\, .
\end{equation}
Only in the classical limit the on-shell condition is satisfied. Similarly, correlations functions and amplitudes can be straightforwardly computed within this formalism through perturbation theory.
Note that there is an infinite-to-one mapping from actions to on-shell scattering amplitudes: infinitely many actions can give rise to the same on-shell amplitude due to field redefinitions $\Phi\rightarrow \Phi'(\Phi)$.
This begs the question: what is the point of introducing fields in the first place?
First and foremost,
they provide us with an organising principle for interactions among particles governed by symmetries.
What is more,
non-perturbative effects, running couplings as well as off-shell correlation functions can be systematically studied.
As Weinberg himself stresses in \cite{Weinberg:1996kw},
quantum fields are ``the only way of satisfying the principles of Lorentz invariance plus quantum mechanics plus cluster decomposition''.\index{Cluster decomposition} Recently, though, there has been much effort towards computing amplitudes directly without the use of Lagrangians.
For details, we refer to \cite{Weinberg:1996kr,Schwartz:2014sze}.

Let us now provide examples of free field theories focussing on spin/helicity states less than one for which the massless and massive states have the same number of degrees of freedom:
\begin{itemize} 
\item {\it Free scalar field {\rm (}spin/helicity $0${\rm )}}. 
In order to create and annihilate spinless particles either massive $|m,j=0;p^\mu,j_3=0\rangle $ or massless $|p^\mu,\lambda=0\rangle $ we introduce a Lorentz scalar field $\phi(x^\mu)$ satisfying classically the \emph{Klein-Gordon equation}\index{Klein-Gordon equation} 
\begin{equation}
(\partial^\mu\partial_\mu +m^2)\phi=0
\end{equation}
which is nothing but the Fourier transformation of the \emph{on-shell} condition $p^\mu p_\mu=m^2$ in momentum space. The expansion of $\phi$ in creation and annihilation operators
obtained from \eqref{eq:FieldExpansionCreationAnnihi} reads
\begin{equation}
\phi(x)=\int\dif p\,\left (  a(p)\, \ee^{\I p_{\mu}x^{\mu}}+  a^{\dagger}(p)\, \ee^{-\I p_{\mu}x^{\mu}}\right )\, .
\end{equation}
The Lagrangian density that reproduces the Klein-Gordon equation by plugging it into the Euler-Lagrange equations \eqref{eq:EulerLagrangeEquation} is
\begin{equation}
\cL= \frac{1}{2}\partial^\mu\phi\partial_\mu \phi - \frac{1}{2}m^2\phi^2\, .
\end{equation}
Notice that the single degree of freedom of a free scalar matches the single one-particle state in both cases, massive and massless. Since the field is real, the corresponding particle will be its own anti-particle. Extending to a complex scalar $\Phi=\phi_1+ i \phi_2$ with $\phi_{1,2}$ real scalar fields, is straightforward and they will correspond to two one-particle states: the particle and its anti-particle.

\item{\it Free spin {\rm (}helicity{\rm )} $\frac{1}{2}$ fermion}.
The one-particle states are now  for the massive case $|m,j=\frac{1}{2};p^\mu,j_3=\pm \frac{1}{2}\rangle$ and $|p^\mu,\lambda=\pm \frac{1}{2}\rangle $ for the massless case. The corresponding field could either be a left-handed  $\psi_L$ or right-handed $\psi_R$ spinor. The free Dirac Lagrangian containing both fields to include a natural mass term is
\begin{equation}
\cL=\overline{\psi}\,\I\cancel{\p}\psi-m\overline{\psi}\psi
=\overline{\psi}_{L}\,\I\cancel{\p}\psi_{L}+\overline{\psi}_{R}\,\I\cancel{\p}\psi_{R}-m\left (\overline{\psi}_{R}\,\psi_{L}+\overline{\psi}_{L}\,\psi_{R}\right )\, .
\end{equation}
Here, for completeness, we also wrote the corresponding Dirac spinor $\psi$ satisfying the standard \emph{Dirac equation}\index{Dirac equation}
\begin{equation}
(\I\cancel{\p}-m)\psi=0\, .
\end{equation}
Again, this can be obtained from the on-shell mass relation $p^\mu p_\mu=m^2$ after using some algebra for the $\gamma$-matrices.

Following the general expression \eqref{eq:FieldExpansionCreationAnnihi} the Dirac field can be written as
\begin{equation}
\psi(x^\mu)=\sum_{\lambda=\pm \frac12}\int\dif p\,\left ( u(p,\lambda)\, a(p,\lambda)\, \ee^{\I p_{\mu}x^{\mu}}+ v(p,\lambda)\,  b^{\dagger}(p,\lambda)\, \ee^{-\I p_{\mu}x^{\mu}} \right )\, .
\end{equation}
We have omitted spinorial indices in $\psi(x^\mu)$ and the momentum-space   wave functions $u(p^\mu),v(p^\mu)$. Here, on-shell $u,v$ satisfy the Dirac equation in momentum space 
$(\cancel{p}-m)u=(\cancel{p}+m)v=0$.
Note that for both massless and massive one-particle states, the corresponding multiplet has two spin states. Adding the same for the corresponding anti-particle, we are left with four independent degrees of freedom (two spin states for each particle and anti-particle) matching the four independent degrees of freedom for a Weyl spinor (two complex-component spinor).

\end{itemize}
Higher spin/helicity states will be discussed below.
More work is required to write down their actions because we need to be careful about additional constraints that have to be imposed to account for the correct number of physical degrees of freedom.
It turns out that the notion of \emph{symmetries} will be crucial for this process which we introduce in the next section.


\section{Symmetries in QFT}\label{sec:symmetriesQFT}

A guiding principle to understand the structure of quantum field theories are \emph{symmetries}, that is, transformations of the fields and spacetime coordinates that leave physics invariant.
Up to this point, we have talked extensively about the Poincar\'e group acting on spacetime coordinates.
On the level of the action, we talk about \emph{Poincar\'e invariance}\index{Poincar\'e invariance} as the statement that Poincar\'e transformations should leave the action invariant (possibly up to total derivatives).
But there are various other notions of symmetries that play a pivotal role in constructing general field theories.
Here, we give a brief summary of the most relevant types for understanding Standard Model physics.

\subsection{Coleman-Mandula theorem}

The celebrated Coleman-Mandula theorem\index{Coleman-Mandula theorem} states that the most general symmetries of the $S$-matrix are of the form
\begin{equation}
\text{Spacetime }\otimes \text{ Internal}\, .
\end{equation}
The left hand side is given by the Poincar{\'e} group\footnote{This symmetry can be extended by introducing anti-commuting generators $Q^{I}_{\alpha}$ with $\lbrace Q_{\alpha}^{I},\overline{Q}_{\dot{\alpha}}^{I}\rbrace=2\sigma^{\mu}_{\alpha\dot{\alpha}}P_{\mu}$ in terms of the anti-commutator $\lbrace\cdot,\cdot\rbrace$. This leads to the concept of \emph{supersymmetry}\index{Supersymmetry}. The corresponding representations lead to multiplets including fields of different spin and the fact that $N_{\text{bosons}}=N_{\text{fermions}}$. The latter has however not (yet) been observed in nature.} with generators $P^{\mu}$, $M^{\mu\nu}$ and has been studied in detail in the previous chapter. We recall that under a general Poincar{\'e} transformation $\lbrace\Lambda |a\rbrace\in\cP(3,1)$ the states $\ket{\psi}$ in our Hilbert space $\cH$ transform with respect to some operator $\mathrm{U}(\Lambda,a)$, $\Lambda\in\mathrm{SO}(3,1)$, $a\in\bR^{3,1}$, such that
\begin{equation}
\ket{\psi}\raw U(\Lambda,a)\ket{\psi}\, .
\end{equation}
Operators $\cO_{i}$ in our theory transform according to
\begin{equation}
\cO_{i}\raw U^{\dagger}(\Lambda,a)\cO_{i}U(\Lambda,a)\, .
\end{equation}
We require that these representations are unitary, i.e.,
\begin{equation}
\braket{\psi_{1}|\psi_{2}}\raw \braket{\psi_{1}|U^{\dagger}U|\psi_{2}}=\braket{\psi_{1}|\psi_{2}}\kom \ket{\psi_{1}},\ket{\psi_{2}}\in \cH\, .
\end{equation}
This can also be realised for the parity operator, while the time reversal operator needs to be anti-unitary, $\cT^{\dagger}\cT=-1$. 
For a field $\Phi_\alpha (x)$, with the index $\alpha$ specifying the corresponding representation of the Lorentz group, this implies that
\begin{equation}
\Phi_\alpha(x)\rightarrow U^{\dagger}(\Lambda,a) \Phi_\alpha (x^{\prime}) U(\Lambda,a)\,= D_{\alpha}\,^{\beta}\, \Phi_\beta(x')
\end{equation}
where $D_{\alpha}\,^{\beta}$ are representation matrices for the Lorentz group. Note the dependence on $x'=\Lambda x+a$ rather than $x$ on the right hand side.

Next, we consider the \emph{internal symmetries}\index{Internal symmetries} for which local operators transform according to
\begin{equation}
\cO_{i}(x)\raw \cO_{i}^{\prime}(x)\, .
\end{equation}
We stress that, contrary to the case of spacetime symmetries, the operators are evaluated \emph{at the same spacetime point} $x^{\mu}$ on both the left and right hand side. If our theory is invariant under such a transformation, then it is called \emph{internal} and the corresponding transformation $U$ commutes with the Hamiltonian
\begin{equation}
\bigl [H,U\bigl ]=0\, .
\end{equation}
Under an internal transformation $g(\alpha_{a})\in G$ with parameters $\alpha_a$, $a=1,\ldots,\text{dim}(G)$, a field $\Phi_\alpha^{i}(x)$ transforms as
\begin{equation}\label{eq:TransformationInternalSymmetry} 
\Phi_\alpha^{i}(x)\rightarrow U^{\dagger}(\alpha_{a}) \Phi_\alpha^{i} (x) U(\alpha_{a})\,= g^{i}\,_{j}\, \Phi_\alpha^{j}(x)
\end{equation}
with $g^{i}\,_{j}$ matrices representing the internal symmetry group $G$.
Above, the indices $i, j,\ldots$ are associated with the representations under $G$ which have nothing to do with the indices $\alpha,\beta,\ldots$ coming from the representation of Lorentz group.
Hence, the left and right hand side of \eqref{eq:TransformationInternalSymmetry} have both the same index $\alpha$ because the internal transformation does not act on the spacetime components of the field.
If the action $S$ is invariant under those transformations, we speak of symmetries of $\cL$ (up to a total derivative).

The Coleman-Mandula theorem is very important in the sense that the fact that the most general symmetry is a direct product (and not a semi-direct or other combination) it forbids non-trivial combinations of internal and spacetime symmetries. In particular it guarantees that particle states are still labeled by mass and  spin/helicity and the internal symmetries could only add extra labels such as electric charge.\footnote{This theorem was generalised by Haag-Lopuszanski and Sohnius  to include supersymmetry which is a spacetime and not internal symmetry. Supersymmetry implies that particles of different spins can be in the same multiplet, see e.g. \cite{Quevedo:2010ui}.}

\subsection{Examples of Internal Symmetries}

In order to be more explicit regarding internal symmetries, let us consider the Lagrangian for
\begin{itemize}
\item A massive Dirac spinor with
\begin{equation}\label{eq:DiracSpinorLag} 
\cL=\overline{\psi}\,\I\cancel{\p}\psi-m\overline{\psi}\psi
\end{equation}
which can be written in terms of the left- and right-handed Weyl spinors as
\begin{equation}
\cL=\overline{\psi}_{L}\,\I\cancel{\p}\psi_{L}+\overline{\psi}_{R}\,\I\cancel{\p}\psi_{R}-m\left (\overline{\psi}_{R}\,\psi_{L}+\overline{\psi}_{L}\,\psi_{R}\right )\, .
\end{equation}
In the limit $m\raw 0$, $\cL$ is invariant under a symmetry group $\mathrm{U}(1)_{L}\times \mathrm{U}(1)_{R}$ since both field can be independently altered by a phase,
\begin{equation}
\psi_{L}\raw \ee^{\I\alpha_{L}}\psi_{L}\kom \psi_{R}\raw \ee^{\I\alpha_{R}}\psi_{R}\, ,
\end{equation}
which is referred to as \emph{chiral symmetry}\index{Chiral symmetry}. In the massive case, the Lagrangian is only invariant under a single $\mathrm{U}(1)_{V}$ with $\alpha_{V}=\alpha_{L}=\alpha_{R}$, i.e., 
\begin{equation}
\psi_{L}\raw \ee^{\I\alpha_{V}}\psi_{L}\kom \psi_{R}\raw \ee^{\I\alpha_{V}}\psi_{R}\, .
\end{equation}
\item A massive complex scalar field with quartic interactions
\begin{equation}
\cL=\p^{\mu}\phi\p_{\mu}\phi^{*}-m^{2}|\phi|^{2}-\lambda|\phi|^{4}\, .
\end{equation}
This Lagrangian is invariant under $\mathrm{U}(1)$-transformations of the form
\begin{equation}
\phi\raw\ee^{\I e\alpha}\phi\, .
\end{equation}
\item A massive scalar in the fundamental representation of $\mathrm{SU}(2)$
\begin{equation}\label{eq:SU2LagrangianScalarExample} 
\cL=\p^{\mu}\phi\p_{\mu}\phi^{\dagger}-m^{2}|\phi|^{2}\kom \phi=\left (\begin{array}{c}
\phi^{1} \\ 
\phi^{2}
\end{array} \right )\, .
\end{equation}
Here, $\phi^{1}$ and $\phi^{2}$ are two complex-valued scalar fields.
In those components, the action reads
\begin{equation}
\cL = \sum_{i=1}^{2}\, \left (\p^{\mu}\phi^{i}\p_{\mu}(\phi^{i})^{*}-m^{2}|\phi^{i}|^{2}\right )\, .
\end{equation}
For an $\mathrm{SU}(2)$ transformation
\begin{equation}
U(\alpha_{a}) = \exp\left (-\dfrac{\I}{2}\alpha_{a}\sigma^{a}\right )\in\mathrm{SU}(2)\kom \alpha_{a}\in\bR\kom a=1,2,3
\end{equation}
in terms of the Pauli matrices $\sigma^{a}$, recall \eqref{eq:PauliMatrices} (without the identity), 
we clearly see that \eqref{eq:SU2LagrangianScalarExample} is invariant since $UU^{\dagger}=\mathds{1}_{2}$.
\end{itemize}

\subsection{Types of symmetries}

There are many types of internal symmetries that we will consider next. In general we distinguish the following types of symmetries:
\begin{enumerate}
\item {\it Spacetime or internal}. As we have already mentioned above, a field $\Phi_\alpha (x)$ transforms under a spacetime transformation $x'=\Lambda x+a$ according to
\begin{equation}
\Phi_\alpha^{i}(x)\rightarrow U^{\dagger}(\Lambda,a) \Phi_\alpha^{i}(x^{\prime}) U(\Lambda,a)\,= D_{\alpha}\,^{\beta}\, \Phi_\beta^{i}(x')
\end{equation}
where $D_{\alpha}\,^{\beta}$ are representation matrices for the Lorentz group. 
Under an internal transformation $g(\alpha_{a})\in G$, $a=1,\ldots,\text{dim}(G)$, a field $\Phi_\alpha^{i}(x)$ transforms instead through
\begin{equation}
\Phi_\alpha^{i}(x)\rightarrow U^{\dagger}(\alpha_{a}) \Phi_\alpha^{i} (x) U(\alpha_{a})\,= g^{i}\,_{j}\, \Phi_\alpha^{j}(x)
\end{equation}
where $g^{i}\,_{j}$ are representation matrices of the internal symmetry group $G$.
\item {\it Continuous or discrete}. E.g., $\phi\raw\ee^{\I\alpha}\phi$ for a $\mathrm{U}(1)$ or $\phi\raw -\phi$ corresponding to a $\bZ_{2}$ symmetry. For instance, consider a real scalar field with
\begin{equation}
\cL=\p_{\mu}\phi\p^{\mu}\phi-m^{2}\phi^{2}+g\phi^{3}-\lambda\phi^{4}\, .
\end{equation}
Imposing the $\bZ_{2}$-symmetry $\phi\raw -\phi$ forbids the presence of the cubic term $\sim g\phi^{3}$. On the other hand, for a $\mathrm{U}(1)$-symmetry $\phi\raw \ee^{\I e \alpha}\phi$ of a complex scalar $\phi$, we have a Lagrangian of the form
\begin{equation}\label{eq:LagGlobalU1} 
\cL=\p_{\mu}\phi\p^{\mu}\phi^{*}-m^{2}|\phi|^{2}-\lambda|\phi|^{4}\, .
\end{equation}
The potential for $m^{2}>0$ is depicted on the left and for the case $m^2<0$ on the right in Fig.~\ref{fig:PotentialCompScalar}.

\begin{figure}[t!]
\centering
\includegraphics[scale=0.8]{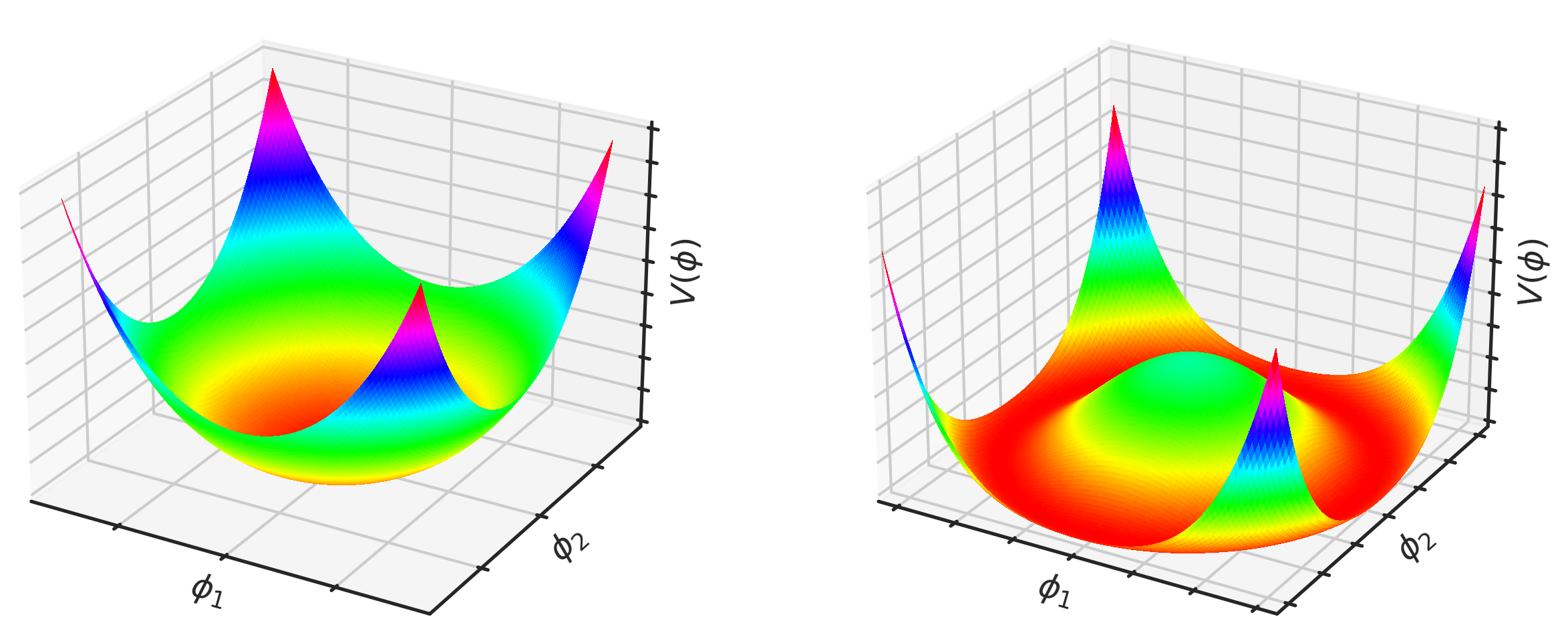}
\caption{\emph{Left:} Plot of the scalar potential \eqref{eq:LagGlobalU1} in the unbroken phase for $m^{2}>0$ with a minimum at the origin. The potential clearly respects the proposed $\mathrm{U}(1)$-symmetry.
\emph{Right:} Spontaneously broken phase for $m^{2}<0$ in \eqref{eq:LagGlobalU1} with a maximum at the origin. Projection onto the $\phi$-plane shows that the $\mathrm{U}(1)$-symmetry is intact even in the ``broken'' phase.}\label{fig:PotentialCompScalar}
\end{figure}

\item {\it Global or local}. In the case of a $\mathrm{U}(1)$-symmetry $\phi\raw \ee^{\I e\alpha}\phi$ , we distinguish 
\begin{equation*}
\begin{cases}
\alpha=\text{const.} &\text{global symmetry}\, ,\\
\alpha=\alpha(x) &\text{local symmetry}\, .
\end{cases}
\end{equation*}
The latter case of a local symmetry leads to a modification of \eqref{eq:LagGlobalU1}, since the original kinetic term is not invariant $\p_\mu\phi \rightarrow  \ee^{\I e\alpha}( \p \phi +\I \p \alpha \phi)$. A possible modification of the Lagrangian is to introduce a new field $A_\mu$ such that its transformation compensates for the lack of invariance of the kinetic term.
That is, we write instead
\begin{equation}
\cL=D_{\mu}\phi D^{\mu}\phi^{*}-m^{2}|\phi|^{2}-\lambda|\phi|^{4}\, .
\end{equation}
where we define the \emph{covariant derivative} with an additional term $\sim A_{\mu}\phi$ as
\begin{equation}
D_{\mu}\phi=\p_{\mu}\phi-\I eA_{\mu}\phi\, .
\end{equation}
Under the $\mathrm{U}(1)$-action, $A_{\mu}$ transforms as
\begin{equation}
A_{\mu}\raw A_{\mu}+\p_{\mu}\alpha
\end{equation}
and $\cL$ is modified in order to add kinetic terms for the $A_\mu$ fields according to
\begin{equation}
\cL'=\cL+F^{\mu\nu}F_{\mu\nu}\kom F_{\mu\nu}=\p_{\mu}A_{\nu}-\p_{\nu}A_{\mu}\, .
\end{equation}
The interactions between $A_{\mu}$ and $\phi$ are hidden in the kinetic term $D_{\mu}\phi D^{\mu}\phi^{*}$ for $\phi$. Also for Dirac fields, the Lagrangian \eqref{eq:DiracSpinorLag} gets modified for a local $\mathrm{U}(1)$ such that
\begin{equation}
\cL=\overline{\psi}\,\cancel{D}\psi-m\overline{\psi}\psi
\end{equation}
where the kinetic term is again modified to
\begin{equation}
\bar{\psi}\,\I\cancel{D}\psi=\bar{\psi}\, \I\gamma^{\mu}D_{\mu}\psi\kom D_{\mu}=\p_{\mu}-\I eA_{\mu}\, .
\end{equation}
This can be seen by considering the transformation property of $\overline{\psi}\,\I\cancel{\p}\psi$, that is,
\begin{equation}
\psi(x)\raw \ee^{\I e\alpha(x)}\psi(x)\quad\Rightarrow\quad \overline{\psi}\,\I\cancel{\p}\psi\raw \overline{\psi}\,\I\cancel{\p}\psi-e(\overline{\psi}\,\gamma^{\mu}\psi)\,\p_{\mu}\alpha
\end{equation}
which is clearly not invariant. On the other hand, the modified kinetic term is indeed gauge invariant,
\begin{equation}
\bar{\psi}\,\I\cancel{D}\psi\raw \overline{\psi}\,\I\cancel{\p}\psi-e(\overline{\psi}\,\gamma^{\mu}\psi)\,\p_{\mu}\alpha+e\overline{\psi}\,\cancel{A}\psi+e(\overline{\psi}\,\gamma^{\mu}\psi)\, \p_{\mu}\alpha=\bar{\psi}\,\I\cancel{D}\psi\, .
\end{equation}
We will see later that these Lagrangians describe interactions of matter fields $\phi, \psi$ to spin 1 fields $A_\mu$ like the photon.
\item {\it Manifest or hidden}. A symmetry is manifest if the vacuum state, or state of minimum energy, shares the same symmetries of the theory. Fig.~\ref{fig:PotentialCompScalar} shows on the left the scalar potential with a $U(1)$ symmetry where the vacuum state is the origin which respects the symmetry.  On the other hand, a symmetry is hidden if the vacuum state does not share the symmetry of the Lagrangian. Hidden symmetries are usually called \emph{spontaneously broken}\index{SSB} which may be a misleading term. The symmetry is not actually broken, but just not respected by the vacuum state. For an observer living in such a vacuum, the symmetry is in that sense hidden. However, even though the observer can hardly detect the symmetry, it remains intact from a global point of view, see the plot on the right of Fig.~\ref{fig:PotentialCompScalar}.
\item {\it Anomalous\index{Anomalous symmetry} or non-anomalous (exact)}. A symmetry is referred to as being anomalous whenever it is realised in the classical theory, but gets broken by quantum corrections. In that sense, the notion of anomalies is crucial in understanding the quantum theory. Not surprisingly, the Standard Model itself is free of (gauge) anomalies, which is a very strong consistency test, see section~\ref{sec:sym_gauge_anom}.
In general, the concept of anomalies is a well established guiding principle in building new theoretical models of particle physics and quantum gravity.


\item {\it Real or accidental}. It may happen that the Lagrangian consistent with a set of symmetries may accidentally have extra symmetries that were not imposed on it. This will happen in the Standard Model with symmetries such as the conservation of lepton or baryon number.
\item {\it Compact or non-compact}. The Poincar{\'e} group is non-compact which is forced upon us by special relativity. Since quantum mechanics tells us to work with unitary representations, we are naturally lead to work with infinite dimensional representations of non-compact groups. On the other hand, we typically restrict to compact internal symmetry groups which allows us to restrict to finite dimensional representations.
\item {\it Abelian or non-Abelian}. E.g., $\mathrm{U}(1)$ or $\mathrm{SU}(N)$, $\mathrm{SO}(N)$, $\mathrm{Sp}(2N)$, $\mathrm{G}_{2}$, $\mathrm{F}_{4}$, $\mathrm{E}_{6}$, $\mathrm{E}_{8}$. All these groups are characterised by Dynkin diagrams in the Cartan classification of simple Lie groups.
\end{enumerate}

\subsection{Noether's theorem}

An important result related to the presence of symmetries in a theory is \emph{Noether's theorem}\index{Noether's theorem}. 
\begin{equ}[Noether's Theorem]
If the action  $S=\int d^4x \, \cL[\Phi_{\alpha},\p\Phi_{\alpha}]$ has a continuous symmetry for $\Phi_{\alpha}\raw \Phi^{\prime}_{\alpha}$, then there exists a current $J^{\mu}$ that is conserved when the field equations are satisfied (sometimes referred to as the current is conserved \emph{on-shell}), that is,
$\p^{\mu}J_{\mu}=0\, $ ,
and the corresponding charge
$Q=\int\dif^{3}x\, J^{0}\, $, 
is a constant of motion ($\dif Q/\dif t=0$).
\end{equ}

Let us review the arguments entering the proof.
We consider a general field transformation
\begin{equation}
\Phi_\alpha\rightarrow \Phi_\alpha+ \Delta_\alpha
\end{equation}
that leaves the action invariant, i.e.,
\begin{equation}
\delta S=0\quad  \Rightarrow \quad \delta \cL=\partial_\mu F^\mu(\Phi_\alpha)
\end{equation}
where $F^\mu$ are some arbitrary functions so that the Lagrangian density transforms as a total derivative.
The variation of the Lagrangian is then given by
\begin{equation}
\delta \cL=\left[\frac{\partial\cL}{\partial\Phi_\alpha}-\partial_\mu\frac{\partial\cL}{\partial(\partial_\mu\Phi_\alpha)}\right]\,  \Delta_\alpha+\partial_\mu\left(\frac{\partial\cL}{\partial(\partial_\mu\Phi_\alpha)}\, \Delta_\alpha\right)
\end{equation}
Then, as long as the equations of motion are satisfied, the first term cancels so that
\begin{equation}
J^\mu=\frac{\partial\cL}{\partial(\partial_\mu\Phi_\alpha)}\, \Delta_\alpha-F^\mu, \qquad \partial_\mu J^\mu=0\, .
\end{equation}
If the Langrangian density (and not only the action) is invariant under the symmetry transformations  $\delta \cL=0$, then
\begin{equation}
J^\mu=\frac{\partial\cL}{\partial(\partial_\mu\Phi_\alpha)}\, \Delta_\alpha
\end{equation}
and 
\begin{equation}\label{eq:NoetherCharge} 
Q\,=\, \int d^3x\, J^0\, =\, \int\dif^{3}x\, \dfrac{\p \cL}{\p\dot{\Phi}_\alpha}\Delta_\alpha\, .
\end{equation}
This charge $Q$ is conserved since it satisfies
\begin{equation}
\dfrac{d Q}{d t}=\int \dif^{3}x\, \p_{t} J^{0}=-\int \dif^{3}x\, \nabla\cdot\mathbf{J}=0
\end{equation}
where in the last step we have assumed that the current $\mathbf{J}$ vanishes at spatial infinity and falls off fast enough.

Noether's theorem expresses the importance of symmetries for physical observables.
It essentially states that the existence of a continuous symmetry implies a conservation law. We may identify the most important physical quantities such as energy, momenta, electric charge, etc., as those that are conserved due to the existence of symmetries: time translations for energy, space translations for momenta, rotations for angular momenta. For electric charge the corresponding symmetry is an internal $U(1)$ symmetry.
Similar observations hold e.g. for baryon and lepton number, although their associated $U(1)$ symmetry is global, whereas for electric charge it is local. We will discuss these symmetries in future chapters.

\subsection{Charges as generators}

The statement of Noether's theorem above is valid in classical physics.
We will see now that QFT adds another layer of importance to Noether's theorem by interpreting the conserved charges as operators.
It is a general fact that Noether charges themselves generate the symmetry underlying their conservation in a quantum theory which is why Noether's theorem remains indispensable even in QFT. The proof relies on the fact that in QFT the fields $\Phi_\alpha$ are operators determined by the creation and annihilation operators.
It further uses the canonically conjugate fields 
\begin{equation}
\Pi_\alpha= \frac{\partial\cL}{\partial \dot\Phi^\alpha}\, .
\end{equation}
The canonical commutation relations among conjugate variables at equal time are
\begin{align}
\left[\Phi_\alpha(\mathbf{x},t),\Pi_\beta(\mathbf{y},t)\right]&=\, \I\delta^3(\mathbf{x}-\mathbf{y})\delta_{\alpha\beta}\, ,\\[0.2em]
 \left[\Phi_\alpha(\mathbf{x},t),\Phi_\beta(\mathbf{y},t)\right]\,=\left[\Pi_\alpha(\mathbf{x},t),\Pi_\beta(\mathbf{y},t)\right]&=0\, .
\end{align}
They determine $Q$ above as an operator since \eqref{eq:NoetherCharge} implies that the canonical momenta $\Pi_\alpha$ are essentially the Noether charges.
We can then extract from this the commutation relations between the conserved charges and the fields $\Phi_\alpha$ through
\begin{equation}
Q\,=\, \int\dif^{3}x\,  \Pi_\alpha \Delta_\alpha\quad \Rightarrow\quad \left[\Phi_\alpha(\mathbf{x},t),Q\right]=\, \I\Delta_{\alpha}(\mathbf{x},t)\, .
\end{equation}
This shows that the conserved charges Q of Noether’s theorem \emph{act as generators of the corresponding symmetry $\Phi_\alpha\rightarrow \Phi_\alpha+ \Delta_\alpha$}.

For a general internal symmetry with Lie group $G$,
we consider the infinitesimal transformation
\begin{equation}
\Phi_{\alpha}^{i}\rightarrow \Phi_{\alpha}^{i}+{\rm i}\alpha^{a}(T_{a})^{i}\,_{j}\Phi^{j}_{\alpha}\, ,
\end{equation}
with parameters $\alpha_a$ and generators $T_{a}$, $a=1,\ldots,\text{dim}(G)$.
The conserved charges are the operators that act on the fields $\Phi^i_{\alpha}$ according to the transformation generated by the generator $T_a$. 
It is easily verified that the charge associated to the transformation by $T_a$ satisfies
\begin{equation}
[\Phi_{\alpha}^{i},Q_{a}]=(T_{a})^{i}\,_{j}\Phi^{j}_{\alpha}
\end{equation}
proving that the conserved charges in Noether's theorem are in one-to-one correspondence with the generators of the symmetry group and act as generators of the corresponding symmetry. 
To see it more explicitly, we may exponentiate the above expression to find
\begin{equation}
U^\dagger \Phi^i_{\alpha}U=\left(e^{\I \alpha^a T_a}\Phi_{\alpha}\right)^{i}\kom U=e^{\I\alpha^a Q_a}\, .
\end{equation}
For an internal $\mathrm{U}(1)$-symmetry with $\psi\raw\ee^{\I e\alpha}\psi$, the conserved current is $j_{\mu}=e\overline{\psi}\gamma_{\mu}\psi$ and the conserved charge is the electric charge $e$ that can be seen as the generator of the $\mathrm{U}(1)$ symmetry.

Consider now spacetime translations $x^{\mu}\raw x^{\mu}+ a^{\mu}$ with current $T^{\mu}\,_{\nu}$ being the stress-energy tensor. The corresponding charges are given by
\begin{equation}
P^{0}=E=\int\dif^{3}x\, T^{00}\kom P^{i}=\int\dif^{3}x\, T^{0i} \, .
\end{equation}
For rotations, one can compute the charges
\begin{equation}
M^{ij}=\int\dif^{3}x\, \left (x^{i}T^{0j}-x^{j}T^{0i}\right )\, .
\end{equation}
Thus, the conserved charges themselves play the role of generators of the Poincar{\'e} group as we have seen previously in section~\ref{sec:PropertiesOfLorentzGroup}.

\section{Effective Field Theories in a nutshell}\label{sec:EFTs}
\index{Effective Field Theory|see {EFT}}

Up to this point,
we collected all the necessary tools to build arbitrary QFTs in $4$ dimensions and compute $S$-matrix elements that lead to observable quantities such as cross sections and decay rates.
Before we come to building up the Standard Model,
let us make general remarks about how QFTs can be used to understand physics at different energy scales by introducing  \emph{Effective Field Theories} (EFTs).

The introduction of EFTs has been one of the most important theoretical developments in the past 50 years. It is only relatively recently that their full power has been fully appreciated. They provide a systematic way to organise our understanding of nature at different energy scales, from low to high energies.
Indeed, we experience the world always through an effective low-energy description.

Even if we know a more fundamental description,
we only need to take into account the behaviour of the degrees of freedom that are accessible at the energy scale of our experiments when describing low-energy phenomena.
To study the properties of water in everyday life, for instance,
we do not need to start from the Lagrangian of QCD or understand quantum gravity.

A more relevant example for the purposes of this lecture is QED where we focus only on photons and electrons, while neglecting all other heavier particles.
This is a valid approximation at energies $E<2m_{e}$ with $m_{e}=511\,$keV.
In this case we know the theory at higher energies, at least in a path integral prescription.
If we are interested only in the low-energy physics, we \emph{integrate out} the heavy states to obtain a theory only for the low-energy states (electrons and photons for instance) defining the corresponding EFT. 

More generally, even if we do not know the theory at high energies, we can identify the relevant low-energy states and write down an EFT. 
This effective description is then capable of accurately describing physical phenomena at low energies.
While these types of calculations can be explicitly performed in e.g. perturbation theory,
in the case of the Standard Model, we do not even know precisely which new degrees of freedom have to be added at higher energies. Said differently, it is almost impossible to backtrack the process of integrating out modes without knowing the full spectrum in the UV.
Luckily, it is irrelevant for many purposes: at low enough energies, theories can be immensely predictive even if we do not have the full information about a given theory at hand.
In other words, we do not need quantum gravity to describe a cup of tea.
It is this basic principle that makes \emph{effective field theories so powerful}.

\subsection{Interactions: organising physics by energy scales}

Let us now describe how physics can be organised by energy scales.
Starting from some theory at high energies described by a path integral of the form \eqref{eq:PathIntegral},
we can integrate out all the heavy degrees of freedom above a given energy scale $\Lambda$ and treat the remaining fields in an effective description.
This typically implies that this description involves \emph{less} degrees of freedom than the theory from which we originally started in the UV.

In practice, when we write down a quantum field theory that should be predictive at energies $E\ll \Lambda$ below some cutoff $\Lambda$, the most basic question we might ask is which operators $\cO_{i}$ to include in the expansion \eqref{eq:LagGen} for the Lagrangian of a QFT.
That is, at low energies, our theory should be represented by an \emph{effective Lagrangian}
\begin{equation}\label{eq:EFTLagGen} 
\cL_{\text{eff}}[\Phi_{\alpha},\p\Phi_{\alpha}]=\sum_{i}\, c_{i}\, \cO_{i}(\Phi_{\alpha},\p\Phi_{\alpha})\, .
\end{equation}
Here, $c_{i}$ are some ``constant'' coefficients, $\cO_{i}$ operators and $\Phi_{\alpha}$ all fields in our theory.
The question about which operators $\cO_{i}$ to include in \eqref{eq:EFTLagGen} can be addressed by noticing that there exists an ordering principle for the operators $\cO_{i}$. Indeed, it turns out that the level of importance of the operators  $\cO_{i}$ depends on their dimensionality and the energies which we are interested in exploring. Since the action $S$ is dimensionless, we can determine the mass dimension of $\cL_{\text{eff}}$ as
\begin{equation}
S_{\text{eff}}[\Phi_{\alpha},\p\Phi_{\alpha}]=\int\,\cL_{\text{eff}}[\Phi_{\alpha},\p\Phi_{\alpha}]\dif^{4}x\quad\Longrightarrow\quad[\cL_{\text{eff}}]=4\, .
\end{equation}
Operators $\cO_{i}$ of dimensions $d_{i}=[\cO_{i}]$ fall into three categories:
\begin{enumerate}
\item \emph{Irrelevant}\index{Irrelevant operators}: $c_{i}$ becomes smaller at lower energies: $d_{i}>4$, $[c_{i}]<0$;
\item \emph{Relevant}\index{Relevant operators}: $c_{i}$ increases at lower energies: $d_{i}<4$, $[c_{i}]>0$;
\item \emph{Marginal}\index{Marginal operators}: $c_{i}$ does not change with the energy scale: $d_{i}=4$, $[c_{i}]=0$.
\end{enumerate}
The coefficients with negative dimensionality would naturally be suppressed by powers of a UV scale $\Lambda$ and would then be less relevant if we are interested in the physics at scales $E\ll \Lambda$. Thus, we call a theory
\begin{itemize}
\item \emph{Renormalisable} if\index{EFT!Renormalisable}
\begin{equation}
[c_{i}]\geq 0\quad\forall i\, .
\end{equation}
This is quite restrictive for the simple reason that
\begin{equation}
d_{i}=[\cO_{i}]=4-[c_{i}]\geq 0\, .
\end{equation}
This implies that in a renormalisable theory only a few $c_{i}$ are non-zero and, hence, the theory is immensely predictive: only those few coefficients $c_{i}$ have to be matched with experiments at energies $E\ll\Lambda$.
\item \emph{Non-renormalisable} if\index{EFT!Non-renormalisable}
\begin{equation}
[c_{i}]\leq 0\quad\text{for some }i\, .
\end{equation}
Then, the coefficients $c_{i}$ scale with the characteristic energy scale $\Lambda$ of our theory as
\begin{equation}
c_{i}\sim \Lambda^{4-d_{i}}\, .
\end{equation}
We distinguish the following scenarios where $E$ is a typical energy of the theory being studied:
\begin{itemize}
\item if $E\ll\Lambda$, it is generically sufficient to keep only a few operators $\cO_{i}$ and the EFT becomes predictive.
\item if $E\sim\Lambda$, we have to include infinitely many operators and we loose predictive power. Thus, we need a UV completion of our theory.
\end{itemize}
Clearly, there can be, in principle, infinitely many such coefficients because $d_{i}=4-[c_{i}]\geq 0$ is always satisfied.
\end{itemize}
Notice that for a non-renormalisable theory, the scale of new physics $\Lambda$ may be very large and therefore the theory may be predictive for a large range of energies $E$ as long as $E\ll \Lambda$.

A typical example that illustrates renormalisable and non-renormalisable theories is to consider the following Lagrangian for a real scalar field
\begin{equation}
{\mathcal L}_{\text{eff}}=\underbrace{\underbrace{\partial^\mu\phi\partial_\mu \phi - m^2\phi^2-g\phi^3-\lambda\phi^4}_{\text{Renormalisable}} + \frac{\alpha}{\Lambda}\phi^5+\frac{\beta}{\Lambda^2}\phi^6+\ldots}_{\text{Non-Renormalisable}}\, .
\end{equation}
The first four terms define a renormalisable theory which has predictive power due to having only $3$ arbitrary couplings $m,g,\lambda$. Beyond that, adding operators of higher dimensionality would make the theory non-renormalisable, while keeping a few terms would define an EFT valid for energies $E\ll \Lambda$. Otherwise the theory breaks down at energies close to $\Lambda$ and would need to be substituted by either a new EFT valid at higher energies or, ultimately, by an ultra-violet (UV) complete theory. 
Let us mention the $4$-Fermi theory as an effective description for the weak interactions at energies $E\ll 80\,$GeV as one prominent example that we will discuss in detail in chapter~\ref{chap:ew}.

\subsection{General relativity as an EFT*}

Einstein's gravity is an example of such theory that needs to be UV completed. Einstein's gravity can be treated quantum mechanically as long as it is an EFT addressing questions at low energies (meaning $E\ll M_{P}\simeq 10^{18}$ GeV).

Einstein's theory of \emph{General Relativity} (GR)\index{General Relativity}\index{GR} is described by the \emph{Einstein-Hilbert (EH) term}\index{Einstein-Hilbert term}
\begin{equation}\label{eq:EHT} 
\cL_{EH}=M_{P}^{2}\, R^{(4)}\sqrt{-g} \kom M_{P}^{2}=\dfrac{\hbar c}{G_{N}}
\end{equation}
in terms of the $4$-dimensional Ricci scalar $R^{(4)}$.
The coupling constant $G_{N}$ has negative mass dimension and is therefore non-renormalisable.
Alternatively,
we may expand the Ricci scalar $R^{(4)}$ in terms of fluctuations of the metric around a constant Minkowski background, that is,
\begin{equation}
g_{\mu\nu}=\eta_{\mu\nu}+\dfrac{1}{M_{P}}h_{\mu\nu}\;\;\Rightarrow\;\; M_{P}^{2}R^{(4)}=(\p h)^{2}+\dfrac{h}{M_{P}}(\p h)^{2}+\dfrac{h^{2}}{M_{P}^{2}}(\p h)^{2}+\ldots\, .
\end{equation}
Thus, an infinite number of counterterms would be necessary in perturbation theory.
Hence,
the theory does not admit a continuum limit, but has an intrinsic cutoff set by the \emph{Planck scale}\index{Planck scale} $M_{P}$.
Having said that,
non-renormalisability does not constitute an obstruction to making reliable perturbative quantum calculations in gravity as long as we limit our considerations to energies $\mu$ well below $M_{P}$,
\begin{equation}
\mu\ll M_{P}=\sqrt{\dfrac{\hbar c}{G_{N}}}\sim 10^{19}\text{GeV}\, .
\end{equation}
In this regime,
we can treat gravity as an EFT which is extraordinarily predictive \cite{Burgess:2003jk}.
In fact, pure gravity is finite at 1-loop \cite{tHooft:1974toh}.

Issues arise, however, once we ask the ``wrong'' questions which can only be answered within a fully consistent quantum theory of gravity.
First and foremost, these questions concern phenomena in the early Universe where energies came close to $M_{P}$.
Similarly,
the quantum nature of black holes might only be fully understood within quantum gravity.
A potential candidate for a theory describing the physics at the Planck scale is \emph{string theory}\index{String theory}.


\chapter{\bf Gauge Theories}
\label{chap:sym}

\vspace{0.5cm}
\begin{equ}[Symmetries and Inevitability]
{\it  There is one common feature that gives both general relativity and the standard model most of their sense of inevitability and simplicity: they obey principles of symmetry.}\\

\rightline{\it Steven Weinberg}
\end{equ}
\vspace{0.5cm}

This chapter presents an overview of local symmetries.
The main message is that the celebrated gauge symmetries are nothing but redundancies of a theory in order to describe massless particles of helicity greater or equal than $1$.
We emphasise that, even though there is some freedom in describing interacting theories for spin/helicity $0,1/2$, there are strong constraints for higher helicities (essentially because the dimension of little group representations is always $2$), recall Sect.~\ref{sec:RepresentationsOfPoincareGroup}.
This is important towards a proper description of the Standard Model: we will see in this chapter that only massless particles of helicities $0,1/2,1,3/2,2$ can exist as interacting theories.
Further, we argue that helicity $1$ only allows QED or Yang-Mills theories, whereas helicities $2$ only gravity.

This limits substantially the options to build the Standard Model. Therefore, once we formulate the Standard Model in terms of these theories for massless fields, it is not because we will make a particular choice of theory, \emph{but it is the only option we have}.
There are no alternatives as long as quantum mechanics and the symmetries of special relativity are valid. In other words, the beauty of the basic principles behind the Standard Model is not because symmetries are beautiful, it is because there is a sense of \emph{inevitability}. Things cannot be otherwise.
 
In this chapter we also present a few concepts which may have been introduced in
other courses, but which are crucial to construct the Standard Model.
This also allows us to set our notation and conventions.  Throughout, we use
natural units, $\hbar = c = 1$.

\vfill

\newpage

\section{The Origin of Gauge (Local) Symmetries}

Initially, let us introduce gauge redundancies.
They are essentially the price that we have to pay when trying to describe massless helicity-1 fields in an \emph{off-shell} formalism.


\subsection{Gauge symmetries from Lorentz invariance}\label{sec:GaugeSymmetriesLorentzInv} 
\index{Gauge Symmetry}\index{Gauge Symmetry!Abelian}

To begin with,
we consider a real spin/helicity-$1$ field described by
\begin{equation}\label{eq:SpinOneField} 
A_{\mu}(x)=\sum_\lambda\int\dif p\left [\epsilon_{\mu}\, a(p^{\nu},\lambda)\ee^{\I px}+\epsilon_{\mu}^{*}\, a^{\dagger}(p^{\nu},\lambda)\ee^{-\I px}\right ]\, .
\end{equation}
Here, $\epsilon_{\mu}$, $\epsilon_{\mu}^{*}$ are the \emph{polarisation vectors}\index{Polarisation vectors} as the objects carrying the Lorentz index and describing the propagation of the fields in spacetime, while $\lambda$ represents helicity for massless particles or spin for massive ones.
Moreover, recall that we use the notation
\begin{equation}
\int\dif p\equiv \int\dfrac{\dif^{4} p}{(2\pi)^{4}}\delta(p^{2}-m^{2})\Theta(p^{0})=\int\dfrac{\dif^{3} p}{E_{p}(2\pi)^{3}}\kom E_{p}^2=\mathbf{p}^{2}+m^{2}\, .
\end{equation}
As it stands, $A_{\mu}$ has in total $4$ degrees of freedom given by $\mu=0,1,2,3$. However, we learnt in section~\ref{sec:RepresentationsOfPoincareGroup} that the corresponding $1$-particle states have either $3$ degrees of freedom for massive particles or $2$ for massless. Hence, we need extra constraints:
\begin{itemize}
\item \emph{Massive case}: we impose the Lorentz invariant condition
\begin{equation}\label{eq:ConstraintMassiveSpinOne} 
p^{\mu}\epsilon_{\mu}=0
\end{equation}
which reduces the number of degrees of freedom to $3$. This is the only Lorentz invariant quantity that can be written to constrain the polarisation vector $\epsilon_\mu$ and the other ingredient at hand, namely the momenta $\p^\mu$.  It successfully reduces the number of degrees of freedom from 4 to 3 agreeing with the degrees of freedom of a massive particle.
\item \emph{Massless case}: apart from \eqref{eq:ConstraintMassiveSpinOne}, there are no more Lorentz invariant constraints. However, since in the massless  case $p^\mu p_\mu=0$,
there is the following ambiguity
\begin{equation}
\epsilon_{\mu}\underbrace{\equiv}_{\text{same state}}\epsilon_{\mu}+\hat{\alpha}(p)p_{\mu}\, ,
\end{equation}
where $\hat{\alpha}(p)$ is an arbitrary function of the momenta. Both sides clearly satisfy \eqref{eq:ConstraintMassiveSpinOne}. This implies that actually $\epsilon_{\mu}$ is \emph{not} a Lorentz vector. The arbitrariness in the parameter $\hat\alpha$ reduces the number of degrees of freedom by 1, leading to the 2 degrees of freedom needed to describe massless particles. This implies a similar equivalence relation for the field in position space \eqref{eq:SpinOneField} of the form
\begin{equation}
A_{\mu}(x)\equiv A_{\mu}(x)+\p_{\mu}\alpha(x)
\end{equation}
where $\alpha(x)$ is the Fourier transform of $\hat{\alpha}(p)$.
This is referred to as \emph{gauge invariance}\index{Gauge invariance} which simply corresponds to a mathematical redundancy in our description of physics. Generally, one can state that
\begin{Boxequ}
\vspace{0.1cm}
A proper Lorentz invariant description of physical amplitudes for massless helicity-$1$ fields \emph{implies} gauge invariance.
\end{Boxequ}
\end{itemize}

Notice that we have already seen gauge transformations $A_{\mu}(x)\equiv A_{\mu}(x)+\p_{\mu}\alpha(x)$ in Sect.~\ref{sec:symmetriesQFT} in the context of local symmetries. For matter fields $\psi(x)$ transforming as $\psi(x)\rightarrow e^{\I \alpha(x)}\psi(x)$, this defines a $U(1)$ transformation. Their gauge invariant kinetic energy is $\bar\psi\I \gamma^\mu D_\mu \psi$ involves the covariant derivative $D_\mu=\p_\mu-\I eA_\mu$ provided that $A_\mu$ transforms through gauge transformations as above. This determines the coupling of the matter field $\psi$ to the gauge field $A_\mu$, namely $A_{\mu}J^{\mu} = \bar\psi\gamma^\mu A_\mu \psi$ with $J^{\mu}$ a conserved current. 

The physical quantity to look at is in fact the field strength tensor (just as in electrodynamics) given by
\begin{equation}
F_{\mu\nu}(x)=\I\sum_{\lambda=\pm 1}\int\dif p\left [\epsilon_{\mu}p_{\nu}-\epsilon_{\nu}p_{\mu}\right ]\, a(p^{\mu},\lambda)\,\ee^{\I px}+\text{h.c.}
\end{equation}
which is invariant under $\epsilon_{\mu}\raw\epsilon_{\mu}+\hat{\alpha}(p) p_{\mu}$ and, after using \eqref{eq:SpinOneField}, amounts to
\begin{equation}
F_{\mu\nu}(x)=\p_{\mu}A_{\nu}(x)-\p_{\nu}A_{\mu}(x)\, .
\end{equation}
In electromagnetism $F_{0i}=E_i$ and $\epsilon_{ijk}F_{jk}=B_i$ with $E_i, B_i$ the components of the electric and magnetic fields, respectively. $F_{\mu\nu}$ carries the degrees of freedom of a helicity 1 field. This is already apparent from the decomposition $(1/2,1/2)\otimes (1/2,1/2)=(0,0)\oplus (1,0)\oplus (0,1)\oplus (1,1)$ where the latter has $2$ indices and $(1,0)\oplus (0,1)$ is the antisymmetric component corresponding to $F_{\mu\nu}$  that carries helicity $\pm 1$ (the $(0,0)$ state is a scalar of helicity 0 and the $(1,1)$ state is a helicity 2 state).

Notice that neither $\epsilon_{\mu}$ nor $A_{\mu}$ are Lorentz vectors since
\begin{equation}
A_{\mu}\raw\Lambda_{\mu}\,^{\nu}A_{\nu}+\p_{\mu}\alpha
\end{equation}
is only a vector up to a gauge transformation.
In contrast, $F_{\mu\nu}$ is a proper Lorentz tensor since it transforms as
\begin{equation}
F_{\mu\nu}\rightarrow  \Lambda_{\mu}\,^{\alpha}\Lambda_{\nu}\,^{\beta}  F_{\alpha\beta}\, .
\end{equation}
and as we saw above, it is invariant under gauge transformations.\footnote{This is special for Abelian gauge theories, while for non-Abelian gauge theories $F_{\mu\nu}$ will only be covariant.}

However, in order to consider interactions of gauge fields with matter fields, we cannot just concentrate on $F_{\mu\nu}$ but need to include also $A_\mu$. For amplitudes, interactions involving helicity-1 fields that do not vanish at $0$-momentum need to be described with $A_{\mu}$ rather than just $F_{\mu\nu}$.
This allows us to introduce the Lagrangian
\begin{equation}
\cL=\int\, \left (-\frac{1}{4} F^{\mu\nu}F_{\mu\nu}+A_{\mu}J^{\mu}+\ldots\right )
\end{equation}
for some current $J^{\mu}$, e.g., $J^{\mu}=\overline{\psi}\gamma^{\mu}\psi$. The equivalence $A_{\mu}\raw A_{\mu}+\p_{\mu}\alpha$ implies that the current must be conserved
\begin{equation}
\p_{\mu}J^{\mu}=0\, .
\end{equation}
In general, the matrix elements involved in the amplitude written in terms of the fields $A_\mu$ and their corresponding polarisations $\epsilon_\mu$ are of the form
\begin{equation}
\cM(p_{i}^{\mu},\lambda_{i})=\epsilon^{\mu}\cM_{\mu}
\end{equation}
In order for the amplitude to be Lorentz invariant, it has to be invariant under the shift of polarisations $\epsilon_\mu\raw \epsilon_\mu +\hat{\alpha}(p) p_\mu$  (gauge redundancy).
Hence, it has to  satisfy the following constraint: 
\begin{equ}[Ward identity]\index{Ward identity}
\vspace*{-0.3cm}
\begin{equation}\label{eq:WardIdentityAbelian} 
p^{\mu}\cM_{\mu}=0\, .
\end{equation}
\end{equ}
This so-called \emph{Ward identity} is an important condition that is solely determined by the requirement of  Lorentz invariance and the fact that $\epsilon_\mu$ is not a Lorentz vector but
enjoys the equivalence relation $\epsilon_{\mu}\sim\epsilon_{\mu}+\hat{\alpha}(p) p_{\mu}$.
We continue our discussion about gauge theories in Sect.~\ref{sec:YMc} after a brief detour into soft theorems.

\subsection{Gravity from helicity-$2$ states}

We consider massless particles of helicity $\lambda=\pm 2$ described by a two-index symmetric field $h_{\mu\nu}$ written in terms of polarisation vectors $\epsilon_{\mu\nu}$ that also have only two degrees of freedom or polarisation states:
\begin{equation}\label{eq:SpinTwoField} 
h_{\mu\nu}(x)=\sum_\lambda\int\dif p\left [\epsilon_{\mu\nu}\, a(p^{\nu},\lambda)\ee^{\I px}+\epsilon_{\mu\nu}^{*}\, a^{\dagger}(p^{\nu},\lambda)\ee^{-\I px}\right ]\, .
\end{equation}

 As for the case of helicity 1, the Lorentz invariant constraint $q^\mu\epsilon_{\mu\nu}=0$ leaves a gauge redundancy
\begin{equation}
\epsilon_{\mu\nu}\raw \epsilon_{\mu\nu}+\hat{\alpha}_{\mu}(q)\, q_{\nu}+\hat{\alpha}_{\nu}(q)\, q_{\mu}
\end{equation}
or equivalently
\begin{equation}
h_{\mu\nu}\raw h_{\mu\nu}+\p_{\mu}\alpha_{\nu}(x)+\p_{\nu}\alpha_{\mu}(x)\, .
\end{equation}
which can be identified with the local version of a general coordinate transformation. This is the underlying symmetry that defines General Relativity.

For completeness, we also need to impose $\epsilon^{\mu}\,_{\mu}=0$ and $q^{\mu}\epsilon_{\mu\nu}=0$. In total, a symmetric tensor has $10$ degrees of freedom. The constraints reduce those by $5=1+4$ in the massive and $8=1+4+3$ in the massless case. Notice that $\epsilon^{\mu}\,_{\mu}=0$ requires $p^{\mu}\hat{\alpha}_{\mu}=0$ resulting in an additional constraint on the components $\hat{\alpha}_{\mu}$ which is why the number of degrees of freedom is only reduced by $3$ for massless spin $2$ particles. Therefore  we can confirm that  a massless helicity-2 particle has only $10-8=2$  independent degrees of freedom.

As discussed in the previous chapter, an invariant action (the Einstein-Hilbert action \eqref{eq:EHT}) can be written in terms of the field $h_{\mu\nu}(x)$ taken as the perturbation of the metric $g_{\mu\nu}(x)=\eta_{\mu\nu}+h_{\mu\nu}(x)$.

\section{Soft Theorems}\label{sec:softtheorems} \index{Charge conservation}

In this section,
we demonstrate the true power of Lorentz invariance in terms of the constraints it imposes on the S-matrix of four dimensional theories.
The idea is to use so-called \emph{soft particles} which simply means considering particles whose  momenta are taken to zero.
As we will see, this approach directly leads to powerful statements like conservation of charge or the absence of long-range interactions involving massless particles with helicities $>2$.
While the results presented here have been derived in the 1960's by Weinberg in \cite{Weinberg:1964ev,Weinberg:1964ew,Weinberg:1965rz},
there are modern derivations available in the literature such as \cite{Benincasa:2007xk} using considerations of \cite{Britto:2004ap,Britto:2005fq}.\footnote{It needs to be stressed that we are working here in four-dimensional Minkowski space. For different backgrounds, higher spin theories have been proposed such as for anti-de-Sitter space in \cite{Bekaert:2004qos}.}

\subsection{Charge conservation}

In this section, we show that charge conservation already follows from Lorentz invariance. Thereto, we consider the following scattering diagram
\begin{equation*}
\begin{tikzpicture}[scale=1.2]
\begin{feynhand}
\node (o) at (-0.75,0) {\large $\mathcal{M}_{0}=$} ; 
\vertex (a0) at (0,0.25); 
\vertex (a1) at (0,1.25); 
\vertex (a2) at (0,-0.25); 
\vertex (a3) at (0,-1.25); 
\vertex [NWblob] (b) at (2,0) {};
\vertex (c0) at (4,0.75); 
\vertex (c1) at (4,0.); 
\vertex (c22) at (4,-0.75); 
\propag [fer] (a0) to (b);
\propag [fer] (a1) to (b);
\propag [fer] (a2) to (b);
\propag [fer] (a3) to (b);
\propag [fer] (b) to (c0);
\propag [fer] (b) to (c1);
\propag [fer] (b) to (c22);
\end{feynhand}
\end{tikzpicture}
\end{equation*}
Let us add a soft photon with momentum $q^{\mu}$ for which we want to take the soft limit $q^{\mu}\raw 0$ before as well as after the interactions encoded in the ``blow''
\begin{equation*}
\begin{tikzpicture}[scale=1.35]
\begin{feynhand}
\node (o) at (-0.75,0) {\large $\mathcal{M}=$} ; 
\vertex (a0) at (0,0.5); 
\vertex (a1) at (0,1.5); 
\vertex (a2) at (0,-0.5); 
\vertex (a3) at (0,-1.5); 
\vertex [NWblob] (b) at (2,0) {};
\vertex (c0) at (4,1.5); 
\vertex (c1) at (4,0.5); 
\vertex [dot] (c2) at (3,-0.25) {}; 
\vertex (c22) at (4,-0.5); 
\vertex (c3) at (2.25,-0.75); 
\vertex (c4) at (3.5,-1.25); 
\node (o1) at (2.8,-0.425) { $\Gamma^{\mu}$} ; 
\propag [fer] (a0) to (b);
\propag [fer] (a1) to (b);
\propag [fer] (a2) to (b);
\propag [fer] (a3) to (b);
\propag [fer] (b) to (c0);
\propag [fer] (b) to (c1);
\propag [fer] (b) to [edge label = $p+q$] (c2);
\propag [fer] (c2) to [edge label = $p$] (c22);
\propag [bos] (c2) to [below,edge label = $q$] (c4);
\node (o) at (5.,0) {\Large $+$} ; 
\vertex (a0) at (6,0.5); 
\vertex (a1) at (6,1.5); 
\vertex (a2) at (6,-0.5); 
\vertex (a3) at (6,-1.5); 
\vertex (a4) at (7.75,-1.25); 
\vertex [NWblob] (b) at (8,0) {};
\vertex (c0) at (10,1.5); 
\vertex (c1) at (10,0.5); 
\vertex (c2) at (10,-0.5); 
\node (o1) at (7.,-0.55) { $\Gamma^{\mu}$} ; 
\node[rotate=40] (o1) at (7.6,-0.55) {$p-q$} ; 
\vertex [dot] (a33) at (7,-0.75) {}; 
\propag [fer] (a0) to (b);
\propag [fer] (a1) to (b);
\propag [fer] (a2) to (b);
\propag [fer] (a33) to (b);
\propag [fer] (a3) to [edge label = $p$] (a33);
\propag [bos] (a4) to [edge label = $q$] (a33);
\propag [fer] (b) to (c0);
\propag [fer] (b) to (c1);
\propag [fer] (b) to (c2);
\end{feynhand}
\end{tikzpicture}
\end{equation*}
The interaction vertex $\Gamma^{\mu} = \Gamma^{\mu}(p,q)$ can be written in general as 
\begin{equation}
\Gamma^{\mu}(p,q)=p^{\mu}F(p^{2},q^{2},p\cdot q)+q^{\mu}G(p^{2},q^{2},p\cdot q)
\end{equation}
Since the final amplitude is of the form $\epsilon^{\mu}\Gamma_{\mu}$ and $\epsilon^{\mu}q_{\mu}=0$, we can forget $G$. Since $p^{2}=m^{2}$ and $q^{2}=0$, we find
\begin{equation}
\Gamma^{\mu}=p^{\mu}F\left (\dfrac{p\cdot q}{m^{2}}\right )
\end{equation}
by dimensionality. If we have one soft photon per external line, one finds
\begin{equation}
\cM=\cM_{0}\left (\sum_{\text{incoming}}\dfrac{p_{i}^{\mu}\epsilon_{\mu}}{2p_{i}^{\mu}q_{\mu}}F_{i}(0)-\sum_{\text{outgoing}} \dfrac{p_{i}^{\mu}\epsilon_{\mu}}{2p_{i}^{\mu}q_{\mu}}F_{i}(0)\right )
\end{equation}
where we used that for $q^{\mu}\raw 0$ the propagators for incoming and outgoing particles can be written as\footnote{Note that, as it stands, the presence of the ${1}/{2 p_{i}^{\mu}q_{\mu}}$ in the propagator seems to indicate an infrared divergence in the limit of $q\raw 0$ for $\cM$. However these diagrams are such that they can be resummed to all orders and  the final result is free of divergences. }
\begin{equation}\label{eq:PropagatorsSimplifyCC} 
\dfrac{1}{(p_{i}+q)^{2}-m^{2}}\sim \dfrac{1}{2 p_{i}^{\mu}q_{\mu}}\kom \dfrac{1}{(p_{i}-q)^{2}-m^{2}}\sim- \dfrac{1}{2 p_{i}^{\mu}q_{\mu}}\, .
\end{equation}
Invariance under $\epsilon_{\mu}\raw\epsilon_{\mu}+\alpha q_{\mu}$ then implies
\begin{equation}
\sum_{\text{incoming}}\, F_{i}(0)-\sum_{\text{outgoing}}\, F_{i}(0)=0
\end{equation}
where $F_{i}(0)$ is nothing but the charges of the particles involved. Hence, one finds that charge is conserved:
\begin{equation}
\sum_{\rm incoming}Q_i=\sum_{\rm outgoing}Q_i.
\end{equation}

This is already a remarkable observation as we can see that the well known result that electric charge is a conserved quantity can be derived directly by just arguments of Lorentz invariance.

\subsection{The equivalence principle and (no) helicities $\mathbf{>2}$}

One can play the same game for gravity. That is, consider scattering processes with soft gravitons of the form
\begin{equation*}
\begin{tikzpicture}[scale=1.35]
\begin{feynhand}
\node (o) at (-0.75,0) {\large $\mathcal{M}=$} ; 
\vertex (a0) at (0,0.5); 
\vertex (a1) at (0,1.5); 
\vertex (a2) at (0,-0.5); 
\vertex (a3) at (0,-1.5); 
\vertex [NWblob] (b) at (2,0) {};
\vertex (c0) at (4,1.5); 
\vertex (c1) at (4,0.5); 
\vertex [dot] (c2) at (3,-0.25) {}; 
\vertex (c22) at (4,-0.5); 
\vertex (c3) at (2.25,-0.75); 
\vertex (c4) at (3.5,-1.25); 
\node (o1) at (2.8,-0.425) { $\Gamma^{\mu\nu}$} ; 
\propag [fer] (a0) to (b);
\propag [fer] (a1) to (b);
\propag [fer] (a2) to (b);
\propag [fer] (a3) to (b);
\propag [fer] (b) to (c0);
\propag [fer] (b) to (c1);
\propag [fer] (b) to [edge label = $p+q$] (c2);
\propag [fer] (c2) to [edge label = $p$] (c22);
\propag [glu] (c2) to [below,edge label = $q$] (c4);
\node (o) at (5.,0) {\Large $+$} ; 
\vertex (a0) at (6,0.5); 
\vertex (a1) at (6,1.5); 
\vertex (a2) at (6,-0.5); 
\vertex (a3) at (6,-1.5); 
\vertex (a4) at (7.75,-1.25); 
\vertex [NWblob] (b) at (8,0) {};
\vertex (c0) at (10,1.5); 
\vertex (c1) at (10,0.5); 
\vertex (c2) at (10,-0.5); 
\node (o1) at (7.,-0.55) { $\Gamma^{\mu\nu}$} ; 
\node[rotate=40] (o1) at (7.6,-0.55) {$p-q$} ; 
\vertex [dot] (a33) at (7,-0.75) {}; 
\propag [fer] (a0) to (b);
\propag [fer] (a1) to (b);
\propag [fer] (a2) to (b);
\propag [fer] (a33) to (b);
\propag [fer] (a3) to [edge label = $p$] (a33);
\propag [glu] (a4) to [edge label = $q$] (a33);
\propag [fer] (b) to (c0);
\propag [fer] (b) to (c1);
\propag [fer] (b) to (c2);
\end{feynhand}
\end{tikzpicture}
\end{equation*}
This results in an amplitude
\begin{equation}
\cM=\cM_{0}\left (\sum_{\text{incoming}}\dfrac{p_{i}^{\mu}\epsilon_{\mu\nu}p_{i}^{\nu}}{2p_{i}^{\mu}q_{\mu}}F_{i}(0)-\sum_{\text{outgoing}} \dfrac{p_{i}^{\mu}\epsilon_{\mu\nu}p_{i}^{\nu}}{2p_{i}^{\mu}q_{\mu}}F_{i}(0)\right )\, .
\end{equation}
We then find that
\begin{equation}
\sum_{\text{incoming}}\, F_{i}(0)p_{i}^{\nu}-\sum_{\text{outgoing}}\, F_{i}(0)p_{i}^{\nu}=0
\end{equation}
which can only be satisfied for arbitrary $p_{i}^{\mu}$ if \textbf{all} $F_{i}(0)$ are the same, i.e., 
\begin{equation}
F_i (0)=\kappa  \kom \forall i\, .
\end{equation}
Thus, the coupling to gravity must be \emph{universal} implying the \emph{principle of equivalence}\index{Principle of equivalence}.

It is hard to overemphasise the importance of this result. The principle of equivalence is the basic premise behind Einstein's General theory of relativity. Here it is not assumed but derived from basic principles of quantum mechanics and special relativity. The whole concept of gravity reduces to be the unique theory that describes the interaction of massless particles of helicity $\pm 2$. This somehow enhances the beauty of the theory not because of the symmetries behind (which are only redundancies) but because of its inevitability. It also provides a different perspective on what gravity is. In the search of a more fundamental theory describing gravity at the quantum level, the concrete requirement is to describe interactions of particles of helicity $\pm 2$.
We note however that, as we mentioned above, interactions can be described in terms of an EFT for the corresponding helicity $\pm 2$ field $h_{\mu\nu}(x)$.
 
Considering helicity $\lambda=\pm 3$ particles results in a constraint 
\begin{equation}
\sum_{\rm incoming}F_i(0)p_i^\mu p_i^\nu=\sum_{\rm outgoing}F_i(0)p_i^\mu p_i^\nu
\end{equation}
which is only satisfied by $F_{i}(0)\equiv 0\,  \forall i$. Thus, there are no interacting massless particles of helicity greater than $2$. 
This is again a crucial result eliminating an infinite number of possibilities for interacting elementary particles.\footnote{This result has been used to prove that supersymmetric theories are constrained to a maximum number of 8 supersymmetries and that the highest possible dimensionality of spacetime is $D=10$ for even dimensionality and $D=11$ for odd dimensionality. This also coincides with the critical dimensionalities found in string theory.} 

Therefore we conclude, by only using arguments of Lorentz invariance that any theory that can describe interacting massless particles can only include a handfull of particles:
\begin{equ}[Massless particle content of any interacting QFT]
$\lambda=0,\pm \dfrac{1}{2},\pm 1,\pm\dfrac{3}{2},\pm 2$ are all possible massless particle states in an interacting theory.
\end{equ}

In nature we have examples of particles of precisely these helicities,\footnote{Massless particles are the most relevant if we are asking questions at low energies, as usual in physics. We will see in the subsequent chapters how some massless particles can get a mass.}
the Higgs particle is the example for $\lambda=0$, quarks and leptons for $\lambda=\pm 1/2$, photons, gluons, $W$- and $Z$-particles for $\lambda=\pm 1$ and the graviton for $\lambda=\pm 2$ with the (so far) only exception of $\lambda = \pm 3/2$. A proper interactive theory of these particles (known individually as the {\it gravitino})\index{Gravitino} is only consistent in supersymmetric theories where they have to couple to gravity. Their study and potential impact in nature is beyond the scope of these lectures.

\section{Non-Abelian gauge theories from scattering amplitudes}\index{Gauge Symmetry!Non-Abelian}

Above, we focussed on theories with a single type of field $A_{\mu}$.
Here, we would like to understand how gauge symmetries are modified in the presence of several species of massless helicity-1 particles.
Instead of stating the answer right away (Yang-Mills theory is the proper theoretical description of helicity-1 particles), we provide a brief derivation of the structure underlying non-Abelian gauge theories.

\begin{figure}[t!]
\centering
\begin{tikzpicture}[scale=1.1]
\setlength{\feynhanddotsize}{1.5ex}
\begin{feynhand}
\node (o) at (-0.25,2.9) {$e$} ; 
\node (o1) at (-0.25,1.1) {$e$} ; 
\vertex (a0) at (-1.5,4) {$e^{-}$}; 
\vertex (a1) at (1.5,4) {$\epsilon^{\nu}_{\text{in}}$}; 
\vertex (b0) at (0,3); 
\vertex (b1) at (0,1); 
\vertex (c0) at (-1.5,0) {$e^{-}$}; 
\vertex (c1) at (1.5,0.) {$\epsilon^{\mu}_{\text{out}}$}; 
\propag [fer, mom={$p$}] (a0) to (b0);
\propag [pho, mom={$q$}] (a1) to (b0);
\propag [fer, mom={$p+q$}] (b0) to (b1);
\propag [fer, mom={$p^{\prime}$}] (b1) to (c0);
\propag [pho, mom={$q^{\prime}$}] (b1) to (c1);
\end{feynhand}
\end{tikzpicture}
\hspace*{1.5cm}
\begin{tikzpicture}[scale=1.1]
\setlength{\feynhanddotsize}{1.5ex}
\begin{feynhand} 
\node (o) at (-0.25,2.9) {$e$}; 
\node (o1) at (-0.25,1.1) {$e$};
\vertex (a0) at (-1.5,4) {$e^{-}$}; 
\vertex (a1) at (2.,4) {$\epsilon^{\nu}_{\text{in}}$}; 
\vertex (a11) at (1.,2.5); 
\vertex (b0) at (0,3); 
\vertex (b1) at (0,1); 
\vertex (b00) at (-0.25,2.5); 
\vertex (b11) at (-0.25,1.5); 
\node (o1) at (-0.75,2.0) {$p-q^{\prime}$};
\vertex (c0) at (-1.5,0) {$e^{-}$}; 
\vertex (c1) at (2.,0.) {$\epsilon^{\mu}_{\text{out}}$}; 
\vertex (c11) at (1.,1.5); 
\propag [fer] (b0) to (b1);
\propag [fer, with arrow = 1] (b00) to(b11);
\propag [fer, mom={$p$}] (a0) to (b0);
\propag [pho, mom={$q^{\prime}$}] (c11) to (c1);
\propag [pho] (b0) to (c11);
\propag [fer, mom={$p^{\prime}$}] (b1) to (c0);
\propag [pho, top, mom={$q$}] (a1) to (a11);
\propag [pho, top] (b1) to (a11);
\end{feynhand}
\end{tikzpicture}
\caption{The two Feynman diagrams contributing to Compton scattering $e^{-}\gamma\raw e^{-}\gamma$ in QED.}\label{fig:ComptonScatteringQED} 
\end{figure}
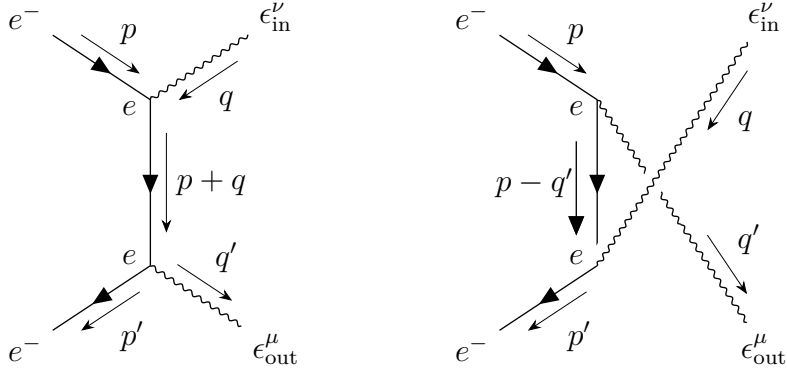

First, we consider \emph{Compton scattering}\index{Compton scattering} in \emph{QED}\index{QED} ($e^{-}\gamma\raw e^{-}\gamma$) as shown in Fig.~\ref{fig:ComptonScatteringQED}.
Using the standard Feynman rules,
the amplitude can be written as
\begin{align}
\cM&=-\I e^{2}\, \bar{u}(p^{\prime},\sigma^{\prime})\left (\dfrac{\gamma_{\mu}(\cancel{p}+\cancel{q}+m)\gamma_{\nu}}{(p+q)^{2}-m^{2}}+\dfrac{\gamma_{\nu}(\cancel{p}-\cancel{q}^{\prime}+m)\gamma_{\mu}}{(p-q^{\prime})^{2}-m^{2}}\right )u(p,\sigma)\epsilon_{\text{in}}^{\nu}\epsilon_{\text{out}}^{\mu}\nn\\
&=\cM_{\mu\nu}\epsilon^{\nu}_{\text{in}}\epsilon_{\text{out}}^{\mu}
\end{align}
where, as usual $\cancel{p}=\gamma^\mu p_\mu,$ etc. Using basic identities like
\begin{equation}
p+q=p^{\prime}+q^{\prime}\kom (\cancel{p}-m)u=0\kom \bar{u}(\cancel{p}^{\prime}-m)=0\, ,
\end{equation}
we can check the Ward identity \eqref{eq:WardIdentityAbelian} by considering how the amplitude changes under $\epsilon^{\nu}_{\text{in}}\rightarrow \epsilon^{\nu}_{\text{in}}+q^\nu$
\begin{align}\label{eq:WardIdentityComptonScatteringQED} 
\cM_{\mu\nu}q^{\nu}\epsilon_{\text{out}}^{\mu}&=-\I e^{2}\, \bar{u}(p^{\prime},\sigma^{\prime})\left (\dfrac{\cancel{\epsilon}_{\text{out}}(\cancel{p}+\cancel{q}+m)\cancel{q}}{(p+q)^{2}-m^{2}}+\dfrac{\cancel{q}(\cancel{p}^{\prime}-\cancel{q}+m)\cancel{\epsilon}_{\text{out}}}{(p^{\prime}-q)^{2}-m^{2}}\right )u(p,\sigma)\nn\\
&= -\I e^{2}\, \bar{u}(p^{\prime},\sigma^{\prime})\, \cancel{\epsilon}_{\text{out}}\, u(p,\sigma)\, \left (\dfrac{2p_{\mu}q^{\mu}}{(p+q)^{2}-m^{2}}+\dfrac{2p^{\prime}_{\mu}q^{\mu}}{(p^{\prime}-q)^{2}-m^{2}}\right )\nn\\
&=0
\end{align}
using \eqref{eq:PropagatorsSimplifyCC} in the last step.

\begin{figure}[t!]
\centering
\begin{tikzpicture}[scale=1.2]
\setlength{\feynhanddotsize}{1.5ex}
\begin{feynhand}
\node (o) at (-0.25,2.9) {$\color{red}e$} ; 
\node (o1) at (-0.25,1.1) {$\color{red}e^{\prime}$} ; 
\vertex (a0) at (-1.5,4) {$e^{-}$}; 
\vertex (a1) at (1.5,4) {$\epsilon^{\nu}_{\text{in}}, {\color{red}\gamma}$}; 
\vertex (b0) at (0,3); 
\vertex (b1) at (0,1); 
\vertex (c0) at (-1.5,0) {$e^{-}$}; 
\vertex (c1) at (1.5,0.) {$\epsilon^{\mu}_{\text{out}}, {\color{red}\gamma^{\prime}}$}; 
\propag [fer, mom={$p$}] (a0) to (b0);
\propag [pho, mom={$q$}] (a1) to (b0);
\propag [fer, mom={$p+q$}] (b0) to (b1);
\propag [fer, mom={$p^{\prime}$}] (b1) to (c0);
\propag [pho, mom={$q^{\prime}$}] (b1) to (c1);
\end{feynhand}
\end{tikzpicture}
\hspace*{1.5cm}
\begin{tikzpicture}[scale=1.2]
\setlength{\feynhanddotsize}{1.5ex}
\begin{feynhand} 
\node (o) at (-0.25,2.9) {$\color{red}e^{\prime}$}; 
\node (o1) at (-0.25,1.1) {$\color{red}e$};
\vertex (a0) at (-1.5,4) {$e^{-}$}; 
\vertex (a1) at (2.,4) {$\epsilon^{\nu}_{\text{in}}, {\color{red}\gamma}$}; 
\vertex (a11) at (1.,2.5); 
\vertex (b0) at (0,3); 
\vertex (b1) at (0,1); 
\vertex (b00) at (-0.25,2.5); 
\vertex (b11) at (-0.25,1.5); 
\node (o1) at (-0.75,2.0) {$p-q^{\prime}$};
\vertex (c0) at (-1.5,0) {$e^{-}$}; 
\vertex (c1) at (2.,0.) {$\epsilon^{\mu}_{\text{out}}, {\color{red}\gamma^{\prime}}$}; 
\vertex (c11) at (1.,1.5); 
\propag [fer] (b0) to (b1);
\propag [fer, with arrow = 1] (b00) to(b11);
\propag [fer, mom={$p$}] (a0) to (b0);
\propag [pho, mom={$q^{\prime}$}] (c11) to (c1);
\propag [pho] (b0) to (c11);
\propag [fer, mom={$p^{\prime}$}] (b1) to (c0);
\propag [pho, top, mom={$q$}] (a1) to (a11);
\propag [pho, top] (b1) to (a11);
\end{feynhand}
\end{tikzpicture}
\caption{Generalised Compton scattering $e^-\gamma \rightarrow e^- \gamma'$ for two types of photons.}\label{fig:ComptonScatteringTwoPhotons} 
\end{figure}
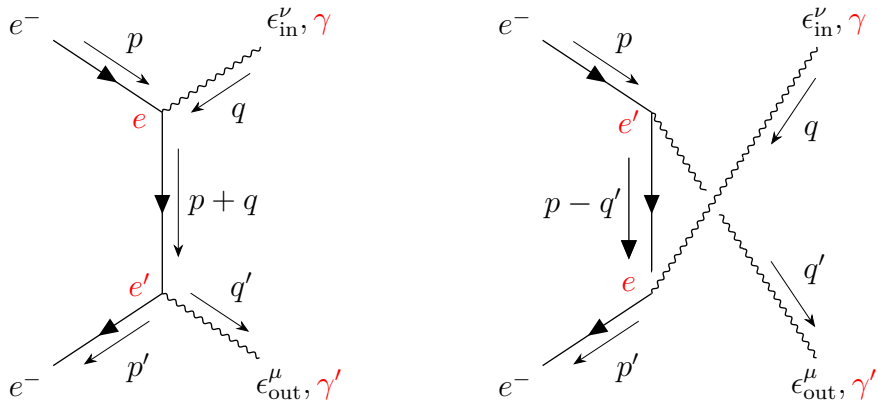

Similarly we may consider having two different types of photons with couplings $e$ and $e'$.
We consider the process $e^-\gamma \rightarrow e^- \gamma' $ shown diagrammatically in Fig.~\ref{fig:ComptonScatteringTwoPhotons} generalising the Compton scattering in Fig.~\ref{fig:ComptonScatteringQED}.
The only difference to the previous amplitude are the two different couplings in each of the vertices.
Going through the algebra from above,
we find that \eqref{eq:WardIdentityComptonScatteringQED} becomes
\begin{equation}
ee^{\prime}-ee^{\prime}=0\, .
\end{equation}
This just means that any value of the two couplings are allowed and hence that the matter fields can be charged under two independent $\mathrm{U}(1)$'s.

\begin{figure}[t!]
\centering
\begin{tikzpicture}[scale=1.2]
\setlength{\feynhanddotsize}{1.5ex}
\begin{feynhand}
\node (o) at (-0.45,2.8) {$\I eT^{a}_{ik}$} ; 
\node (o1) at (-0.45,1.2) {$\I eT^{b}_{kj}$} ; 
\vertex (a0) at (-1.5,4) {$i$}; 
\vertex (a1) at (1.5,4) {$a$}; 
\vertex (b0) at (0,3); 
\vertex (b1) at (0,1); 
\vertex (c0) at (-1.5,0) {$j$}; 
\vertex (c1) at (1.5,0.) {$b$}; 
\propag [fer] (a0) to (b0);
\propag [pho] (a1) to (b0);
\propag [fer] (b0) to [edge label = {$k$}] (b1);
\propag [fer] (b1) to (c0);
\propag [pho] (b1) to (c1);
\end{feynhand}
\end{tikzpicture}
\hspace*{1.5cm}
\begin{tikzpicture}[scale=1.2]
\setlength{\feynhanddotsize}{1.5ex}
\begin{feynhand} 
\node (o) at (-0.45,2.8) {$\I eT^{b}_{ik}$} ; 
\node (o1) at (-0.45,1.2) {$\I eT^{a}_{kj}$} ; 
\vertex (a0) at (-1.5,4) {$i$}; 
\vertex (a1) at (2.,4) {$a$}; 
\vertex (a11) at (1.,2.5); 
\vertex (b0) at (0,3); 
\vertex (b1) at (0,1); 
\vertex (b00) at (-0.25,2.5); 
\vertex (b11) at (-0.25,1.5); 
\vertex (c0) at (-1.5,0) {$j$}; 
\vertex (c1) at (2.,0.) {$b$}; 
\vertex (c11) at (1.,1.5); 
\propag [fer] (b0) to [edge label = {$k$}] (b1);
\propag [fer] (a0) to (b0);
\propag [pho] (c11) to (c1);
\propag [pho] (b0) to (c11);
\propag [fer] (b1) to (c0);
\propag [pho] (a1) to (a11);
\propag [pho, top] (b1) to (a11);
\end{feynhand}
\end{tikzpicture}
\caption{Generalised Compton scattering $e_i\, \gamma^a\rightarrow e_j\, \gamma^b$ for many species of matter and gauge fields with couplings $T^{a}_{ij}$ between two matter and one gauge particle.}\label{fig:ComptonScatteringGeneral} 
\end{figure}
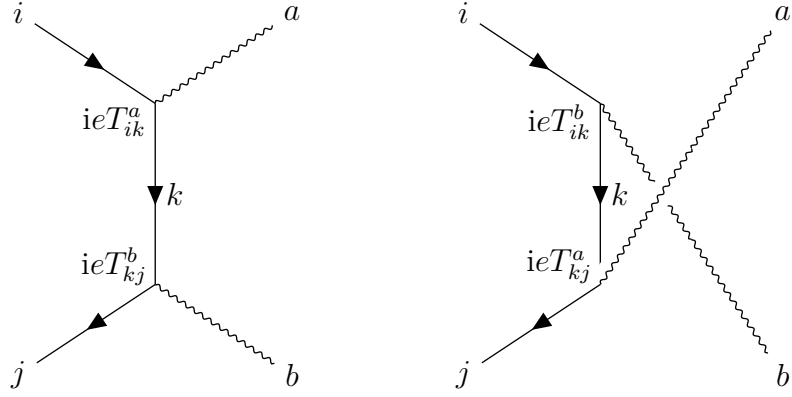

Now, for an arbitrary number of matter particles $e_i$ and gauge particles $\gamma^a$,
we look at the process
\begin{equation}
e_i\, \gamma^a\rightarrow e_j\, \gamma^b\, , \qquad i,j=1, \cdots,  N\qquad a,b=1, \cdots,  D\, .
\end{equation}
The Feynman diagrams are shown in Fig.~\ref{fig:ComptonScatteringGeneral} where we introduced coupling constants $T^{c}_{lk}$ between two matter particles $e^{l}$, $e^{k}$ and one gauge particle $\gamma^{c}$.
We compute the diagrams in Fig.~\ref{fig:ComptonScatteringGeneral}
\begin{align}\label{eq:QEDextManySpinOne} 
\cM^{ab}_{ij}&=-\I  e^{2}\, \bar{u}(p^{\prime},\sigma^{\prime})\left (T_{ik}^{a}\dfrac{\gamma_{\mu}(\cancel{p}+\cancel{q}+m)\gamma_{\nu}}{(p+q)^{2}-m^{2}} T_{kj}^{b}+T_{ik}^{b}\dfrac{\gamma_{\nu}(\cancel{p}-\cancel{q}^{\prime}+m)\gamma_{\mu}}{(p-q^{\prime})^{2}-m^{2}}T_{kj}^{a}\right )u(p,\sigma)\epsilon_{\text{in}}^{\nu}\epsilon_{\text{out}}^{\mu}\nn\\
&=\left(\cM^{\mu\nu}\right)_{ij}^{ab}\epsilon^{\nu}_{\text{in}}\epsilon_{\text{out}}^{\mu}
\end{align}
which gives rise to
\begin{equation}\label{eq:QEDextManySpinOneWard} 
q^{\nu}(\epsilon_{\text{out}}^{\mu})\left(\cM^{\mu\nu}\right)_{ij}^{ab}\propto T_{ik}^{a}T_{kj}^{b}-T_{ik}^{b}T_{kj}^{a}=0\, .
\end{equation}
Said differently, the Ward identity implies the vanishing commutator
\begin{equation}
[T^{a},T^{b}] = 0\, .
\end{equation}
This means we have simply $D$ copies of QED or, more precisely, a $\mathrm{U}(1)^{D}$ gauge theory coupled to $N$ charged scalar fields.

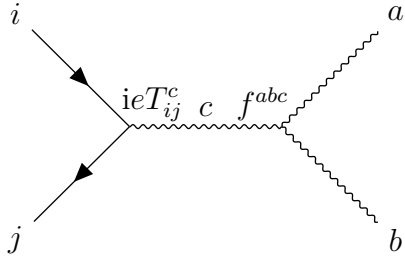
\begin{figure}[t!]
\centering
\begin{tikzpicture}[scale=1.0]
\setlength{\feynhanddotsize}{1.5ex}
\begin{feynhand}
\node (o) at (0.3,2.3) {$\I eT^{c}_{ij}$};
\node (o) at (1.75,2.3) {$f^{abc}$};
\vertex (a0) at (-1.5,3.5) {$i$}; 
\vertex (a1) at (2,2); 
\vertex (d0) at (3.5,3.5) {$a$}; 
\vertex (d1) at (3.5,0.5) {$b$}; 
\vertex (b0) at (0,2); 
\vertex (c0) at (-1.5,0.5) {$j$}; 
\propag [fer] (a0) to (b0);
\propag [pho] (b0) to [edge label = {$c$}] (a1);
\propag [pho] (a1) to (d0);
\propag [pho] (a1) to (d1);
\propag [fer] (b0) to (c0);
\end{feynhand}
\end{tikzpicture}
\caption{Feynman diagram contributing to the process $e_i\, \gamma^a\rightarrow e_j\, \gamma^b$ involving the exchange of a gauge particle and therefore requiring a cubic interaction vertex among the gauge particles.}\label{fig:ComptonScatteringThirdDiagram} 
\end{figure}

This is true unless there exist self-interactions among the gauge bosons.
That is, if there exists an interaction vertex $\sim f^{abc}$ coupling three gauge particles to each other,
then there is a third diagram shown in Fig.~\ref{fig:ComptonScatteringThirdDiagram} that contributes to the process $e_i\, \gamma^a\rightarrow e_j\, \gamma^b$.
Contrary to the diagrams in Fig.~\ref{fig:ComptonScatteringGeneral}, there is now a gauge particle exchanged between the two vertices.
It turns out that the structure of such a three point vertex is highly constrained by e.g. permutation symmetries of external particles. For the sake of brevity, we leave a more detailed discussion to App.~\ref{app:compton_sQED} where we present the full argument that non-abelian Yang-Mills theory is the unique description theories with many helicity-1 fields.
Crucially, this argument works \emph{without having to impose any gauge symmetry to begin with: the underlying Lie-algebra structure arises as a consistency condition of the Ward identity (or Lorentz invariance)}.\footnote{At this point, the reader might wonder why we have to add self-interactions for the gauge fields in the first place. In the case of scalar QED as discussed in App.~\ref{app:compton_sQED}, one finds that the modified $4$-vertex develops a pole in the soft limit suggesting the hidden exchange of a massless gauge field. For QED, one argues that, based on symmetries and charge conservation, there must be an additional diagram if one requires e.g. $[T^{a},T^{b}]\neq 0$. If it sufficed to just add a new vertex involving two fermions and two gauge fields, one would again find that this contribution to Compton scattering must have a pole naturally leading to the diagram in Fig.~\ref{fig:ComptonScatteringThirdDiagram}. Hence, any violation of $[T^{a},T^{b}]=0$ consistent with the Ward identity in QED would require 3-point self-interactions of the gauge fields which can then be fixed using the arguments presented in App.~\ref{app:compton_sQED}.}

To make a long story short,
combining the contribution from the diagram in Fig.~\ref{fig:ComptonScatteringThirdDiagram} with \eqref{eq:QEDextManySpinOne} and testing the Ward identity analogously to \eqref{eq:QEDextManySpinOneWard} amounts to requiring
\begin{equation}
T_{ik}^{a}T_{kj}^{b}-T_{ik}^{b}T_{kj}^{a}=\I f^{abc}T^{c}_{ij}
\end{equation}
or equivalently
\begin{Boxequ}
\begin{equation}\label{eq:ComGenNonAbYM} 
[T^{a},T^{b}]= \I f^{abc}T^{c}\, .
\end{equation}
\end{Boxequ}
This is nothing but a \emph{non-Abelian algebra} which in turn gives rise to the notion of \emph{non-Abelian gauge symmetries} and the associated \emph{Yang-Mills theories}\index{Yang-Mills theories}.
It is important to appreciate the significance of this statement which is sometimes taken for granted when starting from a given gauge group with underlying Lie algebra: the couplings of helicity-1 fields to matter $\sim T_{ij}^{a}$ and among themselves $\sim f^{abc}$ satisfy the non-linear relationship \eqref{eq:ComGenNonAbYM} (an algebra) \emph{because} the Ward identity needs to be imposed. Said differently,
the Lie algebra structure of the underlying theory \emph{emerges from Lorentz invariance and unitarity}. It is in fact inevitable!

To summarise, we conclude that a system with many gauge fields is either
\begin{enumerate}
\item a theory with many photon-like gauge bosons, that is, $G=\mathrm{U}(1)^{n}$
\item or a non-Abelian Yang-Mills system with $G$ being some non-Abelian group $\mathrm{SU}(N)$, $\mathrm{SO}(N)$ etc.
\end{enumerate}
The structure constants\index{Structure constants} $f^{abc}$ appearing in \eqref{eq:ComGenNonAbYM} satisfy the \emph{Jacobi identity}\index{Jacobi identity}
\begin{equation}
f^{abd}f^{dce}+f^{bcd}f^{dae}+f^{cad}f^{dbe}=0
\end{equation}
due to
\begin{equation}
[A,[B,C]]+[B,[C,A]]+[C,[A,B]]=0\, .
\end{equation}

In general this describes the algebra of a Lie group. The group elements are obtained from exponentiating\footnote{We mostly work with compact, simply-connected Lie groups for which this is always the case.}
\begin{equation}
U=\ee^{\I\theta^{a}T^{a}}=\mathds{1}+\I \theta^{a}T^{a}+\ldots
\end{equation}
with $T^{a}$ the generators and $\theta^{a}$ some parameter. The Lie group $G$ itself corresponds to a smooth manifold with coordinates $\theta^{a}$, $a=1,\ldots ,D$. We call $D$ the dimension of $G$, while the rank $r$ corresponds to the number of generators that commute. This is the general structure of Yang-Mills theories that we describe next.

\section{Yang-Mills theory}\label{sec:YMc} \index{Yang-Mills theory}

Above, we derived the structure of scattering amplitudes or rather the properties of couplings between different species of helicity-1 particles from first principles.
Now, let us see how this translates into theories of fields.
We briefly review the Abelian case complementing the treatment of section~\ref{sec:GaugeSymmetriesLorentzInv} before we put the lessons learned in the preceding section about non-Abelian gauge theories to good use.

\subsection{The Abelian case}

Recall that for Abelian gauge fields $A_\mu$ transforming as $A_\mu \rightarrow A_\mu+\partial_\mu \alpha$ we have a Lagrangian of the form
\begin{equation}
\cL=-\frac{1}{4}F^{\mu\nu}F_{\mu\nu}+J^\mu A_\mu
\end{equation}
with $J^\mu$ a conserved current $\partial_\mu J^\mu=0$. Coupling this field to a matter spin $1/2$ field, the current is $J^\mu=e\bar\psi\gamma^\mu\psi$ associated to the symmetry 
$\psi\rightarrow e^{\I e\alpha'}\psi$ with $\alpha^{\prime}\in\mathbb{R}$ some constant. Plugging this current into the Lagrangian together with the kinetic term $\bar\psi\I \cancel{\partial}\psi$ for $\psi$ we can see that the Lagrangian can be written as
\begin{align}
\cL&=-\frac{1}{4}F^{\mu\nu}F_{\mu\nu}+J^\mu A_\mu+\bar\psi\I \cancel{\partial}\psi\nn\\
&=-\frac{1}{4}F^{\mu\nu}F_{\mu\nu}+\bar\psi\I \cancel{D}\psi
\end{align}
with $D_\mu\equiv \partial_\mu-{\rm i}e A_\mu $ the covariant derivative.
Note that the conserved current $J^{\mu}$ and the kinetic term for the fermions together give rise to $D_{\mu}$.
In this form, the Lagrangian is invariant  under the {\bf local} ($\alpha'=\alpha(x)$) gauge transformation
\begin{equation}
\psi\raw\ee^{\I e \alpha}\psi, \qquad D_{\mu}\psi=\left(\p_{\mu}-\I e A_{\mu} \right)\psi\raw \ee^{\I e\alpha}D_{\mu}\psi\kom A_{\mu}\raw A_{\mu}+\p_{\mu}\alpha\, .
\end{equation}
The field strength can be written as
\begin{equation}
F_{\mu\nu}=\frac{\rm i}{e}\, [D_\mu, D_\nu]=\partial_\mu A_\nu -\partial_\nu A_\mu
\end{equation}
and it is invariant under the gauge transformation.

Note that
\begin{equation}
\p_{\mu}\psi\raw\ee^{\I e\alpha}\left (\p_{\mu}\psi+\I e(\p_{\mu}\alpha)\psi\right )
\end{equation}
is not covariant, but 
$D_\mu$ transforms covariantly in the sense that it transforms with a phase like $\psi$ does, i.e., $D_{\mu}\psi\raw \ee^{\I e\alpha}D_{\mu}\psi$.
This brings us to the standard argument (that we did not follow here) that promoting a global symmetry ($\alpha^{\prime} =$ constant) to a local symmetry $\alpha^{\prime}=\alpha(x)$ motivates introducing a gauge field $A_\mu$ to turn the standard derivative into a covariant derivative containing $A_\mu$.
The latter transforms appropriately under gauge transformations such that the Lagrangian is gauge invariant for the matter fields $\psi$ and gauge fields $A_\mu$. This is a simple prescription to follow. However, since this is an arbitrary logic (why to impose that the symmetry is local? Is the introduction of $A_\mu$ unique? etc.), we have preferred to actually {\bf derive}  the existence of the gauge symmetry and the covariant derivative rather than imposing it, in the sense that we asked the question of how to consistently construct a theory for helicity $|\lambda|=1$ massless vector fields coupled to matter fields of spin/helicity $0,1/2$. The coupling of matter to the gauge field is determined by the coupling of the conserved current to the gauge field $A^\mu J_\mu$ which gives precisely the contribution that turns normal derivatives into covariant derivatives.

\subsection{The general non-Abelian case}

Let us now generalise this to the non-Abelian case. For a general\footnote{We will comment on the types of Lie groups that appear frequently in the case of non-Abelian gauge theories at the end of this chapter. For those, the assumptions being made throughout this section do hold.} Lie group with generators $T_{R}^{a}$ in some representation $R$ that we keep implicit here, a field $\psi$ transforms as
\begin{equation}
\psi\raw U\psi\kom U=\ee^{\I\theta^{a}T_{R}^{a}}\, .
\end{equation}
The covariant derivative transforms also covariantly in the sense that
\begin{equation}\label{eq:TrafoCovDerNonAb} 
D_{\mu}\psi\raw U\, D_{\mu}\psi
\end{equation}
where now in terms of the coupling constant $g$
\begin{equation}
D_{\mu}=\p_{\mu}-\I gA_{\mu}
\end{equation}
or, more explicitly, in components
\begin{equation}
(D_{\mu})_{ij}=\p_{\mu}\delta_{ij}-\I g A_{\mu}^{a} (T_{R}^{a})_{ij}
\end{equation}
with the generators being in the same representation $R$ of $\psi$.
Let us now impose that \eqref{eq:TrafoCovDerNonAb} is true.
Then we want to know how $A_{\mu}$ transforms which is why we compute
\begin{align}
D_{\mu}\psi&\raw \p_{\mu}(U\psi)-\I g\, A^{\prime}_{\mu}\, U\psi\nn\\
&=U\p_{\mu}\psi+(\p_{\mu}U)\psi-\I g\, A^{\prime}_{\mu}\, U\psi\nn\\
&=UD_{\mu}\psi+\left (\I g\, UA_\mu+(\p_{\mu}U)-\I g\, A^{\prime}_{\mu}\, U\right )\psi\, .
\end{align}
In order to ensure \eqref{eq:TrafoCovDerNonAb}, we need to impose
\begin{align}
\I g\, UA_\mu+(\p_{\mu}U)-\I g\, A^{\prime}_{\mu}\, U=0
\end{align}
which amounts to the following general and infinitesimal transformations:
\begin{Boxequ}
\vspace*{0.05cm}
\begin{equation}
A^{\prime}_{\mu}=UA_{\mu}U^{-1}-\dfrac{\I}{g}\left (\p_{\mu} U\right )\, U^{-1}\kom A_{\mu}^{a}\raw A_{\mu}^{a}+\dfrac{1}{g}\p_{\mu}\theta^{a}-f^{abc}\theta^{b}A^{c}_{\mu}\, .
\end{equation}
\end{Boxequ}
Here we used \eqref{eq:ComGenNonAbYM} for the commutator for $T^{a}_{R}$.

The field strength for non-Abelian groups is most easily found by considering
\begin{equation}
[D_{\mu},D_{\nu}]\psi(x)=\left (-\I g(\p_{\mu}A_{\nu}-\p_{\nu}A_{\mu})-g^{2}[A_{\mu},A_{\nu}]\right )\psi(x)
\end{equation}
so that
\begin{equation}
F_{\mu\nu}=\dfrac{\I}{g}[D_{\mu},D_{\nu}]=\p_{\mu}A_{\nu}-\p_{\nu}A_{\mu}-\I g [A_{\mu},A_{\nu}]\, .
\end{equation}
We call $A_{\mu}^{a}$ the \emph{gauge connection}\index{Yang-Mills theory!Gauge connection} and $F_{\mu\nu}^{a}$ the \emph{curvature}\index{Yang-Mills theory!Curvature}.\footnote{Let us briefly explain the terminology here. In the language of mathematics, gauge theories are simply built from principal and associated vector bundles over some spacetime manifold $M$.
The former are bundles $\pi:\,P\raw M$ whose fibres are gauge groups $\pi^{-1}(x)=G$, $x\in M$, while the latter are obtained from suitable representations of $G$.
That is, let $\rho:\, G\raw \text{GL}(V)$ be a representation of $G$, then the bundle $\pi_{E}:\, P\times_{\rho}V\raw M$ has fibres $\pi^{-1}(x)=V$, $x\in M$.
A choice of gauge corresponds to a local patch of $P$ in which the connection is defined by the $1$-form $A$ which takes values in the corresponding Lie algebra.
This connection then defines a curvature $2$-form $F$ which we identify with the field strength above.
From this point of view, matter fields are sections of the associated vector bundle in which local trivialisations are the associated choices of gauge.
For a more detailed introduction, see the Part III lecture notes on \emph{Advanced Quantum Field Theory} by D. Skinner \cite{skinneraqft} or the (publicly available) book \cite{hamilton2017mathematical}. } 
The latter transforms under gauge transformations as
\begin{equation}
F^{\prime}_{\mu\nu} = UF_{\mu\nu}U^{-1} \kom F_{\mu\nu}^{a}\raw F_{\mu\nu}^{a}-f^{abc}\theta^{b}F^{c}_{\mu\nu}\, .
\end{equation}
Notice, as the index structure indicates, gauge fields always transform in the {\bf adjoint representation} which is the one for which the generators are the structure constants themselves $T^a_{bc}=f^a_{bc}$. 

The most general gauge invariant, renormalisable Lagrangian takes the form
\begin{equation}\label{eq:GeneralLagrangianYM} 
\cL=-\dfrac{1}{4}g_{ab}F^{a}_{\mu\nu}F^{b,\mu\nu}+\cL_{M}(\psi,D_{\mu}\psi)+\Theta\, F^{a}_{\mu\nu}\tilde{F}^{a,\mu\nu}
\end{equation}
where $g_{ab}$, $a,b=1,\cdots,\text{dim}(G)$, is a metric on the group manifold.
In components, we write
\begin{equation}
F^{a}_{\mu\nu}=\p_{\mu}A^{a}_{\nu}-\p_{\nu}A^{a}_{\mu}+gf^{abc}A^{b}_{\mu}A_{\nu}^{c}\kom \tilde{F}^{a,\mu\nu}=\frac{1}{2}\epsilon^{\mu\nu\rho\sigma}F^{a}_{\rho\sigma}\, .
\end{equation}
In order to have only physically propagating particles (positive kinetic energy), we restrict to groups for which $g_{ab}$ is positive definite. This implies that the group is compact, simple or semi-simple and eliminates all non-compact groups. This again is a strong argument by which we can eliminate an infinite number of potential symmetry groups.
We can thus safely restrict out attention to the compact groups classified by Cartan. These groups (unlike the non-compact ones) allow for finite dimensional unitary representations which makes them suitable to describe physical  interactions.

We typically normalise the metric as  $g_{ab}=\kappa\delta_{ab}$ with $\kappa=1$ since $\kappa$ may be absorbed in a rescaling of $F_{\mu\nu}$ and $A_\mu$. This rescaling allows to move the coupling constant from the kinetic term for gauge fields to its appearance in the definition of the covariant derivative and the $F_{\mu\nu}$ fields. For instance, rescaling $A_\mu\raw A_\mu/g$ amounts to rescale $F_{\mu\nu}\raw F_{\mu\nu}/g$ and have the covariant derivative independent of $g$. But then $g$ appears in the kinetic term for the gauge fields as $F^{\mu\nu}F_{\mu\nu}/g^2$. Therefore the arbitrariness in rescaling the metric $g_{ab}$ amounts to the freedom in where to include the coupling $g$ in the Lagrangian. As long as we are consistent, the physical results are unaffected by this rescaling, but it illustrates the need to have the free parameter $g$.

Notice that the last term in \eqref{eq:GeneralLagrangianYM} can be written as
\begin{equation}
\Theta\, F^{a}_{\mu\nu}\tilde{F}^{a,\mu\nu}=2\Theta\, \p_{\mu}\left (\epsilon^{\mu\nu\rho\sigma}A_{\nu}^{a}F^{a}_{\rho\sigma}\right )\, .
\end{equation}
Hence, being a total derivative, this term has no immediate effect on the classical equations of motion, but is indispensable in a full quantum theory.
In fact, one can show using canonical quantisation that the term associated to $\Theta$ needs to be taken into account when working in a basis of gauge invariant physical states \cite{JackiwTopInv}. $\Theta$ is arbitrary in the sense that no physical principle determines the value of $\Theta$. However, $\Theta$ does not change under local gauge-invariant perturbations nor under time evolution. Hence, $\Theta$ labels different sectors of the theory and it corresponds to a different choice of vacuum. In fact, this can be formulated in terms of a a superselection rule: \emph{quantising non-Abelian gauge theories requires a definite choice of $\Theta$ restricting the Hilbert space of states in a specific way}.
Once $\Theta$ has been fixed, one cannot reach states of the full Hilbert space with another value of $\Theta$. It therefore is a new \emph{fundamental constant} which is required to specify the dynamics of quantum Yang-Mills theory.

We assume here implicitly that $\Theta$ is constant.
We will briefly introduce \emph{axions} in section~\ref{sec:TopDownBSM} in which case $\Theta$ becomes a dynamical field itself.
But even for constant $\Theta$, it can have non-trivial effects as studied in \cite{Witten:1979ey} showing that magnetic monopoles have non-integer valued electric charge. This effect was called \emph{Witten effect}\index{Witten effect}. Beyond that, for the specific value of $\Theta=\pi$, the resulting theory describes a new state of matter corresponding to \emph{topological insulators} \cite{Qi:2008pi}. Such topological states of matters have been constructed in experiments and have surprisingly interesting properties. For instance, if we place an electric charge in a vacuum with $\Theta=0$ opposite of a material with $\Theta=\pi$, the resulting induced electromagnetic field mirrors that of a \emph{magnetic monopole} placed at the same distance to the boundary (rather than that of a mirror electric charge as for a conventional metal), see \cite{Qi:2008pi} for details.

\subsubsection*{Field equations and Bianchi identities}

Ignoring $\Theta$ for the moment, we mostly work with the Lagrangian
\begin{equation}
\cL=-\dfrac{1}{4}\left (F^{a}_{\mu\nu}\right )^{2}+\cL_{M}(\psi,D_{\mu}\psi)
\end{equation}
where $\cL_{M}(\psi,D_{\mu}\psi)$ stands for a general Lagrangian density describing interactions between the gauge fields and some matter fields $\psi$. The field equations are determined as usual from the Euler-Lagrange equations
\begin{equation}
\p_{\mu}\dfrac{\p\cL}{\p (\p_{\mu}A_{\nu}^{a})}=\dfrac{\p\cL}{\p A_{\nu}^{a}}\, .
\end{equation}
One finds that
\begin{equation}
\p_{\mu}\dfrac{\p\cL}{\p (\p_{\mu}A_{\nu}^{a})}=-\p_{\mu}F^{a,\mu\nu}\kom \dfrac{\p\cL}{\p A_{\nu}^{a}}=-gf^{abc} A_{\mu}^{b}F^{c,\nu\mu}-\I\dfrac{\p\cL_{M}}{\p(D_{\nu}\psi)}T^{a}\psi
\end{equation}
and thus
\begin{equation}
\p_{\mu}F^{a,\mu\nu}=-J^{a,\nu}
\end{equation}
in terms of the current
\begin{equation}
J^{a,\nu}=-gf^{abc} A_{\mu}^{b}F^{c,\nu\mu}-\I\dfrac{\p\cL_{M}}{\p(D_{\nu}\psi)}T^{a}\psi\, .
\end{equation}
The current is conserved
\begin{equation}
\p_{\nu}J^{\nu}_{a}=0\, ,
\end{equation}
as expected from Noether's theorem. But this equation is not gauge covariant. In terms of covariant derivatives, one can write this equation as
\begin{equ}[Yang-Mills equations]\index{Yang-Mills equations}\index{Yang-Mills theory!YM equations}
\begin{equation}
D_{\mu}F^{a,\mu\nu}=-j^{a,\nu}\, .
\end{equation}
\end{equ}
The associated current
\begin{equation}
j^{a,\nu}=-\I\dfrac{\p\cL_{M}}{\p(D_{\nu}\psi)}T^{a}\psi
\end{equation}
corresponding to the contribution to the total current $J^{a,\nu} $ coming from the matter fields. In terms of  $j^{a,\nu} $ we can write a  gauge covariant equation
\begin{equation}
D_{\nu}j^{\nu}_{a}=0
\end{equation}
meaning that the matter current $j^{a,\nu} $ is not conserved but only  ``covariantly'' conserved. This is a manifestation of the fact that the total current $J^{a,\nu}$ includes the matter part $j^{a,\nu}$ but also a gauge field dependent part, since for non-Abelian gauge theories the gauge fields transform non-trivially under the gauge transformations (and therefore are self-interacting). This is similar to the situation in General Relativity in which the total energy momentum tensor $T_{\mu\nu}$ is conserved but the matter part (the right hand side of Einstein's equations) is only covariantly conserved due to the fact that gravity is also self-interactive and contributes a term to the total energy momentum tensor.

In addition, the curvature $F^{a}_{\mu\nu}$ also satisfies the \emph{Bianchi identity}\index{Bianchi identity}\index{Yang-Mills theory!Bianchi identity}
\begin{equ}[Bianchi identity]
\begin{equation}
D_{\mu}F^{a}_{\nu\lambda}+D_{\nu}F^{a}_{\lambda\mu}+D_{\lambda}F_{\mu\nu}^{a}=0\, .
\end{equation}
\end{equ}
This can also be written in terms of the dual field strength $\tilde{F}_{\mu\nu}=\frac{1}{2}\epsilon_{\mu\nu\rho\sigma}F^{\rho\sigma}$ as
\begin{equation}
D^\mu \tilde{F}_{\mu\nu}=0\, .
\end{equation}

Notice that Bianchi identities and field equations are interchanged when we interchange $\tilde{F}_{\mu\nu}\leftrightarrow F_{\mu\nu}$.
This has interesting consequences because in the language of differential forms we can construct solutions $F=\pm\star F$ which are called \emph{(anti-)self-dual instanton solutions}.
For them, the Biachi identity automatically implies Yang-Mills equations.
In fact, $F=\pm\star F$ is an even simpler system of equations which is only of first order, whereas Yang-Mills equation is of order $2$.\footnote{This is also why these solutions are particularly interesting from the mathematics point of view and have been used to investigate the topology of four manifolds \cite{donaldson1983self}. In some way, this fact is analogous to the study of manifolds in two dimensions and their topology where holomorphic functions play an important role. The Cauchy-Riemann equations form a system of first order differential equations depending only on the conformal structure determined by the metric, but they also imply Laplace’s equation for real and imaginary part as a second order equation. In this sense, one can think of self-dual solutions $F=\star F$ as the generalisation of holomorphic functions on $2$-dimensional Riemann surfaces to $4$-dimensional manifolds.}
These field configurations are topologically non-trivial (because $\int_{\bR^{4}}\, \tr(F\wedge F)$ is a topological invariant for $4$-dimensional manifolds) and localised both in space and time -- hence the name \emph{instant}on.
Such states cannot be described in the usual perturbative manner through Feynman diagrams, but they are inherently non-perturbative.
While we will not have time to discuss instantons in detail throughout these lectures,
many standard textbooks like \cite{Srednicki:2007qs,Shifman:2012zz,Weinberg:2012pjx} provide detailed introductions to the topic of these solutions.

\subsubsection*{General Comments}

Let us make some general comments about the above:
\begin{enumerate}
\item In comparison with QED, the kinetic term $\sim (F^{a}_{\mu\nu})^{2}$ includes a cubic coupling $\sim g(\p A)AA$ as well as a quartic coupling $g^2 AAAA$. Hence, the gauge fields have self-interactions unlike photons. This is as predicted in the previous section when the need for Yang-Mills was demonstrated. 
In particular, this non-linearity implies that, unlike for light beams as in QED, propagating waves of non-Abelian fields interact among each other.
\item For QED $F_{\mu\nu}$ is invariant, but for Yang-Mills $F_{\mu\nu}^{a}$ is only covariant. The field equations and Bianchi identities are very similar in both cases and the Yang-Mills case reduces to the Maxwell case when the group is Abelian, i.e., a $\mathrm{U}(1)$.
\item For QED, there exists a conserved, gauge invariant current. For Yang-Mills, of both currents $J_{\mu}^{a}$ and $j_{\mu}^{a}$ one is conserved, but not gauge invariant and the other is gauge invariant but not conserved. 

\begin{table}[t!]
\centering
\begin{tabular}{|c|c|c|}
\hline 
 & YM & gravity \\ 
\hline 
\hline 
Connections & $A_{\mu}^{a}$ & $\Gamma^{\rho}_{\mu\nu}$ \\ 
\hline 
Covariant derivative & $D_{\mu}\psi=(\p_{\mu}-\I gA_{\mu})\psi$ &$D_{\mu}V^{\nu}=\p_{\mu}V^{\nu}+\Gamma_{\mu\rho}^{\nu}V^{\rho}$  \\ 
\hline 
Curvature & $\frac{\I}{g}[D_{\mu},D_{\nu}]\psi=F_{\mu\nu}\psi$&$[D_{\mu},D_{\nu}]V^{\rho}=R_{\alpha}\,^{\rho}\,_{\mu\nu}V^{\alpha}$ \\ 
\hline 
\end{tabular}
\caption{Comparison between Yang-Mills theory and General Relativity.}\label{tab:ComparisonYMGR} 
\end{table}

\item \emph{Analogy with gravity}. There is an interesting similarity between the structure of Yang-Mills theories and gravity as summarised in table~\ref{tab:ComparisonYMGR}.
Notice that we can define $F^{a}\,_{b\mu\nu}=F^{c}_{\mu\nu}f_{c}\,^{a}\,_{b}$ which describes the change parallel transported around infinitesimal loops in internal space $V=V^{a}T_{a}$ for the generators $T_{a}$.\footnote{There is an elegant connection between gauge theories and the mathematics of fibre bundles that we mentioned above which is however beyond the scope of this course.}
Also recall that in gravity the stress energy tensors are only covariantly conserved and the Bianchi identity holds. Yang-Mills theories may be considered as symmetries in an extended spacetime, adding compact dimensions to our 4-dimensional spacetime. The simplest realisation is a fifth dimension corresponding to a circle which would naturally have the $\mathrm{U}(1)$ symmetry. Symmetries of higher dimensional compact manifolds would correspond to Yang-Mills groups. This is the idea behind Kaluza-Klein theories\index{Kaluza-Klein theory} of extra dimensions\index{Extra dimensions} \cite{Kaluza:1921tu,Klein:1926tv,Lovelace:1971fa,Schwarz:1972asw}.

\item \emph{Weinberg-Witten theorem \cite{Weinberg:1980kq}}\index{Weinberg-Witten theorem} (stated without proof): There can never be a conserved Lorentz covariant current in a theory with massless $\lambda=1$ particles with non-vanishing charges associated to that current. The version of this theorem for gravity states: A theory with a conserved Lorentz covariant energy momentum tensor cannot have a massless particle of helicity $2$. This illustrates the uniqueness of both Yang-Mills and gravity theories.

\end{enumerate}

\subsubsection*{Properties of compact Lie groups}

\begin{table}[t!]
\centering
\begin{tabular}{|c|c|c|}
\hline 
Group & Rank & Dimension \\ 
\hline 
\hline 
$\mathrm{SU}(N)$ & $N-1$ & $N^{2}-1$ \\ 
\hline 
$\mathrm{SO}(N)$ & $\frac{N}{2}, \frac{N-1}{2}$ & $\frac{N(N-1)}{2}$ \\ 
\hline 
$\mathrm{SP}(N)$ & $N$ & $N(2N+1)$\\ 
\hline 
Exceptional $\mathrm{G}_{2},\mathrm{F}_{4},\mathrm{E}_{6},\mathrm{E}_{7},\mathrm{E}_{8}$ & $2,4,6,7,8$ & $14,52,78,133,248$ \\ 
\hline 
\end{tabular} 
\caption{Compact simple Lie groups}\label{tab:CompactLieGroups}\index{Groups!Classification}
\end{table}

Recall the following about compact Lie groups:
\begin{enumerate}
\item Compact Lie groups are classified as summarised in table~\ref{tab:CompactLieGroups}.
\item A group is called
\begin{itemize}
\item \emph{Simple}\index{Groups!Simple} if there exists no non-trivial ideal (invariant sub-algebra), 
\item \emph{Semi-simple}\index{Groups!Semi-simple} if it can be written as a product of simple groups and $\mathrm{U}(1)$'s.
\end{itemize}
For these cases there is always finite-dimensional irreducible Hermitian representations of the  algebra lifting to unitary representations of the group.
\item Standard representations:\index{Representations}
\begin{itemize}
\item \emph{Fundamental}\index{Representations!Fundamental}: this is the smallest non-trivial representation and denoted as
\begin{equation}
\phi_{i}\raw \phi_{i}+\I\alpha^{a}(T^{a}_{F})_{ij}\phi_{j}
\end{equation}
where $T_{F}^{a}$ are the generators $T^{a}$ in the corresponding representation. For $\mathrm{SU}(N)$, this representation is $N$-dimensional.
\item \emph{Anti-fundamental}\index{Representations!Anti-fundamental}: The anti-fundamental representation is related to the fundamental via
\begin{equation}
T_{AF}^{a}=-(T_{F}^{a})^{*}\, .
\end{equation}

Similarly, one then finds (using the hermiticity of $T^a$)
\begin{equation}
\phi_{i}^{*}\raw \phi_{i}^{*}+\I\alpha^{a}(T^{a}_{AF})_{ij}\phi_{j}^{*}=\phi_{i}^{*}-\I\alpha^{a}(T^{a}_{F})_{ji}\phi_{j}^{*}\, .
\end{equation}
\item \emph{Adjoint}\index{Representations!Adjoint}: For the adjoint representation for $\mathrm{SU}(N)$, one can choose the generators
\begin{equation}
(T_{A}^{a})^{bc}=-\I f^{abc}
\end{equation}
corresponding to a $(N^{2}-1)$-dimensional representation.
\end{itemize}
\item Normalisation\index{Representations!Normalisation}: for $\mathrm{SU}(N)$, we normalise the generators $T^{a}$ such that
\begin{equation}
\tr(T^{a}T^{b})=\dfrac{1}{2}\delta^{ab}
\end{equation}
where
\begin{equation}
T^{a}T^{b}=\dfrac{1}{2N}\delta^{ab}+\dfrac{1}{2}d^{abc}T^{c}+\dfrac{1}{2}\I f^{abc}T^{c}
\end{equation}
in terms of the symmetric coefficients
\begin{equation}\label{eq:coefficientsdabcym} 
d^{abc}=2\tr\left (T^{a}\lbrace T^{b},T^{c}\rbrace\right )\, .
\end{equation}
For a specified representation $R$,
we then have
\begin{equation}
\tr(T_{R}^{a}T_{R}^{b})=T(R)\delta^{ab}
\end{equation}
for some index $T(R)$. $T(R)$ is the \emph{Dynkin index}\index{Dynkin index} of the representation.
E.g., $T(R)$is equal to $\tfrac12$ for generators in the fundamental representation. The quadratic Casimir is\index{Representations!Casimir}
\begin{equation}
C(R)=T_{R}^{a}T_{R}^{a}
\end{equation}
or the identity operator by Schur's lemma\index{Schur's lemma}. For the fundamental representation, one finds
\begin{equation}
C_{F}=\dfrac{N^{2}-1}{2N}
\end{equation}
and for the adjoint
\begin{equation}
C_{A}=N\, .
\end{equation}
\end{enumerate}


\chapter{\bf Broken Symmetries}
\label{chap:ssb}

\vspace{0.5cm}
\begin{equ}[Symmetries we may not see]
{\it  It suddenly came home to us that there is much more symmetry in the laws of nature than one would guess merely by looking at the properties of elementary particles.
The reality we observe in our laboratories is only an imperfect reflection of a deeper and more beautiful reality, the reality of the equations that display all the symmetries of the theory.}\\

\rightline{\it Steven Weinberg}
\end{equ}
\vspace{0.5cm}



In this chapter, we introduce the fundamental concepts of symmetries that may be hidden from an observer -- symmetries upheld by the Lagrangian but not apparent in physical observables. This phenomenon is commonly known as \emph{Spontaneous Symmetry Breaking} (SSB)\index{Spontaneous symmetry breaking}\index{SSB}. As will be explored in later chapters, SSB is a critical component for understanding the weak interactions through the Higgs mechanism. It also plays a significant role in theories with global symmetries, superconducting materials, and the broader comprehension of quantum Yang-Mills theories.

Additionally, we examine \emph{anomalies}, a distinct way symmetries can be broken. Anomalies arise when quantum effects disrupt a symmetry present in classical field theory, and in some cases -- such as gauge symmetries -- they can make the theory inconsistent. This is another crucial factor in ensuring the consistency of the Standard Model, with notable consequences such as charge quantisation.

\vfill

\newpage

\section{Unitarity problems with massive spin-$1$ fields}\label{sec:lossunitarityssb}

Let us see where we stand and provide a brief motivation to study SSB to begin with.
So far, we have been able to identify the following states labelled by spin or helicity:
\begin{enumerate}
\item Helicity or spin  $0,\pm1/2$ particles which can be massless or massive.
\item Helicity $1$ massless particles implying gauge invariance with Abelian (QED) or  Non-Abelian (Yang-Mills) gauge symmetries.
\item Helicity  $2$ gravity.
\end{enumerate}
Historically, Pauli criticised the idea of Yang and Mills: Massless non-Abelian gauge fields should have been seen, like the photon, but had not at the time. Hence, he suggested to forget about Yang-Mills theory right from the beginning since he had found the same result but discarded it as unphysical and did not publish it.

What about massive particles? Massive scalar fields (spin $0$) and fermion fields (spin $1/2$) can be consistently described in QFT, but what about spin-$1$ fields? Recall that these fields possess $3$ polarisation states constrained by $p^{\mu}\epsilon_{\mu}=0$. Furthermore, they satisfy $p_{\mu}p^{\mu}=m^{2}$ and the polarisations can be normalised such that
\begin{equation}
\epsilon^{*}_{\mu}\epsilon^{\mu}=-1\, .
\end{equation}
The field and the equation of motion read
\begin{equation}
(\square+m^{2})A_{\mu}=0\kom A_{\mu}=\int\dfrac{\dif^{3}p}{(2\pi)^{3}}\, a_{i}(p)\epsilon^{i}_{\mu}\ee^{\I px}\, + \rm{h.c.}.
\end{equation}
But what could possibly be wrong with such fields?
The answer is simple and really just depends on what we mean with consistent.
As an EFT below some cutoff scale $\Lambda$, the theory is totally valid and predictive. However, there is a subtle issue when taking the energy to too large values as we now explain.

Let us consider momenta in the $z$-direction
\begin{equation}
p^{\mu}= (E,0,0,p_{z})\kom E^{2}-p_{z}^{2}=m^{2}\, .
\end{equation}
The transverse polarisations can be defined as
\begin{equation}
\epsilon_{1}^{\mu}=(0,1,0,0)\kom \epsilon_{2}^{\mu}=(0,0,1,0)
\end{equation}
and the longitudinal as
\begin{equation}
\epsilon_{L}^{\mu}=\left (\dfrac{p_{z}}{m},0,0,\dfrac{E}{m}\right )\, .
\end{equation}
At high energies $E\gg m$, we may write the latter as
\begin{equation}
\epsilon_{L}^{\mu}\sim \dfrac{E}{m}(1,0,0,1)\, .
\end{equation}
As we have seen before, scattering amplitudes $\cM$ are proportional to the polarisation vectors and hence this implies for an amplitude with two external massive particles of spin $1$
\begin{equation}
g^{2}\epsilon_{L}^{0}\epsilon_{L}^{3}\sim g^{2}\dfrac{E^{2}}{m^{2}}\, .
\end{equation}
What this actually implies is that probabilities (obtained from cross-sections) blow-up at $E\gg m$.

\begin{Boxequ}

\vspace*{0.1cm}

Perturbative unitarity is broken and therefore the theory of massive spin-$1$ fails at $E\gg m$. 

\end{Boxequ}

For instance, having $m\sim 100$GeV with $g\sim 0.1$ results in a scale $E\sim 1$TeV at which the theory fails to make sense. This is a signal that the theory needs to be replaced by a different theory which often involves new degrees of freedom. We then typically speak of a \emph{UV completion}\index{UV completion}. 

To summarise, there are two problems associated with the above:
\begin{enumerate}
\item For massless particles: No massless Yang-Mills field observed.
\item For massive particles: Theories of massive spin-$1$ \emph{fail perturbative unitarity}.
\end{enumerate}
We will see that the the solution to both problems is to consider \emph{spontaneously broken symmetries}. We will next describe this phenomenon in detail.

\section{Spontaneous breaking of a discrete symmetry}
\label{sec:ssb_disc}\index{SSB!Discrete symmetries}

\begin{figure}[t]
\centering
\includegraphics[width = \linewidth]{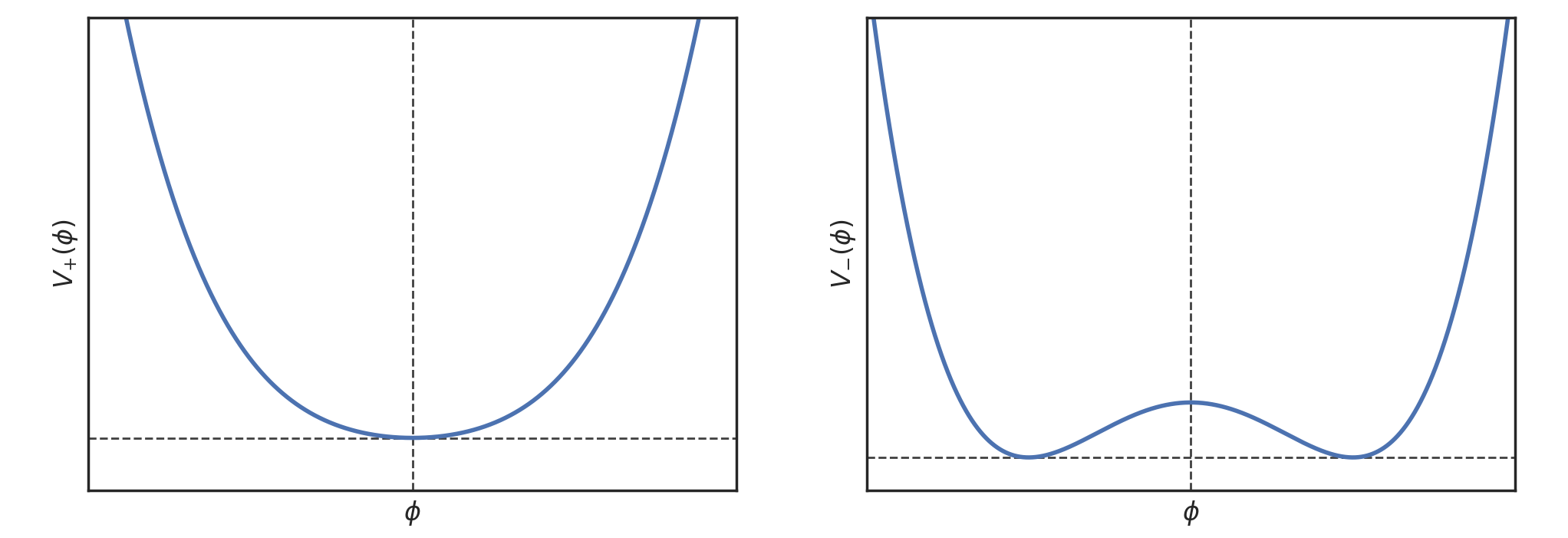}
\caption{\emph{Left:} Scalar potential $V_{+}$ with symmetric minimum. \emph{Right:} Double well potential $V_{-}$ with two degenerate minima exhibiting spontaneous symmetry breaking.
}\label{fig:ssb_doublewell}
\end{figure}

Let us consider the simplest system of a real scalar field $\phi(x)$ with a discrete $\bZ_{2}$ symmetry $\phi\raw -\phi$. The most general renormalisable Lagrangian is of the form
\begin{equation}
\L[\phi] = \frac12 \partial_\mu \phi\,\partial^\mu \phi \;-\; V_{\pm}(\phi)
\end{equation}
where the scalar potential can take one of two forms
\begin{equation}
V_{\pm}(\phi)=\pm \dfrac{1}{2}m^{2}\phi^{2}+\dfrac{\lambda}{4}\phi^{4}+\kappa_{\pm}\, .
\end{equation}
We need $\lambda>0$ for stability, i.e., the scalar potential should be bounded from below. The constant $\kappa_{\pm}$ is chosen such that the potential vanishes at the minimum which in the absence of gravity is not physically important because the relevant quantity is the difference of energies. The main feature of these two potentials is the quadratic piece or mass term for which the sign is not determined. We consider now each case separately since they have different physical implications.
\begin{enumerate}
\item $V_{+}$ (with $m^2 > 0$): The scalar potential $V_{+}(\phi)$ has a classical minimum at $\phi_{0} = 0$, cf. the left panel of Fig.~\ref{fig:ssb_doublewell}. Clearly, the minimum is invariant under $\phi\raw-\phi$ so that the symmetry is manifest. In a quantum theory, we consider the vacuum expectation value (VEV) 
\begin{equation}
\langle\phi\rangle=\braket{0|\phi|0}=\int\D\phi\;\phi\;\exp\left ({\frac{\I}{\hbar}\int\cL\dif^{4}x}\right )
\end{equation}
normalised with respect to $\braket{0|0}=\int\D\phi\ee^{\frac{\I}{\hbar}\int\cL\dif^{4}x} $. In the limit $\hbar \raw 0$, the path integral is dominated by the stationary value of the action and hence
\begin{equation}
\langle\phi\rangle=\phi_0= 0\, .
\end{equation}
Perturbations around the minimum can be written as $\phi=\phi_{0}+\sigma(x)$, $\sigma\ll 1$, so that
\begin{equation}
\cL[\phi]\raw \cL_{\text{pert}}[\sigma] = \dfrac{1}{2}\p_{\mu}\sigma\p^{\mu}\sigma-\dfrac{1}{2}m^{2}\sigma^{2}+\dfrac{\lambda}{4}\sigma^{4}\, .
\end{equation}
In the following, we denote the Lagrangian for the perturbations around the true vacuum as $\cL_{\text{pert}}$.
This Lagrangian describes nothing but a heavy particle of mass $m^{2}>0$ with $p_{\mu}p^{\mu}=m^{2}$ and
\begin{equation}
m^{2}=\dfrac{\p^{2}V}{\p\phi^{2}}\biggl |_{\phi=\phi_{0}}\, .
\end{equation}
So the parameter $m^2$ in the potential is actually the \emph{physical mass-squared of the corresponding particle}.

\item $V_{-}$ (with $m^{2}>0$): in this case the potential can be written as:
\begin{equation}
V_{-}(\phi)=\dfrac{\lambda}{4}\left (\phi^{2}-v^{2}\right )^{2}
\end{equation}
in terms of
\begin{equation}
 v=\sqrt{\dfrac{m^{2}}{\lambda}}\, .
\end{equation}
There are two degenerate vacua $|0_\pm \rangle $ corresponding to  $\phi_{0}=\pm v$ as can be seen from the right panel of Fig.~\ref{fig:ssb_doublewell}. Let us consider perturbations around any one of these minima by setting
\begin{equation}
\phi=\pm v+h(x)
\end{equation}
to write the Lagrangian for the fluctuation $h(x)$
\begin{equation}
\cL[\phi]\raw \cL_{\text{pert}}[h] = \dfrac{1}{2}\p_{\mu}h\p^{\mu}h-V_{\text{pert}}(h)
\end{equation}
where the scalar potential reads
\begin{align}\label{eq:PotHiggsZ2} 
V_{-}(\phi)\raw V_{\text{pert}}(h)=\lambda v^{2}\, h^{2}\pm\lambda v\, h^{3}+\dfrac{\lambda}{4}h^{4}\, .
\end{align}
The first term implies that $h$ describes a massive particle of mass
\begin{equation}
m_h^2=2\lambda v^{2}=2m^{2}>0\, .
\end{equation}
Equivalently, this can be obtained from the original potential $V_{-}$ by taking the second derivative and evaluating at the correct minimum, that is,
\begin{equation}
m_h^{2}=\dfrac{\p^{2}V_{-}(\phi)}{\p\phi^{2}}\biggl |_{\phi^{2}=v^{2}}=2\lambda v^{2}\, .
\end{equation}
Note that the physical mass-squared $m_h^{2}$ differs from the original $m^2$ parameter in the potential. This illustrates the fact that the mass can be read directly from the potential only if the vacuum state corresponds to $\phi=0$ as in the case for $V_+$ above.
Note also that if we had expanded around the symmetric point $\phi=0$, the particle spectrum would consist of a particle of negative mass squared, $-m^{2}$, which is called \emph{tachyon}\index{Tachyon} signalling the wrong expansion. That is, expanding around a maximum instead of a minimum as becomes evident from the right panel in Fig.~\ref{fig:ssb_doublewell}. As we had mentioned in chapter~\ref{chap:stsym} in the context of representations of the Poincar\'e group, tachyon states ($p^\mu p_\mu <0$) are allowed by special relativity, but the apparent presence of a tachyon only indicates an instability of the corresponding expansion point (the maximum) towards the true vacuum of the theory which might be degenerate as in the case above.
 
Let us now comment on the apparent ``breaking'' of the symmetry. The cubic term in \eqref{eq:PotHiggsZ2} seems to break the original $\bZ_{2}$ symmetry which is why it is \emph{hidden}.
Importantly, however, the symmetry is not actually broken, but is still present due to the symmetry under
\begin{equation}
h\pm v\raw -(h\pm v)\quad\Rightarrow\quad h\raw -h\mp 2v\, .
\end{equation}
However once we expand around one of the two vacua, the symmetry is not manifest in the expansion.
In this sense the word hidden is more appropriate.

Quantum mechanically we could have considered the vacuum state to be a superposition of the two vacua  $|0_+\rangle \pm |0_-\rangle$ but for large systems (where the  infinite volume limit can be considered) locality or cluster decomposition implies that any two Hermitian  operators $\mathcal{O}_1(x,t), \mathcal{O}_2(x,t)$ representing physical observables commute at fixed time and large separations and matrix elements factorise $ \langle 0| \mathcal{O}_1(x,t) \mathcal{O}_2(0,t) |0\rangle= \langle  0| \mathcal{O}_1(x,t)|0\rangle \langle  0| \mathcal{O}_2(0,t) |0\rangle$. This condition is satisfied if the system is at one of the two vacuum states, but not at a superposition (see \cite{Weinberg:1996kr} for a detailed discussion of this point).

\end{enumerate}

\subsection{Topological defects}\index{Topological defects}

In the context of spontaneously broken discrete symmetries,
an interesting phenomenon takes place. In physical 3-dimensional space we will find regions in which the system is in vacuum $|0_+\rangle $ and other regions in which the system is in the second vacuum $|0_-\rangle $. Going from a region for which $\langle \phi \rangle =+v$ to the region for which $\langle \phi\rangle =-v$ we need to pass through $\langle \phi\rangle =0$ where the potential is actually non-vanishing. The boundary that separates the two regions is a 2-dimensional wall called a {\emph{domain wall}}\index{Topological defects!Domain wall}\index{Domain wall}. This is a topological defect of the system reflecting the fact that the space of vacua is disconnected, consisting of two points. 
 
\begin{figure}[t!]
\centering
\includegraphics[width=0.55\textwidth]{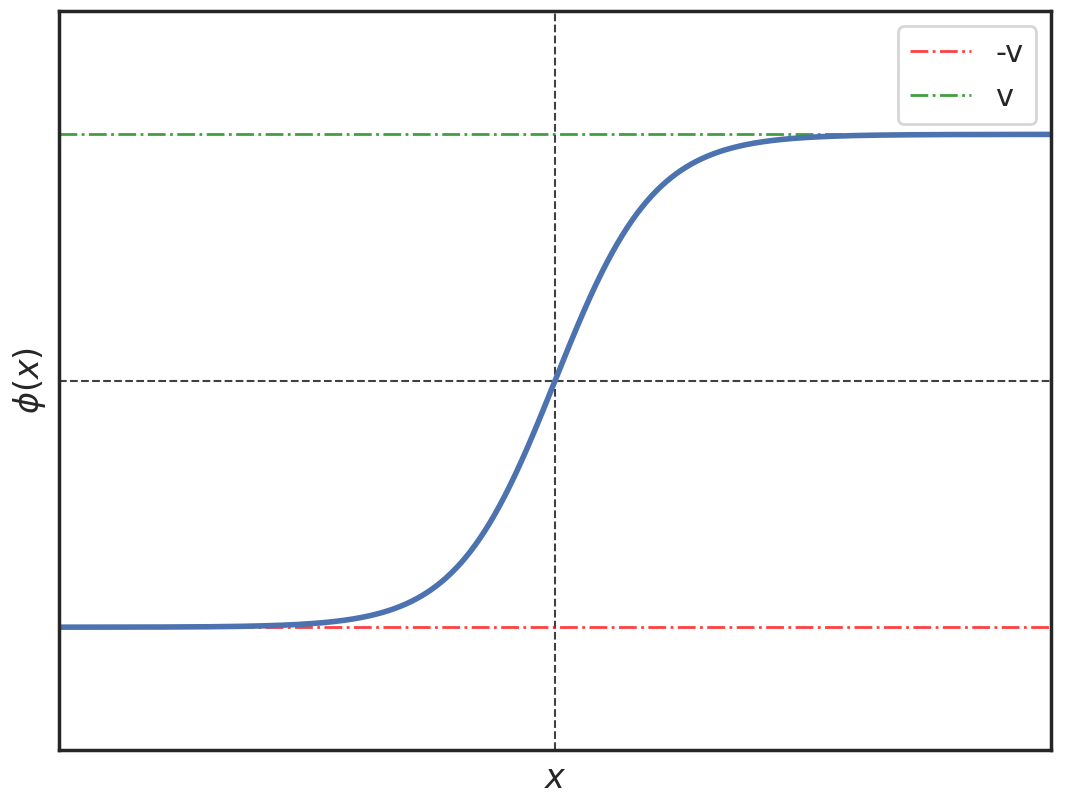}
\caption{Domain wall solution interpolating between the two minima $\phi(x)=\pm v$ of the scalar potential.
}\label{fig:domainwall}
\end{figure}

\begin{figure}[t!]
\centering
\includegraphics[width=0.55\textwidth]{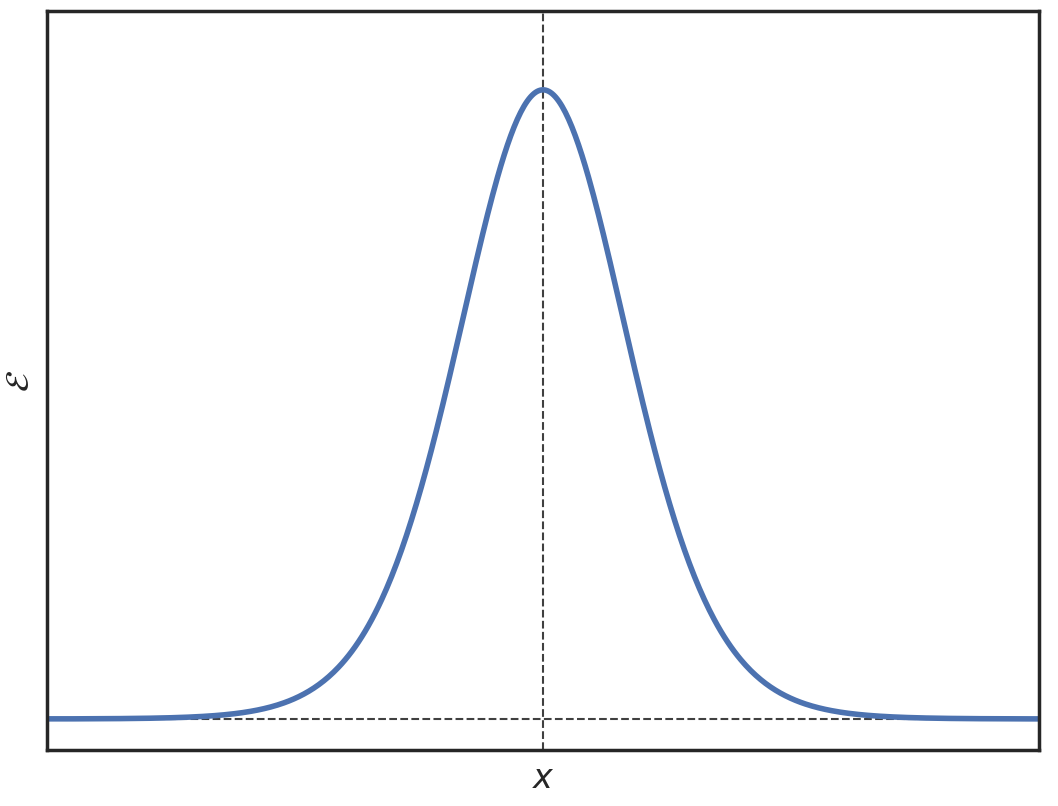}
\caption{Energy density profile for the domain wall solution illustrating that the domain wall solution actually corresponds to a physical object carrying energy (and tension) which is localised in the $x$ direction and arbitrary in the $yz$ directions which is the standard picture of a wall (or 2-brane).
}\label{fig:domainwallenergy}
\end{figure}

Static domain wall solutions can be easily found for the system by solving the field equations 
\begin{equation}
\square \phi +V'(\phi)=0
\end{equation}
for which $\phi(x)$ (independent of $y,z$)  satisfies $\phi''= \lambda \phi(\phi^2-v^2)$ that, even though it is a non-linear equation, it has a closed solution with a profile of the form 
\begin{equation}
\phi(x)=v \tanh\rho x, \qquad \rho^2\equiv {\frac{\lambda v^2}{2}}=\frac{m^2}{2}
\end{equation}
which interpolates between the two vacua in the limits $x\raw \pm \infty$ (see Fig.~\ref{fig:domainwall}). The domain wall would extend through the $y,z$ directions.  
Domain walls are physical entities that carry energy and could play an important role in the dynamics of the system. We can explicitly compute the energy density in this case from the $\phi(x)$ profile above. The energy density $\mathcal E$ of the wall  can be computed by evaluating   it at $\phi(x)=v \tanh \rho x$ for which we get
\begin{equation}
{\mathcal E} ~=~ \frac12 \partial_\mu \phi\,\partial^\mu \phi \;+\; V_{\pm}(\phi) = \frac{\lambda v^4}{2\cosh^4 \rho x}\,. 
\end{equation}
It can be seen in Fig.~\ref{fig:domainwallenergy} that this profile is highly localised at $x\sim 0$ and $y,z$ arbitrary. The value of $\rho$ determines the thickness of the wall (or membrane or 2-brane using current terminology).

\subsection{Symmetry restoration at high temperature}

\begin{figure}[t!]
\centering
\includegraphics[width=0.7\textwidth]{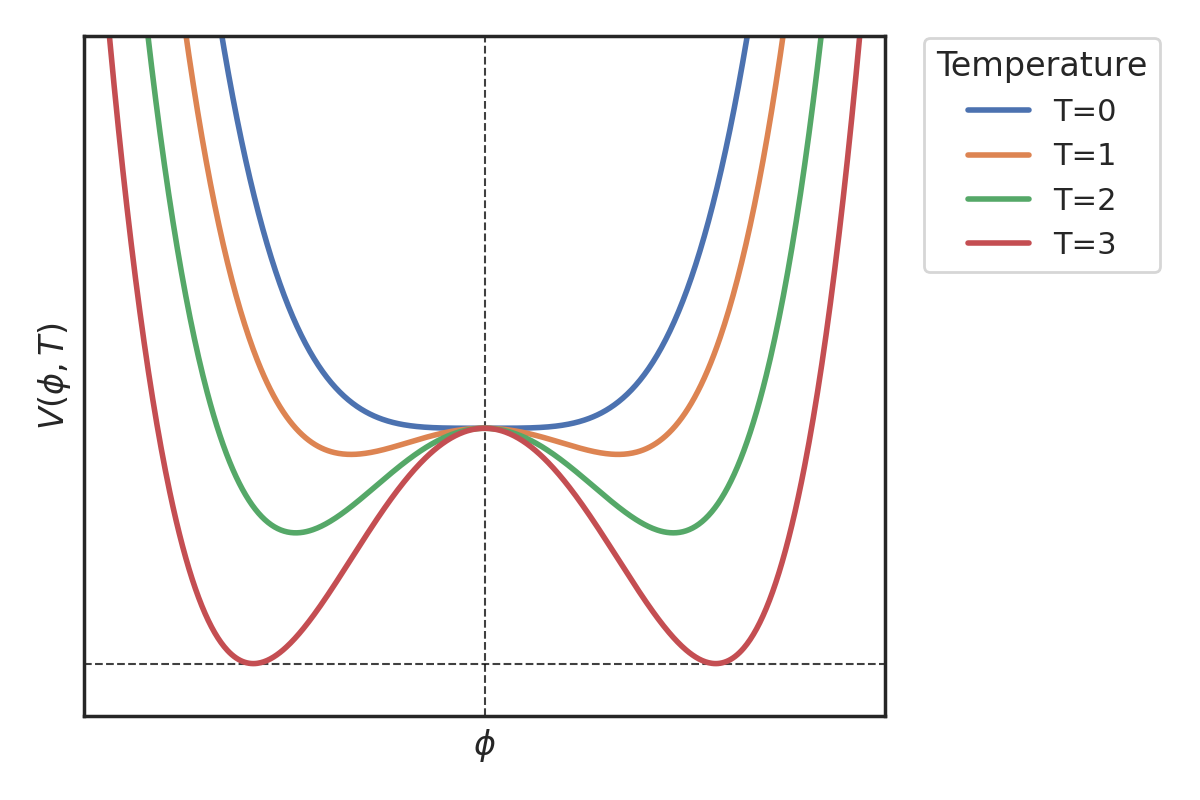}
\caption{Temperature dependent potential. At hight temperature the unbroken phase. When temperature gets reduced (universe expanding gets colder) the broken symmetry phase is realised.}\label{fig:temperature}
\end{figure}

Symmetry breaking can be understood as a phase transition. This model captures the physics of several systems such as ferromagnetic materials. In this case the parameter $m^2$ of the scalar potential corresponds to a temperature difference 
\begin{equation}
m^{2}\propto (T-T_{c})\, .
\end{equation}
That is, for high temperatures $T>T_c$ the potential is $V_+$ with one single minimum and unbroken symmetry. In this case the expectation value of the ferromagnetic material vanishes meaning that all different directions of magnetisation are realised and the average is zero. While the system cools down, we enter the region for which $T<T_c$ where a phase transition occurs and the potential is now of the form $V_-$ exhibiting SSB. The two possible expectation values indicate two opposite directions of the magnets which are polarised pointing at only one direction in one vacuum and the opposite direction in the other vacuum (the average value or expectation value of the field is non-zero now).

In cosmology, the early universe has a high temperature.
If there was a scalar field with a potential of the above type, the quadratic term in the scalar potential would read $-m^2\phi^2 + T^2 \phi^2$. At high temperatures ($T \gg m$) the effective coefficient of $\phi^2$ is positive and the system is in the unbroken phase with a minimum at $\phi=0$. As the universe cools down while expanding, it reaches a critical temperature $T_c=m$ and, for temperatures smaller than $T_c$, the coefficient of $\phi^2$ becomes negative such that the universe enters the symmetry breaking phase, see Fig.~\ref{fig:temperature}.
In this regime, the phase transition may occur towards a broken phase. This is usually referred to as \emph{symmetry restoration}\index{Symmetry restoration} in the early universe. In this case, domain walls, if present, may have a significant impact on the evolution of the universe by contributing a large amount to the energy density of the universe which may over close it.

\section{Spontaneous symmetry breaking (SSB) of continuous global symmetries}\label{sec:ssb_cont}\index{SSB!Continuous (global) symmetries}

\begin{figure}[t!]
\centering
\includegraphics[width=0.99\textwidth]{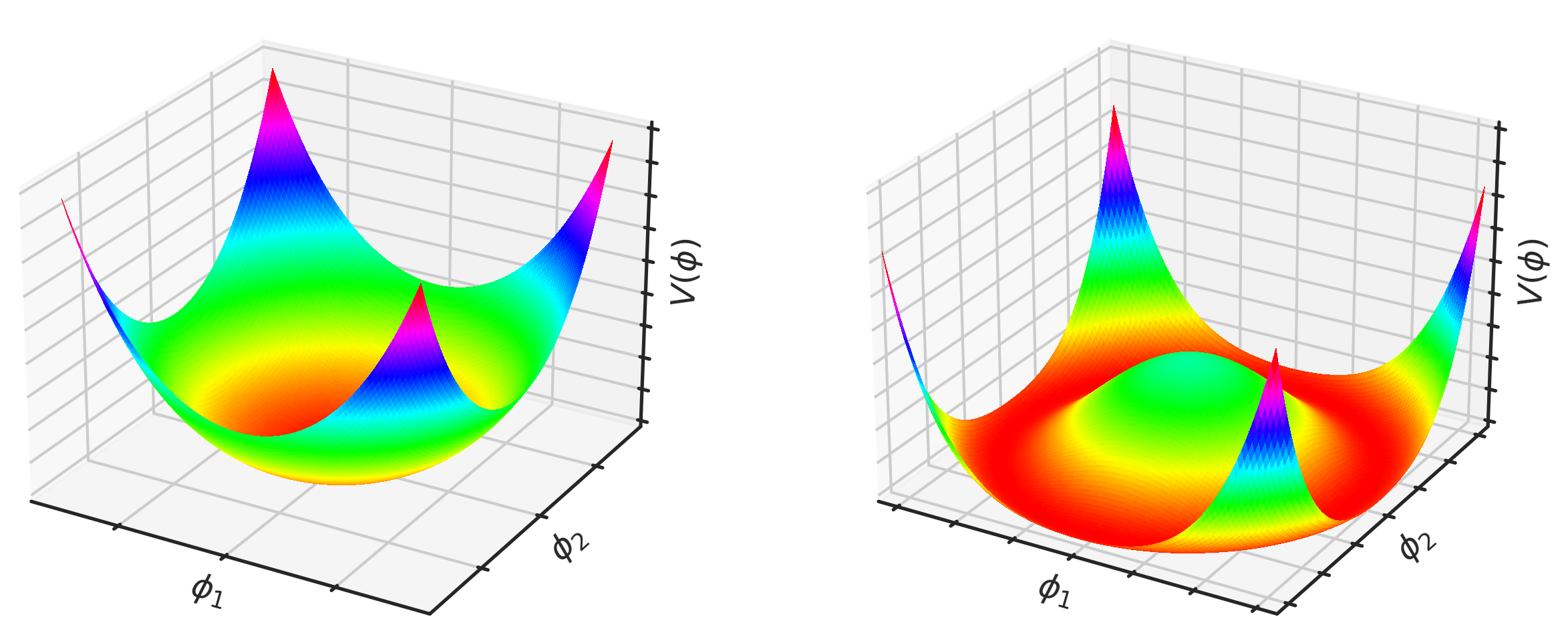}
\caption{Potential in the symmetric or unbroken phase with $m^2>0$ (left) and in the spontaneously broken phase with $m^2<0$ (right).}\label{fig:ssb_V}
\end{figure}

Let us begin our generalisation to the case of spontaneous breaking
of \emph{continuous global symmetries} with a simple example, namely that of an
$N$-component real scalar field $\phi = (\phi_1, \ldots, \phi_N)^T$.
The Lagrangian is given by
\begin{equation}
\L[\phi] = \frac12 \partial_\mu \phi\,\cdot\,\partial^\mu \phi - V_{\pm}(\phi) 
\end{equation}
with
\begin{equation}
V_{\pm}(\phi) =\pm \frac12 m^2\phi^2 + \frac{\lambda}{4}\phi^4 \kom \lambda>0\kom \phi^2 = \phi\cdot\phi \kom\phi^4=(\phi^2)^2 \, .
\end{equation}
The Lagrangian is invariant under global $\mathrm{O}(N)$ transformations of the field which naturally leave the inner product on $\mathbb{R}^{N}$ invariant.

We are primarily interested in the case $V_{-}(\phi)$ because it leads to SSB.
We can replace the potential (up to an irrelevant constant term) by
\begin{equation}
V_-(\phi) ~=~ \frac{\lambda}{4}\left(\phi^2 - v^2\right)^2 \kom v^2 = \frac{m^2}{\lambda} > 0 \,.
\end{equation}
This potential is often called the \textit{Mexican hat} potential and has the shape of a wine bottle as shown on the right of Fig.~\ref{fig:ssb_V}. The vacua are defined by the equation
\begin{equation}\label{eq:VacCondONModel} 
\langle\phi\rangle^{2}=\phi_{0}^{2}=v^{2}
\end{equation}
resembling the defining equation of an $(N-1)$-sphere. Without loss of generality let us pick one possible solution to \eqref{eq:VacCondONModel}
\begin{equation}
\langle\phi\rangle=\phi_{0}=\left (\begin{array}{c}
0 \\ [-0.25cm]
0 \\ [-0.15cm]
\vdots \\ [-0.15cm]
0 \\ [-0.25cm]
v
\end{array} \right )\, .
\end{equation}
This VEV breaks the original symmetry $\mathrm{O}(N)$ to the subgroup $\mathrm{O}(N-1)$ because $\phi_{0}$ is invariant only under rotations in $N-1$ dimensions.

When performing fluctuations around the minimum with fluctuating fields $\pi_i(x)$ and $\sigma(x)$, we write
\begin{equation}
\phi(x)=\left (\begin{array}{c}
\pi_{1}(x) \\ [-0.15cm]
\pi_{2}(x) \\ [-0.15cm]
\vdots \\ [-0.2cm]
\pi_{N-1}(x) \\ [-0.15cm]
v+\sigma(x)\, .
\end{array} \right )\, .
\end{equation}
The Lagrangian for those fluctuations becomes
\begin{equation}
\cL_{\text{pert}}[\pi_{i},\sigma]=\dfrac{1}{2}\p_{\mu}\pi_{i}\p^{\mu}\pi_{i}+\dfrac{1}{2}\p_{\mu}\sigma\p^{\mu}\sigma-V_{\text{pert}}(\pi_{i},\sigma)
\end{equation}
with scalar potential
\begin{equation}
V_{\text{pert}}(\pi_{i},\sigma)=\dfrac{1}{2}m_{\sigma}^{2}\sigma^{2}+\lambda v(\sigma^{2}+\pi_{i}^{2})\sigma+\dfrac{\lambda}{4}\left (\sigma^{2}+\pi_{i}^{2}\right )^{2}\, .
\end{equation}
The (diagonalised) mass matrix schematically looks like
\begin{equation}
\dfrac{\p^{2}V_{-}(\phi)}{\p\phi_{i}\p\phi_{j}}\biggl |_{\phi = \langle\phi\rangle}=\left (\begin{array}{ccc}
0&  &  \\ 
 & \ddots &  \\ 
 &  & 2\lambda v^2
\end{array} \right )
\end{equation}
with the eigenvalues
\begin{equation}
m_{\sigma}=\sqrt{2\lambda v^{2}}\kom  m_{\pi_i}=0\, .
\end{equation}
The only non-zero eigenvalue
corresponds to the mass of $\sigma$. We obtain $N-1$ massless fields $\pi_{i}$, $i=1,\ldots ,N-1$, which are called \emph{Goldstone bosons}\index{SSB!Goldstone bosons}\index{Goldstone bosons} for reasons to become clear below. It turns out that the theory after SSB possesses a manifest $\mathrm{O}(N-1)$ symmetry. This makes sense intuitively from the wine-bottle shape of the potential (Fig.~\ref{fig:ssb_V}): radial excitations come with a large energy penalty, whereas excitations in the field which locally seek to transform the field to another of the equivalent vacua can be made to have arbitrarily small energy difference from the vacuum.

\newpage

\section{Goldstone's theorem}
\label{sec:ssb_goldstone}\index{Goldstone's theorem}

Next, we introduce Goldstone's theorem which makes general statements about the number of massive and massless fields (or occasionally also synonymously referred to as modes/excitations/fluctuations) after SSB for general continuous groups $G$. We distinguish the classical version formulated simply in terms of a scalar potential and the quantum version phrased in terms of quantum mechanical Hilbert spaces and path integrals.

\subsection{The classical version}\index{Goldstone's theorem!Classical}

In general, if a Lagrangian is invariant under $G$, then a non-zero \emph{vacuum expectation value} (VEV)\index{Vacuum expectation value (VEV)} $\phi_{0}=\langle \phi\rangle$ for a field $\phi$ breaks $G\raw H\subset G$.\footnote{The unbroken subgroup $H$ may, but does not need to be a proper subgroup of $G$.} The associated \emph{vacuum manifold}\index{Vacuum manifold} $\cM_{0}$ to $\phi_{0}$ is defined as
\begin{equation}
\cM_{0}=\lbrace \phi_{0}:\, V(\phi_{0})=V_{\text{min}}\rbrace\, .
\end{equation}
In the former case of $\mathrm{O}(N)$, $\cM_{0}$ is determined by condition \eqref{eq:VacCondONModel} implying that $\cM_{0}\cong S^{N-1}$.

\subsubsection*{The invariant group $H$}

The invariant or \emph{stability group}\index{Stability group} $H_{\phi_{0}}$ is the subgroup of $G$ that leaves the vacuum $\phi_0$ invariant, that is, 
\begin{equation}
H_{\phi_{0}}=\lbrace h\in G:\, h\phi_{0}=\phi_{0}\rbrace\, .
\end{equation}
The different vacua themselves are linked by transformations in $G$, i.e., for $\phi_{0},\phi_{0}^{\prime}\in \cM_{0}$ there exists $g\in G$ such that
\begin{equation}
\phi_{0}^{\prime}=g\phi_{0}\, .
\end{equation}

Let us prove that the stability groups $H_{\phi_{0}}$ for different $\phi_{0}\in \cM_{0}$ are isomorphic. As before, let $\phi_{0}^{\prime}\in\cM_{0}$ and $g\in G$ be such that $\phi_{0}^{\prime}=g\phi_{0}$ and define
\begin{equation}
H_{\phi_{0}^{\prime}}=\lbrace h\in G:\, h\phi_{0}^{\prime}=\phi_{0}^{\prime}\rbrace\, .
\end{equation}
Then we can write for $h\in H_{\phi_{0}^{\prime}}$
\begin{equation}
h\phi_{0}^{\prime}=\phi_{0}^{\prime}\quad\Rightarrow\quad h g\phi_{0}=g\phi_{0}\quad\Rightarrow\quad g^{-1}h g\phi_{0}=\phi_{0}
\end{equation}
implying that 
\begin{equation}
g^{-1}hg\in H_{\phi_{0}}\quad\Rightarrow\quad H_{\phi_{0}}=g^{-1}H_{\phi_{0}^{\prime}}g\, .
\end{equation}
Hence, the stability groups $H_{\phi_{0}}$ are indeed isomorphic.
Therefore, in the following, we will denote the stability group of the vacua collectively as $H$. 

\subsubsection*{The coset $G/H$ and Goldstone modes}

The elements $g\in G$ mapping one vacuum to another belong to the coset $G/H$ and fall into equivalence classes $g_{1}\sim g_{2}$ if there exists $h\in H$ such that $g_{1}=g_{2}h$. That is, if there exists two elements $g_{1},g_{2}\in G$ such that
\begin{equation}
\phi_{0}^{\prime}=g_{1}\phi_{0}=g_{2}\phi_{0}\, ,
\end{equation}
then 
\begin{equation}
g_{2}^{-1}g_{1}\in H\, .
\end{equation}
We can associate to $\phi_{0}^{\prime}\in\cM_{0}$ an equivalence class
\begin{equation}
\cM_{0}\cong G/H\, .
\end{equation}

To derive Goldstone's theorem, let us consider infinitesimal transformations around a field $\phi$ in a representation $R$ of a group $G$ of dimension $\dim(R)=N$ so that
\begin{equation}
g\phi=\phi+\delta\phi\kom (g\phi)_{r}=\phi_{r}+\delta\phi_{r}
\end{equation}
with
\begin{equation}
\delta\phi_{s}=\I\alpha^{a}T_{sr}^{a}\phi_{r}\kom a=1,\ldots,\dim(G)\kom r,s=1,\ldots,N\, .
\end{equation}
For a scalar potential invariant under $G$, we have
\begin{equation}
V(g\phi)=V(\phi+\delta\phi)=V(\phi)\, .
\end{equation}
Expanding around $\phi$ amounts to
\begin{equation}\label{eq:ExpGTPot} 
V(\phi+\delta\phi)-V(\phi)=\I\alpha^{a}(T^{a}\phi)_{r}\, \dfrac{\p V}{\p \phi_{r}}=0
\end{equation}
Differentiating \eqref{eq:ExpGTPot} once again leads to
\begin{equation}
\I\alpha^{a}\left[ \left(T^a\right)_{rs}\, \dfrac{\p V}{\p \phi_{r}}+\, (T^{a}\phi)_{r}\,  \dfrac{\p^{2}V}{\p \phi_{r}\p\phi_{s}}\right]=0
\end{equation}
and evaluating at $\phi=\phi_{0}$ gives rise to
\begin{equation}\label{eq:ZeroEigenvalueMassMat} 
(T^{a}\phi_{0})_{r}\,  M_{rs}^{2}=0
\end{equation}
where we defined the mass matrix $M_{rs}^{2}$ as
\begin{equation}
M_{rs}^{2}=\dfrac{\p^{2}V}{\p \phi_{r}\p\phi_{s}}\biggl |_{\phi=\phi_{0}}\, .
\end{equation}
We then distinguish the following cases:
\begin{itemize}
\item if the symmetry is unbroken and the vacuum unique in the sense that $g\phi_{0}=\phi_{0}$ for all $g\in G$, then $\delta\phi=0$ and thus
\begin{equation}
(T^{a}\phi_{0})_{r}=0
\end{equation}
for all $a=1,\ldots,\dim(G)$ and $r=1,\ldots,N$.
\item if there exists $g\in G$ and $a\in\lbrace 1,\ldots,\dim(G)\rbrace$ with $T^{a}\phi_{0}\neq 0$, then by \eqref{eq:ZeroEigenvalueMassMat} $T^{a}\phi_{0}$ is an eigenvector of the mass matrix $M_{rs}^{2}$ with zero eigenvalue.
\end{itemize}

The question is how many of such massless states exist? Let us assume that $G$ is compact and semi-simple and let us split the generators $T^{a}$ as
\begin{equation}
T^{a}=(\tilde{T}^{i},R^{\alpha})
\end{equation}
with $\tilde{T}^{i}\in H$ so $i=1, \ldots, \dim H$ with
\begin{equation}
\tilde{T}^{i}\phi_{0}=0
\end{equation}
and the orthogonality condition
\begin{equation}
\tr(\tilde{T}^{i}R^{\alpha})=0
\end{equation}
with $\alpha = 1, \ldots, \dim(G/H)$. Each vector $R^{\alpha}\phi_{0}$ is a unique eigenvector of eigenvalue zero to $M^{2}_{rs}$ and therefore there are
\begin{equation}
\dim(G/H)=\dim(G)-\dim(H)=\dim(\cM_{0})
\end{equation}
massless modes called \emph{Goldstone bosons}\index{Goldstone bosons}. Since $M^{2}_{rs}$ is a $N\times N$-matrix, then there are at most $N-\dim(G/H)$ massive modes. This is \emph{Goldstone's theorem}:

\begin{equ}[Goldstone's theorem]\index{Goldstone's theorem}
Every spontaneously broken continuous global symmetry of a QFT gives rise to massless states. For an internal symmetry with breaking pattern $G\raw H$ it leads to 
\begin{equation}
\dim(G/H)=\dim(G)-\dim(H)=\dim(\cM_{0})\, \nonumber
\end{equation}
massless particles: the {\it Goldstone bosons}.
\end{equ}

Coming back to the example of the $\mathrm{O}(N)$ model, we found $N-1$ massless fields and that the unbroken symmetry group is $\mathrm{O}(N-1)$. We now understand that
\begin{equation}
\dim(\mathrm{O}(N)/\mathrm{O}(N-1))=\dfrac{N(N-1)}{2}-\dfrac{(N-1)(N-2)}{2}=N-1
\end{equation}
is the number of Goldstone bosons and that there is exactly one massive field $\sigma$.

For spacetime symmetries, their breaking also leads to massless states but the counting is not given by $ \dim(G)-\dim(H) $. Examples include the phonons in condensed matter physics and the domain walls we have just discussed. In this case translation invariance is clearly broken and the free motion of the wall corresponds to a Goldstone mode, but the number of these modes does not match the number of broken symmetries (that include boosts and translations). See \cite{Burgess:2020tbq} for a detailed discussion of this case.

\subsection{Quantum aspects of SSB}\index{Goldstone's theorem!Quantum}

Since this is such an important theorem with far reaching implications, we will now provide a different quantum perspective of Goldstone's theorem and SSB in general. 

\subsubsection*{Order parameter for SSB}\index{SSB!Order parameter}

The Noether charge associated to some symmetry is defined as
\begin{equation}
Q^{a}=\int\dif^{3}x\, J_{0}^{a}\, .
\end{equation}
As discussed in the previous section, 
the charges themselves act as quantum operators such that
\begin{equation}
\left [\phi_{i},Q^{a}\right ]=\I T_{ij}^{a}\phi_{j}\, .
\end{equation}
The order parameter of SSB is given by the VEV of the field operator $\phi$
\begin{equation}
\braket{0|\phi|0}=\langle\phi\rangle\begin{cases}
=0 &\text{unbroken}\, ,\\
\neq 0&\text{broken}\, .
\end{cases}
\end{equation}
If the symmetry is spontaneously broken, i.e., if $\langle\phi\rangle\neq 0$, then it follows that
\begin{equation}
\braket{0|[\phi,Q^{a}]|0}=\braket{0|\phi Q^{a}-Q^{a}\phi|0}\neq 0\, .
\end{equation}
Hence, we deduce that
\begin{equation}
Q^{a}\ket{0}\begin{cases}
=0 &\text{unbroken}\, ,\\
\neq 0&\text{broken}\, .
\end{cases}
\end{equation}

It is then usually stated that the condition for an unbroken symmetry is that the corresponding generator {\bf annihilates the vacuum}. Equivalently, this can be formulated in terms of generators of the underlying symmetry group. That is, let
\begin{equation}
U=\ee^{\I\alpha_{a}T_{a}}
\end{equation}
such that for the vacuum state $\phi_{0}$
\begin{equation}
U_{ij}(\phi_{0})_{j}=\left (\delta_{ij}+\I\alpha^{a}T^{a}_{ij}\right )(\phi_{0})_{j}\equiv (\phi_{0})_{i}\, .
\end{equation}
This implies that
\begin{equation}
T_{ij}^{a}(\phi_{0})_{j}=0\, .
\end{equation}

\subsubsection*{Degenerate energies and Goldstone modes}\index{SSB!Degenerate energies}

In quantum mechanics, if 
\begin{equation}
\ket{\psi}=Q\ket{\chi}
\end{equation}
and $Q$ is conserved,
\begin{equation}
[Q,H]=0\, ,
\end{equation}
then one can show that
\begin{equation}
E_{\psi}\ket{\psi}=H\ket{\psi}=HQ\ket{\chi}=QH\ket{\chi}=E_{\chi}Q\ket{\chi}=E_{\chi}\ket{\psi}\, .
\end{equation}
Thus, the energy levels $E_{\psi}$ and $E_{\chi}$ are the same, 
\begin{equation}
E_{\psi}=E_{\chi}\, ,
\end{equation}
and are degenerate energies.

But in field theories where SSB occurs, the particles need not be degenerate necessarily. Consider for example two fields related by a symmetry transformation, that is,
\begin{equation}
\I\phi_{1}=[\phi_{2},Q]
\end{equation}
for some fields $\phi_{1}$, $\phi_{2}$. Then, for the corresponding particle states, we obtain
\begin{equation}
\ket{1}=a_{1}^{\dagger}\ket{0}=\I[a_{2}^{\dagger},Q]\ket{0}=\I a_{2}^{\dagger}Q\ket{0}-\I Qa_{2}^{\dagger}\ket{0}=-\I Q\ket{2}+\I a_{2}^{\dagger}Q\ket{0}\, .
\end{equation}
So the two particle states corresponding to the two fields are directly related $\ket{1}\leftrightarrow Q\ket{2}$ only if $Q\ket{0}=0$. This is not true for SSB. In this case energy degeneracy of physical states will not hold.

However, for SSB, if $Q^{a}\ket{0}\neq 0$, we find a new result
\begin{equation}
H(Q^{a}\ket{0})=Q^{a}H\ket{0}=E_{0}(Q^{a}\ket{0})\, .
\end{equation}
Hence, $Q^{a}\ket{0}$ is degenerate with the vacuum $\ket{0}$ and both have energy $E_{0}$. 

Let us now define the momentum states
\begin{equation}
\ket{\pi^{a}(p)}=K\int\dif^{3}x\,\ee^{-\I\mathbf{p}\cdot\mathbf{x}}\, J_{0}^{a}(x)\ket{0}
\end{equation}
of energy $E(\mathbf{p})+E_{0}$ with $E^{2}(\mathbf{p})=p^{2}+m^{2}$. Since
\begin{equation}
\ket{\pi^{a}(0)}=KQ^{a}\ket{0}
\end{equation}
has energy $E_{0}$, then $E(\mathbf{p})\raw 0$ for $|\mathbf{p}|\raw 0$. This implies that the states $\ket{\pi^{a}}$ are massless and correspond to the Goldstone modes (one per broken symmetry). This can be seen as a {\bf quantum version of the Goldstone's theorem}.

\subsubsection*{Quantum effective action}\index{Quantum effective action}

Let us now consider the description in terms of path integrals and quantum effective actions. Recall from the path integral formulations of QFT that we can define the functional $W[J]$, usually called \emph{effective action}, as
\begin{equation}
\ee^{\I W[J]}=\int\D\, \phi\,\ee^{\I\int(\cL+J\phi)}
\end{equation}
which is the generating functional for all fully connected Green's functions. We can define
\begin{equation}
\dfrac{\delta W}{\delta J}=\dfrac{\int\D\phi\, \phi\, \ee^{\I\int(\cL+J\phi)}}{\int\D\phi\, \ee^{\I\int(\cL+J\phi)}}=\dfrac{\braket{0|\phi|0}}{\braket{0|0}}=\phi_{c}(x)\, .
\end{equation}
which is nothing but the $1$-point function of a field $\phi(x)$ in the presence of sources. Like in statistical field theory, we define the Legendre transformed functional
\begin{equation}
\Gamma[\phi_{c}]=W[J]-\int\dif^{4}x\, J(x)\phi_{c}(x)\kom \dfrac{\delta\Gamma}{\delta\phi_{c}(x)}=-J(x)
\end{equation}
which is the \emph{$1$PI effective action}\index{$1$PI effective action $\Gamma[\phi_{c}]$} generating $1$PI (one-particle irreducible) connected amputated Green's functions $\Gamma^{(n)}(x_{1},\ldots ,x_{n})$.\footnote{\emph{Warning:} this 1PI effective action should {\it not} be confused with the effective action that we mentioned in chapter \ref{chap:stsym} in the discussion of EFTs. The action appearing in EFTs is effective in the sense that it is the action at low-energies after integrating out all higher momenta, including heavier particles. It is usually called the \emph{Wilsonian effective action} to differentiate it from the 1PI effective action. The Wilsonian action at low energies $E<\mu,$ for a  fixed energy scale $\mu$, can be obtained from the 1PI action after integrating out all states much heavier than $\mu$. }
Expanding in momenta, we can write:
\begin{equation}
\Gamma[\phi_{c}]=\int \dif^{4}x \left [-V_{\text{eff}}(\phi_{c})+\dfrac{1}{2}(\p_{\mu}\phi_{c})^{2}\, Z(\phi_{c})+\ldots\right ]
\end{equation}
with the effective potential $V_{\text{eff}}(\phi_{c})$.
This potential is different from the (classical) potential $V(\phi)$: it knows about the full quantum effects in the theory.
The inverse propagator is
\begin{equation}
\Delta^{-1}=\dfrac{\delta^{2}\Gamma}{\delta\phi\delta\phi}
\end{equation}
which, at vanishing momentum, is the mass matrix
\begin{equation}
\Delta^{-1}\bigl |_{p=0}=\dfrac{\dif^{2}V_{\text{eff}}}{\dif \phi_{c}^{2}}\biggl |_{0}\, .
\end{equation}
This provides the standard interpretation as mass of a particle in terms of the location of the poles of the propagator. 

SSB occurs when the classical field $ \phi_{c}\neq 0$ in the absence of a current  $J=0$, i.e., 
\begin{equation}
\dfrac{\delta\Gamma}{\delta\phi_{c}}=0\kom \phi_{c}\neq 0
\end{equation}
so that at zero momentum
\begin{equation}
\dfrac{\dif V_{\text{eff}}}{\dif \phi_{c}}=0
\end{equation}
for $\phi_{c}\neq 0$. Goldstone modes correspond to
\begin{equation}
\dfrac{\delta^{2}V_{\text{eff}}}{\delta\phi_r\delta\phi_s}\left(T^a\phi\right)_s=0\, .
\end{equation}
This clearly promotes the original claim to the full quantum domain since the effective potential reduces to the classical potential at leading order.

\section{Spontaneous breaking of gauge symmetries}\label{sec:ssb_higgs}\index{SSB!Gauge symmetries}

We started this chapter pointing out two problems: massless Yang-Mills fields have not been observed and massive spin-1 fields fail perturbative unitarity. But rather than making progress in addressing these problems,
what we achieved so far is only adding yet another apparent problem: the non-observation of Goldstone modes. In summary, we now have to deal with three problems:
\begin{enumerate}
\item Yang-Mills fields are massless and have not been observed.
\item Goldstone bosons are massless and have not been observed.
\item Theories with massive spin $1$ fields are not valid at high energies.
\end{enumerate}
In this section, we will understand how all of these problems can be cured at once. The key point is the Higgs mechanism which is based on the simple idea of SSB in the presence of gauge theories.

\subsection{The Abelian Higgs model}\label{sec:AbelianHiggsmodel} 

The first model that we like to consider is the \emph{Abelian Higgs model}\index{Abelian Higgs model} for a complex scalar field $\phi$ coupled to a $\mathrm{U}(1)$ gauge field $A_{\mu}$ with Lagrangian
\begin{equation}
\cL[A_{\mu},\phi]=-\dfrac{1}{4}F_{\mu\nu}F^{\mu\nu}+\dfrac{1}{2}D_{\mu}\phi D^{\mu}\phi^{\dagger}-V(\phi^{*}\phi)
\end{equation}
where
\begin{equation}
D_{\mu}=\p_{\mu}+\I e A_{\mu}\kom F_{\mu\nu}=\p_{\mu}A_{\nu}-\p_{\nu}A_{\mu}\, .
\end{equation}
The Lagrangian is invariant under gauge transformations
\begin{equation}
\phi\raw\ee^{\I\alpha(x)}\phi\kom A_{\mu}\raw A_{\mu}-\dfrac{1}{e}\p_{\mu}\alpha\, .
\end{equation}

The scalar potential is assumed to be of the form
\begin{equation}
V(\phi^{*}\phi)=\dfrac{\lambda}{4}\left (|\phi|^{2}-v^{2}\right )^{2}\, .
\end{equation}
The minimum is clearly given by
\begin{equation}
|\phi_{0}|^{2}=v^{2}\, ,
\end{equation}
so we can choose $\phi_{0}$ to be real by setting
\begin{equation}
\langle\phi\rangle=\phi_{0}=v\, .
\end{equation}
Fluctuations around the minimum can be included by considering
\begin{equation}
\phi(x)=\ee^{\I\xi(x)}\left (\eta(x)+v\right )
\end{equation}
so that the kinetic term for $\phi$ becomes
\begin{align}\label{eq:ExpansionCovKinTermAbelianHiggsModel} 
D^{\mu}\phi D_{\mu}\phi^{*}&=\p^{\mu}\eta\p_{\mu}\eta+(\eta+v)^{2}\left (\p^{\mu}\xi+e A^{\mu}\right )^{2}
\end{align}
and similarly for the potential
\begin{align}
V(\eta)&=\dfrac{\lambda}{4}\left ((\eta+v)^{2}-v^{2}\right )^{2}=\lambda v^{2}\, \eta^{2}+\lambda v\, \eta^{3}+\dfrac{\lambda}{4}\eta^{4}\, .
\end{align}
At this point, we count a single massive field $\eta$ of mass
\begin{equation}
m_{\eta}=\sqrt{2\lambda v^{2}}
\end{equation}
as well as a massless field $\xi$ since there is no quadratic term in $\xi$ in the scalar potential. Actually the only way that $\xi$ appears in the Lagrangian is in the combination $ \p_{\mu}\xi +e A_{\mu }$ as can be seen from \eqref{eq:ExpansionCovKinTermAbelianHiggsModel}. Since we are working with a $\mathrm{U}(1)$ gauge theory with unfixed gauge, we can use this freedom to modify the above result accordingly. That is, we fix the gauge (known as \emph{unitary gauge}\index{Unitary gauge}) by redefining
\begin{equation}
A_{\mu}\raw A_{\mu}-\dfrac{1}{e}\p_{\mu}\xi
\end{equation}
which gives rise to a Lagrangian
\begin{equation}
\cL[A_{\mu},\phi]\raw \cL_{\text{pert}}[A_{\mu},\eta]=\cL^{\text{quadratic}}+\cL^{\text{interaction}}
\end{equation}
with
\begin{align}
\cL^{\text{quadratic}}&=-\dfrac{1}{4}F_{\mu\nu}F^{\mu\nu}+\dfrac{1}{2}\p^{\mu}\eta\p_{\mu}\eta-\lambda v^{2}\, \eta^{2}+\dfrac{1}{2}e^{2}v^{2}A^{\mu}A_{\mu}\, ,\\
\cL^{\text{interaction}}&=-\lambda v\, \eta^{3}-\dfrac{\lambda}{4}\eta^{4}+\dfrac{1}{2}(\eta^{2}+2v\eta)A^{\mu}A_{\mu}\, .
\end{align}
The first line encodes a massive vector field, but no massless Goldstone mode $\xi$! The spectrum is therefore given by
\begin{itemize}
\item a massive gauge field with three degrees of freedom with two from the transverse polarisation from the original massless spin $1$ vector boson and an additional from the ``Goldstone'' mode acting now as longitudinal polarisation
\item and a massive scalar $\eta$ which we call the \emph{Higgs boson}\index{Higgs boson}.
\end{itemize}
This is the \emph{Higgs mechanism.}\index{Higgs mechanism}\index{SSB!Higgs mechanism} 

\begin{equ}[Higgs Mechanism]
For a spontaneously broken gauge symmetry the transverse degrees of freedom of the broken symmetry gauge fields $A_\mu (x)$ combine with the Goldstone bosons $\xi(x)$ to complete the degrees of freedom of a massive vector field with the Goldstone modes becoming the longitudinal components. The gauge fields of the unbroken symmetries remain massless. The remaining scalar degrees of freedom are massive corresponding to the Higgs bosons. 
\end{equ}

\subsubsection{Aside: Cosmic strings}

\begin{figure}[t!]
\centering
\includegraphics[width=0.85\textwidth]{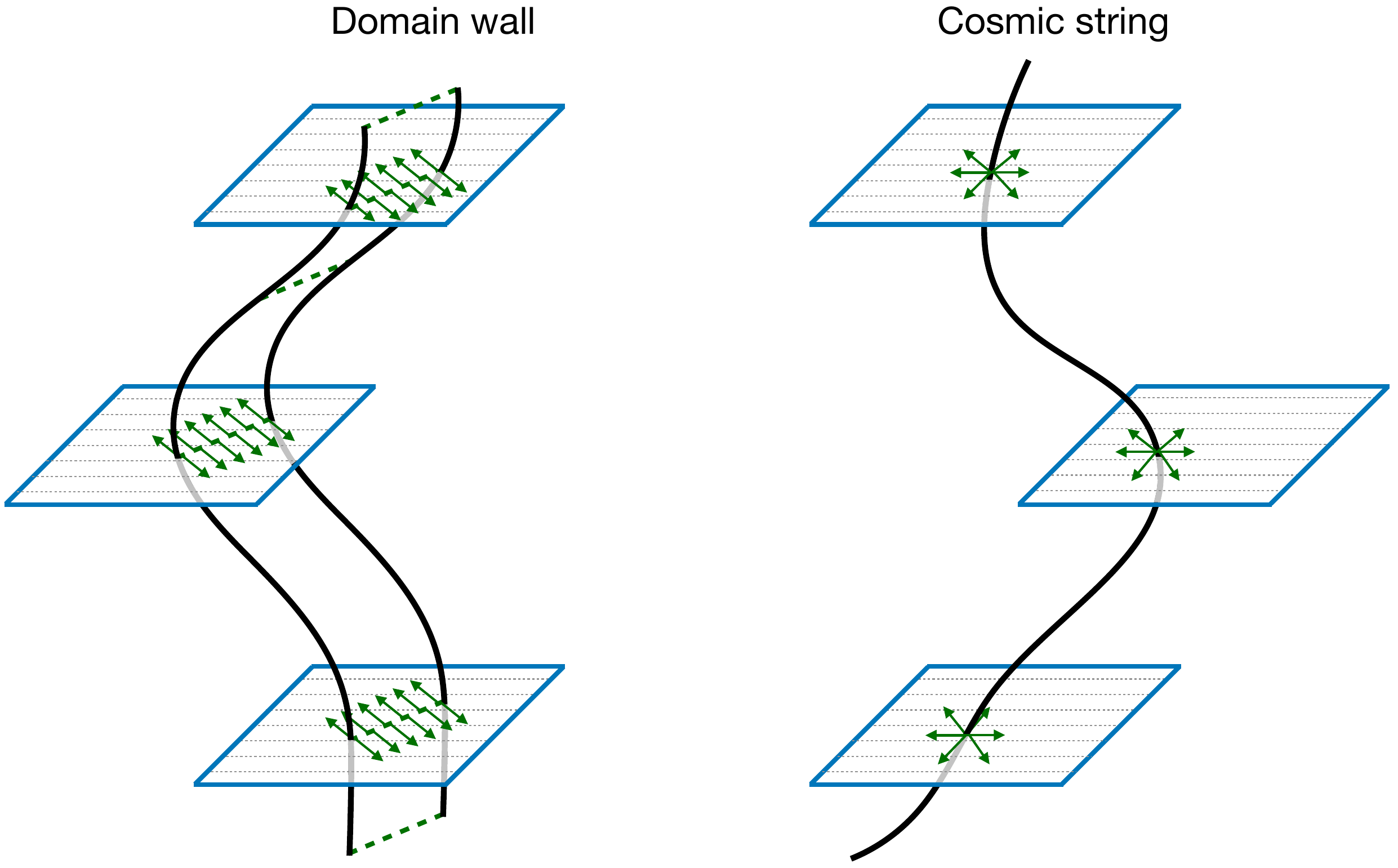}
\caption{In our 3D physical space there appear topological defects separating regions of different vacua (represented here by the arrows as in ferromagnetism). For an SSB potential with two vacua the domain walls separate the different domains.  These are topological defects interpolating between different vacua for a discrete set of vacua. In the figure on the right, we illustrate the case for the continuous symmetry breaking with continuous vacuum degeneracy. The corresponding topological defects are one-dimensional and are known as cosmic strings or vortices.
}\label{fig:cosmicstrings} 
\end{figure}

In general, similar to the case of discrete symmetry breaking,
there may be \emph{topological defects}\index{Topological defects} corresponding not to domain walls, but to what is called as \emph{cosmic strings}\index{Cosmic strings}\index{Topological defects!Cosmic strings}, see Fig.~\ref{fig:cosmicstrings}.
As discussed above, domain walls arise from the presence of two different vacua where some regions in physical 3-dimensional space could be either in one vacuum or the other.
The domain wall then corresponds to the separation between the two phases.
In the current example of a continuous group, the vacuum manifold is not only two points, but a whole circle containing infinitely many vacua.
Then in a given plane we may have vacua in all directions, giving rise to a string since the topology of the vacuum manifold (a circle) is non-trivial. Cosmic strings may be relevant in early universe cosmology \cite{Kibble:1976sj}.

\subsubsection{Aside: Superconductivity}

Even though the Abelian Higgs model provides the simplest concrete example of the Higgs mechanism at work, just by itself it is not yet particularly interesting for particle physics, but it turns out to be relevant for \emph{superconductivity}\index{Superconductivity}. In this case the Abelian Higgs model has been shown to be an effective description of superconductivity.


\begin{equ}[Superconductor]

\emph{Definition:}
A superconductor is a material for which the $\mathrm{U}(1)$ of electromagnetism is spontaneously broken.

\end{equ}

In terms of an effective field theory, known as \emph{Landau-Ginzburg}, a scalar field $\phi$ with $\langle \phi \rangle\neq 0$ is identified with the Cooper pair of two electrons $\psi_e\psi_e$ moving in the material. Even though electrons are fermions, a scalar can be obtained from a pair of electrons. 

At low temperatures this composite field may condense $\langle \phi \rangle =\langle \psi_e \psi_e \rangle \neq 0$, thereby breaking the electromagnetic $\mathrm{U}(1)$ to a discrete
${\mathbb Z}_2$ symmetry (since the electric charge of the condensing field is $2e$).
It then gives a mass to the photon and induces a new phase for the corresponding material in which there is a current with essentially no resistance. In fact, a magnetic field inside the material is energetically unfavourable (since the $A^\mu A_\mu$ term adds a substantial component to the energy so it minimises at $A^\mu=0$ which in turn implies zero magnetic field $\mathbf B=0$). This is known as the \emph{Meissner effect}\index{Meissner effect}. Also, the relevant scales in the superconductor can be interpreted in terms of this model: the physical penetration depth of the magnetic field in the material can be estimated to be proportional to the mass of the photon and the correlation length corresponds to the mass of the Higgs field $\eta$. Cosmic string defects correspond to vortices in the material, etc. 

For a detailed discussion of superconductivity from effective field theories, we refer to \cite{Weinberg:1996kr,Burgess:2020tbq}. Here let us briefly try to see why there is a superconducting behaviour. The Lagrangian can be written as a function of $A^\mu +\partial^\mu \xi/e$. The electric current and charge density are:
\begin{equation}
J^i=\frac{\partial\cL}{\partial A^i}\kom J^0= \frac{\partial\cL}{\partial A^0}=e\, \frac{\partial\cL}{\partial \dot\xi}\, .
\end{equation}
We can see that $\dot\xi$ acts as a conjugate variable to $J^0$. Therefore, we find
\begin{equation}
\dot\xi=e\,  \frac{\partial\cH}{\partial J^0}\propto \rm{voltage}
\end{equation}
in terms of the Hamiltonian $\cH$ since voltage is the variation of energy with respect to the  charge density at a given point. From here we can conclude that we can have a  time-independent configuration with stationary current with $\dot\xi=0$ implying zero voltage which is essentially the definition of superconductivity.

\subsection{A Non-Abelian Example}

Let us briefly see how the Higgs mechanism can be straightforwardly extended to the non-Abelian case. Let us consider the $N=3$ of the $\mathrm{O}(N)$ model discussed above.
We promote the model to a gauge theory with generators $T^i_{jk}=-{\rm i}\epsilon_{ijk}$
\begin{equation}
\L[A_\mu^{i},\phi] = -\frac{1}{4} \left(F^i_{\mu\nu}\right)^2 - \frac12 D_\mu \phi\,\cdot\,D^\mu \phi - V(\phi) 
\end{equation}
with {$\phi^2 = \phi\cdot\phi; ~\phi^4=(\phi^2)^2$ and $\left(D_\mu\right)_{ij}=\delta_{ij}\partial_\mu+g\epsilon_{ijk}A_\mu^k$ with $i,j,k=1,2,3$.
We consider the potential
\begin{equation}
V_{\pm}(\phi) = \pm \frac12 m^2\phi^2 + \frac{\lambda}{4}\phi^4 \kom \lambda>0 \,.
\end{equation}
For $V_+$ the minimum sits at $\phi=(0,0,0)^T$ describing a standard $\mathrm{O}(3)$ Yang-Mills theory with three massless gauge fields $A_\mu^i$ with $i,j=1,2,3$ and three real massive scalars $\phi_i$ with a total of $2\times 3+3=9$ degrees of freedom. 

For $V_-$,
the minimum sits at a non-zero VEV which we can choose as
\begin{equation}
\langle\phi\rangle_{0}=\phi_{0}=\left (\begin{array}{c}
0 \\ 
0 \\ 
v
\end{array} \right )\, .
\end{equation}
This breaks the symmetry to $\mathrm{O}(2)$ or $\mathrm{U}(1)$. Again, we can consider fluctuations around the minimum
\begin{equation}
\phi={\rm exp}\left[\frac{i}{v}\left(\xi_1(x)T_1+\xi_2(x) T_2\right)\right]\left (\begin{array}{c}
0 \\ 
0 \\ 
v+\eta(x)
\end{array} \right )\, .
\end{equation}
We can go to unitary gauge by setting
\begin{equation}
\phi\rightarrow \phi'={\rm exp}\left[-\frac{i}{v}\left(\xi_1T_1+\xi_2T_2\right)\right]\, \phi=\left (\begin{array}{c}
0 \\ 
0 \\ 
v+\eta(x)
\end{array} \right )\, .
\end{equation}
The Lagrangian then becomes
\begin{equation}
\L_{\text{pert}}[A_\mu^{i},\eta] = -\frac{1}{4} \left(F^i_{\mu\nu}\right)^2 - \frac12 \partial^\mu\eta\partial_\mu \eta -m^2\eta^2 - \frac{g^2v^2}{2} \left( A_\mu^1 A^{1\mu} + A_\mu^2 A^{2\mu}  \right)+ \L^{int.}
\end{equation}
with a massive Higgs field $\eta$, massive vector fields $A^1_\mu , A^2_\mu $ and a massless gauge field $A^3_\mu$ corresponding to the unbroken symmetry.

In summary we have seen that the problems of massless gauge bosons and massless Goldstone bosons solve each other by the Higgs mechanism. Furthermore the presence of the physical, massive, Higgs field $\eta(x)$ takes care of the perturbative unitarity problem mentioned before in the sense that given that gauge theories are renormalisable even after symmetry breaking is implemented, the corresponding left-over theory is UV complete. This means that a low-energy theory of massive vector fields obtained after SSB is different from a low-energy theory of massive vector fields not based on gauge invariance. For an explicit calculation to illustrate how perturbative unitarity is recovered by an spontaneously broken gauge theory we have to wait for the next chapter. Before that, we will consider another SSB example and then the crucial aspect of chiral gauge theories known as anomalies.

\subsection{SSB in an $\mathrm{SU}(2)$ gauge theory*}

For illustrative purposes,
let us consider another example, namely $\SUTw$ gauge theory coupled to a two component complex scalar field
\begin{equation}
\phi=\left (\begin{array}{c}
\phi_{1} \\ 
\phi_{2}
\end{array} \right )\, .
\end{equation}
The generators of $\SUTw$ are given by the Pauli matrices,
\begin{equation}
\tau^{a}=\dfrac{1}{2}\sigma^{a}\, .
\end{equation}
The associated Lagrangian is defined as
\begin{equation}
\cL=-\dfrac{1}{4}\tr(F_{\mu\nu}F^{\mu\nu})+\tr((D^{\mu}\phi)^{\dagger}(D_{\mu}\phi))-\dfrac{1}{2}\lambda\left (\phi^{\dagger}\phi-\dfrac{1}{2}v^{2}\right )^{2}\, .
\end{equation}
The bold face notation on the sheet simply corresponds to putting the components $F_{\mu\nu}^{a}$ and $A_{\mu}^{a}$ into a vector with the conventional definition of scalar and cross product. We work in conventions where
\begin{equation}
D_{\mu}=\p_{\mu}\mathds{1}_{2}+\I gA_{\mu}^{a}\tau^{a}
\end{equation}
and
\begin{equation}
F_{\mu\nu}=F_{\mu\nu}^{a}\tau^{a}=\left (\p_{\mu}A_{\nu}^{a}-\p_{\nu}A_{\mu}^{a}-g\epsilon^{abc} A^{b}_{\mu}A_{\nu}^{c}\, \right )\tau^{a}\, .
\end{equation}
The fact that the indices run over $a=1,2,3$ and that $\epsilon^{abc}$ appears in the last term as the structure constants of $\SUTw$ explains the appearance of the cross product on the sheet.

\subsubsection*{Spontaneous symmetry breaking}

Regardless of the notation, we are interested in understanding the breaking pattern of the theory at minima of the scalar potential $V(\phi)$.
Hence, we need to have a closer look at
\begin{equation}\label{eq:ContSymBrPot} 
V(\phi)=\dfrac{\lambda}{2}\left (\phi^{\dagger}\phi-\dfrac{1}{2}v^{2}\right )^{2}\, .
\end{equation}
The first derivative is given by
\begin{equation}\label{eq:Q6GradV} 
\dfrac{\dif V}{\dif \phi_{i}}=\lambda\phi^{*}_{i}\left (|\phi_{1}|^{2}+|\phi_{2}|^{2}-\dfrac{1}{2}v^{2}\right )\, .
\end{equation}
The two stationary points are given by
\begin{equation}
\tilde{\phi}_{0}=0\kom \phi_{0}^{\dagger}\phi_{0}=\dfrac{v^{2}}{2}\, .
\end{equation}
One easily verifies that the Hessian matrix at $\tilde{\phi}_{0}$ has only negative eigenvalues for $\lambda, v>0$ corresponding to a maximum. In contrast,
at $\phi_0$ the Hessian has eigenvalues
\begin{equation}\label{eq:Q6MassEV} 
\nu_{1}=\nu_{2}=\nu_{3}=0\kom \nu_{4}=\lambda |\phi_{0}|^{2}=\dfrac{\lambda v^{2}}{2}\, .
\end{equation}
This implies that the Hessian is positive semi-definite and thus $\phi_{0}$ describes a local minimum.

Now that we have established that the minimum is given by 
\begin{equation}\label{eq:BrokenContSimMinSol} 
\phi_{0}^{\dagger}\phi_{0}=\dfrac{v^{2}}{2}
\end{equation}
we can discuss the breaking pattern in more detail. First, keep in mind that the potential in \eqref{eq:ContSymBrPot} is non-negative for $\lambda>0$ and also
\begin{equation}
V(\phi)\bigl |_{\phi=\phi_{0}}=0
\end{equation}
is the minimal energy. The vacuum manifold is given by
\begin{equation}
\cM_{0}=\biggl \{ \phi_{0}\text{ such that }\phi_{0}^{\dagger}\phi_{0}=\frac{v^{2}}{2}\biggl \}\, .
\end{equation}
Since $\phi^{\dagger}\phi \geq 0$, we must have $v^{2}\geq 0$ for any symmetry breaking to occur. 
At the minimum \eqref{eq:BrokenContSimMinSol}, we e.g. set
\begin{equation}\label{eq:BrokenContSimMinSol1} 
\phi_{0}=\dfrac{1}{\sqrt{2}}\left (\begin{array}{c}
0 \\ 
v
\end{array} \right )\, .
\end{equation}

Before we start any actual calculation, we should first look at what Goldstone's theorem tells us. We can understand the constraint for the vacuum manifold
\begin{equation}
\dfrac{v^{2}}{2}=\phi_{0}^{\dagger}\phi_{0}=a^{2}+b^{2}+c^{2}+d^{2}
\end{equation}
as the defining equation for a $3$-sphere $S^{3}$. Hence,
we deduce
\begin{equation}
\cM_{0}=S^{3}\quad\Rightarrow\quad \dim(\cM_{0})=3=\dim(G)-\dim(H)
\end{equation}
and we expect $3$ massless Goldstone bosons after spontaneous symmetry breaking. We started with $G=\SUTw$ with $\dim(G)=3$ which is why
\begin{equation}
\dim(H)=0
\end{equation}
and the symmetry \emph{is completely broken}. The three massless Goldstone modes become the longitudinal degree of freedom of the new massive gauge bosons.
This means that, whenever SSB occurs in a gauge theory, $\dim(\cM_{0})$ \emph{counts the number of massive gauge bosons after SSB in unitary gauge}.

As it stands, our theory lives in a state expanded around $\tilde{\phi}_{0}=0$ corresponding to a saddle point as shown above. We would like to understand the theory from the perspective of an observer at the minimum \eqref{eq:BrokenContSimMinSol1}. Hence, we consider fluctuations around \eqref{eq:BrokenContSimMinSol1} by defining
\begin{equation}
\phi=\left (\begin{array}{c}
a+\I b \\ 
c+\I d
\end{array} \right )
\end{equation}
as well as some element $U\in\SUTw$ as
\begin{equation}
U=\exp\left (\I\alpha_{i}\tau^{i}\right )\, .
\end{equation}
By definition, $U$ has $3$ degrees of freedom associated with the individual generators. These can be used to \emph{remove} $3$ degrees of freedom in $\phi$ which then fixes a gauge. This allows us to write
\begin{equation}\label{eq:BrokenContSimMinSol2} 
\phi=\dfrac{1}{\sqrt{2}}\left (\begin{array}{c}
0 \\ 
v+f
\end{array} \right )\, .
\end{equation}
for some \emph{real} scalar field $f$. 
It is crucial to keep in mind that \emph{only because we fix a particular gauge, we write $\phi$ in the form \eqref{eq:BrokenContSimMinSol2}}. This gauge is referred to as \emph{unitary gauge}, and is the gauge in which a particular subset of gauge fields absorb the massless Goldstone bosons to become massive. It is also important to remember that the the gauge is completely fixed and we cannot perform any more gauge transformations. This is as it should be: gauge invariance is only the sign for a mathematical redundancy, but does not have any physical significance. That is, to discuss physical phenomena, one first needs to specify a certain gauge.

\subsubsection*{Masses and interactions after SSB}

To obtain the masses explicitly, we examine the quadratic piece of the Lagrangian. First, we compute for \eqref{eq:BrokenContSimMinSol2} the terms $\sim f^{2}$
\begin{equation}
\phi^{\dagger}\phi=\dfrac{1}{2}\left (v^{2}+2vf+f^{2}\right )
\end{equation}
and therefore
\begin{equation}
V(\phi)=\dfrac{\lambda}{2}\left (vf+\dfrac{f^{2}}{2}\right )^{2}=\dfrac{\lambda}{2}v^{2}f^{2}+\dfrac{\lambda}{2}vf^{3}+\dfrac{\lambda}{8}f^{4}\, .
\end{equation}
We find that $f$ has mass
\begin{equation}
m_{f}^{2}=\lambda v^{2}
\end{equation}
as expected from the eigenvalues of the Hessian \eqref{eq:Q6MassEV} (up to factors of 2).
Next, we look at the covariant derivative term for the complex doublet $\phi$ which reads
\begin{equation}
D_{\mu}\phi=\p_{\mu}\phi+\I g A_{\mu}^{a}\tau^{a}\phi \dfrac{1}{2\sqrt{2}}\, \left (\begin{array}{c}
\I g(A_{\mu}^{1}-\I A_{\mu}^{2})(v+f) \\ [0.15em]
2\p_{\mu}f-\I g A_{\mu}^{3}(v+f)
\end{array} \right )\, .
\end{equation}
We are then able to compute (keeping in mind that $v$, $f$ and $A_{\mu}^{a}$ are real-valued)
\begin{align}
(D^{\mu}\phi)^{\dagger} D_{\mu}\phi &=\dfrac{1}{8}\biggl [ g^{2}((A_{\mu}^{1})^{2}+ (A_{\mu}^{2})^{2}+(A_{\mu}^{3})^{2}) (v^{2}+2vf+f^{2})+4(\p_{\mu}f)^{2}\biggl ]\, .
\end{align}
We then read off
\begin{itemize}
\item the kinetic term $\sim (\p_{\mu}f)^{2}$ for the massive real scalar $f$,
\item the mass terms for \textbf{all} the gauge fields $A^{a}_{\mu}$,
\item a cubic interaction vertex $\sim f (A_{\mu}^{a})^{2}$ of coupling strength
\begin{equation}\label{eq:ContBrCubCoup} 
\lambda_{3}=\dfrac{g^{2}v}{2}\, ,
\end{equation}
\item and a quartic interaction vertex $\sim f^{2} (A_{\mu}^{a})^{2}$ of coupling strength
\begin{equation}\label{eq:ContBrQuarCoup} 
\lambda_{4}=\dfrac{g^{2}}{2}\, .
\end{equation}
\end{itemize}
The masses for the gauge particles are
\begin{equation}
m_{A}^{2}=\dfrac{g^{2}v^{2}}{4}\, .
\end{equation}
Each gauge boson becomes massive and the symmetry is completely broken.

\

Even though the symmetry is completely broken in the vacuum, there remains a trace of its original presence. If we look closely at the cubic \eqref{eq:ContBrCubCoup} and quartic coupling \eqref{eq:ContBrQuarCoup}, we obtain
\begin{equation}
\dfrac{\lambda_{3}}{\lambda_{4}}=v=\dfrac{m_{f}}{\sqrt{\lambda}}\, .
\end{equation}
Measuring $m_{f}$ and $\lambda$, i.e., the mass of $f$ and its quartic self interaction determines the ratio of both couplings which would not be true for generic cubic and quartic interactions! Observations of this kind can therefore be seen as a hint for spontaneous symmetry breaking.

\newpage

\section{Anomalies}\label{sec:sym_gauge_anom}

Finally, let us discuss aspects of symmetries in quantum theories which are different from classical theories. It was a profound observation in the 1960's and 1970's that not all classical symmetries survive in a quantum theory \cite{Bell:1969ts,Adler:1969gk,Bardeen:1969md}. This leads us to the notion of \emph{anomalies}\index{Anomalies}:

\begin{equ}[Anomalies]

Classical symmetries are called \emph{anomalous} if they are broken in the quantum theory.

\end{equ}

To understand how anomalies arise,
recall that in the quantum theory we work with path integrals of the form
\begin{equation}
\int\D\Phi_{\alpha}\;\ee^{\I S[\Phi_{\alpha},\p\Phi_{\alpha}]}\, .
\end{equation}
Let us assume that the classical action $S[\Phi_{\alpha},\p\Phi_{\alpha}]$ is invariant under $\Phi_{\alpha}\raw\Phi^{\prime}_{\alpha}$.
For this symmetry to survive at the quantum level, the path integral above also needs to be invariant.
This necessitates in particular that the measure $\D\Phi_{\alpha}$ itself is invariant.
Otherwise, we speak of the presence of an anomaly.
In general, we distinguish two types of anomalies:
\begin{itemize}
\item \emph{Anomalies of a global symmetry} which lead to a breaking of the symmetry in the quantum theory.
This implies that the corresponding classical conservation law does not hold in the quantum theory.
A typical example in the Standard Model is \emph{baryon number} which is not conserved in nature, i.e., it is anomalous,
while the difference of baryon and (total) lepton number is non-anomalous.
We will discuss these anomalies in more detail later in Sect.~\ref{sec:global_anomalies_SM}.
\item \emph{Anomalies for local symmetries}: the associated current is not conserved and the Ward identity correspondingly violated (recall our discussion in Sect.~\ref{sec:GaugeSymmetriesLorentzInv}). This leads to unphysical polarisations of the gauge field and the loss of  Lorentz invariance. In this sense, the theory becomes inconsistent.
\end{itemize}
As we will see, the notion of anomalies is most relevant for chiral theories where so-called ``\emph{chiral anomalies}''\index{Chiral anomalies} appear.

In the subsequent section, we explain in detail how anomalies arise in the presence of gauge fields by studying QED as an example.
We explicitly derive the anomaly there from first principles.
We then generalise anomalies to non-Abelian gauge theories.
We apply our insights to the Standard Model later in Sect.~\ref{sec:anomalies_SM} proving that, despite being a chiral theory, the Standard Model is gauge anomaly free. As another proof of how powerful the machinery of anomalies actually is, we also show in Sect.~\ref{sec:anomalies_SM} that the charges of electron and proton are exactly equal implying charge quantisation in nature which also guarantees the existence of electrically neutral atoms -- everything that we are made of.

\subsection{Abelian gauge theories -- derivation of the anomaly}\index{Anomalies!Abelian}\index{Anomalies!QED}

Let us consider QED as an illustrative example\index{QED}
\begin{equation}
\cL_{QED}[A,\psi]=-\dfrac{1}{4}F_{\mu\nu}F^{\mu\nu}+\overline{\psi}\left (\I\cancel{\p}-e\cancel{A}-m\right )\psi\, .
\end{equation}
The Dirac spinor $\psi$ can be separated into left- and right-handed components such that
\begin{equation}
\overline{\psi}\left (\I\cancel{\p}-e\cancel{A}-m\right )\psi=\overline{\psi}_{L}\left (\I\cancel{\p}-e\cancel{A}\right )\psi_{L}+\overline{\psi}_{R}\left (\I\cancel{\p}-e\cancel{A}\right )\psi_{R}-m\overline{\psi}_{L}\psi_{R}-m\overline{\psi}_{R}\psi_{L}\, .
\end{equation}
In the limit $m\raw 0$, the theory is invariant under the two transformations
\begin{equation}
\psi\raw\ee^{\I\alpha}\psi\kom \psi\raw \ee^{\I\beta\gamma_{5}}\psi
\end{equation}
or equivalently for $\psi_{L},\psi_{R}$
\begin{equation}
\psi_{L}\raw\ee^{\I(\alpha-\beta)}\psi_{L}\kom \psi_{R}\raw\ee^{\I(\alpha+\beta)}\psi_{R}\, .
\end{equation}
The conserved currents associated with these two symmetries are the \emph{vector current}\index{QED!Vector current}
\begin{equation}
J^{\mu}_{v}=\overline{\psi}\gamma^{\mu}\psi
\end{equation}
and the \emph{axial current}\index{QED!Axial current}
\begin{equation}
J^{\mu}_{ax}=\overline{\psi}\gamma^{\mu}\gamma^{5}\psi\, .
\end{equation}
In the limit $m=0$, both currents are conserved
\begin{equation}
\p_{\mu}J^{\mu}_{v}=\p_{\mu}J^{\mu}_{ax}=0\, ,
\end{equation}
while for $m\neq 0$
\begin{equation}
\p_{\mu}J^{\mu}_{v}=0\kom \p_{\mu}J^{\mu}_{ax}=2\I m\overline{\psi}\gamma^{5}\psi\, .
\end{equation}
So, only the vector current is actually conserved in both limits. We will see now that even in the massless limit the \emph{axial current conservation will be broken in the quantum theory.} 
This was originally seen by considering the triangle diagrams shown in Fig.~\ref{fig:TriangleDiagramsAnomalyQED} which, when computed, lead to $ \p_{\mu}J^{\mu}_{ax}\neq 0$. For gauge currents the corresponding Ward identities fail to cancel in a way proportional to the divergence of the axial current, signalling the breaking of gauge symmetry and inconsistency of the theory. However there is a very general way to understand the anomaly directly from the path integral. We choose to present this proof which illustrates the generality of the problem and the power of path integral methods. This was done by Fujikawa in 1979 \cite{Fujikawa:1979ay}.

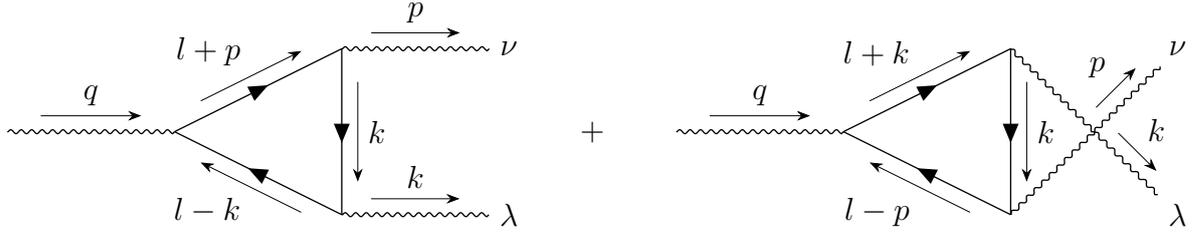
\begin{figure}[t!]
\centering
\begin{tikzpicture}[scale=1.1]
\setlength{\feynhanddotsize}{1.5ex}
\begin{feynhand}
\vertex (a0) at (0,0); 
\vertex (b0) at (2,0); 
\vertex (b1) at (4,1); 
\vertex (b2) at (4,-1); 
\vertex (c1) at (6,1) {$\nu$}; 
\vertex (c2) at (6,-1) {$\lambda$};
\node (0) at (7,0) {+};
\vertex (a00) at (8,0); 
\vertex (b00) at (10,0); 
\vertex (b11) at (12,1); 
\vertex (b22) at (12,-1); 
\vertex (c11) at (14,1) {$\nu$}; 
\vertex (c22) at (14,-1) {$\lambda$}; 
\vertex (c33) at (13,0); 
\propag [pho, mom={$q$}] (a0) to (b0);
\propag [pho, mom={$p$}] (b1) to (c1);
\propag [pho, mom={$k$}] (b2) to (c2);
\propag [fer, mom={$l-k$}] (b2) to (b0);
\propag [fer, mom={$l+p$}] (b0) to (b1);
\propag [fer, mom={$k$}] (b1) to (b2);
\propag [pho, mom={$q$}] (a00) to (b00);
\propag [pho] (b11) to (c33);
\propag [pho] (b22) to (c33);
\propag [pho, mom={$k$}] (c33) to (c22);
\propag [pho, mom={$p$}] (c33) to (c11);
\propag [fer, mom={$l-p$}] (b22) to (b00);
\propag [fer, mom={$l+k$}] (b00) to (b11);
\propag [fer, mom={$k$}] (b11) to (b22);
\end{feynhand}
\end{tikzpicture}
\caption{Triangle diagrams of a three photon process containing a single loop of fermions.}\label{fig:TriangleDiagramsAnomalyQED} 
\end{figure}

To derive the anomaly,
let us consider the path integral for massless QED
\begin{equation}\label{eq:ActionQED} 
\mathcal{Z}_{QED}=\int\D\psi\D\overline{\psi}\D A\,\ee^{\I S_{QED}[A,\psi]}\kom S_{QED}[A,\psi] = \int\dif^{4}x(-\frac{1}{4}F_{\mu\nu}^{2}+\I\overline{\psi}\cancel{D}\psi)
\end{equation}
with the classical symmetry
\begin{equation}
\psi\raw\ee^{\I\alpha}\psi\kom \psi\raw \ee^{\I\beta\gamma_{5}}\psi\kom S_{QED}\raw S_{QED}\, .
\end{equation}
To understand the transformation properties of the measure, we consider the general transformations
\begin{equation}
\psi\raw\Delta\psi\kom\overline{\psi}\raw \overline{\psi}{\Delta}_{c} 
\end{equation}
such that
\begin{equation}\label{eq:TrafoMeasureAnomaly} 
\D\overline{\psi}\D\psi\raw(\cJ_{c}\cJ)^{-1}\D\overline{\psi}\D\psi 
\end{equation}
with Jacobians
\begin{equation}
\cJ=\det(\Delta)\kom \cJ_{c}=\det(\Delta_{c}) \, .
\end{equation}
More explicitly, the Jacobian for $\Delta$ can be written as
\begin{equation}
\cJ=\det(\Delta)=\ee^{\tr(\ln(\Delta))}=\ee^{\int\dif^{4}x\bra{x}\tr(\ln(\Delta(x)))\ket{x}}\, . 
\end{equation}
For $\Delta=\ee^{\I\alpha}$,
\eqref{eq:TrafoMeasureAnomaly} remains invariant since the phases from $\cJ$ and $\cJ_c$ cancel, while for the axial transformations $\Delta=\ee^{\I\beta\gamma_{5}}$ the phases add up.
Therefore, under the axial transformation the measure transforms as
\begin{equation}
\D\overline{\psi}\D\psi\rightarrow\D\overline{\psi}\D\psi {\ee}^{i\int d^4x \beta(x)\mathcal{A}(x)} 
\end{equation}
where the anomaly coming from the Jacobian is:
\begin{equation}
{\mathcal A}=-2{\rm Tr}\, \left[\gamma_5 \delta^4(x-x)\right] \, .
\end{equation}
Note that even though $\tr(\gamma_5)=0$, the $\delta^4(x-x)$ factor is not well defined and we will need to regularise this term. 
But before we get to this, let us first understand what this implies.

The path integral transforms as
\begin{equation}
\mathcal{Z}_{QED}\raw \int\D\psi\D\overline{\psi}\D A\,\exp\left({\I}\int \dif^{4}x \left [\cL_{{\rm{QED}}} - J^{\mu}_{ax}\p_{\mu}\beta+  \beta(x){\mathcal A}(x)\right]\right)
\end{equation}
This means that infinitesimally the integral only over the fermions transforms as
\begin{equation}
\delta\int  \D\psi\D\overline{\psi}\,{\ee}^{{\I} S_{{\rm{QED}}}}\raw \I \int\D\psi\D\overline{\psi}\beta(x)\left(\p_\mu J^\mu_{ax}+{\mathcal A}(x)\right)\,\exp\left({\I}\int d^{4}x \left [\cL_{{\rm{QED}}} \right]\right)
\end{equation}
Therefore, instead of having the classical Noether current conservation law for the axial current $\p_\mu J^\mu_{ax}=0$, we have\footnote{In terms of the integral, the chiral transformation amounts to a change of variables and once the measure is properly considered the value of the integral should not change.}
\begin{equation}
\braket{\p_\mu J^\mu_{ax}}=-{\mathcal A}(x)
\end{equation}

In order to compute the anomaly let us introduce a regulator
\begin{equation}
{\mathcal A}=-2{\rm Tr}\, \left[\gamma_5 \delta^4(x-x)\right]=-2\lim_{y\raw x}\, {\rm Tr}\, \left\{\gamma_5 F\left(-\frac{\cancel D^2}{\Lambda^2}\right)\right\}\, \delta^4(x-y)
\end{equation}
where $F$ is a well-behaved function such as a Gaussian. Concretely we  normalise it to $F(0)=1$ and impose that $F$ and its derivatives vanish at infinity. $D$ is the covariant derivative $  D_\mu\equiv \partial_\mu-ieA_\mu(x)$\footnote{Note that this dependence on the covariant derivative is natural since $\cancel D^2$ is gauge invariant. Furthermore, a way to explicitly compute the path integral is to expand the fermions $\psi$ in terms of eigenvectors of the Dirac operator $D$. Please note also that having $\cancel D^2$ as an argument of the function $F$ is not trivial since for instance the differential operator can act on the gauge field inside $D$.} and $\Lambda$ a cut-off scale. Now,
we apply a Fourier transformation
\begin{align}
{\mathcal A} &=-2\int \frac{d^4k}{(2\pi)^4}\, \lim_{y\raw x}\, {\rm Tr}\left\{\gamma_5 F\left(-\frac{\cancel D^2}{\Lambda^2}\right)\right\}\, {\ee}^{{\I}k\cdot(x-y)} \nonumber \\
& = -2\int \frac{d^4k}{(2\pi)^4}\, {\rm Tr}\left\{\gamma_5 F\left(-\frac{(i\cancel{k}+\cancel D)^2}{\Lambda^2}\right)\right\}\, \nonumber \\
& = -2 \Lambda^4 \int \frac{d^4k}{(2\pi)^4}\, {\rm Tr}\left\{\gamma_5 F\left(-\left(i\cancel{k}+\frac{\cancel D}{\Lambda}\right)^2\right)\right\}\, .
\end{align}
We used  the fact that $D$ is a differential operator and rescaled  $k\raw \Lambda k$. The argument of the function $F$ is 
\begin{equation}
-\left(i\cancel{k}+\frac{\cancel D}{\Lambda}\right)^2=k^2-\frac{ik \cdot D}{\Lambda}-\left(\frac{\cancel{D}}{\Lambda}\right)^2\, .
\end{equation}
In the Taylor expansion we can see that terms with less than four powers of $\gamma$ matrices vanish because their trace vanishes. This takes care of the positive powers of $\Lambda$ in the limit $\Lambda\raw \infty$. Also higher factors than $4$ powers of $1/\Lambda $ will vanish in this limit. We are then left with
\begin{equation}
{\mathcal A}=- \int \frac{d^4k}{(2\pi)^4}\, F''(k^2) {\rm Tr}\left(\gamma_5 \, \cancel D\, ^4\right) \nn
\end{equation}
We can now evaluate the ordinary integral
\begin{equation}
\int d^4kF''(k^2)={\I}\, \int_0^\infty 2\pi^2 k^3 F''(k^2) dk={\I} \pi^2 \int_0^\infty ds s F''(s)=-{\I} \pi^2 \int_0^\infty ds  F'(s)={\I} \pi^2 \nn
\end{equation}
where we have used the assumption that the function $F$ is such that $sF''(s)$ and $F'(s)$ vanish at $s=\infty$. Now, before calculating the trace we will need the following result
\begin{equation}
\cancel D^2=\frac{1}{4}\, \left\{D^\mu,D^\nu\right\}\, \left\{\gamma_\mu,\gamma_\nu\right\}+\frac{1}{4}\, \left[D^\mu,D^\nu\right]\, \left[\gamma_\mu,\gamma_\nu\right]=D^2-\frac{{\I}e}{4}\, F^{\mu\nu}\left[\gamma_\mu,\gamma_\nu\right]\, .
\end{equation}
Using the identity
\begin{equation}
{\rm Tr}\,\left\{\gamma_5 \left[\gamma_\mu,\gamma_\nu\right]\left[\gamma_\rho,\gamma_\sigma\right]\right\}\, =\, 16\, {\I}\, \epsilon_{\mu\nu\rho\sigma}\, ,
\end{equation}
we finally arrive at
\begin{equation}
{\mathcal A}(x) =-\frac{e^2}{16\pi^2}\, \epsilon_{\mu\nu\rho\sigma} F^{\mu\nu}(x) F^{\rho\sigma}(x)
\end{equation}

Therefore, from the path integral above, we find that within  the QED background the axial anomaly is given by
\begin{equ}[Axial Anomaly]
\begin{equation}\label{eq:AxialAnomalyQED} 
\braket{\partial_\mu J_{ax}^\mu}=-\frac{e^2}{16\pi^2}\epsilon^{\alpha\beta\gamma\delta}F_{\alpha\beta}F_{\gamma\delta}\, .
\end{equation}
\end{equ}
More precisely, this is called the \emph{Adler-Bell-Jackiw anomaly}\index{Adler-Bell-Jackiw anomaly} \cite{Bell:1969ts,Adler:1969gk}.
It can be shown that this result is valid to all loop orders (Adler, Bardeen \cite{Adler:1969er}) so the one-loop triangle diagrams above happen to capture the whole structure of the anomaly. In general we can write the anomaly as
\begin{equation}
\p_{\mu}\langle J^{\mu}_{ax}\cO\rangle=-\dfrac{e^{2}}{16\pi^{2}} \langle\epsilon^{\alpha\lambda\beta\nu}F_{\alpha\beta}F_{\lambda\nu}\, \cO\rangle\, .
\end{equation}

What does this result imply?
Essentially the anomaly tells us that the axial current is not conserved. The anomaly happens to be a total derivative itself. That is, defining the quantity $G^\mu=\epsilon^{\mu\nu\rho\sigma}A_\nu F_{\rho\sigma}$, known as the {\bf Chern-Simons term}, we can see that $\partial_\mu G^\mu=\frac{1}{2}\epsilon^{\mu\nu\rho\sigma}F_{\mu\nu}F_{\rho\sigma}$ and so we may construct a conserved quantity
\begin{equation}
\partial^\mu K_\mu=0\kom  K^\mu=\braket{J^\mu_{ax}}+\frac{e^2}{8\pi^2}G^\mu\, .
\end{equation}
However, notice that since $G^\mu$ depends explicitly on $A_\nu$ it is not gauge invariant. Also if we had used $F(-\cancel \partial^2)$ instead of $F(-\cancel D^2)$ as a regulator, we would have  obtained a vanishing anomaly term. But again, this would not have been gauge invariant. We may then say that either the anomaly breaks the chiral symmetry or the gauge symmetry, but we cannot find a way to preserve both.

The study of anomalies provides useful techniques to eliminate inconsistent and identify consistent theories.
Anomalies are relevant particularly for chiral theories such as the Standard Model.
This will be explained momentarily in more detail below.
Beyond that, anomaly cancellation was crucial to identify the consistent string theories in ten dimensions for which an anomaly cancelling-term in the action, known as the \emph{Green-Schwarz term}, cancels the gravitational anomaly determined by the change in the measure \cite{Green:1984sg,Green:1984ed,Green:1984qs}. This opened the way towards considering (chiral) string theories as the best candidates for a consistent theory of gravity and all other interactions at the quantum level in 1984.

As a last comment, let us mention that in some sense the Renormalisation Group\index{Renormalisation Group} (RG) flow can be seen as an anomaly for scale invariance, which is called the trace anomaly for which a non-vanishing trace of the stress energy tensor indicates that scale invariance is broken $\langle T^{\mu}\,_{\mu}\rangle\neq 0$ and therefore the couplings in QFT can change with the energy scale.

\subsection{Anomalies in non-Abelian gauge theories}\index{Anomalies!Non-Abelian}

\begin{figure}[t!]
\centering
\begin{tikzpicture}[scale=1.1]
\setlength{\feynhanddotsize}{1.5ex}
\begin{feynhand}
\vertex (a0) at (0,0) {$a$}; 
\vertex (b0) at (2,0); 
\vertex (b1) at (4,1); 
\vertex (b2) at (4,-1); 
\vertex (c1) at (6,1) {$b$}; 
\vertex (c2) at (6,-1) {$c$};
\node (0) at (7,0) {+};
\vertex (a00) at (8,0) {$a$}; 
\vertex (b00) at (10,0); 
\vertex (b11) at (12,1); 
\vertex (b22) at (12,-1); 
\vertex (c11) at (14,1) {$b$}; 
\vertex (c22) at (14,-1) {$c$}; 
\vertex (c33) at (13,0); 
\propag [pho] (a0) to (b0);
\propag [pho] (b1) to (c1);
\propag [pho] (b2) to (c2);
\propag [fer] (b2) to (b0);
\propag [fer] (b0) to (b1);
\propag [fer] (b1) to (b2);
\propag [pho] (a00) to (b00);
\propag [pho] (b11) to (c33);
\propag [pho] (b22) to (c33);
\propag [pho] (c33) to (c22);
\propag [pho] (c33) to (c11);
\propag [fer] (b22) to (b00);
\propag [fer] (b00) to (b11);
\propag [fer] (b11) to (b22);
\end{feynhand}
\end{tikzpicture}
\caption{Triangle diagrams with}\label{fig:TriangleDiagramNonAbelian}
\end{figure}
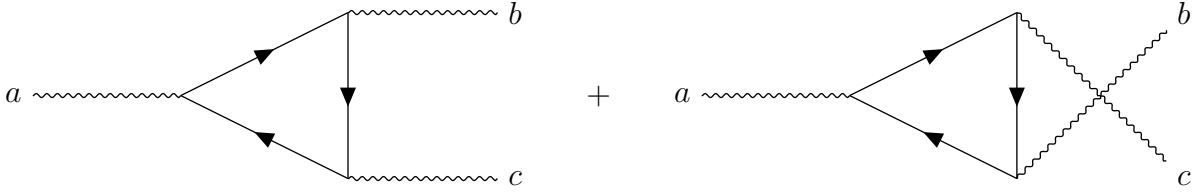

Let us next look at the generalisation of the above results to non-Abelian gauge theories.
Now, we consider the diagrams in Fig.~\ref{fig:TriangleDiagramNonAbelian} with different gauge fields as external legs.
The resulting amplitude $\mathcal{A}(a,b,c)$ is proportional to the divergence of the current as mentioned above, but with constant of proportionality depending on the couplings.
In gauge theories, the latter are proportional to the charges for Abelian theories and the generators of the corresponding gauge group for the non-Abelian case, that is,
\begin{equation}
\mathcal{A}(a,b,c)\sim A^{abc}\cdots
\end{equation}
where we defined the proportionality constants
\begin{equation}
A^{abc} = \tr(T^{a}T^{b}T^{c})+\tr(T^{a}T^{c}T^{b}) = \tr(T^{a}\lbrace T^{b},T^{c}\rbrace)=A(R)d^{abc}\, .
\end{equation}
Here, $d^{abc}$ is the group invariant defined in the previous chapter (cf.~\eqref{eq:coefficientsdabcym}),
but normalised for the fundamental representation, i.e.,
$A(R)=1$ for the fundamental and $A(R)=-1$ for the anti-fundamental representation.
Altogether, the non-Abelian generalisation of Eq.~\eqref{eq:AxialAnomalyQED} reads
\begin{equ}[Anomaly in non-Abelian gauge theories]
\begin{equation}\label{eq:AxialAnomalyNonAbelian} 
\p_{\mu}J^{a,\mu}_{ax}=\left (\sum_{\text{left}}A(R_{l})-\sum_{\text{right}}A(R_{r})\right )\dfrac{g^{2}}{128\pi^{2}} d^{abc}\epsilon^{\alpha\lambda\beta\nu}F^{b}_{\alpha\beta}F^{c}_{\lambda\nu}\, .
\end{equation}
\end{equ}

The structure of this result reveals an important fact mentioned several times before:
if left- and right-handed representations within a given theory are the same, i.e., the theory is non-chiral, the anomaly cancels automatically. 
Vice versa, this means that cancellation of anomalies puts non-trivial constraints on chiral theories like the Standard Model or its possible extensions.
But there is more:
if $R$ is (pseudo) real  $(T^{a})^{*}=-ST^{a}S^{-1}$, we find that the right hand side of \eqref{eq:AxialAnomalyNonAbelian} vanishes because
\begin{align}
A^{abc}=\tr(\lbrace (T^{c})^{*},(T^{b})^{*}\rbrace (T^{a})^{*})=-A^{abc}=0\, .
\end{align}
It so happens that most groups arising in physics exhibit pseudo-real representations and therefore the associated gauge theories have no anomalies. This applies to the following compact groups: $\mathrm{SO}(2n+1)$, $\mathrm{SO}(4n)$ for $n \geq 2$, $\mathrm{Usp}(2n)$ for $n\geq 3$, $G_2, F_4, E_7, E_8$.  Other groups, such as $E_6$ also have $A_{abc}=0$. This leaves only $\mathrm{SU}(n)$ and products of $\mathrm{U}(1)$'s as potentially dangerous. These are precisely the groups relevant for the Standard Model. In some way, nature likes to live dangerously! But for physicists this is actually great news: \emph{the spectrum of the Standard Model is highly constrained and far from arbitrary}.

Let us mention two examples before arguing that the Standard Model is free of (gauge) anomalies later in Sect.~\ref{sec:anomalies_SM}. First,
for $\mathrm{U}(1)^{3}$ the resulting constraint reads:
\begin{equation*}
\begin{tikzpicture}[scale=1.]
\setlength{\feynhanddotsize}{1.5ex}
\begin{feynhand}
\vertex (a0) at (0,0) {$\gamma$}; 
\vertex (b0) at (2,0); 
\vertex (b1) at (4,1); 
\vertex (b2) at (4,-1); 
\vertex (c1) at (6,1) {$\gamma$}; 
\vertex (c2) at (6,-1) {$\gamma$};
\node (0) at (10,0) {$\sim A\sim \tr(Q^{3})\begin{cases}
=0 & \text{consistent}\\
\neq 0 & \text{inconsistent}
\end{cases}$};
\propag [pho] (a0) to (b0);
\propag [pho] (b1) to (c1);
\propag [pho] (b2) to (c2);
\propag [fer] (b2) to (b0);
\propag [fer] (b0) to (b1);
\propag [fer] (b1) to (b2);
\end{feynhand}
\end{tikzpicture}
\end{equation*}
Hence, the sum over the cubed charges has to vanish.
For gravity, we get
\begin{equation*}
\begin{tikzpicture}[scale=1.]
\setlength{\feynhanddotsize}{1.5ex}
\begin{feynhand}
\vertex (a0) at (0,0) {$\gamma$}; 
\vertex (b0) at (2,0); 
\vertex (b1) at (4,1); 
\vertex (b2) at (4,-1); 
\vertex (c1) at (6,1) {$\text{graviton}$}; 
\vertex (c2) at (6,-1) {$\text{graviton}$};
\node (0) at (10,0) {$\sim A\sim  \tr(Q)\begin{cases}
=0 & \text{consistent}\\
\neq 0 & \text{inconsistent}
\end{cases}$};
\propag [pho] (a0) to (b0);
\propag [glu] (b1) to (c1);
\propag [glu] (b2) to (c2);
\propag [fer] (b2) to (b0);
\propag [fer] (b0) to (b1);
\propag [fer] (b1) to (b2);
\end{feynhand}
\end{tikzpicture}
\end{equation*}
This means that the sum over charges needs to vanish, i.e., it implies charge conservation.

\chapter{\bf Electroweak Interactions}
\label{chap:ew}
\index{Electroweak interactions}

\vspace{0.5cm}
\begin{equ}[Electroweak interactions]
{\it  The search for a renormalizable theory of weak interactions was the right strategy but, as it turned out, not for the reasons we originally thought.}\\

\rightline{\it Steven Weinberg}
\end{equ}
\vspace{0.5cm}


We now apply the theoretical framework developed in previous chapters to describe weak interactions. Up to this point, we have mainly relied on the fact that the underlying theories -- quantum mechanics and special relativity -- define the fundamental constituents of matter and their interactions through irreducible representations of the Poincar\'e group. To describe the interactions among these elementary particles, the principles of locality and unitarity guided us to quantum field theory as the fundamental framework. We have argued that, to ensure interactions consistent with Lorentz invariance, local gauge symmetries are necessary: abelian symmetries lead to the well-known case of QED, while non-abelian or Yang-Mills theories have a more complex structure. When coupled to scalar fields, renormalisable Lagrangians allow for at least two distinct phases of the corresponding gauge theory: unbroken gauge theories or those with spontaneous symmetry breaking (SSB). In the SSB phase, Yang-Mills gauge fields can acquire mass through the Higgs mechanism and mediate short-range interactions.


The goal here is to build a gauge theory for the weak interactions of nature, using the Higgs mechanism to give mass to the weak gauge bosons. Specifically, we will emphasise some of the key decisions that were essential in shaping the Standard Model as we understand it today. To make tangible progress, we must now rely not only on theoretical consistency but also on crucial experimental input. After all, physics is fundamentally an experimental science.

\vfill

\newpage

\section{Weak, but powerful}

As mentioned in the introduction, weak interactions were discovered essentially once radioactivity was found. The prime example of a process involving the weak interactions is $\beta$-decay ($n\raw p+e^-+{\overline\nu}_e$). The name {\it weak} was given to differentiate them from the strong interaction. We know that the electromagnetic interactions are responsible to keep electrons bound to nuclei to make atoms and the strong interactions  are responsible to keep the quarks inside hadrons together and indirectly the protons and neutrons bound in the nuclei of all the elements. Weak interactions  do not lead to bound states. However they are crucial for our existence since they are responsible for the leading interactions that give rise to the thermonuclear fusion in stars, including the Sun, from which we receive the energy to live.

As a side note, this process is a beautiful illustration of the 4 interactions at play: gravity dominating at large distances and being attractive induces the formation of stars and galaxies, the electromagnetic and strong interactions compete in the process of having two protons close enough to then allow the weak interaction to start the fusion chain reaction.

Starting with the fusion of two protons (nuclei of Hydrogen) $p+p \raw d+e^-+{\nu}_e$ with $d=pn$ the deuteron or heavy Hydrogen nucleus. Two of these processes produce an $\alpha$ particle (2 protons and 2 neutrons) or nucleus of Helium, releasing energy in the form of neutrinos and photons. This starts the proton chain reaction of fusion interactions that make the stars burn and also create the heavy elements like Oxygen, Carbon, Nitrogen, etc. from which we are all made. Weak interactions are the source of any interaction involving neutrinos and also play an important role in the early history of the Universe. Describing weak interactions within a consistent QFT was very challenging and the success in achieving it will remain as one of the greatest scientific achievements of all time.

The electroweak theory is attributed primarily to Steven Weinberg and Abdus Salam with important work earlier by Sheldon Glashow and John Ward. They all built from the original work of Enrico Fermi, George Sudarshan, Robert Marshak, Murray Gell-Mann, Richard Feynman and others, cf. Sect.~\ref{sec:histSM} for references and the historical development.



\section{Electromagnetic and weak processes}

Before we study the weak interactions, let us briefly summarise some basic facts about QED and compare with the weak interactions that were already established at the time of developing the electroweak theory.

\subsection{Electromagnetic interactions}

As we said several times before,
QED is based on an unbroken $\mathrm{U}(1)$ gauge theory, leading to long-range interactions mediated by the corresponding gauge field: the \emph{photon} $\gamma$.
The basic interaction vertex between photons and matter fields of spin $1/2$ like electrons
\begin{equation*}
\begin{tikzpicture}[scale=1.]
\setlength{\feynhanddotsize}{1.5ex}
\begin{feynhand}
\vertex (a2) at (0.25,-4) {$\gamma$}; 
\vertex (b2) at (-1.25,-4); 
\vertex (c2) at (-2.5,-3) {$\psi$}; 
\vertex (d2) at (-2.5,-5) {$\psi$}; 
\propag [pho] (a2) to (b2);
\propag [fer] (b2) to (d2);
\propag [fer] (c2) to (b2);
\end{feynhand}
\end{tikzpicture}
\end{equation*}
involves two fermions and one photon associated with the conserved current
\begin{equation}
J^{\mu}=\overline{\psi}\gamma^{\mu}\psi\, .
\end{equation}
In the action \eqref{eq:ActionQED}, the interaction term is as usual represented by $A_\mu J^\mu$.
Concrete physical processes include 
electron-positron annihilation
\begin{align*}
\begin{tikzpicture}[scale=0.8]
\setlength{\feynhanddotsize}{1.5ex}
\begin{feynhand}
\node (o) at (-3.5,2) {$e^{-}e^{+}\raw e^{-}e^{+}=$};
\vertex (a0) at (-1.5,3) {$e^{-}$}; 
\vertex (a1) at (2,2); 
\vertex (d0) at (3.5,3) {$e^{-}$}; 
\vertex (d1) at (3.5,1) {$e^{+}$}; 
\vertex (b0) at (0,2); 
\vertex (c0) at (-1.5,1) {$e^{+}$}; 
\propag [fer] (a0) to (b0);
\propag [pho] (b0) to [edge label = $\gamma$] (a1);
\propag [fer] (a1) to (d0);
\propag [fer] (d1) to (a1);
\propag [fer] (b0) to (c0);
\end{feynhand}
\end{tikzpicture}
\end{align*}
or Compton scattering
\begin{align*}
\begin{tikzpicture}[scale=0.8]
\setlength{\feynhanddotsize}{1.5ex}
\begin{feynhand}
\node (o) at (-3.5,2) {$e^{-}\gamma\raw e^{-}\gamma=$};
\vertex (a0) at (-1.5,3) {$e^{-}$}; 
\vertex (a1) at (2,2); 
\vertex (d0) at (3.5,3) {$e^{-}$}; 
\vertex (d1) at (3.5,1) {$\gamma$}; 
\vertex (b0) at (0,2); 
\vertex (c0) at (-1.5,1) {$\gamma$}; 
\propag [fer] (a0) to (b0);
\propag [fer, mom={$e^-$}] (b0) to (a1);
\propag [fer] (a1) to (d0);
\propag [pho] (d1) to (a1);
\propag [pho] (b0) to (c0);
\end{feynhand}
\end{tikzpicture}
\end{align*}
The power of QED is simply that it successfully describes all these different interactions observed in nature with calculations that can be contrasted with experiments.

\subsection{Weak interactions}

Next, let us try to answer the following question: \emph{can the weak interactions be described in a way similar to QED with one or more mediating fields playing the role of the photons?}

\subsubsection*{Fermi, V-A theory and chirality}

For the Weak Interactions, let us first take on a historical perspective.
In the early 1900’s, many physical processes had been observed that guided physicists towards a proper description of a consistent theory of the weak interactions.
One typical interaction was the $\beta$-decay $n\raw p + \nu_e+\bar\nu_e$
\begin{equation}
\centering
\begin{tikzpicture}[scale=1.]
\setlength{\feynhanddotsize}{1.5ex}
\begin{feynhand}
\vertex (a00) at (6.,2) {$n$}; 
\vertex (a11) [dot] at (8.,2) {}; 
\node (o1) at (8.,1.55) {\small $G_{F}$}; 
\vertex (d00) at (10.,2) {$p$}; 
\vertex (d11) at (9.5,0.5) {$\bar{\nu}_{e}$}; 
\vertex (c00) at (8.95,3.5) {$\nu_{e}$}; 
\propag [fer] (a00) to (a11);
\propag [antfer] (d00) to (a11);
\propag [antfer] (a11) to (d11);
\propag [fer] (a11) to (c00);
\end{feynhand}
\end{tikzpicture}
\end{equation}
Fermi proposed a concrete formalism to describe these interactions. This is the famous \emph{Fermi-interaction}\index{Fermi-interaction}\index{EFT!Fermi-interaction} with effective Lagrangian description
\begin{equation}
\cL_{\text{Fermi}}=G_{F}\psi_1\psi_2\psi_3\psi_4\, .
\end{equation}
Here, the $\psi_i$ represent each of the particles in the interaction and $G_F$ determines the coupling. The fermion fields $\psi_i$ have mass dimension $[\psi]=3/2$ which is why the mass dimension of $G_{F}$ is
\begin{equation}
[G_{F}]=-2\, .
\end{equation}
Experimentally the coupling had been found to be $G_F= 1.164\times 10^{-5}$ GeV$^{-2}$. Clearly, this interaction is \emph{non-renormalisable}\index{EFT!Non-renormalisable} which is why this is only a good description at energies 
$E\ll G_F^{{-1/2}}$.

Given the absence of an underlying theory at the time, people considered  the most general Lorentz invariant 4-fermion interactions which take the form
\begin{equation}\label{eq:FourFermiCurrent} 
{\mathcal L}=\sum_i g_i J^i J_i\kom J^i =  \overline{\psi}\cO^i\psi\kom g_i\sim G_F
\end{equation}
with the currents $J^i$ written in terms of operators $\cO^i$. Here $i$ is a generic index that labels the operators listed below. These operators are classified depending on how they transform under Lorentz transformations as follows:
\begin{itemize}
\item $\cO=1$  scalar
\item $\cO=\gamma^{5}$  pseudo-scalar
\item $\cO=\gamma^{\mu\nu}$  tensor
\item $\cO=\gamma^{\mu}$ vector (V)
\item $\cO=\gamma^{\mu}\gamma^{5}$ axial vector (A)
\end{itemize}
A detailed analysis of several experiments in the $1950$'s led Marshak and Sudarshan to identify the correct combination that describes all the weak interaction processes as $V-A$ \cite{Sudarshan:1958vf}. 
This included interactions, e.g., for the $\beta$-decay
\begin{equation}
g\overline{\psi}_{p}\gamma^{\mu}(1-\gamma^{5})\psi_{n}\,\overline{\psi}_{e}\gamma^{\mu}(1-\gamma^{5})\psi_{\nu}+h.c.\, .
\end{equation}
The $V-A$ theory describes an  important concept for the weak interactions, namely \emph{chirality}\index{Chirality} (parity violation). This difference between left- and right-handed fermions, as can be seen from the presence of only $1-\gamma^5$ and not $1+\gamma^5$ in the operators,  is a very important property of the weak interactions that comes from observations.
Any theory describing weak interactions has to have this property, as Lee aand Yang had observed earlier. Therefore, from the dependence on $1-\gamma^5$, it is usually said that weak interactions are left-handed and so chiral.
This was the first successful description of weak interactions at \emph{low energies}.
But it became apparent pretty quickly that this cannot be the right theory to describe weak interactions at all energies, because
the theory is non-renormalisable.
Moreover, even though the calculation  of physical quantities, like cross sections and decay rates fit well with experiments at energies $E\ll G_F^{-1/2}$, for higher energies the theory gives diverging results which are clearly against experiments.
This suggested that the four-fermion vertex with dimensionful coupling $G_F$ vertex should be replaced by a three-point interaction and propagator for mediator particles as in QED.

We need a consistent description that unlike the Fermi theory, is valid at all energies. This will turn out to be unique: \emph{a spontaneously broken gauge theory}.

\subsection*{Mediators for weak interactions}

As in QED, we expect a mediator and a basic interaction vertex of the form
\begin{equation*}
\begin{tikzpicture}[scale=1.]
\setlength{\feynhanddotsize}{1.5ex}
\begin{feynhand}
\vertex (a2) at (1.25,-4) {$W^{\pm},Z^{0}\; ?$}; 
\vertex (b2) at (-1.25,-4); 
\vertex (c2) at (-2.5,-3) {$\psi$}; 
\vertex (d2) at (-2.5,-5) {$\psi$}; 
\propag [pho] (a2) to (b2);
\propag [fer] (b2) to (d2);
\propag [fer] (c2) to (b2);
\end{feynhand}
\end{tikzpicture}
\end{equation*}
Contrary to QED the mediating fields may have an electric charge in order to have charge conservation at each vertex and we may name them $W^+, W^-$ and $Z^0$ with the superscript representing the electric charge. The $\beta$-decay process may then be seen as:
\begin{equation*}
\centering
\begin{tikzpicture}[scale=1.]
\setlength{\feynhanddotsize}{1.5ex}
\begin{feynhand}
\node (o) at (4.75,2) {$\xrightarrow{E\ll m_{W}}$};
\vertex (a0) at (-2.,2) {$n$}; 
\vertex (a1) at (2,1); 
\vertex (d0) at (3.5,2) {$e^-$}; 
\vertex (d1) at (3.5,0) {$\bar{\nu}_{e}$}; 
\vertex (b0) at (0,2); 
\vertex (c0) at (0.95,3.5) {$p$}; 
\vertex (a00) at (6.,2) {$n$}; 
\vertex (a11) [dot] at (8.,2) {}; 
\node (o1) at (8.,1.55) {\small $G_{F}$}; 
\vertex (d00) at (10.,2) {$e^-$}; 
\vertex (d11) at (9.5,0.5) {$\bar{\nu}_{e}$}; 
\vertex (c00) at (8.95,3.5) {$p$}; 
\propag [fer] (a0) to (b0);
\propag [pho, mom={$W^{-}$}] (b0) to (a1);
\propag [antfer] (d0) to (a1);
\propag [antfer] (a1) to (d1);
\propag [fer] (b0) to (c0);
\propag [fer] (a00) to (a11);
\propag [antfer] (d00) to (a11);
\propag [antfer] (a11) to (d11);
\propag [fer] (a11) to (c00);
\end{feynhand}
\end{tikzpicture}
\end{equation*}
Similarly, other weakly interacting processes may be considered.  We distinguish processes according to the particles involved:
\begin{enumerate}
\item \emph{Leptonic}: processes that include only leptons in the initial and final states that include {\it charged current} processes such as
\begin{align*}
\begin{tikzpicture}[scale=0.8]
\setlength{\feynhanddotsize}{1.5ex}
\begin{feynhand}
\node (o) at (-4.25,2) {$\mu^{-}\raw e^{-}\nu_{\mu}\bar{\nu}_{e}=$};
\vertex (a0) at (-2.,2) {$\mu^{-}$}; 
\vertex (a1) at (2,1); 
\vertex (d0) at (3.5,2) {$e^{-}$}; 
\vertex (d1) at (3.5,0) {$\bar{\nu}_{e}$}; 
\vertex (b0) at (0,2); 
\vertex (c0) at (0.95,3.5) {$\nu_{\mu}$}; 
\propag [fer] (a0) to (b0);
\propag [pho, mom={$W^{-}\; ?$}] (b0) to (a1);
\propag [fer] (a1) to (d0);
\propag [antfer] (a1) to (d1);
\propag [fer] (b0) to (c0);
\end{feynhand}
\end{tikzpicture}
\end{align*}
and
\begin{align*}
\begin{tikzpicture}[scale=0.8]
\setlength{\feynhanddotsize}{1.5ex}
\begin{feynhand}
\node (o) at (-3.5,2) {$e^{-}\nu_e\raw e^{-}\nu_e=$};
\vertex (a0) at (-1.5,3) {$e^{-}$}; 
\vertex (a1) at (2,2); 
\vertex (d0) at (3.5,3) {$e^{-}$}; 
\vertex (d1) at (3.5,1) {$\nu_e$}; 
\vertex (b0) at (0,2); 
\vertex (c0) at (-1.5,1) {$\nu_e$}; 
\propag [fer] (a0) to (b0);
\propag [pho, mom={$W^{-}\; ?$}] (b0) to (a1);
\propag [fer] (a1) to (d0);
\propag [fer] (d1) to (a1);
\propag [fer] (b0) to (c0);
\end{feynhand}
\end{tikzpicture}
\end{align*}
and
\begin{align*}
\begin{tikzpicture}[scale=0.8]
\setlength{\feynhanddotsize}{1.5ex}
\begin{feynhand}
\node (o) at (-4.25,2) {$\nu_{\mu}e^{-}\raw \nu_{e}\mu^{-}=$};
\vertex (a0) at (-2.,2) {$\nu_{\mu}$}; 
\vertex (a1) at (2,1); 
\vertex (d0) at (-0.3,-0.2) {$e^{-}$}; 
\vertex (d1) at (3.8,1.4) {${\nu}_{e}$}; 
\vertex (b0) at (0,2); 
\vertex (c0) at (0.95,3.5) {$\mu^{-}$}; 
\propag [fer] (a0) to (b0);
\propag [pho, mom={$W^{+}\; ?$}] (b0) to (a1);
\propag [fer] (d0) to (a1);
\propag [fer] (a1) to (d1);
\propag [fer] (b0) to (c0);
\end{feynhand}
\end{tikzpicture}
\end{align*}
and non-charged or {\it neutral current} processes as
\begin{align*}
\begin{tikzpicture}[scale=0.8]
\setlength{\feynhanddotsize}{1.5ex}
\begin{feynhand}
\node (o) at (-4.25,2) {$\nu_{\mu}e^{-}\raw \nu_{\mu}e^{-}=$};
\vertex (a0) at (-2.,2) {$\nu_\mu$}; 
\vertex (a1) at (2,1); 
\vertex (d0) at (-0.3,-0.2) {$e^{-}$}; 
\vertex (d1) at (3.8,1.4) {$e^{-}$}; 
\vertex (b0) at (0,2); 
\vertex (c0) at (0.95,3.5) {$\nu_\mu$}; 
\propag [fer] (a0) to (b0);
\propag [pho, mom={$Z^{0}\; ?$}] (b0) to (a1);
\propag [fer] (d0) to (a1);
\propag [fer] (a1) to (d1);
\propag [fer] (b0) to (c0);
\end{feynhand}
\end{tikzpicture}
\end{align*}

\item \emph{Semi-leptonic}, that is weak interactions that involve leptons and hadrons in initial or final states, e.g., $\beta$-decay.

\item \emph{Non-leptonic}, that is weak interactions that do not involve leptons in the initial or final state, e.g.,
\begin{align*}
\begin{tikzpicture}[scale=0.9]
\setlength{\feynhanddotsize}{1.5ex}
\begin{feynhand}
\node (o) at (-4.25,0.5) {$\Lambda^{0}\raw p+\pi^{-}=$};
\vertex (a0) at (-2.,1) {$u$}; 
\vertex (b0) at (2.,1) {$u$}; 
\vertex (a1) at (-2.,0.5) {$d$}; 
\vertex (b1) at (2.,0.5) {$d$}; 
\vertex (a2) at (-2.,0) {$s$}; 
\vertex (b2) at (2.,0) {$u$}; 
\vertex (c2) at (0.5,0); 
\vertex (c3) at (1.5,-2.); 
\vertex (c4) at (3.,-1.75) {$d$}; 
\vertex (c5) at (2.35,-3.15) {$\bar{u}$}; 
\propag [fer] (a0) to (b0);
\propag [fer] (a1) to (b1);
\propag [fer] (a2) to (b2);
\propag [pho, mom={\small $W^{-}\;?$}] (c2) to (c3);
\propag [fer] (c3) to (c4);
\propag [antfer] (c3) to (c5);
\end{feynhand}
\end{tikzpicture}
\end{align*}
Here we write the hadrons ($\Lambda^0$ and proton) in terms of their component quarks. Again, neutral and charged mediators are needed.
\end{enumerate}

Some comments are in order.
At the time when these processes were first hypothesised or even measured,
it remained unclear what the mediators of the weak force would be.

Charge conservation suggests that there should be three types of particles with electric charge $0$ or $\pm 1$.
But then, as we already discussed in Sect.~\ref{sec:lossunitarityssb},
we face problems with the loss of unitarity due to the presence of massive spin-1 fields.
As we will see in this chapter, this is where spontaneous symmetry breaking enters the stage.
E.g. at the perturbative level, we can show explicitly how the aforementioned unitarity problem is resolved through SSB in Sect.~\ref{sec:pertuniWZ}.

In all of these processes, we identify a potential mediating particle that we called $W^\pm, Z^0$, all with question marks since up to this point we cannot specify their nature but their name already indicates the corresponding value of their electric charge, assuming charge is conserved on each interaction. We immediately observe that, contrary to electromagnetic interactions in which there is only one mediating particle, the photon, weak interactions require at least three particles. Interactions mediated by $W^\pm$ are called \emph{charged current}\index{Charged current} interactions and those mediated by $Z^0$ are called \emph{neutral current}\index{Neutral current} interactions (only charged currents were observed before the Glashow-Weinberg-Salam theory was developed).

Since the weak interactions are short-ranged (and the decay rates are such that the corresponding decaying particles have long lifetimes), the mediating particles $W^{\pm},Z^{0}$ are expected to be very massive (recall Yukawa theory in which a mediating particle of mass $m$ would give rise to a force of the type $\sim\ee^{- mr}/r^{2}$ which reduces to the standard $1/r^2$ when $m=0$ as in electromagnetism. For $m\neq 0$ the interaction decays exponentially fast with distance and would therefore be short-range as observed for the weak interactions).

At small energies, that is energies much smaller than the mass of the corresponding mediating particle, propagators involving heavy gauge bosons can be replaced by an effective $4$-fermion interaction as shown above.

\newpage

\section{Identifying the model for SSB}\label{sec:SSBmodelEW}

In this section, we turn to the question of what the structure of the gauge theory needs to be. Rather than simply writing down the answer given by the \emph{Electroweak Theory}\index{Electroweak Theory}, we try to discuss and justify the proper framework for the gauge fields and fermions in that sector. For simplicity, we consider only the neutrino-electron system with $4$-component Weyl spinors
\begin{itemize}
\item electron $e_{L}$, $e_{R}$
\item neutrino $\nu_{L}$.
\end{itemize}
Here, the right- and left-handed electron fields are defined as
\begin{equation}
e_{L,R}=\dfrac{1}{2}\left (1\mp\gamma^{5}\right )e
\end{equation}
in terms of the Dirac spinor $e$. To identify the gauge group, Lorentz invariance requires to put those fields with the same Lorentz transformation properties into a single representation. Hence, we split the fields into left- and right-handed content
\begin{equation}
\left (\begin{array}{c}
\nu_{L} \\ 
e_{L}
\end{array} \right )\kom e_{R}\, .
\end{equation}
The largest possible group allowing for such representations is
\begin{equation}
G=\mathrm{SU}(2)_{L}\times \mathrm{U}(1)_{L}\times \mathrm{U}(1)_{R}
\end{equation}
with associated generators $T^{a}$, $a=1,2,3$, $Q_{L}$ and $Q_{R}$. The actions of the individual generators on the fields can be described as follows:
\begin{itemize}
\item $T^{a}=\sigma^{a}/2$ acts on the doublet field $(\nu_{L},e_{L})^{T}$ in the fundamental of $\mathrm{SU}(2)$, but \emph{not} on $e_{R}$ which is only a singlet of $\mathrm{SU}(2)$, i.e., in the trivial representation.
\item The individual $\mathrm{U}(1)$-generators act on the fields in the following way
\begin{align}
&Q_{L}\left (\begin{array}{c}
\nu_{L} \\ 
e_{L}
\end{array} \right )=\dfrac{1}{2}\left (\begin{array}{c}
\nu_{L} \\ 
e_{L}
\end{array} \right )\kom Q_{L}e_{R}=0\nn\\
&Q_{R}\left (\begin{array}{c}
\nu_{L} \\ 
e_{L}
\end{array} \right )=0\kom Q_{R}e_{R}=e_{R}\, \nn.
\end{align}
\end{itemize}
We define \emph{hypercharge}\index{Hypercharge} $Y$ as the combination
\begin{equation}
Y=-Q_{R}-Q_{L}
\end{equation}
and the \emph{electron lepton number}\index{Electron lepton number}\index{Lepton number} $L_{e}$ as
\begin{equation}
L_{e}=2Q_{L}+Q_{R}\, .
\end{equation}
These two operators act on the fields in the following way
\begin{align}
&Y\left (\begin{array}{c}
\nu_{L} \\ 
e_{L}
\end{array} \right )=-\dfrac{1}{2}\left (\begin{array}{c}
\nu_{L} \\ 
e_{L}
\end{array} \right )\kom Ye_{R}=-e_{R}\nn\\
&L_{e}\left (\begin{array}{c}
\nu_{L} \\ 
e_{L}
\end{array} \right )=\left (\begin{array}{c}
\nu_{L} \\ 
e_{L}
\end{array} \right )\kom L_{e}e_{R}=e_{R}\, .
\end{align}
Since the eigenvalues of $L_e$ are all equal to $1$, it means that $L_e$ counts the number of leptons. Finally, we observe that the combination
\begin{equ}[Electric charge]\index{Electric charge}
\begin{equation}\label{eq:ElectricChargeU1} 
Q=T^{3}+Y 
\end{equation}
\end{equ}
gives
\begin{equation}
Q\left (\begin{array}{c}
\nu_{L} \\ 
e_{L}
\end{array} \right )=\left (\begin{array}{c}
0 \\ 
- e_{L}
\end{array} \right )\kom Qe_{R}=-e_{R}
\end{equation}
where the action of $Q$ on the fields is understood in the associated representation of the fields. Since both left- and right-handed electrons have eigenvalue $-1$ and the neutrino zero eigenvalue under $Q$, it is identified with the electric charge.

Altogether, we found the group
\begin{equation}
G=\mathrm{SU}(2)_{L}\times \mathrm{U}(1)_{Y}\times \mathrm{U}(1)_{L_{e}}\, .
\end{equation}
Since there is no evidence for the existence of a gauge field associated with $\mathrm{U}(1)_{L_{e}}$, we will forget about it for the moment, but come back to it later, cf.~section~\ref{sec:TopDownBSM}.

\section{Glashow-Weinberg-Salam Model}\index{Glashow-Weinberg-Salam model}\index{GSW model}

Let us now start with the core part of this chapter which is the \emph{Weinberg-Salam model}\index{Weinberg-Salam model} with useful earlier work of Glashow, who shared the Nobel prize with them and hence the name \emph{GSW model}. This is the description of the weak interactions by means of a gauge symmetry with SSB. This is probably the most complex component of the Standard Model. We will construct it by following several steps to make it more comprehensible. We promise that the effort to follow all the details pays off by the impressive success of this model which is theoretically sound and experimentally tested with great precision.

Following the previous section, we concentrate on building-up a gauge theory based on the group
\begin{equation}
G_{EW}=\mathrm{SU}(2)_{L}\times \mathrm{U}(1)_{Y}
\end{equation}
with general group elements $U\in G_{EW}$ defined as
\begin{equation}
U=\ee^{\I\alpha^{a}T^{a}}\ee^{\I\beta Y}\, \qquad a=1,2,3\, .
\end{equation}
The gauge fields are defined as 
\begin{itemize}
\item $W_\mu=W_{\mu}^{a}T^a$ for $\mathrm{SU}(2)_{L}$ with field strength
\begin{equation}
W_{\mu\nu}=\p_{\mu}W_{\nu}-\p_{\nu}W_{\mu}-\I g[W_{\mu},W_{\nu}]\, .
\end{equation}
Under infinitesimal gauge transformations,
they behave as
\begin{equation}
\delta W_{\mu}^{a}=\dfrac{1}{g}\p_{\mu}\alpha^{a}-\epsilon^{abc}\alpha^{b}W_{\mu}^{c}\, .
\end{equation}
\item $B_{\mu}$ for $\UO_{Y}$ with field strength
\begin{equation}
B_{\mu\nu}=\p_{\mu}B_{\nu}-\p_{\nu}B_{\mu}
\end{equation}
transforming as
\begin{equation}
\delta B_{\mu}=\dfrac{1}{g^{\prime}}\p_{\mu}\beta\, .
\end{equation}
\end{itemize}
Keep in mind that $g$ and $g^{\prime}$ are two independent gauge couplings associated with either $\SUTw_{L}$ or $\UO_{Y}$. These are free parameters of the theory that eventually have to be determined experimentally.

Now, let us remind ourselves what requirements we want the above gauge theory to satisfy:
\begin{itemize}
\item SSB phase with $3$ massive gauge bosons.
\item Chirality: only left-handed fields feel the weak interactions.
\item Massless gauge field corresponding to the photon.
\item $\ldots$
\end{itemize}

\subsection{Bosonic Lagrangian and SSB}

In order to consider the possibility of spontaneous symmetry breaking, we add to the theory a scalar field transforming non-trivially under the action of $G_{EW}$. We then introduce a complex scalar $H$ as a doublet\footnote{Of course, historically other options were also contemplated such as  $H$ being an $SU(2)$ triplet rather than doublet, with no success.} under $\SUTw_{L}$ and hypercharge conventionally chosen\footnote{The choice of $Y_H=1/2$ is at the moment arbitrary but we will see how well it fits when we discuss the couplings of $H$ to matter fields.}
as  $Y_{H}=1/2$ so that\index{Electroweak Theory!Bosonic Lagrangian}
\begin{equation}
H=\left (\begin{array}{c}
H_{+} \\ 
H_{0}
\end{array} \right )_{Y_{H}=1/2}
\end{equation}
with $H_+$ and $H_0$ complex components of the scalar field $H$\index{Higgs field}. We then start with six degrees of freedom coming from $W_\mu^a$ (the two polarisation degrees of freedom for each value of $a$) and four degrees of freedom from $H$ (a doublet with complex entries).

The purely bosonic part of the renormalisable Lagrangian is then\footnote{We will postpone the discussion of the $\Theta$-term (which is also renormalisable) to section~\ref{sec:ThetaTermDiscussionProbs}.}
\begin{align}
\cL_{B}[W_{\mu}^{a},B_{\mu},H]=-\dfrac{1}{4}\left (W^{a}_{\mu\nu}\right )^{2}-\dfrac{1}{4}B_{\mu\nu}^{2}+D_{\mu}H\, (D^{\mu}H)^{\dagger}-V(H)
\end{align}
with scalar potential
\begin{equation}\label{eq:HiggsScalarPotential} 
V(H)=\lambda\left (H^{\dagger}H-\dfrac{v^{2}}{2}\right )^{2}\kom v^{2}=\dfrac{m^{2}}{\lambda}\, .
\end{equation}
The gauge covariant derivative acting on $H$ is here given by
\begin{equation}
D_{\mu}H=\p_{\mu}H-\I g W_{\mu}^{a}T^{a}H-\dfrac{\I}{2}g^{\prime}B_{\mu}H
\end{equation}
where now the two last  terms appear because of  the product structure of $G$ and the $1/2$ in the last term corresponds to the hypercharge of $H$.

The gauge symmetry is broken for a non-trivial VEV $\langle H\rangle\neq 0$.
For the scalar potential in \eqref{eq:HiggsScalarPotential}, we have extrema at
\begin{equation}
\p_{H}V=2\lambda H^{\dagger}\left (HH^{\dagger}-\dfrac{v^{2}}{2}\right )=0\, .
\end{equation}
For SSB, the terms in bracket have to cancel.
As usual, let us pick one direction
\begin{equation}
\langle H\rangle=\dfrac{1}{\sqrt{2}}\left (\begin{array}{c}
0 \\ 
v
\end{array} \right )
\end{equation}
with $v$ positive and expand around the vacuum as
\begin{equation}\label{eq:HiggsVEVPlusFluctuation} 
H=\dfrac{1}{\sqrt{2}}\ee^{-\I\xi^{a}(x)T^{a}}\left (\begin{array}{c}
0 \\ 
v+h(x)
\end{array} \right )
\end{equation}
with $\xi^{a}(x)$ the 3 fields that will correspond to the Goldstone modes and one Higgs boson $h(x)$\index{Higgs boson}. Plugging this back into the Lagrangian (do it!),
we find, as in the previous chapter, that the $\xi^{a}(x)$ only appear in the combination
\begin{equation}
\p_{\mu}\xi^{a}T^{a}+gW^{a}_{\mu}T^{a}+\frac{1}{2}g^{\prime}B_{\mu}
\end{equation}
which is why we can redefine the gauge fields using gauge transformations such that the Goldstone modes $\xi^{a}$ are being absorbed. This is nothing but fixing the gauge to be the \emph{unitary gauge}\index{Unitary gauge} where the massless fields $\xi^{a}$ give rise to the longitudinal polarisation of the massive spin-$1$ particles. 

After SSB, the bosonic Lagrangian can be separated into two pieces
\begin{equation}
\cL_{B}[W_{\mu}^{a},B_{\mu},H]\xrightarrow{\;\; \text{SSB}\;\;}\cL_{B}^{\text{quadratic}}[W_{\mu}^{a},B_{\mu},h]+\cL^{\text{interaction}}_{B}[W_{\mu}^{a},B_{\mu},h]\, .
\end{equation}
Below, we carefully analyse the different contributions to this Lagrangian.

\subsubsection*{Bosonic Lagrangian -- Quadratic Pieces}

Let us begin with the discussion of the quadratic terms, namely
\begin{align}\label{eq:BosLagEWSQ} 
\cL_{B}^{\text{quadratic}}[W_{\mu}^{a},B_{\mu},h]
&=-\dfrac{1}{4}\left (W^{a}_{\mu\nu}\right )^{2}-\dfrac{1}{4}B_{\mu\nu}^{2}+\dfrac{1}{2}\p^{\mu}h\p_{\mu}h-m^{2}h^{2}\nn\\[0.75em]
&\quad+\dfrac{g^{2}v^{2}}{8}\left [(W_{\mu}^{1})^{2}+(W_{\mu}^{2})^{2}+\left (\dfrac{g^{\prime}}{g}B_{\mu}-W_{\mu}^{3}\right )^{2}\right ]\, .
\end{align}
We diagonalise the mass matrix for the gauge fields by defining\footnote{$Z_\mu$ is usually referred to as $Z^0_\mu$ to specify that it has $0$ electric charge. We will not write explicitly the superscript $0$ for ease of notation.}
\begin{align}
W_{\mu}^{\pm}&=\dfrac{1}{\sqrt{2}}\left (W_{\mu}^{1}\mp \I W_{\mu}^{2}\right )\\[0.75em]
Z_{\mu}&=W_{\mu}^{3}\cos(\theta_{W})-B_{\mu}\sin(\theta_{W})\\[0.75em]
A_{\mu}&=W_{\mu}^{3}\sin(\theta_{W})+B_{\mu}\cos(\theta_{W}) 
\end{align}
in terms of the \emph{Weinberg angle}\index{Weinberg angle} or \emph{weak mixing angle}\index{Weak mixing angle} $\theta_{W}$ defined as
\begin{equation}
\cos(\theta_{W})=\dfrac{g}{\sqrt{g^{2}+(g^{\prime})^{2}}}\kom \sin(\theta_{W})=\dfrac{g^{\prime}}{\sqrt{g^{2}+(g^{\prime})^{2}}}\, .
\end{equation}
The mass spectrum can be identified from \eqref{eq:BosLagEWSQ} to be
\begin{align}
m_{h}&=\sqrt{2\lambda}v&\text{Higgs boson mass, experimentally }m_{h}\approx125.2\text{GeV}\nn\\[0.75em]
m_{W_{\mu}^{\pm}}&=\dfrac{vg}{2}&W\text{-mass, experimentally }m_{W}\approx80.38\text{GeV}\nn\\[0.75em]
m_{Z_{\mu}}&=\dfrac{v}{2}\sqrt{g^{2}+(g^{\prime})^{2}}&Z\text{-mass, experimentally }m_{Z}\approx 91.19\text{GeV}\nn\\[0.75em]
m_{A_{\mu}}&=0&\text{Photon, experimentally }m_{\gamma}<10^{-18}\text{eV}\, .
\end{align}
Notice that
\begin{equation}
m_{W}=m_{Z}\cos(\theta_{W})<m_{Z}
\end{equation}
is a prediction of the theory which is indeed confirmed by experiment.

The first question that comes to mind is why is $A_\mu$ massless? And why does it correspond to the photon? We know the answer to the first question: the symmetry group is not completely broken by $\langle H\rangle$. To see this, we write $U=\mathds{1}+\I\alpha^{a}T^{a}+\I\beta Y+\ldots$ for $U\in G_{EW}$
and consider
\begin{align}
\delta \langle H\rangle=U\langle H\rangle -\langle H\rangle =\dfrac{\I v}{2\sqrt{2}}\left (\begin{array}{c}
\alpha^{1}-\I\alpha^{2} \\ 
\beta-\alpha^{3}
\end{array} \right )\, .
\end{align}
The unbroken symmetry group is then defined as $\delta \langle H\rangle=0$ and so $\alpha^{1}=\alpha^{2}=0,\, \alpha^{3}=\beta$.
That is, $U\in H\subset G_{EW}$ can be written as $U=\ee^{\I\alpha^{a}T^{a}}\ee^{\I\beta Y}=\ee^{\I\beta (T^{3}+Y)}=\ee^{\I\beta Q}\in \mathrm{U}(1)$
in terms of the electric charge \eqref{eq:ElectricChargeU1}.
This then allows us to identify the unbroken gauge group as the electromagnetic $\mathrm{U}(1)_{EM}$. The breaking pattern is given by
\begin{equ}[SSB of the Electroweak theory]
\begin{equation}
\SUTw_{L}\times \UO_{Y}\xrightarrow{\;\langle H\rangle\neq 0\;}\UO_{EM}\, .
\end{equation}
\end{equ}
This also fits the predictions of Goldstone's theorem since
\begin{equation}
\# \text{ Goldstone modes }\xi^{a}=\dim(\SUTw_{L}\times \UO_{Y})-\dim(\UO_{EM})=3\, .
\end{equation}

Next, we need to work out the charges of the physical fields. Thereto, we consider global rotations $U\in\UO_{EM}$ with $U=\ee^{\I\beta Q}\, \sim \mathds{1} + \I\beta Q$.
For the Higgs, we have then $\delta H = \I\beta QH$ implying
\begin{equation}
\delta H=\left (\dfrac{\I\beta}{2}\left (\begin{array}{cc}
1 & 0 \\ 
0 & -1
\end{array} \right )+\dfrac{\I\beta}{2}\left (\begin{array}{cc}
1 & 0 \\ 
0 & 1
\end{array} \right )\right )\left (\begin{array}{c} H_{+} \\ H_{0}\end{array} \right )=\I\beta \left (\begin{array}{c}
H_{+} \\ 
0
\end{array} \right )\, .
\end{equation}
We deduce that $H_{+}$ has electric charge $+1$ and $H_{0}$ charge $0$ justifying their names. For the gauge fields, we obtain
\begin{align}
\delta W_\mu^1 & = \dfrac{\partial_\mu \alpha ^1}{g}-\epsilon^{132}\alpha^3 W_\mu^2=\beta W_\mu^2 \kom \delta W_\mu^2 & = \dfrac{\partial_\mu \alpha^2}{g}-\epsilon^{231}\alpha^3 W_\mu^1=-\beta W_\mu^1
\end{align}
where we used $\alpha^1=\alpha^2=0$ and $\alpha^3=\beta$. This implies
\begin{equation}
\delta W_\mu^\pm=\pm \I \beta W_\mu^\pm
\end{equation}
and therefore
the charges of $W_\mu^\pm$ are $Q W_{\mu}^{\pm}=\pm W_{\mu}^{\pm}$ justifying their definition.
Also, since $QW_{\mu}^{3}=QB_{\mu}=0$, the charge of $Z_{\mu}$ and $A_{\mu}$ is zero.

In order to define experimentally meaningful parameters, we observe that in $D_{\mu}H$ the terms $gW_{\mu}^{3}$ and $g^{\prime}B_{\mu}$ lead to terms of the form $g\sin(\theta_{W})A_{\mu}+\ldots$ and $g^{\prime}\cos(\theta_{W})A_{\mu}+\ldots$. This motivates the definition of the \emph{electromagnetic coupling}\index{Electromagnetic coupling}
\begin{equation}\label{eq:defEMCoupling} 
e=g\sin(\theta_{W})=g^{\prime}\cos(\theta_{W})\, .
\end{equation}
We now treat the original parameters $m,\lambda,g,g^{\prime}$ in the original Lagrangian \eqref{eq:BosLagEWSQ} for $e,\theta_{W},m_{h},m_{W}$ which need to be measured experimentally and with the remaining observables being predictions of the theory. The 4 free parameters are determined experimentally to have the values
\begin{equation}
e=0.303\kom\sin^{2}(\theta_{W})=0.223\kom g=\dfrac{e}{\sin(\theta_{W})}=0.64\kom g^{\prime}=\dfrac{e}{\cos(\theta_{W})}=0.34\, .
\end{equation}
In terms of the fields $A_{\mu}, Z_{\mu}^{0}$ and $W_{\mu}^{\pm}$, the quadratic Lagrangian reads

\begin{Boxequ}

\vspace*{-0.1cm}

\begin{align}\label{eq:QuadBosLagASSB} 
\cL_{B}^{\text{quadratic}}[W_{\mu}^{a},B_{\mu},h]&=-\dfrac{1}{4}F_{\mu\nu}^{2}-\dfrac{1}{4}Z_{\mu\nu}^{2}-\dfrac{1}{2}\left (\p_{\mu}W^{+}_{\nu}-\p_{\nu}W^{+}_{\mu}\right )\left (\p_{\mu}W^{-}_{\nu}-\p_{\nu}W^{-}_{\mu}\right )\nn\\[0.35em]
&\quad+\dfrac{1}{2}m_{Z}^{2}Z^{\mu}Z_{\mu}+m_{W}^{2}W_{\mu}^{+}W^{-\mu}+\dfrac{1}{2}\p_{\mu}h\p^{\mu}h-\frac{1}{2}m_{h}^{2}h^{2}\, .
\end{align}

\vspace*{-0.2cm}

\end{Boxequ}

\noindent This is a Lagrangian for one massless $A_\mu$ and three massive spin $1$ fields $W_\mu^\pm, Z_\mu^0$ and one massive scalar $h$ as a function of the $4$ arbitrary parameters. Note that here $m_Z$ is not a free parameter, but determined by $m_W$ and $\theta_W$ (recall that $m_W=m_Z \cos \theta_W$ ).  The total number of degrees of freedom now is three for each of the massive vector fields $W_\mu^\pm$ and $Z_\mu,$ two for the photon $A_\mu$ and one for the Higgs $h$ adding up to the total of twelve degrees of freedom which matches our counting for the original fields $W_\mu^a, B_\mu$ and $H$. 

It is actually remarkable that this attempt to describe the physics of weak interactions leads not only to a consistent theory for the weak interactions, but also, as a bonus, the theory includes the electromagnetic interactions in a unified way. Both interactions, mediated either by $A_\mu$ giving rise to QED or by $W_\mu^\pm, Z_\mu^0$ giving rise to the weak interactions, come from one and the same underlying theory, a spontaneously broken $\mathrm{SU}(2)_L\times U(1)_Y$ gauge theory. This unification of two interactions in one single theory is an achievement that may be comparable with Newton's unification of terrestrial and celestial gravitational interactions and the unification of electric and magnetic interactions within electromagnetism by Maxwell and Faraday. For this reason this theory is often referred to as the {\bf electroweak theory}.

\subsubsection*{Propagator for a massive vector field}

Since we have found that both the $Z_{\mu}$ and $W_{\mu}^\pm$ bosons are massive, before we consider the interactions coming from the cubic and quartic terms, let us compute explicitly the propagator of a massive vector field that is constructed from the quadratic piece of the Lagrangian. In order to be as general as possible we will compute the propagator for any massive vector field. It would correspond in particular to the propagators for both $Z_{\mu}$ and $W_{\mu}^\pm$  in {\bf unitary gauge}.

Let us start with the Lagrangian density for the massive vector field $X_\mu$
\begin{equation}
\cL=-\frac{1}{4}X^{\mu\nu} X_{\mu\nu}\, +\, \frac{1}{2}m^2 X^\mu X_\mu\, .
\end{equation}
In order to extract the propagator let us manipulate this expression as follows (in which we use integration by parts)
\begin{align}
\cL &= -\frac{1}{4}\left(\partial^\mu X^\nu-\partial^\nu X^\mu\right)\left(\partial_\mu X_\nu-\partial_\nu X_\mu\right) \, +\, \frac{1}{2}m^2 X^\mu X_\mu\nonumber \\
&= \frac{1}{2} X^\mu\left[\left(\partial^\alpha\partial_\alpha + m^2\right)\eta_{\mu\nu} -\partial_\mu\partial_\nu \right] X^\nu\nn\\
&\equiv  \frac{1}{2} X^\mu \cD_{\mu\nu}X^\nu \, .
\end{align}
Therefore, starting from the matrix $\cD_{\mu\nu} $ and going to momentum space, we can read off the corresponding propagator as
\begin{equation}\label{eq:propagator_massive_spin_one} 
\Delta_{\mu\nu}=D^{-1}_{\mu\nu}(p)=-\frac{1}{p^2-m^2}\left(\eta_{\mu\nu}-\frac{p_\mu p_\nu}{m^2}\right)\, .
\end{equation}
Note that the massive case is in some sense simpler than the massless case since in the massless case the matrix $D_{\mu\nu}$ has one zero eigenvalue and extracting the propagator is more difficult as we know from QED. Also remember that for an arbitrary massive vector this is the propagator but if the massive vector comes from a broken gauge symmetry this propagator is only valid in unitary gauge. This will play an important role later on.

\subsubsection*{Bosonic Lagrangian -- Cubic and Quartic Interactions}

Now that we have full control of the quadratic part of the bosonic Lagrangian, we can consider the interactions which can also be separated into two contributions depending on the number of interacting fields
\begin{equation}
\cL^{\text{interaction}}_{B}[W_{\mu}^{a},B_{\mu},h]=\cL^{\text{cubic}}_{B}[W_{\mu}^{a},B_{\mu},h]+\cL^{\text{quartic}}_{B}[W_{\mu}^{a},B_{\mu},h]
\end{equation}
where the cubic interactions are (written in terms of the physical fields $W_\mu^\pm, Z_\mu, A_\mu, h$):
\begin{align}\label{eq:BosLagCubicInt} 
\cL^{\text{cubic}}_{B}&[W_\mu^\pm, Z_\mu, A_\mu, h]\nn\\[0.75em]
&=\I e\cot(\theta_{W})\left [Z^{\mu\nu}W_{\mu}^{+}W_{\nu}^{-}-\left (\p_{\mu}W^{+}_{\nu}-\p_{\nu}W^{+}_{\mu}\right )Z^{\mu}W^{-\nu}+\left (\p_{\mu}W_\nu^{-}-\p_{\nu}W^{-}_{\mu}\right )Z^{\mu}W^{+\nu}\right ]\nn\\[0.75em]
&\quad +\I e\left [F^{\mu\nu}W_{\mu}^{+}W_{\nu}^{-}-\left (\p_{\mu}W^{+}_{\nu}-\p_{\nu}W^{+}_{\mu}\right )A^{\mu}W^{-\nu}+\left (\p_{\mu}W^{-}_{\nu}-\p_{\nu}W^{-}_{\mu}\right )A^{\mu}W^{+\nu}\right ]\nn\\[0.75em]
&\quad -g\dfrac{m_{h}^{2}}{4m_{W}}h^{3}+\dfrac{2h}{v}\left [m_{W}^{2}W_{\mu}^{+}W^{-\mu}+\dfrac{1}{2}m_{Z}^{2}Z_{\mu}^{2}\right ]\, .
\end{align}
The first line encodes the interactions between the massive gauge bosons, the second line the interactions between the massless photon with the charged, massive bosons $W_{\mu}^{\pm}$ and the third row the interactions involving the Higgs.

The quartic interactions are given by
\begin{align}\label{eq:BosLagQuarticInt} 
\cL^{\text{quartic}}_{B}&[W_{\mu}^{a},B_{\mu},h]=\dfrac{1}{2}\dfrac{e^{2}}{\sin^{2}(\theta_{W})}\left [W^{+\mu}W^{+}_{\mu}W^{-\nu}W^{-}_{\nu}-W^{+\mu}W^{-}_{\mu}W^{+\nu}W^{-}_{\nu}\right ]\nn\\[0.75em]
&+e^{2}\left [A^{\mu}W_{\mu}^{+}A^{\nu}W_{\nu}^{-}-A_{\mu}^{2}W^{+\nu}W^{-}_{\nu}\right ]+e^{2}\cot^{2}(\theta_{W})\left [Z^{\mu}W_{\mu}^{+}Z^{\nu}W_{\nu}^{-}-Z_{\mu}^{2}W^{+\nu}W^{-}_{\nu}\right ]\nn\\[0.75em]
&+e^{2}\cot(\theta_{W})\left [A^{\mu}Z^{\nu}W^{+}_{\mu}W_{\nu}^{-}+A^{\mu}Z^{\nu}W^{-}_{\mu}W_{\nu}^{+}-2W^{+\mu}W_{\mu}^{-}A^{\nu}Z_{\nu}\right ]\nn\\[0.75em]
&-\dfrac{g^{2}}{32}\dfrac{m^{2}_{h}}{m^{2}_{W}}\, h^{4}+\left (\dfrac{h}{v}\right )^{2}\left [m_{W}^{2}W^{+\mu}W^{-}_{\mu}+\dfrac{1}{2}m_{Z}^{2}Z_{\mu}^2\right ]\, .
\end{align}
As before, the first row encodes quartic interactions among the $W_{\mu}^{\pm}$, the second interactions of the $W_{\mu}^{\pm}$ with one of the gauge fields $A_{\mu}$, $Z_{\mu}$, the third interactions involving all gauge fields and the last row interactions between the Higgs itself as well as $W_{\mu}^{\pm}$, $Z_{\mu}$.

Notice that despite the length of the Lagrangians and the presence of many couplings between the individual fields, there are \textbf{only four arbitrary parameters}, i.e., $e,\theta_{W},m_{h},m_{W}$ which is why many of the predictions of this theory can be tested.
This is completely general observation: in an arbitrary theory involving e.g. a scalar field $h$, its cubic and quartic interactions need not be related to each other, while in the case of SSB (as in \eqref{eq:BosLagCubicInt} and \eqref{eq:BosLagQuarticInt}) one observes that their ratio
\begin{equation}
\dfrac{\lambda_{3}^{2}}{\lambda_{4}}=2m_{h}^{2}
\end{equation}
is proportional to the mass of $h$.

Let us work out some of the Feynman rules, cf. e.g. appendix D in \cite{Burgess:2007zi} for a full list,
\begin{itemize}\index{Electroweak Theory!Feynman rules}
\item Cubic interactions $\sim W_{\mu}^{+}W_{\nu}^{-}Z_{\lambda}$ originating from the first row in \eqref{eq:BosLagCubicInt}
\begin{align*}
\begin{tikzpicture}[scale=1.2]
\setlength{\feynhanddotsize}{1.5ex}
\begin{feynhand}
\node (o) at (5,-4) {\small$=-\I e\cot(\theta_{W})\left [\eta^{\mu\nu}(p_{1}-p_{2})^{\lambda}+\eta^{\nu\lambda}(p_{2}-p_{3})^{\mu}+\eta^{\lambda\mu}(p_{3}-p_{1})^{\nu}\right ]\, .$};
\vertex (a2) at (0.25,-4) {$Z_{\lambda}$}; 
\vertex (b2) at (-1.25,-4); 
\vertex (c2) at (-2.5,-3) {$W_{\mu}^{+}$}; 
\vertex (d2) at (-2.5,-5) {$W_{\nu}^{-}$}; 
\propag [pho, mom={$p_{3}$}] (a2) to (b2);
\propag [pho, mom={$p_{1}$}] (d2) to (b2);
\propag [pho, mom={$p_{2}$}] (c2) to (b2);
\end{feynhand}
\end{tikzpicture}
\end{align*}
\item Quartic interactions among the heavy gauge fields $\sim W_{\mu}^{+}W_{\nu}^{-}Z_{\alpha}Z_{\beta}$ as obtained from the second term in the second row of \eqref{eq:BosLagQuarticInt}
\begin{align*}
\begin{tikzpicture}[scale=1.2]
\setlength{\feynhanddotsize}{1.5ex}
\begin{feynhand}
\node (o) at (8.5,-4) {\small$=\I e^{2}\cot^{2}(\theta_{W})\left [\eta^{\alpha\mu}\eta^{\beta\nu}+\eta^{\alpha\nu}\eta^{\beta\mu}-2\eta^{\alpha\beta}\eta^{\mu\nu}\right ]\, .$};
\vertex (a31) at (4.5,-3) {$W_{\mu}^{+}$}; 
\vertex (a32) at (4.5,-5) {$W_{\nu}^{-}$}; 
\vertex (b3) at (3.25,-4); 
\vertex (c3) at (2.,-3) {$Z_{\beta}$}; 
\vertex (d3) at (2.,-5) {$Z_{\alpha}$}; 
\propag [pho] (a31) to (b3);
\propag [pho] (a32) to (b3);
\propag [pho] (d3) to (b3);
\propag [pho] (c3) to (b3);
\end{feynhand}
\end{tikzpicture}
\end{align*}
\item Cubic interactions involving the massive gauge bosons and the Higgs implied by the second term in the third row of \eqref{eq:BosLagCubicInt}. We have to distinguish between the interactions $\sim hW_{\mu}^{+}W_{\nu}^{-}$
\begin{align*}
\begin{tikzpicture}[scale=1.2]
\setlength{\feynhanddotsize}{1.5ex}
\begin{feynhand}
\node (o) at (2.25,-4) {\small$=\I\dfrac{e}{\sin(\theta_{W})}m_{W}\eta^{\mu\nu}$};
\vertex (a2) at (0.25,-4) {$h$}; 
\vertex (b2) at (-1.25,-4); 
\vertex (c2) at (-2.5,-3) {$W_{\mu}^{+}$}; 
\vertex (d2) at (-2.5,-5) {$W_{\nu}^{-}$}; 
\propag [chasca] (b2) to (a2);
\propag [pho] (d2) to (b2);
\propag [pho] (c2) to (b2);
\end{feynhand}
\end{tikzpicture}
\end{align*}
and $\sim hZ_{\mu}Z_{\nu}$
\begin{align*}
\begin{tikzpicture}[scale=1.2]
\setlength{\feynhanddotsize}{1.5ex}
\begin{feynhand}
\node (o) at (4.25,-4) {\small$=\I\dfrac{e}{\sin(\theta_{W})}\dfrac{m_{Z}^{2}}{m_{W}}\eta^{\mu\nu}=\I\dfrac{e}{\sin(\theta_{W})\cos^{2}(\theta_{W})}m_{W}\eta^{\mu\nu}\, .$};
\vertex (a2) at (0.25,-4) {$h$}; 
\vertex (b2) at (-1.25,-4); 
\vertex (c2) at (-2.5,-3) {$Z_{\mu}$}; 
\vertex (d2) at (-2.5,-5) {$Z_{\nu}$}; 
\propag [chasca] (b2) to (a2);
\propag [pho] (d2) to (b2);
\propag [pho] (c2) to (b2);
\end{feynhand}
\end{tikzpicture}
\end{align*}
\end{itemize}

\newpage

\subsection{$WZ\rightarrow WZ$ Scattering and Perturbative Unitarity}\label{sec:pertuniWZ}
\index{Electroweak Theory!WZ to WZ scattering}\index{Electroweak Theory!Perturbative Unitarity}

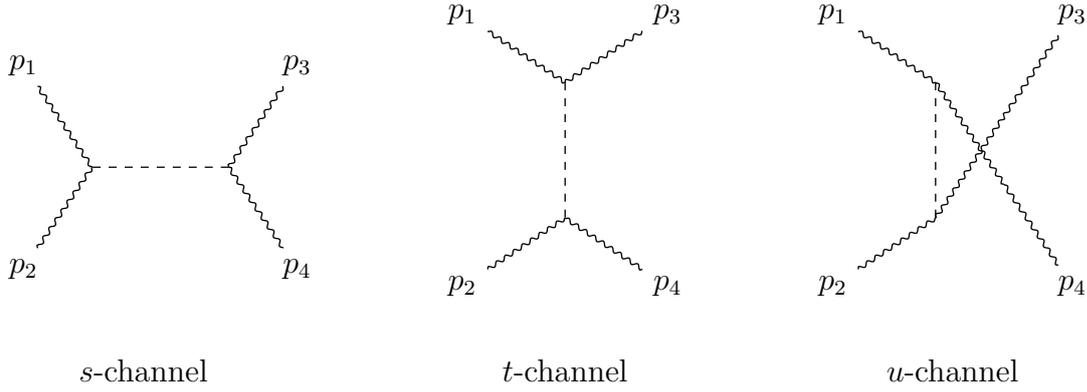
\begin{figure}[t]
\begin{center}
\begin{tikzpicture}[scale=0.9]
\setlength{\feynhanddotsize}{1.5ex}
\begin{feynhand}
\vertex (a0) at (4,-1.5) {$p_{4}$}; 
\vertex (a1) at (4,1.5) {$p_{3}$}; 
\vertex (b0) at (3,0); 
\vertex (b1) at (1,0); 
\vertex (c0) at (0,-1.5) {$p_{2}$}; 
\vertex (c3) at (0,-2.) {}; 
\vertex (c4) at (1.75,-3.) {$s$-channel}; 
\vertex (c1) at (0.,1.5) {$p_{1}$}; 
\propag [bos] (a0) to (b0);
\propag [bos] (a1) to (b0);
\propag [scalar] (b0) to (b1);
\propag [bos] (b1) to (c0);
\propag [bos] (b1) to (c1);
\end{feynhand}
\end{tikzpicture}
\hspace*{1.25cm}
\begin{tikzpicture}[scale=0.9]
\setlength{\feynhanddotsize}{1.5ex}
\begin{feynhand}
\vertex (a0) at (-1.5,4) {$p_{1}$}; 
\vertex (a1) at (1.5,4) {$p_{3}$}; 
\vertex (b0) at (0,3); 
\vertex (b1) at (0,1); 
\vertex (c0) at (-1.5,0) {$p_{2}$}; 
\vertex (c1) at (1.5,0.) {$p_{4}$}; 
\vertex (c4) at (0,-1.25) {$t$-channel}; 
\propag [bos] (a0) to (b0);
\propag [bos] (a1) to (b0);
\propag [scalar] (b0) to (b1);
\propag [bos] (b1) to (c0);
\propag [bos] (b1) to (c1);
\end{feynhand}
\end{tikzpicture}
\hspace*{1.25cm}
\begin{tikzpicture}[scale=0.9]
\setlength{\feynhanddotsize}{1.5ex}
\begin{feynhand} 
\vertex (a0) at (-1.5,4) {$p_{1}$}; 
\vertex (a1) at (2.,4) {$p_{3}$}; 
\vertex (a11) at (1.,2.5); 
\vertex (b0) at (0,3); 
\vertex (b1) at (0,1); 
\vertex (b00) at (-0.25,2.5); 
\vertex (b11) at (-0.25,1.5); 
\vertex (c0) at (-1.5,0) {$p_{2}$}; 
\vertex (c1) at (2.,0.) {$p_{4}$}; 
\vertex (c4) at (0.25,-1.25) {$u$-channel}; 
\vertex (c11) at (1.,1.5); 
\propag [sca] (b0) to (b1);
\propag [bos] (a0) to (b0);
\propag [bos] (c11) to (c1);
\propag [bos] (b0) to (c11);
\propag [bos] (b1) to (c0);
\propag [bos] (a1) to (a11);
\propag [bos] (b1) to (a11);
\end{feynhand}
\end{tikzpicture}
\end{center}
\vspace*{-0.5cm}
\caption{\label{fig:mandelstam}Mandelstam variables and the corresponding $s,t,u$ channels for any 2-2 scattering amplitude.
}\label{fig:Mandelstam4PointScattering} 
\end{figure}

Before we get to fermions, let us consider the amplitude for scattering longitudinally polarised gauge bosons $W^{\pm}$ and $Z$.
First, we introduce \emph{Mandelstam variables}.\index{Mandelstam variables}
Recall that for any 2-2 scattering with external momenta $p_1,p_2, p_3, p_4$ it is convenient to work with the Mandelstam variables $s,t,u$
\begin{align}
s&=(p_1+p_2)^2=(p_3+p_4)^2\, ,\nn\\
t&=(p_1-p_3)^2=(p_2-p_4)^2\, ,\nn\\
u&=(p_1-p_4)^2=(p_2-p_3)^2
\end{align}
satisfying the useful identity
\begin{equation}
s+t+u=\sum_{k=1}^4 m_k^2\, .
\end{equation}
The corresponding Feynman diagrams, as depicted in Fig.~\ref{fig:Mandelstam4PointScattering}, follow the $s,t,u$ channels respectively.

Let us now consider all the channels that contribute to the $W_LZ_L\raw W_LZ_L$ scattering where here the sub-index $L$ stands for longitudinal. More explicitly,
we compute the amplitude
\begin{equation}\label{eq:ScatteringLongPolGB} 
\cM (W_{L}Z_{L}\raw W_{L}Z_{L}) = \cM_{s}+\cM_{t}+\cM_{u}+\ldots
\end{equation}
channel by channel.\footnote{Note that the relevant interaction vertex is the $W^+ W^- Z$ coupling and the internal line can only be  a $W^\pm$. So in particular there is no $t$-channel contribution to the amplitude.}
We will narrow down the missing pieces contributing to $\ldots$ in \eqref{eq:ScatteringLongPolGB} below.

Since the amplitudes depend explicitly on the polarisation vectors, let us try to choose a suitable basis that captures the fact that we are interested only on the longitudinal modes. Recall that for a massive particle we can pick a frame such that $p^\mu=(E,0,0,p_z)$  ($E^2-p_z^2=m^2$) and so the polarisation vectors are
\begin{equation}
\epsilon_{T1}^\mu=(0,1,0,0), \qquad \epsilon_{T2}^\mu= (0,0,1,0), \qquad \epsilon_L^\mu=\left(\frac{E}{m},0,0,\frac{p_z}{m}\right).
\end{equation} 
where the subscripts $T$ and $L$ stand for transverse and longitudinal respectively.  In the limit $E\gg m$ the longitudinal vector is approximately $\epsilon_L^\mu\sim \frac{E}{m}(1,0,0,1)$. But since this gives $\epsilon_L\cdot p\neq 0$ we have to have an expression that is valid beyond leading order in a $m/E$ expansion. A convenient set of approximate (unnormalised) longitudinal polarisation vectors for each particle satisfying $\epsilon_k\cdot p_k=0$ can be constructed as
\begin{align}\label{eq:PolarisationWZWZ} 
\epsilon_1^\mu &= \frac{1}{m_W}\left(p_1^\mu+\frac{2m_W^2}{t-2m_W^2}p_3^\mu\right)\kom  \epsilon_2^\mu = \frac{1}{m_Z}\left(p_2^\mu+\frac{2m_Z^2}{t-2m_Z^2}p_4^\mu\right)\nonumber \\
\epsilon_3^\mu &= \frac{1}{m_W}\left(p_3^\mu+\frac{2m_W^2}{t-2m_W^2}p_1^\mu\right)\kom  \epsilon_4^\mu = \frac{1}{m_Z}\left(p_4^\mu+\frac{2m_Z^2}{t-2m_Z^2}p_2^\mu\right)
\end{align}
where $t$ is the Mandelstam variable defined above. Now we can compute the contribution of each channel to the amplitude \eqref{eq:ScatteringLongPolGB}:

\begin{figure}[t!]
\centering
\begin{tikzpicture}[scale=1.1]
\setlength{\feynhanddotsize}{1.5ex}
\begin{feynhand}
\vertex (a5) at (-2.5,2) {$\cM_{s}=$}; 
\vertex (a0) at (-1.5,3) {$W$}; 
\vertex (a1) at (2,2); 
\vertex (d0) at (3.5,3) {$W$}; 
\vertex (d1) at (3.5,1) {$Z$}; 
\vertex (b0) at (0,2); 
\vertex (c0) at (-1.5,1) {$Z$}; 
\vertex (c5) at (0,0) {}; 
\propag [chabos, mom={$p_{1}$}] (a0) to (b0);
\propag [chabos, mom={$W$}] (b0) to (a1);
\propag [chabos, mom={$p_{3}$}] (a1) to (d0);
\propag [pho, mom={$p_{4}$}] (a1) to (d1);
\propag [pho, mom={$p_{2}$}] (c0) to (b0);
\end{feynhand}
\end{tikzpicture}
\hspace*{1.25cm}
\begin{tikzpicture}[scale=1.1]
\setlength{\feynhanddotsize}{1.5ex}
\begin{feynhand}
\vertex (a5) at (-2.65,2) {$\cM_{u}=$}; 
\vertex (a0) at (-1.5,4) {$W$}; 
\vertex (a1) at (2.,4) {$W$}; 
\vertex (a11) at (1.,2.5); 
\vertex (b0) at (0,3); 
\vertex (b1) at (0,1); 
\vertex (c0) at (-1.5,0) {$Z$}; 
\vertex (c1) at (2.,0.) {$Z$}; 
\vertex (c11) at (1.,1.5); 
\propag [chabos] (b0) to (b1);
\propag [chabos, mom={$p_{1}$}] (a0) to (b0);
\propag [pho, mom={$p_{4}$}] (c11) to (c1);
\propag [pho] (b0) to (c11);
\propag [pho, mom={$p_{2}$}] (c0) to (b1);
\propag [chabos, top, mom={$p_{3}$}] (a1) to (a11);
\propag [pho, top] (b1) to (a11);
\end{feynhand}
\end{tikzpicture}
\caption{\emph{Left:} $s$-channel contribution $\cM_{s}$ to $WZ\raw WZ$. \emph{Right:} $u$-channel contribution $\cM_{u}$ to $WZ\raw WZ$.
}\label{fig:WZWZContributionsDiags}
\end{figure}

\begin{itemize}
\item{\it s-channel}:
Looking at the diagram on the left in Fig.~\ref{fig:WZWZContributionsDiags},
we can use the Feynman rules derived above as well as the expressions for the polarisation tensors in \eqref{eq:PolarisationWZWZ} to write down the expression for the $s$-channel contribution, namely
\begin{align}
\cM_{s}&=({\I}e\cot\theta_W)^2\epsilon^\mu_1\epsilon^\nu_2 \epsilon^\alpha_3 \epsilon^\beta_4 \frac{1}{s-m_W^2}\left(-\eta_{\lambda\kappa}+\frac{(p_1+p_2)_\lambda (p_1+p_2)_\kappa}{m_W^2}\right)\nonumber \\[0.35em]
&\quad  \times \left[\eta_{\mu\nu}(p_1-p_2)_\lambda+\eta_{\nu\lambda}(p_1+2p_2)_\mu-\eta_{\lambda\mu}(2p_1+p_2)_\nu\right]\nonumber \\[0.35em]
&\quad \times\left[\eta_{\alpha\beta}(p_3-p_4)_\kappa+\eta_{\beta\kappa}(p_3+2p_4)_\alpha-\eta_{\kappa\alpha}(2p_3+p_4)_\beta\right]\nonumber \\[0.5em]
&= \frac{e^2\cot^2\theta_W}{4m_W^2m_Z^2}\biggl [2su+s^2-2m_W^2\frac{3su+u^2}{s+u}+2m_Z^2\frac{s^2-3su-2u^2}{s+u}\nn\\
&\hphantom{= \frac{e^2\cot^2\theta_W}{4m_W^2m_Z^2}\biggl [} -\frac{m_Z^4}{m_W^2}s+\cO(1)\biggl ]\, .
\end{align}
Specifically, we used the concrete expression for the $W^+ W^- Z$ vertex and also the propagator of the massive vector as computed above, cf.~\eqref{eq:propagator_massive_spin_one}. Note that we kept only the leading order terms in an $m/E$ expansion. Since both $s,u\sim E^2$ we can see that the first two terms diverge like $\cO(E^4/m^4)$ where $m$ stands for either $m_W,m_Z$, the next three terms are $\cO(E^2/m^2)$ and we have not computed the finite terms which are $\cO(1)$ and below. This is because we are interested in the potentially divergent contributions to the amplitude that would imply the break down of perturbative unitarity.

\item{\it t-channel}:
This contribution vanishes since it requires a cubic vertex for the $Z_\mu$ boson that does not exist,
\begin{equation}
\cM_{t}=0\, .
\end{equation}

\item{\it u-channel}:
The $u$-channel contribution is the same as the $s$-channel by just changing $s\leftrightarrow u$ in the expression for the $s$-channel,
\begin{equation}
\cM_{u}(s,u)=\cM_{s}(u,s)\, .
\end{equation}
It will then have also quartic and quadratic divergences in $E/m$.
\end{itemize}
Notice that the combination of both processes above gives
\begin{align}
\cM_s+\cM_u&=\frac{e^2\cot^2\theta_W}{4m_W^2m_Z^2}\biggl [4su+s^2+u^2-2\left(m_W^2+m_Z^2\right)\frac{6su+s^2+u^2}{s+u}\nn\\
&\hphantom{=\frac{e^2\cot^2\theta_W}{4m_W^2m_Z^2}\biggl [}-\frac{m_Z^4}{m_W^2}(s+u)+\cO(1)\biggl ]\, .
\end{align}
If these were the only contributions, the amplitude \eqref{eq:ScatteringLongPolGB} would diverge in the limit $s,u\raw\infty$ 
\begin{equation}
\cM (W_{L}Z_{L}\raw W_{L}Z_{L})\xrightarrow{s,u\raw \infty}\infty\, .
\end{equation}

\begin{figure}[t!]
\centering
\begin{tikzpicture}[scale=1.2]
\setlength{\feynhanddotsize}{1.5ex}
\begin{feynhand}
\node (o) at (-2.5,1.5) {$\cM_{4}=$} ; 
\vertex (a0) at (-1.5,3) {$W$}; 
\vertex (a1) at (1.5,3) {$W$}; 
\vertex (b0) at (0,1.5); 
\vertex (c0) at (-1.5,0) {$Z$}; 
\vertex (c1) at (1.5,0.) {$Z$}; 
\vertex (c5) at (1.5,-0.5) {}; 
\propag [chabos, mom={$p_{1}$}] (a0) to (b0);
\propag [chabos, mom={$p_{3}$}] (b0) to (a1);
\propag [pho, mom={$p_{2}$}] (c0) to (b0);
\propag [pho, mom={$p_{4}$}] (b0) to (c1);
\end{feynhand}
\end{tikzpicture}
\hspace*{1.1cm}
\begin{tikzpicture}[scale=1.1]
\setlength{\feynhanddotsize}{1.5ex}
\begin{feynhand}
\node (o) at (-2.5,2) {$\cM_{h}=$} ; 
\vertex (a0) at (-1.5,4) {$W$}; 
\vertex (a1) at (1.5,4) {$W$}; 
\vertex (b0) at (0,3); 
\vertex (b1) at (0,1); 
\vertex (c0) at (-1.5,0) {$Z$}; 
\vertex (c1) at (1.5,0.) {$Z$}; 
\propag [chabos, mom={$p_{1}$}] (a0) to (b0);
\propag [chabos, mom={$p_{3}$}] (b0) to (a1);
\propag [sca, mom={$h$}] (b0) to (b1);
\propag [pho, mom={$p_{2}$}] (b1) to (c0);
\propag [pho, mom={$p_{4}$}] (b1) to (c1);
\end{feynhand}
\end{tikzpicture}
\caption{\emph{Left:} $4$-point vertex contribution $\cM_{4}$ to $WZ\raw WZ$. \emph{Right:} Higgs contribution $\cM_{h}$ to $WZ\raw WZ$.
}\label{fig:WZWZTwoAddDiags} 
\end{figure}

However, we are clearly not done yet.
Remembering \eqref{eq:BosLagQuarticInt}, there are two additional diagrams that contribute to $WZ\raw WZ$, see Fig.~\ref{fig:WZWZTwoAddDiags}.
The contribution on the left stems from the $4$-point vertex $\sim W^{2}Z^{2}$ in \eqref{eq:BosLagQuarticInt} which can be computed as
\begin{align}\label{eq:WZWZ4PointVertexContr} 
\cM_4 &=e^2\cot^2\theta_W\epsilon^\mu_1\epsilon^\nu_2 \epsilon^\alpha_3 \epsilon^\beta_4 \left(\eta_{\mu\nu}\eta_{\alpha\beta}+\eta_{\mu\beta}\eta_{\nu\alpha}-2\eta_{\mu\alpha}\eta_{\nu\beta}\right)\nonumber\\[0.5em]
&=\frac{e^2\cot^2\theta_W}{4m_W^2m_Z^2}\left[-s^2-4su-u^2+2(m_W^2+m_Z^2)\, \frac{s^2+6su+u^2}{s+u}+\cO(1)\right]
\end{align}
which  has quartic and quadratic divergences (quartic from the first three terms (recall, $s,t,u$ scale like ${\mathcal O}(E^2)$) and quadratic from the next term). 

However, we notice something remarkable: if we compute the combination $\cM_s+\cM_u+\cM_4$, the most dangerous quartic divergences in \eqref{eq:WZWZ4PointVertexContr} precisely cancel those of the $s$- and $u$-channels.
While non-trivial at first, it is an important consequence of SSB in the presence of gauge fields.
As we mentioned several times before,
there are remain traces of the original gauge invariance even after SSB which manifest themselves through relations among the quartic and cubic vertices of the theory.
This is an important test for our theory since amplitudes for generic massive vectors would have quartic divergences because there are no such relationships between the various quartic and cubic vertices.

Notice that, even though the quartic divergences cancel, the quadratic divergences are only partially cancelled and remain of order
\begin{equation}
\cM_s+\cM_u+\cM_4= \frac{t}{v^2}=\cO(E^2)
\end{equation}
where we have used the relation of the Higgs VEV $v$ in terms of $m_W,m_z,\theta_W$.
It might not really come as a surprise that
this last dangerous piece is cancelled by an additional contribution coming from interactions involving the Higgs.
Specifically,
we have a Higgs channel diagram shown on the right in Fig.~\ref{fig:WZWZTwoAddDiags} which amounts to
\begin{align}
\cM_h &= -\frac{e^2}{\sin^2\theta_w\cos^2\theta_W}\, \epsilon^\mu_1\epsilon^\nu_2 \epsilon^\alpha_3 \epsilon^\beta_4 \left(\eta_{\alpha\mu}\eta_{\beta\nu}\right)\, \frac{m_W^2}{t-m_h^2}\nonumber \\[0.35em]
& =  -\frac{e^2}{4m_Z^2\sin^2\theta_w\cos^2\theta_W}\,\frac{t^2(t-4m_W^2)(t-4m_Z^2)}{(t-m_h^2)(t-2m_W^2)(t-2m_Z^2)}\nn\\[0.35em]
&=-\frac{t}{v^2}
\end{align}
This contribution clearly diverges as $\cO(E^2/m^2)$, but it exactly cancels the quadratic divergence of the sum of the other amplitudes!

To summarise,
putting all the pieces together, we find that
\begin{Boxequ}
\begin{equation}
\cM(WZ\raw WZ)=\cM_s+\cM_u+\cM_4+\cM_h=\cO(1)\, .
\end{equation}
\end{Boxequ}
Therefore, the total amplitude is \textbf{finite} and perturbative unitarity is recovered from the Higgs.
Let us pause here for a moment to appreciate how important this result is. Recall that when we discussed massive spin-1 fields we pointed out the perturbative unitarity problem of increasing probability amplitudes with energy, that at high energies would give rise to probabilities bigger than one breaking unitarity, cf.~section~\ref{sec:lossunitarityssb}. Here we see explicitly how SSB theories solve this problem by having the Higgs field recovering unitarity. Before the Higgs discovery, this was the main argument to justify LHC and its energy range since, in order to recover unitarity the Higgs mass should be smaller than $1$ TeV. This was referred to as the \emph{no-loose argument}\index{No-loose argument} when arguments in favour of the construction of LHC were given. The discovery of the Higgs at the $125$ GeV spectacularly confirmed theoretical expectations.

\begin{equ}[Higgs particle and perturbative unitarity]
The dangerous potential loss of unitarity in the scattering of massive vector fields is precisely cancelled by their interaction with the Higgs field. This illustrates that the Higgs mechanism provides a UV completion for an effective theory of massive vectors.
\end{equ}

This perturbative unitarity problem is present for a generic massive vector field, despite the naive appearance that a theory for a massive vector seems renormalisable by simple power counting argument (canonical kinetic term plus a mass term and a quartic term).
In fact, the theory is non-renormalisable in the following sense: for high energies or momenta $p$,  as we have seen, the propagator of a generic massive vector field goes like 
$\Delta \sim \cO(1/m^2)$ for $p\gg m,$ which is very different from the massless case in which $\Delta \sim \cO(1/p^2)$.
Even for massive fermions $\Delta \sim \cO(1/p)$ and massive scalars $\Delta \sim \cO(1/p^2)$,
the propagator decreases with increasing $p$. Given a particular non-divergent loop diagram and adding an internal line of the vector field to add an extra loop,
the level of the divergence may increase because of the extra integral corresponding to the new loop without any falloff at high energies.
In contrast, if the propagator vanishes for large momenta $p$ such as for massive fermions or scalars, it compensates the level of the divergence and keeps the theory safe of divergences. But if the propagator does not vanish with increasing $p$, as for the case of the massive vectors, then the divergence is not cancelled. This will generically make the theory behave as a non-renormalisable theory in which an infinite number of terms would be required to renormalise the theory.\footnote{The very particular case of the Proca theory corresponding to a Lagrangian as in QED plus a mass term  for the gauge field avoids this problem since there is a conserved current and the dangerous term in the propagator cancels when contracted with the current.}

A typical example is that of a box diagram with four external legs. If the four internal lines correspond to fermions, the loop will have an integral over momenta with the four propagators contributing an $\cO(p^{-4})$ in total that combine with the line element $p^3 dp$ to contribute a harmless logarithmic behaviour. This situation remains the same if we add a new internal line corresponding to a massless vector, like the photon, since the two-loop diagram now will have two integrals over momenta with six internal fermion lines $\cO(p^{-6})$ that combine with the $1/p^2$ behaviour of the photon propagator to keep the integral logarithmic. But if instead of a photon we add a massive vector internal line, the contribution of the propagator would be $1/m^2$ instead of $1/p^2$ and the integral would become quadratically divergent.

Massive vectors coming from spontaneously broken gauge theories avoid this problem since in that case the $\Delta \sim \cO(1/m^2)$ behaviour  is just an artefact of the unitary gauge. Going to other gauges the behaviour $\Delta \sim \cO(1/p^2)$ is recovered and the divergences disappear. This explains why in the unitary gauge that we used above we needed several cancellations to obtain a finite result but these are not actually miraculous cancellations, but a reflection of the underlying gauge symmetry. In summary, the only renormalisable theories of massive vector fields are those derived from gauge theories with spontaneous symmetry breaking. They include not only the massive vectors, but also a massive scalar, the Higgs, that can be seen as providing the UV completion of theories with only massive vectors.

A final comment: as stressed many times before, the Goldstone modes provide the longitudinal degrees of freedom of the massive vector field. Therefore, computing scattering of longitudinal gauge bosons is therefore the same as computing the scattering of Goldstone bosons (using a gauge different from the unitary gauge).  It is often useful to make directly the calculation in terms of the Goldstone bosons since calculations with scalars are usually easier than calculations with vectors.
This is known as the \emph{Goldstone equivalence theorem}\index{Goldstone equivalence theorem}. A complete discussion of this is given in \cite{Peskin:1995ev}.

\subsection{Lagrangian for boson and fermion couplings}\label{sec:fermioncouplingsEW} 

So far, we have focussed exclusively on the bosonic fields in the GSW model.
But we know that matter is formed by fermions like the electron and up-/down-type quarks.
In this section, we collect all of the couplings involving fermions and their corresponding interactions with the Higgs before discussing their properties after SSB.

We denote the fermionic fields as follows:
\begin{itemize}
\item Left-handed leptons transforming as doublets under $\mathrm{SU}(2)_L$:
\begin{equation}
L_{L}^{i}=\biggl \{\left (\begin{array}{c}
\nu_{e,L} \\ 
e_{L}
\end{array} \right )\, ,\, \left (\begin{array}{c}
\nu_{\mu,L} \\ 
\mu_{L}
\end{array} \right )\, ,\, \left (\begin{array}{c}
\nu_{\tau,L} \\ 
\tau_{L}
\end{array} \right )\biggl \}_{Y_{L}=-1/2}
\end{equation}
with $\nu_{e,L},\nu_{\mu,L},\nu_{\tau,L}$  and $e_{L},\mu_{L},\tau_{L}$ left-handed {\bf Weyl spinors} written as 4-component Dirac spinors with only the top two components non-vanishing as we saw in chapter \ref{chap:stsym}. The index $i$ labels the \emph{families} or \emph{generations}\index{Generations}.
\item Right-handed leptons\footnote{The right-handed neutrinos $\nu_R$ were usually not included in the Standard Model spectrum since neutrinos were thought to be massless, but there is now a compelling evidence for neutrinos to have a mass and the natural objects to consider are the right-handed neutrinos that can pair with the left-handed to have a mass term in the Lagrangian as we will see later.} transforming as singlets under $\mathrm{SU}(2)_L$
\begin{equation}
e_{R}^{i}=\lbrace e_{R},\mu_{R},\tau_{R}\rbrace_{Y_{e}=1}\kom \left( \nu_{R}^{i}=\lbrace \nu_{e,R},\nu_{\mu,R},\nu_{\tau,R}\rbrace_{Y_{\nu}=0} \right)\, .
\end{equation}
\item Left-handed quarks\footnote{As mentioned in the introduction, quarks transform also as triplets of the strong interactions gauge group $\mathrm{SU}(3)_c$ to be introduced properly in the next chapter.} transforming as doublets under $\mathrm{SU}(2)_L$
\begin{equation}
Q_{L}^{i}=\biggl \{\left (\begin{array}{c}
u_{L} \\ 
d_{L}
\end{array} \right )\, ,\, \left (\begin{array}{c}
c_{L} \\ 
s_{L}
\end{array} \right )\, ,\, \left (\begin{array}{c}
t_{L} \\ 
b_{L}
\end{array} \right )\biggl \}_{Y_{Q}=1/6}\, .
\end{equation}
The index $i$ labels again the three families or generations. There are $6$ quark flavours $u,d,c,s,t,b$ (two flavours per family).
\item Right-handed quarks transforming as singlets under $\mathrm{SU}(2)_L$
\begin{equation}
u_{R}^{i}=\lbrace u_{R},c_{R},t_{R}\rbrace_{Y_{u}=2/3}\kom d_{R}^{i}=\lbrace d_{R},s_{R},b_{R}\rbrace_{Y_{d}=-1/3}\, .
\end{equation}
\end{itemize}
The assignments for the hypercharges are not arbitrary, in the absence of right-handed neutrinos, they are essentially uniquely specified by anomaly cancellation.
We will show this in great detail in section~\ref{sec:anomalies_SM}.

The weak interactions for the fermions can be split into two terms
\begin{equation}\label{eq:CouplingsFermions} 
\cL_{F}=\cL_{F}^{\text{kin}}+\cL_{F}^{\text{Yukawa}}
\end{equation}
where the kinetic terms for the fermions are\index{Electroweak Theory!Fermionic Lagrangian}
\begin{align}\label{eq:KinTermsFermionsCov} 
\cL_{F}^{\text{kin}}[D]&=\I \overline{L}_{L}^{i}\cancel{D}L_{L}^{i}+\I \overline{Q}_{L}^{i}\cancel{D}Q_{L}^{i}+\I \overline{e}_{R}^{i}\cancel{D}e_{R}^{i} +\I \overline{\nu}_{R}^{i}\cancel{D}\nu_{R}^{i}+\I \overline{u}_{R}^{i}\cancel{D}u_{R}^{i}+\I \overline{d}_{R}^{i}\cancel{D}d_{R}^{i}\, .
\end{align}
The covariant derivative for our gauge group $G_{EW}=\mathrm{SU}(2)_{L}\times\mathrm{U}(1)_{Y}$ can be written as (recall \eqref{eq:defEMCoupling})
\begin{align}
D_{\mu}&=\p_{\mu}-\I g W_{\mu}^{a}T^{a}-\I g^{\prime}B_{\mu} Y {\mathbb{1}}\\[0.45em]
&=\p_{\mu}-{\I g}\left (W_{\mu}^{+}T^{+}+W_{\mu}^{-}T^{-}\right ) -\I Z_{\mu} \left ( g \cos(\theta_W)T^{3}-g^{\prime} \sin(\theta_{W})Y {\mathbb{1}}\right )-\I eA_{\mu}Q{\mathbb{1}}\nn
\end{align}
where
\begin{equation}
T^{\pm}=\dfrac{T^{1}\pm\I T^{2}}{\sqrt{2}}\, .
\end{equation}
As always, the action of $D_{\mu}$ on a field is understood to be in the corresponding representation. Hence, $D_{\mu}$ acts differently for each field, e.g., for $d_{R}^{i}$ we have
\begin{equation}
D_{\mu}d_{R}^{i}=\p_{\mu}d_{R}^{i}-\I g^{\prime} B_{\mu}Yd_{R}^{i}\, =\p_{\mu}d_{R}^{i}-\I e\left(A_\mu
-\tan(\theta_W) Z_\mu\right) Q  d_{R}^{i}\, .
\end{equation}
It is occasionally useful to separate the interactions terms between fermions and gauge fields from the kinetic terms by writing
\begin{align}\label{eq:currentsWZFermions} 
\cL_{F}^{\text{kin}}[D]= \cL_{F}^{\text{kin}}[\p]+eA_{\mu}J_{\mu}^{EM}+\dfrac{e}{\sin(\theta_{W})}Z_{\mu}J_{\mu}^{Z}
+\dfrac{e}{\sqrt{2}\sin(\theta_{W})}\left(W^+_\mu J_+^\mu+W^-_\mu J_-^\mu\right)
\end{align}
where $\cL_{F}^{\text{kin}}[\p]$ is given by the kinetic terms in \eqref{eq:KinTermsFermionsCov} with the covariant derivatives $D_{\mu}$ replaced by ordinary derivatives $\p_{\mu}$.
In \eqref{eq:currentsWZFermions}, we extract easily the expressions for the various current interactions involving combinations of the fermion fields.
The \emph{electromagnetic current}\index{Electroweak Theory!Electromagnetic current} from couplings to the photon $A_{\mu}$ are given by
\begin{equation}\label{eq:EMCurrent} 
J_{\mu}^{EM}=\sum_{i}\, Q_{i}\left (\overline{\psi}_{i}^{L}\gamma_{\mu}\psi_{i}^{L}+\overline{\psi}_{i}^{R}\gamma_{\mu}\psi_{i}^{R}\right )\, .
\end{equation}
Here, $Q_{i}$ denotes the electric charge under \eqref{eq:ElectricChargeU1} and $\psi_{i}^{L,R}$ denotes all left- and right-handed fermions from above.
The \emph{neutral current}\index{Neutral current}\index{Neutral current}\index{Electroweak Theory!Neutral current} from couplings to $Z_{\mu}$ reads
\begin{equation}\label{eq:ZCurrent} 
J_{\mu}^{Z}=\dfrac{1}{\cos(\theta_{W})}\left (\sum_{i}\overline{\psi}_{i}^{L}\gamma_{\mu}T^{3}\psi_{i}^{L}-\sin^{2}(\theta_{W})J_{\mu}^{EM}\right )\, .
\end{equation}
The \emph{charged currents}\index{Charged current}\index{Electroweak Theory!Charged current} $J_\mu^\pm$ from the couplings to $W_\mu^\pm$ in the Lagrangian above are
\begin{equation}\label{eq:ChargedCurrents} 
J_{\mu}^{+}=\overline{\nu}^{i}_{L}\gamma_{\mu}e_{L}^{i}+\overline{u}^{i}_{L}\gamma_{\mu}d_{L}^{i}\kom J_{\mu}^{-}=\overline{e}^{i}_{L}\gamma_{\mu}\nu_{L}^{i}+\overline{d}^{i}_{L}\gamma_{\mu}u_{L}^{i}\, .
\end{equation}

Lastly, we consider the interactions involving fermions which can be written as
\begin{align}
\cL_{F}^{\text{Yukawa}}&=\cL_{F}^{\text{Higgs-leptons}}[L^{i}_{L},e_{R}^{i},H]+\cL_{F}^{\text{Higgs-quarks}}[Q_{L}^{i},u_{R}^{i},d_{R}^{i},H]\, .
\end{align}
The quark contribution is\index{Electroweak Theory!Yukawa interactions}
\begin{equation}\label{eq:YukawaQuarks} 
\cL_{F}^{\text{Higgs-quarks}}[Q_{L}^{i},u_{R}^{i},d_{R}^{i},H]=-y_{ij}^{d}\, \overline{Q}_{L}^{i}Hd_{R}^{j}-y_{ij}^{u}\, \overline{Q}_{L}^{i}\tilde{H}u_{R}^{j}+\text{h.c.}
\end{equation}
where $y^{d}$, $y^{u}$ are free parameters called \emph{Yukawa couplings}\index{Yukawa couplings} and
\begin{equation}
\tilde{H}=\I\sigma^{2}H^{*}\, .
\end{equation}
Notice that these combinations are all gauge invariant, since they are uncharged under $\UO_{Y}$ where
\begin{align}
\overline{Q}_{L}^{i}Hd_{R}^{j}&\raw -\dfrac{1}{6}+\dfrac{1}{2}-\dfrac{1}{3}=0\kom \overline{Q}_{L}^{i}\tilde{H}u_{R}^{j}\raw -\dfrac{1}{6}-\dfrac{1}{2}+\dfrac{2}{3}=0\, .
\end{align}
They are also $\mathrm{SU}(2)_L$ singlets (recall the product of two $\mathrm{SU}(2)$ doublets gives a triplet plus a singlet).
Further, knowing that left-handed quarks transform as $\bf{3}$ of $\mathrm{SU}(3)_c$ (as mentioned in chapter \ref{chap:intro} and to be properly introduced in the next chapter) and right-handed quarks as $\bf \bar{3}$, these couplings are also $\mathrm{SU}(3)_c$ invariant. It is remarkable that the hypercharge assignments imposed by anomaly cancellations are precisely what are needed in order to have non-zero Yukawa couplings.\footnote{Note that assigning a hypercharge $1/2$ to the Higgs field could be justified to guarantee a non-zero Yukawa coupling for electrons. The remarkable fact is that given this assignment and anomaly cancellation, all left- and right-handed quarks have non-vanishing Yukawa couplings and therefore mass terms generated after higgsing, see also section~\ref{sec:anomalies_SM}.}
This is important because these couplings are the ones responsible to give the quarks a mass. Note that the masses for the fermions arise only after the Higgs gets a VEV $\langle H\rangle\neq 0$. There are {\bf no direct mass terms for the quarks which are forbidden by gauge (chiral) symmetry}. Therefore, it is the Yukawa couplings and the Higgs VEV that give mass to fermions.

Without right-handed neutrinos, the Lagrangian for the Yukawa interactions between the leptons and the Higgs includes only the following couplings
\begin{equation}\label{eq:Higgs_leptons_no_RHNeutrinos} 
\cL_{F}^{\text{Higgs-leptons}}[L^{i}_{L},e_{R}^{i},H]= -y_{ij}^{e}\overline{L}^{i}_{L}He_{R}^{j}+\text{h.c.}\, .
\end{equation}
Contrary to \eqref{eq:YukawaQuarks} for the quarks involving two different types of Yukawa couplings for $Q_L^i $ to $u_R^i$ and $d_R^i$ respectively, here there is only a single Yukawa term. As we will see below, this implies that there is only a mass term for the electrons, muons and tauons, but \emph{not to their neutrinos}. Since there are only left-handed neutrinos inside $L^i$, but no right-handed neutrinos, the neutrinos remain massless in this case after SSB.

\subsection{Quarks: mass matrix and weak couplings}\label{sec:quarks_masses_couplings}

Let us come back to the Yukawa couplings between the Higgs and the quarks, cf.~\eqref{eq:YukawaQuarks}.
As mentioned before, these couplings give rise to the masses for the quarks after SSB.
Indeed, we can see this explicitly by replacing the Higgs by \eqref{eq:HiggsVEVPlusFluctuation} in the Yukawa couplings above which leads to
\begin{equation}
\cL_{F}^{\text{Higgs-quarks}}[Q_{L}^{i},u_{R}^{i},d_{R}^{i},H] = \cL_{F}^{\text{Higgs-boson-quarks}}[Q_{L}^{i},u_{R}^{i},d_{R}^{i},h]+\cL_{F}^{\text{mass}}[Q_{L}^{i},u_{R}^{i},d_{R}^{i}]
\end{equation}
in terms of the Higgs-boson-quark interactions
\begin{equation}\label{eq:interactionsHiggsBosonQuarksGeneral} 
\cL_{F}^{\text{Higgs-boson-quarks}}[Q_{L}^{i},u_{R}^{i},d_{R}^{i},h]=-\dfrac{h}{\sqrt{2}}\left [\overline{d}_{L}^{i}y^{d}_{ij}d^{j}_{R}+\overline{u}_{L}^{i}y^{u}_{ij}u^{j}_{R}\right ]+\text{h.c.}
\end{equation}
and the mass term for the quarks
\begin{equation}
\cL_{F}^{\text{mass}}[Q_{L}^{i},u_{R}^{i},d_{R}^{i}]=-\dfrac{v}{\sqrt{2}}\left [\overline{d}_{L}^{i}y^{d}_{ij}d^{j}_{R}+\overline{u}_{L}^{i}y^{u}_{ij}u^{j}_{R}\right ]+\text{h.c.}
\end{equation}
with the mass matrices encoded by $y^{d}_{ij}$ and $y^{u}_{ij}$. This is the source of the standard claim that the Higgs field is responsible to give a mass to all the other particles since the masses are proportional to the Higgs VEV $v$. This claim will be reconsidered in the next chapter.

Note that the mass matrices given by $vy^{d}_{ij}$ and $v y^{u}_{ij}$ are in general non-diagonal. In order to read the masses of the physical particles, we need to diagonalise these matrices. In general, any matrix can be diagonalised using two unitary matrices $U,K$\footnote{A way to see this is using the polar decomposition of the matrix $y=HS$ with $H$ hermitian and $S$ unitary which is possible for any matrix as we have seen before. Now use the fact that $H=yS^\dagger$ being hermitian can be diagonalised to a real diagonal matrix $M$ by means of a single unitary matrix $U$: $M=U^\dagger H U= U^\dagger yS^\dagger U$. Defining $K=S^\dagger U$ implies that $M=U^\dagger y K$ is real and diagonal as required.}
\begin{equation}
y^{d}=U_{d}M^{d}K^{\dagger}_{d}\kom y^{u}=U_{u}M^{u}K^{\dagger}_{u}
\end{equation}
with $M^{d}$, $M^{u}$ real and diagonal leads to a new basis of fields
\begin{equation}
d_{L}^{\prime}=U^\dagger_{d}d_{L}\kom  d_{R}^{\prime}=K^\dagger_{d}d_{R}\kom  u_{L}^{\prime}=U^\dagger_{u}u_{L}\kom  u_{R}^{\prime}=K^\dagger_{u}u_{R}\, .
\end{equation}
The Lagrangian for the mass terms in this new basis reads
\begin{equation}
\cL_{F}^{\text{mass}}=-\dfrac{v}{\sqrt{2}}\left [\overline{d}_{L}^{\prime,i}M^{d}_{ii}d^{\prime,i}_{R}+\overline{u}_{L}^{\prime,i}M^{u}_{ii}u^{\prime,i}_{R}\right ]+\text{h.c.}
\end{equation}
implying that the quark masses are given by
\begin{equation}
m_{d^{\prime}_{i}}=\dfrac{v}{\sqrt{2}}M^{d}_{ii}\kom m_{u^{\prime}_{i}}=\dfrac{v}{\sqrt{2}}M^{u}_{ii}\, .
\end{equation}
Therefore, we managed to define a basis for the up and down quarks for which the mass matrix is diagonal leading to six free parameters $m_{d^{\prime}_i}$ and $m_{u^{\prime}_i}$.
In this basis, the interactions \eqref{eq:interactionsHiggsBosonQuarksGeneral} of this Higgs bosons to the quarks reads
\begin{equation}\label{eq:interactionsHiggsBosonQuarksMassBasis} 
\cL_{F}^{\text{Higgs-boson-quarks}}[Q_{L}^{i},u_{R}^{i},d_{R}^{i},h]=-\dfrac{h}{v}\left [m_{d^{\prime}_{i}}\overline{d}_{L}^{\prime,i} d^{\prime,i}_{R}+m_{u^{\prime}_{i}}\overline{u}_{L}^{\prime,i}u^{\prime,i}_{R}\right ]+\text{h.c.}\, .
\end{equation}
Thus, the coupling scales proportional to the masses of the quarks which is why the decay $h\raw b\bar{b}$ has the highest probability.\footnote{The top quark is too heavy which is why $h\raw t\bar{t}$ is forbidden.}
From now-on, we will drop the primes on $u_i$ and $d_i$ and work in this basis unless stated otherwise.

It is, however, important to keep in mind that the basis that diagonalises the mass matrix is such that in general the couplings to the gauge fields coming from the kinetic terms are \textbf{not} diagonal in this basis. To see this, let us write the couplings \eqref{eq:currentsWZFermions} in the \emph{\bf mass eigenstate}\index{Electroweak Theory!Mass eigenstates}\index{Mass eigenstates} basis
\begin{align}\label{eq:currentsWZFermionsMass} 
\cL_{F}^{\text{kin+Higgs}}&= \cL_{F}^{\text{kin}}[\p]+eA_{\mu}J_{\mu}^{EM}+\dfrac{e}{\sin(\theta_{W})}Z_{\mu}J_{\mu}^{Z}
+\dfrac{e}{\sqrt{2}\sin(\theta_{W})}\left(W^+_\mu J_+^\mu+W^-_\mu J_-^\mu\right) 
\nn \\
&\quad  -m_{j}^{d}\left (\overline{d}_{L}^{j}d_{R}^{j}+\overline{d}_{R}^{j}d_{L}^{j}\right )
-m_{j}^{u}\left (\overline{u}_{L}^{j}u_{R}^{j}+\overline{u}_{R}^{j}u_{L}^{j}\right )
\end{align}
Both $J_\mu^{EM}$ and $J_\mu^Z$ defined in \eqref{eq:EMCurrent} and \eqref{eq:ZCurrent} are manifestly diagonal in this new basis since they come from diagonal generators of the gauge group that do not mix up and down quarks. But the \emph{charged currents} $J_\mu^\pm$ in \eqref{eq:ChargedCurrents} are not diagonal since they mix up- and down-type quarks in a generation.
This is because the different quark flavours are diagonalised by different matrices $U_u$ and $U_d$ which implies that in the mass eigenbasis the $J_\mu^\pm$ read (for the quarks only)
\begin{equation}\label{eq:ChargedCurrentsQuarksMassEigenbasis} 
J_{\mu}^{+}=\overline{u}^{i}_{L}\gamma_{\mu}V^{ij}d_{L}^{j} \kom J_{\mu}^{-}=\overline{d}^{i}_{L}\gamma_{\mu}(V^{\dagger})^{ij}u_{L}^{j}\, .
\end{equation}
Here, the mixing between the different quark flavours is encoded in the \emph{Cabibbo-Kobayashi-Maskawa (CKM) matrix}\index{CKM matrix}
\begin{equation}\label{eq:CKM_matrix} 
V=V_{CKM}=U_{u}^{\dagger}U_{d}=\left (\begin{array}{ccc}
V_{ud} & V_{us} & V_{ub} \\ 
V_{cd} & V_{cs} & V_{cb} \\ 
V_{td} & V_{ts} & V_{tb}
\end{array} \right )\, .
\end{equation}
This matrix being unitary has $9$ free parameters which in principle need to be determined experimentally ($3$ real angles and $6$ phases). However, we can still reduce the number of independent parameters as follows.

Note that there is a remnant $\UO^{6}$ global symmetry of the mass terms
\begin{equation}
d_{R,L}^{i}\raw\ee^{\I\alpha_{i}}d_{R,L}^{i}\kom u_{R,L}^{i}\raw\ee^{\I\beta_{i}}u_{R,L}^{i}\, .
\end{equation}
This symmetry can be used to eliminate $5$ phases (only phase difference work, since the overall $\UO$ for which all the parameters $\alpha_i$ and $\beta_i$ are equal is a symmetry of the whole Lagrangian and corresponds to baryon number). There remain $9-5=4$ free parameters within $V_{CKM}$: $3$ real parameters (the three standard rotation angles in three dimensions $\theta_{12}$, $\theta_{13}$, $\theta_{23}$) and $1$ phase. One parametrisation of this matrix is as follows (writing $c_{ij}=\cos(\theta_{ij}$, $s_{ij}=\sin(\theta_{ij})$)
\begin{equation}
V_{CKM}= \begin{pmatrix} c_{12} c_{13} & s_{12}c_{13} & s_{13} e^{-i\delta} \\
-s_{12}c_{23}-c_{12}s_{23}s_{13}e^{i\delta} & 
c_{12}c_{23}-s_{12}s_{23}s_{13}e^{i\delta} & s_{23}c_{13} \\
s_{12}s_{23}-c_{12}c_{23}s_{13}e^{i\delta} & 
-c_{12}s_{23}-s_{12}c_{23}s_{13}e^{i\delta} & c_{23}c_{13} \end{pmatrix} \, .
\end{equation}
Another parametrisation (known as \emph{Wolfenstein's parametrisation}\index{Wolfenstein's parametrisation}) can be written as
\begin{equation}
V_{CKM}= \left (\begin{array}{ccc}
1-\dfrac{\lambda^{2}}{2} & \lambda & A\lambda^{3}(\rho-i\eta) \\[0.25em]
-\lambda & 1-\dfrac{\lambda^{2}}{2} & A\lambda^{2} \\[0.25em] 
A\lambda^{3}(1-\rho-i\eta) & -A\lambda^{2} & 1
\end{array} \right )
\end{equation}
with parameters $A,\rho,\lambda,\eta$.
Approximately, $s_{12}=\lambda$ is determined by the Cabibbo angle\index{Cabibbo angle} $\theta_{12}$, $\sin(\theta_{12})\approx 0.22$ so for small $\lambda$ and $A,\rho, \eta$ of order one
we can see that there is a hierarchy in the matrix elements showing that to leading order in $\lambda$ the matrix is diagonal and the mixing between the first two families is stronger than the mixing of each of them with the third family.

\subsubsection*{Mass eigenstates $\neq$ ``weak or flavour eigenstates''}
\index{Electroweak Theory!Weak eigenstates}\index{Weak eigenstates}

This statement simply implies that if we work with quarks from different families, and diagonalise the mass matrix in flavour space (a fixed mass for each up and down quark), then the couplings to the $W^\pm_\mu$ fields are not diagonal.
This means that the charged currents $J_{\mu}^{\pm}$ connect fermions of different flavours. This generation mixing is responsible for weak interactions that change flavour. This suggests in particular that initial states from one family can decay into final states from a different family as in the muon decay, see section~\ref{sec:4FermiMuonDecay} below.

\subsubsection*{Glashow-Iliopoulos-Maiani (GIM) Mechanism}
\index{GIM mechanism}

Unlike the charged currents $J_{\mu}^{\pm}$ that mix different generations, the neutral current $J_\mu^Z$ is flavour diagonal. This implies that there are no flavour changing neutral currents (FCNC). In the 1960's and early 1970's only three quarks were known $u,d,s$. If we go back and compute the neutral current with only these three quarks then it would not be diagonal as it can be easily checked.
But interactions that would be mediated by a neutral current such as $s\raw d+e^++ e^-$ were not observed and there was no explanation. This was the reason for Glashow, Iliopoulos and Maiani  to predict the existence of a fourth quark (charm $c$) that leads to the absence of FCNC and forbids such decays. This was spectacularly confirmed with the discovery of the $J/\psi$ resonance in 1974 \cite{Augustin:1974xw,Aubert:1974js}.
From the current perspective there is a compelling need for the existence of $c$ once $s$ was discovered. Anomaly cancellations require both members of a family for consistency. 

\subsubsection*{CP-violation}

The presence of a phase $\delta$ in $V_{CKM}$ implies CP-violation\index{CP-violation}. This can be shown explicitly by analysing the behaviour of the currents under time reversal $T$ (since $CPT$ is an exact symmetry non-invariance under $T$ is equivalent to $CP$ violation).
If there were only two families, the CKM matrix would have only one real parameter (the \emph{Cabibbo angle}\index{Cabibbo angle}) and no phases; therefore no CP-violation. So the observation of CP-violation led Kobayashi and Maskawa to predict a third family \cite{Kobayashi:1973fv}.
Experimentally, the four parameters in the CKM matrix were measured to be
\begin{align}
&\theta_{12}=13.02\pm 0.04\kom \theta_{23}=2.56\pm 0.03\, ,\nn\\
&\theta_{13}=0.20\pm 0.02\kom \delta=69\pm 5\, .
\end{align}

\subsubsection*{Baryon number}
\index{Baryon number}

The couplings in $\cL_{F}$ defined in \eqref{eq:CouplingsFermions} have an accidental $\mathrm{U}(1)_{B}$ symmetry
\begin{equation}
B(Q_{L}^{i})=B(u_{R}^{i})=B(d_{R}^{i})=\dfrac{1}{3}\kom B(\overline{Q}_{L}^{i})=B(\overline{u}_{R}^{i})=B(\overline{d}_{R}^{i})=-\dfrac{1}{3}
\end{equation}
and all others $B=0$. This $\mathrm{U}(1)_{B}$ symmetry corresponds to baryon number (charge $B=+1$ for baryons and $B=-1$ for anti-baryons whereas leptons have charge $B=0$).
We will discuss this baryon number symmetry later in section~\ref{sec:globalsymmetries}.

\subsubsection*{Unitarity Triangle}\index{Unitarity Triangle}

The fact that the CKM matrix is unitary $VV^\dagger={1} $ can be expressed nicely in terms of its components as the statement
\begin{equation}
\frac{V_{ud}V^*_{ub}}{V_{cd}V^*_{cb}}+\frac{V_{td}V^*_{tb}}{V_{cd}V^*_{cb}}+1=0
\end{equation}
where the indices refer to the generations and where we divided by the factor $V_{cd}V^*_{cb} $. This expression is in general the sum of three numbers of wich two can be complex. If so, they would produce a triangle in the complex plane. If these numbers end up being real, the triangle collapses. Therefore a measure of the existence of CP violation is to measure the area of this triangle and a test of unitarity of the CKM matrix is to determine experimentally the sides of the triangle and check that it closes. 

Using these components of the CKM matrix we can construct the invariant quantity known as the \emph{Jarlskog invariant}\index{Jarlskog invariant} \cite{Beyer:2002tj} which is a measure of the area of the unitarity triangle.
More specifically, it is a convenient way to express the non-vanishing area of the triangle by the existence of a non-vanishing phase in the CKM matrix
\begin{equation}
J\equiv{\rm Im}\left( V_{ud}V^*_{ub}V_{tb}V^*_{td}\right)= s_{12}s_{23}s_{31}c_{12}c_{23}c_{31}^2\, \sin\delta =    (2.96\pm 0.20)\times 10^{-5}\, .
\end{equation}
Measuring $J\neq 0$ guarantees the presence of CP violation in weak interactions.  $J$ is invariant under any of the phase rotations that gave rise to the CKM matrix. A simple way to see it is that each of the indices $u,d,t,b$ appear in one $V_{ij}$ and in one $V^*_{ij}$ component so the phase rotation of the corresponding quark cancels. It can be easily shown that $J$ is twice the area of the unitarity triangle (using that the area generated by two vectors can be computed by computing their cross product).

\subsection{Leptons: mass matrix and weak couplings}

We now repeat the same analysis of the weak couplings of leptons to the gauge fields and the Higgs as we just did for the quarks. Since right-handed neutrinos have not been observed, even though it has been confirmed that neutrinos have a non-zero mass, we will consider first the possibility of no right-handed neutrinos $\nu_{R}$.
In the subsequent section, we add right-handed neutrinos and discuss the implications.

Without right-handed neutrinos, the Higgs-lepton interactions are given by \eqref{eq:Higgs_leptons_no_RHNeutrinos}, namely
\begin{equation}\label{eq:YukawaCouplingsHiggsDecayLag} 
\cL_{F}^{\text{Higgs-leptons}}=-y_{ij}^{L}\, \overline{L}_{L}^{i}He_{R}^{j}-(y^{\dagger})_{ij}^{L}\, \overline{e}_{R}^{i}HL_{L}^{j}\, .
\end{equation}
After SSB, we set the Higgs field in unitary gauge to
\begin{equation}
H=\dfrac{1}{\sqrt{2}}\left (\begin{array}{c}
0 \\ 
v+h
\end{array} \right )
\end{equation}
in terms of the Higgs boson $h$. Plugging this into \eqref{eq:YukawaCouplingsHiggsDecayLag}, we find
\begin{align}
\cL_{F}^{\text{Higgs-leptons}}&=-\dfrac{y_{ij}^{L}}{\sqrt{2}}\, \bar{e}_{L}^{i}(v+h)e_{R}^{j}-\dfrac{(y^{\dagger})_{ij}^{L}}{\sqrt{2}}\, \overline{e}_{R}^{i}(v+h)e_{L}^{j}\, .
\end{align}
As before, we can find a field basis in which the Yukawa couplings $y_{ij}^{L}$ are diagonal. The eigenvalues are non-negative and will be denoted $m_{i}\sqrt{2}/v$ in the following.
In this basis, the interaction Lagrangian can be written as
\begin{align}
\cL_{F}^{\text{Higgs-leptons}}&=- \dfrac{m_{i}}{v}\, \bar{e}_{L}^{i}(v+h)e_{R}^{i}-\dfrac{m_{i}}{v} \, \overline{e}_{R}^{i}(v+h)e_{L}^{i}\nn\\
&=-m_{i}\left (\bar{e}_{L}^{i}e_{R}^{i}+\bar{e}_{R}^{i}e_{L}^{i}\right )-\dfrac{m_{i}}{v}h\left (\bar{e}_{L}^{i}e_{R}^{i}+\bar{e}_{R}^{i}e_{L}^{i}\right )\, .
\end{align}
The first term is a Dirac mass term for the leptons $e,\mu,\tau$. Let us define the Dirac spinors
\begin{equation}
\ell^{i}=e_{L}^{i}+e_{R}^{i}=\left (\begin{array}{c}
\tilde{e}_{L}^{i} \\ 
\tilde{e}_{R}^{i}
\end{array} \right )
\end{equation}
in terms of $2$-component Weyl spinors $\tilde{e}_{L,R}^{i}$ so that
\begin{equation}\label{eq:YukawaCouplingsHiggsDecayLagMassCoup}
\cL_{F}^{\text{Higgs-leptons}}=-m_{i}\bar{\ell}^{i}\ell^{i}-\lambda_{i}\, h\bar{\ell}^{i}\ell^{i}
\end{equation}
in terms of the couplings
\begin{equation}\label{eq:CoupYukLep} 
\lambda_{i}^{\text{Higgs-leptons}}=\dfrac{m_{i}}{v}\, .
\end{equation}
As in the case of the Yukawa couplings for the quarks, a non-zero VEV for the Higgs field, $\langle H\rangle=v\neq 0$, will provide mass terms for the leptons. But because there is only a single Yukawa coupling, there are only mass terms for the electrons, muons and tauons, but \emph{not for their neutrinos}. This is because there are only left-handed neutrinos inside $L^i$, but no right-handed neutrinos which is why the neutrinos remain massless even after SSB.

Further, this implies that the mass matrix for $e,\mu,\tau$ can be diagonalised without affecting the mixing of the couplings coming from the kinetic terms. Therefore, in the absence of right-handed neutrinos, the weak and mass eigenstates of leptons are the same and there is no analogue of the CKM matrix. This diagonal property of the quadratic Lagrangian for leptons also implies separate conservations of lepton numbers for each family: three independent global $\mathrm{U}(1)$ symmetries corresponding to conservation of 
 $L_{e}$, $L_{\mu}$, $L_{\tau}$, e.g., $L_{e}(e)=L_{e}(\nu_{e})=1$, other zero, etc. 

Since neutrinos are massive, we know this cannot be the full story. The natural way to proceed is to introduce the right-handed neutrinos that we will do next. However, the fact that there is no mass term for neutrinos may be just a limitation of the fact that we are imposing the Lagrangian to be renormalisable. Once we relax this condition there will be neutrino masses generated from couplings such as $HHLL$ which is of dimension five and therefore non-renormalisable. Once the Higgs gets a VEV this will induce neutrino masses. We will discuss this in chapter \ref{chap:probs}.

If neutrinos were massless as above, then the fact that the mass matrix can be diagonalised implies that there would be independently conserved family lepton numbers  $ L_{e}, L_{\mu}, L_{\tau}$ corresponding to three accidental $\mathrm{U}(1)$ global symmetries  acting as independent phase rotations $\psi_k\raw e^{\I \alpha_k}\psi_k$ where $\psi_k$ $k=1,2,3$ represents teach of the three families of leptons. However, since neutrinos are massive, these symmetries are not actually there, but there remains the overall accidental symmetry $L=L_{e}+L_{\mu}+L_{\tau}$ which counts the overall lepton number which is similar to baryon number for quarks.
Except that, if there is a Majorana mass for right-handed neutrinos, it also breaks the overall lepton number $L$, see below.
Note that both lepton and baryon number are accidental global symmetries.

\subsection{$Z$-boson decay and the number of light neutrinos}

One of the big successful results from the LEP\footnote{Nicknamed the $Z$-factory for this reason.} experiment in the 1990s was the study of the decay modes of the $Z$-boson \cite{ParticleDataGroup:2006fqo}. The total decay rate (width of the $Z$-boson resonance) $\Gamma_Z$ is well determined given all the possible  decay channels as:
\begin{equation}\label{eq:DecayZFull} 
\Gamma_Z=\Gamma_{ee}+\Gamma_{\mu\mu}+\Gamma_{\tau\tau}+\Gamma_{\text{hadrons}}+ N_\nu\Gamma_{\nu\nu}
\end{equation}
Since the width was measured very precisely, from this the number of neutrinos lighter than the $Z$ boson can be determined accurately as we now demonstrate.

To this end, let us compute the decay of the $Z$-boson into leptons.
We initially have to identify the interactions in the Lagrangian relevant for the decay.
These are obtained from the kinetic terms for the leptons after SSB, namely
\begin{align}
\I\overline{L}_L\,  \gamma^{\mu}{D}_{\mu}L_L+\I\overline{e}_R\,  \gamma^{\mu}{D}_{\mu}e_R&=-\dfrac{g}{2\cos(\theta_{W})} J^{\mu}_{n}\, Z_{\mu} 
\end{align}
in terms of the neutral current
\begin{align}\label{eq:neutral_current_leptons} 
J^{\mu}_{n}&=\overline{L}_{L} \gamma^{\mu}\left (\cos(\theta_{W})^{2}\sigma_{3}+\sin(\theta_{W})^{2}\mathds{1}_{2} \right )\,  L_{L}+2\sin(\theta_{W})^{2}\, \bar{e}_{R}\gamma^{\mu}e_{R} \, . 
\end{align}
To treat all decays into leptons simultaneously, it is convenient to write
\begin{align}
J^{\mu}_{n}&=\sum_{\ell\in\lbrace e,\nu_{e},\mu,\nu_{\mu},\tau,\nu_{\tau}\rbrace}  \bar{\ell}\gamma^{\mu}(v_{\ell}\mathds{1}_{4}-a_{\ell}\gamma^{5})\ell \, . 
\end{align}
Here, we distinguish between vector couplings $v_{\ell}$ and axial couplings $a_{\ell}$ where
\begin{equation}
v_{i}=2\sin(\theta_{W})^{2}-\dfrac{1}{2}\kom a_{i}=-\dfrac{1}{2}\kom v_{\nu_{i}}=a_{\nu_{i}}=\dfrac{1}{2}\kom i\in\lbrace e,\mu,\tau\rbrace\, .
\end{equation}

Diagrammatically, the decay of the $Z$-bosons into leptons can be visualised as
\begin{align*}
\begin{tikzpicture}[scale=1.2]
\setlength{\feynhanddotsize}{1.5ex}
\begin{feynhand}
\vertex (a2) at (-0.5,-4) {$Z, \epsilon_{\mu}$}; 
\vertex (b2) at (1.25,-4); 
\node (o) at (1.25,-4.3) {}; 
\vertex (c2) at (2.5,-3) {$\bar{\ell}^{i}$}; 
\vertex (d2) at (2.5,-5) {$\ell^{i}$}; 
\propag [bos, mom={$p$}] (a2) to (b2);
\propag [fer, mom={$k$}] (b2) to (d2);
\propag [antfer, mom={$q$}] (b2) to (c2);
\end{feynhand}
\end{tikzpicture}
\end{align*}
The associated tree level S-matrix element for this decay is
\begin{equation}
\cM=\bra{\ell(k)\bar{\ell}(q)}\,\cL_{Z\raw\bar{\ell}^{i}\ell^{i}}\ket{Z(p,\epsilon)}
\end{equation}
where $\epsilon_{\mu}$ is the polarisation of the $Z$-boson.
Momentum conservation implies
\begin{equation}\label{eq:DecZMC} 
m_{Z}^{2}=p\cdot k+p\cdot q\kom p\cdot k=m_{\ell_{i}}^{2}+k\cdot q=p\cdot q \kom 2k\cdot q=m_{Z}^{2}-m_{\ell_{i}}^{2}\, .
\end{equation}
To compute $\cM$, we recall the Feynman rules for in and out states listed in App.~\ref{app:drc}.
Hence, we have
\begin{align}
\cM&=\mathcal{N}\, \bar{u}_{\ell}(k) \gamma^{\mu}(v_{\ell}\mathds{1}_{4}-a_{\ell}\gamma^{5})v_{\bar{\ell}}(q)\,\epsilon_{\mu}(p)\kom \mathcal{N}=\dfrac{g}{2\cos(\theta_{W})}\, .
\end{align}
As usual, we compute the square of the matrix element and sum over all spins of the leptons and polarisations for the $Z$-boson,
\begin{align*}
 \sum_{\text{spins},\ldots}|\cM|^{2}&=\mathcal{N}^{2}\, \sum_{\text{spins},\ldots}\bar{u}_{\ell}(k) \gamma^{\mu}(v_{\ell}\mathds{1}_{4}-a_{\ell}\gamma^{5})v_{\bar{\ell}}(q)\,\epsilon_{\mu}(p)\epsilon_{\nu}^{*}(p) \bar{v}_{\bar{\ell}}(q) \gamma^{\nu}(v_{\ell}\mathds{1}_{4}-a_{\ell}\gamma^{5})u_{{\ell}}(k)\nn\\
 &=\mathcal{N}^{2}\, \tr\left [(\cancel{k}+m_{\ell})\gamma^{\mu}(v_{\ell}\mathds{1}_{4}-a_{\ell}\gamma^{5})(\cancel{q}-m_{\ell})\gamma^{\nu}(v_{\ell}\mathds{1}_{4}-a_{\ell}\gamma^{5}) \right ]\left (-g_{\mu\nu}+\dfrac{p_{\mu}p_{\nu}}{m_{Z}^{2}}\right )
\end{align*}
where we used
\begin{equation}
 \sum_{\text{polarisations}}\, \epsilon_{\mu}(p)\epsilon_{\nu}^{*}(p) =-g_{\mu\nu}+\dfrac{p_{\mu}p_{\nu}}{m_{Z}^{2}}\, .
\end{equation}
This identity gives exactly the numerator of the propagator of a massive spin-1 boson, cf.~\eqref{eq:propagator_massive_spin_one}.
In a first approximation, one can neglect the lepton masses which are small compared to the mass of the $Z$-boson so that
\begin{align}
 \sum_{\text{spins},\ldots}|\cM|^{2}&=\mathcal{N}^{2}\, \tr\biggl [ (v_{\ell}^{2}+a_{\ell}^{2})\cancel{k}\gamma^{\mu}\cancel{q}\gamma^{\nu}-2a_{\ell}v_{\ell}\cancel{k}\gamma^{\mu}\cancel{q}\gamma^{\nu}\gamma^{5} \biggl ]\left (-g_{\mu\nu}+\dfrac{p_{\mu}p_{\nu}}{m_{Z}^{2}}\right )
\end{align}
where we used various commutation relations for the $\gamma$-matrices listed in App.~\ref{app:drc}.
Next, we apply \eqref{eq:TrGam5Gam4} to argue that the term $\sim a_{\ell}v_{\ell}$ vanishes due to antisymmetry of $\varepsilon^{\mu\nu\rho\sigma}$.
Then, we utilise \eqref{eq:TrGam4} to write
\begin{align}
 \sum_{\text{spins},\ldots}|\cM|^{2}&=\dfrac{g^{2}}{\cos(\theta_{W})^{2}}\, (v_{\ell}^{2}+a_{\ell}^{2})\, \left [k\cdot q+\dfrac{2}{m_{Z}^{2}}(k\cdot p)(q\cdot p)\right ]\, .
\end{align}
Ignoring the lepton masses, the conditions \eqref{eq:DecZMC} yield
\begin{align}
 \sum_{\text{spins},\ldots}|\cM|^{2}&=\dfrac{g^{2}}{\cos(\theta_{W})^{2}}\, (v_{\ell}^{2}+a_{\ell}^{2})\, m_{Z}^{2}\, .
\end{align}

Using \eqref{eq:decayrate} for the partial decay rate and including an averaging factor of $1/3$ for the initial spins of the $Z$-bosons, we have to compute
\begin{align}
\Gamma(Z\raw \ell\bar{\ell})&=\dfrac{1}{2m_{Z}}\,\int\, \dfrac{\dif^{3}k}{(2\pi)^{3}2k^{0}}\, \dfrac{\dif^{3}q}{(2\pi)^{3}2q^{0}}\,  (2\pi)^{4}\delta^{(4)}\left (p-k-q\right )\dfrac{1}{3} \sum_{\text{spins},\ldots}|\cM_{\alpha\beta}|^{2}\nn\\
&=\dfrac{g^{2}\, (v_{\ell}^{2}+a_{\ell}^{2})\, m_{Z}}{96\pi^{2}\cos(\theta_{W})^{2}}\int\, \dfrac{\dif^{3}k}{k^{0}}\, \dfrac{\dif^{3}q}{q^{0}}\,  \delta^{(4)}\left (p-k-q\right )\, .
\end{align}
In the rest frame of the $Z$-boson and for massless leptons, we find
\begin{align}
\Gamma(Z\raw \ell\bar{\ell})&=\dfrac{g^{2}\, (v_{\ell}^{2}+a_{\ell}^{2})\, m_{Z}}{96\pi^{2}\cos(\theta_{W})^{2}}\int\, \dfrac{\dif^{3}k}{|\mathbf{k}|^{2}}\delta\left (m_{Z}-2|\mathbf{k}|\right )\, .
\end{align}
After evaluating the final integral, this becomes
\begin{align}
\Gamma(Z\raw \ell\bar{\ell})&=\dfrac{G_{F}}{\sqrt{2}}\, \dfrac{m_{Z}^{3}(v_{\ell}^{2}+a_{\ell}^{2})}{16\pi}\kom \dfrac{G_{F}}{\sqrt{2}}=\dfrac{g^{2}}{8m_{Z}^{2}\cos(\theta_{W})^{2}}
\end{align}
where we introduced \emph{Fermi's constant}\index{Fermi's constant} $G_F$.

Looking back at \eqref{eq:DecayZFull}, we find that in this approximation
\begin{equation}
\Gamma_{ee}=\Gamma_{\mu\mu}=\Gamma_{\tau\tau}\kom \dfrac{\Gamma_{\nu\nu}}{\Gamma_{ee}} \approx 1.977\, .
\end{equation}
Experimentally, it was found that
\begin{equation}
\dfrac{\Gamma_Z}{\Gamma_{ee}} \approx 29.7366 \kom \dfrac{\Gamma_{\text{hadrons}}}{\Gamma_{ee}} \approx 20.80\, .
\end{equation}
Thus, one estimates that the number of light neutrino species is
\begin{equation}
N_{\nu} \approx \dfrac{29.74 \Gamma_{ee}-20.80\Gamma_{ee}-3\Gamma_{ee}}{1.977\Gamma_{ee}} \approx 3.005\, .
\end{equation}
While we made crude approximations here to arrive at this result,
this is strikingly close to the experimentally measured value.
The various partial decay rates for the different sectors have been measured very precisely from which the number of neutrinos lighter than the $Z$ boson was determined to be \cite{ALEPH:2005ab}
\begin{equation}
N_\nu=2.9840\pm 0.0082\, .
\end{equation}
This is a very strong indication that there are no more than three families of quarks and leptons although it does not rule out the possibility of extra neutrinos heavier than the $Z$ boson. This result is beautifully complemented with cosmological observations of the cosmic microwave background which limit the number of neutrino-like particles to be no more than 3.
In both measurements, the precision was such that in order to fit the experiments it would not be possible to have more decay channels including more than the three known neutrinos.

\subsection{Neutrino masses and lepton flavour mixing}

\subsubsection*{Right-handed neutrinos and neutrino masses}
\index{Right-handed neutrinos}

Let us now include right handed neutrinos $\nu_{R}^i$. But before we start, we stress that these particles \emph{have not been detected} and strictly speaking do not have to be part of the Standard Model. However, the fact that all other matter particles have a left-handed and a right-handed part and that neutrinos have been found to be  massive, hints at their existence. These fields are {\it sterile} in the sense that they are singlets under all non-Abelian gauge groups and also have vanishing hypercharges $Y_{\nu_{R}}=0$ making them electrically neutral $Q=0$.\footnote{This can be understood from anomaly cancellation conditions discussed in Sect.~\ref{sec:anomalies_SM}.}
This means that they only feel gravitational interactions, explaining why they might have escaped detection so far.
However, as we will see momentarily, they can play an important role in generating neutrino masses. 

The mass terms for the leptons include the two Yukawa terms similar to the quark case plus an extra term that is possible only for neutrinos corresponding to a direct mass term for right-handed neutrinos
 \begin{equation}\label{eq:LagrangianLeptonsWithRHNeutrinos} 
\cL_{F}^{\text{leptons}}= \cL_{F}^{\text{kinetic, leptons}}+ \cL_{F}^{\text{Higgs-leptons}}+\cL_{F}^{\text{Majorana-mass}}
\end{equation}
where $ \cL_{F}^{\text{kinetic, leptons}}$ encodes the standard kinetic terms for the leptons and
\begin{equation}
\cL_{F}^{\text{Higgs-leptons}}=-y_{ij}^{e}\overline{L}^{i}_{L}He_{R}^{j}-y_{ij}^{\nu}\underbrace{\overline{L}^{i}_{L}\tilde{H}\nu_{R}^{j}}_{\text{Dirac}}+\text{h.c.} \kom \cL_{F}^{\text{Majorana-mass}}= M_{ij}^{\nu}\underbrace{\nu_{R}^{i}\nu_{R}^{j}}_{\text{Majorana}}\, .
\end{equation}

In principle the number of right-handed neutrinos is not limited to coincide with the number of families of the other particles due to the fact that they are not charged under any of the gauge symmetries. So there could be more or less  than three $\nu_{R}$'s since they are sterile. The Majorana mass $M^\nu$ may be very large and these particles may have only indirect implications at low energies. Note that this Majorana mass is the only mass term allowed for the fermions of the Standard Model. So $M^\nu$ are the only mass parameters together with the Higgs mass.

Another question that we should address concerns the nature of neutrinos, namely Dirac or Majorana neutrinos\index{Dirac or Majorana neutrinos}.
Recall that, if $\nu$ were massless then the state with helicity $\lambda=+1/2$ is called neutrino $\nu_{L}$ and the state with helicity $\lambda=-1/2$ is the anti-neutrino $\overline{\nu}_{L}$ (or $\nu_{R}$). The CPT action is such that $\ket{\nu}\neq CPT\ket{\nu}$ but gives $\overline{\nu}_{L}$.
If however $\nu$ is massive, then it has spin $j=1/2$ with the two states $j_{3}=\pm 1/2$. If $\ket{\nu}\neq CPT\ket{\nu}$ the corresponding particle is a \emph{Dirac neutrino}, while $\ket{\nu}= CPT\ket{\nu}$ it is called a \emph{Majorana neutrino}, described by a Majorana spinor and the reality condition implies the particle is the same as the anti-particle. So far there is not enough information to settle if the neutrinos are best represented by Dirac or Majorana spinors.

\subsubsection*{Dirac and Majorana masses}\index{Dirac mass term}\index{Majorana mass term} 

The first two terms in $\cL_F^{\text{leptons}}$ are Yukawa couplings for both electrons and neutrinos coupled to the Higgs. Again once the Higgs gets a VEV ($\langle H\rangle \neq 0$) they will give rise to mass terms for both electron and neutrinos proportional to the electron and neutrino Yukawa couplings and the VEV of the Higgs $v$. This gives rise to standard Dirac mass $m\overline{\psi}_{L}\psi_{R}$.
 The third term is the only explicit mass term allowed in the Standard Model. This is due to the fact that the right-handed neutrinos are by themselves invariant under any gauge transformation and therefore a Lorentz invariant quadratic term is allowed by the gauge symmetries and gives an explicit mass to the right-handed neutrinos. Since $\nu_R$ are written as a 2-component Weyl spinor, the Lorentz invariant combination is then $ M_{ij}^{\nu}{\nu_{R}^{i}\nu_{R}^{j}}$. Expressed in terms of 4-component spinors this term is of the type $M\overline \psi_R\psi_R$ (contrary to the $\overline\psi_L\psi_R$ coming from the Yukawa coupling).
 
A simple way to understand Majorana masses as opposed to Dirac masses is as follows. Recall our discussion of spinor types in chapter~\ref{chap:stsym}. Let us consider two 2-component Weyl spinors $\psi^\alpha$ and $ \chi^\alpha$ and their conjugates. From them we may construct Lorentz invariant quadratic or mass terms of the form $\psi\chi,\psi\psi,\chi\chi$ and their complex conjugates. The first term $\psi\chi+h.c.$ is the standard Dirac mass $\overline\Psi_D \Psi_D$ once the two Weyl spinors are combined in a Dirac fermion  $\Psi_D=\left(\psi^\alpha,\, \overline \chi_{\dot\alpha}\right)^T$. The $\psi\psi$ and $\chi\chi$ terms may  be directly obtained if we combine $\psi$ and $\chi$ in  two Majorana spinors. Namely $\Psi_M^1=(\psi^\alpha,\, \overline \psi_{\dot\alpha})^T$ and $\Psi_M^2=(\chi^\alpha,\, \overline \chi_{\dot\alpha})^T$ and so the Majorana mass terms can be obtained from ${\overline\Psi}_M^1 \Psi_M^1$ and $\overline\Psi_M^2\Psi_M^2$ and the Dirac mass term from $\overline\Psi_M^1 \Psi_M^2$. Notice that if $\psi$ or $\chi$ carry some charge (e.g. electric charge), we could not combine them in Majorana spinors. Right-handed neutrinos are unique in that sense. From the discussion above we may identify the left-handed Weyl spinor $\psi$ with $\psi_L$ and the right-handed conjugate of $\chi$ with $\psi_R$.

\subsubsection*{See-saw mechanism}\index{See-saw mechanism}

In the presence of right-handed neutrinos, the mass matrix for $\nu$'s can be written as
\begin{equation}
m\overline{\psi}_{L}\psi_{R}+\dfrac{1}{2}M \overline{\psi}_{R}\psi_{R}\quad\Rightarrow\quad \left (\begin{array}{cc}
0 & m \\ 
m & M
\end{array} \right )
\end{equation}
focussing here on a single generation. The mass matrix has eigenvalues
\begin{equation}
\lambda=\left (1\pm\sqrt{1+\dfrac{m^{2}}{4M^{2}}}\right )\dfrac{M}{2}, \qquad |\lambda|\cong M, \dfrac{m^{2}}{M}\quad\text{ for }M\gg m\, .
\end{equation}
Therefore, if $M$ is large (say close to the Planck scale, $M_{Planck}\sim 10^{19}$ GeV), this may be an explanation for the smallness of neutrino masses since $m^2/M$ may be naturally very small. Note that a negative mass eigenvalue (corresponding to the negative root) for a fermion can always be rotated away to get a positive physical mass. This mechanism is called see-saw since if we increase $M$ then $m$ decreases and vice versa.


\subsubsection*{Flavour Mixing}\index{Flavour Mixing}

In the presence of right-handed neutrinos, we have a mixing similar to the quark case in the sense that both mass and weak eigenstates do not coincide. Diagonalising the mass matrix leads to the equivalent of the CKM matrix for the lepton sector, known as the \emph{PMNS} matrix. To understand the flavour mixing in the mass eigenstate basis, we have to study the current interactions:
\begin{itemize}
\item The neutral currents are given by
\begin{equation}
\cL_{nc}=Z^{\mu}J^{Z}_{\mu}\kom J^{Z}_{\mu} = -\dfrac{g}{2\cos(\theta_{W})} J^{\mu}_{n}
\end{equation}
with $J^{\mu}_{n}$ previously defined in \eqref{eq:neutral_current_leptons}.
This shows that they remain flavour diagonal similar to the case of quarks.
\item In contrast, the charged currents coupling to $W_{\mu}^{\pm}$ are \emph{not} flavour diagonal.
Indeed, they can be written as
\begin{equation}
\cL_{cc}=-\dfrac{g}{\sqrt{2}}U_{PMNS}^{ij}\left (\overline{e}^{i}_{L}\gamma^\mu W_{\mu}^- \nu_{L}^{j}+\text{h.c.}\right )
\end{equation}
and are determined in terms of the \emph{PMNS matrix}\index{PMNS matrix} (named after Pontecorvo-Maki-Nakagawa-Sakata)
\begin{align}
U_{PMNS} =& \begin{pmatrix} c_{12} c_{13} & s_{12}c_{13} & s_{13} e^{-i\delta} \\
-s_{12}c_{23}-c_{12}s_{23}s_{13}e^{i\delta} & 
c_{12}c_{23}-s_{12}s_{23}s_{13}e^{i\delta} & s_{23}c_{13} \\
s_{12}s_{23}-c_{12}c_{23}s_{13}e^{i\delta} & 
-c_{12}s_{23}-s_{12}c_{23}s_{13}e^{i\delta} & c_{23}c_{13} \end{pmatrix} 
\nn\\
&\qquad\times \mathrm{diag}\left(1, e^{i\alpha_{21}/2}, e^{i\alpha_{31}/2}\right) 
\label{eq:ew_PMNSmatrix}
\end{align}
where $c_{ij}=\cos\beta_{ij}$ and $s_{ij}=\sin\beta_{ij}$, for the 3
angles $\beta_{12}$, $\beta_{23}$, and $\beta_{13}$, and $3$ phases $\delta$, $\alpha_{21}$, $\alpha_{31}$. Note that $\alpha_{21}$, $\alpha_{31}$ appear if there are right-handed neutrinos. The reason being that contrary to the quark case, here we can reduce the original $9$ parameters of the matrix to $6$ and \emph{not} $4$ because there are only three $\mathrm{U}(1)$'s that can be used to reduce the number of parameters, corresponding to a phase rotation to the electron fields. There are no $\mathrm{U}(1)$ symmetries for neutrinos because  the Majorana mass term is not symmetric. Therefore the number of free parameters of the \emph{PMNS} matrix is $9-3=6$. Measurements show that
\begin{align}
\sin^{2}(2\beta_{12})=0.857(4)\kom \sin^{2}(2\beta_{23})=0.95\kom \sin^{2}(2\beta_{13})=0.098(13)\, ,
\end{align}
but the phases have not been measured yet.
\end{itemize}
The value of neutrino masses is still unknown, the only information available experimentally is the difference between their squared masses:
\begin{align}
m_{2}^{2}-m_{1}^{2}=(7.5\pm 0.2)\cdot 10^{-5}\text{eV}^{2}\kom m_{3}^{2}-m_{2}^{2}=0.00252\pm 0.00012\text{eV}^{2}\, .
\end{align}
This range of masses hints at extremely small neutrino masses and can be compared with the top quark mass $173$GeV, illustrating the huge range of masses within the Standard Model fermions.

\subsubsection*{Neutrino oscillations}\index{Neutrino oscillations}
 
If neutrinos are massive, they can oscillate. This means that they can be produced in a reaction as ``weak'' eigenstates $\ket{\nu_{\alpha}}$ as opposed to mass eigenstates $\ket{\nu_{i}}$ and, while they travel large enough distances, they may change flavours, oscillating between different flavours.
Both are related by the mixing matrix $U_{\alpha i}$
\begin{equation}
\ket{\nu_{\alpha}}=\sum_{i=1}^3 \, U_{\alpha i}^{*}\ket{\nu_{i}},  \qquad  \ket{\nu_{i}}=\sum_{\alpha=1}^3 \, U_{\alpha i}\ket{\nu_{\alpha}}
\end{equation}
while they travel for a time $t$
\begin{equation}
\ket{\nu_{\alpha}(t)}=\sum \, U_{\alpha i}^{*}\ket{\nu_{i}(t)}\, .
\end{equation}
The mass eigenstates $\ket{\nu_i}$ have plane wave solutions (like standard solutions of Klein-Gordon equations) that determine easily the time dependence
\begin{equation}
\ket{\nu_i(t)}= {\ee}^{-{\I} (E_it-p_i\cdot x)} \ket{\nu_i(0)}\, .
\end{equation}
Since neutrinos are very light they move close to the speed of light and we can take the ultra-relativistic limit
 \begin{equation}
p_i\gg m_i, \qquad 
E_i=\sqrt{p_i^2+m_i^2}\simeq p_i+\frac{m_i^2}{2p_i}\simeq E+\frac{m_i^2}{2E}\, .
\end{equation}
Setting the speed of light $c=1$ we can approximate 
\begin{equation}
t\simeq L, \qquad p_i\cdot x\simeq EL
\end{equation}
with $L$ the distance travelled by the neutrino at time $t$. Then we can write
\begin{equation}
\ket{\nu_i(L)}= {\ee}^{-{\I} m_i^2L/(2E)} \ket{\nu_i(0)}\, .
\end{equation}

At some point when they are  detected they will also be in terms of a weak interaction and therefore the relevant question is: What is the probability of a neutrino being produced at a flavour eigenstates $\nu_{\alpha}$ and detected in $\nu_{\beta}$ at a distance $L$? Using the equations above we can make an estimate of this probability
\begin{equation}
P_{\alpha\beta}=|\braket{\nu_{\beta}|\nu_{\alpha}(t)}|^{2}=\sum_{ij}\, \ee^{-\I (m_{i}^{2}-m_{j}^{2})L/2E}U_{\beta i}U_{\beta j}^{*}U_{\alpha j}U_{\alpha i}^{*}
\end{equation}
where $L$ is the distance travelled and  $E$ the  energy. Then studying neutrinos travelling long distances would allow us to determine if they oscillate between different flavours.  Oscillations can be observed if the neutrinos travel distances $L$ large enough so that the argument of the exponential is relevant. Concretely, if $\Delta m^2\neq 0, L\geq \lambda(E)$ where
\begin{equation}
\lambda(E)=\frac{2E}{\Delta m^2}=500 m\left(\frac{E}{1{\rm GeV}}\right)\, \left(\frac{1 {\rm eV^2}}{\Delta m^2}\right)\, .
\end{equation}
This means that for energies in the GeV region and mass differences in the eV$^2$ region oscillations can be detected after traveling distances of order of kilometers. This has been observed in many experiments and it is the \emph{best evidence we have for neutrino masses different from zero}. The evidence comes from three sources:
\begin{itemize}
\item \emph{Solar $\nu$-problem.}\index{Solar $\nu$-problem} The thermo-nuclear reactions in the Sun in which Hydrogen nuclei (protons) fuse to produce Helium followed by fusion of Helium to heavier elements produce $\nu_{e}$. The main source of electron neutrinos is the $p+p\raw d+\bar\nu+e^+$ reaction (with $d=p+n$ the deuteron) which account for more than $90\%$ of the neutrinos produced in the Sun. The flux of neutrinos can be computed with much confidence from our understanding of the weak interactions, but the amount of electron neutrinos detected on earth was only $35$\%. This was known as the solar neutrino problem for many years which was a mystery since only $\nu_e$ where possible to detect. Now it is understood in terms of neutrino oscillations in the sense that the other $65$\% is a combination of $\nu_{\mu}$ and $\nu_{\tau}$). It was a big triumph to identify that this problem was a fundamental physics issue rather than an astrophysical issue regarding the physics of the solar model and that the Standard Model with oscillating massive neutrinos actually matches the observations.\footnote{The solution actually requires the study of neutrino oscillations in media with high matter density as the core of the Sun and is known as the \emph{Mikheyev-Smirnov-Wolfenstein (MSW) effect}. Describing the MSW effect is beyond the scope of these lectures.} 
\item\emph{Atmospheric neutrinos.}\index{Atmospheric neutrinos}  Also neutrino oscillations from neutrinos produced in the atmosphere from pions produced by cosmic rays $\nu$: $\pi^{-}\raw \mu^{-}\overline{\nu}_{\mu}\raw (e^{-}\nu_{e}\nu_{\mu})\overline{\nu}_{\mu}$ not $2:1$ ratio. Illustrates the oscillation
\item \emph{Neutrino oscillations in the laboratory.}  On earth, neutrinos can be produced in nuclear reactors and particle accelerators $\nu$'s can give rise to oscillations as long as they are allowed to travel substantial distances. For this the analysis of neutrino oscillations mentioned above applies and need to have detectors very far away from the source. There are several experimental facilities being developed that can detect neutrino oscillations.
\end{itemize}
The experimental evidence for neutrino masses comes from neutrino oscillations. From the equations above, it becomes clear though that the only information we can extract is not about the neutrino masses directly, but only about mass differences, explaining the experimental limits quoted above.

\subsection{4-Fermi Theory and muon decay*}\label{sec:4FermiMuonDecay} 
\index{4-Fermi Theory}\index{EFT!4-Fermi Theory}

We recall from the beginning of this chapter that the 4-Fermi theory was the first successful description of the weak interactions at low energies through an interaction vertex involving four fermions $G_F\, \psi\psi\psi\psi$.
After having written down the full electroweak theory, we are now ready to reconsider how this effective description arises from the GSW model at low energies.

We know that the GSW model contains charge current interactions of the form (recall \eqref{eq:currentsWZFermions})
\begin{equation}
\cL_W=\frac{e}{\sqrt{2}\sin\theta_W}\left(W^{+\mu}J_\mu^++W^{-\mu}J_\mu^-\right)+\frac{e}{\sin\theta_W}Z^\mu J^Z_\mu\, .
\end{equation}
For instance, the lepton sector contributes to $J_\mu^\pm$
\begin{align}
J_\mu^+&= \bar{\nu}_{eL}\gamma_\mu e_L+\bar{\nu}_{\mu L}\gamma_\mu \mu_L+\bar{\nu}_{\tau L}\gamma_\mu \tau_L\, ,\\
J_\mu^{-}&= \bar{e}_{L}\gamma_\mu \nu_{e L}+\bar{\mu}_{L}\gamma_\mu \nu_{\mu L}+\bar{\tau}_{L}\gamma_\mu \nu_{\tau L}\, ,
\end{align}
where for concreteness  we are assuming massless neutrinos and no-mixing issues. From here and knowing the propagator for massive vectors such as $W^\pm$ we can see that for energies $E\ll m_W$ these interactions give rise to amplitudes that can be directly obtained from a 4-Fermi interaction.

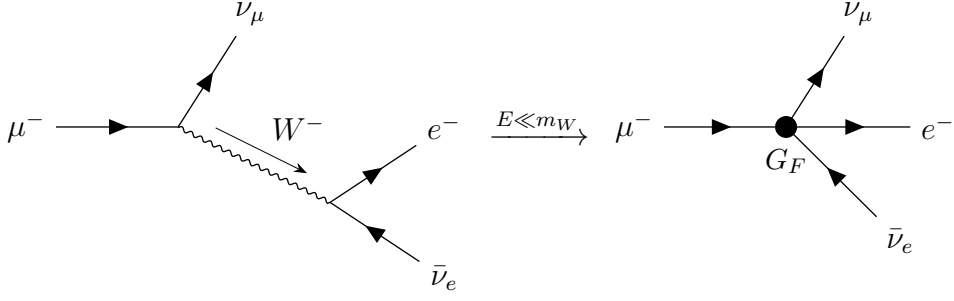
\begin{figure}[t!]
\centering
\begin{tikzpicture}[scale=1.]
\setlength{\feynhanddotsize}{1.5ex}
\begin{feynhand}
\node (o) at (4.75,2) {$\xrightarrow{E\ll m_{W}}$};
\vertex (a0) at (-2.,2) {$\mu^{-}$}; 
\vertex (a1) at (2,1); 
\vertex (d0) at (3.5,2) {$e^{-}$}; 
\vertex (d1) at (3.5,0) {$\bar{\nu}_{e}$}; 
\vertex (b0) at (0,2); 
\vertex (c0) at (0.95,3.5) {$\nu_{\mu}$}; 
\vertex (a00) at (6.,2) {$\mu^{-}$}; 
\vertex (a11) [dot] at (8.,2) {}; 
\node (o1) at (8.,1.55) {\small $G_{F}$}; 
\vertex (d00) at (10.,2) {$e^{-}$}; 
\vertex (d11) at (9.5,0.5) {$\bar{\nu}_{e}$}; 
\vertex (c00) at (8.95,3.5) {$\nu_{\mu}$}; 
\propag [fer] (a0) to (b0);
\propag [pho, mom={$W^{-}$}] (b0) to (a1);
\propag [antfer] (d0) to (a1);
\propag [antfer] (a1) to (d1);
\propag [fer] (b0) to (c0);
\propag [fer] (a00) to (a11);
\propag [antfer] (d00) to (a11);
\propag [antfer] (a11) to (d11);
\propag [fer] (a11) to (c00);
\end{feynhand}
\end{tikzpicture}
\caption{The muon decay $\mu^-\raw e^-+\nu_\mu+\bar\nu_e$ can be approximated at low energies ($E\ll m_W$) through a 4-point fermion vertex in the 4-Fermi effective theory. }\label{fig:muon_decay_4Fermi} 
\end{figure}

Let us describe this in more detail by computing the cross section for muon decay depicted in Fig.~\ref{fig:muon_decay_4Fermi}.
In this case, we can focus on interactions for the first two families of leptons (muons, electrons and their associated neutrinos).
From the form of the propagator for massive spin-$1$ fields in \eqref{eq:propagator_massive_spin_one}, we can see that for energies $E\ll m_W$ the contribution from the $W^{\pm}_{\mu}$-propagator reduces to 
\begin{equation}
\Delta_{\mu\nu}^{W^{\pm}}=-\frac{1}{p^2-m_{W}^2}\left(\eta_{\mu\nu}-\frac{p_\mu p_\nu}{m_{W}^2}\right)\approx \frac{\eta_{\mu\nu}}{m_{W}^2}\, .
\end{equation}
Thus, the process can be written as a 4-Fermi interaction coming from a Lagrangian
 \begin{equation}
\cL_{4F}=-\frac{4G_F}{\sqrt{2}}\left( \bar{e}\gamma_\mu P_{L} \nu_e+\bar{\mu}\gamma_\mu P_L \nu_\mu\right)\,\left(\bar{\nu}_e\gamma^\mu P_L e+\bar{\nu}_\mu\gamma^\mu P_L \mu\right) 
\end{equation}
with 
\begin{equation}
\frac{4G_F}{\sqrt{2}}=\frac{e^2}{2m_W^2\sin^2\theta_W}=\frac{2}{v^2}\kom P_{L}=\dfrac{1-\gamma^{5}}{2}\, .
\end{equation}
Here we have written the left handed fermions in the Dirac notation including the projector $P_L$.
This allows us to see directly that the coupling we get after setting the limit $E\ll m_W$ is the $V-A$ combination mentioned at the beginning of the chapter as proposed before the GSW model by Marshak and Sudarshan in 1958 \cite{Sudarshan:1958vf}, see in particular Eq.~\eqref{eq:FourFermiCurrent}. This can be seen as another success of the GSW model in the sense that it {\it explains} why the 4-Fermi model was successful to describe weak interactions at low energies. It also can relate the Fermi coupling $G_F$ to fundamental parameters of the GSW model.  In this sense the GSW model is a UV completion of the 4-Fermi model. 

As an application of the above observation and in order to illustrate how physical amplitudes are computed and to be as explicit as possible, let us now compute in full detail the decay rate for the muon decay shown in Fig.~\ref{fig:muon_decay_4Fermi}.
More specifically, we will consider the decay $\mu^-\raw e^-+\nu_\mu+\bar\nu_e$ at low energies ($E\ll m_W$) and then start from the 4-Fermi effective theory. This is justified since $m_W (80.385 \, {\rm GeV}) \gg m_\mu (105.6583745\,  {\rm MeV}) $. The decay rate $\Gamma(\alpha\raw\beta)$ from the initial $\alpha$ to final $\beta$ states depend on the interaction matrix $\cM$\footnote{See Appendix~\ref{app:drc} for a review of decay rates and cross sections.} as
\begin{equation}\label{eq:decayratemuon} 
\Gamma(\alpha\raw\beta)=\dfrac{1}{2m_{\alpha}}\,\int\,  \sum_{\text{spins},\ldots}|\cM_{\alpha\beta}|^{2}\dif\rho_{\beta}\, ,
\end{equation}
where $d\rho_f$ is the phase space measure
\begin{equation}
\dif\rho_{\beta}=(2\pi)^{4}\delta^{(4)}\left (p_{\alpha}-\sum_{r\in\beta}\, p_{r}\right )\, \prod_{r\in\beta}\, \dfrac{\dif^{3}p_{r}}{(2\pi)^{3}}\, \dfrac{1}{2p_{r}^{0}}\, .
\end{equation}
where as usual we sum over final moment and the delta function imposes energy-momentum conservation. For this 4-Fermi case the interaction matrix in terms of the particle ($u$) and antiparticle ($v$) wave functions is
\begin{equation}
\cM=-\frac{G_F}{\sqrt{2}}\bar u_e(k)\gamma^\alpha(1-\gamma^5)v_{\nu_e}(q)\bar u_{\nu_\mu}(q')\gamma_\alpha (1-\gamma^5)u_\mu(p)\, .
\end{equation}
Next we need to compute the square of the amplitude  summing over spin states of the final particles
\begin{align}
\frac{1}{2}\sum_{\rm spins}|\cM|^2 & = \frac{G_F^2}{4}\sum_{\rm spins}\left[\bar u_e(k)\gamma^\alpha(1-\gamma^5)v_{\nu_e}(q)\bar v_{\nu_e}(q)\gamma^\beta(1-\gamma^5)u_e(k)\right]\nonumber \\
& \times \left[\bar u_{\nu_\mu}(q')\gamma_\alpha (1-\gamma^5)u_\mu(p)\bar u_\mu(p)\gamma_\beta(1-\gamma^5)u_{\nu_\mu}(q')\right]\nn\\
&\equiv \frac{G_F^2}{4} S_e^{\alpha\beta}S_{\mu,\alpha\beta}\, .
\end{align}
To compute $S_e^{\alpha\beta}$ depending on the electrons and the muon dependent $S_{\mu,\alpha\beta}$, we need to use the fact that the wave functions $u,v$ solve the Dirac equation in momentum space and use the completeness conditions for the sum over spins
\begin{equation}
\sum u^s(p) \bar u^s(p)=\cancel p +m, \qquad \sum v^s(p) \bar v^s(p)=\cancel p - m\, .
\end{equation}
We will also assume that the neutrino masses are negligibly small and set them to zero. Then we can write
\begin{align}
S_e^{\alpha\beta} & =   {\rm Tr}\left[(\cancel k+m_e)\gamma^\alpha (1-\gamma^5) \cancel q \gamma^\beta (1-\gamma^5) \right]\, , \nn \\
S_{\mu,\alpha\beta} & =   {\rm Tr}\left[\cancel q'\gamma_\alpha (1-\gamma^5) (\cancel p + m_\mu)\gamma_\beta (1-\gamma^5) \right] \, .
\end{align}
Now we can use the standard gamma matrix identities \eqref{eq:TrOdd}, \eqref{eq:TrGam4} and \eqref{eq:TrGam5Gam4} to find
\begin{align}
S_e^{\alpha\beta} &= 8\left(k^\alpha q^\beta+k^\beta q^\alpha - k\cdot q \, \eta^{\alpha\beta}-{\I}\epsilon^{\alpha\beta\sigma\rho}k_\sigma q_\rho\right)\, ,\nn \\
S_{\mu,\alpha\beta} & =  8\left( p_\alpha q'_\beta+ p_\beta q'_\alpha - p\cdot q'\,   \eta_{\alpha \beta}-{\I}\epsilon_{\alpha\beta\lambda\tau}q'^\lambda p^\tau \right)\, .
\end{align}
Finally, from here we can easily  compute the contraction $S_1^{\alpha\beta}S_{2,\alpha\beta}$ to get
\begin{equation}\label{eq:SumM2Muon} 
\frac{1}{2}\sum_{\rm spins}|\cM|^2= 64 G_F^2 \left(p\cdot q \right) \left( k \cdot q' \right)\, .
\end{equation}
This finishes the calculation of the integrand in \eqref{eq:decayratemuon}. 

The remaining task is to compute the integral.
We start by plugging the expression \eqref{eq:SumM2Muon} into the partial decay rate \eqref{eq:decayratemuon}
\begin{align}\label{eq:decayratemuon2} 
\Gamma & =  \frac{1}{2m_\mu (2\pi)^5}\int \frac{d^3k}{2 k^0}\int \frac{d^3q}{ 2 q^0}\int \frac{d^3q'}{2 q'^0}  \,\delta^{(4)}(p-k-q-q')\frac{1}{2}\sum_{\rm spins}|\cM|^2\nn \\
& =  \frac{G_F^2}{8\pi^5 m_\mu}\int \frac{d^3 k}{k^0} \frac{d^3 q}{q^0}\frac{d^3 q'}{q'^0}\delta^{(4)}(p-k-q-q')\, (p\cdot q) (k\cdot q')
\end{align}
where we are using that the electron's momentum is $k^\mu$ and the neutrinos have momenta $q^\mu,q'^\mu$. In order to perform the integral, let us consider the following object
\begin{equation}
I_{\mu\nu}(Q)= \int \frac{d^3 q}{|\vec{q}|} \frac{d^3 q'}{|\vec{q'}|}\, \delta^{(4)}(Q-q-q') q_\mu q'_\nu\, .
\end{equation}
Given that the final result should be a function of $Q_\mu$, the only possible tensorial dependence of $I_{\mu\nu}$ is the following
\begin{equation}
I_{\mu\nu}= a Q_\mu Q_\nu + b Q^2  \eta_{\mu\nu} 
\end{equation}
with unknown coefficients $a,b$. In order to determine them, we can contract this equation with $\eta^{\mu\nu}$ and also with $Q^\mu Q^\nu$. Using the fact that $q^2=q'^2=0$ and that inside the integral we can use the delta function condition $Q=q+q'$, we get
\begin{equation}
a+4b=\frac{I}{2}, \qquad a+b= \frac{I}{4}
\end{equation}
where
\begin{equation}
I \equiv \int \frac{d^3 q}{|\vec{q}|} \frac{d^3 q'}{|\vec{q'}|}\delta^{(4)}(Q-q-q') = \int \frac{d^3 q}{|\vec{q}|^2}\delta(Q^0-2|\vec{q}|) =4\pi\int d|\vec{q}|\delta(Q^0-2|\vec{q}|)=2\pi\, .
\end{equation}
Here we used the fact that $I$ is Lorentz invariant and could evaluate it for $Q=(Q^0,0,0,0)$ with no loss of generality.
Therefore, the coefficients $a,b$ lead to
\begin{equation}
I_{\mu\nu}= \frac{\pi}{3}Q_\mu Q_\nu + \frac{\pi}{6} Q^2 \eta_{\mu\nu}\, .
\end{equation}
Plugging this into the integral for $\Gamma$ in \eqref{eq:decayratemuon2}, we find
\begin{equation}
\Gamma=\frac{G_F^2}{3m_\mu (2\pi)^4} \int \frac{d^3k}{k^0}\left[2p\cdot (p-k)\, k\cdot (p-k)+ p\cdot k\, (p-k)^2\right]\, .
\end{equation}
Now we choose the rest frame of the muon where $p=(m_\mu,0,0,0)$.
Further, we can safely take $k=(E, \vec{k})$ with $|\vec{k}| \sim E$ since the electron is much lighter than the muon. Therefore, $\Gamma$ can be approximated by
\begin{equation}\label{eq:decay_muon_4Fermi} 
\Gamma=\frac{2G_F^2 m_\mu}{3m_\mu (2\pi)^3} \int_0^{m_\mu/2} \dif E \, E^2 (3m_\mu -4E)= \frac{G_F^2 m_\mu^5}{192\pi^3}
\end{equation}
which is our final result. The limits of integration correspond to the two extreme cases: the electron at rest and neutrinos in opposite direction $E=0$ and the neutrinos moving in the same direction and the electron in the opposite for which momentum conservation implies $E=m_\mu/2$.

Experimentally the muon lifetime has been well measured to be \cite{ParticleDataGroup:2016lqr} 
\begin{equation}
\tau_{\mu}=\dfrac{1}{\Gamma}\approx 2.1970\times 10^{-6}\text{ seconds}\, .
\end{equation}
From this and the knowledge of the muon mass we can obtain the value of
\begin{equation}
G_F=1.164\times 10^{-5}\text{ GeV}^{-2}\, .
\end{equation}
Of course, the value of $G_F$ can be determined by other transitions.
The fact that they all agree is a further confirmation of the validity of the theory.
Knowing the value of $G_F$ also fixes the VEV of the Higgs 
\begin{equation}
v^2=\dfrac{1}{\sqrt{2}G_F} \quad \Rightarrow\quad v=247\text{ GeV}
\end{equation}
which is very close to the experimentally measured value
\begin{equation}
\text{experimentally: } v=\dfrac{2m_{W}}{g} \approx 246.22\text{ GeV}\, .
\end{equation}
Also, even without knowing the value of $\theta_W$, already having this value for $v$ and using its relations to $m_W$ and $m_Z$ would then require
\begin{equation}
m_W=\dfrac{ev}{2\sin\theta_W}>37.4\text{ GeV}\kom m_Z=\dfrac{m_W}{\cos\theta_W}>m_W\, .
\end{equation}
This information was important to know how heavy the $W$ and $Z$ particles could be when experiments were designed to search them.

\

This concludes this chapter which is at the core of the physics of the Standard Model, in which two of the fundamental interactions known in nature are successfully described in terms of a simple unified gauge theory with many experimental successes: proper explanation of a huge number of accumulated experimental evidence regarding weak interactions, spectacularly confirmed predictions (neutral currents, $W^\pm, Z$, Higgs, etc.), plus precision experimental tests -- an outstanding triumph for fundamental physics.

\chapter{\bf Strong Interactions}\index{Strong Interactions}
\label{chap:qcd}

\vspace{0.5cm}
\begin{equ}[Confinement]
{\it Color particles like quarks and gluons can never be isolated. This has never been proved $\ldots$, but since it is true I for one am happy to leave the proof to the mathematicians.}\\

\rightline{\it Steven Weinberg}
\end{equ}
\vspace{0.5cm}

In the previous chapters, we have seen that interacting theories of helicity $\pm 1$ particles give rise to gauge theories. So far, we investigated the following two phases:
\begin{enumerate}
\item \emph{Coulomb phase}\index{Coulomb phase}: QED with $G=\UO_{EM}$ mediated by a massless field $A_{\mu}$ leading to long range interactions.
\item \emph{Higgs phase}\index{Higgs phase}: GSW model\index{GSW model} (Glashow-Salam-Weinberg) with breaking pattern $G=\SUTw_{L}\times\UO_{Y}\raw\UO_{EM}$ with massive mediators $W^{\pm},Z$ and short range interactions at low energies (in addition to the Coulomb phase).
\end{enumerate}
But there still remain several questions unanswered:
\begin{enumerate}
\item Can we describe strong interactions in terms of a gauge theory? 
\item Can a similar theory to the electroweak theory describe strong interactions that are also short range?
\item Are there other phases of gauge theories that can describe short range interactions (different from the Higgs phase)? 
\end{enumerate}
Recall that in the 1930's Yukawa predicted strong interactions to be mediated by the later discovered $\pi$-mesons\index{$\pi$-mesons} or pions\index{Pions} that having a mass (intermediate between protons and electron, hence the name mesons\index{Mesons}) could explain the fact that the interactions are short ranged.
This is because the scalar potential would decay exponentially with the mass $V\propto \ee^{-mr}/r$ compared with the long-range interactions in QED that give $V\propto 1/r$. However, even though non-relativistic descriptions of the force were partially successful, the efforts to describe it in terms of relativistic QFT failed, mostly due to the need to deal with strong interactions which rendered the standard perturbative expansions used in QED and the electroweak theory useless. The answer to the second question above is then \textbf{No}. Actually, when Weinberg discovered his model for the electroweak interactions, he was trying unsuccessfully to describe the strong interactions.
In this chapter, we will see however that the answer to the first and third questions is \textbf{Yes}.

\section{Motivation for $\text{SU}(3)$}

Let us list a couple of arguments why an $\SUTh_{c}$ gauge theory is an appropriate ansatz to model the string interactions:
\begin{itemize}
\item \emph{Colour and Eightfold way}\index{Eightfold way}. As mentioned in the introduction, cf.~Fig.~\ref{fig:EFW}, the eightfold way model was very successful in classifying the strongly interacting particles: mesons and baryons (hadrons) in terms of representations of an approximate symmetry $\SUTh_f$ (not to be confused with colour $\SUTh_c$) and the proposal of Gell-Mann and Zweig to consider the fundamental representations of this group to be the fundamental components of all hadrons hit the spot by the prediction and subsequent detection of the $\Omega^-$ particle. However this particle also poses a problem in the sense that its composition is $\ket{\Omega^{-}}=\ket{s}\otimes \ket{s}\otimes \ket{s}$ for the strange quark $\ket{s}$. The problem is (again!) the Pauli exclusion principle since this would require three identical quarks in the same quantum state.
The solution (Greenberg \cite{Greenberg:1964pe}, Nambu and Han \cite{Han:1965pf}) was to introduce a new quantum number \emph{colour}\index{Colour} such that 
$$\ket{\Omega^{-}}=\epsilon_{ijk}\ket{s}^{i}\otimes \ket{s}^{j}\otimes \ket{s}^{k} \qquad i,j,k=r,g,b$$
 with $r,g,b$ three colours (for red, green, blue, but of course this has nothing to do with real colours except for the name).
\item \emph{QCD}. Later on it was suggested by Harald Fritzsch, Gell-Mann \cite{Fritzsch:1972jv} and, independently, Julius Wess that this new quantum number would be due to an exact gauge symmetry corresponding to $\SUTh_c$ which was coined Quantum Chromodynamics (QCD) in \cite{Fritzsch:1973pi}. Given the fact the there were three and only three colours the options for a gauge group are limited to those that acted on a three-dimensional space. $\SUTh$ is the only viable choice of gauge group since:
\begin{itemize}
\item $\mathrm{SO}(3)$ cannot distinguish quarks and anti-quarks so, if $\ket{q}\otimes \ket{\bar{q}}$ exists, then also $\ket{q}\otimes \ket{q}$ which are states of fractional charge, which is not how nature behaves,
\item $\mathrm{U}(3)$ could not work either since $\UO\subset\mathrm{U}(3)$ would mediate long range interactions just like in QED. That singled out $\SUTh_c$.
\end{itemize}
\item \emph{Colour blindness}. However, the fact that the theory is formulated in terms of fields (gluons\index{Gluons} and quarks) that are not the observed degrees of freedom in nature (hadrons\index{Hadrons}) made it difficult to see how this gauge theory could make contact with the observed spectrum. 
\item \emph{Leptons}. Leptons do not carry colour which is why it seems unnatural to assign to them a gauge theory with gauge group $\SUTh_{c}$ to describe strong interactions that are not experienced by leptons. So they should be singlets under this symmetry.

\begin{figure}[t!]
\centering
 \includegraphics[scale=0.75]{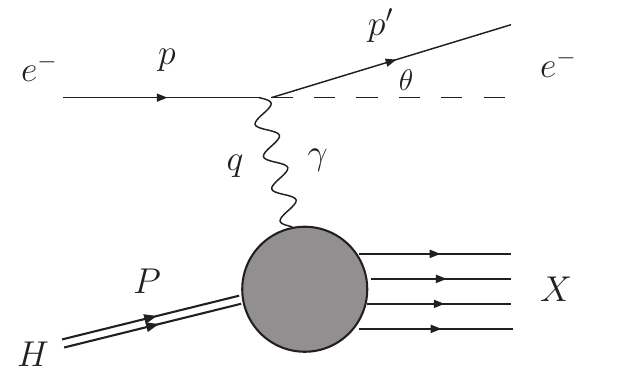}
\caption{\emph{Deep inelastic scattering} (DES). Scattering of electrons off protons exposed the composite structure of the protons. Analogous to the old Rutherford experiment to uncover the structure of atoms. In this case a high energy electron $e^-$ of momentum $p$ scatters off a hadron $H$ (which in this case is a proton) with centre of mass momentum $P$. The interaction is mediated by a photon $\gamma$ of momentum $q$. The electron is scattered at an angle $\theta$. The proton absorbs a good part of the energy and emits further hadrons $X$ (hence the inelastic nature of the event). This event can be explained if both $H$ and $X$ are composed of quarks and gluons.
}\label{fig:DIS}
\end{figure}

\item \emph{Experimental support}. But the proposal gained support experimentally. 
\begin{itemize}
\item \emph{Deep Inelastic Scattering (DIS)}\index{Deep Inelastic Scattering (DIS)}. In 1969 experiments at SLAC probed the structure of protons by bombarding them with electrons at high energies. This was the modern version of the Rutherford experiments that uncovered the structure of the atom. Instead of $\alpha$-particles scattering off gold atoms, this time high energy electrons bombarded protons and the result was equally spectacular. The experiments showed that \emph{hadrons, like protons, behave as composed particles experimentally}, cf.~Fig.~\ref{fig:DIS}. Not only that, these experiments also showed that the proton's components are weakly interacting once the energy is increased. This is manifested by the fact that the structure of the proton remained the same independent of how hard it was struck. In the graph in Fig.~\ref{fig:BjorkenScaling}, it can be illustrated by the presence of an approximate scale invariance (flat curves) of the proton structure function. It hinted at the protons being composed of more fundamental particles that interact very weakly (to explain the approximate scaling, known as \emph{Bjorken scaling}\index{Bjorken scaling}). These components were called \emph{Partons}\index{Partons} and only several years later were identified with the quarks and gluons of QCD.

\begin{figure}[t!]
\centering
 \includegraphics[scale=0.55]{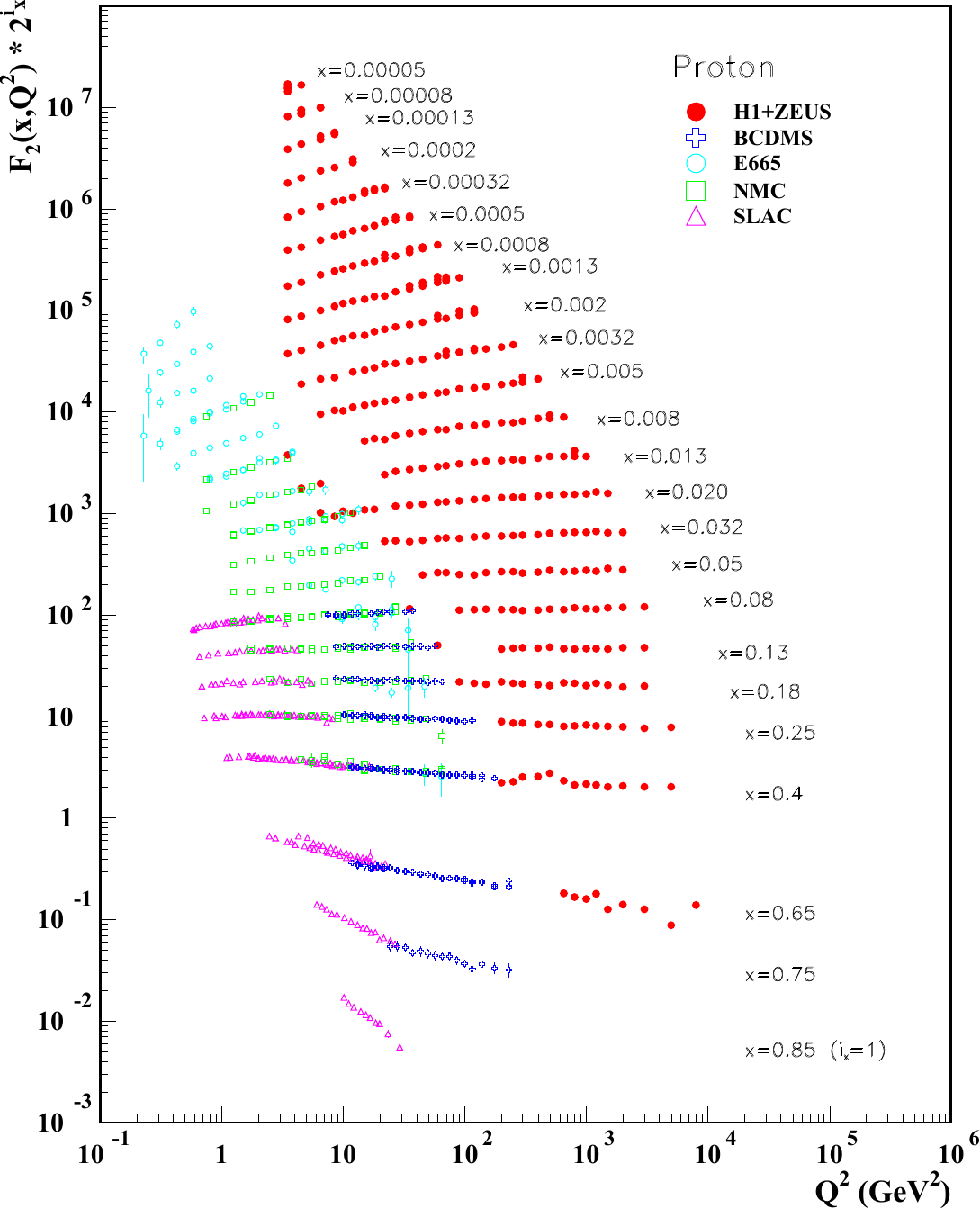}
\caption{\emph{Bjorken scaling}. The structure of the proton remains the same no matter how hard it is struck. Figure taken from the particle data group, see \href{https://pdg.lbl.gov/2020/reviews/rpp2020-rev-structure-functions.pdf}{this link}.}\label{fig:BjorkenScaling} 
\end{figure}
\item \emph{Three colours}. An impressive source of support for the three colour hypothesis can be obtained by considering the following ratio of two processes.
The first is the annihilation of an electron-positron into $\mu^{-}\mu_{+}$,
\begin{align*}
\begin{tikzpicture}[scale=1.]
\setlength{\feynhanddotsize}{1.5ex}
\begin{feynhand}
\node (o) at (-3.5,2) {$e^{-}e^{+}\raw \mu^{-}\mu^{+}=$};
\vertex (a0) at (-1.5,3) {$e^{-}$}; 
\vertex (a1) at (2,2); 
\vertex (d0) at (3.5,3) {$\mu^{-}$}; 
\vertex (d1) at (3.5,1) {$\mu^{+}$}; 
\vertex (b0) at (0,2); 
\vertex (c0) at (-1.5,1) {$e^{+}$}; 
\propag [fer] (a0) to (b0);
\propag [pho, mom={$\gamma$}] (b0) to (a1);
\propag [fer] (a1) to (d0);
\propag [fer] (d1) to (a1);
\propag [fer] (b0) to (c0);
\end{feynhand}
\end{tikzpicture}
\end{align*}
and the second corresponds to the decay into hadrons
\begin{align*}
\begin{tikzpicture}[scale=1.]
\setlength{\feynhanddotsize}{1.5ex}
\begin{feynhand}
\node (o) at (-3.5,2) {$e^{-}e^{+}\raw q\bar{q}=$};
\vertex (a0) at (-1.5,3) {$e^{-}$}; 
\vertex (a1) at (2,2); 
\vertex (d0) at (3.5,3) {$q\;\text{ (hadrons)}$}; 
\vertex (d1) at (3.5,1) {$\bar{q}\;\text{ (hadrons)}$}; 
\vertex (b0) at (0,2); 
\vertex (c0) at (-1.5,1) {$e^{+}$}; 
\propag [fer] (a0) to (b0);
\propag [pho, mom={$\gamma$}] (b0) to (a1);
\propag [fer] (a1) to (d0);
\propag [fer] (d1) to (a1);
\propag [fer] (b0) to (c0);
\end{feynhand}
\end{tikzpicture}
\end{align*}
These are simple electromagnetic interactions with the vertices just differing by the electric charge of the outgoing particles. Therefore, when taking the ratio, one can probe the electric charges involved in the two different processes.
That is, we can define the \emph{R-factor}
\begin{equation}\label{eq:DefRFactor} 
R=\dfrac{\sigma(e^{-}e^{+}\raw\text{ hadrons})}{\sigma(e^{-}e^{+}\raw\mu^{-}\mu^{+})}\sim N_{c}\sum Q_{q}^{2}
\end{equation}
with $N_{c}$ the number of colours and $Q_{q}^{2}$ the charges of quarks below the top mass which are the energies available for electron-positron colliders so that
\begin{equation}
3\sum Q_{q}^{2}=2\left (\dfrac{2}{3}\right )^{2}+3\left (-\dfrac{1}{3}\right )^{2}=\dfrac{11}{3}\, .
\end{equation}
More generally, we have depending on the probed energies
\begin{equation}
  N_c\sum Q_{q}^{2}=
    \begin{cases}
      \frac{2}{3}\, N_c & uds\quad  \text{light}\, ,\\
      \frac{10}{9}\, N_c  & udsc \quad \text{light}\, ,\\
      \frac{11}{9}\, N_c   & udscb \quad \text{light}\, .
    \end{cases}       
\end{equation}
This fits experiment with $N_{c}=3$ to high precision. These numbers are such that it is impossible to have $N_{c}=2,4,\ldots$ or any other integer when compared to experiments. Furthermore, as illustrated in Fig.~\ref{fig:Rfactor}, the experiments show a series of plateaus with $R$ increasing once the threshold for the mass of one of the quarks is reached (charm, bottom for instance). As it can be observed in Fig.~\ref{fig:Rfactor}, one finds three regimes:
\begin{enumerate}
\item the horizontal green line is the plateau corresponding to contributions coming only from the $u,d,s$ quarks ($R=2$), 
\item then at energies of order 4 GeV (after the $\psi(2s)$ resonance), there is a plateau in which the $c$ quark can be produced $R=10/3$,
\item and then at energies close to 10 GeV, after the $\Upsilon$ resonance there is the next plateau $R=11/3$ when the $b$ quarks can also be produced. 
\end{enumerate}
This includes 5 quarks, the top quark threshold (top mass $\sim 174$ GeV) needs higher energies.
However, at higher energies there is also the $Z$ resonance that can also be seen in Fig.~\ref{fig:Rfactor}, that opens a new decay channel with the $Z$ boson as the mediator instead of the photon.
If we compute the above ratio $R$ for $\gamma$ and $Z$ as mediators, we find theoretically $R=20.09$ which agrees well with the experimental value $R=20.79\pm 0.04\, $ and the agreement improves when loop corrections are included. This is compelling evidence of the three colour hypothesis and therefore for a $\mathrm{SU}(3)_c$ gauge theory of the strong interactions.
\end{itemize}
\end{itemize}

\begin{figure}[t!]
\centering
 \includegraphics[scale=0.75]{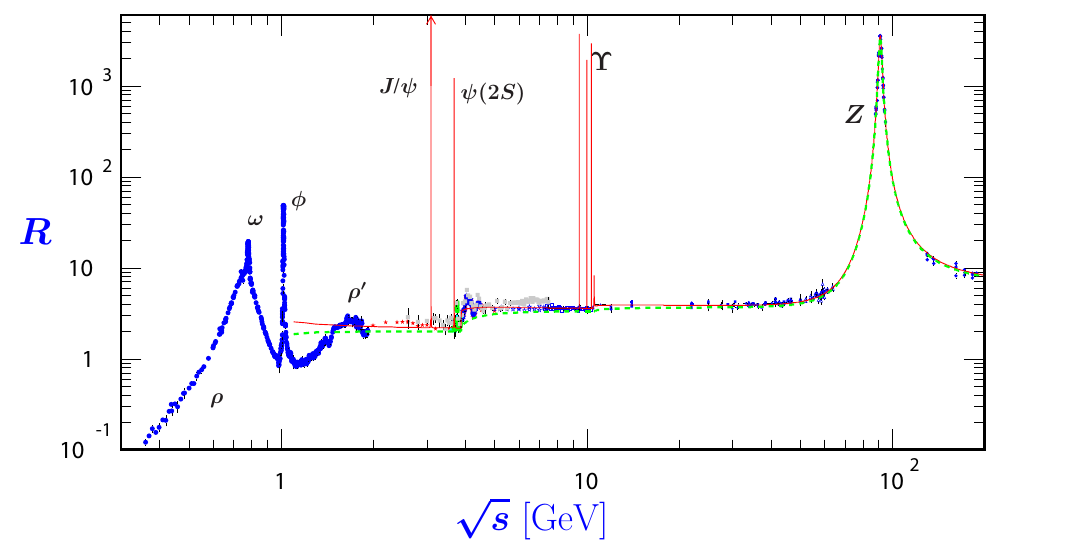}
\caption{Evidence for three colours. Electron-positron scattering to muons and hadrons allows the definition of the $R$ factor as in Eq.~\eqref{eq:DefRFactor}. Depending on the masses of quarks there are different plateaus (besides the resonances) at values corresponding to the masses of the quarks that fit below the threshold. Notice that at the energies shown the top quark does not contribute. Furthermore once the $S$ particle threshold is reached there is the new channel corresponding to $Z$ boson exchange together with the original photon-exchange. The value of $R$ fits only if $N_c=3$.
Figure from the particle data group, see \href{https://pdg.lbl.gov/2012/hadronic-xsections/rpp2012-sigma_R_ee_plots.pdf}{this link}.
}\label{fig:Rfactor} 
\end{figure}

\section{Quantum Chromodynamics (QCD)}\index{Quantum Chromodynamics (QCD)}

Having established the $\mathrm{SU}(3)_c$ structure of strong interactions and given our knowledge of gauge theories so far, we will then start studying QCD as a gauge theory of the symmetry group $\mathrm{SU}(3)_c$ .

The most general Lagrangian for the $\SUTh_{c}$ gauge theory with fermions in the fundamental representation is
\begin{equation}
\cL=-\dfrac{1}{4}\left (G_{\mu\nu}^{A}\right )^{2}+\I\bar{q}_{i}\, \cancel{D}_{ij}\, q_{j}-m_{i}\bar{q}_{i}q_{i}+\theta G_{\mu\nu}^{A}\tilde{G}_{\mu\nu}^{A}\, .
\end{equation}
The indices $i,j$ are color indices. The field strength is defined as
\begin{equation}
G_{\mu\nu}^{A}=\p_{\mu}G_{\nu}^{A}-\p_{\nu}G_{\mu}^{A}+g_{s}f^{ABC}G_{\mu}^{B}G_{\nu}^{C}
\end{equation}
and the covariant derivative reads as usual
\begin{equation}
(D_{\mu})_{ij}=\p_{\mu}\delta_{ij}-\I g_{s}G_{\mu}^{A}T^{A}_{ij}\, .
\end{equation}
The mediators $G_{\mu}^{A}$, $A=1,\ldots ,8$, $\dim(\SUTh)=8$, are called \emph{gluons}\index{Gluons}. A convenient choice of generators for $\SUTh$ is
\begin{equation}
T_{ij}^{A}=\dfrac{1}{2}\lambda_{ij}^{A}
\end{equation}
where the matrices $\lambda^{A}$ are the \emph{Gell-Mann matrices}\index{Gell-Mann matrices} which are the analogues of the Pauli matrices of $\mathrm{SU}(2)$ for $\SUTh$:
\begin{align}
&\lambda^{1}={\begin{pmatrix}0&1&0\\1&0&0\\0&0&0\end{pmatrix}}\kom	
\lambda^{2}={\begin{pmatrix}0&-\I&0\\\I&0&0\\0&0&0\end{pmatrix}}\kom
\lambda^{3}={\begin{pmatrix}1&0&0\\0&-1&0\\0&0&0\end{pmatrix}}\kom \lambda^{4}={\begin{pmatrix}0&0&1\\0&0&0\\1&0&0\end{pmatrix}}\\[0.5em]
&\lambda^{5}={\begin{pmatrix}0&0&-\I\\0&0&0\\ \I&0&0\end{pmatrix}}\kom
\lambda^{6}={\begin{pmatrix}0&0&0\\0&0&1\\0&1&0\end{pmatrix}}	\kom \lambda^{7}={\begin{pmatrix}0&0&0\\0&0&-\I\\0&\I &0\end{pmatrix}}\kom
\lambda^{8}={\frac {1}{\sqrt {3}}}{\begin{pmatrix}1&0&0\\0&1&0\\0&0&-2\end{pmatrix}}\, .\nn
\end{align}
These matrices satisfy
\begin{equation}
\tr(T^{A}T^{B})=\dfrac{1}{2}\delta^{AB}\, .
\end{equation}
Notice that their structure is very similar to that of the Pauli matrices, the first three are just the Pauli matrices illustrating an $\SUTw$ embedding within $\SUTh$. Also, contrary to $\SUTw$ in which only $\sigma_3$ is diagonal, here both $\lambda_3$ and $\lambda_8$ are diagonal, illustrating the fact that $\SUTh$ has rank two. This implies that two of the generators can be diagonalised simultaneously and the weight and root diagrams are 2-dimensional as we have seen for the eightfold way.

The QCD Lagrangian\index{QCD Lagrangian} is a particular case of the general Yang-Mills case. As we have discussed, the kinetic term for the gluons gives rise to cubic and quartic self-couplings.The vertices involved are
\begin{align*}
\begin{tikzpicture}[scale=1.2]
\setlength{\feynhanddotsize}{1.5ex}
\begin{feynhand}
\node (o) at (5,-4) {\small$=\I g_{s}f^{ABC}\left [\eta^{\mu\nu}(p_{1}-p_{2})^{\rho}+\eta^{\nu\rho}(p_{2}-p_{3})^{\mu}+\eta^{\rho\mu}(p_{3}-p_{1})^{\nu}\right ]$};
\vertex (a2) at (0.25,-4) {$B,\nu$}; 
\vertex (b2) at (-1.25,-4); 
\vertex (c2) at (-2.5,-3) {$A,\mu$}; 
\vertex (d2) at (-2.5,-5) {$C,\rho$}; 
\propag [glu, mom={$p_{2}$}] (a2) to (b2);
\propag [glu, mom={$p_{3}$}] (d2) to (b2);
\propag [glu, mom={$p_{1}$}] (c2) to (b2);
\end{feynhand}
\end{tikzpicture}
\end{align*}
\begin{align*}
\begin{tikzpicture}[scale=1.2]
\setlength{\feynhanddotsize}{1.5ex}
\begin{feynhand}
\node (o) at (9.75,-4) {\small$=\I g^{2}_{s}f^{ABE}f^{CDE}\left (\eta^{\mu\rho}\eta^{\nu\sigma}-\eta^{\mu\sigma}\eta^{\nu\rho}\right )+ \I g_{s}^{2}f^{ACE}f^{BDE}\left (\eta^{\mu\nu}\eta^{\sigma\rho}-\eta^{\mu\sigma}\eta^{\nu\rho}\right )$};
\node (1) at (7.75,-5) {\small$+ \I g_{s}^{2}f^{ADE}f^{BCE}\left (\eta^{\mu\nu}\eta^{\sigma\rho}-\eta^{\mu\rho}\eta^{\nu\sigma}\right )$};
\vertex (a31) at (4.5,-3) {$B,\nu$}; 
\vertex (a32) at (4.5,-5) {$C,\rho$}; 
\vertex (b3) at (3.25,-4); 
\vertex (c3) at (2.,-3) {$A,\mu$}; 
\vertex (d3) at (2.,-5) {$D,\sigma$}; 
\propag [glu] (a31) to (b3);
\propag [glu] (a32) to (b3);
\propag [glu] (d3) to (b3);
\propag [glu] (c3) to (b3);
\end{feynhand}
\end{tikzpicture}
\end{align*}
Also the coupling of the gauge field to fermions coming from the covariant derivative takes the form
\begin{align*}
\begin{tikzpicture}[scale=1.2]
\setlength{\feynhanddotsize}{1.5ex}
\begin{feynhand}
\node (o) at (2.25,-4) {\small$=\I g_{s}\gamma^{\mu}T^{A}_{ij}$};
\vertex (a2) at (0.25,-4) {$A,\mu$}; 
\vertex (b2) at (-1.25,-4); 
\vertex (c2) at (-2.5,-3) {$q_{i}$}; 
\vertex (d2) at (-2.5,-5) {$\bar{q}_{j}$}; 
\propag [glu] (b2) to (a2);
\propag [fer] (b2) to (d2);
\propag [antfer] (b2) to (c2);
\end{feynhand}
\end{tikzpicture}
\end{align*}

\section{Interaction potentials}

Let us next try to get a better grasp of what the strong interactions actually are. It is always helpful to compare with the well known case of QED.

We know in QED the study of the scattering of $e^{-}p^{+}\raw e^{-} p^{+}$ (time running upwards)
\begin{align*}
\begin{tikzpicture}[scale=1.]
\setlength{\feynhanddotsize}{1.5ex}
\begin{feynhand}
\node (o) at (-3.5,2) {$e^{-}p^{+}\raw e^{-} p^{+}=$};
\vertex (a0) at (-1.5,3) {$e^{-}$}; 
\vertex (a1) at (2,2); 
\vertex (d0) at (3.5,3) {$p^{+}$}; 
\vertex (d1) at (3.5,1) {$p^{+}$}; 
\vertex (b0) at (0,2); 
\vertex (c0) at (-1.5,1) {$e^{-}$}; 
\propag [fer] (b0) to (a0);
\propag [pho] (b0) to (a1);
\propag [fer] (a1) to (d0);
\propag [fer] (d1) to (a1);
\propag [fer] (c0) to (b0);
\end{feynhand}
\end{tikzpicture}
\end{align*}
can be used to determine the Coulomb interaction potential.
Indeed, one can show that
\begin{equation}
|\cM|\propto -\frac{e^2}{{|\mathbf p}|^2}=\tilde V(|\mathbf{p}|^2)
\end{equation}
that (after a Fourier transform) in position space reads
\begin{equation}
V(r)\propto -\frac{e^2}{4\pi r}
\end{equation}

For QCD, a typical process for quark-antiquark scattering looks like 
\begin{align*}
\begin{tikzpicture}[scale=1.]
\setlength{\feynhanddotsize}{1.5ex}
\begin{feynhand}
\node (o) at (4.25,-4) {\small$\sim g_s^2T^{A}_{ji}T^{A}_{kl}$};
\vertex (a2) at (1,-4); 
\vertex (b2) at (-1,-4); 
\vertex (c2) at (-2.5,-3) {$q_{j}$}; 
\vertex (d2) at (-2.5,-5) {${q}_{i}$};
\vertex (e2) at (2.5,-3) {$\bar{q}_{l}$}; 
\vertex (f2) at (2.5,-5) {$\bar{q}_{k}$}; 
\propag [glu] (b2) to (a2);
\propag [antfer] (b2) to (d2);
\propag [fer] (b2) to (c2);
\propag [antfer] (f2) to (a2);
\propag [fer] (e2) to (a2);
\end{feynhand}
\end{tikzpicture}
\end{align*}
This process is slightly more involved than the analogue for QED above: the incoming quarks and anti-quarks are in $\mathbf{3}\otimes\bar{\mathbf{3}}$ representations of $\mathrm{SU}(3)$ that couple naturally to the gluons who are in the adjoint representation (and produce gauge singlets) and the same for the outgoing quark-antiquark pair. For each case incoming and outgoing quark-antiquark pair, we know from group theory that
\begin{equation}
\mathbf{3}\otimes\bar{\mathbf{3}}={\mathbf{8}}+{\mathbf{1}}
\end{equation}
where the octet is just equivalent to the adjoint representation given by the Gell-Mann matrices and the singlet would be the combination proportional to the trace of the matrix $q_i\bar q^j $: ($\ket{1}\propto \ket{1\bar 1}+ \ket{2\bar 2}+\ket{3\bar 3} $).
Using the Gell-Mann matrices above, it is easy to check that for the octets $T^{A}T^{A}<0$ meaning a repulsive force, whereas for the singlet combination $T^{A}T^{A}>0$ attractive. This is similar to the Coulomb interactions. 
More explicitly, we can compute the potential as follows:
\begin{itemize}
\item \emph{\textbf{Singlet state}}. 

Let us consider first the original quark/anti-quark pair to be the singlet
\begin{equation}
\ket {q\bar q}_S=\frac{1}{\sqrt{3}}\left(\ket{R\bar R}+\ket{G\bar G}+\ket{B\bar B}\right)\qquad R,G,B=1,2,3\, .
\end{equation}
That is, the incoming quarks have the same colour $i=k$ (red and anti-red, green and anti-green, blue and anti-blue) and using the explicit representation of the Gell-Mann matrices we have for each
\begin{equation}
T^A_{j1}T^A_{1l}=\begin{pmatrix} \frac{1}{3} & 0 & 0 \\ 0 & \frac{1}{2} & 0 \\ 0 & 0 & \frac{1}{2} \end{pmatrix}\kom T^A_{j2}T^A_{2l}=\begin{pmatrix} \frac{1}{2} & 0 & 0 \\ 0 & \frac{1}{3} & 0 \\ 0 & 0 & \frac{1}{2} \end{pmatrix}\kom T^A_{j3}T^A_{3l}=\begin{pmatrix} \frac{1}{2} & 0 & 0 \\ 0 & \frac{1}{2} & 0 \\ 0 & 0 & \frac{1}{3} \end{pmatrix}\, .
\end{equation}
The strength of the interaction can be computed by taking the trace (summing over all possibilities) of each of these matrices as indicated by the expression of $\ket {q\bar q}_S$ above, which  gives for each colour $+(4/3)$ (times 3 for the 3 colours  and times $1/(\sqrt{3})^2$ from the normalisation) to give
\begin{equation}
V(r)=-\frac{4}{3} \frac{g_s^2}{4\pi r}=-\frac{4}{3} \frac{\alpha_s}{r}\, .
\end{equation}
The minus sign indicates that, similar to the electromagnetic case, this is an attractive force.
It means in particular that bound states can exist (just like an atom or positronium for electromagnetism). 
\item\emph{\textbf{Octet state}}. 

Now for the octet $j\neq k$, it is easy to show
\begin{equation}
T^A_{ij}T^A_{kl}=-\frac{1}{6}\delta_{ij} \delta_{kl}\kom j\neq k\, .
\end{equation}
This illustrates that the colours of the outgoing particles are the same as the colours of the incoming particles as it should. The interaction potential is
\begin{equation}
V(r)=+\frac{1}{6} \frac{g_s^2}{4\pi r}=+\frac{1}{6} \frac{\alpha_s}{r}\, .
\end{equation}
Since the sign is positive, the interaction is repulsive. This means we cannot create bound states from the octets. This goes a long way towards explaining why mesons are colour singlets. 
\end{itemize}

This result also can be extended for bound states of three quarks. Since
\begin{equation}
{\bf 3}\otimes {\bf 3} \otimes {\bf 3}={\bf 10}\oplus {\bf 8}\oplus {\bf 8}\oplus {\bf 1}
\end{equation}
it can be seen again that only the singlet combination gives an attractive interaction. This is expected since recall that ${\bf 3}\otimes {\bf 3} ={\bf 6}\oplus {\bar {\bf 3}} $ so the singlet in 
${\bf 3}\otimes {\bf 3} \otimes {\bf 3}$ also comes from the singlet in the ${\bf 3}\otimes {\bar {\bf 3}} $ product. In this case it corresponds to the invariant antisymmetric combination
\begin{equation}
\ket {q q q}_S=\frac{1}{\sqrt{6}}\left(\ket{RGB}+\ket{BRG}+\ket{GBR}-\ket{BGR}-\ket{RBG}-\ket{GRB}\right)\, .
\end{equation}
This is a first indication that quarks and anti-quarks attract each other in colour singlets that can be identified with mesons, whereas the colour octets being repulsive do not form bound states.
Similarly out of three quarks we can form another colour singlet $\epsilon_{ijk}\psi_i\psi_j\psi_k$ which are the baryons. In summary, colour singlets are \emph{hadrons}\index{Hadrons} which can be distinguished in
\begin{itemize}
\item $q_{i}\bar{q}_{i}$: \emph{mesons}\index{Mesons}
\item $\varepsilon_{ijk}q_{i}q_{j}q_{k}$: \emph{baryons}\index{Baryons}
\end{itemize}
Notice that this is consistent with the fact that only colour singlets are observed in nature, but it does not {\it explain it}. Contrary to QED, in which both bound states and isolated fundamental states, such as free electrons exist; in QCD quarks on the other hand only live in bound states.
Below, we will present arguments towards understanding why this is the case.

\section{Asymptotic freedom}\label{sec:AsymptoticFreedom}\index{Asymptotic freedom}

The most important property of QCD and many Yang-Mills systems is \emph{asymptotic freedom} (Gross-Wilczek \cite{Gross:1973ju}, Politzer \cite{Politzer:1973fx}, 1973) that we will discuss now.
It is known from general QFT that proper renormalisation techniques lead to the fact that coupling constants actually change with energy.\footnote{This subject is covered in detail in the advanced quantum field theory (AQFT) course.}
For instance, quantum contributions to vacuum polarisation\index{Vacuum polarisation} for the gauge fields include loops of matter and gauge fields which implies that the gauge coupling $g_{s}$ becomes energy dependent. This is well familiar from e.g. QED. For general Yang-Mills, due to the additional self interactions of the gauge fields, the one-loop diagram involves not only fermions and scalars in the loops but also the gauge fields themselves.
The QCD $\beta$-function\index{$\beta$-function!QCD}\index{QCD!$\beta$-function} is given by
\begin{equation}
\beta(\alpha_{s})=\mu\dfrac{\dif}{\dif\mu}\alpha_{s}\kom \alpha_{s}=\dfrac{g_{s}^{2}}{4\pi}\, .
\end{equation}
To leading order, one finds
\begin{equation}
\beta(\alpha_{s})=-\dfrac{\alpha_s^2}{4\pi} b\, .
\end{equation}
Here the coefficient $b$ and higher order coefficients are determined by group theoretical numbers. For $\mathrm{SU}(N_{c})$ with $N_{f}$ flavours,  
\begin{equation}
b=\dfrac{11}{3}N_{c}-\dfrac{2}{3}N_{f}\, ,
\end{equation}
and the running coupling is given by
\begin{equ}[Running coupling for $\mathrm{SU}(N_{c})$ with $N_{f}$ flavours]
\begin{equation}
\dfrac{1}{g_{s}^{2}(\mu)}=\dfrac{1}{g_{s}^{2}(\Lambda)}-\dfrac{1}{(4\pi)^{2}}\left [\dfrac{11}{3}N_{c}-\dfrac{2}{3}N_{f}\right ]\log\left (\dfrac{\Lambda^{2}}{\mu^{2}}\right )
\end{equation}
\end{equ}
where $\Lambda$ is some reference scale. In comparison, recall that the running of the QED coupling is given by
\begin{equation}
\dfrac{1}{e^{2}(\mu)}=\dfrac{1}{e^{2}(\Lambda)}+\dfrac{1}{12\pi^{2}}\log\left (\dfrac{\Lambda^{2}}{\mu^{2}}\right )\, .
\end{equation}
It leads to a Landau pole\index{Landau pole} at $\mu\sim 10^{286}$eV at which point the theory ceases to be valid.
However, this energy is so high that it is clear the theory will need modifications at much smaller energies, knowing for instance that gravity cannot be neglected at the Planck scale. 

\begin{figure}[t!]
\centering
 \includegraphics[width=0.9\linewidth]{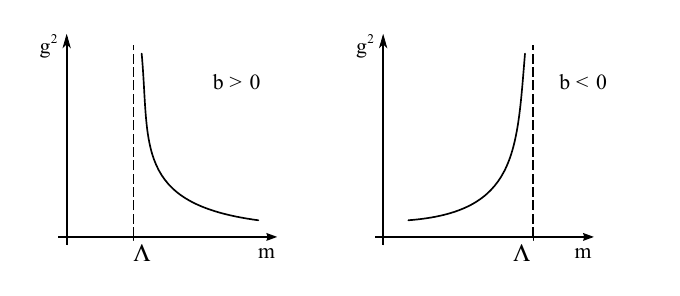}
\caption{The running of the coupling constants for asymptotically free ($b>0$) and for asymptotically slave ($b<0$) theories.}\label{fig:RunCouplingSU3}
\end{figure}

Yang-Mills is much richer than QED. In QCD, we need to distinguish the different behaviours depending on whether $N_{c}$ is greater or smaller than $2 N_{f}/11$, cf. Fig. \ref{fig:RunCouplingSU3}.
For QCD, we have $N_{c}=3$ and $N_{f}=6$ which implies that the theory is \emph{asymptotically free}! This means that, despite the coupling being relatively strong at low energies, it decreases logarithmically with energy and the theory is well behaved in the ultra-violet. In this sense it is better behaved than QED. The main difference is the factor $-11N_c/3$ due to the self interactions of gauge bosons.
This contribution has a definite sign opposite from that due to matter fields. If we insist that a given gauge theory is well defined in the UV, this puts a bound on the number of matter fields, in this case on $N_f<11N_c/2$ which is well satisfied by QCD.

\begin{figure}[t!]
\centering
\includegraphics[width=0.9\linewidth]{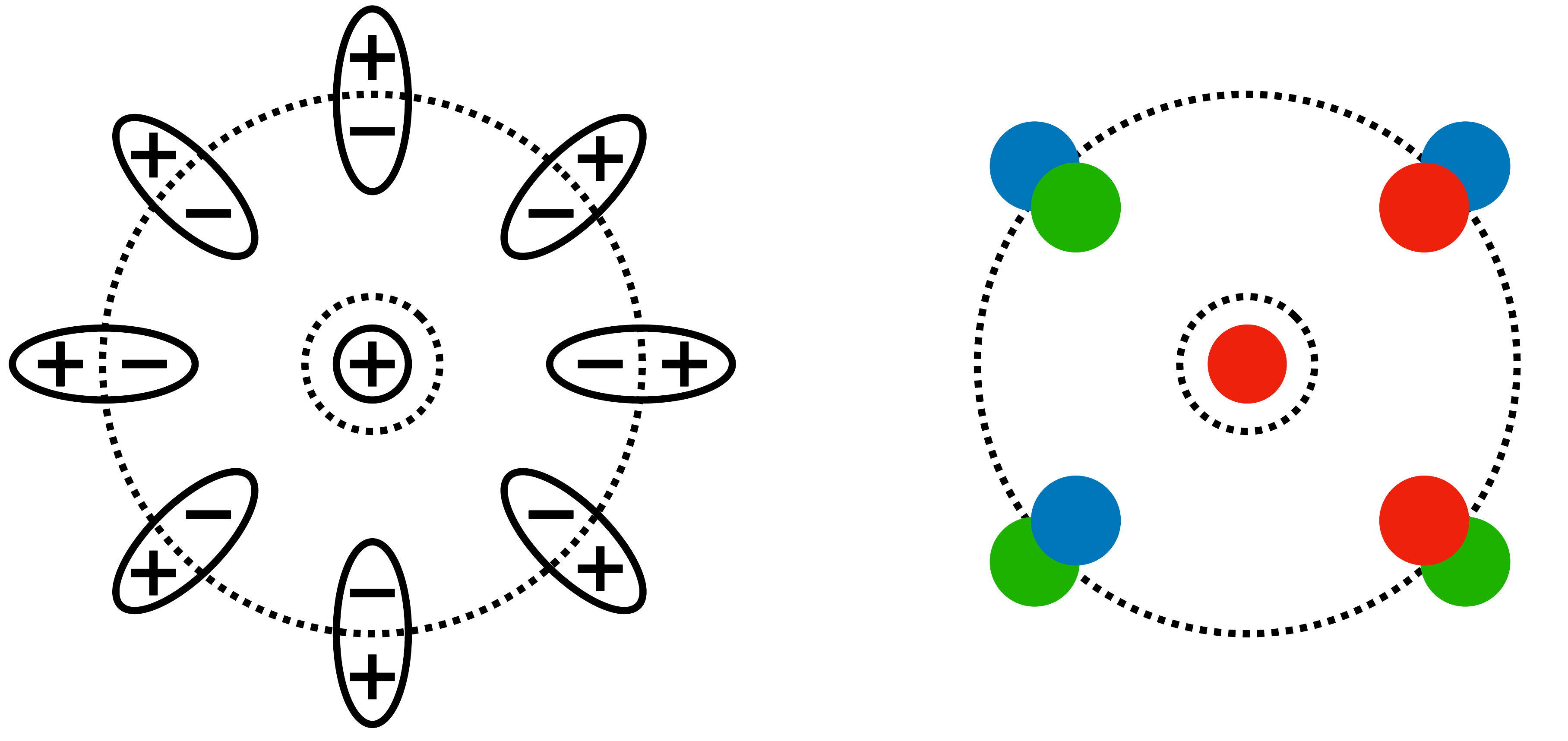}
\caption{Screening vs Anti-screening. A cartoon representation of the screening effect in QED ($b<0$) for which the effective electric charge increases closer to the bare charge and the anti-screening effect for asymptotically free theories ($b>0$) such as QCD in which the net colour (red in this case) decreases closer to the bare particle.}
\label{fig:Screening}
\end{figure}

This difference in sign between QED and QCD is crucial. In QED it leads to what is known as the \emph{screening effect} in which the contributions to the vacuum polarisation screen the value of the bare electric charge. We can imagine a vacuum with pairs of particle and anti-particles being produced in the vicinity of an electric charge in which the electromagnetic attraction somehow screens the value of the bare charge in such a way that the effective charge decreases with distance (see Fig.~\ref{fig:Screening}). For QCD the opposite happens, i.e., there is an \emph{anti-screening} effect in which the effective colour decreases closer to the coloured particle.

Since QCD is asymptotically free, we can see that it is a weakly interacting theory at high energies, explaining the observational fact that in deep inelastic scattering (DIS) the components of the  proton behaved as free particles. It would also imply that, in the early universe at high temperatures, quarks and gluons were essentially free, interacting in a quark-gluon plasma soup together with leptons and photons. Once the universe cools down and particles interacted less violently, the coupling becomes stronger and at some point {\it hadronisation} happens, that means the observable spectrum becomes hadrons (and leptons) and no longer quarks and gluons. Knowing how the coupling changes with energies, we may estimate the value of this scale. The scale $\Lambda_{QCD}$ is called the \emph{confining scale}\index{Confining scale} corresponding to the QCD Landau pole. It is usually estimated by computing the value of $\mu$ for which $1/g_{s}^{2}\raw 0$ \footnote{It is clear this is only an indicative estimate since the running coupling receives corrections to all loops that become more important with the coupling getting stronger. Furthermore perturbation theory ceases to be valid at the coupling of order one, that is $1/g_s^2 \sim 1$ rather than $1/g_s^2 \sim 0$, but this does not change substantially the estimate of $\Lambda_{QCD}$ as it can be easily verified. } and is given by
\begin{equation}
\Lambda_{QCD}=\Lambda_{UV}\exp\left ({-\dfrac{8\pi^{2}}{bg_{s}^{2}}}\right )\, .
\end{equation}
Notice the non-perturbative nature of this expression (the function $e^{-1/x^2}$ is such that the function and all its derivatives vanish at $x=0$ and therefore it does not have a proper Taylor expansion). The above relation is known as \emph{dimensional transmutation}\index{Dimensional transmutation}: the dimensionless coupling in the action $g_s$ can be traded for an energy scale at which perturbation theory breaks down $\Lambda_{QCD}$. In the case of QCD, it is the natural way of getting a smaller scale, namely $\Lambda_{QCD}\approx 200$MeV, from high energy scales. 

The changing of the strong coupling with energy has been checked experimentally with the best fit giving $\alpha_s= 0.1184\pm 0.0007$ at the $Z$ mass scale $\mu=M_Z$. This can be compared with the value of the QED fine structure 'constant' $\alpha_{EM}\sim 1/129\sim 0.0077$ at the same scale. This explains why we observe QED as a weakly coupled theory and QCD as strong, but the relation changes at higher energies since QCD gets weaker and QED gets stronger. The values of $\alpha_s$ and $\Lambda_{QCD}$ have been computed using corrections up to 4-loops to the beta function with excellent agreement with experiment, see e.g. \cite{Chetyrkin:1997sg}.

\begin{figure}[t!]
\centering
 \includegraphics[width=0.8\linewidth]{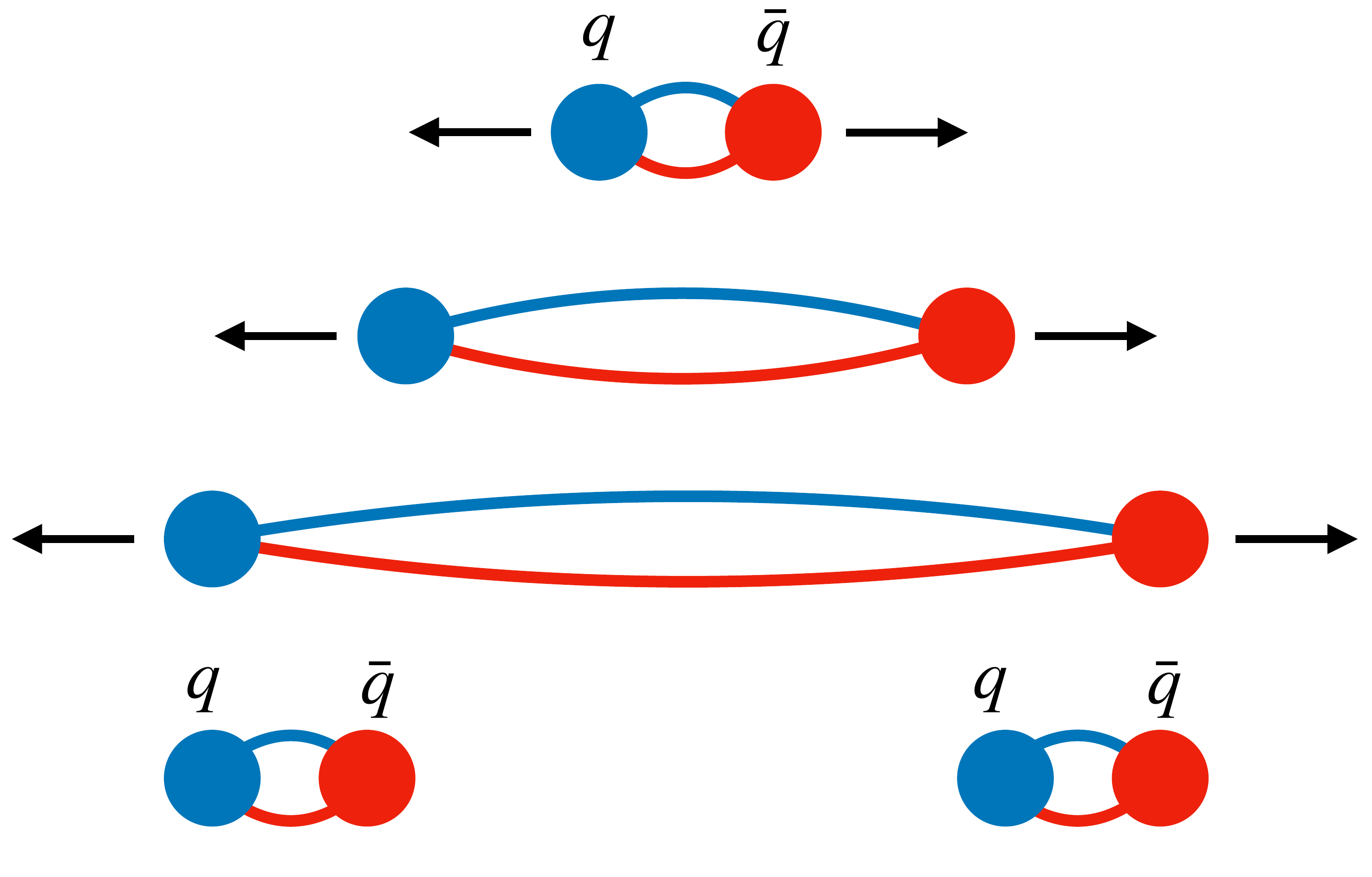}
\caption{Trying to split the quark and anti-quark from a meson. The gluons act as providing a constant force. Flux tubes are formed and at some point energetics favoured the creation of new quark-anti-quark pair, thereby preventing the generation of a single quark. This is a potential visualisation of quark and gluons confinement.}
\end{figure}

\subsubsection*{Quark Confinement and Hadron Masses*}

We argued in the preceding section that only colour singlets are allowed as bound states: mesons and baryons. Now, asymptotic freedom allows us to have an intuitive understanding of why we cannot see quarks and gluons in isolation. At  low energies the interaction is strong enough as to keep quarks confined into the singlet bound states. This is equivalent to have electrons and protons bound in a Hydrogen atom. But for atoms, if we apply enough energy we can eject the electron from the atom, which is possible since the attractive interaction decreases with the separation ($V(r)\propto -1/r$). However, due to the anti-screening implied by asymptotic freedom, applying energy to try to separate a quark from a hadron, the farther apart they are the stronger the attractive interaction. At some point it is energetically preferred to create a quark-anti-quark pair than to break-up the hadron and create an isolated quark. Experimentally this would appear as jets of mesons being produced. 

An effective linear interaction ($V(r) \propto r$) can describe this behaviour. Even though this is a compelling picture, due to our technical limitations to explicitly compute observable quantities at strong coupling, a proper proof of confinement is still an open question. Over several decades a technique known as lattice QCD has been developed in which quarks and gluons are described in a lattice discretisation of 3-dimensional space.
With the help of huge computer facilities, this approach has partially succeeded in computing some amplitudes and quark masses. There are several techniques to extract information about the physics of quarks inside hadrons.

In general since $m_u,m_d,m_s<\Lambda_{QCD}<m_c,m_b,m_t$,
we can see that hadrons made-up of $u,d,s$ quarks may have different properties than those made-up of the $c,b,t$ quarks. For instance let us roughly estimate the values of hadron masses. Since $\Lambda_{QCD}$ determines the energy of confinement, we may roughly say, using the uncertainty principle, that the momentum of quarks inside a hadron is of order $p\sim 1/r_c\sim \Lambda_{QCD}$ with $r_c$ the confinement radius. The mass of the hadron is determined by the total energy $E=\sqrt{p^2+m_q^2}$ with $m_q$ the mass of the constituent quarks. For $u,d,s$ we know $m_q<\Lambda_{QCD}$ and therefore the corresponding hadrons would have a mass of order $m_{hadron}\sim E\sim\Lambda_{QCD}$, whereas for $c,b,t$ the total energy is dominated by the quark mass and so $m_{hadron}\sim E\sim m$. This essentially agrees with the pattern of the hundreds of hadrons known so far. With very few notable exceptions, such as pions, which are hierarchically lighter than $\Lambda_{QCD}$.
This will be explained more thoroughly in the next section.

Before we finish this section, let us go one step further and use these simple techniques to make a more precise numerical estimate of the size of the proton. We know that in a Hydrogen atom the total energy is actually $E=m_p+m_e+E_b$ where $E_b\sim -13.6 $eV  the binding energy which is the work that needs to be done to eject the electron from the atom. For QCD, we know that a proton is a bound state of the form $uud$. The total energy is then $E=E_{quarks}+E_b$ where now $E_{quarks}\sim 3 p\sim 3/r$ and the binding energy is of the form $E_b=kr$ with $k$ a constant known as the \emph{string tension}. The total energy is then
\begin{equation}
E\sim \frac{3}{r}+kr\, .
\end{equation}
This function has an extremum at $k=3/r_m^2$ with $r_m$ the minimum distance between quarks. Plugging this back into the expression for the energy gives
\begin{equation}
E(r_m)=m_{proton}=\frac{3}{r_m}+\frac{3}{r_m}=\frac{6}{r_m}\sim 1\, {\rm GeV}
\end{equation}
where we have used the known value of the proton mass ($\sim 1$ GeV). This implies that the minimum quark separation is $r_m\sim 6\, $ GeV$^{-1} \sim 10^{-15}{\rm cm}\sim 1.2 $ Fermi. The string tension is then $k=3/r_m^2=1/12\, $GeV$^{-2}$.

Finally the proton radius is determined by assuming the three quarks form an equilateral triangle of side $r_m$ and the radius is the radius of the circle embedding the corresponding triangle which by trigonometry we know it is $r_m/\sqrt{3}$. Therefore the proton radius is $R_{proton}\sim r_m/\sqrt{3}\sim 0.7$ Fermi which fits well with experiments which suggest that $R_{proton}^{experiment}\sim 0.84-0.87$ Fermi.

\section{Effective Chiral Lagrangian}\label{sec:chiraltheory} 
\index{Effective Chiral Lagrangian}\index{EFT!Chiral Lagrangian}

Asymptotic freedom is somehow a present from nature to physicists that can now use the perturbative techniques based on Feynman diagrams and loop expansions for QCD despite being a strongly coupled theory at low energies, as long as the calculations are done for processes at high enough energies to justify the weak coupling approximation. It also shows that QCD is by itself UV complete which is a major achievement. However, we still need to be able to describe the physics at scales lower than $\Lambda_{QCD}$ in which it should describe the confinement of quarks and gluons into hadrons. It is important to emphasise that a formal proof of confinement is not yet available and it is considered as one of the top challenges in mathematics. Physicists are not only convinced it is true given all the evidence from experiment and asymptotic freedom. 

The challenge is then to perform reliable calculations in the strong coupling regime of QCD. For this the most powerful tool available at the moment is Lattice QCD in which for more than 40 years a large community has been able to make progress by putting QCD in a discretised space, a lattice, with the lattice size as an inverse cut-off scale and together with heavy computational power concrete calculations such as hadron masses and couplings have been obtained.

\begin{table}[t!]
\centering
\begin{tabular}{|c|cc||c||cc||c|}
\hline 
quark & $u$ & $d$ & $s$ & $c$ & $b$ & $t$ \\ 
\hline 
mass & $1.7-3.3$MeV & $4.1-5.8$MeV & $104$MeV & $1270$MeV & $4$GeV & $173$GeV \\ 
\hline 
\end{tabular} 
\caption{List of all the quark masses.}\label{tab:QuarkMasses} 
\end{table}

Other approaches have been proposed over the years to extract some information in the strong coupling regime. Here we will concentrate on a general approach known as \emph{chiral perturbation theory}.\index{Chiral perturbation theory}
This is a way of treating QCD in terms of an effective field theory. Recall that EFT's are good descriptions at given energies. We can organise physics by energies by integrating out high energy modes. For this, notice the hierarchy in the quark masses in Table~\ref{tab:QuarkMasses} where the \emph{confining scale}\index{Confining scale} $\Lambda_{QCD}$ is located between strange- and charm-quark mass
\begin{equation}
m_{s}<\Lambda_{QCD}<m_{c}\, .
\end{equation}
It is clear from the table that the $u$ and $d$ quarks are substantially lighter than the rest.
We also learn that the $s$ quark may still be considered in an effective theory below $\Lambda_{QCD}$, but the $c,b, t$ quarks are clearly in another regime heavier than $\Lambda_{QCD}$ with the top quark far heavier than the rest. Let us then consider QCD with only $u$ and $d$ quarks as an effective low-energy theory so that
\begin{align}
\cL&=-\dfrac{1}{4}\left (G_{\mu\nu}^{A}\right )^{2}+\I\bar{u}^{L}\,\cancel{D}\, u^{L}+\I\bar{u}^{R}\,\cancel{D}\, u^{R}+\I\bar{d}^{L}\,\cancel{D}\, d^{L}+\I\bar{d}^{R}\,\cancel{D}\, d^{R}\nn\\
&\quad -m_{u}(\bar{u}_{L}u_{R}+\bar{u}_{R}u_{L})-m_{d}(\bar{d}_{L}d_{R}+\bar{d}_{R}d_{L})\, .
\end{align}
The quark masses are generated originally by the Yukawa couplings to the Higgs, but for our purposes here they are just free small parameters. In the limit $m_{u},m_{d}\raw 0$, the Lagrangian clearly has the following global symmetry
\begin{equation}
G=\SUTw_{L}\times\SUTw_{R}\times\UO_{V}\times\UO_{A}
\end{equation}
where $\SUTw_{L}$ acts on $(u_{L},d_{L})$ and $\SUTw_{R}$ on $(u_{R},d_{R})$ building together the \emph{chiral symmetry}\index{Chiral symmetry}.
$\UO_{V}$ is associated with baryon number, whereas $\UO_{A}$ is anomalous as we discussed in chapter \ref{chap:ssb}. Since $\UO_V$ is already a symmetry even in the massive case and $\UO_A$ is clearly anomalous we will concentrate on the approximate $\SUTw_{L}\times\SUTw_{R}$ chiral symmetry.
Let us stress here that the symmetry is approximate in the sense that is only a symmetry if the quark masses were identically zero. 
Nonetheless, its existence will help us understand a lot about the spectrum of hadrons as we now explain.

Let us define the generators as follows
\begin{equation}
T_{V}^{a}=T_{L}^{a}+T^{a}_{R}\kom T_{A}^{a}=T_{L}^{a}-T_{R}^{a}\kom a=1,2,3
\end{equation}
generating $\SUTw_{V}$ and $\SUTw_{A}$ respectively. Then
\begin{equation}
\left (\begin{array}{c}
u \\ 
d
\end{array} \right )\raw \exp\left (\I(\theta_{V}^{a}T^{a}+\gamma_{5}\theta_{A}^{a}T^{a})\right )\left (\begin{array}{c}
u \\ 
d
\end{array} \right )\, .
\end{equation}
Notice that as the $\gamma_5$ presence indicates, $\SUTw_{A}$ can map a hadron $\ket{h}$ to another hadron $\ket{h^{\prime}}$ of opposite chirality, but all other quantum numbers the same. If this symmetry was not spontaneously broken, it would imply that both hadrons would be degenerate in mass. Since such a pair of degenerate hadrons does not exist, this means that this symmetry should be spontaneously broken. The natural order parameter would be\footnote{Recall that only scalar fields can have a non-vanishing VEV because any other field having a VEV would break Poincar\'e invariance. However condensates of fermions like $\bar{u}_{L}u_{R}$ being scalars can have a VEV. In QCD also gluon condensates $G_{\mu\nu}^A G^{A\mu\nu}$ could have a VEV. All of these condensates would define the QCD vacuum\index{QCD vacuum}\index{QCD!Vacuum}.}
\begin{equation}
\langle \bar{u}_{L}u_{R}\rangle=\langle\bar{d}_{L}d_{R}\rangle\neq 0
\end{equation}
breaking partially the chiral symmetry to the diagonal (or vector) $\SUTw_V$
\begin{equation}
\SUTw_{V}\otimes\SUTw_{A}\raw \SUTw_{V}\, .
\end{equation}

The unbroken vector symmetry acts equally on $u_L$ and $u_R$ quarks as
\begin{equation}
\left (\begin{array}{c}
u_{L} \\ 
d_{L}
\end{array} \right )\raw g\left (\begin{array}{c}
u_{L} \\ 
d_{L}
\end{array} \right )\kom \left (\begin{array}{c}
u_{R}\\ 
d_{R}
\end{array} \right )\raw g\left (\begin{array}{c}
u_{R} \\ 
d_{R}
\end{array} \right )
\end{equation}
and so
\begin{equation}
\left (\begin{array}{c}
u \\ 
d
\end{array} \right )\raw g\left (\begin{array}{c}
u \\ 
d
\end{array} \right )\, .
\end{equation}
The associated quantum number is \emph{isospin}\index{Isospin}. This is the same isospin introduced by Heisenberg \cite{Heisenberg:1932dw} when he proposed that strong interactions would make no difference between protons and neutrons and that they would be related as different spin states of the rotation group given by an internal $\SUTw$. Notice $p=uud$ and $n=udd$ have the same isospin and therefore
\begin{equation}
\left (\begin{array}{c}
p \\ 
n
\end{array} \right )\raw g \left (\begin{array}{c}
p \\ 
n
\end{array} \right )
\end{equation}
under $g\in \SUTw_{V}$. This is remarkable since now we can \emph{explain} this isospin symmetry as just an approximate symmetry derived from QCD and the symmetry is only approximate as long as the $u$ and $d$ quarks are considered massless. Once their mass is taken into account the proton and neutron do not have to be degenerate in mass. Note that the mass of proton and neutron are of order $1$GeV, but their mass difference is only of order $1$MeV which is of the same order as the mass of $u$ and $d$ quarks.

From the early universe perspective, we may say that in the early universe the $\SUTw_L\times \SUTw_R$ symmetry was manifest, but once the universe cools down to temperatures of order $T<T_{c}\cong \Lambda_{QCD}$ the binding energy among quarks is strong enough to confine them (and the gluons) into hadrons and the quark-anti-quark pair can condense (get a VEV) to break the symmetry.

We may now rightfully ask: How can we study the quark condensate phase? 
Recall that a quark condensate is similar to the EFT of superconductivity. In that case, the abelian Higgs model was an appropriate EFT to describe the main aspects of superconductors, cf. Section~\ref{sec:AbelianHiggsmodel}.
Even though the corresponding scalar field is not a physical field in the spectrum, it captures the physics of Cooper pairs of electrons in the medium.
Let follow here a similar logic and introduce scalar fields $\Sigma_{ij}(x)$ transforming under $G=\SUTw_{L}\times\SUTw_{R}$ via
\begin{equation}
\Sigma\raw g_{L}\Sigma g_{R}^{\dagger}\kom \Sigma^{\dagger}\raw g_{R}\Sigma^{\dagger}g_{L}\, .
\end{equation}
We then consider an effective Lagrangian with the same global symmetries as the original Lagrangian, namely $\SUTw_{L}\times\SUTw_{R}$. The most general such a renormalisable Lagrangian is
\begin{equation}
\cL=\tr\left [\p_{\mu}\Sigma(\p^{\mu}\Sigma)^{\dagger}\right ]+m^{2}\tr(\Sigma\Sigma^{\dagger})-\dfrac{\lambda}{4}\tr\left [\Sigma\Sigma^{\dagger}\Sigma\Sigma^{\dagger}\right ]\, .
\end{equation}
For positive $m^2$ and $\lambda$ this gives rise to spontaneous symmetry breaking as we have seen already several times. The VEV of $\Sigma$ that keeps the diagonal $\SUTw$ unbroken is
\begin{equation}
\langle \Sigma_{ij}\rangle=\dfrac{v}{\sqrt{2}}\left (\begin{array}{cc}
1 & 0 \\ 
0 & 1
\end{array} \right )\kom v=\dfrac{2m}{\sqrt{\lambda}}\, .
\end{equation}
Then $\SUTw_{L}\times\SUTw_{R}\raw\SUTw_{V}$ as required.
We can make contact with quark condensate as follows
\begin{equation}
v\sim\Lambda_{QCD}\sim\langle \bar{u}u\rangle^{\frac{1}{3}}\, .
\end{equation}
Notice the power $1/3$ is due to dimensional analysis and $[u]=[d]=3/2$. 

Now we can expand around the vacuum
\begin{equation}
\Sigma(x)=\dfrac{v+\sigma(x)}{\sqrt{2}}\exp\left (\dfrac{2\I T^{a}\pi^{a}(x)}{(F_{\pi})}\right ) = \dfrac{v+\sigma(x)}{\sqrt{2}}\, U(x)
\end{equation}
where the constant $F_\pi=v$ is introduced to keep standard notation in the literature. Here, as usual, $\sigma(x)$ is a massive Higgs field, invariant under $\SUTw_{V}$ and $\pi^{a}(x)$ are the massless Goldstone modes that transform in the adjoint of the unbroken  group. More generally, one finds
\begin{equation}
\Sigma\raw g_{L}\Sigma g_{R}^{\dagger}\implies \delta\pi^a=\frac{F_\pi}{2} \left(\theta^a_L-\theta^a_R\right)-\frac{1}{2}f^{abc}\left(\theta^b_L+\theta^b_R\right)\pi^c+\cdots
\end{equation}

Concentrating only on the low-energy theory, we can integrate out the massive $\sigma(x)$ field (mass of order $\Lambda_{QCD}$) and work only with the effective theory for the low energy Goldstone modes $\pi^{a}$ in terms of the field $U(x)$ defined as
\begin{equation}
U(x)=\exp\left (\dfrac{2\I T^{a}\pi^{a}(x)}{F_{\pi}}\right )=\exp\left (\dfrac{\I}{F_{\pi}}\left (\begin{array}{cc}
\pi^{0} & \sqrt{2}\pi^{-} \\ 
\sqrt{2}\pi^{+} & -\pi^{0}
\end{array} \right )\right )
\end{equation}
with $\pi^{0}=\pi^{3}$ and $\pi^{\pm}=(\pi^{1}\pm\I \pi^{2})/\sqrt{2}$. The invariant Lagrangian for Goldstone modes can then be written as an expansion in derivatives know as the \emph{Chiral Lagrangian}\index{Chiral Lagrangian}
\begin{equation}
\cL_{\chi}=\dfrac{F_{\pi}^{2}}{4}\tr\left [(D^{\mu}U)(D_{\mu}U)^{\dagger}\right ]+\lambda_{1}\tr\left [(D^{\mu}U)(D_{\mu}U)^{\dagger}\right ]^{2}+\ldots\, .
\end{equation}
Expanding the exponentials
\begin{equation}
\cL_{\chi}\raw \dfrac{1}{2}(\p_{\mu}\pi^{0})(\p^{\mu}\pi^{0})+(D^{\mu}\pi^{+})(D_{\mu}\pi^{-})^{\dagger}+\dfrac{1}{F_{\pi}^{2}}\left [-\dfrac{1}{3}\pi^{0}\pi^{0}D_{\mu}\pi^{+}D^{\mu}\pi^{-}+\ldots\right ]+\ldots
\end{equation}
where in general $D_{\mu}$ is the electroweak covariant derivative (but without including the QCD gauge fields since the field $U(x)$ is a colour singlet). Notice that the normalisation factor $F_\pi^2/4$ implies that the pion fields have canonical kinetic terms. The higher derivative terms have arbitrary coefficients $\lambda_i$ that as usual should be bound by experiment. This is an expansion in powers of $E/F_{\pi}$, since an expansion in derivatives correspond to an expansion in momenta and then energy $E$, so the above is only predictive for energies $E\ll F_{\pi}$. This formalism is called \emph{chiral-perturbation theory} ($\chi PT$)\index{Chiral perturbation theory}. It can easily be seen that $\pi^{0}$, $\pi^{\pm}$ have the same quantum numbers as the well known pion fields! So these fields can be seen as both
\begin{itemize}
\item hadrons made out of quarks, or as
\item (pseudo-)Goldstone bosons of the approximate $SU(2)_L\times SU(2)_R$ chiral symmetry breaking.
\end{itemize}
This is remarkable. We have now an effective field theory obtained from QCD, but not in terms of the fundamental degrees of freedom (quarks and gluons), but in terms of the lightest hadrons, the pions. Furthermore, this makes contact with the original proposal of Yukawa treating pions\index{Pions} as mediators of strong interactions \cite{Yukawa:1935xg}. However, Yukawa's concept of strong interactions was based on direct interactions among hadrons, whereas we know that this is only a low-energy behaviour in the hadronic phase of QCD and not the fundamental degrees of freedom. Therefore, $\chi$PT, though not fundamental, is an appropriate effective field theory to describe interactions among hadrons.

The Lagrangian above is only based on derivatives of the $U(x)$ field. We may also add 'mass terms' of the form
\begin{equation}
\delta \cL=C \tr \left( MU+M^\dagger U^\dagger\right)
\end{equation}
where $M$ is the $u,d$ mass matrix $M= \rm{diag}\, (m_u, m_d)$ and $C$ a dimension $3$ constant of order $C\sim \Lambda_{QCD}^3$. Expanding the exponential in $U$ we can get the quadratic terms in the $\pi$ fields proportional to
\begin{equation}
\delta \cL_{quadratic}=\dfrac{C}{F_\pi^2}\left(m_u+m_d\right)\left(\pi_0^2+\pi_1^2+\pi_2^2\right)
\end{equation}
implying that the mass$^2$ of the pions is of order
\begin{equation}
m_\pi^2=\dfrac{C}{F_\pi^2}(m_u+m_d)\sim \Lambda_{QCD}(m_u+m_d)
\end{equation}
fitting well with the experimental results with $m_{\pi^{0}}\sim 135$ GeV and $m_{\pi^{\pm}}\sim 139.6$ GeV.

Note however that we considered only the lightest quarks $u$ and $d$ and the chiral symmetry is only approximate due to the lightness of $u,d$ quarks. We may move on and extend this formalism to include the other quark whose mass is lighter than $\Lambda_{QCD}$, namely the $s$ quark. In this case the approximate chiral symmetry and its breaking would be extended to
\begin{equation}
\SUTh_{L}\times \SUTh_{R}\raw \SUTh_{V}
\end{equation}
with $16-8=8$ Goldstone bosons, 
\begin{equation}
\left (\begin{array}{ccc}
\frac{\pi^{0}}{\sqrt{2}}+\frac{\eta^{0}}{\sqrt{6}} & \pi^{+} & K^{+} \\[0.1em]
\pi^{-} & -\frac{\pi^{0}}{\sqrt{2}}+\frac{\eta^{0}}{\sqrt{6}} & K^{0} \\[0.1em] 
\bar{K}^{-} & \bar{K}^{0} & -\sqrt{\frac{2}{3}}\eta^{0}
\end{array} \right )
\end{equation}
that can be identified with the known mesons $\eta, K, \tilde{K}$ besides the pions. The remaining unbroken symmetry $\SUTh_V$ is nothing else but the flavour $\SUTh$ of the \emph{Eightfold way}\index{Eightfold way}, recall Fig.~\ref{fig:EFW}. Again this allows us to \emph{explain} the original approximate global symmetry proposed to classify hadrons as a consequence of an approximate symmetry coming from the more fundamental theory that is QCD. Furthermore, the approximate nature of this symmetry is under less solid grounds since the mass of the $s$ quark is smaller than $\Lambda_{QCD}$, but not much smaller and it is substantially heavier than $u$ and $d$. Therefore, the extra Goldstone modes are hierarchically heavier than the pions, as observed in nature. As in the case of $\SUTw$ an effective Lagrangian can be written describing the interactions of all these mesons.

Baryons can also be introduced as
\begin{equation}
B=\varepsilon_{ijk}q^{i}q^{j}q^{k}
\end{equation}
using the well known $\SUTh_V$ product
\begin{equation}
\mathbf{3}\otimes{\mathbf{3}}\otimes{\mathbf{3}}= {\mathbf{10}}+{\mathbf{8}}+{\mathbf{8}}+{\mathbf{1}}\, .
\end{equation}
We stress that these are representations of $\SUTh_V=\SUTh_f$ which is not $\SUTh_c$, in particular the indices $i,j,k$ above are flavour indices corresponding to $u,d,s$.
Therefore, for example, the $\SUTw$ isospin doublet containing the proton and the neutron can be embedded into a full $\SUTh$ multiplet, the following octet of baryons
\begin{equation}
\left (\begin{array}{ccc}
\frac{\Sigma^{0}}{\sqrt{2}}+\frac{\Lambda}{\sqrt{6}} & \Sigma^{+} & p \\[0.1em]
\Sigma^{-} & -\frac{\Sigma^{0}}{\sqrt{2}}+\frac{\Lambda}{\sqrt{6}} & n \\[0.1em] 
\Xi^{-} & \Xi^{0} & -\sqrt{\frac{2}{3}}\Lambda
\end{array} \right )\, .
\end{equation}
Couplings of $B$ to $U$ can be introduced to have then a low-energy theory of interacting mesons and baryons. Once again, we stress that this approach can really only be used for the $u,d,s$ quarks and their corresponding hadrons. For hadrons made out of $c,b,t$ this is not suitable since these quarks are heavier than $\Lambda_{QCD}$ and there would be no justification for a low-energy effective action. Other approaches (such as heavy quark effective theory) should be used in this case.

Note that somehow we closed a circle. We started with the historical introduction of isospin and the eightfold way that lead to the proposal of quarks as fundamental degrees of freedom and then to colour as the appropriate gauge symmetry to treat strong interactions. Now we started with QCD as the fundamental theory and ended up \emph{explaining} why there are these approximate symmetries of isospin and the eightfold way. The existence of these symmetries is explained by the fact that there is a hierarchy of quark masses and $2$ or $3$ of them can be taken to be approximately zero compared to $\Lambda_{QCD}$ and the other heavier quarks. Why this hierarchy of masses exists, is still an open question.

\chapter{\bf The Standard Model}
\label{chap:SM}

\begin{equ}[Confusion before clarity]
{\it  I think it is important to understand how confusing these things seemed back then, and no one knows better than I do how confused I was.}\\

\rightline{\it Steven Weinberg}
\end{equ}
\vspace{0.5cm}




The journey toward the complete Standard Model has been a fascinating one, full of theoretical challenges, experimental breakthroughs, and conceptual shifts. The theory as we know it today is the culmination of decades of collaborative effort and ingenuity, unifying the fundamental forces -- except gravity -- under a single quantum framework. The Standard Model represents a major triumph of modern physics, explaining not only the electromagnetic and weak forces but also the strong interactions. In this final chapter, we will collect and synthesise all the key concepts developed in the previous discussions of electroweak and strong interactions to arrive at the full picture.

In the preceding chapters, we introduced the two major components that together form the backbone of the Standard Model. The electroweak interactions, described by a spontaneously broken gauge theory, have shown us how the electromagnetic and weak forces can be unified. Meanwhile, the strong force, governed by quantum chromodynamics, demonstrates the phenomenon of confinement and explains the behaviour of hadrons. These two branches of fundamental interactions display very different characteristics, but their unification under the Standard Model reflects the depth and consistency of modern theoretical physics.
Together, they illustrate two distinct phases of gauge theories, which account for the existence of short-range interactions -- either through the Higgs mechanism or quark confinement.

In this chapter, we bring together all the elements we have developed throughout these lectures. However, the result is more than just a straightforward combination of the two Lagrangians. We will present the full Standard Model as a gauge theory for the group $SU(3)_c\times SU(2)_L\times U(1)_Y$, which is spontaneously broken to $SU(3)_c\times U(1)_{em}$, and provide the following key insights:
\begin{itemize}
\item Count the total number of free parameters that can be determined through experiments.
\item Explain why all gauge symmetries are anomaly-free.
\item Identify the global symmetries present, including approximate, accidental, and anomalous ones.
\item Highlight a non-trivial parameter, $\Theta$, which requires the consideration of both weak and strong interactions since it is related to the phases in the CKM matrix \eqref{eq:CKM_matrix}.
This parameter is central to one of the most significant puzzles in the Standard Model: the \emph{strong CP problem}.
\end{itemize}
Though this chapter consolidates the material discussed in earlier sections, it also introduces new perspectives and presents the completion of our understanding of fundamental interactions. The structure of the Standard Model, with its symmetry-breaking mechanisms, anomalies, and free parameters, serves as the best-tested theory in physics, and yet it leaves us with profound open questions. While straightforward in some aspects, the content of this chapter reflects the culmination of the deepest theoretical insights of 20th-century physics.




\section{The Standard Model -- all at once}

In this section, we synthesise the results from previous discussions on electroweak and strong interactions to present the complete Standard Model. Both interactions, formulated as gauge theories, are naturally unified into a single Lagrangian invariant under local transformations from the group
\begin{equation}
G_{\text{SM}}=\SUTh_{c}\times\SUTw_{L}\times\UO_{Y}\, .
\end{equation}
As before, we denote the fundamental gauge fields as $G_{\mu}^{A}$, $W_{\mu}^{a}$, $B_{\mu}$, while after SSB we write $G_{\mu}^{A}$, $W_{\mu}^{\pm}$, $Z_{\mu}$, $A_{\mu}$. The Lagrangian for the kinetic and self-interaction terms of the gauge bosons is
\begin{equation}
\cL^{\text{gauge}}=-\dfrac{1}{4}\left (G_{\mu\nu}^{A}\right )^{2}-\dfrac{1}{4}\left (W_{\mu\nu}^{a}\right )^{2}-\dfrac{1}{4}\left (B_{\mu\nu}\right )^{2}-\Theta_{G}G_{\mu\nu}^{A}\tilde{G}_{\mu\nu}^{A}-\Theta_{W}W_{\mu\nu}^{a}\tilde{W}_{\mu\nu}^{a}-\Theta_{B}B_{\mu\nu}\tilde{B}_{\mu\nu}\, .
\end{equation}
In principle, the three topological $\theta$-terms coming with different angles $\Theta_{G}$, $\Theta_{W}$ and $\Theta_{B}$ are allowed. Even though classically and in perturbation theory they correspond to a total derivative that does not affect the equations of motion, quantum mechanically they can play a role once non-perturbative effects are included.\footnote{Non-perturbative effects\index{Non-perturbative effects} are explicit effects that are not captured by the standard perturbation expansion in terms of Feynman diagrams in which the amplitudes are written as a Taylor expansion in terms of couplings $g$ such as the electromagnetic coupling. There are effects, such as instantons\index{Instantons}, that are not captured by these expansions even at weak coupling. An example of a non-perturbative quantity is $\Lambda_{QCD}$ that as we discussed can be written in terms of expressions such as $e^{-1/g^2}$ which is a function that has no non-trivial Taylor expansion around $g=0$ since the function and all of its derivatives vanish at $g=0$. A detailed description of non-perturbative effects in the Standard Model is beyond the scope of these lectures.}
We will discuss later their importance in QCD.

The fermionic matter content can be summarised as follows\footnote{Recall that when we we write $Q_L$ as a doublet here it is a $\mathrm{SU}(2)_L$ doublet, each entry  can be seen as a two-component left-handed Weyl spinor (or as a projected four-component Dirac spinor). Also, the right-handed particles are also right-handed Weyl spinors that can be written as projections of 4-component Dirac spinors. }
\begin{align}
Q_{L}^{i}&=\biggl \{\left (\begin{array}{c}
 u_{L} \\[-4pt]
 d_{L}
 \end{array}\right ),\left (\begin{array}{c}
 c_{L} \\[-4pt]
 s_{L}
 \end{array}\right ),\left (\begin{array}{c}
 t_{L} \\[-4pt]
 b_{L}
 \end{array}\right ) \biggl \}\kom &\left (\mathbf{3},\mathbf{2},\frac{1}{6}\right )&\nn \\[0.3em]
u_{R}^{i}&=\bigl \{u_{R},c_{R},t_{R}\bigl \}\kom &\left (\bar{\mathbf{3}},\mathbf{1},\frac{2}{3}\right )&\nn \\[0.3em]
d_{R}^{i}&=\bigl \{d_{R},s_{R},b_{R}\bigl \}\kom &\left (\bar{\mathbf{3}},\mathbf{1},-\frac{1}{3}\right )&\nn \\[0.3em]
L_{L}^{i}&=\biggl \{\left (\begin{array}{c}
\nu_{e,L} \\ 
e_{L}
\end{array} \right ),\left (\begin{array}{c}
\nu_{\mu,L} \\ 
\mu_{L}
\end{array} \right ),\left (\begin{array}{c}
\nu_{\tau,L} \\ 
\tau_{L}
\end{array} \right )\biggl \}\kom &\left (\mathbf{1},\mathbf{2},-\frac{1}{2}\right )&\nn \\[0.3em]
e_{R}^{i}&=\bigl \{e_{R},\mu_{R},\tau_{R}\bigl \}\kom &\left (\mathbf{1},\mathbf{1},-1\right )&\nn \\[0.3em]
\biggl (\nu_{R}^{i}&=\bigl \{\nu_{e,R},\nu_{\mu,R},\nu_{\tau,R}\bigl \} \kom &\left (\mathbf{1},\mathbf{1},0\right )&\biggl )^{*}
\end{align}
The Lagrangian for the fermions can be split into two pieces so that
\begin{equation}
\cL_{F}=\cL_{F}^{\text{kinetic}}+\cL_{F}^{\text{Yukawa}}
\end{equation}
where the kinetic term for the fermions is of the form
\begin{align}
\cL_{F}^{\text{kinetic}}&=\I \overline{L}_{L}^{i}\cancel{D}L_{L}^{i}+\I \overline{Q}_{L}^{i}\cancel{D}Q_{L}^{i}+\I \overline{e}_{R}^{i}\cancel{D}e_{R}^{i}\nn\\[0.5em]
&\quad +\I \overline{\nu}_{R}^{i}\cancel{D}\nu_{R}^{i}+\I \overline{u}_{R}^{i}\cancel{D}u_{R}^{i}+\I \overline{d}_{R}^{i}\cancel{D}d_{R}^{i}\, .
\end{align}
The covariant derivative can be written as
\begin{align}
D_{\mu}&=\p_{\mu}-\I g_{s}G_{\mu}^{A}T^{A}-\I g W_{\mu}^{a}T^{a}-{\I}g^{\prime}Y B_{\mu}\, .
\end{align}
The Yukawa couplings are defined as
\begin{equation}
\cL_{F}^{\text{Yukawa}}=-y_{ij}^{d}\, \overline{Q}_{L}^{i}Hd_{R}^{j}-y_{ij}^{u}\, \overline{Q}_{L}^{i}\tilde{H}u_{R}^{j}+y_{ij}^{e}\overline{L}^{i}_{L}He_{R}^{j}(+y_{ij}^{\nu}\overline{L}^{i}_{L}H\nu_{R}^{j})^{*}+\text{h.c.}\, .
\end{equation}
The Higgs field is a complex scalar doublet
\begin{equation}
H=\left (\begin{array}{c}
H_{+} \\ 
H_{0}
\end{array} \right )\kom \left (\mathbf{1},\mathbf{2},\dfrac{1}{2}\right )
\end{equation}
with Lagrangian
\begin{equation}\label{eq:HiggsLag} 
\cL^{\text{Higgs}}=D_{\mu}H (D^{\mu}H)^{\dagger}+V(H)\kom V(H)=m^{2}|H|^{2}-\lambda |H|^{4}\, .
\end{equation}

\begin{table}[t!]
\centering
\begin{tabular}{|c|c|c|c|}
\hline 
Sector & Parameters & Physical & Number \\ 
\hline 
Gauge & $g_{s},g,g^{\prime},\theta_{G},\theta_{W},\theta_{B}$ & $g_s, e,\cos(\theta_{W}),\bar{\theta}$ & 4 \\ 
\hline 
Higgs & $m^{2},\lambda$ & $m_{h},m_W$ & 2 \\ 
\hline 
 & & $m_{i}^{u},m_{i}^{d},m_{i}^{e}$ & 9 \\ 
Yukawa & $y_{ij}^{u},y_{ij}^{d},y_{ij}^{e}(,y_{ij}^{\nu})^{*}$ & $V_{CKM}$ & 4 \\ 
& & $V_{PMNS}$ & 6? \\ 
& & $m_{i}^{\nu},M_{i}^{\nu_{R}}$ & $3^+$? \\ 
\hline 
\hline 
Total & & & $25^+$ \\ 
\hline
\end{tabular} 
\caption{Parameter count for the Standard Model.}\label{tab:SMParameterCount} 
\end{table}

Altogether, the Standard Model Lagrangian is given by
\begin{equ}[Standard Model Lagrangian]\index{Standard Model Lagrangian}
\begin{equation}\label{eq:SMLag} 
\cL_{SM}=\cL^{\text{gauge}}+\cL_{F}^{\text{kinetic}}+\cL_{F}^{\text{Yukawa}}+\cL^{\text{Higgs}}\, .
\end{equation}
\end{equ}
Let us perform a parameter count for the Standard Model summarised in Table~\ref{tab:SMParameterCount}.
We have more than $25$ free parameters in the Standard Model. We do not specify the number of parameters coming from the right-handed neutrinos since it is yet not known how right-handed neutrinos will appear and couple to the remaining fields in the Standard Model.
In particular, their number does not have to be restricted to the number of families as for the other fields since right-handed neutrinos are simply fermions that do not couple to any of the gauge fields of the Standard Model.

The Standard Model Lagrangian \eqref{eq:SMLag} is renormalisable and can be expanded in terms of operators of different dimensions. Let us write
\begin{equation}
\cL_{SM}=\sum_{i}\, c_{i}\cO_{i}\kom [c_{i}]+[\cO_{i}]=4\, .
\end{equation}
The dimensions of the individual operators $\cO_{i}$ are
\begin{itemize}
\item $[\cO_{i}]=0$: $c_{0}=\Lambda$ is the constant term in the scalar potential ($\lambda v^4/2$ in the Higgs potential). Once coupled to gravity, this term would correspond to the \emph{cosmological constant}\index{Cosmological constant}.
\item $[\cO_{i}]=2$: $c_{2}=m^{2}$ with $m^2$ the coefficient in the quadratic term $m^2H^2$ of the HIggs potential.
\item $[\cO_{i}]=3$: there is no dimension $3$ operator in the Standard Model Lagrangian. But if right-handed neutrinos are involved then the corresponding Majorana mass $c_{3}=M^\nu$ multiplying $\nu_R\nu_R=\cO_{3}$.
\item $[\cO_{i}]=4$: all the other terms implying the coefficients (gauge couplings, Yukawa couplings, $\theta$ terms are dimensionless).
\end{itemize}
Notice that
\begin{itemize}
\item no mass terms are allowed for gauge fields because of gauge invariance, and
\item no mass terms are allowed for fermions, again from (chiral) gauge symmetry. The fermion masses arise from Yukawa couplings and $\langle H\rangle=v\neq 0$. 
\end{itemize}
Then, unlike the case of the Higgs, fermion and gauge field masses are only generated after symmetry breaking and quantum corrections to the Lagrangian cannot induce masses for gauge fields and fermions.

The only missing piece in the Standard Model Lagrangian is gravity. To include gravity, we have to introduce the metric degrees of freedom and make the Lagrangian invariant under general coordinate transformations.
We then arrive at an non-renormalisable EFT with Lagrangian
\begin{equation}
\cL_{SM}\raw\sqrt{-g}\left (\cL^{\prime}_{SM}+\Lambda+M_P^2R+\ldots\right )
\end{equation}
where $g$ is the determinant of the metric $\Lambda$ the cosmological constant and $R$ the Ricci scalar.
We also defined
\begin{equation}
\cL^{\prime}_{SM}=\cL_{SM}[D_{\mu}\raw\cD_{\mu}]
\end{equation}
in terms of the covariant derivative $\cD_{\mu}$ suitably adjusted for gravity. Here, we wrote the gravity component as an expansion in powers of the curvature (which is a derivative expansion) with leading order term the cosmological constant $\Lambda$, next order the Einstein-Hilbert action in terms of the Ricci scalar and then higher powers of curvature terms ${\mathcal{O}}(R^2)$.

\begin{figure}[t!]
\centering
\includegraphics[scale=0.4]{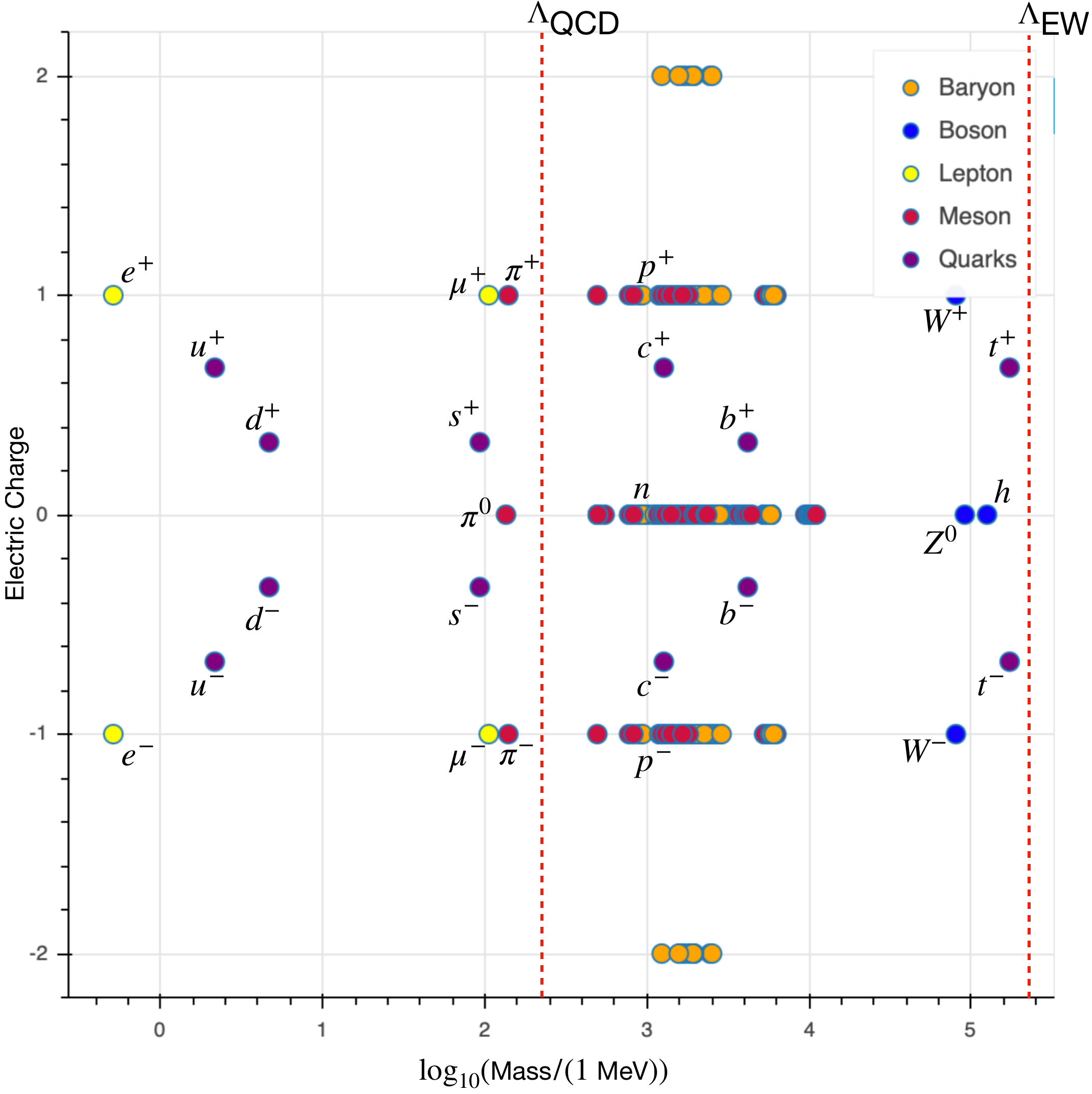}
\caption{Summary of massive particles in the Standard Model.
We highlight the types of particles with different colours and assign labels to the most prominent ones.
The QCD scale $\Lambda_{\text{QCD}}\approx 250\,  {\rm MeV} $ and electroweak scale $\Lambda_{\text{EW}}=v=246 \, {\rm GeV}$ are also indicated as red lines.}\label{fig:MPSM}
\end{figure}

This is \emph{the} model (or theory or framework) that explains and describes the world we observe. It is
\begin{itemize}
\item Mathematically consistent in the sense that the Lagrangian is consistent with all spacetime and internal symmetries, it is the most general renormalisable Lagrangian consistent with these symmetries and all gauge symmetries are free from anomalies.
\item Explains \emph{all} the experiments done before the formulation of the model (all interactions, decay rates, etc.).
\item Made predictions that were spectacularly confirmed over the years (neutral currents, $W^{\pm},Z^{0}$, Higgs).
\item Precision tests: theory $\leftrightarrow$ experiment with agreeing precision at many decimal figures.
\item Explains the observed ``approximate'' and accidental symmetries such as
\begin{itemize}
\item Baryon number (accidental)
\item Lepton number (accidental) 
\item Isospin (approximate)
\item $\SUTh_{f}$ (eightfold way) (approximate)
\end{itemize}
\item Consistent coupling to gravity as an EFT at energies
\begin{equation}
E\ll M_{P}=\sqrt{\dfrac{\hbar c}{G}}\sim 10^{19}\, \text{GeV}\, .
\end{equation}
It is important to compare the magnitude of this scale compared to the relevant scales in the Standard Model namely the VEV of the Higgs $v=\langle H \rangle = 246 \, {\rm GeV}$ and 
the QCD scale: $\Lambda_{\text{QCD}}\sim 250\,  {\rm MeV} $.
We summarise the massive particle content of the Standard Model together with the relevant scales in Fig.~\ref{fig:MPSM}.
\end{itemize}

\section{Global symmetries in the Standard Model}\label{sec:globalsymmetries} 

Throughout these lectures, we mentioned various types of global symmetries of the Standard Model.
Having completed the full structure of the Standard Model in the previous section,
we are in the right position to actually collect all of the global symmetries in the Standard Model.
We refer to section 2.5 of \cite{Burgess:2007zi} for a more detailed discussion.

Let us start with the Lagrangian in the absence of any interaction (including self-interactions of the gauge fields) or mass terms. This determines the maximum possible symmetry group of the theory. It can be determined by counting the number of real degrees of freedom for each spin:
\begin{itemize}
\item the Higgs $H$ as a complex doublet corresponds to $4$ real degrees of freedom,
\item there are three generations of fermions and $15$ different species, namely one for $e_{R}^{i}$, two from $L_{L}^{i}$, three from each $u_{R}^{i}$ and $d_{R}^{i}$ and six from $Q_{L}^{i}$. Altogether, this leads to $45$ complex degrees of freedom, and
\item the number of independent helicity-$1$ fields is obtained by adding the ranks of all gauge groups. In total, we have $8+3+1=12$ real degrees of freedom.
\end{itemize}
We can use orthogonal rotations to transform the various bosonic fields of the same spin into each other and similarly unitary transformations for the fermions. This allows us to define the maximum possible global symmetry group as
\begin{equation}
G_{\text{max}}=G_{0}\times G_{1/2}\times G_{1}=\mathrm{O}(4)\times \mathrm{U}(45)\times \mathrm{O}(12)\, .
\end{equation}

The next step is to consider individual interactions in the Standard Model Lagrangian in order to determine the subgroup of $G_{\text{max}}$ that is preserved. Clearly, the gauge group of the Standard Model
\begin{equation}
G_{SM}=\SUTh_{c}\times \SUTw_{L}\times \UO_{Y}\subset G_{\text{max}}
\end{equation}
must be preserved by construction.
We distinguish the following couplings:
\begin{itemize}
\item \emph{gauge self-interactions}: the only transformations that leave the structure constants invariant by virtue of Jacobi's identity are the transformations associated with $G_{SM}$. As a result, there are no additional global symmetries in this sector. In fact, this is a general statement about Lie groups meaning that there will never arise accidental symmetries in this sector.
\item \emph{scalar self-couplings}: these terms leave the full $G_{0}=\mathrm{O}(4)$ intact since the scalar potential $V(H)$ in the Higgs Lagrangian \eqref{eq:HiggsLag} is invariant under rotations (considering the Higgs as a real $4$-vector).
\item \emph{scalar-gauge couplings}: the Standard Model only includes a single irreducible representation of scalar fields and there is no other subgroup than $\SUTw_{L}\times \UO_{Y}$ that commutes with $\mathrm{O}(4)$. So only the $\SUTw_{L}\times \UO_{Y}$ subgroup survives and no additional global symmetries arise. However, for small gauge coupling $g^{\prime}$ associated with $\UO_{Y}$, the $\mathrm{O}(4)$ symmetry is a good \emph{approximate} symmetry.
\item\emph{fermion-gauge couplings}: if we pick a basis for the fermions in the corresponding representations of $G_{SM}$, then there are no larger subgroup than $G_{SM}$ transforming particles into each other in a single generation. However, there are $3$ generations and we can perform unitary rotations relating a fermion in one to another in a different generation. Thus, the subgroup of $G_{1/2}$ giving rise to an accidental global symmetry is
\begin{equation}\label{eq:FermGlSymL} 
G_{f}=\UTh_{Q_{L}}\times \UTh_{u_{R}}\times \UTh_{d_{R}}\times \UTh_{L_{L}}\times \UTh_{e_{R}}\subset G_{1/2}\, .
\end{equation}
Each factor corresponds to a unitary transformation in the space of generations for the individual representations of $G_{SM}$.
\item \emph{Yukawa interactions}: It is convenient to work in a basis in which the mass matrix and therefore also the Yukawa couplings are diagonal. In this basis, the couplings are diagonal and real. Using the experimental observations that the eigenvalues of the Yukawa coupling matrix are all non-zero and non-degenerate, we can restrict the five choices of $\UTh$ matrices from \eqref{eq:FermGlSymL} to be also diagonal with pure phases along the diagonal. This breaks each factor to an a priori independent $\UO$. However, since the Yukawa couplings couple left- and right-handed fields to each other, the $\UO$'s associated to the RH fields equal the LH $\UO$'s. So for the leptons there are three independent $\UO$'s for each generation so that $ \UTh_{L_{L}}\times \UTh_{e_{R}}\raw \UO_{e}\times \UO_{\mu}\times \UO_{\tau}$. For the quarks, we also need to guarantee that the transformations leave the charge current interactions invariant. As we have seen in \eqref{eq:ChargedCurrentsQuarksMassEigenbasis},
these currents involve the CKM-matrix in the basis of mass eigenstates. For a generic such unitary CKM-matrix, the only way of guaranteeing the invariance is by having a diagonal matrix which is why $\UTh_{Q_{L}}\times \UTh_{u_{R}}\times \UTh_{d_{R}}\raw \UO_{B}$ breaks to a single $\UO$ corresponding to the choice of one phase on the diagonal of the $3\times 3$-matrix.

Interestingly, the observed fermion masses and, hence, the measured Yukawa couplings are small in comparison to, e.g., $m_{W}$ and $m_{Z}$. Ignoring the Yukawa couplings results in a larger \emph{approximate} flavour symmetry
\begin{equation}
\text{leptons: }\, \mathrm{U}(3)_{L_{L}}\times  \mathrm{U}(3)_{e_{R}}\kom \text{quarks: }\, \mathrm{U}(3)_{Q_{L}}\times \mathrm{U}(3)_{u_{R}}\times \mathrm{U}(3)_{d_{R}}\, .
\end{equation}
For more details on approximate symmetries in the Standard Model, see chapter 8 in \cite{Burgess:2007zi}.
\item \emph{chiral symmetry:} in the absence of electroweak interactions, we find the approximate chiral symmetry
\begin{equation}
G_{\text{chiral}}=\mathrm{U}(3)_{L}\times \mathrm{U}(3)_{R}\, .
\end{equation}
It becomes relevant when studying the low energy behaviour of strongly interacting quarks, especially the lightest particles as we discussed in section~\ref{sec:chiraltheory}.
\end{itemize}

All in all, we find that the accidental global symmetry group of the Standard Model is given by
\begin{equation}
G_{ac}=\UO_{e}\times \UO_{\mu}\times \UO_{\tau}\times \UO_{B}\, .
\end{equation}
Each generator of these groups is associated with a quantum number that is confirmed to be conserved experimentally. The associated quantum numbers are
\begin{enumerate}
\item As discussed in Sect.~\ref{sec:SSBmodelEW},
the \emph{electron number} $L_{e}$ can be defined in terms of the generators $Q_{L}$ and $Q_{R}$ so that
\begin{equation}
L_{e}=2Q_{L}+Q_{R}
\end{equation}
and hence
\begin{equation}
L_{e}L_{L}^{1}=L_{L}^{1}\kom L_{e}e_{R}^{1}=e_{R}^{1}\kom L_{e}\nu_{R}^{1}=\nu_{R}^{1}\, .
\end{equation}
Acting on any of the other fields, the charges are zero.
Similarly, one can define \emph{muon number} $L_{\mu}$ as
\begin{equation}
L_{\mu}L_{L}^{2}=L_{L}^{2}\kom L_{\mu}e_{R}^{2}=e_{R}^{2}\kom L_{\mu}\nu_{R}^{2}=\nu_{R}^{2}
\end{equation}
and \emph{tau number} $L_{\tau}$ as
\begin{equation}
L_{\tau}L_{L}^{3}=L_{L}^{3}\kom L_{\tau}e_{R}^{3}=e_{R}^{3}\kom L_{\tau}\nu_{R}^{3}=\nu_{R}^{3}\, .
\end{equation}
It is again understood that $L_{\mu}, L_{\tau}$ acting on all other fields vanishes.
The number
\begin{equation}
L_{\text{tot}}=L_{e}+L_{\mu}+L_{\tau}
\end{equation}
is called \emph{lepton number}.
\item \emph{baryon number:}
Baryon number $B$ is defined as
\begin{equation}
BQ_{L}^{i}=\dfrac{1}{3}Q_{L}^{i}\kom Bu_{R}^{i}=\dfrac{1}{3}u_{R}^{i}\kom Bd_{R}^{i}=\dfrac{1}{3}d_{R}^{i}\, .
\end{equation}
\end{enumerate}

The conservation of these quantum numbers has direct physical implications. First, it ensures the stability of the lightest particles carrying a non-zero charge in each generation. Thus, all neutrinos, the electron and the lightest baryon (corresponding to the \textbf{proton}!) are predicted to be absolutely stable. The neutron in an isolated environment is unstable with a lifetime of approximately $880$ second, albeit being stable when bound in a nucleus. Otherwise, life as we know it would not be possible. Moreover, the above observations suggest that processes like $\mu\raw e\gamma$ are forbidden by conservation of $L_{e}$ and $L_{\mu}$. Due to the fact that neutrinos are not perfectly massless, there is evidence that the separate lepton numbers are \textbf{not} conserved, but the violation is so tiny that they are hardly detectable in particle experiments. Hence, the last statement is true only approximately and these so called \emph{charged lepton flavour violations} are expected to appear in nature. 

Baryon and lepton number are symmetries of nature not because of an ad-hoc assumption, but simply because the Lagrangian written as the most general renormalisable Lagrangian consistent with the gauge and spacetime symmetries of the theory happens to be also invariant under baryon and lepton number. We have to bear in mind though that these symmetries are anomalous, as we show below, and hence would be broken in the quantum theory, which for global symmetries is not a sign of inconsistency. However, as we will see in chapter~\ref{chap:probs}, once we add non-renormalisable couplings to the Standard Model Lagrangian, there is no reason for them to exist and they would naturally be broken.

\section{Anomalies in the Standard Model}\label{sec:anomalies_SM}

Let us recall our discussion about anomalies from section~\ref{sec:sym_gauge_anom}.
The general idea is to understand the potential consequences of breaking classical symmetries in quantum theories.
Noether's theorem states that every global continuous symmetry has an associated conserved current.
The conservation is violated in the presence of anomalies.
We then distinguish between anomalies of local and global symmetries.
\begin{enumerate}
\item The former constitute a violation of Ward identities invalidating the quantum theory.
This is because a massless helicity-$1$ particle coupling to a non-conserved current gives rise to an unphysical longitudinal polarisation violating unitarity.
In general, the absence of such \emph{gauge anomalies}\index{Gauge anomalies} is a very strong consistency requirement on any QFT.
\item In contrast, anomalies of global symmetries are not dangerous for the existence of a theory.
They are nonetheless essential to single out conserved quantities.
The prime example is \emph{baryon number} that we introduced in section~\ref{sec:quarks_masses_couplings} which assigns a quantum number $B=1/3$ to all quarks and $B=0$ to all leptons.
As we show below, the associated Noether current $J_{B}^{\mu}=\sum_{i}\,\bar{q}_{i}\gamma^{\mu}q_{i}$ is anomalous.
The violation of baryon number is a crucial ingredient to explain the matter/anti-matter asymmetry in the observed Universe.
\end{enumerate}

In section~\ref{sec:sym_gauge_anom}, we argued that anomalies can be derived from triangle diagrams of the form
\begin{equation*}
\begin{tikzpicture}[scale=1.]
\setlength{\feynhanddotsize}{1.5ex}
\begin{feynhand}
\vertex (a0) at (0,0) {$a$}; 
\vertex (b0) at (2,0); 
\vertex (b1) at (4,1); 
\vertex (b2) at (4,-1); 
\vertex (c1) at (6,1) {$b$}; 
\vertex (c2) at (6,-1) {$c$};
\node (0) at (7,0) {+};
\vertex (a00) at (8,0) {$a$}; 
\vertex (b00) at (10,0); 
\vertex (b11) at (12,1); 
\vertex (b22) at (12,-1); 
\vertex (c11) at (14,1) {$b$}; 
\vertex (c22) at (14,-1) {$c$}; 
\vertex (c33) at (13,0); 
\propag [pho] (a0) to (b0);
\propag [pho] (b1) to (c1);
\propag [pho] (b2) to (c2);
\propag [fer] (b2) to (b0);
\propag [fer] (b0) to (b1);
\propag [fer] (b1) to (b2);
\propag [pho] (a00) to (b00);
\propag [pho] (b11) to (c33);
\propag [pho] (b22) to (c33);
\propag [pho] (c33) to (c22);
\propag [pho] (c33) to (c11);
\propag [fer] (b22) to (b00);
\propag [fer] (b00) to (b11);
\propag [fer] (b11) to (b22);
\end{feynhand}
\end{tikzpicture}
\end{equation*}
which leads to a contribution
\begin{equation}
A^{abc}=\tr(T^{a}\lbrace T^{b},T^{c}\rbrace)=A(R)d^{abc}
\end{equation}
for some representation $R$ of the fermions running in the loop. These are the so-called \emph{anomaly-coefficients} which need to vanish to guarantee the absence of anomalies. The trace implies that we need to sum over all types of fermions that can contribute to these processes which in the case of the Standard Model typically means summing over every colour, flavour and generation.
Overall, the result of summing over the two triangle diagrams leads to (recall \eqref{eq:AxialAnomalyNonAbelian})
\begin{equation}\label{eq:ViolConsCurGGG} 
\p^{\alpha}J_{\alpha}^{a}=\left (\sum_{\text{left}}\, A(R_{l})-\sum_{\text{right}}\, A(R_{r})\right )\, \dfrac{g^{2}}{128\pi^{2}}\, d^{abc}\,\varepsilon^{\mu\nu\alpha\beta}F^{b}_{\mu\nu}F^{c}_{\alpha\beta}
\end{equation}
with a sum over all left-handed and over all right-handed particles.
If the right hand side is non-vanishing, the current is clearly not conserved.
As we argued already section~\ref{sec:sym_gauge_anom}, a non-chiral theory, i.e., a theory with an equal amount of left-handed and right-handed fields is automatically free of anomalies which is obvious from the expression \eqref{eq:ViolConsCurGGG}.
Since the SM is a chiral gauge theory with several gauge group factors, the cancellation of all gauge anomalies is not obvious at all.

\subsection{Local anomaly cancellation in the Standard Model}

Let us get started and show that Standard Model is free of gauge anomalies. Let us denote the anomaly coefficients $A(n_{1},n_{2},n_{3})$ for $n_{i}\in \lbrace 1,2,3\rbrace$ denoting a choice of gauge group factor in $G=\SUTh_{c}\times\SUTw_{L}\times \UO$. We use the $\SUTh$ generators
\begin{equation}
T^{\alpha}=\dfrac{\lambda^{\alpha}}{2}\kom \alpha=1,\ldots,8
\end{equation}
in terms of the $3\times 3$-matrices $\lambda^{\alpha}$ called the \emph{Gellmann-matrices} and similarly the $\SUTw$ generators
\begin{equation}
T^{a}=\dfrac{\sigma^{a}}{2}\kom a=1,2,3
\end{equation}
in terms of the Pauli-matrices $\sigma^{a}$. 

By virtue of \eqref{eq:AxialAnomalyNonAbelian}, the coefficient of the anomaly will be proportional to $\tr\left[\{T_i.T_j\},T_k\right]_{L} -\tr\left[\{T_i.T_j\},T_k\right]_{R}$ with $T_i$ the generators of the corresponding groups and $L,R$ meaning left- and right-handed fermions respectively. 
For example, the diagrams
\begin{equation*}
\begin{tikzpicture}[scale=1.]
\setlength{\feynhanddotsize}{1.5ex}
\begin{feynhand}
\vertex (a0) at (0,0) {$\SUTh$}; 
\vertex (b0) at (2,0); 
\vertex (b1) at (4,1); 
\vertex (b2) at (4,-1); 
\vertex (c1) at (6,1) {$\SUTh$}; 
\vertex (c2) at (6,-1) {$\SUTh$};
\node (0) at (7,0) {+};
\vertex (a00) at (8,0) {$\SUTh$}; 
\vertex (b00) at (10,0); 
\vertex (b11) at (12,1); 
\vertex (b22) at (12,-1); 
\vertex (c11) at (14,1) {$\SUTh$}; 
\vertex (c22) at (14,-1) {$\SUTh$}; 
\vertex (c33) at (13,0); 
\propag [pho] (a0) to (b0);
\propag [pho] (b1) to (c1);
\propag [pho] (b2) to (c2);
\propag [fer] (b2) to (b0);
\propag [fer] (b0) to (b1);
\propag [fer] (b1) to (b2);
\propag [pho] (a00) to (b00);
\propag [pho] (b11) to (c33);
\propag [pho] (b22) to (c33);
\propag [pho] (c33) to (c22);
\propag [pho] (c33) to (c11);
\propag [fer] (b22) to (b00);
\propag [fer] (b00) to (b11);
\propag [fer] (b11) to (b22);
\end{feynhand}
\end{tikzpicture}
\end{equation*}
correspond to the coefficients $A(3,3,3)$ given by
\begin{equation}
A(3,3,3)_{\alpha\beta\gamma}=\tr(T^{\alpha}\lbrace T^{\beta},T^{\gamma}\rbrace)\, .
\end{equation}
Before we start, we can make a few simplifications. Notice that anomaly coefficients associated with e.g. $\UO^{2}\SUTw$ are
\begin{equation}
A(1,1,2)_{a}=\tr\left (Y^{2}T_{a}\right )=\tr(Y^{2})\tr\left (\dfrac{\sigma_{a}}{2}\right )=0
\end{equation}
because the Pauli-matrices are traceless. In general, the generators of $\SUN$ are all traceless which is why we only need to consider the anomaly coefficients associated with $\SUN^{3}$ or $\SUN^{2}\UO$. This means that the only non-trivial anomalies could arise from
\begin{equation}
\UO^{3}\kom\UO\SUTw^{2}\kom\UO\SUTh^{2}\kom \SUTw^{3}\kom \SUTh^{3}\, .
\end{equation}

Let us now go through all these possibilities and show that each anomaly vanishes individually:
\begin{itemize}
\item $A(1,1,1)$: we need to sum over all fermions so that
\begin{align}
A(1,1,1)&=\sum_{\text{fermions}}\, Y^{3}\nn\\[0.2em]
&=3\cdot \left (3\cdot 2\cdot Y_{Q}^{3}+2\cdot Y_{L}^{3}-(Y_{e_{R}}^{3}+3\cdot Y_{u_{R}}^{3}+ 3\cdot Y_{d_{R}}^{3}\right )\nn\\[0.2em]
&=3\cdot \left (3\cdot 2\cdot \left (\dfrac{1}{6}\right )^{3}+2\cdot  \left (-\dfrac{1}{2}\right )^{3}-(-1)^{3}-3\cdot \left (\dfrac{2}{3}\right )^{3}- 3\cdot\left (-\dfrac{1}{3}\right )^{3}\right )\nn\\[0.2em]
&=3\left (\dfrac{1}{36}-\dfrac{1}{4}+1-\dfrac{8}{9}+\dfrac{1}{9}\right )\nn\\[0.2em]
&=0\, .
\end{align}
\item $A(1,2,2)$: Using the fact that
\begin{equation}
\lbrace\sigma^{a},\sigma^{b}\rbrace= 2\delta^{ab}\mathds{1}_{2}\, ,
\end{equation}
we find by summing over all fermions in non-trivial representations of $\SUTw$ (effectively summing over all doublets)
\begin{align}
A(1,2,2)^{ab}&=\tr\left (Y\left \{\dfrac{\sigma^{a}}{2},\dfrac{\sigma^{b}}{2}\right \}\right )\nn\\[0.2em]
&=\delta^{ab}\sum_{\text{doublets}}\, Y\nn\\[0.2em]
&=\delta^{ab}\, 3\left (Y_{L}+3Y_{Q}\right )\nn\\[0.2em]
&=\delta^{ab}\, 3\left (-\dfrac{1}{2}+3\left (\dfrac{1}{6}\right )\right )\nn\\[0.2em]
&=0\, .
\end{align}
\item $A(1,3,3)$: Using the fact that
\begin{equation}
\lbrace\lambda^{\alpha},\lambda^{\beta}\rbrace= \dfrac{4}{3}\delta^{\alpha\beta}\mathds{1}_{3}+2d^{\alpha\beta\gamma}\lambda^{\gamma}\, ,
\end{equation}
we find by summing over all fermions in non-trivial representations of $\SUTh$ (effectively summing over all triplets)
\begin{align}
A(1,3,3)^{\alpha\beta}&=\tr\left (Y\left \{\dfrac{\lambda^{\alpha}}{2},\dfrac{\lambda^{\beta}}{2}\right \}\right )\nn\\[0.2em]
&=\delta^{\alpha\beta}\sum_{\text{triplets}}\, Y\nn\\[0.2em]
&=\delta^{\alpha\beta}\, 3\left (-Y_{u_{R}}-Y_{d_{R}}+2Y_{Q}\right )\nn\\[0.2em]
&=\delta^{\alpha\beta}\, 3\left (-\dfrac{2}{3}+\dfrac{1}{3}+2\left (\dfrac{1}{6}\right )\right )\nn\\[0.2em]
&=0\, .
\end{align}
\item $A(2,2,2)$: Here, we have
\begin{align}
A(2,2,2)^{cab}&=\tr\left (\dfrac{\sigma^{c}}{2}\left \{\dfrac{\sigma^{a}}{2},\dfrac{\sigma^{b}}{2}\right \}\right )\nn\\[0.2em]
&=\dfrac{1}{4}\tr\left (\sigma^{c}\delta^{ab}\mathds{1}_{2}\right )\nn\\[0.2em]
&=0\, .
\end{align}
We could also have used the fact that
\begin{equation}
\sigma_{a}^{*}=-\sigma_{2}\sigma_{a}\sigma_{2}
\end{equation}
which is why the corresponding representations are pseudo-real. This means that the anomaly-coefficients vanish identically.
\item $A(3,3,3)$: Here, we can use that the $\SUTh$ gauge theory is non-chiral and the corresponding coefficients must vanish by the the fact that in general  pseudo-real representations are anomaly free.

\end{itemize}

In addition to these constraints, we also need to discuss potential \emph{gravitational anomalies}. Although we always think about the Standard Model in Minkowski spacetime, we would like to be able to couple it to gravity. In this case, the Poincar{\'e} group should not be viewed as a global, but as a local group since we work with a dynamical, massless spin-$2$ field. Notice that there is nothing wrong with quantising gravity perturbatively and treating it as an effective field theory below the Planck scale. We can compute all kinds of observables in analogy to Yang-Mills theory with only a few modifications. A full non-perturbative treatment of quantum gravity is, however, a serious problem which is for instance addressed by string theory, but we do not care about these subtleties here.

Now, what are gravitational anomalies? Computing anomalies for one gauge boson and two gravitons\footnote{Recall that fermions are the only particles being in complex representations of the Lorentz group. The associated generators are essentially equivalent to $\SUTw$ generators which is why we can apply similar arguments as for $\SUTw$.} leads to
\begin{equation}
\p^{\alpha}J_{\alpha}^{a}\sim \tr(T^{a})\; \varepsilon^{\mu\nu\alpha\beta}R_{\mu\nu}\,^{\gamma\delta}R_{\alpha\beta\gamma\delta}\, .
\end{equation}
Since $\mathrm{SU}(N)$ generators are traceless, the anomalies of $\mathrm{grav}^{2}\mathrm{SU}(N)$ are automatically zero.
The only non-vanishing coefficient therefore is $A(1,J,J)$ for some Lorentz generators $J$. This coefficient corresponds to the sum over all fermions
\begin{align}
A(1,J,J)&=3\left (2Y_{L}-Y_{e_{R}}+6Y_{Q}-3Y_{u_{R}}-3Y_{d_{R}}\right )\nn\\[0.3em]
&=3\left (2\left (-\dfrac{1}{2}\right )-(-1)+6\left (\dfrac{1}{6}\right )-3\left (\dfrac{2}{3}\right )-3\left (-\dfrac{1}{3}\right )\right )\nn\\[0.3em]
&=0\, .
\end{align}

All in all, this implies the the SM as a chiral gauge theory is free of gauge and gravitational anomalies and, as such, is well defined.
Although historically anomaly cancellation was not an initial assumption (simply because they were discovered quite late in the development of the SM, namely around 1969 by Adler \cite{Adler:1969gk} as well as Bell, Jackiw \cite{Bell:1969ts}), the final theory turns out to be anomaly free.
Furthermore, anomaly cancellation conditions give strict guidelines on possible extensions of the SM as will be exemplified with right-handed neutrinos below.

In total, we have the following four anomaly cancellation conditions 
\begin{align}
\label{eq:AnomCanc1} 2 Y_{L}^{3}-Y_{e_{R}}^{3}+6 Y_{Q}^{3}-3 Y_{u_{R}}^{3}- 3 Y_{d_{R}}^{3}&=0\, ,\\[0.1em]
\label{eq:AnomCanc2}Y_{L}+3Y_{Q}&=0\, ,\\[0.1em]
\label{eq:AnomCanc3}-Y_{u_{R}}-Y_{d_{R}}+2Y_{Q}&=0\, ,\\[0.1em]
\label{eq:AnomCanc4}2Y_{L}-Y_{e_{R}}+6Y_{Q}-3Y_{u_{R}}-3Y_{d_{R}}&=0\, .
\end{align}
Since we have in total $5$ independent hypercharges for the fermions, we can fix them uniquely by choosing the value for one of them. 
So choosing a certain value of the hypercharge for, say, the electron leads to a unique assignment of hypercharges for all matter fields by requiring the absence of gauge anomalies.

\subsubsection*{Charge of electron and proton}

From the anomaly cancellation conditions derived above, the most interesting one is almost certainly \eqref{eq:AnomCanc2}, i.e.,
\begin{equation}\label{eq:AnCancIM} 
Y_{L}+3Y_{Q}=0\, .
\end{equation}
This conditions arises due to the chirality of the electroweak sector relating (left-handed) leptons and quarks. There is no contribution from the RH fields and the multiplicities are determined from the associated representation with respect to $\SUTh_{c}$.

The electric charge of the proton is
\begin{equation}
Q_{P}=2Q_{u}+Q_{d}
\end{equation}
where the electric charges of up- and down-quark are determined from
\begin{equation}
QQ_{L}^{i}=\left (T^{3}+Y_{Q}\mathds{1}_{2}\right )Q_{L}^{i}=\left (\begin{array}{cc}
\dfrac{1}{2}+Y_{Q} & 0 \\ 
0 & -\dfrac{1}{2}+Y_{Q}
\end{array} \right )Q_{L}^{i}=\left (\begin{array}{cc}
Q_{u} & 0 \\ 
0 & Q_{d}
\end{array} \right )Q_{L}^{i}\, .
\end{equation}
Thus, the electric charge is given by (using \eqref{eq:AnomCanc2})
\begin{equation}
Q_{P}=2\left (\dfrac{1}{2}+Y_{Q}\right )-\dfrac{1}{2}+Y_{Q}=\dfrac{1}{2}-Y_{L}\, .
\end{equation}
On the other hand, the electric charge of the leptons is determined from
\begin{equation}
QL_{L}^{i}=\left (\begin{array}{cc}
\dfrac{1}{2}+Y_{L} & 0 \\ 
0 & -\dfrac{1}{2}+Y_{L}
\end{array} \right )L_{L}^{i}=\left (\begin{array}{cc}
Q_{\nu} & 0 \\ 
0 & Q_{e}
\end{array} \right )L_{L}^{i}
\end{equation}
and, in particular, of the electron
\begin{equation}
Q_{e}=-\dfrac{1}{2}+Y_{L}\, .
\end{equation}
This immediately implies that
\begin{equation}
Q_{P}=-Q_{e}\, .
\end{equation}

This is remarkable since this allows us to understand a very basic property of matter, that is that the magnitude of the electric charge of protons is exactly the same as that of the electrons with opposite sign.
On the Lie algebra level, there is no argument of why hypercharge or electric charge should be quantised in the SM, cf. the discussion in Sect.~8.5.7 in \cite{Hamilton:2017gbn}. However, the anomaly cancellation condition \eqref{eq:AnCancIM} implies that, whatever the actual value of the electric charges is, being it $\pm 1$ or $\pm\pi$, the electron $e^{-}$ and the proton $p$ have \textbf{exactly opposite charges} and not just approximately.
This is crucial to build stable atoms and hence any type of matter!

For the neutron, we obtain
\begin{equation}
Q_{n}=Q_{u}+2Q_{d}=-\dfrac{1}{2}-Y_{L}=-Q_{\nu}\, .
\end{equation}
Hence, \eqref{eq:AnCancIM} on its own implies
\begin{equation}
Q_{P}=-Q_{e}\kom Q_{n}=-Q_{\nu}
\end{equation}
\emph{whatever values we assign to the hypercharges}.
Only after imposing $Y_{L}=-1/2$,
we recover
\begin{equation}
Q_{P}=-Q_{e}=1\kom Q_{n}=Q_{\nu}=0\, .
\end{equation}
We conclude that charges must always be quantised given that these are the lightest particles in nature out of which matter is being formed.
While the lightest mesons such as the pions $\pi^{0},\pi^{\pm}$ decay into electrons, neutrinos and photons,
the proton is in fact believed to be stable or at least to have an huge lifetime of more than $10^{34}$ years (compare this to the age of the Universe which is $\sim 10^{10}$ years).

\subsubsection*{Gauge anomalies including right-handed neutrinos}

Finally, let us add the right-handed neutrinos to the Standard Model. The new terms in the Lagrangian are
\begin{equation}
\cL^{\text{RH }\nu}=\bar{\nu}_{R}^{i}\I\gamma^{\mu}D_{\mu}\nu_{R}^{i}+(-y_{ij}^{\nu}\bar{L}_{L}^{i}\tilde{H}\nu_{R}^{j}+\text{h.c.})\, .
\end{equation}
The anomaly cancellation conditions become
\begin{align}
\label{eq:AnomCancNR1} 2 Y_{L}^{3}-Y_{e_{R}}^{3}-Y_{\nu_{R}}^{3}+6 Y_{Q}^{3}-3 Y_{u_{R}}^{3}- 3 Y_{d_{R}}^{3}&=0\, ,\\[0.1em]
\label{eq:AnomCancNR2}Y_{L}+3Y_{Q}&=0\, ,\\[0.1em]
\label{eq:AnomCancNR3}-Y_{u_{R}}-Y_{d_{R}}+2Y_{Q}&=0\, ,\\[0.1em]
\label{eq:AnomCancNR4}2Y_{L}-Y_{e_{R}}-Y_{\nu_{R}}+6Y_{Q}-3Y_{u_{R}}-3Y_{d_{R}}&=0\, .
\end{align}
These are still $4$ equations, but now in $6$ variables. Our previous argument about charge quantisation remains true. However, there are now more than one solution. Clearly, nothing changes when setting $Y_{\nu}=0$ which leads to well known hypercharges discussed before.

The most general solutions for the above equations are (cf. Sect.~30.4 in \cite{Schwartz:2013pla})
\begin{align}
Y_{L}&=-\dfrac{a}{2}-b\kom Y_{e_{R}}=-a-b\kom Y_{\nu_{R}}=-b\, ,\nn\\[0.3em]
 Y_{Q}&=\dfrac{a}{6}+\dfrac{b}{3}\kom Y_{u_{R}}=\dfrac{2a}{3}+\dfrac{b}{3}\kom Y_{d_{R}}=-\dfrac{a}{3}+\dfrac{b}{3}
\end{align}
for any $a,b$ and
\begin{align}
Y_{Q}=Y_{L}=0\kom Y_{u_{R}}=c\kom Y_{d_{R}}=-c\kom Y_{e_{R}}=d\kom Y_{\nu_{R}}=-d\, ,
\end{align}
for any $c,d$ which are the only two solutions up to $u_{R}\leftrightarrow e_{R}$ and $e_{R}\leftrightarrow \nu_{R}$. The Standard Model is obtained from the first solutions with the assignments $a=1$ and $b=0$. Again, we only need to impose for instance $Y_{\nu_{R}}=0$ to obtain $b=0$. Then $a=1$ can be obtained by rescaling the $\UO$ coupling constant so that the Standard Model hypercharge is again uniquely determined from the first. The second solution spoils this argument, but it does not match with observations.
Taking into account $\nu_{R}^{i}$, we find another solution $B-L$ (difference between baryon and lepton number).

\subsubsection*{Hypercharge of the Higgs}

One way to constrain the hypercharge for the Higgs is by demanding a non-trivial Yukawa coupling with the electron giving rise to the mass of the electron. We would like the Yukawa couplings to involve only hypercharge neutral terms in order to guarantee gauge invariance. Recall that for the leptons
\begin{equation}
\cL^{\text{Yukawa, leptons}}=-y_{ij}^{e}\overline{L}^{i}_{L}He_{R}^{j}+\text{h.c.}\, .
\end{equation}
To ensure the vanishing of the net hypercharge, we require
\begin{equation}
-Y_{L}+Y_{H}+Y_{e_{R}}=0
\end{equation}
which results in
\begin{equation}
Y_{H}=\dfrac{1}{2}
\end{equation}
as expected. The Lagrangian for all Yukawa couplings is given by
\begin{align}
\cL^{\text{Yukawa}}&=-y_{ij}^{e}\overline{L}^{i}_{L}He_{R}^{j}-y_{ij}^{d}\, \overline{Q}_{L}^{i}Hd_{R}^{j}-y_{ij}^{u}\, \overline{Q}_{L}^{i}\tilde{H}u_{R}^{j}+\text{h.c.}
\end{align}
which gives rise to the following three conditions
\begin{align}
-Y_{Q}+Y_{H}+Y_{d_{R}}&=0\, ,\\
-Y_{Q}-Y_{H}+Y_{u_{R}}&=0\, ,\\
-Y_{L}+Y_{H}+Y_{e_{R}}&=0\, .
\end{align}
It is a remarkable outcome that, with this assignment in combination with anomaly cancellation, not only does the electron get a mass, but also all left- and right-handed quarks due to the presence of non-vanishing Yukawa couplings.

\subsection{Anomalies of global symmetries in the Standard Model}\label{sec:global_anomalies_SM}

In this section, we discuss the possible anomalies associated with the global symmetries $G_{ac}$ found in the previous section. We begin with the baryon number for which
\begin{itemize}
\item $A(B,B,B)$: This coefficient is {\color{blue}non-anomalous}
\begin{align}
A(B,B,B)&=\sum_{\text{all}}\, 2B^{3}=2\dfrac{36-18-18}{27}=0\, .
\end{align} 
\item $A(B,B,1)$: This coefficient is {\color{blue}non-anomalous}
\begin{align}
A(B,B,1)&=\sum_{\text{all}}\, 2YB^{2}\nn\\[0.2em]
&=36\left (\dfrac{1}{6}\right )\left (\dfrac{1}{3}\right )^{2}-18\left (\dfrac{2}{3}\right )\left (-\dfrac{1}{3}\right )^{2}+18\left (\dfrac{1}{3}\right )\left (-\dfrac{1}{3}\right )^{2}\nn\\[0.2em]
&=0\, .
\end{align} 
\item $A(B,1,1)$: This coefficient is {\color{red}anomalous}
\begin{align}
A(B,1,1)&=\sum_{\text{all}}\, 2Y^{2}B\nn\\[0.2em]
&=36\, \left (\dfrac{1}{6}\right )^{2}\dfrac{1}{3}+18\left (\dfrac{2}{3}\right )^{2}\left (-\dfrac{1}{3}\right )+18\left (\dfrac{1}{3}\right )^{2}\left (-\dfrac{1}{3}\right )\nn\\[0.2em]
&=-3\, .
\end{align} 
\item $A(B,2,2)$: This coefficient is {\color{red}anomalous}
\begin{align}
A(B,2,2)&=\sum_{\text{doublets}}\, B=9\left (\dfrac{1}{3}\right )=3\, .
\end{align} 
\item $A(B,3,3)$: This coefficient is {\color{blue}non-anomalous}
\begin{align}
A(B,3,3)&=\sum_{\text{quarks}}\, B\nn\\[0.2em]
&=6\left (\dfrac{1}{3}\right )+3\left (-\dfrac{1}{3}\right )+3\left (-\dfrac{1}{3}\right )\nn\\[0.2em]
&=0\, .
\end{align} 
\item $A(B,J,J)$: This coefficient is {\color{blue}non-anomalous}
\begin{align}
A(B,J,J)&=\sum_{\text{all}}\, B=12-6-6=0\, .
\end{align} 
\end{itemize}

For the individual lepton numbers, we find with $L\in\lbrace L_{e},L_{\mu},L_{\tau}\rbrace$
\begin{itemize}
\item $A(L,L,L)$: This coefficient is {\color{red}anomalous}
\begin{align}
A(L,L,L)&=\sum_{\text{all}}\, 2L^{3}\nn\\[0.2em]
&=4\left (1\right )^{}+2\left (-1\right )^{3}\nn\\[0.2em]
&= 2\, .
\end{align}
\item $A(L,L,1)$: This coefficient is {\color{blue}non-anomalous}
\begin{align}
A(L,L,1)&=\sum_{\text{all}}\, 2YL^{2}\nn\\[0.2em]
&=4\left (-\dfrac{1}{2}\right )\left (1\right )^{2}+2\left (1\right )\left (-1\right )^{2}\nn\\[0.2em]
&= 0\, .
\end{align}
\item $A(L,1,1)$: This coefficient is {\color{red}anomalous}
\begin{align}
A(L,1,1)&=\sum_{\text{all}}\, 2Y^{2}L\nn\\[0.2em]
&=4\left (-\dfrac{1}{2}\right )^{2}\left (1\right )+2\left (1\right )^{2}\left (-1\right )\nn\\[0.2em]
&= -1\, .
\end{align}
\item $A(L,2,2)$: This coefficient is {\color{red}anomalous}
\begin{align}
A(L,2,2)&=\sum_{\text{doublets}}\, L\nn\\[0.2em]
&= 1\, .
\end{align}
\item $A(L,J,J)$: This coefficient is {\color{red}anomalous}
\begin{align}
A(L,J,J)&=\sum_{\text{all}}\, L\nn\\[0.2em]
&=2\left (1\right )+1\left (-1\right )\nn\\[0.2em]
&= 1\, .
\end{align}
\end{itemize}

Some of the anomaly coefficients above are non-zero which implies that the Standard Model suffers from global anomalies. Thus, these symmetries do not survive in the quantum theory. However, this should be understood as a scale dependent observation.
At energies well below $m_{W}$, the violation of the associated conservation laws due to quantum effects is exponentially suppressed. Hence, most of these symmetries are approximately realised in the Standard model, although those involving $\SUTh_{c}$ are strongly broken.

The only anomaly free global symmetries of the Standard Model are obtained by taking appropriate linear combinations, namely
\begin{equation}
L_{e}-L_{\mu}\kom L_{e}-L_{\tau}
\end{equation}
with the third $L_{\mu}-L_{\tau}$ being linearly dependent on the first two. To see this, let us consider any linear combination $L=L_{1}-L_{2}$ with $L_{1},L_{2}\in\lbrace L_{e},L_{\mu},L_{\tau}\rbrace$. Then for example
\begin{align}
A(L,1,1)&= A(L_{1},1,1)-A(L_{2},1,1)=0\, .
\end{align}
The only non-trivial term is
\begin{align}
A(L,L,L)&=\sum_{\text{all}}\, 2(L_{1}-L_{2})^{3}\nn\\[0.2em]
&=\sum_{\text{all}}\, 2(L_{1}^{3}-3L_{1}^{2}L_{2}+3L_{1}L_{2}^{2}-L_{2}^{3})\nn\\[0.2em]
&=A(L_{1},L_{1},L_{1})-3A(L_{1},L_{2},L_{2})+3A(L_{1},L_{2},L_{2})-A(L_{2},L_{2},L_{2})\nn\\[0.2em]
&=0
\end{align}
which is zero because $A(L_{1},L_{2},L_{2})=A(L_{1},L_{2},L_{2})$ by the equality of the associated charges.

One can show that the total lepton number
\begin{equation}
L_{\text{tot}}=L_{e}+L_{\mu}+L_{\tau}
\end{equation}
has the same gauge anomalies as $B$, only the anomaly coefficients for $B^{3}$, $L_{\text{tot}}^{3}$ and the gravitational anomaly disagree. In the presence of right handed neutrinos, all anomaly coefficients agree so that the combination
\begin{equation}
B-L_{\text{tot}}
\end{equation}
is anomaly free.

This makes this symmetry suitable for a potential extension of the Standard Model in which $U(1)_{B-L}$ could be promoted to a full gauge symmetry with a massive $Z'$ gauge field. 


\section{The $\theta$ parameter and quark masses}\label{sec:ThetaTermDiscussionProbs} 
\index{$\theta$ angle}

Most of the structure of the Standard Model Lagrangian is simply adding the electroweak terms of chapter \ref{chap:ew} to the strong interactions Lagrangian of chapter \ref{chap:qcd} and just taking care of the full covariant derivatives. However, by putting them together, there is a term that becomes more relevant once the strong and electroweak interactions are combined in the full Standard Model Lagrangian. This is the $\theta_3$ term for the strong interactions
\begin{equation}
\frac{ g_s^2\theta_3}{32\pi^2} \epsilon^{\mu\nu\rho\sigma}G^A_{\mu\nu} G^A_{\rho\sigma}\kom 
\theta_3= \frac{64\pi^2}{g_s^2} \Theta_G\, .
\end{equation}
First, let us recall three important properties of this term:
\begin{itemize}
\item{\it Total derivative}. This term is a total derivative since $\epsilon^{\mu\nu\rho\sigma}G^A_{\mu\nu} G^A_{\rho\sigma}=\partial^\mu K_\mu$ with $K_\mu$ the Chern-Simons current defined as 
\begin{equation}
K^\mu=\epsilon^{\mu\nu\rho\sigma}\left(G^A_\nu G^A_{\rho\sigma}-\frac{g_s}{3}f^{ABC}G^A_\nu G^B_\rho G^C_\sigma \right)\, .
\end{equation}
This means that this term is not relevant in perturbation theory. However it is important once non-perturbative effects are included.\footnote{Being topological, it so happens that there are field configurations $G^A_\mu\neq 0$ for which the field strength $G^A_{\mu\nu}=0$ and  the  charge $N_{CS}$ associated to the Chern-Simons current\index{Chern-Simons current} $K_\mu$ is non-vanishing $N_{CS}=\int d^3x K_0 \neq 0$. $N_{CS}$ probes the different theta vacua discussed below. A gauge field  configuration such that $\Delta N_{CS} =1$ is called an \emph{instanton}.}

\item{\it CP violation}\index{CP violation}. This term violates CP (see example sheet for full details). The easiest way to see it is that this term has the same structure as the QED case (abelian) for which it is just ${\mathbf E}\cdot {\mathbf B}$ which changes sign under time reversal (and parity) since ${\mathbf E}$ being a polar vector does not change sign under time reversal, but ${\mathbf B}$ does. By the CPT theorem, if it violates T, it violates CP.

\item{\it Quantisation}. In fact, it needs to be taken into account to ensure a proper quantum description. This is discussed in detail in \cite{JackiwTopInv} where it is shown that a proper canonical quantisation of Yang-Mills theory in terms of a complete set of gauge invariant states necessitates the presence of the $\theta$-term. It is therefore important to keep in mind that it is not some mathematical obscurity, but quintessential for the quantisation process itself.

\end{itemize}

For $\mathrm{SU}(2)$ and $\mathrm{U}(1)$ the corresponding $\theta$ parameter can be redefined away as we will explain at the end of this section. We will show now  why for QCD the $\theta_3$ parameter is physical. 
Let us consider the mass term for the six quark flavours ($u,d,s,c,b,t$) in the mass basis
\begin{equation}
\cL_{mass}=m_j \bar\psi_L^j\psi_R^j \kom j=1, \cdots, 6
\end{equation}
Then, we consider a chiral field redefinition for all quark fields
\begin{equation}
\psi_L^j\raw e^{\I \alpha_j} \psi_L^j \kom  \psi_R^j\raw e^{-\I \alpha_j} \psi_R^j\, .
\end{equation}
If the quark masses $m_j$ were real, these rotations can be compensated by transforming the masses as
\begin{equation}
m_j \raw {\ee}^{2\I \alpha_j} m_j\kom  j=1,2,\cdots, 6
\end{equation}
at the price of making the quark masses complex.
Also, as we discussed in chapter \ref{chap:ssb}, each of these $\mathrm{U}(1)$ axial transformations is anomalous and would induce a $\theta$ term for the corresponding gauge field
\begin{equation}
\cD \psi \cD \bar\psi\raw {\rm exp}\left[-\frac{\I g_s^2\sum_j\alpha_j}{32\pi^2}\int d^4x \epsilon^{\mu\nu\rho\sigma}G^A_{\mu\nu} G^A_{\rho\sigma}\right]\cD \psi \cD \bar\psi\, .
\end{equation}
Therefore, the above transformations shift the $\theta_3$ parameter as
\begin{equation}
\theta_3\raw \theta_3 +2\sum_{j=1}^6\alpha_j\, .
\end{equation}
Thus, by a suitable choice of the parameters $\alpha_j$, we can rotate away the $\theta_3$ setting it to zero. However, this will be done at the price of including a phase in the mass matrix and an extra parameter beyond those of the CKM matrix. This is important because this parameter is a phase that breaks CP. This means that we only moved the CP violating parameter $\theta_3$ into another CP violating parameter, but now in the quark masses. 
There is one physically relevant parameter which is the invariant combination
\begin{equation}\label{eq:thetabarQCD} 
\bar\theta \equiv \theta_3-\sum_{j=1}^6\arg m_j=\theta_3-\arg \prod m_j=\theta_3-\arg \det(M)
\end{equation}
where $M$ denotes the quarks mass matrix, recall the discussion in Sect.~\ref{sec:fermioncouplingsEW}.
This is the effective $\theta$-parameter that cannot be rotated away: the chiral anomaly shifts $\theta_3$ and $\arg \det(M)$ by exactly the same constant making $\bar\theta$ invariant.

\subsection*{Theta vacua*}\index{$\theta$ vacua}

This parameter $\bar\theta$ illustrates a particular property of the QCD vacuum in the following sense.
We know from the previous chapter that in the effective chiral Lagrangian for the Goldstone modes $\pi^a$ we can have  a 'mass term'
\begin{equation}
\delta \cL=C \tr \left( MU+M^\dagger U^\dagger\right)
\end{equation}
With  $M$  the $u,d$ mass matrix $M= \rm{diag}\, (m_u, m_d)$, assumed to be real, and $C$ a dimension $3$ constant of order $C\sim \Lambda_{QCD}^3$. Expanding the exponential in $U$ we can get the quadratic terms in the $\pi$ fields proportional to
\begin{equation}
\delta \cL = \dfrac{C}{F_\pi^2}\left(m_u+m_d\right)\left(\pi_0^2+\pi_1^2+\pi_2^2\right)+\ldots
\end{equation}
implying that the mass$^2$ of the pions is of order
\begin{equation}
m_\pi^2=\dfrac{C}{F_\pi^2}(m_u+m_d)\sim \Lambda_{QCD}(m_u+m_d)\, .
\end{equation}
as we have seen above.

Now, if the quark mass matrices include the $\bar\theta$ phase 
\begin{equation}
M={\rm diag}(m_u, m_d) e^{\I\bar\theta}\, ,
\end{equation}
then $\delta \cL$ acquires  a dependence on $\bar\theta$. Taking the leading term in the expansion of $U$ (the identity) implies 
\begin{equation}
E(\bar\theta)=-2C(m_u+m_d)\cos\bar\theta=-F_\pi^2m_\pi^2 \cos\bar\theta\, .
\end{equation}
This illustrates the fact that the vacuum structure of QCD depends on $\bar\theta$ and each minimum of this energy function is a vacuum with degenerate energies. These are known as \emph{theta vacua} or $\theta$-vacua, see Fig.~\ref{fig:theta_vacua}.

\begin{figure}[t!]
\centering
\includegraphics[width=0.7\linewidth]{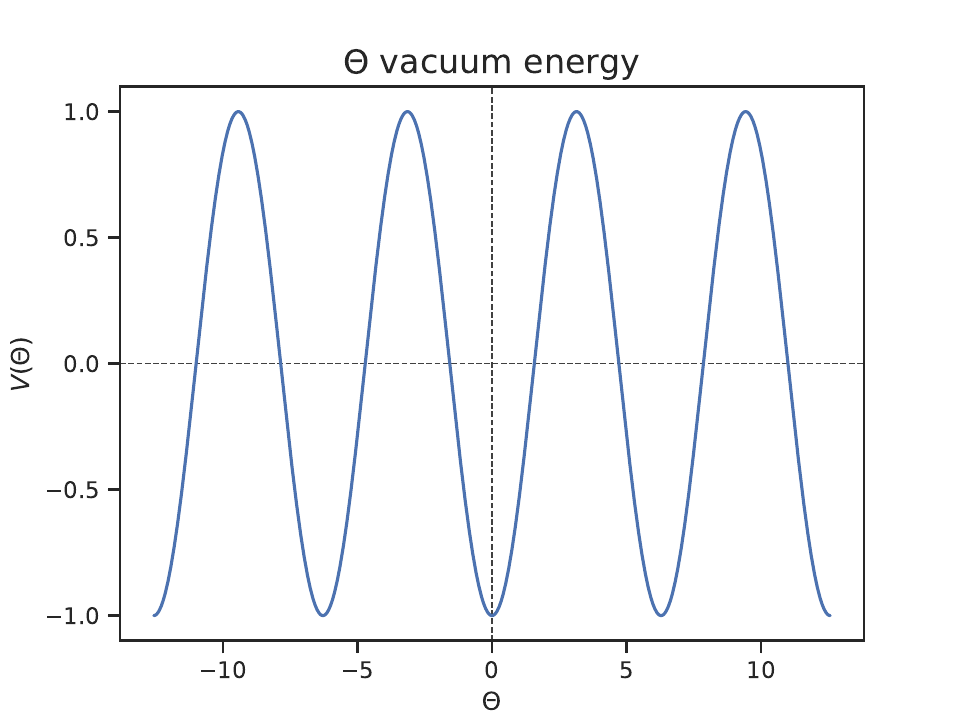}
\caption{The vacuum in QCD has a degeneracy in the sense that it depends on $\bar\theta$ as a periodic function on $\bar\theta$. Each of the minima of this function is an allowed vacuum.}\label{fig:theta_vacua} 
\end{figure}

\subsection*{What about $\theta$ for electroweak interactions?*}



We have seen that a chiral transformation on fermions, being anomalous, generates an effective $\theta$ term. In the case of QCD, we can perform a quark transformation to rotate away the $\theta$ parameter, but this introduces an additional phase in the mass matrix. The physical parameter is $\bar{\theta} = \theta - \arg \det M$. For weak interactions, once the corresponding $\theta$ parameter is shifted to the mass matrix, for instance through a chiral left-handed transformation, we can rotate only the right-handed fermions to remove the phase from the mass matrix. Since right-handed fields are invariant under $\mathrm{SU}(2)_L$, they do not reintroduce a $\theta$ term for $\mathrm{SU}(2)_L$. Therefore, unlike in QCD, the weak interaction $\theta$ term has no physical significance.

A similar argument applies to electromagnetism, where neutrinos can be used to rotate away the corresponding $\theta$ parameter, as they are neutral with respect to electromagnetic interactions. However, in condensed matter systems, electromagnetic interactions involving boundary effects can render the associated $\theta$ term relevant. This is an area of active research and falls beyond the scope of these lectures.

\chapter{\bf Beyond the Standard Model}
\label{chap:probs}

\begin{equ}[In all seriousness]
{\it  Our mistake is not that we take our theories too seriously, but that we do not take them seriously enough. It is always hard to realize that these numbers and equations we play with at our desks have something to do with the real world.}\\

\rightline{\it Steven Weinberg}
\end{equ}
\vspace{0.5cm}



It is no exaggeration to say that the Standard Model ranks among humanity's greatest achievements, and the summary presented in these lectures is merely a brief overview of the remarkable success of this theory. However, as discussed in this chapter, there are still unresolved issues. Over the years, the key limitations of the Standard Model have been identified, and numerous proposals have been put forward to address them. It is essential to outline the most prominent ideas that aim to go \emph{beyond the Standard Model} (BSM).

A word of caution: unlike the preceding chapters, which culminated in the Standard Model -- validated by all experiments up to the TeV scale -- this chapter should be regarded as informed speculation about what might lie beyond the Standard Model at higher energies. None of these ideas currently have experimental support. Nevertheless, since the Standard Model’s development in the 1970's, there has been no shortage of ideas to tackle its shortcomings. Below, we summarise the most significant concepts worth considering in the quest for the next major breakthrough: a high-energy generalisation of the Standard Model.


\vfill

\newpage

\section{Open Questions}

While the Standard Model provides a theoretical description of nature to astonishing accuracy,
it leaves many open questions that need to be addressed by its future extensions.
This section summarises a collection of the most pressing issues.

\subsection{Fundamental}

The UV completion of the gravitational part remains an open question. It is well known that a fully consistent quantum theory describing gravity at all energies does not exist. This is the most fundamental problem in physics. However, as emphasised several times in the lectures, this does not mean that quantum aspects of gravity cannot be addressed at low energies compared to the Planck scale.

\subsection{Strong coupling regimes}

Even though the Standard Model is a consistent quantum theory, performing calculations of physical observables requires essentially solving path integrals that in practice can only be performed in approximation schemes.
We usually resort to perturbative expansions in small parameters like coupling constants, thereby giving rise to Feynman diagrams. Going beyond perturbation theory, even for weak coupling requires techniques not discussed in these lectures such as instantons, monopoles, etc. The most pressing challenge is without a doubt describing gauge theories in strong coupling regimes. As we saw in the previous chapter, QCD at energies below $\Lambda_{QCD}$ is strongly coupled and even though techniques have been developed to address this regime (lattice QCD, $\chi$ PT, large $N_c$ expansion, QCD strings, etc.), this continues to be a major challenge for the wider research landscape. A formal proof of confinement is considered one of the most important questions in mathematical physics. Clearly, this is not a failure of the theory, but it is rather the limitation of physicists to extract all the information from a well formulated theory.

\subsection{Naturalness}

\begin{itemize}
\item $[\cO_{i}]=0$: \emph{The cosmological constant problem}. 
The cosmological constant $\Lambda$ corresponds to energy of the vacuum
\begin{equation}
R_{\mu\nu}-\dfrac{1}{2}Rg_{\mu\nu}=8\pi G\langle T_{\mu\nu}\rangle_{\text{vac}}\sim \Lambda g_{\mu\nu}\, .
\end{equation}
Observations of the current acceleration of the Universe have shown that
 \cite{SupernovaSearchTeam:1998fmf}
\begin{equ}[\hspace{-0.1cm}Cosmological constant problem]
\vspace*{-0.25cm}
\begin{equation}
\Lambda\approx (10^{-3}\text{eV})^{4}\quad\Rightarrow\quad \dfrac{\Lambda}{M_{P}^{4}}\sim 10^{-123}\ll 1\, .
\end{equation}
\end{equ}

\vspace*{-0.3cm}

\noindent However, quantum corrections to $\Lambda$ in the form of vacuum amplitudes are quartically divergent and would naturally lead to $\Lambda\sim M_P^4$.
Therefore, remarkable cancellations are required in order to keep $\Lambda$ small requiring a (doable) fine-tuning of $123$ decimal figures.
The problem becomes even more dramatic since the vacuum energy receives contributions from all sectors in the SM, that is, this issue arises at all scales.
For instance, for the Higgs potential, this would require a tuning of $60$ decimal figures (since $\Lambda/\langle H\rangle^4 \sim 10^{-60}$), similarly for the QCD vacuum all the way to the electron mass. This has been {\bf the biggest puzzle in physics} for the past 50 years given the huge amount of fine tuning required.
While it was originally believed that $\Lambda=0$, after the discovery \cite{SupernovaSearchTeam:1998fmf} of the current acceleration of the Universe, the problem became even more difficult since \emph{explaining} such a small number from first principles looks hopeless.
More generally, the source of this acceleration has been coined \emph{dark energy}\index{Dark energy} \cite{Huterer:1998qv}.
A non-vanishing cosmological constant is the simplest explanation,
but others have been proposed such as \emph{quintessence}\index{Quintessence} \cite{Ratra:1987rm,Wetterich:1987fm}.

\item $[\cO_{i}]=2$: \emph{The hierarchy problem}\index{Hierarchy problem}. Only the Higgs field has an allowed mass term in the Lagrangian $m^2 |H|^2$. At tree-level this can be seen as an insertion on a Higgs line
\begin{align*}
\begin{tikzpicture}[scale=1.2]
\setlength{\feynhanddotsize}{1.5ex}
\begin{feynhand}
\vertex (a2) at (-2,0); 
\vertex (b2) at (2,0); 
\vertex [crossdot] (c2) at (0,0) {};
\propag [sca] (a2) to (c2);
\propag [sca] (c2) to (b2);
\end{feynhand}
\end{tikzpicture}
\end{align*}
Experimentally we know that $m_{h}\sim 125$GeV \cite{ATLAS:2012yve,CMS:2012qbp}.\footnote{Recall that the physical mass of the Higgs $m_h$ is not identical to the parameter $m$ in the Lagrangian but it is proportional to it.}
Also, contrary to gauge and fermion fields, there are quantum corrections to the Higgs mass
\begin{align*}
\begin{tikzpicture}[scale=1.2]
\setlength{\feynhanddotsize}{1.5ex}
\begin{feynhand}
\vertex (a2) at (-2,0); 
\vertex (b2) at (2,0); 
\node (o) at (2.75,0) {$+\, \ldots$};
\vertex (c2) at (-0.75,0);
\vertex (d2) at (0.75,0);
\propag [sca] (a2) to (c2);
\propag[fer] (c2) to [half left] (d2);
\propag[fer] (d2) to [half left] (c2);
\propag [sca] (d2) to (b2);
\end{feynhand}
\end{tikzpicture}
\end{align*}
These diagrams are quadratically divergent ($\sim \int d^4k/k^2$) and therefore give a correction to the Higgs mass of order $\sim M^{2}_{\text{cutoff}}$. Since the Standard Model is renormalisable, the only known cut-off at the moment is the Planck scale. This would then imply that a fine tuning has to be made to quantum corrections up to $15$ decimal figures which leads us to the \emph{hierarchy problem}\index{Hierarchy problem}
\begin{equ}[\hspace{-0.1cm}Hierarchy problem]
\begin{equation}
\dfrac{m_{h}}{M_{P}}\sim 10^{-15}
\end{equation}
\end{equ}

\vspace*{-0.3cm}

\noindent This problem has played an important role in the past 30 years since, to prevent the Higgs mass to become higher than its measured value, the natural expectation is new physics at scales close to the Higgs mass (so that we can replace the cut-off scale by something one or two orders higher than the Higgs mass but not much more). Expectations that the energies explored by the LHC would uncover that new physics are still on, but nothing has been detected.
This implies already that the level of fine tuning is of order one percent. This is not dramatic, but still without a proper explanation.

\item $[\cO_{i}]=4$: \emph{The Strong CP problem}\index{Strong CP problem}. 
We know that the $\theta$ term in the QCD Lagrangian
\begin{equation}
\cL_\theta=\theta_3 \frac{g_s^2}{64\pi^2}\epsilon^{\mu\nu\rho\sigma}G^A_{\mu\nu} G^A_{\rho\sigma}
\end{equation}
can be rotated away by suitable field redefinitions for the quark fields at the prize of introducing a phase in the quarks mass matrix. 
This means that we can change back and forth the phase $\theta_3$ from the $\theta$ term to the quark mass terms. But there is an invariant physical phase $\bar\theta $ defined in \eqref{eq:thetabarQCD}.
This means that we cannot rotate away the $\theta_3$ term by chiral rotations since this would move the CP violating phase from the $\theta$ term to the mass matrix.

What can experiments tell us about the size of $\bar\theta $? Experimentally, the effective dipole moment of the neutron $N$, in an EFT would come from  a CP violating term of the form
\begin{equation}
\cL_{edm}= d_n \epsilon^{\mu\nu\rho\sigma} \bar N \gamma_{\mu\nu} N F_{\rho\sigma}\, .
\end{equation}
The origin of this term is the CP violating part of QCD and is therefore proportional to $\bar\theta$. In a EFT of hadrons it can be induced by a loop of $\pi^-$ and proton coupled to external lines of two neutrons and one photon. The Feynman diagram calculation gives
\begin{equation}
d_n\sim \frac{em_\pi^2}{m_N^3}\bar\theta \sim 10^{-16}e \bar\theta
\end{equation}
and experimentally \cite{Baird:1969px,Crewther:1979pi,Shifman:1979if,
Baker:2006ts,
Pendlebury:2015lrz,
Abel:2020pzs}
\begin{equation}
d_n\leq 10^{-26}e\quad\Rightarrow\quad\bar\theta \leq 10^{-10}\, .
\end{equation}
Explaining why $\bar\theta$ is such a small number is the \emph{strong CP problem}.

\end{itemize}

\subsection{Flavour Problems}

\subsubsection*{Why questions}
The flavour sector is the least elegant part of the Standard Model with three families (six flavours) of matter particles. It actually leads to several puzzles:
\begin{itemize} 
\item \emph{Existence of additional families}. First, matter we know is made only of up and down quarks as well as the electron. Why are there two more families of essentially identical particles then differing only in mass with the first family (and decaying to them by different interactions)? Naively,
it seems to be unnecessary to have such a complicated zoo of particles when the low energy physics is essentially characterised by only a small fraction of them.
\item \emph{Number of families}. Relatedly, why are there exactly three families and no less or more? The only hint we have is that we need at least three families in order to have CP violation. But why is this relevant and chosen by nature?
\item \emph{Mass hierarchies}. This sector is the main source of independent parameters of the Standard Model coming from the Yukawa couplings, including the masses for all particles and the components of the CKM and PMNS matrices \eqref{eq:CKM_matrix} and \eqref{eq:ew_PMNSmatrix}. Furthermore, these parameters differ substantially from each other through huge hierarchies of masses, from $1.7 \times 10^5$MeV for top quark to $0.5$MeV for the electron, not to mention neutrino masses. These parameters are dimensionless (the dimension of masses is given by the Higgs VEV), but there is no explanation of why they have to take the values they do. 

\end{itemize}

\subsubsection*{Neutrino masses}

Neutrino masses have been even less understood than the rest of the flavour sector. This is partly because of the nature of these particles which are so weakly interacting.
Moreover, evidence that their mass is not zero appeared only relatively recently, see \cite{Capozzi:2016rtj} for a review.
If right-handed neutrinos $\nu_{R}$ exist, we can add a term $M\nu_{R}\nu_{R}$ with $[\cO]=3$, recall \eqref{eq:LagrangianLeptonsWithRHNeutrinos}. $M$ may be very large 
since then $\nu_{R}$ should be integrated out.

\subsection{Cosmology*}

One of the most important successes of the Standard Model is the fact that its formulation fits very well with the current picture of the early Universe we have, right after the Big-Bang. But even more importantly, it is equally successful in describing the subsequent history of the Universe including nucleosynthesis\index{Nucleosynthesis}, recombination, matter-radiation equality, large scale structure and the composition, formation and dead of stars, etc. The subfield of \emph{Astro-Particle Physics} has been shaped due to the close connection between particle physics and cosmology questions.

It is within astro-particle physics that the most compelling open questions of the Standard Model can be formulated:
\begin{itemize}
\item \emph{The Big-Bang}. Understanding the Big-Bang is a major problem that requires a full understanding of a fully consistent quantum theory of gravity that is not available.

\item\emph{Dark energy}. As mentioned before the acceleration of the Universe requires an explanation that is not available within the Standard Model unless the cosmological constant is extremely tuned.

\item \emph{Baryogenesis}\index{Baryogenesis}. It is clear that the existence of anti-particles is one of the handful predictions of relativistic QFTs and all particles known have their corresponding anti-particle also discovered. The question is then why did not all particle-anti-particle pairs annihilated themselves in the early Universe and left an empty universe behind. It can be stated that the excess of particles over anti-particles is 1 in $10^{10}$. Andrei Sakharov\index{Sakharov conditions} came with the three conditions that need to be satisfied in order to achieve a mismatch of particles and antiparticles \cite{Sakharov:1967dj}:
\begin{itemize}
\item[$\ast$] \emph{Baryon number violation}: Needed to have an asymmetry between the number of baryons and anti-baryons (core of matter).
\item[$\ast$] \emph{$C$ and $CP$ violation}: Needed so that interactions that produced more baryons are not counter-balanced by interactions that produce more anti-baryons.
\item[$\ast$] \emph{Out of thermal equilibrium}: otherwise, in thermal equilibrium, $CPT$-invariance would restore any generated asymmetry.
\end{itemize}
In principle all of these conditions are satisfied in the Standard Model since baryon number is violated non-perturbatively, $P$ is clearly violated in weak interactions and CP is violated by the CKM matrix. Also there are moments in the early Universe, like phase transitions from unbroken to broken phase of the electroweak theory that are out of thermal equilibrium. However, detailed calculations show that all this is not enough and we may need to go beyond the Standard Model in order to address this question. 

\item \emph{Dark matter}\index{Dark matter}. Since the 1930's evidence has been accumulated at different scales that there is an extra component of matter whose effects are only detected gravitationally \cite{Zwicky:1933gu,Zwicky:1937zza,Freeman:1970mx,Rubin:1970zza}, but are not seen from other interactions. This means that this extra matter, like the neutron or neutrino, is not charged under electromagnetic interactions and for lack of a better name is collectively known as \emph{dark matter}. There are many candidates for this matter including \emph{weakly interacting massive particles} (WIMPs)\index{WIMPs} \cite{Copi:1994ev,deSwart:2017heh}, the axions \cite{Chadha-Day:2021szb} that solve the strong CP problem all the way to primordial black holes \cite{Frampton:2010sw,Lacki:2010zf,Kashlinsky:2016sdv,Espinosa:2017sgp,Clesse:2017bsw}. The search has been going on for several decades already with no success so far in detecting them. The progress has been limited to restrict the large parameter space for these particles which limits substantially the number of possibilities.
Just like for dark energy,
explaining the origin and nature of dark matter remains one of the key targets for fundamental cosmology.

\end{itemize}

\section{Beyond the Standard Model}\label{sec:BSM}

We know that the SM cannot be the final theory. Just to emphasise the aforementioned open questions
again, there is no description of some key principles such as baryogenesis, the nature of dark matter and dark energy and, most importantly, a full quantum description of gravity. But whatever physics will replace it, the Standard Model will remain as the valid description of the world at low energies.
Below,
we mention a few of the ground breaking ideas to extend the Standard Model where we distinguish top-down and bottom-up approaches to \emph{Beyond the Standard Model}\index{Beyond the Standard Model}\index{BSM} (BSM) physics.

\subsection{Top-down}\label{sec:TopDownBSM} 

We begin with a top-down perspective where we construct theories at arbitrarily high energies that have certain desirable properties and subsequently try to derive the Standard Model at low energies by gradually integrating out modes.
Let us mention a couple of the more successful attempts below:

\begin{enumerate}

\item {\bf Quantum Gravity}. \emph{Idea: address the consistency problem.}

The ultimate hope remains having a concrete, consistent theory of quantum gravity. This is arguably the most important problem in all physics and there is no lack of proposals.
It is important to keep in mind that the real problem is to formulate a consistent theory of \emph{all} interactions.
Said differently, it does not suffice to come up with an original way to quantise gravity forgetting about all the other interactions and matter.
At the moment string theory is a concrete promising proposal, but our level of understanding is not yet adequate to claim success. In particular, besides the fundamental questions of its proper formulation, there is no explicit scenario coming from string theory or otherwise that solves all the open questions mentioned before. 
Clearly, this is an ambitious endeavour providing fascinating insights into potential UV completion of the Standard Model coupled to gravity.

\item 
{\bf Supersymmetry}. \emph{Idea: expand basic principles (quantum mechanics and special relativity).}

For this there is a compelling theory: \emph{supersymmetry}\index{Supersymmetry}. This is not an internal symmetry but a spacetime symmetry that transforms fermions into bosons and bosons into fermions.  Briefly, supersymmetry expands the Poincar\' e algebra by including anti-commuting generators enhancing the generators to include the standard $P^{\mu}$, $M^{\mu \nu}$ of Poincar\' e together with the spinor generators $Q_{\alpha}$, $\bar{Q}_{\dot{\alpha}}$, satisfying the algebra
\begin{equation}
 \left[Q_{\alpha} \ , \ M^{\mu \nu} \right] =  (\sigma^{\mu \nu})_{\alpha}\,^{\beta} \, Q_{\beta} , \qquad  \left\{Q_{\alpha} \ , \ \bar{Q}_{\dot{\beta}} \right\} = 2 \,(\sigma^{\mu})_{\alpha \dot{\beta}} \, P_{\mu} 
\end{equation}
where curly brackets stand for anti-commutators.
Notice that from the second equation we can see that two symmetry transformations $Q_{\alpha} \bar{Q}_{\dot{\beta}}$ have the effect of a translation generated by $P^\mu$. That is, let $|B \rangle$ be a bosonic state and $|F \rangle$ a fermionic one, then
\begin{equation}
    Q_{\alpha} \, |F \rangle = |B \rangle , \quad \bar{Q}_{\dot{\beta}}  \, |B \rangle = |F \rangle \implies Q\bar{Q}: \ |B \rangle \ \ \mapsto \ \ |B \ {\rm (translated)} \rangle \ .
\end{equation}
This makes clear that supersymmetry is a spacetime symmetry. Contrary to the Poincar\' e group, one single multiplet includes states with different spins/helicities
\begin{equation}
\left\{ |p^{\mu} , \pm \lambda \rangle , \quad |p^{\mu} , \pm \left(\lambda + \tfrac{1}{2} \right) \rangle \right\}\ .
\end{equation}
This provides a loophole to the Coleman-Mandula theorem\index{Coleman-Mandula theorem} \cite{Coleman:1967ad} mentioned in chapter \ref{chap:stsym}.

Supersymmetry provides a well defined extension of the Standard Model by at least doubling the number of particles. There are, for example, chiral multiplets with $\lambda = 0,\frac{1}{2}$, vector- or gauge multiplets ($\lambda = \frac{1}{2},1$ gauge and gaugino)
    \[\begin{array}{r|l} \lambda = 0 \ {\rm scalar} & \lambda = \frac{1}{2} \ {\rm fermion} \\\hline
    {\rm squark} & {\rm quark} \\ {\rm slepton} & {\rm lepton} \\ {\rm Higgs} & {\rm Higgsino} \end{array} \ \ \ \ \ \ \ \begin{array}{r|l} \lambda = \frac{1}{2} {\rm fermion} & \lambda = 1 \ 
    {\rm boson} \\\hline
    {\rm photino} & {\rm photon} \\ {\rm gluino} & {\rm gluon} \\ W {\rm ino} , \ Z{\rm ino} & W , \ Z \end{array} \ ,
\]
as well as the graviton with its partner
    \[\begin{array}{r|l} \lambda = \frac{3}{2} \ {\rm fermion} & \lambda = 2 \ {\rm boson} \\\hline
    {\rm gravitino} & {\rm graviton} \end{array}
\]
Supersymmetry has several unique properties:
\begin{itemize}
\item[$\ast$] It is at the level of the Poincar\'e symmetry in the sense that it is a spacetime symmetry.
\item[$\ast$] It is singled out as the consistent way to incorporate the only missing allowed states which are helicity $3/2$ particles (the gravitini)\index{Gravitino} \cite{Grisaru:1977kk}. 
\item[$\ast$] One of the predictions of supersymmetry is that superpartners in a single supermultiplet must have the same mass which is in stark contradiction to our observations.
SUSY is clearly not manifest in nature since each particle of the SM should have a supersymmetric partner with the same mass.
However,
not all hope is lost because, the Higgs mechanism demonstrates that there may be many more symmetries in nature that we simply do not observe because they are spontaneously broken.
Similar to the EW vacuum, supersymmetry is not necessarily preserved by the vacuum in which case we say that supersymmetry is spontaneously broken\index{Supersymmetry!Breaking}\index{SUSY breaking}.
In this way, the masses of supersymmetric particles are much higher than those in the Standard Model.
More specifically, they would need to be above $1$ TeV which is the current experimental reach.

\item[$\ast$] It may ameliorate the hierarchy problem\index{Hierarchy problem}. This has been the main argument in favour of supersymmetry. The reason is that the quadratically divergent diagrams that are at the source of the hierarchy problem are cancelled by the contribution of the supersymmetric particles running in the loops. Essentially, fermion loops contribute the same to boson loops, but with the opposite sign such that
\begin{align*}
\begin{tikzpicture}[scale=1.]
\setlength{\feynhanddotsize}{1.5ex}
\begin{feynhand}
\vertex (a2) at (-2,0) {$H$}; 
\vertex (b2) at (2,0) {$H$}; 
\node (o) at (2.75,0) {$+$};
\node (o3) at (8.5,0) {$=\; 0$};
\node (o1) at (0,-1) {fermions};
\node (o2) at (5.5,-1) {bosons};
\vertex (c2) at (-0.75,0);
\vertex (d2) at (0.75,0);
\vertex (a3) at (3.5,0) {$H$}; 
\vertex (b3) at (7.5,0) {$H$}; 
\vertex (c3) at (4.75,0);
\vertex (d3) at (6.25,0);
\propag [sca] (a2) to (c2);
\propag[fer] (c2) to [half left] (d2);
\propag[fer] (d2) to [half left] (c2);
\propag [sca] (d2) to (b2);
\propag [sca] (a3) to (c3);
\propag[bos] (c3) to [half left] (d3);
\propag[bos] (d3) to [half left] (c3);
\propag [sca] (d3) to (b3);
\end{feynhand}
\end{tikzpicture}
\end{align*}
As long as supersymmetry is exact these contributions to the Higgs mass would cancel. But since supersymmetry has to be broken the cancellation happens up to the scale of supersymmetry breaking. If the mass of the superpartners were of order TeV scale it would be enough to protect the Higgs mass and solve the hierarchy problem. However, by the fact that LEP and LHC have not found supersymmetric particles there is already a tension between this solution and experiments.

\item[$\ast$] Supersymmetry may also lead to gauge unification, in the sense that running couplings of the three Standard Model groups join at the same point at a scale close to the Planck scale if there is supersymmetry, but they do not meet without supersymmetry. This may signal that at such energies the three interactions would become one and the same (see the Fig.~\ref{fig:unification}).

\item[$\ast$] Supersymmetry predicts at  least a doubling of the Standard Model particles. Some of the new particles have all the properties to be (at least part of) dark matter candidates known as WIMPS.

\item[$\ast$] Supersymmetry is required for the consistency of string theories and also supersymmetric theories offer simple controllable theories for which non-perturbative effects can be studied under much better control than standard QFTs.

\item[$\ast$] So far supersymmetry has not been observed indicating that, if the symmetry exists, the scale of breaking is probably beyond the reach of LHC. This already affects the argument in favour of supersymmetry solving the hierarchy problem, also the lack of discovery of WIMPS so far put bounds on the supersymmetric candidates for WIMPS. Another weak point is that supersymmetry may alleviate but certainly not solve the cosmological constant problem which is a more pressing problem. Whatever mechanism that solves the cosmological constant problem may also affect the solution of the hierarchy problem. This is then a very open question.

\end{itemize}

\begin{figure}[t!]
\centering
\includegraphics[width=0.9\linewidth]{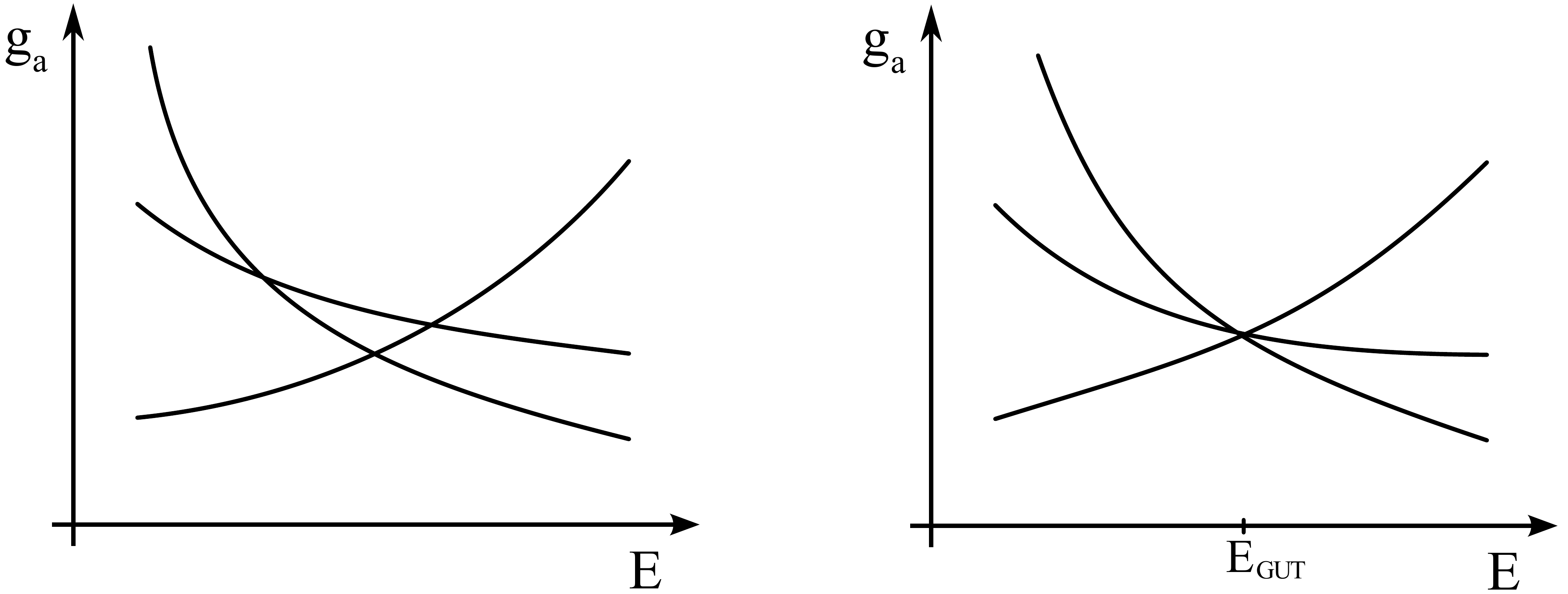}
\caption{The running of the coupling constants for the three interactions in the Standard Model. Without supersymmetry they cross each other at different points. With supersymmetry they cross at the same point hinting at a unified theory at scales of order $10^{16}$ GeV. The running after they meet most probably be different since they may unify to a single simple group such as $\mathrm{SO}(10)$ or directly to the fundamental theory since the scale is close to the Planck scale.
}\label{fig:unification}
\end{figure}

\item{\bf Grand Unification}.\index{Grand Unification} \emph{Idea: embed Standard Model in new gauge theory.}

Following the historical spirit of searching for models that are capable of going beyond the Standard Model and addressing the open questions, a lot of effort has been dedicated to build models beyond the Standard Model (BSM)\index{BSM}.
A natural starting point is enlarging the particle spectrum of the SM in Tab.~\ref{tab:SM}.
However, we know the options are limited since we can only add particles of helicities $0,1/2,1$ (ignoring the gravitino for the moment).
We then have the following options:
\begin{itemize}
\item $\lambda=1/2$: here is certainly room to add more matter particles as long as they do not spoil e.g. anomaly cancellation \eqref{eq:AnomCanc1} - \eqref{eq:AnomCanc4}.
Since the latter is satisfied for each generation separately, adding new generations of quarks and leptons with the same hypercharges is always possible as long as consistent with observational bounds \cite{Carpenter:2010dt,Dighe:2012dz,Nektarijevic:2013hlr}.
But other options are imaginable and might even provide candidates for \emph{dark matter}\index{Dark matter} such as sterile neutrinos \cite{Merle:2013gea}.

\item $\lambda=1$: 
The simplest possibility is to add new abelian gauge particles (commonly referred to $Z'$) such as a local version of the anomaly free $B-L$ symmetry. More ambitious proposals include higher non-abelian groups. The first attempt was the Pati-Salam model \cite{Pati:1974yy} with gauge group
$\mathrm{SU}(4)\times \SUTw_L\times \SUTw_R$ that breaks to $G_{\text{SM}}$ \eqref{eq:GaugeGroupSM} at high energies. This idea played an important role since it was the first concrete realisation of potential baryon number violation with its implications for baryogenesis and proton decay.
Another noteworthy example is the \emph{left-right symmetric model}\index{Left-right symmetric model} \cite{Pati:1974yy,Mohapatra:1974hk,Senjanovic:1975rk} with group $\mathrm{SU}(3)_{C}\times \SUTw_L\times \SUTw_R\times \mathrm{U}(1)_{B-L}$.
It introduces a left-right symmetry which leads to many phenomenologically attractive properties such as an explanation of parity violation \cite{Mohapatra:1974gc,Senjanovic:1975rk} and light neutrinos \cite{Mohapatra:1980qe}.
Moreover, it arises as an intermediate sector in the breaking cascade of $\mathrm{SO}(10)$ to $G_{\text{SM}}$.
Speaking of which, further generalisations of the SM include \emph{Grand Unified Theories (GUTs)}\index{Grand Unified Theories (GUTs)} with groups $\mathrm{SU}(5)$ \cite{Georgi:1974sy},
$\mathrm{SO}(10)$ \cite{Georgi:1974my,Fritzsch:1974nn} and $\mathrm{E}_{6}$ \cite{Gursey:1975ki,Achiman:1978vg}. Here, $\mathrm{SO}(10)$ is particularly interesting since a single representation (the spinorial $\bf 16$) accommodates all particles of one generation of the SM with all the right quantum numbers where
\begin{equation}
{\mathbf {16}}=\left({\mathbf 3}, {\mathbf 2}\right)_{\frac{1}{6}}+\left({\mathbf {\bar 3}}, \mathbf{1}\right)_{\frac{2}{3}}+\left({\mathbf {\bar 3}}, \mathbf{1}\right)_{-\frac{1}{3}}+\left({\mathbf 1}, \mathbf{2}\right)_{-\frac{1}{2}}+\left({\mathbf 1}, \mathbf{1}\right)_{-1}+\left({\mathbf 1}, \mathbf{1}\right)_{0}\, .
\end{equation}
These theories allow the possibility of gauge coupling unification at a scale close to the Planck scale, at least when supplemented by supersymmetry.
That is, using the standard expression for the running of gauge couplings
\begin{equation}
\frac{1}{\alpha_i(\mu)}=\frac{1}{\alpha_i(M)}-\frac{b_i}{4\pi} \log\left(\frac{M^2}{\mu^2}\right)\, ,
\end{equation}
one finds a scale $\mu=\Lambda\sim M_{P}$ for which
\begin{equation}
\alpha_1(\Lambda)=\alpha_2(\Lambda)=\alpha_3(\Lambda)
\end{equation}
where $\alpha_i=g_i^2/(4\pi)$ and $g_i$, $i=1,2,3$ correspond to the gauge couplings for each of the Standard Model gauge groups $\mathrm{U}(1), SU(2)_L, SU(3)_c$ respectively.

\item $\lambda=0$:
After the discovery of the Higgs particle,
it is natural to ask if there are other fundamental scalars in nature. Obviously, if there are higher gauge symmetries, more Higgs-like fields would be needed to mediate the breaking to $G_{\text{SM}}$.
Finally,
a compelling approach to explain the causal structure of our Universe is a period of exponential expansion in the early Universe known as \emph{inflation}\index{Inflation}.
Most models of inflation require a scalar, the \emph{inflaton}\index{Inflaton}, to be the source of this acceleration by slowly rolling down its potential at high energies.

\end{itemize}
These are all interesting approaches in their own right, but there is currently no clear direction due to the lack of experimental guidance, at least when it comes to finding new elementary particles.

\item{\bf Axions}.\index{Axions} \emph{Idea: introduce new fields.}

The strong CP  problem would be resolved if there were a single massless quark because then the determinant of the mass matrix would vanish automatically and $\theta_3$ can be fully rotated away. However there is strong evidence that all quarks have non vanishing mass.

A concrete proposal to address the strong CP problem is the Peccei-Quinn-Weinberg-Wilczek mechanism \cite{Peccei:1977hh,Peccei:1977ur,Wilczek:1977pj,Weinberg:1977ma}.
This is achieved by introducing an anomalous global $\mathrm{U}(1)$ which is broken spontaneously, thereby giving rise to a Goldstone boson field, the \emph{axion} $a(x)$.
That is, the axion $a$ can be seen as the phase of a complex scalar field
\begin{equation}
\Phi(x)=R(x)e^{{\I} a(x)}\rightarrow e^{\I\alpha}\Phi(x)
\end{equation}
which implies that $a$ is equipped with a shift symmetry 
\begin{equation}
a\rightarrow a+ \lambda\kom \lambda \in \bR\, .
\end{equation}
descending from the original $\mathrm{U}(1)$ symmetry.
If the $\mathrm{U}(1)$ symmetry is anomalous (depending on how it couples to fermion fields which we do not need to specify), it would induce a transformation of the Lagrangian
\begin{equation}
\delta\cL=-\frac{\alpha}{64\pi^2}\epsilon^{\mu\nu\rho\sigma}G^A_{\mu\nu} G^A_{\rho\sigma}\, .
\end{equation}
The Lagrangian for this axion field thus takes the form
\begin{equation}
\cL=-\frac{1}{2}\partial^\mu a\, \partial_\mu a -\frac{1}{64\pi^2} \left(\theta+\frac{a}{f_a}\right)\epsilon^{\mu\nu\rho\sigma}G^A_{\mu\nu} G^A_{\rho\sigma}+\cdots
\end{equation}
Here, $f_{a}$ is the axion decay constant which sets the scale of the spontaneous symmetry breaking.
The term $\sim a\, G\wedge G$ arises as a result of the anomalous $\mathrm{U}(1)$.

Non-perturbative effects like instantons satisfying the self-duality condition $F=\star_{4}F$ will generically induce a potential for the axion field given by \cite{Callan:1976je,Peccei:1977hh}
\begin{equation} 
V(a)=-\Lambda_{a}^{4} \cos\left(\frac{a}{f_a}+\Theta \right)
\end{equation} 
with a minimum at $\Theta+\frac{a}{f_a}=2\pi k$, $k\in\bZ$, thereby explaining why the effective $\Theta$ angle is so small. 
Since this is crucial, let us be a bit more explicit.
Using the chiral Lagrangian with the term $MU+M^\dagger U^\dagger$ and using $ M={\rm diag}(m_u, m_d) e^{\I\bar\theta}$ we get a potential of the form
\begin{equation} 
V(a)=E(a(x),\bar\theta)\sim -F_\pi^2m_\pi^2 \cos\left(\frac{a}{f_a}+\bar\theta \right)
\end{equation} 
with minimum at $\bar\theta+\frac{a}{f_a}=0$ (modulo $2\pi $) which means that the effective $\bar\theta$ angle is small.

The above field is the special example of the QCD axion\index{QCD axion} introduced specifically to solve the strong CP problem.
More generally, it may also be argued that, if nature has allowed for fundamental scalar fields like the Higgs, there is definitely room for pseudo-scalars like axions to also exist. So, independent of the strong CP problem, there is a very strong motivation for the existence of axion fields.
They have a rich cosmology \cite{Marsh:2015xka} which makes them attractive for a variety of model building purposes such as for inflation.
The systematic search for axions has been going on for several decades already. Axions have not been detected yet, but they are one of the best motivated particles beyond the Standard Model.

\end{enumerate}

\subsection{Bottom-up}

\begin{itemize}

\item \emph{Standard Model EFT}.

We know that even though the Standard Model is renormalisable, which keeps it consistent and predictive, once it couples to gravity it becomes an EFT with cut-off of order or smaller than $M_P$. Therefore a systematic way to study BSM physics is to consider the particle content of the Standard Model and construct non-renormalisable terms in terms of higher dimension operators which are Lorentz and gauge invariant from the Standard Model particles. 
\begin{equation}
\cL=\cL_{SM}+\dfrac{1}{M}\cL_{5}+\dfrac{1}{M^{2}}\cL_{6}+\cO\left (\dfrac{1}{M^{3}}\right )\, .
\end{equation}
This is known as the \emph{Standard Model effective field theory} or \emph{SMEFT}\index{SMEFT}.
We know that keeping a few of these terms is still predictive as long as the cut-off scale is large enough so that higher order operators can be safely neglected. The importance of this approach is that it is model independent.
By studying the operators themselves, we can put bounds on the magnitude of the couplings and scales which automatically constrains all models that generate these operators under the RG flow at low energies.\index{Renormalisation group flow}\index{RG flow}
For instance, dimension-five operators in $\cL_{5}$ are a source for neutrino masses \cite{Weinberg:1979sa}
\begin{equation}
\cL_5= \left (\dfrac{\lambda_{\nu}}{M}\right ) HH\nu_{L}\nu_{L}\kom M\gg m_{W}
\end{equation}
 This is a direct source of neutrino masses (with no need to introduce right-handed neutrinos at this scale).
For $\langle H\rangle=v\neq 0$
\begin{equation}
\dfrac{\lambda\langle H\rangle^{2}}{M}\sim (50\text{meV})^{2}\quad\Rightarrow\quad M\sim 10^{14}\text{GeV}
\end{equation}
This means that to give rise to neutrino masses at the observed scale ($\sim 50$ meV), the new physics that generates this dimension-five operator has to come at a scale as large as $10^{14}$GeV, assuming the coefficients $\lambda$ of order one.

Similarly, out of a total 63 operators\footnote{The classification of operators up to dimension 8 \cite{Murphy:2020rsh,Li:2020gnx} and 9 \cite{Li:2020xlh} has been achieved quite recently,
see also \cite{Graf:2020yxt,Li:2022abx} for a systematic approach.} of dimension-six \cite{Buchmuller:1985jz,Grzadkowski:2010es} in $\cL_{6}$, there are 4 that violate baryon number.
These operators are of the schematic type
\begin{equation}
\cL_6= \left (\dfrac{\beta}{M^2}\right ) qqql
\end{equation}
where $qqql$ represents three quarks and one lepton like $Q_LQ_LQ_LL_L, Q_LQ_Lu_R e_R$, $Q_LL_Lu_Rd_R, u_Ru_Rd_Re_R$. They all violate baryon number by one unit and therefore allow the proton to decay through processes like $p\rightarrow e^+ + \pi^0$. Knowing the limit on the lifetime of the proton $\tau>1.67\times 10^{34}$ years \cite{Super-Kamiokande:2009yit,Bajc:2016qcc} imply that the new physics that can give rise to these operators has to be at scales $M\geq 10^{15}$GeV. It is interesting to notice that two completely different physical processes, proton decay and neutrino masses hint at a fundamental scale of similar order.
If for some reason the coefficients cancel, then there are dimension $11$ operators $qqq\ell\ell\ell hh/M^{7}$, that would imply the fundamental scale to be $M\gtrsim 10^{5}$GeV.

\item \emph{Amplitudes*}

Another bottom-up approach to address physics BSM is the on-shell amplitudes programme. In this approach, all perturbative aspects of the Standard Model and beyond can be studied by just describing directly the amplitudes of interactions among the corresponding particles without the use of an underlying Lagrangian. One of the motivations of this approach is that in many cases starting from a Lagrangian and computing the amplitudes leads to lengthy calculations that at the end collapse to very simple expressions. Part of the problem is the redundancy generated by gauge invariance. Working directly with the physical on-shell states skips this procedure and amplitudes can be obtained by general requirements of unitarity, locality and causality that are enough to obtain explicit expressions for the amplitudes with much less effort than starting from a Lagrangian, see \cite{Eden:1966dnq,Elvang:2013cua,Benincasa:2013faa,Cheung:2017pzi} for reviews.

This approach has been used to obtain general results such as the ones mentioned in these lectures regarding the possible interacting particles. So far, it reproduces the uniqueness of helicities $\lambda=0,\pm 1/2,\pm 1,\pm 3/2,\pm 2$ as well as provides the general proof for the need of Yang-Mills as an output rather than an input (as also discussed in these lectures).\footnote{In theories with spontaneously broken Lorentz invariance like in cosmology, studying scattering amplitudes requires a modified treatment of on-shell states, see in particular \cite{Pajer:2020wnj}.}
It is within this approach that it has been argued that interactions among massless helicity $2$ particles should be UV completed by introducing an infinite tower of massive states with arbitrarily high spins $j$ as observed in string theory, cf.~\cite{Camanho:2014apa,Caron-Huot:2016icg,Arkani-Hamed:2017jhn,Christensen:2018zcq,Alonso:2019ptb,Arkani-Hamed:2020blm,Cheung:2022mkw,Arkani-Hamed:2023jwn,Cheung:2024obl}.
Ironically, it is a generic theme that theories with IR poles due to long range interactions -- like gravity -- are most challenging to complete in the UV.

This approach has recently been used to derive all Standard Model amplitudes and also the effective operators from SMEFT. A powerful tool for this approach is the so-called ``spinor-helicity'' formalism, which uses explicitly the whole formalism of Weyl spinors developed in chapter \ref{chap:stsym} of these lectures. In particular writing the vector field as a $(1/2,1/2)$ object in therms of $(A,B)$ representations of the Lorentz group (including $\alpha$ and $\dot{\alpha}$ indices instead of a 4d vector. A full description of this formalism is beyond the scope of these lectures, but with the basis learned in chapter \ref{chap:stsym} students can easily follow it. 
For introductory reviews of this subject, we refer to \cite{Elvang:2013cua,Cheung:2017pzi} and references therein.

\end{itemize}

\chapter{Final Remarks}\label{chap:final_remarks}

\begin{equ}[Good ideas take time]
{\it  I am emphasizing here that it took a long time before we realized what these ideas were good for partly because I want to encourage today’s string theorists, who I think also have good ideas that are taking a long time to mature.}\\

\rightline{\it Steven Weinberg}
\end{equ}
\vspace{0.5cm}

This concludes our series of lectures. The primary goal has been to convey the conceptual foundations of the Standard Model, allowing for a deeper understanding and appreciation of the remarkable achievements that have culminated in this comprehensive framework of how nature operates. The structure of these lectures was designed to provide a modern perspective on the logical principles that define the Standard Model. We aimed to illustrate how both inevitable and compelling this model is for describing the world, based on first principles and basic experimental input.

Given the constraints of only 25 lectures, each lasting 50 minutes, and the vast range of material related to the Standard Model, we had to be selective in our approach. The focus was on presenting an overarching view, starting with a historical context, followed by a first-principles approach, now that the subject is mature and well-established. Coordination with the Quantum Field Theory (QFT), Symmetries, Fields, and Particles (SFP), and Advanced QFT (AQFT) courses was necessary to ensure coverage of topics not addressed elsewhere but vital to the Standard Model, while avoiding unnecessary repetition. For instance, we assumed prior knowledge of QED, including the calculation of scattering amplitudes and decay rates. Anomalies were briefly introduced since they are not covered in other courses, though we had to restrict our discussion to basic concepts and tools due to the subject's breadth. The quantisation of Yang-Mills theory was deferred to the AQFT course, and the discussion of running couplings awaited being covered in the concurrent AQFT course.

By bringing together this material, we hope to provide a broad understanding of the subject, equipping students with the knowledge to engage with advanced textbooks or reproduce more detailed calculations of decay rates and scattering amplitudes.



\

Let us finally wrap these notes-up by summarising the main lines of arguments followed in this course:
\begin{enumerate}

\item[1.] The fundamental theories we take as basic postulates are just Special Relativity and Quantum Mechanics.
\item [2.] The symmetries of Special Relativity given by the Poincar\'e group include translations and Lorentz transformations. The basic representations of the Lorentz group correspond to left-handed and right-handed Weyl spinors obtained from the $2$ to $1$ homomorphism between the $\mathrm{SL}(2,\bC)$ and $\mathrm{SO}(3,1)$ groups (indicating that rotations by $4\pi$ rather than $2\pi$ are the ones that are identical to the identity). Out of these two independent representations all other representations of the Lorentz group (vectors, tensors, etc.) can be obtained. This also includes the more standard Dirac spinors which are a reducible representation composed of left- and right-handed Weyl spinors.
\item[3.] Physical states correspond to unitary irreducible representations of the Poincar\'e group. They have very different properties for massive and massless cases.
\begin{itemize} 
\item For the massive case, they are labelled by the eigenvalues of the Casimir operators $C_1=P^\mu P_\mu$ and $C_2=W^\mu W_\mu$ with $P^\mu$ momentum generators and $W_\mu $ the Pauli-Ljubanski vector. Each state within a representation is labelled by the eigenvalues of momenta, chosen in a frame as $p^\mu=(m, 0,0,0)$, and the corresponding Little group associated with the generators that leave invariant $p^\mu$.
In the massive case, the Little group is $\mathrm{SO}(3)$ and so the states are $\ket{m,j;p^\mu, j_3}$ with $m$ the mass, $j=0,1/2, 1, \cdots $ the spin and $j_3=-j,\cdots, j$ the component of the spin in the direction of motion. These states describe massive particles.
\item For massless states the same procedure leads to $C_1=C_2=0$ and the Little group, after imposing finite dimensional representations, reduces to $\mathrm{SO}(2)$ so the states are only labelled by momenta that in a particular frame is $p^\mu=(E,0,0,E)$ and helicity $\lambda=0,\pm 1/2, \pm 1 \cdots $ as $\ket{p^\mu,\lambda}$. 
\end{itemize}
These are the basic quantities to consider to be the building blocks for all matter.
\item[4.] To study interactions among the elementary particle states described above, we impose extra criteria of locality and unitarity which require us to assign a field to every particle state. Then interactions are described by the interaction Hamiltonian and captured by a Lagrangian. Further conditions of stability guarantee the existence of a vacuum state and renormalisability to guarantee predictability, thereby restricting the possible Lagrangians. This condition is understood as only an approximation to a more general Lagrangian written in an Effective Field Theory (EFT) for which the Lagrangian is expanded as $\cL=c_i\cO_i$ with coefficients $c_i$ of higher dimensional operators
$\cO_i$ suppressed with respect to a energy scale $M$ by $c_i=k_i/M^{i-4} $ and $k_i$ dimensionless constants.
\item[5.]
Field theories for particles of spin/helicity $0,1/2$ (for which both massless and massive states have the same number of degrees of freedom) do not offer particular challenges, even though they are very constrained to have Lagrangians with only a few terms.
However, for higher values of spin/helicity constraints become much more severe:
\begin{itemize}
\item Massless states of helicity $\lambda=\pm 1$ have only two polarisation states, whereas the massive particles of spin $j=1$ have $2j+1=3$ degrees of freedom. The polarisation vectors $\epsilon_\mu$, with originally 4-components, in both cases are constrained by $p^\mu\epsilon_\mu=0$ reducing the number of degrees of freedom to $3$.
This is already sufficient for the massive case, but not for the massless case. In this case, since the on-shell condition reads $p^\mu p_\mu=0$, we have to take into account that there is a redundancy in the polarisation vectors.
That is, $\epsilon_\mu+\alpha(p) p_\mu$ is equivalent to the polarisation $\epsilon_\mu$ for any arbitrary function $\alpha(p)$. Moving to position space from this momentum space condition leads to $A_\mu (x)+\partial_\mu \alpha(x)$ being equivalent to $A_\mu$. This manifests gauge invariance for a Lagrangian built to describe the interactions of a field $A_\mu(x)$. Given this redundancy, the polarisation vectors for a helicity $1$ field do not transform as vectors under Lorentz transformations, but they transform as vectors up to $\epsilon_{\mu} \rightarrow \epsilon_\mu+\alpha(p) p_\mu$. In particular an amplitude of the form $\cM=\cM_\mu \epsilon^\mu$ would not be Lorentz invariant unless $p^\mu \cM_\mu=0$. This is the famous Ward identity.

\begin{itemize}
\item[$\ast$] Using the Ward identity and considering a general scattering process with arbitrary incoming and outgoing particles and attaching a ''soft photon'' to each line, we proved that the charges of the particles (defined as the interaction of the photon to the corresponding particle) are conserved, that is: $\sum_{in}Q_i=\sum_{out}Q_i$.

\item[$\ast$]
Using a process equivalent to Compton scattering with arbitrary couplings at the vertex between the $a$ helicity $1$ particle and the matter fields $i$ and $j$ as $T^a_{ij}$. We proved that contributions from the $s$ and $t$ channels satisfy the Ward identity if the couplings satisfy $[T^a,T^b]=0$ (like charges satisfy in QED) unless there is a self-interaction among the helicity $1$ particles with vertex $f_{abc}$ for a cubic (three particle) interaction. In this case the couplings satisfy $[T^a,T^b]=f^{abc}T^c$ which is Yang-Mills theory.

\end{itemize}
This means that only using Lorentz invariance already implies the Ward identity and this in turn implies abelian and non-abelian gauge theories. The corresponding gauge symmetries are not an input but an output of any theory involving helicity $\pm 1$ particles. So they are unavoidable.
 
 \item For massive spin $j=1$ particles, in order to construct a properly defined theory we proved that amplitudes will increase with energy. Since amplitudes are after all probability amplitudes, having an arbitrary large amplitude would break unitarity. Therefore theories of massive spin $1$ particles are not well defined at high energies and need a UV completion.
 
 \item For massless particles of helicity\footnote{Massless particles of helicity $\lambda = \pm 3/2$ can be consistently constructed, but only if they couple to gravity and in a way determined by supersymmetry.} $\lambda=\pm 2$, following the same steps as for charge conservation for helicity $\pm 1$ particles, we established that the equivalent condition to charge conservation is of the form $\sum_{in}\kappa_ip_i^\mu=\sum_{out}\kappa_ip_i^\mu$ with $\kappa_i$ the couplings and $p_i^\mu$ the momenta of the matter particles. This linear condition on momenta is on top of the standard momentum conservation that reads $\sum_{in}p_i^\mu=\sum_{out}p_i^\mu$. Having an extra linear constraint on momenta would not be possible (unless restricting motion to lower dimensional surfaces) and therefore these two conditions should be equivalent implying all the $\kappa_i$ are the same. That means that the interaction mediated by the helicity $2$ particle is the same for all particles. This is the principle of equivalence allowing us to identify this interaction with gravity.
 
 \item Doing the same for higher helicities, say $\lambda=3$, the condition would extend to $\sum_{in}\gamma_ip_i^\mu p_i^\nu=\sum_{out}\gamma_ip_i^\mu p_i^\nu$ for which there are no solutions (keeping in mind momentum conservation) unless the couplings vanish $\gamma_i=0$. This implies that there are no interacting theories for helicity $\lambda>2$ massless particles. This is a very powerful result limiting the possible interactions to be mediated by helicity $1$ particles with abelian or non-abelian gauge symmetries or helicity $\pm 2$ corresponding to gravity and no more. We may add interactions mediated by scalar particles if we prefer to interpret the Higgs as a mediator of interactions rather than a matter particle but this is just a question of semantics. It is then no surprise that the interactions we have observed are precisely of this type. All this is derived only from Lorentz invariance and quantum mechanics.

 \item For massive particles of spin $j \geq 2$ as for spin $j=1$, they do not have perturbative unitarity and need an UV completion which is not known. Notice that there are known composite particles of high spin, but not massive elementary particles of spin $j\geq 2$.\footnote{It has been claimed that a proper UV completion of particles with spin $j\geq 2$ needs an infinite tower of massive states with arbitrarily high spins $j$ (see e.g. \cite{Camanho:2014apa,Caron-Huot:2016icg,Arkani-Hamed:2017jhn,Christensen:2018zcq,Alonso:2019ptb,Arkani-Hamed:2020blm,Cheung:2022mkw,Arkani-Hamed:2023jwn,Cheung:2024obl}), similar to the spectra observed in string theory. But this is an active area of research at the moment with no conclusive results. }
\end{itemize}

\item[6.] Having singled out the theories to those with spin/helicity $0, 1/2, 1, 3/2, 2$, we then concentrated on the Yang-Mills case which also includes QED as the abelian case. Yang-Mills theories are very rich and the number of gauge groups is infinite. First we argued that in order to have positive kinetic terms, we are limited to gauge groups which are compact and simple or semi-simple, that means the groups classified by Cartan.
Specifically, the relevant groups are $\mathrm{SO}(N), \mathrm{SU}(N), \mathrm{Sp}(N), \mathrm{G}_2, \mathrm{F}_4, \mathrm{E}_6, \mathrm{E}_8$ which eliminates the infinite number of non-compact groups, but still leaves a large number of options for gauge symmetries, including all representations for matter fields. Fortunately most of the results are independent of which symmetry group we worked with. We wrote the most general renormalisable Lagrangian coupling gauge fields to matter fields through covariant derivatives and found general properties of these theories.
\begin{itemize} 
\item \emph{Spontaneous symmetry breaking}. Once coupled to scalar fields, the potential for the scalars is of two types depending on the sign of the quadratic term
$m^2|H|^2$. If positive, the minimum of the potential is at $\langle H\rangle =0$ and the symmetry is manifest. If negative, the minimum is at $\langle H\rangle \neq 0$ implying spontaneous symmetry breaking. 
\begin{itemize}
\item[$\ast$] We proved Goldstone's theorem stating that, once a continuous symmetry is broken, there are dim$(G/H)$ massless particles, the Goldstone bosons.
\item[$\ast$] We described the Higgs mechanism in which gauge symmetry is spontaneously broken and the originally massless gauge boson acquires a mass by absorbing the degrees of freedom of the original Goldstone bosons. This solves several problems at once: no massless Goldstone bosons are seen, no massless Yang-Mills fields have been seen but together they become massive spin-1 fields with the extra bonus that the couplings to the massive Higgs field restores perturbative unitarity. Therefore, we ended up with a UV complete theory describing interactions of massive spin $j=1$ particles with matter fields.
\end{itemize}

\item \emph{Asymptotic freedom}. Yang-Mills theories are unique in the sense that the corresponding gauge couplings evolve with energy in a way that the strength of the interactions tends to decrease with increasing energy (asymptotic freedom), unless the theory is abelian (QED) or has a substantial number of matter fields. This allows for the possibility of having fundamental degrees of freedom confined at lower energies. But at high energies the theories are UV complete.

\end{itemize}
\item[7.] \emph{Electroweak theory}. Finally we considered concrete examples of Yang-Mills theory. First an example of spontaneous symmetry breaking describing the weak and electromagnetic interactions. We justified, based on the critical experimental evidence for chirality that the natural group for weak interactions is $\SUTw_L\times \UO_Y$ with $\SUTw_L$ acting only on left-handed fermions. The scalar Higgs field breaks the symmetry to $\UO_{EM}$ and so includes electromagnetism automatically within the model. We described this model in detail including all couplings of gauge fields among themselves, their couplings to fermions (quarks and leptons), to the Higgs particle and the couplings of the Higgs to matter.

\item [8.] \emph{Quantum Chromodynamics}. Strong interactions are an example of asymptotically free Yang-Mills theories explaining the interactions among quarks and gluons and why at low energies it is natural to see only the composite states, namely hadrons. We also proved that well known symmetries such as baryon number are only accidental symmetries in the sense that the most general renormalisable Lagrangian for QCD is automatically symmetric under a global symmetry corresponding to the conservation of baryon number and higher order corrections to the Standard Model Lagrangian would break these symmetries. Furthermore, in describing QCD at energies below $\Lambda_{QCD}$, we uncover the well known approximate symmetries corresponding to isospin (relating protons and neutrons) as well as the eightfold way (the flavour $\SUTh$ symmetry that historically gave rise to the proposal of quarks as the basic building blocks of matter and colours as the symmetry behind the strong interactions). Explaining in this way the origin of these symmetries and their approximate nature due to the fact that there is a hierarchy of masses for the quarks.

\item[9.] \emph{Standard Model and beyond}. We ended up summarising the Standard Model with all its triumphs and limitations. This should serve as a motivation on how to go beyond. In particular the SMEFT naturally adds higher order terms to the renormalisable Lagrangian, including dimension 5 operators that give mass to neutrinos and dimension 6 operators that violate baryon number.

\end{enumerate}

\noindent It is crucial to understand the fundamental principles of the Standard Model in order to gain insight into how we might extend it, particularly in these times when the optimal path forward remains uncertain. Notably, we have made significant progress without direct experimental input, allowing us to conclude that the basic constituents of matter must transform as scalars, fermions, vectors, or symmetric tensors -- and nothing more. We know that spin-1 theories require an underlying symmetry, whether abelian or non-abelian, that such theories are typically asymptotically free and lead to confinement, and that when coupled to scalar fields (whether elementary or composite), they result in symmetry breaking via the Higgs mechanism. Remarkably, all of this follows from the principles of relativity and quantum mechanics alone, making the core features of the Standard Model largely predictable.

The primary choice we face is selecting the appropriate symmetry group and representations. However, not all options are viable, as consistency conditions -- such as the requirement anomaly cancellation -- impose significant constraints (leading, for example, to charge quantisation). Thus, although the development of the Standard Model involved many confusing paths, we can now confidently assert that its structure is highly robust. It is likely that the gauge symmetries and matter content will be modified at higher energies, with the current framework representing a minimal case. Fortunately, significant challenges remain, such as the nature of dark matter, which will help guide us toward the next stage of understanding and bring us closer to a complete fundamental theory of the Universe.

A final note on symmetries: the only true symmetry we have assumed is Poincar\'e invariance from special relativity and CPT symmetry, which is a consequence of relativistic quantum field theories. All gauge symmetries, by contrast, are simply redundancies rather than real symmetries, and global symmetries are only approximate.\footnote{There are also general arguments suggesting that exact global symmetries cannot exist in a fully consistent quantum theory of gravity, see for instance \cite{Kallosh:1995hi}.} Additionally, all spacetime symmetries that can be broken, such as $P$ and $CP$ (or $T$), are broken. Therefore, although symmetries offer a powerful tool for systematically constructing theories through invariant Lagrangians, they may not, in the end, form part of the fundamental principles underlying these theories -- aside from Poincar\'e or more general spacetime symmetries.

We hope that the techniques introduced in this course will help you fully grasp the core principles of the Standard Model and guide you in your future research projects, including those that explore theories beyond the Standard Model.

\

\textbf{Acknowledgements.}
We thank all our colleagues who helped us shape our understanding of this subject and all of our students who asked many interesting questions and found many typos in previous versions of the notes that helped improving the presentation. 
We thank Steven Weinberg for inspiration and for providing the quotes at the beginning of each chapter.
FQ also thanks Cliff Burgess for explaining over many years what Weinberg actually meant.
FQ is grateful to the CERN theory department for providing the perfect environment to finish these lectures.
AS thanks the Department of Applied Mathematics and Theoretical Physics at Cambridge University for support and hospitality where most chapters of these lecture notes have been completed.

\begin{appendices}

\hphantom{Nothing to see here}

\newpage



 
\chapter{Cross sections and decay rates}\label{app:drc}
 
In this appendix we summarise basic formulas to compute cross sections and decay rates which are the final quantities that can be explicitly computed from the full theoretical formalism and at the end are compared with experiments. This is standard QFT material that was only partially covered in this course.
  
Particle physics experiments are some of the most technically complex machines we ever built.
Yet we ask them very basic questions like ``How frequently does $X$ decay to products 
$\alpha+\beta+\gamma+\ldots$?'' or ``Given $N$ collisions between beams of
$A$ and $B$ particles, how many times do we produce particle $X$?''
From these measurements, we determine the free parameters of the Standard Model or even quantify deviations from it.



\section{From Correlation Functions to Scattering Amplitudes}\index{Correlation functions}\index{Scattering amplitudes}

As we mentioned in the lectures, the relevant quantity is the $S$-matrix for scattering between an
initial state $\alpha$ and final state $\beta$.  In the case of decays just
described, $\alpha=X$ and we are interested in inelastic scattering, where
$\beta$ has different particle content than $\alpha$.  In general, the $S$-matrix
elements are given by Dyson's formula
\begin{equation}
\langle \beta | S | \alpha \rangle = \lim_{t_\pm \to \pm\infty} 
\langle \beta | U(t_+, t_-) | \alpha \rangle 
\end{equation}
with
\begin{equation}
U(t_+, t_-) = \mathcal{T} \exp\left( -i\int_{t_-}^{t_+} \! dt' \,H_I(t')
\right) \,.
\end{equation}
Here, the $S$-matrix itself is defined as
\begin{equation}\label{eq:DefSMatrix} 
S=\cT\exp\left (\I\int\dif^{4}x\cL_{I}\right )\, .
\end{equation}
The $S$ matrix can be separated into a boring part (where nothing
happens) and an interesting part (the $T$-matrix)
\begin{equation}
S = 1 + \I \, T\, .
\end{equation}
Due to momentum conservation, we can define the invariant amplitude $\mathcal{M}$ as
\begin{equation}
\label{equ:invamp}
\langle \beta | S - 1 | \alpha \rangle = (2\pi)^4 \delta^{(4)}(p_\beta- p_\alpha) \, \I\mathcal{M}_{\beta\alpha} \,.
\end{equation}
If we expand the $S$-matrix perturbatively, and perform the spacetime integrals, we are able to pull out the factor of $(2\pi)^{4}\delta(p_{\beta}-p_{\alpha})\I$ on the right hand side ensuring momentum conservation.
The quantity $\cM_{\beta\alpha}$ is defined in momentum space and can be computed to every order in perturbation theory by using the momentum space Feynman rules.

To see this, recall that the LSZ formula\index{LSZ formula} certifies that the relevant information in scattering amplitudes is encoded in fully connected correlation functions $\langle\ldots\rangle^{\text{conn}}$.
As an example, let us write the LSZ formula for scattering of $m$ scalar particles in $|\alpha\rangle$ into $n$ new particles in $\langle \beta|$ as
\begin{align}
&\langle p_{1},\ldots ,p_{n},\text{out}|q_{1},\ldots ,q_{m},\text{in}\rangle=\langle p_{1},\ldots ,p_{n},\text{in}|S|q_{1},\ldots ,q_{m},\text{in}\rangle=\langle f | S | i \rangle \\[0.5em]
&=\text{disconnected terms}+(\mathrm{i})^{n+m}\int\, \mathrm{d}^{d}y_{1}\ldots\int\, \mathrm{d}^{d}y_{n}\int\, \mathrm{d}^{d}x_{1}\ldots\int\, \mathrm{d}^{d}x_{m}\mathrm{e}^{\mathrm{i} \sum_{k=1}^{n}\, p_{k}y_{k}-\mathrm{i} \sum_{l=1}^{m}\, q_{l}x_{l}}\nonumber\\[0.5em]
&(\square_{y_{1}}+m^{2})\ldots (\square_{y_{n}}+m^{2})(\square_{x_{1}}+m^{2})\ldots (\square_{x_{m}}+m^{2})\langle\Omega|\mathcal{T}\phi(y_{1})\ldots \phi(y_{n})\phi(x_{1})\ldots \phi(x_{n})|\Omega\rangle\nonumber
\end{align}
assuming that all particles have the same mass $m$.
The first term is associated with disconnected scattering processes where a subset of particles does not participate in the actual scattering such as
\begin{equation*}
\begin{tikzpicture}[scale=1.]
\begin{feynhand}
\vertex (a0) at (0,0.75) {$q_{2}$}; 
\vertex (a1) at (0,0) {$\mathbf{\vdots}$}; 
\vertex (b0) at (0,1.5) {$q_{1}$}; 
\vertex (b1) at (4,1.5) {$p_{1}$}; 
\vertex (a3) at (0,-0.75){$q_{m}$}; 
\vertex [NWblob] (b) at (2,0) {};
\vertex (c0) at (4,0.75) {$p_{2}$};
\vertex (c1) at (4,0.) {$\vdots$}; 
\vertex (c22) at (4,-0.75) {$p_{n}$}; 
\propag [fer] (a0) to (b);
\propag [fer] (a3) to (b);
\propag [fer] (b) to (c0);
\propag [fer] (b) to (c22);
\propag [fer,black] (b0) to (b1);
\end{feynhand}
\end{tikzpicture}
\end{equation*}
The true information about scattering events is encrypted in the fully connected second term.
After plugging in the Fourier transform for the fields, it can be written as
\begin{align}
&\langle p_{1},\ldots ,p_{n},\text{out}|q_{1},\ldots ,q_{m},\text{in}\rangle^{\text{conn}}\\
&=(\mathrm{i})^{n+m} \prod_{k=1}^{n}( -p_{k}^{2}+m^{2}) \prod_{l=1}^{m}(-q_{l}^{2}+m^{2})\, \langle\Omega|\mathcal{T}\hat{\phi}(p_{1})\ldots\hat{ \phi}(p_{n})\hat{\phi}(q_{1})\ldots \hat{\phi}(q_{n})|\Omega\rangle\nonumber
\end{align}
Crucially, the momenta appearing on both sides are \textbf{on-shell} which means that
\begin{equation}
p_{k}^{2}-m^{2}=0\quad ,\quad q_{l}^{2}-m^{2}=0\, .
\end{equation}
For the left hand side to be non-zero, the contributing correlation function on the right has to have a very specific pole structure.
Indeed, these are precisely the connected correlation functions which have $m+n$ poles at $+m^{2}$ because each external line contributes a propagator.
Due to the simple fact that the $S$-matrix measures quantum mechanical probabilities, the right hand side cannot have more poles than that.
Otherwise, it would be singular.
Thus, our final expression for the LSZ formula is
\begin{align}\label{eq:LSZAmpCon} 
&\langle p_{1},\ldots ,p_{n},\text{out}|q_{1},\ldots ,q_{m},\text{in}\rangle^{\text{conn}}=\langle\Omega|\mathcal{T}\hat{\phi}(p_{1})\ldots\hat{ \phi}(p_{n})\hat{\phi}(q_{1})\ldots \hat{\phi}(q_{n})|\Omega\rangle^{\text{conn}}\biggl |_{\text{amputated}}
\end{align}
where $\bigl |_{\text{amputated}}$ gets rid of all external propagators by cutting off the corresponding lines.
To conclude, we reduced the problem of computing scattering amplitudes to computing amputated, connected correlation functions.
The above results can be shown to hold for more general theories involving different species of particles.
Superficially, this just amounts to introducing new labels for spins, polarisations etc.

As a simple example, we consider $2-2$ scattering of particles in a $\phi^{4}$ theory
\begin{equation}
\cL=\dfrac{1}{2}(\p\phi)^{2}+\dfrac{m^{2}}{2}\phi^{2}+\dfrac{\lambda}{4!}\phi^{4}\, .
\end{equation}
At tree level to order $\cO(\lambda)$, there is a single connected diagram
\begin{equation*}
\begin{tikzpicture}[scale=1.]
\begin{feynhand}
\vertex (a0) at (0,1.) {$q_{1}$}; 
\vertex (a3) at (0,-1.){$q_{2}$}; 
\vertex (b) at (2,0);
\vertex (c0) at (4,1.) {$p_{1}$};
\vertex (c22) at (4,-1.) {$p_{2}$}; 
\propag [chasca] (a0) to (b);
\propag [chasca] (a3) to (b);
\propag [chasca] (b) to (c0);
\propag [chasca] (b) to (c22);
\end{feynhand}
\end{tikzpicture}
\end{equation*}
First, we compute the $4$-point correlation function in position space
\begin{align}
\langle\Omega|\mathcal{T}{\phi}(x_{1}) { \phi}(x_{2}){\phi}(y_{1}) {\phi}(y_{2})|\Omega\rangle\bigl |_{\lambda}&=\prod_{j=1}^{4}\, \int\dfrac{\dif^{4}k_{j}}{(2\pi)^{4}}\, \ee^{-\I k_{1}x_{1}}\ee^{-\I k_{2}x_{2}}\ee^{\I k_{3}x_{3}}\ee^{-\I k_{4}y_{2}}\nn\\
&\quad(-\I\lambda)(2\pi)^{4}\delta^{(4)}(k_{1}+k_{2}-k_{3}-k_{3})\prod_{l=1}^{4}\, \dfrac{\I}{k_{l}^{2}-m^{2}}\, .
\end{align}
To find the S-matrix element, we take the Fourier transform
\begin{align}
&\int\dif y_{1}\mathrm{e}^{-\I y_{1}q_{1}}\int\dif y_{2}\mathrm{e}^{-\I y_{2}q_{2}}\int\dif x_{1}\mathrm{e}^{\I x_{1}p_{1}}\int\dif x_{2}\mathrm{e}^{\I x_{2}p_{2}}\langle\Omega|\mathcal{T}{\phi}(x_{1}) { \phi}(x_{2}){\phi}(y_{1}) {\phi}(y_{2})|\Omega\rangle\bigl |_{\lambda}\nn\\
&=(-\I\lambda)(2\pi)^{4}\delta^{(4)}(p_{1}+p_{2}-q_{1}-q_{2}) \dfrac{\I}{p_{1}^{2}-m^{2}}\dfrac{\I}{p_{2}^{2}-m^{2}}\dfrac{\I}{q_{1}^{2}-m^{2}}\dfrac{\I}{q_{2}^{2}-m^{2}}\, .
\end{align}
Discarding all the propagators from external lines, we recover the connected component of the scattering amplitude as given by the right hand side of \eqref{eq:LSZAmpCon}
\begin{equation}
\langle p_{1},p_{2},\text{out}|q_{1},q_{2},\text{in}\rangle^{\text{conn}}\bigl |_{\lambda}=(-\I\lambda)(2\pi)^{4}\delta^{(4)}(p_{1}+p_{2}-q_{1}-q_{2})\, .
\end{equation}
From \eqref{equ:invamp}, we see that
\begin{equation}\label{eq:TTScatPFSA} 
\cM_{q_{1},q_{2}\raw p_{1},p_{2}}^{\text{tree}}=-\lambda\, .
\end{equation}
Recalling \eqref{eq:DefSMatrix} and \eqref{equ:invamp}, we could have simply used that at leading order in the perturbative expansion
\begin{equation}\label{eq:TreeLevelSMatrixElements} 
\cM_{\beta\alpha}^{\text{tree}}=\braket{f|\cL_{I}|i}
\end{equation}
where $\cL_{I}=\lambda\phi^{4}/4!$, but more generally includes all interaction vertices present in the theory.
It is then only a matter of applying suitable Feynman rules for in and out states which can be summarised as:
\begin{itemize}
\item for scalars:
\begin{align}
\wick{\c1{\phi}\ket{\c1{\phi(k)}}} \raw 1\kom \wick{\langle \c1{\phi(k)}| \c1{{\phi}}}\raw 1\, .
\end{align}
\item for fermions:
\begin{align}\label{eq:FeynRulesFermInOut} 
\wick{\c1{f}\ket{\c1{f(k)}}} \raw u_{s}(k)\kom \wick{\c1{\bar{f}}\ket{\c1{\bar{f}(q)}}}\raw \bar{v}_{s}(q)\kom \wick{\langle \c1{f(k)}| \c1{\bar{f}}}\raw \bar{u}_{s}(k)\kom \wick{\langle\c1{\bar{f}(q)}|\c1{\bar{f}}}\raw v_{s}(q)\, .
\end{align}
\item for vectors:
\begin{equation}
\wick{\c1{A_{\mu}}\ket{\c1{A_{\mu}(k)}}}\raw \epsilon_{\mu}(\lambda,k)\kom \wick{\langle\c1{A_{\mu}(q)}|\c1{A}_{\mu}}\raw \epsilon_{\mu}^{*}(\lambda,q)\, .
\end{equation}
\end{itemize}

\section{Decay Rates}\index{Decay rates}

Next, we may ask what is the probability that a state $\alpha$ decays into $\beta$.
The probability that we measure $\alpha \to \beta$ is given by the relevant
$S$ matrix element squared over the norm-squared for the initial and
final states
\begin{equation}
\mathscr{P}(\alpha\raw\beta) = \frac{|\langle  \beta | S - 1 | \alpha \rangle|^2}{\langle  \beta |  \beta
\rangle  \langle \alpha | \alpha \rangle}
\end{equation}
where
\begin{align}
\langle \alpha |  \alpha \rangle = (2\pi)^3 \, 2p_\alpha^0 \delta^{(3)}(0) =
2 p_\alpha^0 V \kom \langle \beta |  \beta \rangle = \prod_{r\in  \beta} \, (2 p_r^0 V)
\end{align}
Here, we work in finite spatial volume $V$ to avoid dealing with subtleties regarding with non-normalisable 
states. The probability the decay will occur is\footnote{The factor $VT$ comes from one factor of the $\delta$-function squared in $|\langle \beta | S |\alpha\rangle|^2$.}
\begin{equation}
\mathscr{P}(\alpha\raw\beta) = \frac{|\cM_{\beta\alpha}|^2}{2 m_i V} \, (2\pi)^4 \,
\delta^{(4)}\Big(p_i - \sum_{r\in f} p_r\Big) \, VT \,\prod_{r\in f} \frac{1}{2p_r^0 V}\, .
\label{eq:weak_probdecay}
\end{equation}

In experiments, the momentum of the final state is never measured with infinite precision.
This means it needs to be integrated over the region in space corresponding to the precision of the detector
or, alternatively, over all possible values for the momenta.
The \emph{partial decay rate} for the process $\alpha\raw\beta$ is then obtained by dividing
the probability \eqref{eq:weak_probdecay} by $T$ and integrating
over momenta.
Since the number of 1-particle states in the box with momentum in a momentum-space volume $\mathrm{d}^3p$ is $V \,\mathrm{d}^3p/(2\pi)^3$, the partial decay rate is
\begin{equation}
\Gamma(\alpha\raw\beta) = \frac{1}{T} \int \sum_{\text{spins},\text{pol.},\ldots}\mathscr{P}(\alpha\raw\beta) \; \prod_{r \in \beta} \frac{V \, \mathrm{d}^3p_r}{(2\pi)^3} \, .
\end{equation}
Here, we sum over spin and polarisation states which can typically not be measured by detectors directly.
The Lorentz-invariant the integral measure on the phase space for the final state $\beta$ is given by
\begin{equation}
\dif\rho_{\beta}=(2\pi)^{4}\delta^{(4)}\left (p_{\alpha}-\sum_{r\in\beta}\, p_{r}\right )\, \prod_{r\in\beta}\, \dfrac{\dif^{3}p_{r}}{(2\pi)^{3}}\, \dfrac{1}{2p_{r}^{0}}\, .
\end{equation}
The partial decay rate then becomes
\begin{align}\label{eq:decayrate}
\Gamma(\alpha\raw\beta)=\dfrac{1}{2m_{\alpha}}\,\int\,  \sum_{\text{spins},\ldots}|\cM_{\alpha\beta}|^{2}\dif\rho_{\beta}\, .
\end{align}
The total decay rate is then simply
\begin{equation}
\Gamma_\alpha =  \sum_\beta \Gamma(\alpha\raw\beta) = \dfrac{1}{2m_{\alpha}}\,  \sum_\beta \,\int\,  \sum_{\text{spins},\ldots}|\cM_{\alpha\beta}|^{2}\dif\rho_{\beta}\, .
\end{equation}

The following identities might come in handy when computing partial decay rates using \eqref{eq:decayrate}:
\begin{itemize}
\item spin sum rules (with $m^2 = k^2$ and $m^{2}=q^{2}$)
\begin{equation}\label{eq:SpinSumRules} 
\sum_{s}\, u_{s}(k)\bar{u}_{s}(k)=\cancel{k}+m\kom \sum_{s}\, v_{s}(q)\bar{v}_{s}(q)=\cancel{q}-m\, .
\end{equation}
\item Trace identities
\begin{align}
\label{eq:TrOdd}\mathrm{Tr}(\gamma^{\mu_1} \cdots \gamma^{\mu_n}) &= 0 ~~~\mbox{for $n$
  odd}\, , \\
\label{eq:TrGam4} \mathrm{Tr}(\gamma^\mu \gamma^\nu \gamma^\rho \gamma^\sigma) & =
4\left(\eta^{\mu\nu}\eta^{\rho\sigma} \,-\, \eta^{\mu\rho}\eta^{\nu\sigma} \,+\, \eta^{\mu\sigma}
\eta^{\nu\rho}\right)\, ,
 \\
\label{eq:TrGam5Gam4} \mathrm{Tr}(\gamma^5\gamma^\mu \gamma^\nu \gamma^\rho \gamma^\sigma)
& = - 4 i \epsilon^{\mu\nu\rho\sigma}\, , \\
\label{eq:TrKSQ} \mathrm{Tr}\,\pslash{k}\,\pslash{q} &= 4\,k\cdot q\, , \\
\label{eq:TrGamFKSQ} \mathrm{Tr}\,\gamma^5 \,\pslash{k}\,\pslash{q} &= 0\, , \\
\label{eq:TrSGSGGF} \mathrm{Tr}\,\gamma^\mu &= \mathrm{Tr}\,\gamma^\mu\gamma^5 =  0\, .
\end{align}
\end{itemize}

\section{Cross Sections}\index{Cross sections}

\begin{figure}[t!]
\begin{center}
\includegraphics[width=0.9\textwidth]{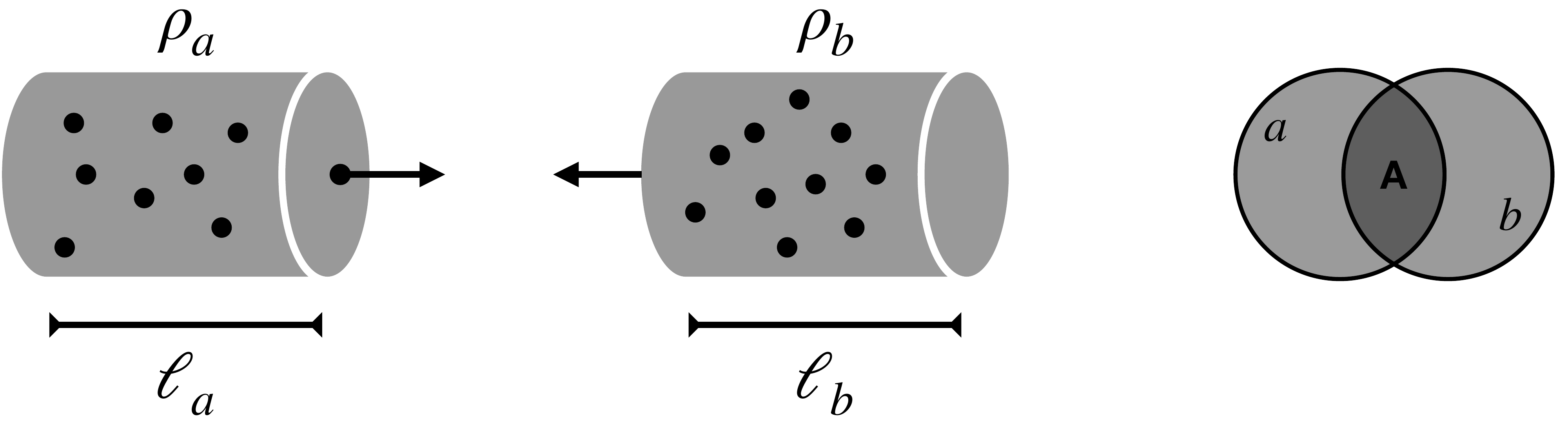}
\end{center}
\caption{Two bunches of particles, with densities $\rho_a, \rho_b$ and lengths $\ell_a, \ell_b$ collide with cross-sectional area $A$.
}\label{fig:weak_Xsection}
\end{figure}

In addition to knowing how often particles decay,
many experiments in particle physics need to quantify how often particle collides.
Suppose we have two beams of particles and collide them as depicted in Fig.~\ref{fig:weak_Xsection}.
The number of collision events can then be estimated through \emph{cross sections}:
they are obtained from the total number of scattering events divided by cross-sectional area of the collision region as well as the densities and size of the beams.
That is, we compute the quantity
\begin{equation}\label{eq:cross_section} 
\sigma = \frac{\#~\mbox{scattering events}}{\rho_a\ell_a\rho_b\ell_b A} = \frac{N}{F \rho_b V}
\end{equation}
where we introduced the number of
\begin{itemize}
\item scattering events per unit time $N$,
\item target particles $\rho_b V$ in volume $V = \ell_b A$, and
\item incoming particles per unit area per unit time $F = |\VEC{v}_a - \VEC{v}_b| \rho_a$ (also referred to as \emph{incident flux}) in terms of the relative velocity of the particles in the two beams $|\VEC{v}_a - \VEC{v}_b|$.\footnote{In our normalisation, we have one particle in volume $V$, i.e, $\rho_a = \rho_b = 1/V$, and so $F = |\VEC{v}_a - \VEC{v}_b| / V$.}
\end{itemize}
The dimension of the cross-section $\sigma$ in \eqref{eq:cross_section} is that of an area.
It is traditionally measured in the unit \textit{barn} with 1 barn $= 10^{-28}
\mathrm{m}^2$.

We are mostly interested in the \emph{differential cross sections}: the beam of outgoing particles will be measured at different angles and we would like to understand the probability distribution of measuring scattered particles e.g. for different solid angle elements.
Specifically, we want to compute the differential probability per unit time of an event $\alpha\to \beta$.
We divide by the flux of particles through the interaction region. 
For the particles moving in the lab frame, 
the prefactor $1/2m_\alpha$ in $\Gamma$ in \eqref{eq:decayrate} becomes $1/2E$ for each beam of particles.
Thus, we find
\begin{align}
\label{eq:weak_dsigma}
d\sigma  &=   \frac{1}{F}\,\frac{1}{4E_a E_b V} \, |\cM_{\beta\alpha}|^2 \, d\rho_\beta \nonumber \\
   &=   \frac{1}{|\VEC{v}_a - \VEC{v}_b|}\,\frac{1}{4E_a E_b} \,|\cM_{\beta\alpha}|^2\,d\rho_f 
\end{align}
in terms of 
\begin{equation}
dN = \frac{1}{4E_a E_b V} \, |\cM_{\beta\alpha}|^2 \, d\rho_\beta \, .
\end{equation}

The notion of differential cross sections is useful as it can reveal the substructures of particles just the atomic nucleus in Rutherford scattering.
In the previous case of $2-2$ scattering in $\phi^{4}$, the leading order expression for the cross section obtained from \eqref{eq:TTScatPFSA} reads
\begin{equation}
\dfrac{\dif\sigma}{\dif\Omega_{3}}=\dfrac{1}{2!}\dfrac{1}{64\pi^{2}}\dfrac{1}{s}\,  \lambda^{2}\sim \dfrac{1}{s}
\end{equation}
in terms of the Mandelstam variable $s=(p_{1}+p_{2})^{2}$.
This result is very characteristic for scattering point-like objects and, in fact, holds more generally:
\begin{Boxequ}
\vspace*{0.1cm}
For a target with no substructure of length $l\geq 1/\sqrt{s}$, the differential cross-section for hard scattering falls off as $1/s$.
\end{Boxequ}
This behaviour was critical to discover that hadrons have a \emph{parton structure} in deep inelastic scattering experiments with hadrons.

\section{$\pi$ decay*}
\label{sec:weak_pidecay}

\begin{figure}[t!]
\begin{center}
\includegraphics[width=0.8\textwidth]{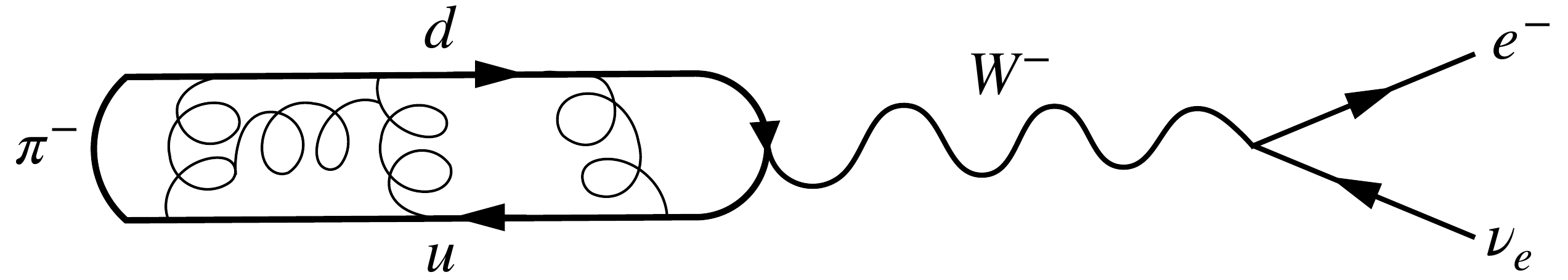}
\end{center}
\caption{Weak decay of a pion to electron and anti-neutrino.
}\label{fig:weak_pi2enu_quark}
\end{figure}

To conclude this appendix, we discuss the $\pi^{-}$ meson decay to electron and anti-neutrino.
The process is similar to the muon's decay discussed in Sect.~\ref{sec:4FermiMuonDecay}
since it is mediated through the charged weak currents in Eq.~\eqref{eq:ChargedCurrents}.
Specifically, a down quark $d$ and an up anti-quark $\bar{u}$ annihilate to a $W^-$ boson,
which then promptly decays into electron and anti-neutrino. 
Crucially, however,the initial state of the $d$-type and $\bar{u}$-type quarks \emph{does not correspond to freely propagating particles};
as we learned in Sect.~\ref{sec:AsymptoticFreedom}, the quark and anti-quark are strongly bound together into a hadronic state which in the above case corresponds to a $\pi^-$ meson, or pion.
The decay is visualised in Fig.~\ref{fig:weak_pi2enu_quark}.

Let us start by collecting some important definitions.
Recall the definition \eqref{eq:ChargedCurrents} for the charges currents, namely 
\begin{equation}
J_{\mu}^{+}=\overline{\nu}^{i}_{L}\gamma_{\mu}e_{L}^{i}+\overline{u}^{i}_{L}\gamma_{\mu}d_{L}^{i}\kom J_{\mu}^{-}=\overline{e}^{i}_{L}\gamma_{\mu}\nu_{L}^{i}+\overline{d}^{i}_{L}\gamma_{\mu}u_{L}^{i}\, .
\end{equation}
To isolate the relevant contributions for the above decay, we first define the \emph{leptonic weak current}
\begin{equation}
J^\mu_{\mathrm{lept}} = \bar{e}\gamma_\alpha(1-\gamma^5)\nu_e\, .
\end{equation}
This can be used to describe the decay $W^{-} \raw e^{-}+\nu_{e}$.
In addition, we need the \emph{hadronic weak current} where it is convenient to separate the current into terms of definite parity, namely
\begin{equation}
J^\mu_{\mathrm{had}} = V^\mu_{\mathrm{had}} - A^\mu_{\mathrm{had}}\, .
\label{eq:weak_LHcurrent}
\end{equation}
Here, the vector and axial-vector currents in the mass eigenbasis for the quarks (recall \eqref{eq:ChargedCurrentsQuarksMassEigenbasis}) are
\begin{align*}
V^\mu_{\mathrm{had}} =& \bar{u}\gamma^\mu (V_{ud} d
+V_{us} s +V_{ub} b)  \\
A^\mu_{\mathrm{had}}=& \bar{u}\gamma^\mu\gamma^5 (V_{ud} d
+V_{us} s +V_{ub} b) \,.
\end{align*}
Here, we keep only the terms relevant for the $\pi^{-}$ decay.
For energies $E\ll m_w$, the effective Lagrangian from integrating out the $W^{\pm}$-boson becomes
\begin{equation}\label{eq:effIntPiEnu} 
\L^{\mathrm{eff}}_W = -\frac{4G_F}{\sqrt{2}} J_{\mu,\,\mathrm{lept}} J^\mu_{\mathrm{had}}\, .
\end{equation}
This is again of the form as in Eq.~\eqref{eq:FourFermiCurrent} as proposed by Marshak and Sudarshan \cite{Sudarshan:1958vf}.

At tree level, the relevant amplitude in the above approximation is simply depicted as
\begin{equation*}
\begin{tikzpicture}[scale=1.1]
\setlength{\feynhanddotsize}{1.75ex}
\begin{feynhand}
\vertex (a00) at (5.,2) {$\pi^{-}$}; 
\vertex (a11) [dot] at (8.,2) {}; 
\vertex (d00) at (10.5,3.5) {$e^{-}$}; 
\vertex (d11) at (10.5,0.5) {$\bar{\nu}_{e}$}; 
\propag [fer] (a00) to (a11);
\propag [antfer] (d00) to(a11);
\propag [antfer] (d11) to (a11);
\end{feynhand}
\end{tikzpicture}
\end{equation*}
The vertex corresponds to the effective interaction in \eqref{eq:effIntPiEnu}.
It is computed as follows
\begin{align}
\mathcal{M} &= \langle e^-(k)\bar{\nu}_e(q) | \L^{\mathrm{eff}}_W |
\pi^-(p)\rangle \nonumber \\
&= -\frac{G_F}{\sqrt{2}}\langle e^-(k)\bar{\nu}_e(q) | 
\bar{e}\gamma_\mu (1-\gamma^5)\nu_e|0\rangle\langle 0|
J^\mu_{\mathrm{had}} |\pi^-(p)\rangle \,.
\end{align}
Since the pseudo-scalar meson $\pi^{-}$ is by definition parity-odd, while the QCD vacuum is parity-even,
the matrix element of $J^\mu_{\mathrm{had}} $ only picks up the piece $A^\mu_{\mathrm{had}}$.
Hence, we obtain
\begin{align}
\mathcal{M} &= \frac{G_F}{\sqrt{2}} \bar{u}_e(k) \gamma_\alpha(1-\gamma^5)
v_{\nu_e}(q)\langle 0 | A^\alpha_{\mathrm{had}}| \pi^-(p)\rangle \,.
\end{align}
QCD is strongly interacting and, in particular, free quarks are forbidden
due to confinement .
Hence, approximating the hadronic matrix element perturbatively is a hopeless task.
Instead, we parametrise our ignorance in a single dimensionful parameter, the so-called pion decay constant $F_\pi$, such that
\begin{equation}
\langle 0 | V_{ud} \bar{u}\gamma^\alpha\gamma^5 d| \pi^-(p)\rangle
= i V_{ud}\sqrt{2}F_\pi p^\alpha \,.
\label{eq:weak_piondc}
\end{equation}
Due to momentum conservation,
we write $p=k+q$ to arrive at (the neutrino is massless)
\begin{equation}
\bar{u}_e(k) \,\pslash{k} = \bar{u}_e(k) m_e \kom
\pslash{q}\,v_{\nu_e}(q) = 0\, .
\end{equation}
Then we find
\begin{equation}
\mathcal{M} = iG_F F_\pi m_e V_{ud} \bar{u}_e(k)(1-\gamma^5) v_{\nu_e}(q) \,.
\end{equation}

Next,
we want to derive the decay rate \eqref{eq:decayrate} for which we need the squared amplitude $|\cM|^{2}$ and sum over all spins.
We therefore compute
\begin{equation}
\sum_{\mathclap{\mathrm{spins}}} |\mathcal{M}|^2 = 2 |G_F F_\pi m_e V_{ud}|^2
\,\mathrm{Tr}\left[(\pslash{k}+m_e)(1-\gamma^5) \,\pslash{q}\right] 
\end{equation}
having used $(1-\gamma^5)\gamma^\mu(1+\gamma^5) = 2(1-\gamma^5)\gamma^\mu$.
Utilising the trace identities \eqref{eq:TrKSQ}, \eqref{eq:TrGamFKSQ} and \eqref{eq:TrSGSGGF}, we obtain
\begin{equation}
\sum_{\mathclap{\mathrm{spins}}} |\mathcal{M}|^2 = 8 |G_F F_\pi m_e V_{ud}|^2 \,
k\cdot q \,.
\end{equation}
Therefore the decay rate in the $\pi$ rest frame is
\begin{align}
\Gamma_{\pi\to e\bar{\nu}} =& \frac{1}{2m_\pi} \int\!\frac{d^3k}{(2\pi)^3 2k^0}
\frac{d^3q}{(2\pi)^3 2q^0}\,(2\pi)^4\,\delta^{(4)}(p-k-q) 
\sum_{\mathclap{\mathrm{spins}}} |\mathcal{M}|^2 \nonumber \\
=& \frac{ |G_F F_\pi m_e V_{ud}|^2}{m_\pi\, (2\pi)^2} \int\!\frac{d^3k}{ k^0}
\frac{d^3q}{ q^0}\,\delta^{(4)}(p-k-q) 
 \,
(k\cdot q)  \nonumber \\
=&  \frac{|G_F F_\pi m_e V_{ud}|^2}{4\pi^2 m_\pi}
\int \!\frac{d^3k}{E|\VEC{k}|}\,\delta(m_\pi - E - |\VEC{k}|)
(E + |\VEC{k}|)|\VEC{k}| \,.  
\end{align}
We now use the composition rule for the $\delta$-distribution
\begin{equation}
\delta(f(k)) = \sum_i \dfrac{\delta(k-k_0^i)}{|f'(k_0^i)|}\, ,
\end{equation}
where $k_0^i$ are the roots of $f(k)=0$.
In our case, we have the roots
\begin{equation}
k_0= \frac{m_\pi^2 - m_e^2}{2m_\pi} \kom |f'(k_0)| = 1+ \frac{k_0}{E}\, .
\end{equation}
Plugging this into the decay rate, we find
\begin{align}
\Gamma_{\pi\to e\bar{\nu}} =& |G_F F_\pi m_e V_{ud}|^2 \, \frac{1}{4\pi^2 m_\pi}
\int_0^\infty\!\frac{4\pi |\VEC{k}|^2 d|\VEC{k}|}{E}(E+|\VEC{k}|)\frac{\delta(|\VEC{k}|-k_0)}{1+\frac{k_0}{E}} 
\nonumber \\
=& \frac{G_F^2 |F_\pi V_{ud}|^2}{4\pi}\, m_e^2 m_\pi 
\left(1 - \frac{m_e^2}{m_\pi^2}\right)^2 \,.
\end{align}

A similar calculation for $\pi \to \mu \bar{\nu}_\mu$ yields
\begin{equation}
\Gamma_{\pi\to \mu\bar{\nu}} = \frac{G_F^2 |F_\pi V_{ud}|^2}{4\pi} \,
 m_\mu^2 m_\pi \left(1 - \frac{m_\mu^2}{m_\pi^2}\right)^2  \,.
\end{equation}
One can take the ratio where the least well-known quantities cancel
\begin{equation}
\frac{\Gamma(\pi\to e\bar{\nu}_e)}{\Gamma(\pi\to \mu\bar{\nu}_\mu)}
\,=\, \frac{m_e^2}{m_\mu^2}\left(\frac{m_\pi^2-m_e^2}{m_\pi^2-m_\mu^2}
\right)^2 \,=\, 1.28\times 10^{-4} \,.
\end{equation}
Experimentally, the ratio is measured to be $1.230(4) \times 10^{-4}$.
The small deviations must come from \emph{quantum effects}, i.e., loop diagrams.


\chapter{Yang-Mills theory from Ward identity and soft limits}\label{app:compton_sQED}


In this appendix, we consider (generalised) scalar electrodynamics in order to derive a consistent interacting theory for massless helicity-$1$ particles.
The objective of this appendix is to prove that Yang-Mills theory is the unique field theory for describing many such states by only imposing Lorentz invariance.
The plan is as follows: 
\begin{enumerate}
\item We start with conventional scalar QED with a single scalar field coupled to a $\UO$ gauge potential to derive a formula for Compton Scattering.
\item Subsequently, we add more and more particles to our scattering formula \emph{without specifying the actual couplings in form of a Lagrangian}.
We rather demand that our scattering amplitude is a Lorentz invariant object by imposing only Ward identities.
\begin{itemize}
\item For additional scalar particles, we will observe that they must appear in the same \emph{mass multiplet}.
\item For additional gauge bosons, we will find that the Ward identity gives rise to a \emph{Lie algebra structure}.
\end{itemize}
\item Without self interactions among the $N$ gauge bosons,
we will show that $[T^{a}, T^{b}]=0$ for matrices $T^{a}_{ij}$ describing the coupling to two scalar $\phi_{i}$, $\phi_{j}$.
This implies that we are considering a gauge theory with $G\cong \mathrm{U}(1)^{N}$, i.e., many disconnected copies of scalar QED.
\item With self interactions among the $N$ gauge bosons,
we derive $[T^{a}, T^{b}]=\I f^{abc}T^{c}$ using nothing but Lorentz invariance and unitarity in form of the Ward identities.
\end{enumerate}
All in all, we end up with Yang-Mills theory for (non-)abelian gauge potentials coupled to any number of scalar fields.

\vfill

\newpage

\section{Scalar QED and Compton Scattering}

\begin{figure}[h]
   \centering
\begin{tikzpicture}[scale=1.]
\setlength{\feynhanddotsize}{1.5ex}
\begin{feynhand}
\node (o) at (4.5,0) {$=\dfrac{\I}{p^{2}-m^{2}+\I\varepsilon}$} ; 
\node (o1) at (5.75,-2) {$=\dfrac{-\I}{p^{2}+\I\varepsilon}\left [\eta_{\mu\nu}-(1-\xi)\dfrac{p_{\mu}p_{\nu}}{p^{2}}\right ]$} ; 
\node (o3) at (4.5,-4) {$=\I e(-p^{\mu}_{1}-p^{\mu}_{2})$} ; 
\node (o3) at (4.,-7) {$=2\I e^{2}\eta^{\mu\nu}$} ; 
\vertex (a1) at (0,-2); 
\vertex (c1) at (2.5,-2); 
\vertex (a2) at (0,-4); 
\vertex (b2) at (1.25,-4); 
\vertex (c2) at (2.5,-3); 
\vertex (d2) at (2.5,-5); 
\vertex (a3) at (0,-6) {$\mu$}; 
\vertex (b3) at (2.5,-6) {$\nu$}; 
\vertex (e3) at (1.25,-7); 
\vertex (c3) at (0,-8); 
\vertex (d3) at (2.5,-8); 
\vertex (a0) at (0,0); 
\vertex (c0) at (2.5,0); 
\propag [chasca] (a0) to (c0);
\propag [pho] (a1) to (c1);
\propag [chasca, mom={$p_{1}$}] (a2) to (b2);
\propag [pho] (b2) to (d2);
\propag [chasca, mom={$p_{2}$}] (b2) to (c2);
\propag [pho] (a3) to (e3);
\propag [pho] (b3) to (e3);
\propag [chasca] (c3) to (e3);
\propag [chasca] (e3) to (d3);
\end{feynhand}
\end{tikzpicture}
\caption{Feynman rules for scalar QED}\label{fig:FRSQED} 
\end{figure}

The Lagrangian for scalar QED is given by
\begin{equation}\label{eq:LagSQED} 
\cL=-\dfrac{1}{4}F_{\mu\nu}F^{\mu\nu}+D_{\mu}\phi\, D^{\mu}\phi^{*}-m^{2}|\phi|^{2}\, .
\end{equation}
Here, the gauge covariant derivatives are given by
\begin{equation}
D_{\mu}\phi=\p_{\mu}\phi+\I eA_{\mu}\phi\kom D^{\mu}\phi^{*}=\p^{\mu}\phi^{*}-\I eA^{\mu}\phi^{*}\, .
\end{equation}
In order to work out the Feynman rules, it is convenient to write out the Lagrangian as
\begin{equation}
\cL=-\dfrac{1}{4}F_{\mu\nu}F^{\mu\nu}-\phi^{*}(\square+m^{2})\phi-\I eA_{\mu}\left [\phi^{*}\p^{\mu}\phi-(\p^{\mu}\phi^{*})\,\phi\right ]+e^{2}A_{\mu}^{2}|\phi|^{2}\, .
\end{equation}
The Feynman rules are summarised in Fig.~\ref{fig:FRSQED}. It is important to keep in mind that the cubic vertex is associated with a derivative operator and comes therefore with a $4$-momentum. Depending on the direction of momentum flow and particle flow, we need to adapt the signs in front of the associated $4$-vector. As discussed in the lecture for QED, we do not need the above Lagrangian formulation to prove charge conservation. All we really need is \emph{Lorentz invariance}.

To begin our endeavour,
let us consider Compton scattering. There are three diagrams contributing to the scattering amplitude, namely
\begin{equation*}
\begin{tikzpicture}[scale=1.2]
\setlength{\feynhanddotsize}{1.5ex}
\begin{feynhand}
\node (o) at (-2.5,2) {\large $\I \cM_{t}=$} ; 
\vertex (a0) at (-1.5,4) {$e^{-}$}; 
\vertex (a1) at (1.5,4) {$\epsilon^{\mu}_{\text{in}}$}; 
\vertex (b0) at (0,3); 
\vertex (b1) at (0,1); 
\vertex (c0) at (-1.5,0) {$e^{-}$}; 
\vertex (c1) at (1.5,0.) {$\epsilon^{\nu}_{\text{out}}$}; 
\propag [chasca, mom={$p_{1}$}] (a0) to (b0);
\propag [pho, mom={$q_{1}$}] (a1) to (b0);
\propag [chasca, mom={$p_{1}+q_{1}$}] (b0) to (b1);
\propag [chasca, mom={$p_{2}$}] (b1) to (c0);
\propag [pho, mom={$q_{2}$}] (b1) to (c1);
\end{feynhand}
\end{tikzpicture}
\hspace*{1.cm}
\begin{tikzpicture}[scale=1.2]
\setlength{\feynhanddotsize}{1.5ex}
\begin{feynhand}
\node (o) at (-2.5,2) {\large $\I \cM_{u}=$} ; 
\vertex (a0) at (-1.5,4) {$e^{-}$}; 
\vertex (a1) at (2.,4) {$\epsilon^{\mu}_{\text{in}}$}; 
\vertex (a11) at (1.,2.5); 
\vertex (b0) at (0,3); 
\vertex (b1) at (0,1); 
\vertex (b00) at (-0.25,2.5); 
\vertex (b11) at (-0.25,1.5); 
\node (o1) at (-0.85,2.0) {$p_{1}-q_{2}$};
\vertex (c0) at (-1.5,0) {$e^{-}$}; 
\vertex (c1) at (2.,0.) {$\epsilon^{\nu}_{\text{out}}$}; 
\vertex (c11) at (1.,1.5); 
\propag [chasca] (b0) to (b1);
\propag [fer, with arrow = 1] (b00) to(b11);
\propag [chasca, mom={$p_{1}$}] (a0) to (b0);
\propag [pho, mom={$q_{2}$}] (c11) to (c1);
\propag [pho] (b0) to (c11);
\propag [chasca, mom={$p_{2}$}] (b1) to (c0);
\propag [pho, top, mom={$q_{1}$}] (a1) to (a11);
\propag [pho, top] (b1) to (a11);
\end{feynhand}
\end{tikzpicture}
\end{equation*}
\begin{equation*}
\begin{tikzpicture}[scale=1.4]
\setlength{\feynhanddotsize}{1.5ex}
\begin{feynhand}
\node (o) at (-2.5,1.5) {\large $\I \cM_{4}=$} ; 
\vertex (a0) at (-1.5,3) {$e^{-}$}; 
\vertex (a1) at (1.5,3) {$\epsilon^{\mu}_{\text{in}}$}; 
\vertex (b0) at (0,1.5); 
\vertex (c0) at (-1.5,0) {$e^{-}$}; 
\vertex (c1) at (1.5,0.) {$\epsilon^{\nu}_{\text{out}}$}; 
\propag [chasca, mom={$p_{1}$}] (a0) to (b0);
\propag [pho, mom={$q_{1}$}] (a1) to (b0);
\propag [chasca, mom={$p_{2}$}] (b0) to (c0);
\propag [pho, mom={$q_{2}$}] (b0) to (c1);
\end{feynhand}
\end{tikzpicture}
\end{equation*}
We can compute the individual contributions using the Feynman rules from Fig.~\ref{fig:FRSQED}
\begin{align}
\I\cM_{t}&=\I e(-p_{1}-(p_{1}+q_{1}))^{\mu}\; \dfrac{\I}{(p_{1}+q_{1})^{2}-m^{2}}\; \I e(-(p_{1}+q_{1})-p_{2})^{\nu}\; \epsilon_{\text{in}}^{\mu}\epsilon_{\text{out}}^{\nu}\\[0.5em]
\I\cM_{u}&=\I e(-p_{1}-(p_{1}-q_{2}))^{\nu}\; \dfrac{\I}{(p_{1}-q_{2})^{2}-m^{2}}\; \I e(-(p_{1}-q_{2})-p_{2})^{\mu}\; \epsilon_{\text{in}}^{\mu}\epsilon_{\text{out}}^{\nu}\\[0.5em]
\I\cM_{4}&=2\I e^{2}\eta_{\mu\nu} \epsilon_{\text{in}}^{\mu}\epsilon_{\text{out}}^{\nu}
\end{align}
Summing over all these processes results in
\begin{equation}
\I\cM=\I\cM_{t}+\I\cM_{u}+\I\cM_{4}=-\I\cM_{\mu\nu}\, \epsilon_{\text{in}}^{\mu}\epsilon_{\text{out}}^{\nu}
\end{equation}
where
\begin{align}
\cM^{\mu\nu}&= e^{2}\biggl \{\dfrac{(2p_{1}^{\mu}+q_{1}^{\mu})(p_{1}^{\nu}+q_{1}^{\nu}+p_{2}^{\nu})}{(p_{1}+q_{1})^{2}-m^{2}}+\dfrac{(p_{1}^{\mu}+p_{2}^{\mu}-q_{2}^{\mu})(2p_{1}^{\nu}-q_{2}^{\nu})}{(p_{1}-q_{2})^{2}-m^{2}}-2\eta^{\mu\nu} \biggl \}\, .
\end{align}

As a consistency check, we want to show that the Ward identity holds, i.e.,
\begin{equation}
q_{1}^{\mu}\cM_{\mu\nu}=0\, .
\end{equation}
As discussed in the lecture, this identity is required in order to guarantee unitarity and Lorentz invariance. Since $p_{1}^{2}=m^{2}=p_{2}^{2}$ and $q_{1}^{2}=q_{2}^{2}=0$ on-shell, we can write
\begin{equation}
(p_{1}+q_{1})^{2}-m^{2}=2 p_{1}^{\mu}q_{1,\mu}\kom (p_{1}-q_{2})^{2}-m^{2}=-2 p_{1}^{\mu}q_{2,\mu}
\end{equation}
so that
\begin{align}
q_{1,\mu}\cM^{\mu\nu}&=e^{2}\biggl \{\dfrac{2q_{1,\mu} p_{1}^{\mu} (p_{1}^{\nu}+q_{1}^{\nu}+p_{2}^{\nu})}{2 p_{1}^{\mu}q_{1,\mu}}-\dfrac{q_{1,\mu}(p_{1}^{\mu}+p_{2}^{\mu}-q_{2}^{\mu})(2p_{1}^{\nu}-q_{2}^{\nu})}{2 p_{1}^{\mu}q_{2,\mu}}-2q_{1,\mu}\eta^{\mu\nu} \biggl \}\nn\\
&=e^{2}\biggl \{(p_{1}^{\nu}+q_{1}^{\nu}+p_{2}^{\nu})-\dfrac{q_{1,\mu}(p_{1}^{\mu}+p_{2}^{\mu}-q_{2}^{\mu})(2p_{1}^{\nu}-q_{2}^{\nu})}{2 p_{1}^{\mu}q_{2,\mu}}-2q_{1}^{\nu} \biggl \}\, .
\end{align}
Using the fact that $p_{1}+q_{1}=p_{2}+q_{2}$, we obtain
\begin{equation}
p_{1}-q_{2}=p_{2}-q_{1}\kom p_{1}^{\nu}q_{2,\nu}=p_{2}^{\nu}q_{1,\nu}
\end{equation}
such that
\begin{align}
q_{1,\mu}\cM^{\mu\nu}&=e^{2}\biggl \{(p_{1}^{\nu}+q_{1}^{\nu}+p_{2}^{\nu})-\dfrac{q_{1,\mu}(2p_{2}^{\mu}-q_{1}^{\mu})(p_{1}^{\nu}+p_{2}^{\nu}-q_{1}^{\nu})}{2 p_{1}^{\mu}q_{2,\mu}}-2q_{1}^{\nu} \biggl \}\nn\\
&=e^{2}\biggl \{(p_{1}^{\nu}+q_{1}^{\nu}+p_{2}^{\nu})-\dfrac{2q_{1,\mu}p_{2}^{\mu} (p_{1}^{\nu}+p_{2}^{\nu}-q_{1}^{\nu})}{2 p_{1}^{\mu}q_{2,\mu}}-2q_{1}^{\nu} \biggl \}\nn\\
&=e^{2}\biggl \{(p_{1}^{\nu}+q_{1}^{\nu}+p_{2}^{\nu})-(p_{1}^{\nu}+p_{2}^{\nu}-q_{1}^{\nu})-2q_{1}^{\nu} \biggl \}\label{eq:SQEDWardIDLastStep1} \\
&=0
\end{align}
as anticipated.

\section{Adding new particle species}

Next,
we include additional scalar fields $\phi_{i}$, $i=1,\ldots,N$ to our original theory. The $t$-channel diagram is modified in such a way that it involves indices for each dashed line, that is,
\begin{equation*}
\begin{tikzpicture}[scale=1.2]
\setlength{\feynhanddotsize}{1.5ex}
\begin{feynhand}
\node (o) at (-2.75,2) {\large $\I \cM_{t}^{k}=$} ; 
\node (o1) at (-0.4,2.) {$k$} ; 
\vertex (a0) at (-1.5,4) {$i$}; 
\vertex (a1) at (1.5,4) {$\epsilon^{\mu}_{\text{in}}$}; 
\vertex (b0) at (0,3); 
\vertex (b1) at (0,1); 
\vertex (c0) at (-1.5,0) {$j$}; 
\vertex (c1) at (1.5,0.) {$\epsilon^{\nu}_{\text{out}}$}; 
\propag [chasca, mom={$p_{1}$}] (a0) to (b0);
\propag [pho, mom={$q_{1}$}] (a1) to (b0);
\propag [chasca, mom={$p_{1}+q_{1}$}] (b0) to (b1);
\propag [chasca, mom={$p_{2}$}] (b1) to (c0);
\propag [pho, mom={$q_{2}$}] (b1) to (c1);
\end{feynhand}
\end{tikzpicture}
\end{equation*}
The resulting contribution reads
\begin{equation}
\I\cM_{t}^{k}=\I e(-p_{1}-(p_{1}+q_{1}))^{\mu}\; \dfrac{\I}{(p_{1}+q_{1})^{2}-m_{k}^{2}}\; \I e(-(p_{1}+q_{1})-p_{2})^{\nu}\; \epsilon_{\text{in}}^{\mu}\epsilon_{\text{out}}^{\nu}\, .
\end{equation}
In contrast to the single particle case, we now have
\begin{equation}
p_{1}^{2}=m_{i}^{2}\kom p_{2}^{2}=m_{j}^{2}
\end{equation}
and hence the denominator in the propagator becomes
\begin{equation}
(p_{1}+q_{1})^{2}-m_{k}^{2}=2p_{1}\cdot q_{1}+m_{i}^{2}-m_{k}^{2}\, .
\end{equation}
Proceeding similarly for the $u$-channel diagram and summing again over all contributions (in particular over all particle propagators) leads to
\begin{align}
q_{1,\mu}\cM^{\mu\nu}_{ij}&=e^{2}\biggl \{\sum_{k}\dfrac{2q_{1,\mu} p_{1}^{\mu} (p_{1}^{\nu}+q_{1}^{\nu}+p_{2}^{\nu})}{2 p_{1}^{\mu}q_{1,\mu}+m_{i}^{2}-m_{k}^{2}}-\sum_{k}\dfrac{q_{1,\mu}(p_{1}^{\mu}+p_{2}^{\mu}-q_{2}^{\mu})(2p_{1}^{\nu}-q_{2}^{\nu})}{2 p_{1}^{\mu}q_{2,\mu}+m_{i}^{2}-m_{k}^{2}}-2q_{1,\mu}\eta^{\mu\nu} \biggl \}\nn\\
&=e^{2}\biggl \{\sum_{k}\dfrac{2q_{1,\mu} p_{1}^{\mu} (p_{1}^{\nu}+q_{1}^{\nu}+p_{2}^{\nu})}{2 p_{1}^{\mu}q_{1,\mu}+m_{i}^{2}-m_{k}^{2}}-\sum_{k}\dfrac{2p_{1}^{\mu}q_{2,\mu} (p_{1}^{\nu}+p_{2}^{\nu}-q_{1}^{\nu})}{2 p_{1}^{\mu}q_{2,\mu}+m_{i}^{2}-m_{k}^{2}}-2q_{1,\mu}\eta^{\mu\nu} \biggl \}\, .
\end{align}
This can only vanish \textbf{for all momenta} if
\begin{equation}
m_{i}^{2}=m_{j}^{2}\, .
\end{equation}
Therefore, \emph{gauge particles can only couple to particles of the same mass}!
In the following, we assume that all scalars have the same mass $m=m_{i}$.

\section{Adding more gauge fields}

As a next step, we would like to include additional gauge particles, i.e., massless vector fields $A_{\mu}^{a}$. The modified $3$-point and $4$-point vertex will be denoted as
\begin{equation}
\begin{tikzpicture}[scale=1.5]
\setlength{\feynhanddotsize}{1.4ex}
\begin{feynhand}
\node (o3) at (4.,-4) {$=-\I e\, \Gamma^{a\mu}_{ij}(p_{1},p_{2},q)$} ; 
\node (o3) at (4.,-7) {$=2\I e^{2} \Gamma^{ab,\mu\nu}_{ij}(p_{1},p_{2},q_{1},q_{2})$} ; 
\vertex (a2) at (0,-4) {$i$}; 
\vertex (b2) [dot] at (1.25,-4) {}; 
\vertex (c2) at (2.5,-3) {$j$}; 
\vertex (d2) at (2.5,-5) {$a,\mu$}; 
\vertex (a3) at (0,-6) {$a,\mu$}; 
\vertex (b3) at (2.5,-6) {$b,\nu$}; 
\vertex (e3) [squaredot] at (1.25,-7) {}; 
\vertex (c3) at (0,-8) {i}; 
\vertex (d3) at (2.5,-8) {j}; 
\propag [chasca, mom={$p_{1}$}] (a2) to (b2);
\propag [pho, mom={$q$}] (d2) to (b2);
\propag [chasca, mom={$p_{2}$}] (b2) to (c2);
\propag [pho] (a3) to (e3);
\propag [pho] (b3) to (e3);
\propag [chasca] (c3) to (e3);
\propag [chasca] (e3) to (d3);
\end{feynhand}
\end{tikzpicture}
\end{equation}
To see how the modified charges can be described, we consider the vertex linear in the gauge fields $A^{a}_{\mu}$. The most general form for this vertex is given by (cf. Sect. 3.3.2 in the \href{https://www.dropbox.com/s/1rbin5mby0cnbyx/StandardModel2020.pdf?dl=0}{SM lecture notes} and notice $p_{2}=p_{1}+q$)
\begin{equation}
\Gamma^{a\mu}_{ij}(p_{1},p_{2},q)=2p_{1}^{\mu}F^{a}_{ij}(p_{1}^{2},q^{2},p_{1}\cdot q)+2q^{\mu}G^{a}_{ij}(p_{1}^{2},q^{2},p_{1}\cdot q)\, .
\end{equation}
For tree level Compton scattering, the external leg associated with the gauge particle will be contracted with a polarisation vector $\epsilon_{\mu}$ which annihilates the second term due to the transversality condition
\begin{equation}
\epsilon_{\mu}q^{\mu}=0\, .
\end{equation}
Moreover, one of the momenta $p_{i}$ of a scalar is going to be on-shell so that
\begin{equation}
\Gamma^{a\mu}_{ij}(p_{1},p_{2},q)=2p_{1}^{\mu}F^{a}_{ij}(\dfrac{p_{1}\cdot q}{2})\, .
\end{equation}
In the soft limit, we get
\begin{equation}
\Gamma^{a\mu}_{ij}(p_{1},p_{2},q)\raw  2p_{1}^{\mu} F^{a}_{ij}(0)\, .
\end{equation}
Therefore, the $3$-vertex must be described by an $N\times N$ matrix
\begin{equation}
T_{ij}^{a}= F^{a}_{ij}(0)\, .
\end{equation}
Clearly, the index structure is to be expected since the vertex connects different particle species.
It should be stressed though that a priori we have no information about the properties of the $T^{a}_{ij}$.
The quartic vertex comes with an (at this point) unknown coupling strength $\Gamma^{ab,\mu\nu}_{ij}$ which is a non-trivial combination of $T^{a}$ and $T^{b}$. We will determine the functional behaviour of this contribution further below.

Let us consider the generalised Compton scattering diagrams
\begin{equation*}
\begin{tikzpicture}[scale=1.4]
\setlength{\feynhanddotsize}{1.5ex}
\begin{feynhand}
\node (o) at (-0.3,2.8) {$\I eT^{a}_{ik}$} ; 
\node (o1) at (-0.3,1.2) {$\I eT^{b}_{kj}$} ; 
\vertex (a0) at (-1.5,4) {$i$}; 
\vertex (a1) at (1.5,4) {$a,\mu$}; 
\vertex (b0) [dot] at (0,3) {}; 
\vertex (b1) [dot] at (0,1) {}; 
\vertex (c0) at (-1.5,0) {$j$}; 
\vertex (c1) at (1.5,0.) {$b,\nu$}; 
\propag [chasca] (a0) to (b0);
\propag [pho] (a1) to (b0);
\propag [chasca] (b0) to [edge label = {$k$}] (b1);
\propag [chasca] (b1) to (c0);
\propag [pho] (b1) to (c1);
\end{feynhand}
\end{tikzpicture}
\hspace*{1.75cm}
\begin{tikzpicture}[scale=1.4]
\setlength{\feynhanddotsize}{1.5ex}
\begin{feynhand} 
\node (o) at (-0.3,2.8) {$\I eT^{b}_{ik}$} ; 
\node (o1) at (-0.3,1.2) {$\I eT^{a}_{kj}$} ; 
\vertex (a0) at (-1.5,4) {$i$}; 
\vertex (a1) at (2.,4) {$a,\mu$}; 
\vertex (a11) at (1.,2.5); 
\vertex (b0) [dot] at (0,3) {}; 
\vertex (b1) [dot] at (0,1) {}; 
\vertex (b00) at (-0.25,2.5); 
\vertex (b11) at (-0.25,1.5); 
\vertex (c0) at (-1.5,0) {$j$}; 
\vertex (c1) at (2.,0.) {$b,\nu$}; 
\vertex (c11) at (1.,1.5); 
\propag [chasca] (b0) to [edge label = {$k$}] (b1);
\propag [chasca] (a0) to (b0);
\propag [pho] (c11) to (c1);
\propag [pho] (b0) to (c11);
\propag [chasca] (b1) to (c0);
\propag [pho] (a1) to (a11);
\propag [pho, top] (b1) to (a11);
\end{feynhand}
\end{tikzpicture}
\end{equation*}
\begin{equation*}
\begin{tikzpicture}[scale=1.4]
\setlength{\feynhanddotsize}{1.5ex}
\begin{feynhand} 
\vertex (a0) at (-1.5,3) {$i$}; 
\vertex (a1) at (1.5,3) {$a,\mu$}; 
\vertex (b0) [squaredot] at (0,1.5) {}; 
\vertex (c0) at (-1.5,0) {$j$}; 
\vertex (c1) at (1.5,0.) {$b,\nu$}; 
\propag [chasca] (a0) to (b0);
\propag [pho] (a1) to (b0);
\propag [chasca] (b0) to (c0);
\propag [pho] (b0) to (c1);
\end{feynhand}
\end{tikzpicture}
\end{equation*}
which amounts to
\begin{align}
\I\cM_{t}&=\I eT^{a}_{ik}(-p_{1}-(p_{1}+q_{1}))^{\mu}\; \dfrac{\I}{(p_{1}+q_{1})^{2}-m^{2}}\; \I eT^{b}_{kj}(-(p_{1}+q_{1})-p_{2})^{\nu}\; \epsilon_{\text{in}}^{\mu}\epsilon_{\text{out}}^{\nu}\nn\\
\I\cM_{u}&=\I e T^{b}_{ik} (-p_{1}-(p_{1}-q_{2}))^{\nu}\; \dfrac{\I}{(p_{1}-q_{2})^{2}-m^{2}}\; \I eT^{a}_{kj}(-(p_{1}-q_{2})-p_{2})^{\mu}\; \epsilon_{\text{in}}^{\mu}\epsilon_{\text{out}}^{\nu}\\[0.5em]
\I\cM_{4}&=2\I e^{2}\Gamma_{ij,\mu\nu}^{ab}\epsilon_{\text{in}}^{\mu}\epsilon_{\text{out}}^{\nu}
\end{align}
As before, we can try to check the Ward identity which is modified to (cf. \eqref{eq:SQEDWardIDLastStep1})
\begin{align}
q_{1,\mu}\cM^{\mu\nu}&=e^{2}(T^{a}_{ik}T^{b}_{kj} (p_{1}^{\nu}+q_{1}^{\nu}+p_{2}^{\nu})-T^{b}_{ik}T^{a}_{kj}(p_{1}^{\nu}+p_{2}^{\nu}-q_{1}^{\nu})-2\Gamma^{ab,\mu\nu}_{ij} q_{1,\mu})\, .
\end{align}
This can be written as
\begin{equation}\label{eq:WardIdeManyGaugeFieldStep1} 
q_{1,\mu}\cM^{\mu\nu}=e^{2}([T^{a},T^{b}]_{ij} (p_{1}^{\nu}+p_{2}^{\nu})+\lbrace T^{a},T^{b}\rbrace_{ij}q_{1}^{\nu}-2\Gamma^{ab,\mu\nu}_{ij} q_{1,\mu})\, .
\end{equation}
Here, we need to sum again over all particles $k$ interchanged via the propagator which implies matrix multiplication in this particular case. 

If we assume for now that there are no pure self interactions between only gauge particles, the vertex contribution $\Gamma^{ab,\mu\nu}_{ij}$ is constant as a function of the momenta which we denote as $\tilde{\Gamma}^{ab,\mu\nu}_{ij}$. This is because $\tilde{\Gamma}^{ab,\mu\nu}_{ij}$ arises from couplings of the form $\phi_{i}\phi_{j}A_{\mu}^{a}A_{\nu}^{b}$ which does not involve any additional derivatives.\footnote{Adding derivatives $\p_{\mu}$ increases the mass dimension.}
In the soft limit $q_{1}\ll p_{1},p_{2}$, we can neglect the terms in $q_{1}$ and write
\begin{align}
q_{1,\mu}\cM^{\mu\nu}&=e^{2}([T^{a},T^{b}]_{ij} (p_{1}^{\nu}+p_{2}^{\nu})+\lbrace T^{a},T^{b}\rbrace_{ij}q_{1}^{\nu}-2\tilde{\Gamma}^{ab,\mu\nu}_{ij} q_{1,\mu})\nn\\
&\xrightarrow{q_{1}\raw 0}e^{2} [T^{a},T^{b}]_{ij}\, (p_{1}^{\nu}+p_{2}^{\nu})\, .
\end{align}
For this to vanish for arbitrary $p_{1},p_{2}$, we require
\begin{equation}\label{eq:ComTATBAB} 
[T^{a},T^{b}]_{ij}=0
\end{equation}
which means that the matrices must commute. This corresponds to a $\UO^{M}$ gauge theory where all the generators commute with each other. 
The simple reason that the individual $\mathrm{U}(1)$-sectors do not talk to each other is obviously the assumption that there are no self interactions which directly implies that $\tilde{\Gamma}^{ab,\mu\nu}_{ij}$ is constant.

\subsection{Aside: The form of the quartic vertex functions}


Let us work out another interesting property of the quartic vertex starting again from \eqref{eq:WardIdeManyGaugeFieldStep1}. We can use our observation \eqref{eq:ComTATBAB} to write
\begin{equation}
q_{1,\mu}\cM^{\mu\nu}= e^{2}(\lbrace T^{a},T^{b}\rbrace_{ij}q^{\nu} -2 \tilde{\Gamma}^{ab,\mu\nu}_{ij} q_{1,\mu}= 2e^{2} \left (T^{a}_{ik}T^{b}_{kj}\eta^{\nu\mu}  -\tilde{\Gamma}^{ab,\mu\nu}_{ik} \right )q_{1,\mu}\, .
\end{equation}
Hence, we identified the unknown coupling as
\begin{equation}\label{eq:VertexPHISQASQ} 
\tilde{\Gamma}^{ab,\mu\nu}_{ij}=T^{a}_{ik}T^{b}_{kj}\eta^{\mu\nu}  \, .
\end{equation}
Here we have worked under the assumption that there are no other couplings than $\phi_{i}\phi_{j}A^{a}_{\mu}A^{b}_{\nu}$ contributing to the quartic vertex.

Before proceeding with the non-abelian scenario, let us try to improve our understanding of ${\Gamma}^{ab,\mu\nu}_{ij}$.
In general, we may split the $4$-point vertex ${\Gamma}^{ab,\mu\nu}_{ij}$ into two contributions
\begin{equation}
{\Gamma}^{ab,\mu\nu}_{ij}=\tilde{\Gamma}^{ab,\mu\nu}_{ij}+\hat{\Gamma}^{ab,\mu\nu}_{ij}
\end{equation}
where the second term $\hat{\Gamma}^{ab,\mu\nu}_{ij}$ encodes all possible self-interactions allowed by Lorentz invariance to be discussed in the next section.
In this case, it turns out that \eqref{eq:ComTATBAB} is \textbf{not true} any more.
The vertex contribution \eqref{eq:VertexPHISQASQ} however is sill partially correct, but guaranteeing symmetry under particles exchange necessitates
\begin{equation}
\tilde{\Gamma}^{ab,\mu\nu}_{ij}=\dfrac{1}{2}\lbrace T^{a},T^{b}\rbrace_{ij}\eta^{\mu\nu} \, .
\end{equation}

Taking into account what we have learned about the vertex \eqref{eq:VertexPHISQASQ}, we may therefore rewrite our general result \eqref{eq:WardIdeManyGaugeFieldStep1} as
\begin{align}
q_{1,\mu}\cM^{\mu\nu}&=e^{2}([T^{a},T^{b}]_{ij} (p_{1}^{\nu}+p_{2}^{\nu})+ \lbrace T^{a},T^{b}\rbrace_{ij} q_{1}^{\nu}-2\left (\tilde{\Gamma}^{ab,\mu\nu}_{ij}+\hat{\Gamma}^{ab,\mu\nu}_{ij}\right ) q_{1,\mu})\nn\\
&=e^{2}([T^{a},T^{b}]_{ij} (p_{1}^{\nu}+p_{2}^{\nu})-2 \hat{\Gamma}^{ab,\mu\nu}_{ij}  q_{1,\mu})\, .
\end{align}
Insisting on $q_{1,\mu}\cM^{\mu\nu}=0$ which needs to be true for \textbf{any} valid Lorentz invariant theory of massless spin-$1$ particles, we get
\begin{equation}\label{eq:QuaVertexSelfInt} 
\hat{\Gamma}^{ab,\mu\nu}_{ij}  q_{1,\mu}=\dfrac{1}{2}[T^{a},T^{b}]_{ij} (p_{1}+p_{2})^{\nu}\, .
\end{equation}
If we had assumed that $[T^{a},T^{b}]=0$, this would mean that there are no additional corrections to Compton scattering. For a non-trivial commutator, we make the following peculiar observation: in the soft limit $q_{1}\raw 0$, the right hand side of \eqref{eq:QuaVertexSelfInt} does not vanish. This means that also the left hand side cannot vanish, i.e.,
\begin{equation}
\hat{\Gamma}^{ab,\mu\nu}_{ij}  q_{1,\mu} \nrightarrow 0\, .
\end{equation}
This implies that $\hat{\Gamma}^{ab,\mu\nu}_{ij}$ \textbf{must have a pole in} $q_{1}$ and so schematically
\begin{equation}
\hat{\Gamma}^{ab,\mu\nu}_{ij}\sim \dfrac{1}{q_{1}}\, .
\end{equation}
A pole of this kind is of course familiar from the exchange of massless particles. This is why we construct the most general cubic self-interaction in the next part. This will allow us to identify the above contribution with an additional diagram of the form
\begin{equation*}
\begin{tikzpicture}[scale=1.2]
\setlength{\feynhanddotsize}{1.5ex}
\begin{feynhand}
\node (o) at (-2.5,2.) {$\hat{\Gamma}^{ab,\mu\nu}_{ij}\sim$};
\node (o) at (0.3,2.3) {$\I  e T^{c}_{ij}$};
\node (o) at (1.75,2.3) {$f^{abc}$};
\vertex (a0) at (-1.5,3) {$i$}; 
\vertex (a1) at (2,2); 
\vertex (d0) at (3.5,3) {$a$}; 
\vertex (d1) at (3.5,1) {$b$}; 
\vertex (b0) at (0,2); 
\vertex (c0) at (-1.5,1) {$j$}; 
\propag [fer] (a0) to (b0);
\propag [pho] (b0) to [edge label = {$c$}] (a1);
\propag [pho] (a1) to (d0);
\propag [pho] (a1) to (d1);
\propag [fer] (b0) to (c0);
\end{feynhand}
\end{tikzpicture}
\end{equation*}

\subsection{Self interactions between gauge bosons}

To compute the above diagram, we include self-couplings between the individual gauge particles. That is, we now consider
\begin{equation}\label{eq:SQEDManySpec3} 
\cL\supset F_{3}(A_{\mu}^{a},\p_{\mu}A_{\nu}^{a})+F_{4}(A_{\mu}^{a},\p_{\mu}A_{\nu}^{a})
\end{equation}
with $F_{3}$ encoding all cubic and $F_{4}$ all quartic interactions. In principle, we could include even higher powers of the gauge particles. However, dimensional analysis tells us that those terms are non-renormalisable and negligible at low energies. Hence, we can also make two simplifications: $F_{4}(A_{\mu}^{a},\p_{\mu}A_{\nu}^{a})=F_{4}(A_{\mu}^{a})$ since each derivative contributes an additional power of mass in the dimensional analysis. For the same reason, $F_{3}(A_{\mu}^{a},\p_{\mu}A_{\nu}^{a})$ can include at most $1$ derivative.

At this point, we do not make any assumptions about the self-interactions for the gauge particles, but simply determine the vertices from considerations in the soft-limit for the associated particles. Let us therefore consider the following vertex:
\begin{equation*}
\begin{tikzpicture}[scale=1.4]
\setlength{\feynhanddotsize}{1.5ex}
\begin{feynhand}
\node (o3) at (4.5,0) {$=-e\Gamma^{abc}_{\mu\nu\rho}(q^{a},q^{b},q^{c})$} ; 
\vertex (a2) at (0,0) {$\mu$,a}; 
\vertex (b2) at (1.25,0); 
\vertex (c2) at (2.5,1) {$\nu$,b}; 
\vertex (d2) at (2.5,-1) {$\rho$,c}; 
\propag [pho, mom={$q^{a}$}] (a2) to (b2);
\propag [pho, mom={$q^{c}$}] (d2) to (b2);
\propag [pho, mom={$q^{b}$}] (c2) to (b2);
\end{feynhand}
\end{tikzpicture}
\end{equation*}
As argued above, $\Gamma_{\mu\nu\rho}^{abc}(q^{a},q^{b},q^{c})$ is at most linear in any of the $q^{a}$, $q^{b}$, $q^{c}$. In fact, this is necessary to ensure the $1/q_{1}$ behaviour of $\hat{\Gamma}^{ab,\mu\nu}_{ij}$ as discussed above, recall \eqref{eq:QuaVertexSelfInt}. Hence, we are led to assume couplings of the form $A^{2}\p A$ which implies that the most general expression for the cubic vertex reads
\begin{align}\label{eq:GamFullExpBefMomCon} 
\Gamma_{\mu\nu\rho}^{abc}(q^{a},q^{b},q^{c})&=f_{1}^{abc}\eta_{\nu\rho}q^{a}_{\mu}+f_{2}^{abc}\eta_{\mu\rho}q^{a}_{\nu}+f_{3}^{abc}\eta_{\mu\nu}q^{a}_{\rho}\nn\\
&\quad+f_{4}^{abc}\eta_{\nu\rho}q^{b}_{\mu}+f_{5}^{abc}\eta_{\mu\rho}q^{b}_{\nu}+f_{6}^{abc}\eta_{\mu\nu}q^{b}_{\rho}\nn\\
&\quad+f_{7}^{abc}\eta_{\nu\rho}q^{c}_{\mu}+f_{8}^{abc}\eta_{\mu\rho}q^{c}_{\nu}+f_{9}^{abc}\eta_{\mu\nu}q^{c}_{\rho}\, .
\end{align}
The appearance of the metric factors together with one factor of the momentum mirrors the fact that in $A^{2}\p A$ the factor $A^{2}$ is symmetric in two spacetime indices, whereas $\p A$ contributes the additional momentum $4$-vector.
A priori, the coefficient functions are unrelated and completely general rank $3$ tensors in the indices $a,b,c$.
Their relationships and symmetry properties can be derived from a few fundamental principles as we now show.

To simplify the above ansatz, we apply the usual rules for vertices such as particle interchange and momentum conservation. First, the vertex should be invariant under cyclic permutations 
\begin{equation}
(\mu, a,q^{a})\raw (\nu, b,q^{b})\raw (\rho, c,q^{c})\raw (\mu, a,q^{a})
\end{equation}
which implies that
\begin{align}
\Gamma_{\mu\nu\rho}^{abc}(q^{a},q^{b},q^{c})&=f_{1}^{abc}\eta_{\nu\rho}q^{a}_{\mu}+f_{2}^{abc}\eta_{\mu\rho}q^{a}_{\nu}+f_{3}^{abc}\eta_{\mu\nu}q^{a}_{\rho}\nn\\
&\quad+f_{3}^{bca}\eta_{\nu\rho}q^{b}_{\mu}+f_{1}^{bca}\eta_{\mu\rho}q^{b}_{\nu}+f_{2}^{bca}\eta_{\mu\nu}q^{b}_{\rho}\nn\\
&\quad+f_{2}^{cab}\eta_{\nu\rho}q^{c}_{\mu}+f_{3}^{cab}\eta_{\mu\rho}q^{c}_{\nu}+f_{1}^{cab}\eta_{\mu\nu}q^{c}_{\rho}\, .
\end{align}
Also, we must ensure momentum conservation at the vertex which implies
\begin{equation}
q^{a}+q^{b}+q^{c}=0
\end{equation}
and so
\begin{align}
\Gamma_{\mu\nu\rho}^{abc}(q^{a},q^{b},q^{c})&=f_{1}^{abc}\eta_{\nu\rho}(-q^{b}_{\mu}-q^{c}_{\mu})+f_{2}^{abc}\eta_{\mu\rho}q^{a}_{\nu}+f_{3}^{abc}\eta_{\mu\nu}q^{a}_{\rho}\nn\\
&\quad+f_{3}^{bca}\eta_{\nu\rho}q^{b}_{\mu}+f_{1}^{bca}\eta_{\mu\rho}(-q^{a}_{\nu}-q^{c}_{\nu})+f_{2}^{bca}\eta_{\mu\nu}q^{b}_{\rho}\nn\\
&\quad+f_{2}^{cab}\eta_{\nu\rho}q^{c}_{\mu}+f_{3}^{cab}\eta_{\mu\rho}q^{c}_{\nu}+f_{1}^{cab}\eta_{\mu\nu}(-q^{a}_{\rho}-q^{b}_{\rho})\nn\\
&=(f_{2}^{abc}-f_{1}^{bca})\eta_{\mu\rho}q^{a}_{\nu}+(f_{3}^{abc}-f_{1}^{cab})\eta_{\mu\nu}q^{a}_{\rho}\label{eq:GamRes1} \\
&\quad+(f_{3}^{bca}-f_{1}^{abc})\eta_{\nu\rho}q^{b}_{\mu}+(f_{2}^{bca}-f_{1}^{cab})\eta_{\mu\nu}q^{b}_{\rho}\label{eq:GamRes2} \\
&\quad+(f_{2}^{cab}-f_{1}^{abc})\eta_{\nu\rho}q^{c}_{\mu}+(f_{3}^{cab}-f_{1}^{bca})\eta_{\mu\rho}q^{c}_{\nu}\, .
\end{align}
The exchange of two states at any of the external lines should not change the physical amplitude. Stated otherwise, if we interchange $(\mu, a,q^{a})$ and $(\nu, b,q^{b})$ in line \eqref{eq:GamRes1}, then this should equal line \eqref{eq:GamRes2}. Therefore, we deduce that
\begin{equation}
(f_{2}^{bac}-f_{1}^{acb})\eta_{\nu\rho}q^{b}_{\mu}+(f_{3}^{bac}-f_{1}^{cba})\eta_{\mu\nu}q^{b}_{\rho}=(f_{3}^{bca}-f_{1}^{abc})\eta_{\nu\rho}q^{b}_{\mu}+(f_{2}^{bca}-f_{1}^{cab})\eta_{\mu\nu}q^{b}_{\rho}
\end{equation}
and therefore
\begin{equation}\label{eq:FConst1} 
f_{2}^{bac}-f_{1}^{acb}=f_{3}^{bca}-f_{1}^{abc}\kom f_{3}^{bac}-f_{1}^{cba}=f_{2}^{bca}-f_{1}^{cab}\, .
\end{equation}
Similarly, if we interchange $(\mu, a,q^{a})$ and $(\nu, b,q^{b})$ in line \eqref{eq:GamRes2}, then equality with line \eqref{eq:GamRes1} demands
\begin{equation}
(f_{3}^{acb}-f_{1}^{bac})\eta_{\mu\rho}q^{a}_{\nu}+(f_{2}^{acb}-f_{1}^{cba})\eta_{\mu\nu}q^{a}_{\rho}=(f_{2}^{abc}-f_{1}^{bca})\eta_{\mu\rho}q^{a}_{\nu}+(f_{3}^{abc}-f_{1}^{cab})\eta_{\mu\nu}q^{a}_{\rho}
\end{equation}
and hence
\begin{equation}\label{eq:FConst2} 
f_{3}^{acb}-f_{1}^{bac}=f_{2}^{abc}-f_{1}^{bca}\kom f_{2}^{acb}-f_{1}^{cba}=f_{3}^{abc}-f_{1}^{cab}\, .
\end{equation}
Clearly, the constraints in \eqref{eq:FConst2} are compatible with \eqref{eq:FConst1} and also redundant as everything boils down to
\begin{equation}
f_{2}^{abc}=f_{3}^{acb}-f_{1}^{bac}+f_{1}^{bca}\, .
\end{equation}
This allows us to eliminate $f_{2}$ in the above expression such that
\begin{align}
\Gamma_{\mu\nu\rho}^{abc}(q^{a},q^{b},q^{c})&=(f_{3}^{acb}-f_{1}^{bac}+f_{1}^{bca}-f_{1}^{bca})\eta_{\mu\rho}q^{a}_{\nu}+(f_{3}^{abc}-f_{1}^{cab})\eta_{\mu\nu}q^{a}_{\rho}\nn\\
&\quad+(f_{3}^{bca}-f_{1}^{abc})\eta_{\nu\rho}q^{b}_{\mu}+(f_{3}^{bac}-f_{1}^{cba}+f_{1}^{cab}-f_{1}^{cab})\eta_{\mu\nu}q^{b}_{\rho}\nn \\
&\quad+(f_{3}^{bac}-f_{1}^{cba}+f_{1}^{abc}-f_{1}^{abc})\eta_{\nu\rho}q^{c}_{\mu}+(f_{3}^{cab}-f_{1}^{bca})\eta_{\mu\rho}q^{c}_{\nu}\nn\\
&=(f_{3}^{acb}-f_{1}^{bac})\eta_{\mu\rho}q^{a}_{\nu}+(f_{3}^{abc}-f_{1}^{cab})\eta_{\mu\nu}q^{a}_{\rho}\nn\\
&\quad+(f_{3}^{bca}-f_{1}^{abc})\eta_{\nu\rho}q^{b}_{\mu}+(f_{3}^{bac}-f_{1}^{cba})\eta_{\mu\nu}q^{b}_{\rho}\nn \\
&\quad+(f_{3}^{bac}-f_{1}^{cba})\eta_{\nu\rho}q^{c}_{\mu}+(f_{3}^{cab}-f_{1}^{bca})\eta_{\mu\rho}q^{c}_{\nu}\, .
\end{align}
It is convenient to define
\begin{equation}
f^{abc}=f_{3}^{abc}-f_{1}^{cab}
\end{equation}
so that
\begin{align}\label{eq:GamALMT} 
\Gamma_{\mu\nu\rho}^{abc}(q^{a},q^{b},q^{c})&=f^{acb}\eta_{\mu\rho}q^{a}_{\nu}+f^{abc}\eta_{\mu\nu}q^{a}_{\rho} +f^{bca}\eta_{\nu\rho}q^{b}_{\mu}+f^{bac}\eta_{\mu\nu}q^{b}_{\rho}\nn \\
&\quad+f^{bac}\eta_{\nu\rho}q^{c}_{\mu}+f^{cab}\eta_{\mu\rho}q^{c}_{\nu}\, .
\end{align}
The constants $f^{abc}$ have certain symmetrisation properties that we need to employ. For instance, we can use the same symmetry as before exchanging $(\mu, a,q^{a})$ and $(\nu, b,q^{b})$ to obtain for the last term in \eqref{eq:GamALMT}
\begin{equation}
f^{cab}\eta_{\mu\rho}q^{c}_{\nu}\raw f^{cba}\eta_{\nu\rho}q^{c}_{\mu}\, .
\end{equation}
This needs to match the second-to last term in \eqref{eq:GamALMT} which is why
\begin{equation}
f^{bac}=f^{cba}\, .
\end{equation}
For other particle interchanges we find
\begin{enumerate}
\item for $(\mu, a,q^{a})$ and $(\rho, c,q^{c})$, then
\begin{equation}
f^{bac}\eta_{\nu\rho}q^{c}_{\mu}\raw f^{bca}\eta_{\nu\mu}q^{c}_{\rho}
\end{equation}
so
\begin{equation}
f^{bca}=f^{abc}
\end{equation}
\item for $(\mu, a,q^{a})$ and $(\nu, b,q^{b})$
\begin{equation}
f^{acb}\eta_{\mu\rho}q^{a}_{\nu}\raw f^{bca}\eta_{\rho\nu}q^{b}_{\mu}
\end{equation}
and then $(\nu, b,q^{b})$ and $(\rho, c,q^{c})$
\begin{equation}
f^{bca}\eta_{\rho\nu}q^{b}_{\mu}\raw f^{cba}\eta_{\rho\nu}q^{c}_{\mu}
\end{equation}
so that
\begin{equation}
f^{cba}=f^{bac}\, .
\end{equation}
\end{enumerate}
Altogether, we find
\begin{equation}
f^{abc}=f^{bca}\kom f^{bac}=f^{cba}
\end{equation}
and thus
\begin{equation}\label{eq:ThreePVertexFinRes} 
\Gamma_{\mu\nu\rho}^{abc}(q^{a},q^{b},q^{c})=f^{abc}\left (\eta_{\mu\rho}q_{\nu}^{c}+\eta_{\mu\nu}q_{\rho}^{a}+\eta_{\rho\nu}q_{\mu}^{b}\right )+f^{bac}\left (\eta_{\nu\rho}q_{\mu}^{c}+\eta_{\mu\rho}q_{\nu}^{a}+\eta_{\mu\nu}q_{\rho}^{b}\right )\, .
\end{equation}

As we will see below, the Ward identity shows that $f^{abc}$ must be anti-symmetric so that
\begin{equation}\label{eq:ThreePVertexFinRes1} 
\Gamma_{\mu\nu\rho}^{abc}(q^{a},q^{b},q^{c})=f^{abc}\left (\eta_{\mu\rho} (q_{\nu}^{c}-q_{\nu}^{a})+\eta_{\mu\nu} (q_{\rho}^{a}-q_{\rho}^{b})+\eta_{\rho\nu}(q_{\mu}^{b}-q_{\mu}^{c})\right )\, .
\end{equation}
However, we have to keep in mind that this cannot be obtained from the above considerations!

\subsection{Commutator for $T^{a}_{ij}$}

Our final task is to show that the constants $f^{abc}$ appearing in \eqref{eq:ThreePVertexFinRes} are nothing but the structure constants of an algebra associated with the matrices $T^{a}_{ij}$. Let us come back to our general expression for the Ward identity from Compton scattering \eqref{eq:WardIdeManyGaugeFieldStep1}
\begin{equation}\label{eq:EqsGenWISIGP} 
q_{1,\mu}\cM^{\mu\nu}=e^{2}([T^{a},T^{b}]_{ij} (p_{1}^{\nu}+p_{2}^{\nu})+\lbrace T^{a},T^{b}\rbrace_{ij}q_{1}^{\nu}-2(\tilde{\Gamma}^{ab,\mu\nu}_{ij}+\hat{\Gamma}^{ab,\mu\nu}_{ij}) q_{1,\mu})\, .
\end{equation}
Due to self-interactions between the gauge particles, the following diagram
\begin{equation*}
\begin{tikzpicture}[scale=1.6]
\setlength{\feynhanddotsize}{1.5ex}
\begin{feynhand}
\node (o) at (0.3,2.25) {\small$\Gamma^{c,\rho}_{ij}$};
\node (o) at (1.8,2.25) {\small$\Gamma_{\mu\nu\rho}^{abc}$};
\vertex (a0) at (-1.5,3) {$i$}; 
\vertex (a1) at (2,2); 
\vertex (d0) at (3.5,3) {$a,\mu$}; 
\vertex (d1) at (3.5,1) {$b,\nu$}; 
\vertex (b0) [dot] at (0,2) {}; 
\vertex (c0) at (-1.5,1) {$j$}; 
\propag [fer, mom={$p_{1}$}] (a0) to (b0);
\propag [pho, mom={$q_{1}-q_{2}$}] (a1) to (b0);
\propag [pho, mom={$q_{1}$}] (d0) to (a1);
\propag [pho, mom={$q_{2}$}] (a1) to (d1);
\propag [fer, mom={$p_{2}$}] (b0) to (c0);
\end{feynhand}
\end{tikzpicture}
\end{equation*}
gives a contribution to the $4$-vertex ${\Gamma}^{ab,\mu\nu}_{ij}$ encoded in $\hat{\Gamma}^{ab,\mu\nu}_{ij}$. We compute the above diagram by applying the usual Feynman rules so that
\begin{align}
\I \cM_{A^{3}}&=-\I e\, \Gamma^{c}_{ij,\lambda}(p_{1},p_{2},p_{1}-p_{2}) \dfrac{\I\eta^{\lambda\rho}}{(q_{1}-q_{2})^{2}}\, (-e)\Gamma_{\mu\nu\rho}^{abc}(q_{1},-q_{2},q_{2}-q_{1})\, \epsilon_{\text{in}}^{\mu}(q_{1})\epsilon_{\text{out}}^{\nu}(q_{2})\, .
\end{align}
Before we plug everything in, let us work out the cubic vertex contribution \eqref{eq:ThreePVertexFinRes}
\begin{align}
\Gamma_{\mu\nu\rho}^{abc}(q_{1},-q_{2},q_{2}-q_{1})&=f^{abc}\left (\eta_{\mu\rho} (q_{2}-q_{1})_{\nu}+\eta_{\mu\nu} q_{1,\rho}-\eta_{\rho\nu} q_{2,\mu}\right )\nn\\
&\quad+f^{bac}\left (\eta_{\nu\rho} (q_{2}-q_{1})_{\mu}+\eta_{\mu\rho}q_{1,\nu}-\eta_{\mu\nu}q_{2,\rho}\right )\, .
\end{align}
Furthermore, we work out that
\begin{align}
\Gamma^{a}_{ij,\lambda}(p_{1},p_{2},p_{1}-p_{2}) &=(p_{1,\lambda}+p_{2,\lambda}) T^{c}_{ij}
\end{align}
so that
\begin{align}\label{eq:CubAAmp} 
\cM_{A^{3}}^{\mu\nu}\epsilon^{\mu}_{a}\epsilon^{*,\nu}_{b}&=-\dfrac{(-e)^{2}\, (p_{1}+p_{2})^{\rho} T^{c}_{ij}}{(q_{1}-q_{2})^{2}}\, \biggl \{f^{abc}\left (\eta_{\mu\rho} (q_{2}-q_{1})_{\nu}+\eta_{\mu\nu} q_{1,\rho}-\eta_{\rho\nu} q_{2,\mu}\right )\\
&\quad+f^{bac}\left (\eta_{\nu\rho} (q_{2}-q_{1})_{\mu}+\eta_{\mu\rho}q_{1,\nu}-\eta_{\mu\nu}q_{2,\rho}\right )\biggl\}\epsilon^{\mu}_{a}\epsilon^{*,\nu}_{b}\nn\\[0.3em]
&=\dfrac{e^{2}\, (p_{1}+p_{2})^{\rho} T^{c}_{ij}}{2q_{1}\cdot q_{2}}\, \biggl \{f^{abc}\left (-\epsilon_{a,\rho}  (q_{1}\cdot\epsilon_{b}^{*})+(\epsilon_{a}\cdot \epsilon_{b}^{*}) q_{1,\rho}-(\epsilon_{a}\cdot q_{2}) \epsilon_{b,\rho}^{*}\right )\\
&\quad+f^{bac}\left ( \epsilon_{b,\rho}^{*} (q_{2} \cdot \epsilon_{a})+\epsilon_{a,\rho} (\epsilon_{b}^{*}\cdot q_{1})-(\epsilon_{a}\cdot\epsilon_{b}^{*}) q_{2,\rho}\right )\biggl\} \nn\, .
\end{align}
using
\begin{equation}
q_{1}\cdot \epsilon_{a}=0\kom  q_{2}\cdot \epsilon_{b}^{*}=0\, .
\end{equation}
The contribution to the Ward identity is obtained by replacing $\epsilon^{\mu}_{a}\raw q_{1}^{\mu}$
\begin{align}\label{eq:WardIDNAGTC1} 
\cM_{A^{3}}^{\mu\nu} q_{1}^{\mu}\epsilon^{*,\nu}_{b}&=\dfrac{e^{2}\, (p_{1}+p_{2})^{\rho} T^{c}_{ij}}{2q_{1}\cdot q_{2}}\, \biggl \{f^{abc}\left (- q_{1,\rho}  (q_{1}\cdot\epsilon_{b}^{*})+(q_{1}\cdot \epsilon_{b}^{*}) q_{1,\rho}-(q_{1}\cdot q_{2}) \epsilon_{b,\rho}^{*}\right )\nn\\
&\quad+f^{bac}\left ( \epsilon_{b,\rho}^{*} (q_{2} \cdot q_{1})+q_{1,\rho} (\epsilon_{b}^{*}\cdot q_{1})-(q_{1}\cdot\epsilon_{b}^{*}) q_{2,\rho}\right )\biggl\} \nn\\
&=\dfrac{e^{2}\, (p_{1}+p_{2})^{\rho} T^{c}_{ij}}{2q_{1}\cdot q_{2}}\, \biggl \{f^{bac} \left (q_{1}-q_{2}\right )_{\rho}  (q_{1}\cdot\epsilon_{b}^{*})  -\left (f^{abc}-f^{bac}\right ) (q_{1}\cdot q_{2}) \epsilon_{b,\rho}^{*} \biggl\} \, .
\end{align}
We use again
\begin{equation}
(p_{1}+q_{1})^{2}=(p_{2}+q_{2})^{2}\quad\Rightarrow\quad p_{1}\cdot q_{1}=p_{2}\cdot q_{2}
\end{equation}
\begin{equation}
(p_{1}-q_{2})^{2}=(p_{2}-q_{1})^{2}\quad\Rightarrow\quad p_{1}\cdot q_{2}=p_{2}\cdot q_{1}
\end{equation}
\begin{equation}
p_{1}\cdot q_{1}+p_{2}\cdot q_{1}=p_{2}\cdot q_{2}+p_{1}\cdot q_{2}\, .
\end{equation}
to deduce that the first term vanishes.
Altogether, we obtain from \eqref{eq:EqsGenWISIGP}
\begin{align}
\cM_{\mu\nu}q_{1}^{\mu}\epsilon^{*,\nu}_{b}&=e^{2}\left ([T^{a},T^{b}]_{ij} (p_{1}+p_{2})_{\nu}+\dfrac{\I}{2} T^{c}_{ij}\, (f^{bac}-f^{abc})\, (p_{1}+p_{2})_{\nu}\right ) \epsilon^{*,\nu}_{b}
\end{align}
which implies
\begin{equation}\label{eq:ComGenYMFCS} 
[T^{a},T^{b}]_{ij}=\dfrac{\I}{2}\left (f^{abc}-f^{bac}\right ) T^{c}_{ij}\, .
\end{equation}
In the soft limit, one can further show
\begin{equation}
f^{bac}=-f^{abc}
\end{equation}
so that we recognise the $f^{abc}$ as structure constants of some algebra with
\begin{equation}
[T^{a},T^{b}]_{ij}=\I f^{abc} T^{c}_{ij}\, .
\end{equation}

\section{Discussion}

The significance of this calculation is that we obtain the expression for the overall contribution to the Ward identity in terms of the commutator.
In fact, it shows that the latter is identified with $\I f^{abc}T_{ij}^{c}$ arising from the two couplings in the above diagram. We did not need to say anything about gauge symmetries or Lie algebras to determine the properties of the generators $T^{a}_{ij}$ for a theory of massless spin-$1$ particles.
Altogether, this appendix shows that we do not have any other choice: \emph{if we want to describe scattering processes involving massless spin-$1$ particles, the underlying field theory is uniquely specified by Yang-Mills theory.}

If you ever doubted the uniqueness of the Yang-Mills Lagrangian,
we just recovered the same result by simply considering scattering with different types of massless spin-$1$ particles and imposing consistency with Lorentz invariance.
The underlying notion of Lie groups as encoding the local symmetries of the theory is again a \emph{derived} concept.
It is nothing that we put in by hand for fun or to make the theory ``beautiful''.
On the contrary, the physical scattering amplitudes together with the Ward identities dictate the correct local Lagrangian description.

\end{appendices}

\backmatter

\cleardoublepage
\phantomsection
\bibliographystyle{utphys}
\renewcommand\bibname{Bibliography}
\bibliography{StandardModel}

\begin{thebibliography}{100}
\bibitem{Zweig:1981pd}
G.~Zweig,
``{An SU(3) model for strong interaction symmetry and its breaking. Version
  1},''.
\end{thebibliography}

\cleardoublepage
\phantomsection
\printindex

\end{document}